# Superconductivity

## The Structure Scale of the Universe

Twenty Fourth Edition – December 08, 2018

Richard D. Saam
525 Louisiana Ave
Corpus Christi, Texas 78404 USA
e-mail: rdsaam@att.net
http://xxx.lanl.gov/abs/physics/9905007

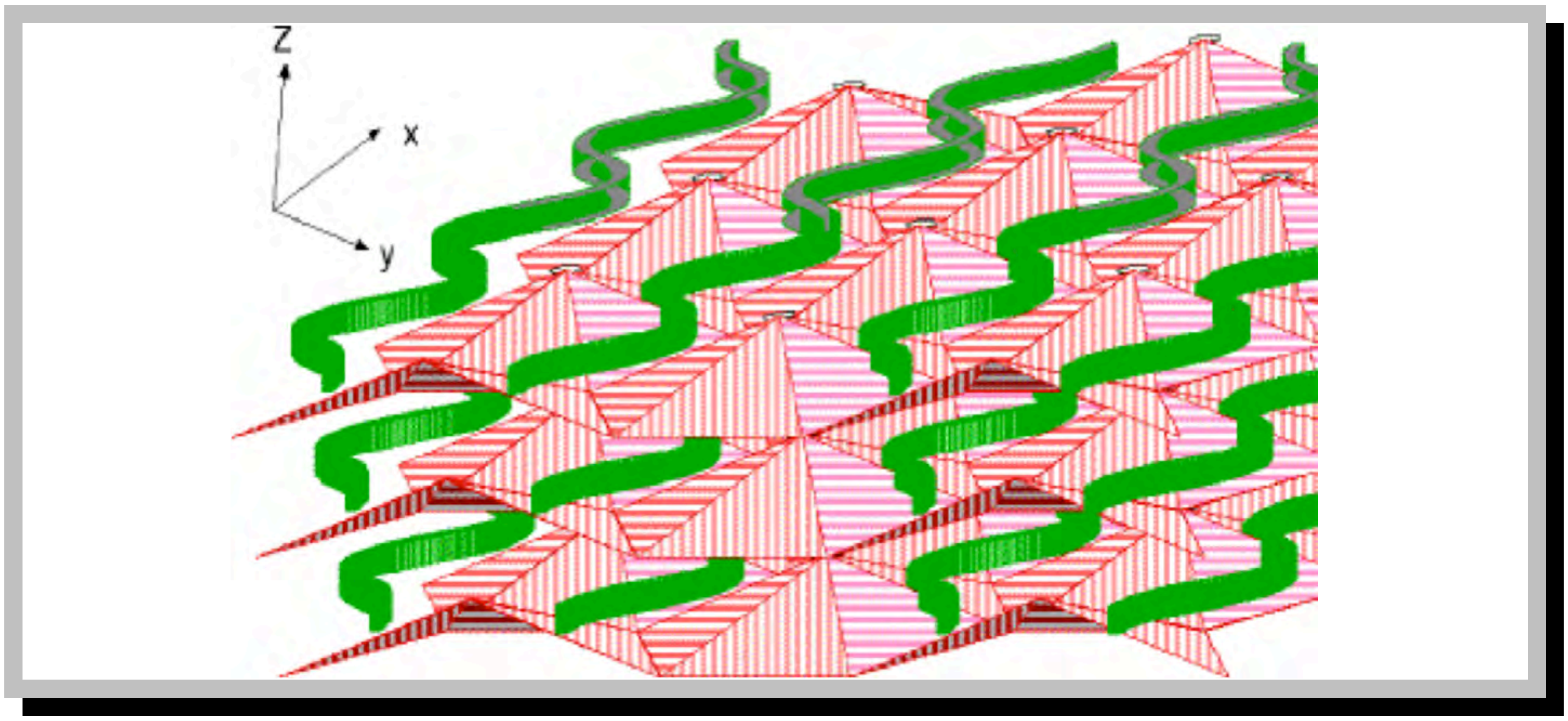

## PREVIOUS PUBLICATIONS

## ABSTRACT

Superconductivity is presented as a resonant space filling cellular 3D tessellating triangular structure (trisine) consistent with universe conservation of energy and momentum 'elastic nature', universal parameters ($G, H, h, c$ and $k_b$ ), Maxwells' permittivity and permeability concepts, Lemaitre Big Bang universe expansion, a Higgs mechanism, de Broglie matter wave hypothesis, Einstein's energy momentum stress tensor and Charge-Conjugation, Parity-Change and Time-Reversal (CPT) theorem (discrete Noether's theorem) without recourse to net matter-antimatter annihilation concepts. These trisine lattice elastic structures scale as visible energies $m_T c^2$/(*permeability* x *permittivity)* and black body energies as $m_T c^2$/(*permittivity)* each related to their reflective energy values $m_T c^2$/(*permeability_reflective* x *permittivity_reflective)* and $m_T c^2$/(*permittivity_reflective)* by their near equality at the nuclear dimension defined by the speed of light($c$). Both visible and reflective trisine lattices are scalable over 15 orders of magnitude from nuclear to universe dimensions with a constant cellular mass ($m_T$) (110.107178208 x electron mass or 56.26465274 MeV/$c^2$) and constant Hubble to Newton gravity ratio '$H_U/G_U$' with a present HU value of 2.31E-18 /sec (71.23 km/(sec million parsec). The standard model weak and strong forces and up, down, strange, charm, bottom and top particles are defined as nuclear trisine entities maintained at the nuclear density 2.34E14 g/cm$^3$, radius 8.75E-14 cm and temperature 8.96E13 K. The trisine lattice scales to the present universe density of 6.38E-30 g/cm$^3$, Hubble radius of curvature of 2.25E28 cm and temperature 8.11E-16 K. The trisine present universe mass and density could be considered a candidate for the 'dark energy' for interpreting present universe data observations. In the proposed trisine dark energy, each particle of mass $m_T$ is contained in a present universe trisine cell volume of 15,733 cm$^3$ resonating at 29,600 seconds. In this context, the universe present cosmic microwave background radiation (CMBR) black body temperature($m_T c^2$/(*permittivity*) = 2.729/$k_b$ K) is linked to an ubiquitous extremely cold ($m_T c^2$/(*permeability* x *permittivity*) = 8.11E-16/$k_b$ K) universe 'dark energy' temperature thermally supporting an equally cold 'dark matter' in the form of colligative nucleosynthetic hydrogen/helium Bose Einstein Condensate(BEC) electromagnetically undetected but gravitationally affecting galactic rotation curves and lensing phenomena. Trisine momentum $m_T c$ space predicts an deceleration value with observed universe expansion and consistent with observed Pioneer 10 & 11 deep space translational and rotational deceleration, flat galactic rotation curves, local universe deceleration relative to supernovae Type 1a standard candles (not the reverse supernovae Type 1a universe expansion acceleration as conventionally reported) and all consistent with the notion that:

**An object moving through momentum space will accelerate.**

The ubiquitous trisine mass 56 MeV/$c^2$ is expressed in Voyager accelerated Anomalous Cosmic Rays, EGRET and FERMI spacecraft data. There is an indication that higher FERMI detected energies falling in intensity by -1/3 power and cosmic ray flux by -23/8 power are an indication of early Big Bang nucleosynthetic processes creating what is known as dark matter dictated by the relationship between reflective dark energy $k_b T_c = m_T c^2$/(*permeability_reflective x permittivity_reflective)* and back round radiation (presently CMBR) $k_b T_b = m_T c^2$/(*permittivity_reflective)*. Trisine resonant time oscillations have been observed in Gravity Probe B gyro, Active Galactic Nuclei (AGN), telluric atmospheric sodium material superconductors and top quark oscillations. Entanglement and 'Action at a Distance' is explained in terms of trisine visible and reflective properties. Nuclear asymptotic freedom as a reflective concept is congruent with the Universe early nucleosynthetic expansion. Also, a reasonable value for the cosmological constant is derived having dimensions of the visible universe. A changing gravitational constant '$G_U$' is consistent with observed galaxy cluster dynamics as a function of '$z$' and also with the recently observed Higg's particle. Also, a 93 K superconductor should lose weight close to reported results by Podkletnov and Nieminen. Also this trisine model predicts 1, 2 and 3 dimensional superconductors, which has been verified by observed magnesium diboride ($MgB_2$) superconductor critical data. Also, the dimensional characteristics of material superconductor stripes are explained. Also dimensional guidelines are provided for design of room temperature superconductors and beyond and which are related to practical goals such as fabricating a superconductor with the energy content equivalent to the combustion energy of gasoline. These dimensional guidelines have been experimentally verified by Homes' Law and generally fit the parameters of a superconductor operating in the "dirty" limit. Confirmation of the trisine approach is represented by correlation to Koide Lepton relation at nuclear dimensions. Also, there is good correlation between, the cosmological constant and a characteristic lattice angle of 22.8 degrees. Also, spacecraft with asymmetrical flyby gravity assist from the earth or other planets would experience modified energy characteristics due to interaction with universe resonant tidal effects or De Broglie matter waves imparted to the earth or planets and subsequently to the spacecraft. Also, the concept of critical volume is introduced for controlled study of characteristics associated with CPT symmetry Critical Optical Volume Energy (COVE) within the context of the Lawson Criterion. A $1.5 billion - 10 year experimental plan is proposed to study and implement this concept under laboratory conditions here on earth.

## Introduction

**1.**

The trisine structure is a geometrical lattice model for the superconductivity phenomena based on Gaussian surfaces (within the context of Maxwell Displacement (D) concept) encompassing superconducting Cooper CPT Charge conjugated pairs (discrete Noether's theorem entities). This Gaussian surface has the same configuration as a particular matrix geometry as defined in reference [1], and essentially consists of mirror image non-parallel surface pairs. Originally, the main purpose of the engineered lattice [1] was to control the flow of particles suspended in a fluid stream and generally has been fabricated on a macroscopic scale to remove particles in industrial fluid streams on the order of $m^3$/sec. This general background is presented in Appendices L and M. In the particular configuration discussed in this report, a more generalized conceptual lattice is described wherein there is a perfect elastic resonant character to fluid and particle flow. In other words there is 100 percent conservation of energy and momentum (elastic or resonant condition), which defines the phenomenon of superconductivity.

The superconducting model presented herein is a logical translation of this geometry [1] in terms of classical and quantum theory to provide a basis for explaining aspects of the superconducting phenomenon and integration of this phenomenon with nuclear, electromagnetic and gravitational forces at all time scales over the age of the universe (from the Big Bang to present) (in terms of *z*) within the context of references[1-160].

This approach is an attempt to articulate a geometrical model for superconductivity in order to anticipate dimensional regimes in which one could look for higher performance materials or photonic media an approach that leads to universal understanding. This approach addresses specific particles such as polarons, exitons, magnons, plasmons and the standard model quarks and gluons as an expression of superconductivity in a given material or photonic medium, but which mask the essence of superconductivity as an expression of virtual particles by Dirac Schwinger pair production/annihilation. Indeed, the dimensionally correct (*mass, length, time*) trisine structure or lattice as presented in this report may be expressed in terms of something very elementary and that is the Charge Conjugate Parity Change Time Reversal (CPT) Theorem as a basis for creating resonant virtual particle structures within the the Heisenberg Uncertainty and consistent with dimensional (*mass, length, time*) correctness or numerical consistency with fundamental constants $(\hbar, G, c, H \ \& \ k_b)$ (encompassing the Planck units) as well as conformance with Maxwell equations[41], Lemaitre cosmology[122] and Einstein's general relativity.

The vacuum energy density as derived by Milonni[45], may not have the extremely high value of 1E114 erg/$cm^3$ due to an undefined boundary condition assumption that is more physically represented by universe parameters $(\hbar, G, c, H \ \& \ k_b)$ describing nature independent of particular '*mass, length and time*' dimensions.

As a corollary:

If a fundamental constant $(\hbar, G, c, H \ \& \ k_b)$ functional relationship describing nature is in terms of *mass, length and time*, then a functional relationship with arbitrary dimensionality $(dim1, dim2, dim3)$ equally describing nature is established with $(dim1, dim2, dim3)$ functionally linked back to (*mass, length, time*).

Model numerical and dimensional coherency is maintained at less than the trisine uncertainty factor $(1-g_s)=.01$ on a Excel spreadsheet and Mathcad dimensional analysis software with name plate calculation precision at 15 significant digits.

Traditionally, superconductivity or resonance is keyed to a critical temperature $(T_c)$, which is representative of the energy at which the superconductive resonant property takes place in a material or photonic medium. The material or photonic medium that is characterized as a superconductor has this property at all temperatures below the critical temperature.

Temperature*(T)* is related to energies $\left(E_{\varepsilon,\,k_m}\right)$ associated with permittivity $(\varepsilon)$ and permeability $(k_m)$ through the Boltzmann constant $(k_b)$ within the the Heisenberg Uncertainty $E_{\varepsilon,\,k_m} TIME = \hbar/2$ at a characteristic mass(*m*).

$$\frac{\hbar}{TIME} = k_b T = E_{\varepsilon,\,k_m} = mv^2 = \varepsilon k_m m c^2$$

Various quantum mechanical separable temperature $(T's)$ such as $T_c, T_b, T_a, T_p, T_r$ *and* $T_i$ are to be defined as various expressions of permittivity $(\varepsilon)$ and permeability $(k_m)$ with the acknowledgement of group and phase velocities.

In terms of the trisine model, any critical temperature $(T_c)$ can be inserted, but for the purposes of this report, eleven universe expansion conditions of critical temperatures $(T_c)$ and associated black body temperatures $(T_b)$ (and other $T's$) are selected as scaled with universe permittivity $(\varepsilon)$ and permeability $(k_m)$ and associated magnetic and electrical fields as well as various thermodynamic properties. This approach can be placed in the context of the universe adiabatic expansion where these particular temperatures $(T_c)$ and associated black body $(T_b)$ existed at a particular universe age from the time of nucleosynthesis to present galactic formation from dark matter within the superconducting 'dark energy' milieu. In this scenario, this cold dark matter would primarily be hydrogen and helium but would not detract from the observed hydrogen and helium abundances as observed in early universe stellar formation.

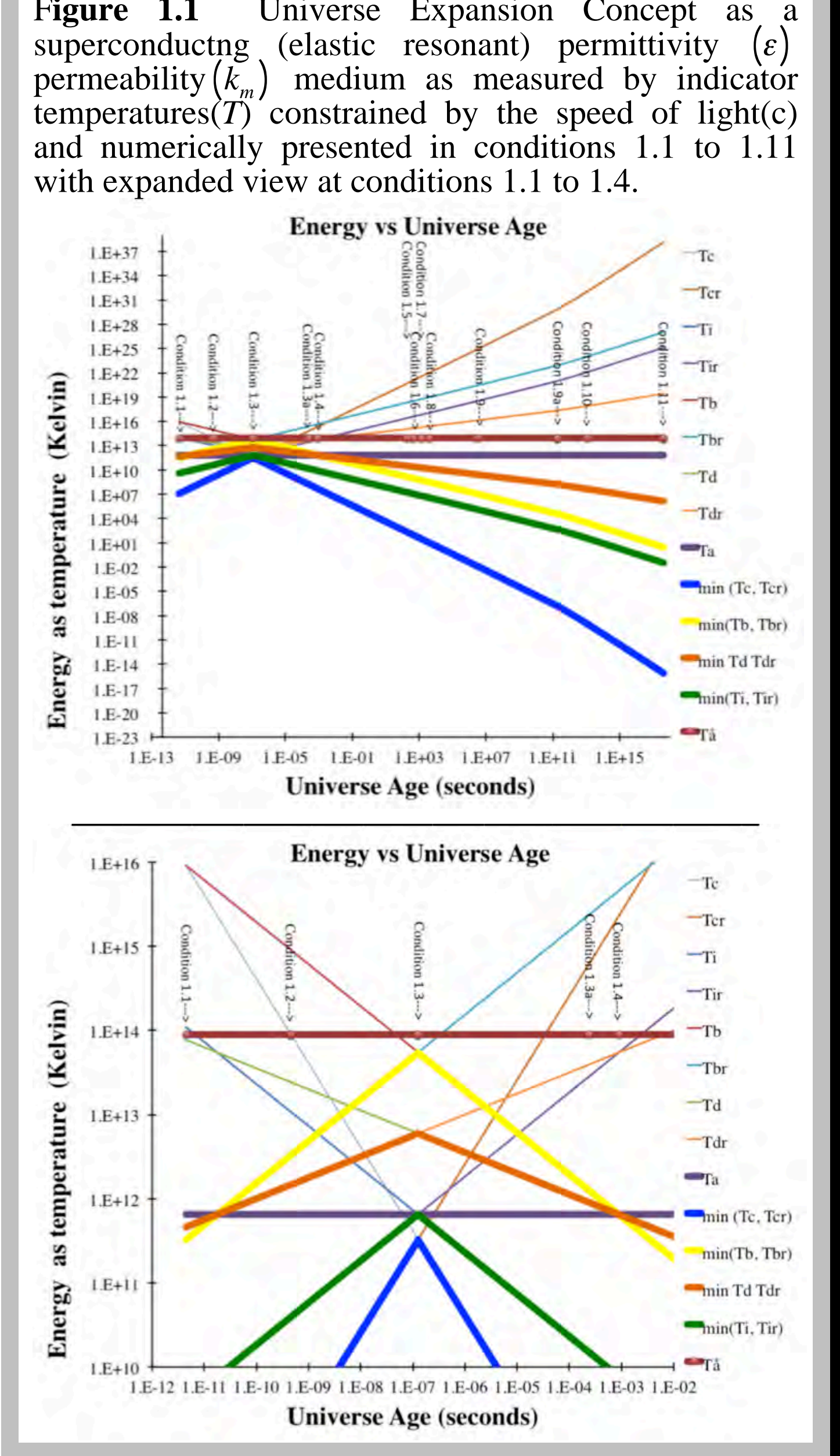

**Figure 1.1** Universe Expansion Concept as a superconductng (elastic resonant) permittivity $(\varepsilon)$ permeability $(k_m)$ medium as measured by indicator temperatures$(T)$ constrained by the speed of light(c) and numerically presented in conditions 1.1 to 1.11 with expanded view at conditions 1.1 to 1.4.

This approach logically indicates that most of the mass of the universe (dark energy) is at the critical temperature $(T_c)$ which is presently extremely cold at 8.11E-16 *K* and in thermal equilibrium with the 'dark baryonic matter' having the chemical nature (mainly hydrogen and helium) of nucleosynthesis period in the early universe. Baryonic 'dark' matter presently at 8.11E-16 *K* (quantum mechanically distinct from the cosmic microwave background radiation $T_b$ presently at 2.73 *K*) could be in a dense (~1 $g/cm^3$) particulate colligative Bose Einstein Condensate (BEC) phase since it has been demonstrated in the laboratory that hydrogen BECs are created at much warmer temperatures on the order of 1E-7 *K* ($BEC_{ST}$ - see Appendix K). Conceptually, the universe expansion can be likened to a $CO_2$ fire extinguisher adiabatic discharge with the resultant $CO_2$ particle cold fog analogous to the universe dark matter. A small portion of this dark matter 'fog' gravitationally collapses into the luminous matter (stars, galaxies etc.) that is observed in the celestial sphere while the substantial residual remains as ~meter sized masses within galactic interstellar space and surrounding halos only observed by its gravitational effects.

More precisely, universe time evolution is dictated by adiabatic process at universe Hubble expansion $H_U$ (initially at an extremely high 8.48E25 $\sec^{-1}$) with the fundamental conservation of energy adiabatic expansion

$$T_{c2}/T_{c1} = \left(\rho_{U2}/\rho_{U1}\right)^{2/3}$$

between Hubble expansion $H_{U1}$ and Hubble expansion $H_{U2}$ where:

$$H_U = \left(1/\sqrt{3}\right)\left(section/cavity\right)\left(k_b T_c/m_T c\right) = 1/Age_U \,.$$

The universe expansion is within the context of solutions to the Einstein field equations with $H_U$ and $G_U$ directly related by the speed of light $(c)$ and the Dirac constant $(\hbar)$:

$$G_{\mu\nu} = \frac{8\pi G_U}{c^4} T_{\mu\nu}$$

$$U = \frac{H_U}{G_U} = \frac{4\pi\rho_U}{H_U} = \frac{4m_T^3 c}{\pi\hbar^2} = 3.46E-11 \quad \left(g\ cm^{-3}\sec\right)$$

where the Newton gravitational parameter $(G_U)$ varies $\sim 1/Age_U$ inversely with universe age coordinate time ($Age_U$) while incremental proper time $(\Delta\tau)$ is approximately zero. The universe expands with Hubble radius of curvature $R_U$ and the volume of lattice component *cavity's* at the universe age $Age_U$. Then lattice dimension $B \sim Age_U^{1/3}$. The different rates of $B$ and $R_U$ change with $Age_U$ may explain the presently observed universe acceleration. We as observers are imbedded in the universe lattice 3D tessellated *cavity* structure and bounded by the gross universe volume $V_U$. All transitions are smooth from the Big Bang (no initial Guth inflationary period is required).

$$\frac{1}{Age_U} = H_U = \frac{1}{\sqrt{3}}\frac{k_b T_c}{cavity}\frac{section}{m_T c} = \sqrt{\frac{8\pi}{2}G_U \rho_U}$$

$$\frac{1}{Age_{Ur}} = H_{Ur} = \frac{1}{\sqrt{3}}\frac{k_b T_c}{cavity_r}\frac{section_r}{m_T c} = \sqrt{\frac{8\pi}{2}G_{Ur} \rho_{Ur}}$$

$$\rho_U = \frac{m_T}{cavity} = \frac{2}{8\pi}\frac{H_U^2}{G_U} = \frac{M_U}{V_U} = \frac{3M_U}{4\pi R_U^3}$$

$$\rho_{Ur} = \frac{m_T}{cavity_r} = \frac{2}{8\pi}\frac{H_{Ur}^2}{G_{Ur}} = \frac{M_{Ur}}{V_{Ur}} = \frac{3M_{Ur}}{4\pi R_{Ur}^3}$$

$$R_U = \sqrt{3}cAge_U = \frac{\sqrt{3}c}{H_U} \qquad \frac{G_U m_T M_U}{R_U} = m_t c^2$$

$$R_{Ur} = \sqrt{3}cAge_{Ur} = \frac{\sqrt{3}c}{H_{Ur}} \qquad \frac{G_{Ur} m_T M_{Ur}}{R_{Ur}} = m_t c^2$$

$$\frac{G_U M_U H_U}{c^3} = \frac{G_{Ur} M_{Ur} H_{Ur}}{c^3} = \frac{\left(G_U G_{Ur} M_U M_{Ur} H_U H_{Ur}\right)^{1/2}}{c^3} = 3^{1/2}$$

$$\left(M_U M_{Ur}\right)^{1/2} = \frac{g_s m_T}{3\alpha}$$

$$\frac{\mathbb{C} e_{\pm}\hbar}{m_t v_{\varepsilon z}} = \left(\frac{cavity}{\mathbb{C}}\right) H_{c1}$$

$$1/\alpha = \hbar c / e^2$$

$$\frac{m_T}{m_e} = 4\pi^{3/4}\sqrt{\frac{2\hbar c}{g_e e^2}} = 4\pi^{3/4}\sqrt{\frac{2}{g_e \alpha}}$$

$$2 g_e g_s \pi^{3/4}\sqrt{\frac{2}{g_e \alpha}}\frac{e_{\pm}\hbar}{2 m_e v_\varepsilon} = cavity\ H_c$$

$$cavity = \frac{4\pi Age_U m_T}{U} = \frac{4\pi m_t}{U H_U} = 2\sqrt{3}AB^2$$

$$section = 2\sqrt{3}B^2 \qquad \frac{section}{cavity} = \frac{1}{A} \qquad \frac{B}{A} = 2.379760996$$

$$sherd = \frac{M_{Upresent}}{M_U} \qquad sherd_{present} = 1$$

(‘*present*’ values are in trisine condition 1.11 green)

$universe\ constant\ mass = M_U sherd = 3.02E+56$ g

$universe\ constant\ energy = M_U c^2 sherd = 2.72E+77\ erg$

Universe Newtonian Gravitational constant G =

$G_U sherd^{-1/2} = G_{Ur} sherd_r^{-1/2} =$ 6.6725985E-8 $cm^3 sec^{-2} g^{-1}$

Universe Total Volume $V_{UT} = M_U sherd \dfrac{cavity}{m_t}$

Most cosmological observation assumes a constant Newtonian gravitational parameter ($G$) with the Universe age making a conversion necessary.

Trisine condition header information have the following universe temperature, size and time definitions:

$$Age_{UG} = Age_U\sqrt{\frac{G_U}{G}} = Age_U\sqrt{\frac{R_U c^2}{GM_U}}$$

$$Age_{UGr} = Age_{Ur}\sqrt{\frac{G_{Ur}}{G}} = Age_U\sqrt{\frac{R_{Ur} c^2}{GM_{Ur}}}$$

$$Age_{UG\nu} = Age_U\sqrt{\frac{R_U c^2}{G\left(M_U + M_\nu\right)}}$$

$$Age_{UG\nu r} = Age_{Ur}\sqrt{\frac{R_{Ur} c^2}{G\left(M_{Ur} + M_{\nu r}\right)}}$$

$Age_{UG\nu} = Age_{UG} = Age_U$ = 13.7E+09 years at the present time

$$T_a = \frac{m_t c^2}{k_b}$$

$$T_{å} = g_d m_t c^2 / \left(\alpha k_b\right)$$

$$T_d = \frac{m_T c^2}{\sqrt{\varepsilon} k_b} \qquad T_{dr} = \frac{m_T c^2}{\sqrt{\varepsilon_r} k_b}$$

$$T_b = \frac{m_T c^2}{\varepsilon k_b} \qquad T_{br} = \frac{m_T c^2}{\varepsilon_r k_b}$$

$$T_i = \frac{m_T c^2}{\left(k_{mx}\varepsilon_x\right)^{1/2} k_b} \qquad T_{ir} = \frac{m_T c^2}{\left(k_{mxr}\varepsilon_{xr}\right)^{1/2} k_b}$$

$$T_c = \frac{m_t c^2}{2k_{mx}\varepsilon_x k_b} = \frac{\hbar^2}{2m_t}\left(\frac{\pi}{B}\right)^2 \frac{1}{k_b} \qquad T_{cr} = \frac{m_t c^2}{2k_{mxr}\varepsilon_{xr} k_b} = \frac{\hbar^2}{2m_t}\left(\frac{\pi}{B_r}\right)^2 \frac{1}{k_b}$$

$$T_j = \frac{1}{4}\frac{m_T c^2}{\left(k_m\varepsilon\right)^2}\frac{1}{k_b} \qquad T_{jr} = \frac{1}{4}\frac{m_T c^2}{\left(k_{mr}\varepsilon_r\right)^2}\frac{1}{k_b}$$

$$k_m = \frac{1}{\varepsilon}\frac{1}{\dfrac{1}{\varepsilon_x k_{mx}} + \dfrac{1}{\varepsilon_y k_{my}} + \dfrac{1}{\varepsilon_z k_{mz}}} \qquad \varepsilon = \frac{3}{\dfrac{1}{\varepsilon_x} + \dfrac{1}{\varepsilon_y} + \dfrac{1}{\varepsilon_z}}$$

$$k_{mr} = \frac{1}{\varepsilon_r}\frac{1}{\dfrac{1}{\varepsilon_{xr} k_{mxr}} + \dfrac{1}{\varepsilon_{yr} k_{myr}} + \dfrac{1}{\varepsilon_{zr} k_{mzr}}} \qquad \varepsilon_r = \frac{3}{\dfrac{1}{\varepsilon_{xr}} + \dfrac{1}{\varepsilon_{yr}} + \dfrac{1}{\varepsilon_{zr}}}$$

$$z_B = \frac{B_{present}}{B} - 1 \qquad z_R = \frac{R_{Upresent}}{R_U} - 1$$

| 1.1 | $z_R$ | 3.79E+46 | $z_R$ | 3.79E+46 | *unitless* |
|---|---|---|---|---|---|
| | $z_B$ | 3.36E+15 | $z_B$ | 3.36E+15 | *unitless* |
| | $T_a$ | 6.53E+11 | $T_a$ | 6.53E+11 | $^oK$ |
| | $T_b$ | 9.14E+15 | $T_{br}$ | 3.27E+11 | $^oK$ |
| | $T_c$ | 9.14E+15 | $T_{cr}$ | 1.17E+07 | $^oK$ |
| | $T_d$ | 7.73E+13 | $T_{dr}$ | 4.62E+11 | $^oK$ |
| | $T_i$ | 1.09E+14 | $T_{ir}$ | 3.90E+09 | $^oK$ |
| | $T_j$ | 7.98E+22 | $T_{jr}$ | 1.30E+05 | $^oK$ |
| | $\varepsilon_x$ | 5.96E-03 | $\varepsilon_{xr}$ | 1.67E+02 | *unitless* |
| | $\varepsilon_y$ | 5.96E-03 | $\varepsilon_{yr}$ | 1.67E+02 | *unitless* |
| | $\varepsilon_z$ | 7.32E-05 | $\varepsilon_{zr}$ | 2.05E+00 | *unitless* |
| | $\varepsilon$ | 7.14E-05 | $\varepsilon_r$ | 2.00E+00 | *unitless* |
| | $k_{mx}$ | 5.99E-03 | $k_{mr}$ | 1.67E+02 | *unitless* |
| | $k_{my}$ | 4.49E-03 | $k_{myr}$ | 1.25E+02 | *unitless* |
| | $k_{mz}$ | 2.16E-02 | $k_{mzr}$ | 6.04E+02 | *unitless* |
| | $k_m$ | 2.00E-02 | $k_{mr}$ | 5.61E+02 | *unitless* |
| | $\rho_U$ | 2.41E+17 | $\rho_{Ur}$ | 1.10E+04 | g/cm$^3$ |
| | $B$ | 6.58E-15 | $B_r$ | 1.84E-10 | *cm* |
| | *cavity* | 4.16E-43 | $cavity_r$ | 9.12E-30 | $cm^3$ |
| | $R_U$ | 5.93E-19 | $R_{Ur}$ | 1.30E-05 | *cm* |
| | $V_U$ | 8.74E-55 | $V_{Ur}$ | 9.24E-15 | $cm^3$ |
| | $V_{Utotal}$ | 1.25E+39 | $V_{Urtotal}$ | 9.24E-15 | $cm^3$ |
| | $M_U$ | 2.11E-37 | $M_{Ur}$ | 1.02E-10 | g |
| | $M_\nu$ | 5.14E-26 | $M_{\nu r}$ | 5.14E-26 | g |
| | $M_{UTotal}$ | 3.03E+56 | $M_{UTotalr}$ | 3.03E+56 | G |
| | $time_\pm$ | 2.63E-27 | $time_{r\pm}$ | 2.06E-18 | *sec* |
| | $1/time_\pm$ | 3.81E+26 | $1/time_{r\pm}$ | 4.86E+17 | */sec* |
| | $H_U$ | 8.75E+28 | $H_{Ur}$ | 3.99E+15 | */sec* |
| | $G_U/G$ | 3.79E+46 | $G_{Ur}/G$ | 1.73E+33 | *unitless* |
| | *sherd* | 1.43E+93 | $sherd_r$ | 2.98E+66 | *unitless* |
| | $Age_U$ | 1.14E-29 | $Age_{Ur}$ | 2.51E-16 | *sec* |
| | $Age_{UG}$ | 2.22E-06 | $Age_{UGr}$ | 1.04E+01 | *sec* |
| | $Age_{UG\nu}$ | 4.50E-12 | $Age_{UG\nu r}$ | 3.53E-03 | *sec* |
| | *BCST* | 2.58E+01 | *BCST* | 2.58E+01 | *sec* |

This energy conjunction at $T_c = T_b$ marks the beginning of the universe where time and space are not defined in a group phase reflection condition where $v_d \sim v_\varepsilon \gg c$. The universe mass $(M_U)$ is nominally at 2.08E-37 gram the radiation mass $\left(M_\nu\right) \sim m_T/2$. This point corresponds to the ankle condition in the observed cosmic ray spectrum as presented in Figure 4.33.1

| 1.2 | $z_R$ | 3.67E+43 | $z_R$ | 3.67E+43 | *unitless* |
|---|---|---|---|---|---|
| | $z_B$ | 3.32E+14 | $z_B$ | 3.32E+14 | *unitless* |
| | $T_a$ | 6.53E+11 | $T_a$ | 6.53E+11 | $^oK$ |
| | $T_b$ | 9.05E+14 | $T_{br}$ | 3.30E+12 | $^oK$ |
| | $T_c$ | 8.96E+13 | $T_{cr}$ | 1.19E+09 | $^oK$ |
| | $T_d$ | 4.80E+13 | $T_{dr}$ | 8.89E+09 | $^oK$ |
| | $T_i$ | 1.08E+13 | $T_{ir}$ | 3.94E+10 | $^oK$ |
| | $T_j$ | 7.67E+18 | $T_{jr}$ | 1.35E+09 | $^oK$ |
| | $\varepsilon_x$ | 6.02E-02 | $\varepsilon_{xr}$ | 1.65E+01 | *unitless* |
| | $\varepsilon_y$ | 6.02E-02 | $\varepsilon_{yr}$ | 1.65E+01 | *unitless* |
| | $\varepsilon_z$ | 7.39E-04 | $\varepsilon_{zr}$ | 2.03E-01 | *unitless* |
| | $\varepsilon$ | 7.21E-04 | $\varepsilon_r$ | 1.98E-01 | *unitless* |
| | $k_{mx}$ | 6.05E-02 | $k_{mr}$ | 1.65E+01 | *unitless* |

| $k_{my}$ | 4.54E-02 | $k_{myr}$ | 1.24E+01 | *unitless* |
|---|---|---|---|---|
| $k_{mz}$ | 2.18E-01 | $k_{mzr}$ | 5.98E+01 | *unitless* |
| $k_m$ | 2.02E-01 | $k_{mr}$ | 5.55E+01 | *unitless* |
| $\rho_U$ | 2.34E+14 | $\rho_{Ur}$ | 1.13E+07 | $g/cm^3$ |
| $B$ | 6.65E-14 | $B_r$ | 1.83E-11 | $cm$ |
| $cavity$ | 4.28E-40 | $cavity_r$ | 8.85E-33 | $cm^3$ |
| $R_U$ | 6.11E-16 | $R_{Ur}$ | 1.26E-08 | $cm$ |
| $V_U$ | 9.57E-46 | $V_{Ur}$ | 8.44E-24 | $cm^3$ |
| $V_{Utotal}$ | 1.29E+42 | $V_{Urtotal}$ | 8.44E-24 | $cm^3$ |
| $M_U$ | 2.24E-31 | $M_{Ur}$ | 9.56E-17 | $g$ |
| $M_v$ | 5.41E-21 | $M_{vr}$ | 5.41E-21 | $g$ |
| $M_{UTotal}$ | 3.03E+56 | $M_{UTotalr}$ | 3.03E+56 | $g$ |
| $time_\pm$ | 2.68E-25 | $time_{r\pm}$ | 2.02E-20 | $sec$ |
| $1/time_\pm$ | 3.73E+24 | $1/time_{r\pm}$ | 4.96E+19 | $/sec$ |
| $H_U$ | 8.49E+25 | $H_{Ur}$ | 4.11E+18 | $/sec$ |
| $G_U/G$ | 3.67E+43 | $G_{Ur}/G$ | 1.78E+36 | *unitless* |
| $sherd$ | 1.35E+87 | $sherd_r$ | 3.16E+72 | *unitless* |
| $Age_U$ | 1.18E-26 | $Age_{Ur}$ | 2.43E-19 | $sec$ |
| $Age_{UG}$ | 7.14E-05 | $Age_{UGr}$ | 3.24E-01 | $sec$ |
| $Age_{UGv}$ | 4.59E-10 | $Age_{UGvr}$ | 3.46E-05 | $sec$ |
| $BCST$ | 2.29E+02 | $BCST$ | 2.29E+02 | $sec$ |

This $T_c$ defines superconductive phenomenon as dictated by the proton density 2.34E14 g/cm$^3$ and radius $B$ of 6.65E-14 cm (8.750E-14 cm NIST reported charge radius value) which is defined at a critical temperature $T_c$ of

$$T_c = T_{å} = g_d m_T \hbar c^3 \big/ \left(e^2 k_b\right) = g_d m_T c^2 \big/ \left(\alpha k_b\right)$$

that is a consequence of a group phase reflection condition:

$$v_{dx} = \left(\frac{2}{\alpha}\right)^{\frac{1}{2}} c$$

that is justifiable in elastic resonant condition (see equations 2.1.5 and 2.1.6 and follow-on explanation) and

$$\frac{\hbar c}{e^2} = \frac{m_r}{m_T} = \frac{v_{dx}^2}{2c^2} = \text{fine structure constant} = \frac{1}{\alpha}$$

This condition 1.2 defines the quark masses in the context of an space filling elastic concept developed in section 2:

Conservation Of Momentum

$$\sum_{n=B,C,Ds,Dn} \Delta p_n \Delta x_n = 0$$

$$g_s\left(\pm K_B \pm K_C\right) \rightleftarrows \pm K_{Ds} \pm K_{Dn}$$

$$g_s\left(\pm K_{Br} \pm K_{Cr}\right) \rightleftarrows \pm K_{Dsr} \pm K_{Dnr}$$

Conservation Of Energy

$$\sum_{n=B,C,Ds,Dn} \Delta E_n \Delta t_n = 0$$

$$K_B^2 + K_C^2 = g_s\left(K_{Ds}^2 + K_{Dn}^2\right)$$

Where

$$K_{Br}^2 + K_{Cr}^2 = g_s\left(K_{Dsr}^2 + K_{Dnr}^2\right)$$

$$2\frac{K_A}{K_B} = \frac{B}{A} = 2\frac{K_{Ar}}{K_{Br}} = \frac{B_r}{A_r} = 2.379329652$$

and

$m_T = 1.003150329E-25 g$ $\left(56.27261166\ MeV/c^2\right)$

*Superconductory gyro number* $g_s = 1.009662156$

*Dirac number* $g_d = 1.001159652180760$

(As explained in section 2.3)

The top quark is associated with energy

$$\left(\frac{4g_s}{\alpha^2}\right)\frac{\hbar^2 K_{Ar}^2}{2m_T} = \frac{\hbar^2 K_A^2}{2m_T} = 174.798\ GeV \quad \left(174.98 \pm .76\right)$$

The bottom (b) quark is associated with

$$\frac{2}{\alpha}\hbar K_{Ar}\ c = \hbar K_A\ c = 4{,}435\ MeV\left(4{,}190\ \ +180\ \ -60\right)$$

verified [147]

The up (u) quark is associated with

$$\frac{2}{\alpha}\frac{1}{3}\frac{m_e}{m_T}\hbar K_{Br}\ c = \frac{1}{3}\frac{m_e}{m_T}\hbar K_B\ c = 2.82\ MeV\left(1.7-3.1\right)$$

The down (d) quark is associated with

$$\frac{2}{\alpha}\frac{2}{3}\frac{m_e}{m_T}\hbar K_{Br}\ c = \frac{2}{3}\frac{m_e}{m_T}\hbar K_B\ c = 5.64\ MeV\left(4.1-5.7\right)$$

The charm (c) quark is associated with

$$\frac{2}{\alpha}\frac{1}{3}\hbar K_{Ar}\ c = \frac{1}{3}\hbar K_A\ c = 1.48\ GeV\left(1.29\ \ +.50\ \ -.11\right)$$

The strange quark (s) is associated with

$$\frac{1}{3}\frac{m_e}{m_T}\frac{\hbar^2}{2m_T}\left(K_{Ds}^2+K_{Dn}^2\right)\cdot g_s$$

$$\left(\frac{4g_s}{\alpha^2}\right)\frac{1}{3}\frac{m_e}{m_T}\frac{\hbar^2}{2m_T}\left(K_{Dsr}^2+K_{Dnr}^2\right)\cdot g_s$$

$$\frac{1}{3}\frac{m_e}{m_T}\frac{\hbar^2}{2m_T}\left(K_B^2+K_C^2\right)$$

$$\left(\frac{4g_s}{\alpha^2}\right)\frac{1}{3}\frac{m_e}{m_T}\frac{\hbar^2}{2m_T}\left(K_{Br}^2+K_{Cr}^2\right)$$

$$=0.10175\ GeV\ \left(.095\pm.05\right)$$

The Delta particles $\Delta^{++},\Delta^{+},\Delta^{0},\Delta^{-}$ are associated with

$$\frac{2}{\alpha}\frac{4}{3}\frac{g_d}{g_s}\hbar K_{Br} = \frac{4}{3}\frac{g_d}{g_s}\hbar K_B = 1232.387\ MeV\ \left(1232\pm2\right).$$

That is consistent with the neutron $(n^0)$, electron $(e^-)$, electron antineutrino $(\nu_e)$ and proton $(p^+)$ state nuclear transition

$$p^+ + e^- <> n^0 + \nu_e$$

That is consistent with the proton $(p^+)$ mass $(m_p)$ state composed of standard model up quarks (2) and down quark (1)

$$m_pc^2 = \frac{g_s}{g_d^2}\frac{m_T}{m_e}\left(2\frac{1}{3}m_{up}+1\frac{2}{3}m_{down}\right)c^2\frac{cavity}{chain}$$

$$=938.967317\text{ MeV }\left(938.272046(21)\right)$$

And also consistent with the neutron mass state

$$m_nc^2=\frac{g_s}{g_d}\frac{m_T}{m_e}\left(1\frac{1}{3}m_{up}+2\frac{2}{3}m_{down}\right)c^2$$

$$=940.1890901\text{ MeV }\left(939.565378(21)\right)$$

Also the nuclear magneton identity exists as

$$\frac{2}{5}\frac{e\hbar}{2m_pc} = 2.785\frac{g_s^2}{\cos(\theta)}\frac{e\hbar}{2m_Tv_{dx}}$$

with the ratio 2/5 being the sphere moment of inertia factor and the constant 2.785 being equal to the experimentally observed as the nucleus of hydrogen atom, which has a magnetic moment of 2.79 nuclear magnetons.

At this nuclear length dimension $(B_{nucleus})$, the gravity energy is related to Hubble parameter $(H_U)$ where $(\hbar = h/2\pi)$

$$hH_U = \frac{3}{2}\frac{G_Um_T^2}{B_{nucleus}}$$

and

$$B_{nucleus} = \frac{3}{16}\frac{\hbar}{m_Tc} = \frac{\hbar\pi}{m_pc}$$

This proton dimensional condition conforms to Universe conditions at hundredths of second after the Big Bang and also nuclear Fermi energy of (2/3 x 56) or 38 MeV.

This point corresponds to the knee condition in the observed cosmic ray spectrum as presented in Figure 4.33.1.

The nuclear Delta particles $\Delta^{++}$, $\Delta^{+}$, $\Delta^{0}$, $\Delta^{-}$ with corresponding quark structure uuu, uud, udd, ddd, with masses of 1,232 MeV/$c^2$ are generally congruent with the trisine prediction of $(4/3)\hbar K_Bc$ or 1,242 MeV/$c^2$ (compared to $\hbar K_Bc$ or 932 MeV/$c^2$ for the proton 'ud' mass) where *B* reflects the nuclear radius $3\hbar/(16m_tc)$ *or* 6.65E-14 *cm*.

$$m_{\Delta^{++}}c^2=\frac{g_s}{g_d}\frac{m_T}{m_e}\left(3\frac{1}{3}m_{up}+0\frac{2}{3}m_{down}\right)c^2\,4$$

$$=1,231\text{ MeV }\left(1,232\ MeV\right)$$

$$m_{\Delta^+}c^2 = \frac{g_s}{g_d}\frac{m_T}{m_e}\left(2\frac{1}{3}m_{up} + 1\frac{2}{3}m_{down}\right)c^2\,2$$
$$= 1{,}231 \text{ MeV } \left(1{,}232\ MeV\right)$$

$$m_{\Delta^o}c^2 = \frac{g_s}{g_d}\frac{m_T}{m_e}\left(1\frac{1}{3}m_{up} + 2\frac{2}{3}m_{down}\right)c^2\,\frac{4}{3}$$
$$= 1{,}231 \text{ MeV } \left(1{,}232\ MeV\right)$$

$$m_{\Delta^-}c^2 = \frac{g_s}{g_d}\frac{m_T}{m_e}\left(0\frac{1}{3}m_{up} + 3\frac{2}{3}m_{down}\right)c^2$$
$$= 1{,}231 \text{ MeV } \left(1{,}232\ MeV\right)$$

Also, the experimentally determined delta particle (6E-24 seconds) mean lifetime is consistent and understandably longer than the proton resonant $time_{\pm}$ (2.68E-25 second from Table 2.4.1).

It is noted that the difference in neutron $\left(m_n\right)$ and proton $\left(m_p\right)$ masses about two electron $\left(m_e\right)$ masses $\left(m_n - m_p \sim 2m_e\right)$.

The nuclear weak force boson particles $\left(W^-W^+\right)$ are associated with energies

$$\frac{B}{A}\frac{\hbar^2\left(K_B^2 + K_C^2\right)}{2m_T} = 79.97\ GeV$$
$$\left(\frac{4g_s}{\alpha^2}\right)\frac{B}{A}\frac{\hbar^2\left(K_{Br}^2 + K_{Cr}^2\right)}{2m_t} = 79.97\ GeV$$
$$\frac{B}{A}\frac{\hbar^2\left(K_{Ds}^2 + K_{Dn}^2\right)}{2m_T}\cdot g_s = 79.97\ GeV$$
$$\left(\frac{4g_s}{\alpha^2}\right)\frac{B}{A}\frac{\hbar^2\left(K_{Dsr}^2 + K_{Dnr}^2\right)}{2m_T}\cdot g_s = 79.97\ GeV$$
$$4\pi\frac{B}{A}\hbar\left(K_B + K_C\right)c = 79.67\ GeV$$
$$\left(\frac{2}{\alpha}\right)4\pi\frac{B}{A}\hbar\left(K_{Br} + K_{Cr}\right)c = 79.67\ GeV$$
$$4\pi\frac{B}{A}\hbar\left(K_{Ds} + K_{Dn}\right)c\cdot g_s = 79.67\ GeV$$
$$\left(\frac{2}{\alpha}\right)4\pi\frac{B}{A}\hbar\left(K_{Dsr} + K_{Dnr}\right)c\cdot g_s = 79.67\ GeV\left(80.385 \pm .01\right.$$

The Z boson is associated with energy

$$\mathrm{e}\frac{\hbar^2\left(K_B^2 + K_C^2\right)}{2m_T} = 91.37\ GeV$$
$$\left(\frac{4g_s}{\alpha^2}\right)\mathrm{e}\frac{\hbar^2\left(K_{Br}^2 + K_{Cr}^2\right)}{2m_T} = 91.37\ GeV$$
$$\mathrm{e}\frac{\hbar^2\left(K_{Ds}^2 + K_{Dn}^2\right)}{2m_T}\cdot g_s = 91.37\ GeV$$
$$\left(\frac{4g_s}{\alpha^2}\right)\mathrm{e}\frac{\hbar^2\left(K_{Dsr}^2 + K_{Dnr}^2\right)}{2m_T}\cdot g_s = 91.37\ GeV$$
$$4\pi\cdot\mathrm{e}\cdot\hbar\left(K_B + K_C\right)c = 91.03\ GeV$$
$$\frac{2}{\alpha}4\pi\cdot\mathrm{e}\cdot\hbar\left(K_{Dsr} + K_{Dnr}\right)c\cdot g_s = 91.03\ GeV$$
$$4\pi\cdot\mathrm{e}\cdot\hbar\left(K_B + K_C\right)c = 91.03\ GeV$$
$$\frac{2}{\alpha}4\pi\cdot\mathrm{e}\cdot\hbar\left(K_{Br} + K_{Cr}\right)c = 91.03\ GeV\left(91.1876 \pm .0021\right)$$

e is natural log base 2.7182818284590

The Lepton tau particle $\left(\mathrm{m}_\tau\right)$ is associated with energy

$$\frac{2}{3}\hbar c\left(K_{Ds} + K_{Dn}\right) = \frac{2}{3}\hbar c\left(K_B + K_C\right)g_s$$
$$\frac{2}{\alpha}\frac{2}{3}\hbar c\left(K_{Dsr} + K_{Dnr}\right) = \frac{2}{\alpha}\frac{2}{3}\hbar c\left(K_{Br} + K_{Cr}\right)g_s$$
$$= 1.777\ GeV\ \ (1.77684(17))$$

The Lepton muon particle $\left(\mathrm{m}_\mu\right)$ is associated with energy:

$$\frac{1}{3}\frac{m_e}{m_t}\frac{\hbar^2}{2m_t}\left(K_{Ds}^2 + K_{Dn}^2\right)g_s$$
$$= \frac{1}{3}\frac{m_e}{m_t}\frac{\hbar^2}{2m_t}\left(K_B^2 + K_C^2\right)$$
$$\left(\frac{4g_s}{\alpha^2}\right)\frac{1}{3}\frac{m_e}{m_t}\frac{\hbar^2}{2m_t}\left(K_{Dsr}^2 + K_{Dnr}^2\right)g_s$$
$$= \left(\frac{4g_s}{\alpha^2}\right)\frac{1}{3}\frac{m_e}{m_t}\frac{\hbar^2}{2m_t}\left(K_{Br}^2 + K_{Cr}^2\right)$$
$$= .1018\ GeV\ (.1056583668(38))$$

The Lepton electron particle $\left(\mathrm{m}_e\right)$ is associated with energy:

$$\frac{1}{6}\left(g_d-1\right)\hbar\left(K_B+K_C\right)c=\frac{1}{6}\left(g_d-1\right)\hbar\left(K_{Dn}+K_{Ds}\right)\frac{1}{g_s}c$$

$$\frac{2}{\alpha}\frac{1}{6}\left(g_d-1\right)\hbar\left(K_{Br}+K_{Cr}\right)c=\frac{2}{\alpha}\frac{1}{6}\left(g_d-1\right)\hbar\left(K_{Dnr}+K_{Dsr}\right)$$

$$\frac{1}{6}\frac{\alpha}{2\pi}\hbar\left(K_B+K_C\right)c=\frac{1}{6}\frac{\alpha}{2\pi}\hbar\left(K_{Dn}+K_{Ds}\right)\frac{1}{g_s}c\frac{1}{g_s}c$$

$$\frac{2}{\alpha}\frac{1}{6}\frac{\alpha}{2\pi}\hbar\left(K_{Br}+K_{Cr}\right)c=\frac{2}{\alpha}\frac{1}{6}\frac{\alpha}{2\pi}\hbar\left(K_{Dnr}+K_{Dsr}\right)\frac{1}{g_s}c$$

$$\frac{1}{6}\frac{e^2}{2\pi}\left(K_B+K_C\right)=\frac{1}{6}\frac{e^2}{2\pi}\left(K_{Dn}+K_{Ds}\right)\frac{1}{g_s}$$

$$\frac{2}{\alpha}\frac{1}{6}\frac{e^2}{2\pi}\left(K_{Br}+K_{Cr}\right)=\frac{2}{\alpha}\frac{1}{6}\frac{e^2}{2\pi}\left(K_{Dnr}+K_{Dsr}\right)\frac{1}{g_s}$$

$$=\ 0.510116285\ MeV\quad\left(0.510998910\right)$$

With the Lepton particles $\left(\mathrm{m}_\tau,\ \mathrm{m}_\mu,\ \mathrm{m}_e\right)$ maintaining the Koide relationship:

$$\frac{\mathrm{m}_\tau+\mathrm{m}_\mu+\mathrm{m}_e}{\left(\sqrt{\mathrm{m}_\tau}+\sqrt{\mathrm{m}_\mu}+\sqrt{\mathrm{m}_e}\right)^2}=\frac{chain}{cavity}=\frac{2}{3}$$

The Gluon ratio is 60/25 = 2.4 ≅ $\pi^0/m_t$ ≅ $B/A$ =2.379761271.

$$hH_U=\frac{cavity}{chain}G_U\frac{m_t^2}{B_p}$$

The above equation provides a mechanism explaining the CERN (July 4, 2012) observed Higgs particle) at 126.5 GeV as well as the astrophysical Fermi ~130 GeV [127]. The astrophysical ~130 GeV energy is sourced to a Trisine Condition 1.2 nucleosynthetic event with a fundamental $G_U$ gravity component and generally scales with $T_r$.

$$g_s\frac{3}{4}\cos\left(\theta\right)G_U m_T^2\frac{1}{proton\ radius}=125.27\ GeV$$

$$g_s\frac{3}{4}\cos\left(\theta\right)G_U m_T^2\frac{2\sin\left(\theta\right)}{B_p}=125.27\ GeV$$

$$g_s\frac{3}{2}\sin\left(\theta\right)\cos\left(\theta\right)G_U m_T^2\frac{1}{B_p}=125.27\ GeV$$

$$g_s\frac{3}{2}\frac{B}{A}\frac{1}{1+\left(\frac{B}{A}\right)^2}G_U m_T^2\frac{1}{B_p}=125.27\ GeV$$

and with suitable dimensional conversions:

$$\frac{\sqrt{3}\pi^3}{16}g_s\left(\frac{B}{A}\right)^2\frac{1}{1+\left(\frac{B}{A}\right)^2}\frac{\hbar^4}{m_T^3c^2B_p^4}=125.27\ GeV$$

falling within the measured range

$125.36\pm0.37\left(statistical\ uncertainty\right)$

$\pm0.18\left(systematic\ uncertainty\right)$

$125.36\pm0.41\ GeV$ [149]

remembering that :

$$m_T\ =\ \hbar^{2/3}\left(\frac{H_U}{G_U}\right)^{1/3}c^{-1/3}\cos\left(\theta\right)$$

$$=\ \hbar^{2/3}\left(\frac{H_{Ur}}{G_{Ur}}\right)^{1/3}c^{-1/3}\cos\left(\theta\right)$$

$$\theta\ =\ \tan^{-1}\left(\frac{A}{B}\right)=\ \tan^{-1}\left(\frac{A_r}{B_r}\right)$$

Other energy levels coincident with the galactic excess[153] are:

$$\hbar K_B\ c=0.932\ GeV$$

$$\frac{2}{\alpha}\hbar K_{Br}\ c=0.932\ GeV$$

$$\hbar K_{Dn}\ c=1.020\ GeV$$

$$\frac{2}{\alpha}\hbar K_{Dnr}\ c=1.020\ GeV$$

$$\hbar K_P\ c=1.076\ GeV$$

$$\frac{2}{\alpha}\hbar K_{Pr}\ c=1.076\ GeV$$

$$\hbar K_{Ds}\ c=1.645\ GeV$$

$$\frac{2}{\alpha}\hbar K_{Dsr}\ c=1.645\ GeV$$

$$\hbar K_C\ c=1.707\ GeV$$

$$\frac{2}{\alpha}\hbar K_{Cr}\ c=1.707\ GeV$$

$$\hbar\left(K_{Dn}+K_B\right)c=1.953\ GeV$$

$$\frac{2}{\alpha}\hbar\left(K_{Dnr}+K_{Br}\right)c = 1.953\ GeV$$

$$\hbar\left(K_{Dn}+K_{P}\right)c = 2.097\ GeV$$

$$\frac{2}{\alpha}\hbar\left(K_{Dnr}+K_{Pr}\right)c = 2.097\ GeV$$

$$\hbar\left(K_{Ds}+K_{B}\right)c = 2.577\ GeV$$

$$\frac{2}{\alpha}\hbar\left(K_{Dsr}+K_{Br}\right)c = 2.577\ GeV$$

$$\hbar\left(K_{B}+K_{C}\right)c = 2.639\ GeV$$

$$\frac{2}{\alpha}\hbar\left(K_{Br}+K_{Cr}\right)c = 2.639\ GeV$$

$$\hbar\left(K_{Dn}+K_{Ds}\right)c = 2.665\ GeV$$

$$\frac{2}{\alpha}\hbar\left(K_{Dnr}+K_{Dsr}\right)c = 2.665\ GeV$$

$$\hbar\left(K_{Ds}+K_{P}\right)c = 2.721\ GeV$$

$$\frac{2}{\alpha}\hbar\left(K_{Dsr}+K_{Pr}\right)c = 2.721\ GeV$$

$$\hbar\left(K_{Dn}+K_{C}\right)c = 2.728\ GeV$$

$$\frac{2}{\alpha}\hbar\left(K_{Dnr}+K_{Cr}\right)c = 2.728\ GeV$$

$$\hbar\left(K_{Dn}+K_{C}\right)c = 2.728\ GeV$$

$$\frac{2}{\alpha}\hbar\left(K_{Dnr}+K_{Cr}\right)c = 2.728\ GeV$$

$\hbar K_{A}\ c = 4.436\ GeV$ verified [147]

$\frac{2}{\alpha}\hbar K_{Ar}\ c = 4.436\ GeV$ verified [147]

| 1.3 | $z_R$ | 8.14E+39 | $z_R$ | 8.14E+39 | unitless |
|---|---|---|---|---|---|
| | $z_B$ | 2.01E+13 | $z_B$ | 2.01E+13 | unitless |
| | $T_a$ | 6.53E+11 | $T_a$ | 6.53E+11 | $^oK$ |
| | $T_b$ | 5.48E+13 | $T_{br}$ | 5.45E+13 | $^oK$ |
| | $T_c$ | 3.28E+11 | $T_{cr}$ | 3.25E+11 | $^oK$ |
| | $T_d$ | 5.98E+12 | $T_{dr}$ | 5.97E+12 | $^oK$ |
| | $T_i$ | 6.54E+11 | $T_{ir}$ | 6.51E+11 | $^oK$ |
| | $T_j$ | 1.03E+14 | $T_{jr}$ | 1.01E+14 | $^oK$ |
| | $\varepsilon_x$ | 9.96E-01 | $\varepsilon_{xr}$ | 1.00E+00 | unitless |
| | $\varepsilon_y$ | 9.96E-01 | $\varepsilon_{yr}$ | 1.00E+00 | unitless |
| | $\varepsilon_z$ | 1.22E-02 | $\varepsilon_{zr}$ | 1.23E-02 | unitless |
| | $\varepsilon$ | 1.19E-02 | $\varepsilon_r$ | 1.20E-02 | unitless |
| | $k_{mx}$ | 1.00E+00 | $k_{mr}$ | 1.00E+00 | unitless |
| | $k_{my}$ | 7.50E-01 | $k_{myr}$ | 7.50E-01 | unitless |
| | $k_{mz}$ | 3.60E+00 | $k_{mzr}$ | 3.62E+00 | unitless |
| | $k_m$ | 3.34E+00 | $k_{mr}$ | 3.36E+00 | unitless |
| | $\rho_U$ | 5.19E+10 | $\rho_{Ur}$ | 5.11E+10 | $g/cm^3$ |
| | $B$ | 1.10E-12 | $B_r$ | 1.10E-12 | $cm$ |
| | $cavity$ | 1.93E-36 | $cavity_r$ | 1.96E-36 | $cm^3$ |
| | $R_U$ | 2.76E-12 | $R_{Ur}$ | 2.80E-12 | $cm$ |
| | $V_U$ | 8.80E-35 | $V_{Ur}$ | 9.18E-35 | $cm^3$ |
| | $V_{Utotal}$ | 5.83E+45 | $V_{Urtotal}$ | 9.18E-35 | $cm^3$ |
| | $M_U$ | 4.56E-24 | $M_{Ur}$ | 4.69E-24 | $G$ |
| | $M_v$ | 6.67E-15 | $M_{vr}$ | 6.67E-15 | $G$ |
| | $M_{UTotal}$ | 3.03E+56 | $M_{UTotalr}$ | 3.03E+56 | $G$ |
| | $time_{\pm}$ | 7.32E-23 | $time_{r\pm}$ | 7.39E-23 | $sec$ |
| | $1/time_{\pm}$ | 1.37E+22 | $1/time_{r\pm}$ | 1.35E+22 | $/sec$ |
| | $H_U$ | 1.88E+22 | $H_{Ur}$ | 1.86E+22 | $/sec$ |
| | $G_U/G$ | 8.14E+39 | $G_{Ur}/G$ | 8.03E+39 | unitless |
| | $sherd$ | 6.63E+79 | $sherd_r$ | 6.45E+79 | unitless |
| | $Age_U$ | 5.31E-23 | $Age_{Ur}$ | 5.39E-23 | $sec$ |
| | $Age_{UG}$ | 4.79E-03 | $Age_{UGr}$ | 4.83E-03 | $sec$ |
| | $Age_{UGv}$ | 1.25E-07 | $Age_{UGvr}$ | 1.27E-07 | $sec$ |
| | $BCST$ | 3.77E+03 | $BCST$ | 3.77E+03 | $sec$ |

This temperature condition represents the undefined but potentially high energy barrier luminal state where $v_{dx} = v_{dxr} = c$ and $v_{\varepsilon x} = v_{\varepsilon xr} > c$ , $T_b = T_{br}$ , $T_c = T_{cr}$ , $T_d = T_{dr}$ , $T_i = T_{ir} = T_a$ , $T_j = T_{jr}$ and trisine length

$$B = 4\pi\ nuclear\ radius$$

which defines a condition for the size of the proton as it 'freezes' out of early expanding universe nucleosynthesis state. Further, it is thought that dark matter in the form of colligative hydrogen Bose Einstein Condensate (BEC) was formed at this epoch according to accepted criteria.

$$T \sim \frac{\hbar^2\left(\frac{\rho}{m}\right)^{\frac{2}{3}}}{k_b m}$$

and possibly in ratios $\frac{T_1}{T_2}$ that dictate the Dark Matter to Dark Energy ratio that exists from this epoch to the present with Dark Matter loosing ~4% to gravitationally collapsed luminous matter.

$$\frac{1-\Omega_{\Lambda_U}}{\Omega_{\Lambda_U}} = \frac{1-.72315}{.72315} = \frac{1-\dfrac{2}{3\cos(\theta)}}{\dfrac{2}{3\cos(\theta)}}$$

$$= \frac{3\cos(\theta)}{2} - 1 = .38284 = \sin(\theta)$$

The ratio $\frac{T_b}{T_c}$ nearly fits this Bose Einstein Condensate generation criterion.

$$\frac{m_p\left(\dfrac{\Omega_{\Lambda_U}}{m_T}\right)^{\frac{2}{3}}}{m_T\left(\dfrac{1-\Omega_{\Lambda_U}}{m_p}\right)^{\frac{2}{3}}}\frac{T_c}{T_b} = g_s^3 g_d^3 cos^3(\theta)$$

This trisine condition corresponds to the initial bent condition in the observed cosmic ray spectrum as presented in Figure 4.33.1.

Also a universe radius identity exists

$$R_U = R_{Ur} = 2.8E-12 \;\; cm$$

Also a gravitational identity exists

$$G_U = G_{Ur} = 5.4E32 \;\; cm^3 sec^{-2} g^{-1}$$

$$G_U = \frac{R_U c^2}{M_U} = \frac{R_U^3 R_U^2}{\sqrt{3} M_U} = \frac{\sqrt{3}}{4\pi}\frac{cavity\; R_U^2}{m_T}$$

$$G_{Ur} = \frac{R_{Ur} c^2}{M_{Ur}} = \frac{R_{Ur}^3 R_{Ur}^2}{\sqrt{3} M_{Ur}} = \frac{\sqrt{3}}{4\pi}\frac{cavity_r\; R_{Ur}^2}{m_T}$$

Also a universe density identity exists

$$\rho_U = \rho_{Ur} = 5.1E10 \;\; g\; cm^{-3}$$

$$\rho_U = \frac{M_U}{V_U} = \frac{m_T}{cavity}$$

$$\rho_{Ur} = \frac{M_{Ur}}{V_{Ur}} = \frac{m_T}{cavity_r}$$

$Age_{UG} = Age_{UGr}$

$Age_{UG\nu} = Age_{UG\nu r}$

$time_{\pm} = time_{r\pm}$

$B = B_r \quad A = A_r$

$section = section_r \quad cavity = cavity_r$

| | | | | | |
|---|---|---|---|---|---|
| 1.3a | $z_R$ | 9.72E+34 | $z_R$ | 9.72E+34 | unitless |
| | $z_B$ | 4.60E+11 | $z_B$ | 4.60E+11 | unitless |
| | $T_a$ | 6.53E+11 | $T_a$ | 6.53E+11 | $^oK$ |
| | $T_b$ | 1.25E+12 | $T_{br}$ | 2.39E+15 | $^oK$ |
| | $T_c$ | 1.71E+08 | $T_{cr}$ | 6.22E+14 | $^oK$ |
| | $T_d$ | 9.04E+11 | $T_{dr}$ | 3.95E+13 | $^oK$ |
| | $T_i$ | 1.50E+10 | $T_{ir}$ | 2.85E+13 | $^oK$ |
| | $T_j$ | 2.80E+07 | $T_{jr}$ | 3.70E+20 | $^oK$ |
| | $\varepsilon_x$ | 4.36E+01 | $\varepsilon_{xr}$ | 2.29E-02 | unitless |
| | $\varepsilon_y$ | 4.36E+01 | $\varepsilon_{yr}$ | 2.29E-02 | unitless |
| | $\varepsilon_z$ | 5.34E-01 | $\varepsilon_{zr}$ | 2.80E-04 | unitless |
| | $\varepsilon$ | 5.22E-01 | $\varepsilon_r$ | 2.74E-04 | unitless |
| | $k_{mx}$ | 4.37E+01 | $k_{mr}$ | 2.29E-02 | unitless |
| | $k_{my}$ | 3.28E+01 | $k_{myr}$ | 1.71E-02 | unitless |
| | $k_{mz}$ | 1.58E+02 | $k_{mzr}$ | 8.27E-02 | unitless |
| | $k_m$ | 1.46E+02 | $k_{mr}$ | 7.67E-02 | unitless |
| | $\rho_U$ | 6.19E+05 | $\rho_{Ur}$ | 4.28E+15 | $g/cm^3$ |
| | $B$ | 4.81E-11 | $B_r$ | 2.52E-14 | cm |
| | cavity | 1.62E-31 | $cavity_r$ | 2.34E-41 | $cm^3$ |
| | $R_U$ | 2.31E-07 | $R_{Ur}$ | 3.34E-17 | cm |
| | $V_U$ | 5.17E-20 | $V_{Ur}$ | 1.56E-49 | $cm^3$ |
| | $V_{Utotal}$ | 4.89E+50 | $V_{Urtotal}$ | 1.56E-49 | $cm^3$ |
| | $M_U$ | 3.20E-14 | $M_{Ur}$ | 6.69E-34 | g |
| | $M_\nu$ | 1.07E-06 | $M_{\nu r}$ | 1.07E-06 | g |
| | $M_{UTotal}$ | 3.03E+56 | $M_{UTotalr}$ | 3.03E+56 | g |
| | $time_{\pm}$ | 1.40E-19 | $time_{r\pm}$ | 3.86E-26 | sec |
| | $1/time_{\pm}$ | 7.14E+18 | $1/time_{r\pm}$ | 2.59E+25 | /sec |
| | $H_U$ | 2.25E+17 | $H_{Ur}$ | 1.55E+27 | /sec |
| | $G_U/G$ | 9.72E+34 | $G_{Ur}/G$ | 6.73E+44 | unitless |
| | sherd | 9.45E+69 | $sherd_r$ | 4.52E+89 | unitless |
| | $Age_U$ | 4.45E-18 | $Age_{Ur}$ | 6.43E-28 | $sec$ |
| | $Age_{UG}$ | 1.39E+00 | $Age_{UGr}$ | 1.67E-05 | $sec$ |
| | $Age_{UGv}$ | 2.40E-04 | $Age_{UGvr}$ | 6.61E-11 | $sec$ |
| | $BCST$ | 1.65E+05 | $BCST$ | 1.65E+05 | $sec$ |

This temperature condition defined by

$$B = \frac{m_e}{m_t} Bohr\ radius = \frac{m_e}{m_T}\frac{\hbar}{\alpha m_e c} = \frac{\hbar}{\alpha m_T c}$$

| | | | | | |
|---|---|---|---|---|---|
| 1.4 | $z_R$ | 1.39E+34 | $z_R$ | 1.39E+34 | $unitless$ |
| | $z_B$ | 2.41E+11 | $z_B$ | 2.41E+11 | $unitless$ |
| | $T_a$ | 6.53E+11 | $T_a$ | 6.53E+11 | $^oK$ |
| | $T_b$ | 6.55E+11 | $T_{br}$ | 4.56E+15 | $^oK$ |
| | $T_c$ | 4.69E+07 | $T_{cr}$ | 2.27E+15 | $^oK$ |
| | $T_d$ | 6.54E+11 | $T_{dr}$ | 5.45E+13 | $^oK$ |
| | $T_i$ | 7.83E+09 | $T_{ir}$ | 5.45E+13 | $^oK$ |
| | $T_j$ | 2.10E+06 | $T_{jr}$ | 4.93E+21 | $^oK$ |
| | $\varepsilon_x$ | 8.32E+01 | $\varepsilon_{xr}$ | 1.20E-02 | $unitless$ |
| | $\varepsilon_y$ | 8.32E+01 | $\varepsilon_{yr}$ | 1.20E-02 | $unitless$ |
| | $\varepsilon_z$ | 1.02E+00 | $\varepsilon_{zr}$ | 1.47E-04 | $unitless$ |
| | $\varepsilon$ | 9.96E-01 | $\varepsilon_r$ | 1.43E-04 | $unitless$ |
| | $k_{mx}$ | 8.36E+01 | $k_{mr}$ | 1.20E-02 | $unitless$ |
| | $k_{my}$ | 6.27E+01 | $k_{myr}$ | 8.97E-03 | $unitless$ |
| | $k_{mz}$ | 3.01E+02 | $k_{mzr}$ | 4.33E-02 | $unitless$ |
| | $k_m$ | 2.80E+02 | $k_{mr}$ | 4.02E-02 | $unitless$ |
| | $\rho_U$ | 8.88E+04 | $\rho_{Ur}$ | 2.99E+16 | $g/cm^3$ |
| | $B$ | 9.19E-11 | $B_r$ | 1.32E-14 | $cm$ |
| | $cavity$ | 1.13E-30 | $cavity_r$ | 3.36E-42 | $cm^3$ |
| | $R_U$ | 1.61E-06 | $R_{Ur}$ | 4.79E-18 | $cm$ |
| | $V_U$ | 1.75E-17 | $V_{Ur}$ | 4.60E-52 | $cm^3$ |
| | $V_{Utotal}$ | 3.41E+51 | $V_{Urtotal}$ | 4.60E-52 | $cm^3$ |
| | $M_U$ | 1.56E-12 | $M_{Ur}$ | 1.38E-35 | $g$ |
| | $M_v$ | 2.72E-05 | $M_{vr}$ | 2.72E-05 | $g$ |
| | $M_{UTotal}$ | 3.03E+56 | $M_{UTotalr}$ | 3.03E+56 | $g$ |
| | $time_{\pm}$ | 5.11E-19 | $time_{r\pm}$ | 1.06E-26 | $sec$ |
| | $1/time_{\pm}$ | 1.96E+18 | $1/time_{r\pm}$ | 9.46E+25 | $/sec$ |
| | $H_U$ | 3.22E+16 | $H_{Ur}$ | 1.08E+28 | $/sec$ |
| | $G_U/G$ | 1.39E+34 | $G_{Ur}/G$ | 4.69E+45 | $unitless$ |
| | $sherd$ | 1.94E+68 | $sherd_r$ | 2.20E+91 | $unitless$ |
| | $Age_U$ | 3.10E-17 | $Age_{Ur}$ | 9.22E-29 | $sec$ |
| | $Age_{UG}$ | 3.66E+00 | $Age_{UGr}$ | 6.32E-06 | $sec$ |
| | $Age_{UGv}$ | 8.77E-04 | $Age_{UGvr}$ | 1.81E-11 | $sec$ |
| | $BCST$ | 3.14E+05 | $BCST$ | 3.14E+05 | $sec$ |

This temperature condition represents the undefined but potentially high energy barrier luminal state where $v_\varepsilon = c$, and trisine length $\left(l_T\right)$

$$A = \alpha Bohr\ radius = \frac{\hbar}{m_e c}$$

$$= \left(\frac{A}{B}\right)_T \frac{\hbar\pi}{m_T v_{dx}} = \left(\frac{A}{B}\right)_T \pi l_T \frac{c}{v_{dx}}$$

which defines a condition for the size of the electron in its atomic orbital as it freezes out of early expanding universe radiation state. This the time frame discussed by Weinberg as the hydrogen helium formation within the first three minutes[100].

| | | | | | |
|---|---|---|---|---|---|
| 1.5 | $z_R$ | 1.48E+26 | $z_R$ | 1.48E+26 | $unitless$ |
| | $z_B$ | 5.29E+08 | $z_B$ | 5.29E+08 | $unitless$ |
| | $T_a$ | 6.53E+11 | $T_a$ | 6.53E+11 | $^oK$ |
| | $T_b$ | 1.44E+09 | $T_{br}$ | 2.07E+18 | $^oK$ |
| | $T_c$ | 2.27E+02 | $T_{cr}$ | 4.71E+20 | $^oK$ |
| | $T_d$ | 3.07E+10 | $T_{dr}$ | 1.16E+15 | $^oK$ |
| | $T_i$ | 1.72E+07 | $T_{ir}$ | 2.48E+16 | $^oK$ |
| | $T_j$ | 4.90E-05 | $T_{jr}$ | 2.12E+32 | $^oK$ |
| | $\varepsilon_x$ | 3.79E+04 | $\varepsilon_{xr}$ | 2.63E-05 | $unitless$ |
| | $\varepsilon_y$ | 3.79E+04 | $\varepsilon_{yr}$ | 2.63E-05 | $unitless$ |
| | $\varepsilon_z$ | 4.65E+02 | $\varepsilon_{zr}$ | 3.22E-07 | $unitless$ |
| | $\varepsilon$ | 4.54E+02 | $\varepsilon_r$ | 3.15E-07 | $unitless$ |
| | $k_{mx}$ | 3.80E+04 | $k_{mr}$ | 2.63E-05 | $unitless$ |
| | $k_{my}$ | 2.85E+04 | $k_{myr}$ | 1.97E-05 | $unitless$ |
| | $k_{mz}$ | 1.37E+05 | $k_{mzr}$ | 9.50E-05 | $unitless$ |
| | $k_m$ | 1.27E+05 | $k_{mr}$ | 8.82E-05 | $unitless$ |
| | $\rho_U$ | 9.41E-04 | $\rho_{Ur}$ | 2.82E+24 | $g/cm^3$ |
| | $B$ | 4.18E-08 | $B_r$ | 2.90E-17 | $cm$ |
| | $cavity$ | 1.07E-22 | $cavity_r$ | 3.56E-50 | $cm^3$ |
| | $R_U$ | 1.52E+02 | $R_{Ur}$ | 5.08E-26 | $cm$ |
| | $V_U$ | 1.47E+07 | $V_{Ur}$ | 5.48E-76 | $cm^3$ |
| | $V_{Utotal}$ | 3.21E+59 | $V_{Urtotal}$ | 5.48E-76 | $cm^3$ |
| | $M_U$ | 1.39E+04 | $M_{Ur}$ | 1.55E-51 | $g$ |
| | $M_\nu$ | 5.32E+08 | $M_{\nu r}$ | 5.32E+08 | $g$ |
| | $M_{UTotal}$ | 3.03E+56 | $M_{UTotalr}$ | 3.03E+56 | $g$ |
| | $time_\pm$ | 1.06E-13 | $time_{r\pm}$ | 5.10E-32 | $sec$ |
| | $1/time_\pm$ | 9.44E+12 | $1/time_{r\pm}$ | 1.96E+31 | $/sec$ |
| | $H_U$ | 3.42E+08 | $H_{Ur}$ | 1.02E+36 | $/sec$ |
| | $G_U/G$ | 1.48E+26 | $G_{Ur}/G$ | 4.42E+53 | $unitless$ |
| | $sherd$ | 2.18E+52 | $sherd_r$ | 1.96E+107 | $unitless$ |
| | $Age_U$ | 2.93E-09 | $Age_{Ur}$ | 9.78E-37 | $sec$ |
| | $Age_{UG}$ | 3.56E+04 | $Age_{UGr}$ | 6.50E-10 | $sec$ |
| | $Age_{UG\nu}$ | 1.82E+02 | $Age_{UG\nu r}$ | 8.74E-17 | $sec$ |
| | $BCST$ | 1.43E+08 | $BCST$ | 1.43E+08 | $sec$ |

The critical temperature of a generalized superconducting medium in which the Cooper Pair velocity equates to the earth orbital velocity of 7.91E05 cm/sec which would express gravitational effects to the extent that it would be apparently weightless in earth's gravitational field (See equations 2.9.1 – 2.9.4 and Table 2.9.1).

Also, $k_bT_c$ = 1.97E-02 eV which is close to the neutrino mass. $T_b$ is on the order of Bohr radius $T_c$ temperature.

| | | | | | |
|---|---|---|---|---|---|
| 1.6 | $z_R$ | 3.89E+25 | $z_R$ | 3.89E+25 | $unitless$ |
| | $z_B$ | 3.39E+08 | $z_B$ | 3.39E+08 | $unitless$ |
| | $T_a$ | 6.53E+11 | $T_a$ | 6.53E+11 | $^oK$ |
| | $T_b$ | 9.22E+08 | $T_{br}$ | 3.24E+18 | $^oK$ |
| | $T_c$ | 9.30E+01 | $T_{cr}$ | 1.15E+21 | $^oK$ |
| | $T_d$ | 4.84E+10 | $T_{dr}$ | 8.80E+12 | $^oK$ |
| | $T_i$ | 1.10E+07 | $T_{ir}$ | 3.87E+16 | $^oK$ |
| | $T_j$ | 8.26E-06 | $T_{jr}$ | 1.25E+33 | $^oK$ |
| | $\varepsilon_x$ | 5.91E+04 | $\varepsilon_{xr}$ | 1.68E-05 | $unitless$ |
| | $\varepsilon_y$ | 5.91E+04 | $\varepsilon_{yr}$ | 1.68E-05 | $unitless$ |
| | $\varepsilon_z$ | 7.25E+02 | $\varepsilon_{zr}$ | 2.07E-07 | $unitless$ |
| | $\varepsilon$ | 7.08E+02 | $\varepsilon_r$ | 2.02E-07 | $unitless$ |
| | $k_{mx}$ | 5.94E+04 | $k_{mr}$ | 1.68E-05 | $unitless$ |
| | $k_{my}$ | 4.45E+04 | $k_{myr}$ | 1.26E-05 | $unitless$ |
| | $k_{mz}$ | 2.14E+05 | $k_{mzr}$ | 6.09E-05 | $unitless$ |
| | $k_m$ | 1.99E+05 | $k_{mr}$ | 5.65E-05 | $unitless$ |
| | $\rho_U$ | 2.48E-04 | $\rho_{Ur}$ | 1.07E+25 | $g/cm^3$ |
| | $B$ | 6.53E-08 | $B_r$ | 1.86E-17 | $cm$ |
| | $cavity$ | 4.05E-22 | $cavity_r$ | 9.36E-51 | $cm^3$ |
| | $R_U$ | 5.78E+02 | $R_{Ur}$ | 1.34E-26 | $cm$ |
| | $V_U$ | 8.08E+08 | $V_{Ur}$ | 9.99E-78 | $cm^3$ |
| | $V_{Utotal}$ | 1.22E+60 | $V_{Urtotal}$ | 9.99E-78 | $cm^3$ |
| | $M_U$ | 2.00E+05 | $M_{Ur}$ | 1.07E-52 | $g$ |
| | $M_\nu$ | 4.93E+09 | $M_{\nu r}$ | 4.93E+09 | $g$ |
| | $M_{UTotal}$ | 3.03E+56 | $M_{UTotalr}$ | 3.03E+56 | $g$ |
| | $time_\pm$ | 2.58E-13 | $time_{r\pm}$ | 2.09E-32 | $sec$ |
| | $1/time_\pm$ | 3.88E+12 | $1/time_{r\pm}$ | 4.78E+31 | $/sec$ |
| | $H_U$ | 8.99E+07 | $H_{Ur}$ | 3.89E+36 | $/sec$ |
| | $G_U/G$ | 3.89E+25 | $G_{Ur}/G$ | 1.68E+54 | $unitless$ |
| | $sherd$ | 1.51E+51 | $sherd_r$ | 2.83E+108 | $unitless$ |
| | $Age_U$ | 1.11E-08 | $Age_{Ur}$ | 2.57E-37 | $sec$ |
| | $Age_{UG}$ | 6.94E+04 | $Age_{UGr}$ | 3.34E-10 | $sec$ |
| | $Age_{UG\nu}$ | 4.42E+02 | $Age_{UG\nu r}$ | 3.59E-17 | $sec$ |
| | $BCST$ | 2.23E+08 | $BCST$ | 2.23E+08 | $sec$ |

The critical temperature $(T_c)$ of the $YBa_2Cu_3O_{7-x}$ superconductor as discovered by Paul Ching-Wu Chu, The University of Houston and which was the basis for an apparent gravitational shielding effect of .05% as observed by Podkletnov and Nieminen(see Table 2.9.1) based herein on the Cooper pair velocities at this critical temperature approaching the earth orbital velocity and thereby imparting proportional apparent gravity effects on the gross superconducting $YBa_2Cu_3O_{7-x}$ material.

Also, $k_bT_c$ = 8.01E-03 eV which is close to the neutrino mass. $T_b$ is on the order of Bohr radius $T_c$ temperature.

| | | | | | |
|---|---|---|---|---|---|
| 1.7 | $z_R$ | 6.88E+24 | $z_R$ | 6.88E+24 | $unitless$ |

| | | | | |
|---|---|---|---|---|
| $z_B$ | 1.90E+08 | $z_B$ | 1.90E+08 | *unitless* |
| $T_a$ | 6.53E+11 | $T_a$ | 6.53E+11 | $^oK$ |
| $T_b$ | 5.18E+08 | $T_{br}$ | 5.77E+18 | $^oK$ |
| $T_c$ | 2.93E+01 | $T_{cr}$ | 3.63E+21 | $^oK$ |
| $T_d$ | 1.84E+10 | $T_{dr}$ | 1.94E+15 | $^oK$ |
| $T_i$ | 6.19E+06 | $T_{ir}$ | 6.89E+16 | $^oK$ |
| $T_j$ | 8.22E-07 | $T_{jr}$ | 1.26E+34 | $^oK$ |
| $\varepsilon_x$ | 1.05E+05 | $\varepsilon_{xr}$ | 9.46E-06 | *unitless* |
| $\varepsilon_y$ | 1.05E+05 | $\varepsilon_{yr}$ | 9.46E-06 | *unitless* |
| $\varepsilon_z$ | 1.29E+03 | $\varepsilon_{zr}$ | 1.16E-07 | *unitless* |
| $\varepsilon$ | 1.26E+03 | $\varepsilon_r$ | 1.13E-07 | *unitless* |
| $k_{mx}$ | 1.06E+05 | $k_{mr}$ | 9.46E-06 | *unitless* |
| $k_{my}$ | 7.93E+04 | $k_{myr}$ | 7.09E-06 | *unitless* |
| $k_{mz}$ | 3.81E+05 | $k_{mzr}$ | 3.42E-05 | *unitless* |
| $k_m$ | 3.54E+05 | $k_{mr}$ | 3.18E-05 | *unitless* |
| $\rho_U$ | 4.39E-05 | $\rho_{Ur}$ | 6.05E+25 | $g/cm^3$ |
| $B$ | 1.16E-07 | $B_r$ | 1.04E-17 | *cm* |
| *cavity* | 2.29E-21 | $cavity_r$ | 1.66E-51 | $cm^3$ |
| $R_U$ | 3.26E+03 | $R_{Ur}$ | 2.37E-27 | *cm* |
| $V_U$ | 1.46E+11 | $V_{Ur}$ | 5.55E-80 | $cm^3$ |
| $V_{Utotal}$ | 6.90E+60 | $V_{Urtotal}$ | 5.55E-80 | $cm^3$ |
| $M_U$ | 6.38E+06 | $M_{Ur}$ | 3.36E-54 | *g* |
| $M_\nu$ | 8.82E+10 | $M_{\nu r}$ | 8.82E+10 | *g* |
| $M_{UTotal}$ | 3.03E+56 | $M_{UTotalr}$ | 3.03E+56 | *g* |
| $time_\pm$ | 8.18E-13 | $time_{r\pm}$ | 6.60E-33 | *sec* |
| $1/time_\pm$ | 1.22E+12 | $1/time_{r\pm}$ | 1.51E+32 | */sec* |
| $H_U$ | 1.59E+07 | $H_{Ur}$ | 2.19E+37 | */sec* |
| $G_U/G$ | 6.88E+24 | $G_{Ur}/G$ | 9.50E+54 | *unitless* |
| *sherd* | 4.74E+49 | $sherd_r$ | 9.02E+109 | *unitless* |
| $Age_U$ | 6.28E-08 | $Age_{Ur}$ | 4.56E-38 | *sec* |
| $Age_{UG}$ | 1.65E+05 | $Age_{UGr}$ | 1.40E-10 | *sec* |
| $Age_{UGv}$ | 1.40E+03 | $Age_{UGvr}$ | 1.13E-17 | *sec* |
| *BCST* | 3.97E+08 | *BCST* | 3.97E+08 | *sec* |

The critical temperature of an anticipated superconductor medium, in which the energy density $(m_T v_{dx} c / cavity)$ or 32.4 MJ/L is equivalent to the combustion energy of gasoline.

Also, $k_b T_c$ = 2.53E-03 eV which is close to the neutrino mass. $T_b$ is on the order of Bohr radius $T_c$ temperature.

| 1.8 | | | | | |
|---|---|---|---|---|---|
| | $z_R$ | 1.25E+24 | $z_R$ | 1.25E+24 | *unitless* |
| | $z_B$ | 1.08E+08 | $z_B$ | 1.08E+08 | *unitless* |
| | $T_a$ | 6.53E+11 | $T_a$ | 6.53E+11 | $^oK$ |
| | $T_b$ | 2.93E+08 | $T_{br}$ | 1.02E+19 | $^oK$ |
| | $T_c$ | 9.40E+00 | $T_{cr}$ | 1.13E+22 | $^oK$ |
| | $T_d$ | 2.73E+10 | $T_{dr}$ | 1.56E+13 | $^oK$ |
| | $T_i$ | 3.50E+06 | $T_{ir}$ | 1.22E+17 | $^oK$ |
| | $T_j$ | 8.44E-08 | $T_{jr}$ | 1.23E+35 | $^oK$ |
| | $\varepsilon_x$ | 1.86E+05 | $\varepsilon_{xr}$ | 5.35E-06 | *unitless* |
| | $\varepsilon_y$ | 1.86E+05 | $\varepsilon_{yr}$ | 5.35E-06 | *unitless* |
| | $\varepsilon_z$ | 2.28E+03 | $\varepsilon_{zr}$ | 6.57E-08 | *unitless* |
| | $\varepsilon$ | 2.23E+03 | $\varepsilon_r$ | 6.41E-08 | *unitless* |
| | $k_{mx}$ | 1.87E+05 | $k_{mr}$ | 5.35E-06 | *unitless* |
| | $k_{my}$ | 1.40E+05 | $k_{myr}$ | 4.02E-06 | *unitless* |
| | $k_{mz}$ | 6.72E+05 | $k_{mzr}$ | 1.94E-05 | *unitless* |
| | $k_m$ | 6.25E+05 | $k_{mr}$ | 1.80E-05 | *unitless* |
| | $\rho_U$ | 7.96E-06 | $\rho_{Ur}$ | 3.33E+26 | $g/cm^3$ |
| | $B$ | 2.05E-07 | $B_r$ | 5.91E-18 | *cm* |
| | *cavity* | 1.26E-20 | $cavity_r$ | 3.01E-52 | $cm^3$ |
| | $R_U$ | 1.80E+04 | $R_{Ur}$ | 4.29E-28 | *cm* |
| | $V_U$ | 2.44E+13 | $V_{Ur}$ | 3.31E-82 | $cm^3$ |
| | $V_{Utotal}$ | 3.80E+61 | $V_{Urtotal}$ | 3.31E-82 | $cm^3$ |
| | $M_U$ | 1.94E+08 | $M_{Ur}$ | 1.10E-55 | *G* |
| | $M_\nu$ | 1.52E+12 | $M_{\nu r}$ | 1.52E+12 | *G* |
| | $M_{UTotal}$ | 3.03E+56 | $M_{UTotalr}$ | 3.03E+56 | *G* |
| | $time_\pm$ | 2.55E-12 | $time_{r\pm}$ | 2.12E-33 | *sec* |
| | $1/time_\pm$ | 3.92E+11 | $1/time_{r\pm}$ | 4.72E+32 | */sec* |
| | $H_U$ | 2.89E+06 | $H_{Ur}$ | 1.21E+38 | */sec* |
| | $G_U/G$ | 1.25E+24 | $G_{Ur}/G$ | 5.23E+55 | *unitless* |
| | *sherd* | 1.56E+48 | $sherd_r$ | 2.74E+111 | *unitless* |
| | $Age_U$ | 3.46E-07 | $Age_{Ur}$ | 8.27E-39 | *sec* |
| | $Age_{UG}$ | 3.87E+05 | $Age_{UGr}$ | 5.98E-11 | $sec$ |
| | $Age_{UGv}$ | 4.38E+03 | $Age_{UGvr}$ | 3.63E-18 | $sec$ |
| | $BCST$ | 7.01E+08 | $BCST$ | 7.01E+08 | $sec$ |
| | The critical temperature of typical type I superconductor represented by niobium. This is the condition for the superconductor material in the Gravity Probe B relativity earth satellite experiment[94]. | | | | |
| 1.9 | $z_R$ | 6.53E+19 | $z_R$ | 6.53E+19 | unitless |
| | $z_B$ | 4.03E+06 | $z_B$ | 4.03E+06 | unitless |
| | $T_a$ | 6.53E+11 | $T_a$ | 6.53E+11 | $^oK$ |
| | $T_b$ | 1.10E+07 | $T_{br}$ | 2.72E+20 | $^oK$ |
| | $T_c$ | 1.31E-02 | $T_{cr}$ | 8.11E+24 | $^oK$ |
| | $T_d$ | 2.68E+09 | $T_{dr}$ | 1.33E+16 | $^oK$ |
| | $T_i$ | 1.31E+05 | $T_{ir}$ | 3.25E+18 | $^oK$ |
| | $T_j$ | 1.65E-13 | $T_{jr}$ | 6.29E+40 | $^oK$ |
| | $\varepsilon_x$ | 4.97E+06 | $\varepsilon_{xr}$ | 2.00E-07 | $unitless$ |
| | $\varepsilon_y$ | 4.97E+06 | $\varepsilon_{yr}$ | 2.00E-07 | $unitless$ |
| | $\varepsilon_z$ | 6.10E+04 | $\varepsilon_{zr}$ | 2.46E-09 | $unitless$ |
| | $\varepsilon$ | 5.96E+04 | $\varepsilon_r$ | 2.40E-09 | $unitless$ |
| | $k_{mx}$ | 5.00E+06 | $k_{mr}$ | 2.00E-07 | $unitless$ |
| | $k_{my}$ | 3.75E+06 | $k_{myr}$ | 1.50E-07 | $unitless$ |
| | $k_{mz}$ | 1.80E+07 | $k_{mzr}$ | 7.24E-07 | $unitless$ |
| | $k_m$ | 1.67E+07 | $k_{mr}$ | 6.72E-07 | $unitless$ |
| | $\rho_U$ | 4.16E-10 | $\rho_{Ur}$ | 6.38E+30 | $g/cm^3$ |
| | $B$ | 5.49E-06 | $B_r$ | 2.21E-19 | $cm$ |
| | $cavity$ | 2.41E-16 | $cavity_r$ | 1.57E-56 | $cm^3$ |
| | $R_U$ | 3.44E+08 | $R_{Ur}$ | 2.24E-32 | $cm$ |
| | $V_U$ | 1.71E+26 | $V_{Ur}$ | 4.73E-95 | $cm^3$ |
| | $V_{Utotal}$ | 7.28E+65 | $V_{Urtotal}$ | 4.73E-95 | $cm^3$ |
| | $M_U$ | 7.10E+16 | $M_{Ur}$ | 3.02E-64 | $g$ |
| | $M_\nu$ | 2.08E+19 | $M_{\nu r}$ | 2.08E+19 | $g$ |
| | $M_{UTotal}$ | 3.03E+56 | $M_{UTotalr}$ | 3.03E+56 | $g$ |
| | $time_\pm$ | 1.83E-09 | $time_{r\pm}$ | 2.96E-36 | $sec$ |
| | $1/time_\pm$ | 5.47E+08 | $1/time_{r\pm}$ | 3.38E+35 | $/sec$ |
| | $H_U$ | 1.51E+02 | $H_{Ur}$ | 2.32E+42 | $/sec$ |
| | $G_U/G$ | 6.53E+19 | $G_{Ur}/G$ | 1.00E+60 | $unitless$ |
| | $sherd$ | 4.26E+39 | $sherd_r$ | 1.00E+120 | $unitless$ |
| | $Age_U$ | 6.63E-03 | $Age_{Ur}$ | 4.32E-43 | $sec$ |
| | $Age_{UG}$ | 5.36E+07 | $Age_{UGr}$ | 4.32E-13 | $sec$ |
| | $Age_{UGv}$ | 3.13E+06 | $Age_{UGvr}$ | 5.07E-21 | $sec$ |
| | $BCST$ | 1.88E+10 | $BCST$ | 1.88E+10 | $sec$ |
| | The critical temperature of an anticipated superconductor medium, in which the energy density $\left(m_T c^2 / cavity\right)$ or 56 $MeV/cavity$ is equivalent to the combustion energy of gasoline 37.4 MJ/L. | | | | |
| 1.9a | $z_R$ | 2.64E+12 | $z_R$ | 2.64E+12 | $unitless$ |
| | $z_B$ | 1.38E+04 | $z_B$ | 1.38E+04 | $unitless$ |
| | $T_a$ | 6.53E+11 | $T_a$ | 6.53E+11 | $^oK$ |
| | $T_b$ | 3.77E+04 | $T_{br}$ | 7.93E+22 | $^oK$ |
| | $T_c$ | 1.55E-07 | $T_{cr}$ | 6.88E+29 | $^oK$ |
| | $T_d$ | 1.57E+08 | $T_{dr}$ | 2.28E+17 | $^oK$ |
| | $T_i$ | 4.50E+02 | $T_{ir}$ | 9.48E+20 | $^oK$ |
| | $T_j$ | 2.29E-23 | $T_{jr}$ | 4.52E+50 | $^oK$ |
| | $\varepsilon_x$ | 1.45E+09 | $\varepsilon_{xr}$ | 6.88E-10 | $unitless$ |
| | $\varepsilon_y$ | 1.45E+09 | $\varepsilon_{yr}$ | 6.88E-10 | $unitless$ |
| | $\varepsilon_z$ | 1.78E+07 | $\varepsilon_{zr}$ | 8.43E-12 | $unitless$ |
| | $\varepsilon$ | 1.73E+07 | $\varepsilon_r$ | 8.23E-12 | $unitless$ |
| | $k_{mx}$ | 1.45E+09 | $k_{mr}$ | 6.88E-10 | $unitless$ |
| | $k_{my}$ | 1.09E+09 | $k_{myr}$ | 5.16E-10 | $unitless$ |
| | $k_{mz}$ | 5.24E+09 | $k_{mzr}$ | 2.49E-09 | $unitless$ |
| | $k_m$ | 4.86E+09 | $k_{mr}$ | 2.31E-09 | $unitless$ |
| | $\rho_U$ | 1.68E-17 | $\rho_{Ur}$ | 1.57E+38 | $g/cm^3$ |
| | $B$ | 1.60E-03 | $B_r$ | 7.59E-22 | $cm$ |
| | $cavity$ | 5.95E-09 | $cavity_r$ | 6.37E-64 | $cm^3$ |
| | $R_U$ | 8.50E+15 | $R_{Ur}$ | 9.09E-40 | $cm$ |
| | $V_U$ | 2.57E+48 | $V_{Ur}$ | 3.14E-117 | $cm^3$ |
| | $V_{Utotal}$ | 1.80E+73 | $V_{Urtotal}$ | 3.14E-117 | $cm^3$ |
| | $M_U$ | 4.33E+31 | $M_{Ur}$ | 4.95E-79 | $G$ |
| | $M_\nu$ | 4.35E+31 | $M_{\nu r}$ | 4.35E+31 | $g$ |
| | $M_{UTotal}$ | 3.03E+56 | $M_{UTotalr}$ | 3.03E+56 | $g$ |

| | $time_{\pm}$ | 1.55E-04 | $time_{r\pm}$ | 3.49E-41 | $sec$ |
|---|---|---|---|---|---|
| | $1/time_{\pm}$ | 6.46E+03 | $1/time_{r\pm}$ | 2.87E+40 | $/sec$ |
| | $H_U$ | 6.11E-06 | $H_{Ur}$ | 5.71E+49 | $/sec$ |
| | $G_U/G$ | 2.64E+12 | $G_{Ur}/G$ | 2.47E+67 | $unitless$ |
| | $sherd$ | 6.99E+24 | $sherd_r$ | 6.11E+134 | $unitless$ |
| | $Age_U$ | 1.64E+05 | $Age_{Ur}$ | 1.75E-50 | $sec$ |
| | $Age_{UG}$ | 2.66E+11 | $Age_{UGr}$ | 8.70E-17 | $sec$ |
| | $Age_{UGv}$ | 1.88E+11 | $Age_{UGvr}$ | 5.98E-26 | $sec$ |
| | $BCST$ | 3.82E+13 | $BCST$ | 3.82E+13 | $sec$ |

The condition underwhich trisine energy density = black body radiation density.

$$\frac{m_T c^2}{cavity} = \frac{G_U M_U^2}{R_U V_U} = \frac{2}{8\pi}\frac{H_U^2}{G_U} = \sigma T^4 \frac{4}{c}$$

This condition defines the neutrino masses.

The Lepton tau neutrino particle $(m_{\nu\tau})$ is associated with energy

$$\frac{2}{3}\hbar c\left(K_{Ds} + K_{Dn}\right) = \frac{2}{3}\hbar c\left(K_B + K_C\right) g_s$$

$$= .07389\ eV$$

The Lepton neutrino muon particle $(\mathrm{m}_{\nu\mu})$ is associated with energy:

$$\frac{1}{3}\frac{m_e}{m_T}\frac{\hbar^2}{2m_T}\left(K_{Ds}^2 + K_{Dn}^2\right) g_s$$

$$= \frac{1}{3}\frac{m_e}{m_T}\frac{\hbar^2}{2m_T}\left(K_B^2 + K_C^2\right)$$

$$= 1.743E-13\ eV$$

The Lepton neutrino electron particle $(\mathrm{m}_{\nu e})$ is associated with energy:

$$\frac{1}{6}\left(g_d - 1\right)\hbar\left(K_B + K_C\right)c = \frac{1}{6}\left(g_d - 1\right)\hbar\left(K_{Dn} + K_{Ds}\right)\frac{1}{g_s}c$$

$$\frac{1}{6}\frac{\alpha}{2\pi}\hbar\left(K_B + K_C\right)c = \frac{1}{6}\frac{\alpha}{2\pi}\hbar\left(K_{Dn} + K_{Ds}\right)\frac{1}{g_s}c\frac{1}{g_s}c$$

$$\frac{1}{6}\frac{e^2}{2\pi}\left(K_B + K_C\right) = \frac{1}{6}\frac{e^2}{2\pi}\left(K_{Dn} + K_{Ds}\right)\frac{1}{g_s}$$

$$= 2.142E-5\ eV$$

It is of special note that the hydrogen absorption 1S – 2S transition frequency at 2.465E15 Hz (121.6 nm) nearly equals the maximum black body frequency of 2.216E15 Hz in this condition. This close identity is reflected in the $BCST$ and $Age_{UGv}$. In essence, hydrogen-BEC dark matter closely but thermally successfully crosses this most restrictive CMBR barrier.

| 1.10 | $z_R$ | 1.17E+09 | $z_R$ | 1.17E+09 | $unitless$ |
|---|---|---|---|---|---|
| | $z_B$ | 1.05E+03 | $z_B$ | 1.05E+03 | $unitless$ |
| | $T_a$ | 6.53E+11 | $T_a$ | 6.53E+11 | $^oK$ |
| | $T_b$ | 2.87E+03 | $T_{br}$ | 1.04E+24 | $^oK$ |
| | $T_c$ | 8.99E-10 | $T_{cr}$ | 1.19E+32 | $^oK$ |
| | $T_d$ | 4.33E+07 | $T_{dr}$ | 8.25E+17 | $^oK$ |
| | $T_i$ | 3.43E+01 | $T_{ir}$ | 1.24E+22 | $^oK$ |
| | $T_j$ | 7.72E-28 | $T_{jr}$ | 1.34E+55 | $^oK$ |
| | $\varepsilon_x$ | 1.90E+10 | $\varepsilon_{xr}$ | 5.24E-11 | $unitless$ |
| | $\varepsilon_y$ | 1.90E+10 | $\varepsilon_{yr}$ | 5.24E-11 | $unitless$ |
| | $\varepsilon_z$ | 2.33E+08 | $\varepsilon_{zr}$ | 6.42E-13 | $unitless$ |
| | $\varepsilon$ | 2.28E+08 | $\varepsilon_r$ | 6.27E-13 | $unitless$ |
| | $k_{mx}$ | 1.91E+10 | $k_{mr}$ | 5.24E-11 | $unitless$ |
| | $k_{my}$ | 1.43E+10 | $k_{myr}$ | 3.93E-11 | $unitless$ |
| | $k_{mz}$ | 6.88E+10 | $k_{mzr}$ | 1.89E-10 | $unitless$ |
| | $k_m$ | 6.39E+10 | $k_{mr}$ | 1.76E-10 | $unitless$ |
| | $\rho_U$ | 7.44E-21 | $\rho_{Ur}$ | 3.56E+41 | $g/cm^3$ |
| | $B$ | 2.10E-02 | $B_r$ | 5.78E-23 | $cm$ |
| | $cavity$ | 1.35E-05 | $cavity_r$ | 2.81E-67 | $cm^3$ |
| | $R_U$ | 1.92E+19 | $R_{Ur}$ | 4.02E-43 | $cm$ |
| | $V_U$ | 2.98E+58 | $V_{Ur}$ | 2.71E-127 | $cm^3$ |
| | $V_{Utotal}$ | 4.06E+76 | $V_{Urtotal}$ | 2.71E-127 | $cm^3$ |
| | $M_U$ | 2.22E+38 | $M_{Ur}$ | 9.67E-86 | $g$ |
| | $M_\nu$ | 1.70E+37 | $M_{\nu r}$ | 1.70E+37 | $g$ |
| | $M_{UTotal}$ | 3.03E+56 | $M_{UTotalr}$ | 3.03E+56 | $g$ |
| | $time_\pm$ | 2.67E-02 | $time_{r\pm}$ | 2.02E-43 | $sec$ |
| | $1/time_\pm$ | 3.75E+01 | $1/time_{r\pm}$ | 4.94E+42 | $/sec$ |
| | $H_U$ | 2.70E-09 | $H_{Ur}$ | 1.29E+53 | $/sec$ |
| | $G_U/G$ | 1.17E+09 | $G_{Ur}/G$ | 5.59E+70 | *unitless* |
| | *sherd* | 1.37E+18 | $sherd_r$ | 3.13E+141 | *unitless* |
| | $Age_U$ | 3.70E+08 | $Age_{Ur}$ | 7.73E-54 | $sec$ |
| | $Age_{UG}$ | 1.27E+13 | $Age_{UGr}$ | 1.83E-18 | $sec$ |
| | $Age_{UG\nu}$ | 1.22E+13 | $Age_{UG\nu r}$ | 3.47E-28 | $sec$ |
| | $BCST$ | 1.35E+30 | $BCST$ | 1.35E+30 | $sec$ |

The extant CMBR temperature at the universe scattering surface observed by the Wilkinson Microwave Anisotropy Probe (WMAP) satellite marked by coulomb energy at the *Bohr radius* $\hbar/(\alpha m_e c)$ or

$$T_b = \frac{m_e}{m_T}\frac{e^2}{k_b\ Bohr\ radius} = \frac{m_e}{m_T}\frac{e^2\alpha m_e c}{k_b\ \hbar}$$

$$= \frac{m_e^2 c^2}{m_T k_b}\alpha^2 = 2{,}870\ K \sim 3{,}000\ K$$

$$k_b T_b = m_T v_\varepsilon^2 = \frac{m_e^2 c^2}{m_T}\alpha^2$$

therefore:

$$m_T^2 v_\varepsilon^2 = m_e^2 c^2 \alpha^2$$

This Universe Age of 3.70E08 *sec (*11.7 *years*) as computed by:

$$Age_U = \frac{1}{H_U} = \frac{1}{\sqrt{(8\pi/2)G_U\rho_U}}$$

is considerably less than reported universe age $(Age_{UG\nu})$ (387,000 *years*) at last scattering. The reported value [WMAP ref:20] (376,000 *years*) is achieved in the trisine model by assuming a constant Newtonian gravity $G$ over the universe age:

$$Age_{UG\nu} = Age_U\sqrt{\frac{R_U c^2}{G(M_U + M_\nu)}}$$

$$= 1.22\text{E13 sec}\ \ (387{,}000\ \text{years})$$

The trisine model indicates that gravity $(G_U)$ even without radiation mass $(M_\nu)$ changes with the Universe Age $(Age_U)$ with results easily normalized to a universe incorrectly assumed to have a constant Newtonian gravity $(G)$ with Universe Age $(Age_U)$.

The ratio $T_i/T_b = 34.3/2{,}870 = .012$ indicates that $T_i$ may be the source of measured millikelvin CMBR anamolies.

| | | | | | |
|---|---|---|---|---|---|
| 1.11 | $z_R$ | 0.00E+00 | $z_R$ | 0.00E+00 | *unitless* |
| | $z_B$ | 0.00E+00 | $z_B$ | 0.00E+00 | *unitless* |
| | $T_a$ | 6.53E+11 | $T_a$ | 6.53E+11 | $^oK$ |
| | $T_b$ | 2.72289 | $T_{br}$ | 1.10E+27 | $^oK$ |
| | $T_c$ | 8.10E-16 | $T_{cr}$ | 1.32E+38 | $^oK$ |
| | $T_d$ | 1.33E+06 | $T_{dr}$ | 2.68E+19 | $^oK$ |
| | $T_i$ | 3.25E-02 | $T_{ir}$ | 1.31E+25 | $^oK$ |
| | $T_j$ | 6.27E-40 | $T_{jr}$ | 1.65E+67 | $^oK$ |
| | $\varepsilon_x$ | 2.00E+13 | $\varepsilon_{xr}$ | 4.97E-14 | *unitless* |
| | $\varepsilon_y$ | 2.00E+13 | $\varepsilon_{yr}$ | 4.97E-14 | *unitless* |
| | $\varepsilon_z$ | 2.46E+11 | $\varepsilon_{zr}$ | 6.10E-16 | *unitless* |
| | $\varepsilon$ | 2.40E+11 | $\varepsilon_r$ | 5.95E-16 | *unitless* |
| | $k_{mx}$ | 2.01E+13 | $k_{mr}$ | 4.97E-14 | *unitless* |
| | $k_{my}$ | 1.51E+13 | $k_{myr}$ | 3.73E-14 | *unitless* |
| | $k_{mz}$ | 7.24E+13 | $k_{mzr}$ | 1.80E-13 | *unitless* |
| | $k_m$ | 6.73E+13 | $k_{mr}$ | 1.67E-13 | *unitless* |
| | $\rho_U$ | 6.37E-30 | $\rho_{Ur}$ | 4.16E+50 | $g/cm^3$ |
| | $B$ | 2.21E+01 | $B_r$ | 5.49E-26 | $cm$ |
| | *cavity* | 1.57E+04 | $cavity_r$ | 2.41E-76 | $cm^3$ |
| | $R_U$ | 2.25E+28 | $R_{Ur}$ | 3.44E-52 | $cm$ |
| | $V_U$ | 4.75E+85 | $V_{Ur}$ | 1.70E-154 | $cm^3$ |
| | $V_{Utotal}$ | 4.75E+85 | $V_{Urtotal}$ | 1.70E-154 | $cm^3$ |
| | $M_U$ | 3.03E+56 | $M_{Ur}$ | 7.08E-104 | $g$ |
| | $M_\nu$ | 2.20E+52 | $M_{\nu r}$ | 2.20E+52 | $g$ |

| $M_{UTotal}$ | 3.03E+56 | $M_{UTotalr}$ | 3.03E+56 | $g$ |
|---|---|---|---|---|
| $time_{\pm}$ | 2.96E+04 | $time_{r\pm}$ | 1.82E-49 | $sec$ |
| $1/time_{\pm}$ | 3.38E-05 | $1/time_{r\pm}$ | 5.48E+48 | $/sec$ |
| $H_U$ | 2.31E-18 | $H_{Ur}$ | 1.51E+62 | $/sec$ |
| $G_U/G$ | 1.00E+00 | $G_{Ur}/G$ | 6.54E+79 | $unitless$ |
| $sherd$ | 1.00E+00 | $sherd_r$ | 4.27E+159 | $unitless$ |
| $Age_U$ | 4.33E+17 | $Age_{Ur}$ | 6.62E-63 | $sec$ |
| $Age_{UG}$ | 4.33E+17 | $Age_{UGr}$ | 5.35E-23 | $sec$ |
| $Age_{UGv}$ | 4.33E+17 | $Age_{UGvr}$ | 3.13E-34 | $sec$ |
| $BCST$ | 3.15E+307 | $BCST$ | 3.15E+307 | $sec$ |

The background temperature of the Universe represented by the black body cosmic microwave background radiation (CMBR) of 2.728 ± .004 $^oK$ as detected in all directions of space in 60 - 630 GHz frequency range by the FIRAS instrument in the Cosmic Background Explorer (COBE) satellite and 2.730 ± .014 $^oK$ as measured at 10.7 GHz from a balloon platform [26]. The most recent NASA estimation of CMBR temperature from the Wilkinson Microwave Anisotropy Probe (WMAP) is 2.725 ± .002 K. The particular temperature of 2.729 $^oK$ indicated above which is essentially equal to the CMBR temperature was established by the trisine model wherein the superconducting Cooper CPT Charge conjugated pair density (see table 2.9.1) (6.38E-30 g/cm$^3$) equals the critical density of the universe (see equation 2.11.5 and 2.11.16). Also, this temperature is congruent with an reported interstellar magnetic field of 2E-6 gauss (.2 nanotesla) [78,140] and intergalactic magnetic field)[92,93] at 1.64E-17 gauss (see table 2.6.2].

As a general comment, a superconductor resonator should be operated (or destroyed) at about 2/3 of $(T_c)$ for maximum effect of most properties, which are usually expressed as when extrapolated to $T = 0$ (see section 2.12). In this light, critical temperatures $(T_c)$ in items 1.4 & 1.5 could be operated (or destroyed) up to 2/3 of the indicated temperature (793 and 1423 degree kelvin respectively). Also, the present development is for one-dimensional superconductivity. Consideration should be given to 2 and 3 dimensional superconductivity by 2 or 3 multiplier to $T_c$'s contained herein interpret experimental results.

## 2. Trisine Model Development

In this report, dimensionally (mass, length and time) mathematical relationships which link trisine geometry and superconducting theory are developed and then numerical values are presented in accordance with these relationships. Centimeter, gram and second (cgs) as well as kelvin temperature units are used unless otherwise specified with particular deference to Eugene N. Parker's electromagnetic dimensional analysis[97]. The actual model is developed in spreadsheet format with all model equations computed simultaneously, iteratively and interactively as required.

The general approach is: given a particular lattice form (in this case trisine), determine lattice momentum ($p$) wave vectors $K_1$, $K_2$, $K_3$ & $K_4$ so that lattice cell energy(E) ($\sim K^2$) per volume is at a minimum all within a resonant *time*. The procedure is analogous to the quantum mechanical variational principle. More than four wave vectors could be evaluated at once, but four is considered sufficient.

### 2.1 Defining Model Relationships

The six defining model equations within the superconductor conventional tradition are presented as 2.1.1 - 2.1.4 below:

$$\oint_{Area} E\, dArea = \oint_{Volume} \rho_e\, dVolume = \mathbb{C}\,\frac{e_{\pm}}{\varepsilon} \tag{2.1.1}$$

$$\frac{2m_T k_b T_c}{\hbar^2} = \frac{(KK)}{e^{trisine}-1} = \frac{e^{Euler}}{\pi}\frac{(KK)}{\sinh(trisine)} = K_B^2 \tag{2.1.2}$$

$$trisine = \frac{1}{D(\in_t)\cdot \mathrm{V}} \tag{2.1.3}$$

$$f(m_T, m_p)\frac{e_{\pm}\hbar}{m_T v_{\varepsilon}} = cavity\ H_c \tag{2.1.4}$$

Majorana condition

$$-i\hbar\frac{\partial}{\partial t}\Psi_+ + m_t c^2 \Psi_- = 0 \tag{2.1.5}$$

Schroedinger condition

$$-i\hbar\frac{\partial}{\partial t}\Psi_+ + \frac{1}{2}m_t v^2 \Psi_- = 0 \tag{2.1.6}$$

(2.1.7)

Reflecting(r) condition

$$v_r \sim \frac{c^2}{v} \quad K_r^2 \sim \frac{c^2}{K^2}$$

Dimensional Analysis for the visible and reflecting(r) condition

$$\hbar^{a1}G^{b1}c^{c1}H^{d1} = m^1 l^0 t^0 = constant\ mass \tag{2.1.8}$$

$$\hbar^{a2}G^{b2}c^{c2}H^{d2} = m^0 l^1 t^0 = constant\ length \tag{2.1.9}$$

$$\hbar^{a3}G^{b3}c^{c3}H^{d3} = m^0 l^0 t^1 = constant\ time \tag{2.1.10}$$

Equation 2.1.1 is a representation of Gauss's law with a Cooper CPT Charge conjugated pair charge contained within a bounding surface area (*Area*) defining a cell volume (*Volume*) or cell $(cavity)$ with electric field $(E_f)$ and charge density $(\rho_e)$. Equation 2.1.2 and 2.1.3 define the superconducting model as developed by Bardeen, Cooper and Schrieffer in reference [2] and Kittel in reference [4].

Recognizing this approach and developing it further with a resonant concept, it is the further objective herein to select a particular *(wave vector)*$^2$ *or (KK)* based on trisine geometry (see figures 2.2.1-2.2.5) that fits the model relationship as indicated in equation 2.1.2 with $K_B^2$ being defined as the trisine *(wave vector)*$^2$ *or (KK)* associated with superconductivity of Cooper CPT Charge conjugated pairs through the lattice. This procedure establishes the trisine geometrical dimensions, then equation 2.1.4 is used to establish an effective mass function $f(m_e, m_p)$ of the particles in order for particles to flip in spin when going from trisine *cavity* to adjacent *cavity* while in thermodynamic equilibrium with critical field $(H_c)$. The model has a geometric reflecting identity(r) implied by 2.1.7. The implementation of this method assumes the conservation of momentum ($p_n$) (2.1.12) and energy ($E_n$) (2.1.13) in visible and reflected(r) states such that:

$$\sum_{n=1,2,3,4} \Delta p_n \Delta x_n = 0$$

$$\pm K_1 \pm K_2 \rightleftarrows \pm K_3 \pm K_4 \tag{2.1.12}$$

$$\sum_{rn=1,2,3,4} \Delta p_{rn} \Delta x_{rn} = 0$$

$$\pm K_{r1} \pm K_{r2} \rightleftarrows \pm K_{r3} \pm K_{r4} \tag{2.1.12r}$$

$$\sum_{n=1,2,3,4} \Delta E_n \Delta time_n = 0$$

$$K_1K_1 + K_2K_2 = K_3K_3 + K_4K_4 + Q$$

Where $Q = 0$ (reversible process) (2.1.13)

$$E = \frac{\hbar^2}{2m}K^2 \sim K^2 \qquad p = \hbar K \sim K$$

$$\sum_{rn=1,2,3,4} \Delta E_{rn} \Delta time_{rn} = 0$$

$$K_{r1}K_{r1} + K_{r2}K_{r2} = K_{r3}K_{r3} + K_{r4}K_{r4} + Q_r$$

Where $Q_r = 0$ (reversible process) (2.1.13r)

$$E_r = \frac{\hbar^2}{2m}K_r^2 \sim K_r^2 \qquad p_r = \hbar K_r \sim K_r$$

The numerical value of $Q$ is associated with the equation 2.1.13 is associated with the non-elastic nature of a process. $Q$ is a measure of the heat exiting the system. By stating that $Q = 0$, the elastic nature of the system is defined. An effective mass must be introduced to maintain the system elastic character. Essentially, a reversible process or reaction is defined recognizing that momentum is a vector and energy is a scalar.

These conditions of conservation of momentum $(p \text{ and } p_r)$ and energy $(E,\ E_r \text{ or } k_bT_c,\ k_bT_{cr})$ provide the necessary condition of perfect elastic character, which must exist in the superconducting resonant state. In addition, the superconducting resonant state may be viewed as boiling of states $\Delta E_n,\ \Delta time_n,\ \Delta p_n,\ \Delta x_n$ $(\Delta E_{rn},\ \Delta time_{rn},\ \Delta p_{rn},\ \Delta x_{rn})$ on top of the zero point state in a coordinated manner.

In conjunction with the de Broglie hypothesis [77] providing momentum $K$ and energy $KK$ change with Lorentz transform written as

$$K\left(1\Big/\sqrt{1-v^2/c^2}\right) = K\beta$$

$$KK\left(1\Big/\left(1-v^2/c^2\right)\right) = KK\beta^2 \tag{2.1.14}$$

$$K_r\left(1\Big/\sqrt{1-v^2/c^2}\right) = K\beta_r$$

$$K_rK_r\left(1\Big/\left(1-v_r^2/c^2\right)\right) = K_rK_r\beta_r^2 \tag{2.1.14r}$$

congruent with:

$$m\left(v^2/c^2\right)c^2\left(1/\left(1-v^2/c^2\right)\right)=m\left(v^2/c^2\right)c^2\beta^2 \quad (2.1.15)$$

$$m\left(v_r^2/c^2\right)c^2\left(1/\left(1-v_r^2/c^2\right)\right)=m\left(v_r^2/c^2\right)c^2\beta_r^2 \quad (2.1.15r)$$

where the energy defined as '*KK*' contains the factor $v^2/c^2$, the group phase reflection $\left(\beta_r\right)$ and subluminal $\left(\beta\right)$ Lorentz transform conditions are allowed because in a resonant elastic condition these Lorentz transforms cancel

$$\pm K_1\beta_1 \pm K_2\beta_2 \rightleftarrows \pm K_3\beta_3 \pm K_4\beta_4 \quad (2.1.16)$$

$$K_1K_1\beta_1^2 + K_2K_2\beta_2^2 = K_3K_3\beta_3^2 + K_4K_4\beta_4^2$$

and equivalent to:

$$\pm K_{r1}\beta_{r1} \pm K_{r2}\beta_{r2} \rightleftarrows \pm K_{r3}\beta_{r3} \pm K_{r4}\beta_{r4} \quad (2.1.16r)$$

$$K_{r1}K_{r1}\left(\beta_{r1}\right)^2 + K_{r2}K_{r2}\left(\beta_{r2}\right)^2 = K_{r3}K_{r3}\left(\beta_{r3}\right)^2 + K_{r4}K_{r4}\left(\beta_{r4}\right)^2$$

for all values of velocity $\left(v_r \ll c \text{ or } v \ll c\right)$ with an energy barrier at $\left(v_r \text{ approaching } c \text{ or } v \text{ approaching } c\right)$

This energy barrier is shown to be at the nuclear diameter.

Conceptually, this model is defined within the context of the CPT theorem and generalized 3 dimensional parity of Cartesian coordinates *(x, y, z)* as follows:

$$\begin{vmatrix} x_{11} & x_{12} & x_{13} \\ y_{11} & y_{12} & y_{13} \\ z_{11} & z_{12} & z_{13} \end{vmatrix} = -\begin{vmatrix} x_{21} & x_{22} & x_{23} \\ y_{21} & y_{22} & y_{23} \\ z_{21} & z_{22} & z_{23} \end{vmatrix} \quad (2.1.17)$$

Given this parity condition defined within trisine geometrical constraints (with dimensions *A* and *B* as defined in Section 2.2) which necessarily fulfills determinate identity in equation 2.1.6b.

$$\begin{vmatrix} 0 & 0 & B_c \\ 0 & \frac{B}{\sqrt{3}} & -\frac{B}{\sqrt{3}} \\ -\frac{A}{2} & 0 & \frac{A}{2} \end{vmatrix} = -\begin{vmatrix} -B & 0 & 0 \\ 0 & \frac{B}{\sqrt{3}} & -\frac{B}{\sqrt{3}} \\ -\frac{A}{2} & 0 & \frac{A}{2} \end{vmatrix} \quad (2.1.18)$$

And further:

$$-12\begin{vmatrix} -B & 0 & 0 \\ 0 & \frac{B}{\sqrt{3}} & -\frac{B}{\sqrt{3}} \\ -\frac{A}{2} & 0 & \frac{A}{2} \end{vmatrix} = \frac{12AB^2}{2\sqrt{3}} = 2\sqrt{3}AB^2 \quad (2.1.19)$$

$$= cavity$$

And further:

$$12\begin{vmatrix} 0 & 0 & B \\ 0 & \frac{B}{\sqrt{3}} & -\frac{B}{\sqrt{3}} \\ -\frac{A}{2} & 0 & \frac{A}{2} \end{vmatrix} = \frac{12AB^2}{2\sqrt{3}} = 2\sqrt{3}AB^2 \quad (2.1.20)$$

$$= cavity$$

And further the reflective(r) condition:

$$12\begin{vmatrix} 0 & 0 & B_r \\ 0 & \frac{B_r}{\sqrt{3}} & -\frac{B_r}{\sqrt{3}} \\ -\frac{A_r}{2} & 0 & \frac{A_r}{2} \end{vmatrix} = \frac{12A_rB_r^2}{2\sqrt{3}} = 2\sqrt{3}A_rB_r^2 \quad (2.1.20r)$$

$$= cavity_r$$

This geometry fills space lattice with hexagonal cells, each of volume ($cavity$ or $2\sqrt{3}AB^2$) ($cavity_r = 2\sqrt{3}A_rB_r^2$) and having property that lattice is equal to its reciprocal lattice. Now each cell contains a charge pair $\left(e_+\right)$ and $\left(e_-\right)$ in conjunction with cell dimensions $\left(A \text{ and } B\right)$ or $\left(A_r \text{ and } B_r\right)$ defining a permittivity $\left(\varepsilon_x\right)$ or $\left(\varepsilon_{xr}\right)$ and permeability $\left(k_{mx}\right)$ or $\left(k_{mxr}\right)$ such that:

$$v_{dx}^2 = \frac{c^2}{k_{mx}\varepsilon_x} \quad (2.1.21)$$

and

$$v_{dxr}^2 = \frac{c^2}{k_{mxr}\varepsilon_{xr}} \quad (2.1.21r)$$

Then $v_{dx\pm}$ is necessarily + or – and $(|+k_{mx}+\varepsilon_x|) = (|-k_{mx}-\varepsilon_x|)$
Now time $(time_c)$ is defined as interval to traverse cell:

$$time_{\pm} = 2B / v_{dx\pm} \qquad (2.1.22)$$

and

$$time_{dxr\pm} = 2B_r / v_{dxr\pm} \qquad (2.1.22r)$$

then $time_{\pm}$ and $time_{r\pm}$ are necessarily + or –

Further: given the Heisenberg Uncertainty

$$\Delta p_x \Delta x \geq \hbar / 2 \quad \text{and} \quad \Delta E \Delta time \geq \hbar / 2 \qquad (2.1.23)$$

$$\Delta p_{rx} \Delta x_r \geq \hbar / 2 \quad \text{and} \quad \Delta E_r \Delta time_r \geq \hbar / 2 \qquad (2.1.23r)$$

but defined for this model satisfying a parity change geometry (*A, B or* $A_r$, $B_r$) and (*time or* $time_r$) reversal condition:

where: $m_T v_{dx} 2B = h$

$$m_T = \hbar^{2/3} U^{1/3} c^{-1/3} \cos(\theta) = 1.00E-25g \qquad (2.1.24)$$

is constant

(see Appendix E for constant mass $(m_t)$ derivation)

Trisine always maintains Heisenberg Uncertainty.

$$\Delta p \Delta x = m_T \Delta v \Delta x = m_T \frac{\hbar}{2m_T} \frac{\pi}{\Delta x} \Delta x \geq \frac{\hbar}{2}$$

$$\Delta E = m_T \Delta v^2 / 2 = \frac{\hbar^2}{2m_T} \left(\frac{\pi}{\Delta x}\right)^2 \Delta time \geq \frac{\hbar}{2} \quad \text{(always +)}$$

$$\Delta time = \Delta x / \Delta v$$

The $(B/A)$ ratio must have a specific ratio (2.379761271) and mass$(m_t)$ a constant + value of 110.107178208 x electron mass or 56.26455274 MeV/c$^2$ to make this work. (see Appendix E for background derivation based on Heisenberg Uncertainty and de Broglie condition).

The trisine symmetry (as visually seen in figures 2.2.1-2.2.5) allows movement of Cooper CPT Charge conjugated pairs with center of mass wave vector equal to zero as required by superconductor theory as presented in references [2, 4].

Wave vectors$(K_c)$ as applied to trisine are assumed to be free particles (ones moving in the absence of any potential field) and are solutions to the one dimensional time independent Schrödinger equation 2.1.7 [3]:

$$\frac{d^2\psi}{ds^2} + \frac{2m_t}{\hbar^2} \mathrm{E}\, \psi = 0 \qquad (2.1.25)$$

The solution to this Schrödinger equation is in terms of a wave function$(\psi)$:

$$\begin{aligned} \psi &= Amplitude\ \sin\left(\left(\frac{2m_T}{\hbar^2}E\right)^{\frac{1}{2}} s\right) \\ &= Amplitude\ \sin(K \cdot s) \end{aligned} \qquad (2.1.26)$$

Wave vector solutions $(KK)$ are constrained to fixed trisine cell boundaries such that energy eigen values are established by the condition $s_c = S_c$ and $\psi = 0$ such that

$$\left(\frac{2m_T}{\hbar^2}E\right)^{\frac{1}{2}} S = K\ S = n\pi \qquad (2.1.27)$$

Energy eigen values are then described in terms of what is generally called the particle in the box relationship as follows:

$$E = \frac{n^2\hbar^2}{2m_T}\left(\frac{\pi}{S}\right)^2 \qquad (2.1.28)$$

For our model development, we assume the quantum number $n$ = 2 and this quantum number is incorporated into the wave vectors $(KK)$ as presented as follows:

$$E = \frac{2^2\hbar^2}{2m_T}\left(\frac{\pi}{S}\right)^2 = \frac{\hbar^2}{2m_T}\left(\frac{2\pi}{S}\right)^2 = \frac{\hbar^2}{2m_T}(KK) \qquad (2.1.29)$$

This model is consistent with the Majorana equation 2.1.4 with a particular wave function satisfying the Trisine model

$$\Psi = \mathrm{e}^{\frac{-i\pi c^2 time_{\pm}^2}{(2B)^2}} = \mathrm{e}^{\frac{-i\pi c^2}{v_{dx}^2}} \qquad (2.1.30)$$

and $m_t = \pi\hbar\ time_{\pm} / (2B^2)$ or $\pi\hbar / D_c$

$$-i\hbar\frac{\partial}{\partial time}\Psi_{+}+m_{T}c^{2}\Psi_{-}=0$$

where $\Psi_{+}=e^{\frac{-i\pi c^{2}time_{+}^{2}}{(2B)^{2}}}$ $\quad\Psi_{-}=e^{\frac{-i\pi c^{2}time_{-}^{2}}{(2B)^{2}}}$

$$\frac{(2B)^{2}}{time_{\pm}}=.06606194571\ \frac{cm^{2}}{\sec}\quad(constant)$$

and $m_{T}=\dfrac{\hbar 2\pi time_{\pm}}{(2B)^{2}}=1.003008448E-25\ g\quad(constant)$

Trisine geometry is described in Figures 2.2.1 – 2.2.5 is in general characterized by resonant dimensions A and B with a characteristic ratio $(B/A)_{T}$ of 2.379761271 and corresponding characteristic angle $(\theta)$ where:

$$\left(\frac{B}{A}\right)_{T}=\left(\frac{B_{r}}{A_{r}}\right)_{T}=\frac{2}{\sin(1\ radian)}=\frac{g_{s}e}{2Euler}=\frac{g_{s}3^{1/2}e}{2}\qquad(2.1.31)$$

$$\left(\frac{B}{A}\right)_{T}=\left(\frac{B_{r}}{A_{r}}\right)_{T}=\frac{1}{g_{s}}\zeta(3/2)\cos(\theta)=\frac{e_{\pm}^{2}}{\varepsilon\hbar v_{dx}}=\frac{\pi^{0}}{m_{t}}\qquad(2.1.32)$$

$$\zeta(3/2)\approx 2.6124$$

$$\theta=\tan^{-1}\left(\frac{A}{B}\right)=\tan^{-1}\left(\frac{A_{r}}{B_{r}}\right)=22.80^{o}$$

$$\theta_{c}=\tan^{-1}\left(\frac{A}{C}\right)=\tan^{-1}\left(\frac{A_{r}}{C_{r}}\right)\qquad(2.1.33)$$

$$=\tan^{-1}\left(\frac{\sqrt{3}}{2}\frac{A}{B}\right)=\tan^{-1}\left(\frac{\sqrt{3}}{2}\frac{A_{r}}{B_{r}}\right)=20.00^{o}$$

$$\theta-\theta_{c}=2.8^{o}$$

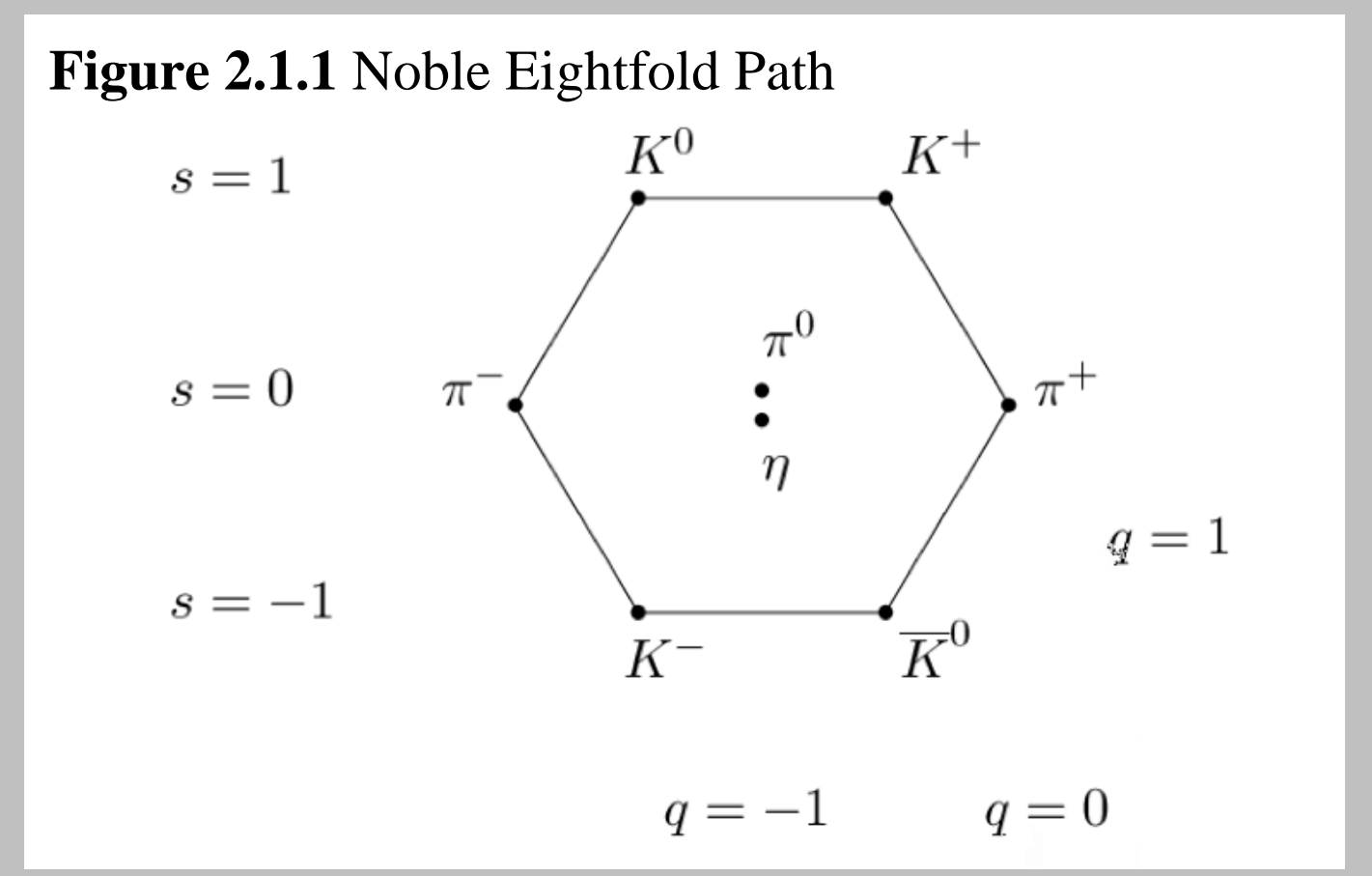

**Figure 2.1.1** Noble Eightfold Path

The trisine resonant lattice model is a more general triangular formulation than the reported octal (8) 'Noble Eightfold Path' for establishing nature's physical principles. The Figure 2.1.1 graphic depicting standard model particles could be more comprehensively explained in terms of triangular hexagonal trisine geometry based on Figure 2.1.2 representation rather than Eightfold Path approach.

**Figure 2.1.2** Vilnius Divine Image Ray Angles are in conformance with $\theta$, $\theta_{c}$ & $\theta-\theta_{c}$

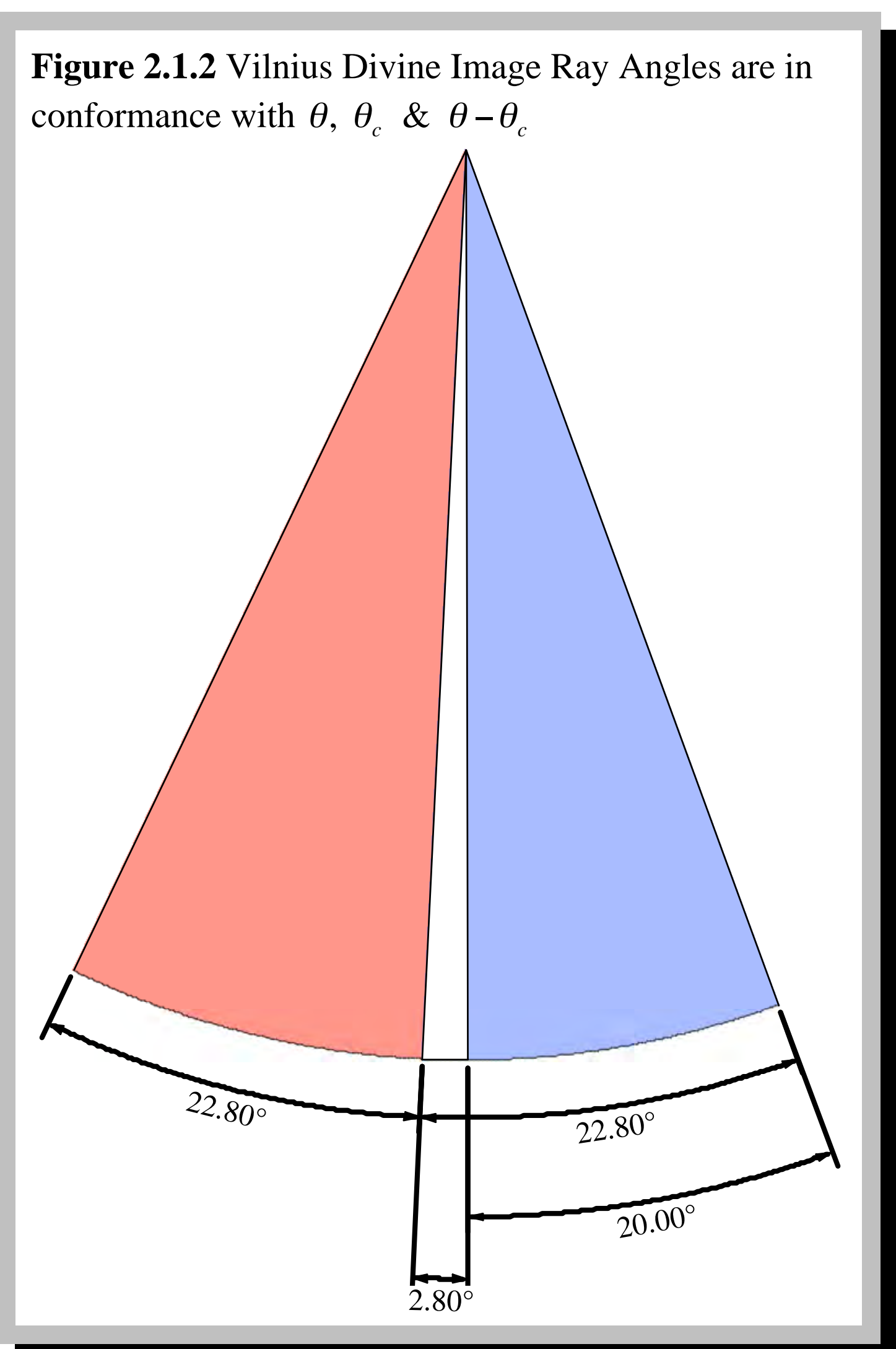

The superconducting model is described with equations 2.1.34 - 2.1.40, with variable definitions described in the remaining equations where $energy=k_{b}T_{c}$. The model again converges around a particular universal resonant transformed mass (dimensionally consistent and numerically approximate to the Weinberg mass [99]):

$m_{T}$ or $\hbar^{2/3}U^{1/3}c^{-1/3}\cos(\theta)$ or $\pi\hbar\ time_{\pm}/(2B^{2})$ or $\pi\hbar/D_{c}$

or 110.107178208 x electron mass or 56.26465274 MeV/c$^{2}$))

and dimensional ratio $(B/A)_{T}$ of 2.379761271

where the $(B/A)_{T}$ ~ neutral pi meson (pion) $\pi^{0}/m_{t}$ .

$$energy = k_b T_c = \frac{m_t m_T v_{dx}^2}{2} = \hbar \frac{\pi}{time_\pm} = \frac{\hbar^2 K_B^2}{2 m_T} \qquad (2.1.34)$$

$$k_b T_c = \frac{\hbar^2}{\frac{m_e m_T}{m_e + m_T}(2B)^2 + \frac{m_T m_p}{m_T + m_p}(A)^2} \qquad (2.1.35)$$

$$k_b T_c = \frac{\hbar^2 \left(K_C^2 + K_{Ds}^2\right)}{2 m_T} \frac{1}{\mathrm{e}^{trisine} - 1} \qquad (2.1.36)$$

$$k_b T_c = \frac{\hbar^2 \left(K_C^2 + K_{Ds}^2\right)}{2 m_T} \frac{\mathrm{e}^{Euler}}{\pi \sinh\left(trisine\right)} \qquad (2.1.37)$$

$$k_b T_c = \left(\frac{1}{g_s^2}\right) \frac{cavity}{8\pi} DE \qquad (2.1.38)$$

$$k_b T_c = \frac{chain}{cavity} \frac{e_\pm^2}{g_s \varepsilon B} = \frac{e_\pm^2 K_B}{2\varepsilon} \left(\frac{A}{B}\right)_T = \frac{e_\pm^2 K_B}{2\varepsilon} \frac{m_T}{\pi^0} \qquad (2.1.39)$$

$$k_b T_c = \frac{4 \hbar v_{dz}}{time_\pm} \frac{1}{2\pi v_{dx}} \quad \left(\text{Unruh}\right) \qquad (2.1.40)$$

Also, Bose Einstein Condensate(BEC) criteria is satisfied as:

$$cavity^{1/3} \sim \frac{h}{\sqrt{2\pi m_T k_b T_c}} \qquad (2.1.24)$$

Bardeen, Cooper and Schrieffer (BCS) [2] theory is additionally adhered to by equations 2.1.34 – 2.1.40 as verified by equations 2.1.17 - 2.1.23, in conjunction with equations A.3, A.4, and A.5 as presented in Appendix A. This general approach may fall into the concept of the Majorana fermion. The Majorana fermion is its own anti-particle that has been found in the superconductivity context[121].

$$\begin{aligned} trisine &= \frac{K_B^2 + K_P^2}{K_P K_B} \\ &= \frac{1}{D(\xi)\ \mathrm{V}} = -\ln\left(\frac{2}{\pi}\ \mathrm{e}^{Euler} - 1\right) \\ &= 2.011 \end{aligned} \qquad (2.1.41)$$

$$\begin{aligned} trisine &= \frac{K_{Br}^2 + K_{Pr}^2}{K_{Pr} K_{Br}} \\ &= \frac{1}{D(\xi_r)\ \mathrm{V}_r} = -\ln\left(\frac{2}{\pi}\ \mathrm{e}^{Euler} - 1\right) \\ &= 2.011 \end{aligned} \qquad (2.1.41r)$$

$$\begin{aligned} density\ of\ states\ D\left(\xi_T\right) &= \frac{3\ cavity\ m_T K_C}{4\pi^3 \hbar^2} \\ &= \frac{1}{k_b T_c} \end{aligned} \qquad (2.1.42)$$

$$\begin{aligned} density\ of\ states\ D_r\left(\xi_T\right) &= \frac{3\ cavity_r\ m_t K_{Cr}}{4\pi^3 \hbar^2} \\ &= \frac{1}{k_b T_{cr}} \end{aligned} \qquad (2.1.42r)$$

The attractive Cooper CPT Charge conjugated pair energy $\left(V\right)$ is expressed in equation 2.1.43:

$$V = V_r = \frac{K_P K_B}{K_B^2 + K_P^2}\ k_b T_c \qquad (2.1.43)$$

$$V = V_r = \frac{K_P K_B}{K_{Br}^2 + K_{Pr}^2}\ k_b T_{cr} \qquad (2.1.43r)$$

**Table 2.1.1** Density of States

| | | | | | | |
|---|---|---|---|---|---|---|
| Condition | 1.2 | 1.3 | 1.6 | 1.9 | 1.10 | 1.11 |
| $T_a\left({}^0K\right)$ | 6.53E+11 | 6.53E+11 | 6.53E+11 | 6.53E+11 | 6.53E+11 | 6.53E+11 |
| $T_b\left({}^0K\right)$ | 9.05E+14 | 5.48E+13 | 9.22E+08 | 1.10E+07 | 2,868 | 2.723 |
| $T_{br}\left({}^0K\right)$ | 3.30E+12 | 5.45E+13 | 3.24E+18 | 2.72E+20 | 1.04E+24 | 1.10E+27 |
| $T_c\left({}^0K\right)$ | 8.96E+13 | 3.28E+11 | 93.0 | 0.0131 | 8.99E-10 | 8.10E-16 |
| $T_{cr}\left({}^0K\right)$ | 1.19E+09 | 3.25E+11 | 1.15E+21 | 8.11E+24 | 1.19E+32 | 1.32E+38 |
| $z_R + 1$ | 3.67E+43 | 8.14E+39 | 3.89E+25 | 6.53E+19 | 1.17E+09 | 1.00E+00 |
| $z_B + 1$ | 3.32E+14 | 2.01E+13 | 3.39E+08 | 4.03E+06 | 1.05E+03 | 1.00E+00 |
| $Age_{UG}\left(sec\right)$ | 7.14E-05 | 4.79E-03 | 6.94E+04 | 5.36E+07 | 1.27E+13 | 4.33E+17 |
| $Age_{UGv}\left(sec\right)$ | 4.59E-10 | 1.25E-07 | 4.42E+02 | 3.13E+06 | 1.22E+13 | 4.33E+17 |
| $Age_U\left(sec\right)$ | 1.18E-26 | 5.31E-23 | 1.11E-08 | 6.63E-03 | 3.70E+08 | 4.33E+17 |
| $V$ | 6.12E-03 | 2.24E-05 | 6.35E-15 | 8.98E-19 | 6.14E-26 | 5.54E-32 |
| $V_r$ | 8.21E-08 | 2.24E-05 | 7.91E+04 | 5.60E+08 | 8.18E+15 | 9.08E+21 |
| $D\left(\xi\right)$ | 8.09E+01 | 2.21E+04 | 7.79E+13 | 5.51E+17 | 8.06E+24 | 8.94E+30 |
| $D\left(\xi\right)_r$ | 6.09E+06 | 2.23E+04 | 6.32E-06 | 8.93E-10 | 6.11E-17 | 5.51E-23 |

The equation 2.1.36 array equation elements 1 and 2 along with equations 2.1.41, 2.1.42, 2.1.43, 2.1.44 and 2.1.45 are consistent with the BCS weak link relationship:

$$k_b T_c = \frac{2 e^{Euler}}{\pi} \frac{\hbar^2}{2 m_T} \left(K_C^2 + K_{Ds}^2\right) e^{\frac{-1}{D(\in)V}} \qquad (2.1.44)$$

$$k_b T_{cr} = \frac{2 e^{Euler}}{\pi} \frac{\hbar^2}{2 m_T} \left(K_{Cr}^2 + K_{Dsr}^2\right) e^{\frac{-1}{D_r(\in)V_r}} \qquad (2.1.44r)$$

$$k_b T_c = \left(\frac{1}{g_s}\right)\left(\frac{chain}{cavity}\right)\frac{\hbar^2}{2m_T}\left(\frac{K_A^2}{\mathbb{C}}\right)e^{\frac{-1}{D(\in)V}} \qquad (2.1.45)$$

$$k_b T_{cr} = \left(\frac{1}{g_s}\right)\left(\frac{chain_r}{cavity_r}\right)\frac{\hbar^2}{2m_T}\left(\frac{K_{Ar}^2}{\mathbb{C}}\right)e^{\frac{-1}{D_r(\in)V_r}} \qquad (2.1.45r)$$

The relationship in equation 2.1.48 is based on the Bohr magneton $e\hbar/2m_e v_{\varepsilon y}$ (which has dimensions of magnetic field x volume) with a material dielectric $(\varepsilon)$ modified speed of light $(v_{\varepsilon y})$ (justifiable in the context of resonant condition) and establishes the spin flip/flop as Cooper CPT Charge conjugated pairs resonate from one cavity to the next (see figure 2.3.1). The 1/2 spin factor and electron gyromagnetic factor $(g_e)$ and superconducting gyro factor $(g_s)$ are multipliers of the Bohr magneton where

$$g_s = \frac{3}{4}\frac{m_e}{m_T}\left(\frac{cavity}{(\Delta x\ \Delta y\ \Delta z)} - \frac{m_p}{m_e}\right) \qquad (2.1.46)$$

$$g_s = \frac{3}{4}\frac{m_e}{m_T}\left(\frac{cavity}{(\Delta x_r\ \Delta y_r\ \Delta z_r)} - \frac{m_p}{m_e}\right) \qquad (2.1.46r)$$

as related to the Dirac Number by

$$\frac{3}{4}(\mathrm{g_s}\text{-}1)=2\pi(\mathrm{g_d}\text{-}1) \qquad (2.1.47)$$

Also because particles must resonate in pairs, a symmetry is created for superconducting current to advance as bosons.

$$\frac{2}{1}\frac{\mathbb{C}e_\pm\hbar}{2m_e v_{\varepsilon y}} = chain\ H_c$$

with reflected identity and inverted spin 1 / 2 vs 2 / 1 (2.1.48)

$$\frac{1}{2}\frac{\mathbb{C}e_\pm\hbar}{2m_e v_{\varepsilon yr}} = chain_r\ H_{cr}$$

The proton to electron mass ratio $(m_p/m_e)$ is verified through the Werner Heisenberg Uncertainty Principle volume $(\Delta x\cdot\Delta y\cdot\Delta z)$ as follows in equations 2.1.33 - 2.1.38 (2.1.33r - 2.1.38r):

$$\Delta x = \frac{\hbar}{2}\frac{1}{\Delta p_x} = \frac{\hbar}{2}\frac{1}{\Delta(\hbar K_B)} = \Delta\left(\frac{B}{2}\frac{1}{\pi}\right) \qquad (2.1.49)$$

$$\Delta x_r = \frac{\hbar}{2}\frac{1}{\Delta p_{xr}} = \frac{\hbar}{2}\frac{1}{\Delta(\hbar K_{Br})} = \Delta\left(\frac{B_r}{2}\frac{1}{\pi}\right) \qquad (2.1.49r)$$

$$\Delta y = \frac{\hbar}{2}\frac{1}{\Delta p_y} = \frac{\hbar}{2}\frac{1}{\Delta(\hbar K_P)} = \Delta\left(\frac{3B}{2\sqrt{3}}\frac{1}{2\pi}\right) \qquad (2.1.50)$$

$$\Delta y_r = \frac{\hbar}{2}\frac{1}{\Delta p_{yr}} = \frac{\hbar}{2}\frac{1}{\Delta(\hbar K_{Pr})} = \Delta\left(\frac{3B_r}{2\sqrt{3}}\frac{1}{2\pi}\right) \qquad (2.1.50r)$$

$$\Delta z = \frac{\hbar}{2}\frac{1}{\Delta p_z} = \frac{\hbar}{2}\frac{1}{\Delta(\hbar K_A)} = \Delta\left(\frac{A}{2}\frac{1}{2\pi}\right) \qquad (2.1.51)$$

$$\Delta z_r = \frac{\hbar}{2}\frac{1}{\Delta p_{zr}} = \frac{\hbar}{2}\frac{1}{\Delta(\hbar K_{Ar})} = \Delta\left(\frac{A_r}{2}\frac{1}{2\pi}\right) \qquad (2.1.51r)$$

$$(\Delta x\cdot\Delta y\cdot\Delta z) = \frac{1}{(2K_B)(2K_P)(2K_A)} \qquad (2.1.52)$$

$$(\Delta x_r\cdot\Delta y_r\cdot\Delta z_r) = \frac{1}{(2K_{Br})(2K_{Pr})(2K_{Ar})} \qquad (2.1.52r)$$

$$\frac{m_T}{m_e} = \frac{3}{4}\frac{1}{g_s}\left(\frac{cavity}{(\Delta x\ \Delta y\ \Delta z)} - \frac{m_p}{m_e}\right) \qquad (2.1.53)$$

$$\frac{m_T}{m_e} = \frac{3}{4}\frac{1}{g_s}\left(\frac{cavity_r}{(\Delta x_r\ \Delta y_r\ \Delta z_r)} - \frac{m_p}{m_e}\right) \qquad (2.1.53r)$$

Equations 2.1.49 - 2.1.53 (2.1.49r - 2.1.53r) provides us with the confidence ( $\Delta x\cdot\Delta y\cdot\Delta z$ is well contained within the *cavity* ) ( $\Delta x_r\cdot\Delta y_r\cdot\Delta z_r$ is well contained within the $cavity_r$ ) that we can proceed in a semi-classical manner outside the envelop of the uncertainty principle with the resonant transformed mass $(m_t)$ and de Broglie velocities $(v_d)$ used in this trisine superconductor model or in other words equation 2.1.54 holds.

$$\mathbb{C}\left(\frac{\hbar^2(KK)}{2m_T}\right) = \mathbb{C}\left(\frac{1}{2}m_T v_d^2\right) = \frac{(m_T v_d)^2}{m_T} \qquad (2.1.54)$$

$$\mathbb{C}\left(\frac{\hbar^2(K_r K_r)}{2m_T}\right) = \mathbb{C}\left(\frac{1}{2}m_T v_{dr}^2\right) = \frac{(m_T v_{dr})^2}{m_T} \qquad (2.1.54)$$

The Meissner condition as defined by total magnetic field exclusion from trisine chain as indicated by diamagnetic susceptibility $(\mathrm{X}) = -1/4\pi$, as per the following equation:

$$\mathrm{X} = -\frac{\mathbb{C}\ k_b T_c}{chain\ H_c^2} = -\frac{1}{4\pi} \qquad (2.1.55)$$

Equation 2.1.57 provides a representation of the trisine geometry in terms of relativistic Lorentz invariant relationship that is consistent with the Majorana condition at near relativity trisine conditions 1.3 and 1.4.

$$\left(\frac{1}{2}mv^2\right)^2 = \left(m\varsigma c\right)^2 + \left(mc^2\right)^2$$

generally if $\left(m \text{ and } c\right) \neq 0$ then:

$$\varsigma = \left\{ \begin{matrix} \dfrac{+i\sqrt{2c^2 - v^2}\sqrt{2c^2 + v^2}}{2c} \\ \dfrac{-i\sqrt{2c^2 - v^2}\sqrt{2c^2 + v^2}}{2c} \end{matrix} \right\} \tag{2.1.56}$$

and $m$ compatible with

$m_T \; = \; \pi\hbar \; time_{\pm} / \left(2B^2\right)$ or $\pi\hbar / D_c$

and $v$ compatible with

$v_{dx}, \; v_{dy}, \; v_{dz}, \; v_{\varepsilon x}, \; v_{\varepsilon y}, \; v_{\varepsilon z}$

to the trisine resonant condition:

$$\frac{1}{2}m_T v_{dx}^2 = \frac{1}{2}m_T v_{dx}^2$$

and $m$ compatible with

$m_T \; = \; \pi\hbar \; time_{r\pm} / \left(2B_r^2\right)$ or $\pi\hbar / D_c$

and $v$ compatible with

$v_{dxr}, \; v_{dyr}, \; v_{zr}, \; v_{\varepsilon x}, \; v_{\varepsilon y}, \; v_{\varepsilon z}$

to the trisine resonant condition:

$$\frac{1}{2}m_T v_{dxr}^2 = \frac{1}{2}m_T v_{dxr}^2$$

The relativity relationship is further explained in terms of the Einstein tensor $G_{\mu\nu}$ and stress energy tensor $T_{\mu\nu}$ relationship, equation 3.22.

**Table 2.1.2** with $\left(m_T\right)$, $\left(m_r\right)$. Angular frequency $\left(2k_b T_c / \hbar\right)$ $\left(\omega\right)$ & Wave Length $\left(2\pi c/\omega\right)$ or $\left(c \cdot time\right)$ with plot (Conceptually this is the Schwinger limit for virtual particles).

| Condition | 1.2 | 1.3 | 1.6 | 1.9 | 1.10 | 1.11 |
|---|---|---|---|---|---|---|
| $T_a\left(^0K\right)$ | 6.53E+11 | 6.53E+11 | 6.53E+11 | 6.53E+11 | 6.53E+11 | 6.53E+11 |
| $T_b\left(^0K\right)$ | 9.05E+14 | 5.48E+13 | 9.22E+08 | 1.10E+07 | 2,868 | 2.723 |
| $T_{br}\left(^0K\right)$ | 3.30E+12 | 5.45E+13 | 3.24E+18 | 2.72E+20 | 1.04E+24 | 1.10E+27 |
| $T_c\left(^0K\right)$ | 8.96E+13 | 3.28E+11 | 93.0 | 0.0131 | 8.99E-10 | 8.10E-16 |
| $T_{cr}\left(^0K\right)$ | 1.19E+09 | 3.25E+11 | 1.15E+21 | 8.11E+24 | 1.19E+32 | 1.32E+38 |
| $z_R + 1$ | 3.67E+43 | 8.14E+39 | 3.89E+25 | 6.53E+19 | 1.17E+09 | 1.00E+00 |
| $z_B + 1$ | 3.32E+14 | 2.01E+13 | 3.39E+08 | 4.03E+06 | 1.05E+03 | 1.00E+00 |
| $Age_{UG}\left(sec\right)$ | 7.14E-05 | 4.79E-03 | 6.94E+04 | 5.36E+07 | 1.27E+13 | 4.33E+17 |
| $Age_{UGv}\left(sec\right)$ | 4.59E-10 | 1.25E-07 | 4.42E+02 | 3.13E+06 | 1.22E+13 | 4.33E+17 |
| $Age_U\left(sec\right)$ | 1.18E-26 | 5.31E-23 | 1.11E-08 | 6.63E-03 | 3.70E+08 | 4.33E+17 |
| $m_T \; (g)$ | 1.00E-25 | 1.00E-25 | 1.00E-25 | 1.00E-25 | 1.00E-25 | 1.00E-25 |
| $m_r \; (g)$ | 1.38E-23 | 5.04E-26 | 1.43E-35 | 2.02E-39 | 1.38E-46 | 1.24E-52 |
| $m_{rr} \; (g)$ | 1.83E-28 | 4.99E-26 | 1.76E-16 | 1.25E-12 | 1.82E-05 | 2.02E+01 |
| $\Delta x\,(cm)$ | 1.06E-14 | 6.62E-09 | 1.75E-08 | 8.13E-07 | 3.20E-03 | 3.52 |
| $\Delta y\,(cm)$ | 9.17E-15 | 5.73E-09 | 1.52E-08 | 7.04E-07 | 2.77E-03 | 3.05 |
| $\Delta z\,(cm)$ | 2.22E-15 | 1.39E-09 | 3.68E-09 | 1.71E-07 | 6.73E-04 | 0.74 |
| $\Delta x\,(cm)$ | 2.90E-12 | 1.76E-13 | 2.96E-18 | 3.52E-20 | 9.20E-24 | 8.74E-27 |
| $\Delta y\,(cm)$ | 2.52E-12 | 1.52E-13 | 2.56E-18 | 3.05E-20 | 7.97E-24 | 7.57E-27 |
| $\Delta z\,(cm)$ | 6.10E-13 | 3.69E-14 | 6.22E-19 | 7.39E-21 | 1.93E-24 | 1.84E-27 |
| $\omega$ (rad/sec) | 2.35E+25 | 8.59E+22 | 2.44E+13 | 3.44E+09 | 2.35E+02 | 2.12E-04 |
| Wave Length (cm) | 8.03E-15 | 2.19E-12 | 7.74E-03 | 5.48E+01 | 8.00E+08 | 8.88E+14 |
| $\omega$ (rad/sec) | 3.12E+20 | 8.51E+22 | 3.00E+32 | 2.12E+36 | 3.10E+43 | 3.44E+49 |
| Wave Length (cm) | 6.05E-10 | 2.21E-12 | 6.28E-22 | 8.87E-26 | 6.07E-33 | 5.47E-39 |

Where $\left(v_d\right)$ is the de Broglie velocity as in equation 2.1.41 and also related to resonant CPT $time_{\pm}$ by equation 2.1.42.

$$v_{dx} = \frac{\hbar}{m_T}\frac{\pi}{B} \qquad \left(1+z_R\right)^{1/3} \text{ or } \left(1+z_B\right)^{1} \tag{2.1.57}$$

$$v_{dxr} = \frac{\hbar}{m_T}\frac{\pi}{B_r} \qquad \left(1+z_R\right)^{-1/3} \text{ or } \left(1+z_B\right)^{-1} \tag{2.1.57r}$$

$$time = \frac{2 \cdot B}{v_{dx}} \qquad \left(1+z_R\right)^{-2/3} \text{ or } \left(1+z_B\right)^{-2} \tag{2.1.58}$$

$$time_r = \frac{2 \cdot B_r}{v_{dxr}} \qquad \left(1+z_R\right)^{2/3} \text{ or } \left(1+z_B\right)^{2} \tag{2.1.58r}$$

The relationship between $\left(v_{dx}\right)$ and $\left(v_{xr}\right)$ is per the group/phase velocity relationship presented in equation 2.1.60.

$$\left(group\ velocity\right) \cdot \left(phase\ velocity\right) = c^2$$

$$\begin{gathered} v_{dx}v_{dxr} = v_{dx}\frac{c^2}{v_{dx}} = c^2 \\ v_{dy}v_{dyr} \neq v_{dy}\frac{c^2}{v_{dy}} \neq c^2 \\ v_{dz}v_{dzr} \neq v_{dz}\frac{c^2}{v_{dz}} \neq c^2 \end{gathered} \tag{2.1.59}$$

dielectric (*group velocity*)·(*phase velocity*) = $c^2$

$$\begin{aligned} v_{ex}v_{exr} &= v_{ex}\frac{c^2}{v_{ex}} = c^2 \\ v_{ey}v_{eyr} &= v_{ey}\frac{c^2}{v_{ey}} = c^2 \\ v_{ez}v_{ezr} &\neq v_{ez}\frac{c^2}{v_{ez}} \neq c^2 \end{aligned} \qquad (2.1.60)$$

And the Phase velocity $v_{dxr}$ is related to universal Hubble radius of curvature dimension $R_U$ over cavity resonance time as well as the speed of light $c$ over the *cavity* resonance velocity $v_{dx}$. At present, the phase velocity $v_{dxr}$ is much greater than $(v_{dx})$ and $(v_{\varepsilon x},\ v_{\varepsilon y}\ \&\ v_{\varepsilon z})$ but in the Big Bang nucleosynthetic super luminal condition, the phase velocity $v_{dxr}$ was less than $(v_{dx})$ and $(v_{\varepsilon x},\ v_{\varepsilon y}\ \&\ v_{\varepsilon z})$.

**Table 2.1.3** Velocity $v_{dx}$ phase velocity $v_{dxr}$

| | | | | | | |
|---|---|---|---|---|---|---|
| Condition | 1.2 | 1.3 | 1.6 | 1.9 | 1.10 | 1.11 |
| $T_a\left({}^0K\right)$ | 6.53E+11 | 6.53E+11 | 6.53E+11 | 6.53E+11 | 6.53E+11 | 6.53E+11 |
| $T_b\left({}^0K\right)$ | 9.05E+14 | 5.48E+13 | 9.22E+08 | 1.10E+07 | 2,868 | 2.723 |
| $T_{br}\left({}^0K\right)$ | 3.30E+12 | 5.45E+13 | 3.24E+18 | 2.72E+20 | 1.04E+24 | 1.10E+27 |
| $T_c\left({}^0K\right)$ | 8.96E+13 | 3.28E+11 | 93.0 | 0.0131 | 8.99E-10 | 8.10E-16 |
| $T_{cr}\left({}^0K\right)$ | 1.19E+09 | 3.25E+11 | 1.15E+21 | 8.11E+24 | 1.19E+32 | 1.32E+38 |
| $z_R+1$ | 3.67E+43 | 8.14E+39 | 3.89E+25 | 6.53E+19 | 1.17E+09 | 1.00E+00 |
| $z_B+1$ | 3.32E+14 | 2.01E+13 | 3.39E+08 | 4.03E+06 | 1.05E+03 | 1.00E+00 |
| $Age_{UG}(sec)$ | 7.14E-05 | 4.79E-03 | 6.94E+04 | 5.36E+07 | 1.27E+13 | 4.33E+17 |
| $Age_{UGv}(sec)$ | 4.59E-10 | 1.25E-07 | 4.42E+02 | 3.13E+06 | 1.22E+13 | 4.33E+17 |
| $Age_U(sec)$ | 1.18E-26 | 5.31E-23 | 1.11E-08 | 6.63E-03 | 3.70E+08 | 4.33E+17 |
| $v_{dx}(cm/\sec)$ | 4.97E+11 | 3.00E+10 | 5.06E+05 | 6.01E+03 | 1.57E+00 | 1.49E-03 |
| $v_{dxr}(cm/\sec)$ | 1.81E+09 | 2.99E+10 | 1.78E+15 | 1.49E+17 | 5.71E+20 | 6.02E+23 |

## 2.2 Trisine Geometry

The characteristic trisine resonant dimensions A and B are indicated as a function of temperature in table 2.2.1 as computed from the equation 2.2.1

$$k_bT_c = \frac{\hbar^2K_B^2}{2m_T} = \frac{\hbar^2K_A^2}{2m_T}\left(\frac{B}{A}\right)_T^2 \qquad (2.2.1)$$

$$k_bT_{cr} = \frac{\hbar^2K_{Br}^2}{2m_T} = \frac{\hbar^2K_{Ar}^2}{2m_T}\left(\frac{B_r}{A_r}\right)_T^2 \qquad (2.2.1r)$$

Trisine geometry is subject to Charge conjugation Parity change Time reversal (CPT) symmetry which is a fundamental symmetry of physical laws under transformations that involve the inversions of charge, parity and time simultaneously. The CPT symmetry provides the basis for creating particular resonant structures with this trisine geometrical character.

**Table 2.2.1** Dimensions $B$ and $A$ with $B$ vs. $T_c$ plot

| | | | | | | |
|---|---|---|---|---|---|---|
| Condition | 1.2 | 1.3 | 1.6 | 1.9 | 1.10 | 1.11 |
| $T_a\left({}^0K\right)$ | 6.53E+11 | 6.53E+11 | 6.53E+11 | 6.53E+11 | 6.53E+11 | 6.53E+11 |
| $T_b\left({}^0K\right)$ | 9.05E+14 | 5.48E+13 | 9.22E+08 | 1.10E+07 | 2,868 | 2.723 |
| $T_{br}\left({}^0K\right)$ | 3.30E+12 | 5.45E+13 | 3.24E+18 | 2.72E+20 | 1.04E+24 | 1.10E+27 |
| $T_c\left({}^0K\right)$ | 8.96E+13 | 3.28E+11 | 93.0 | 0.0131 | 8.99E-10 | 8.10E-16 |
| $T_{cr}\left({}^0K\right)$ | 1.19E+09 | 3.25E+11 | 1.15E+21 | 8.11E+24 | 1.19E+32 | 1.32E+38 |
| $z_R+1$ | 3.67E+43 | 8.14E+39 | 3.89E+25 | 6.53E+19 | 1.17E+09 | 1.00E+00 |
| $z_B+1$ | 3.32E+14 | 2.01E+13 | 3.39E+08 | 4.03E+06 | 1.05E+03 | 1.00E+00 |
| $Age_{UG}(sec)$ | 7.14E-05 | 4.79E-03 | 6.94E+04 | 5.36E+07 | 1.27E+13 | 4.33E+17 |
| $Age_{UGv}(sec)$ | 4.59E-10 | 1.25E-07 | 4.42E+02 | 3.13E+06 | 1.22E+13 | 4.33E+17 |
| $Age_U(sec)$ | 1.18E-26 | 5.31E-23 | 1.11E-08 | 6.63E-03 | 3.70E+08 | 4.33E+17 |
| $A\ (cm)$ | 2.80E-14 | 4.62E-13 | 2.74E-08 | 2.31E-06 | 8.82E-03 | 9.29 |
| $B\ (cm)$ | 6.65E-14 | 1.10E-12 | 6.53E-08 | 5.49E-06 | 2.10E-02 | 2.21E+01 |
| $A_r\ (cm)$ | 7.67E-12 | 4.64E-13 | 7.82E-18 | 9.29E-20 | 2.43E-23 | 2.31E-26 |
| $B_r\ (cm)$ | 1.83E-11 | 1.10E-12 | 1.86E-17 | 2.21E-19 | 5.78E-23 | 5.49E-26 |

These structures have an analog in various resonant structures postulated in the field of chemistry to explain properties of delocalized $\pi$ electrons in benzene, ozone and a myriad of other molecules and incorporate the Noether theorem concepts of energy – time and momentum – length symmetry.

Trisine characteristic volumes with variable names *cavity* and *chain* as well as characteristic areas with variable names *section*, *approach* and *side* are defined in equations 2.2.7 - 2.2.9. See figures 2.2.1 - 2.2.4 for a visual description of these parameters. The mirror image forms in figure 2.2.1 are in conformance with parity requirement in Charge conjugation Parity change Time reversal (CPT) theorem as established by determinant identity in equation 2.1.6b and trivially obvious in cross – product expression in equation 2.2.2 in which case an orthogonal coordinate system is right-handed, or left-handed for result 2*AB* which is later defined as trisine *cavity side* in equation 2.2.9.

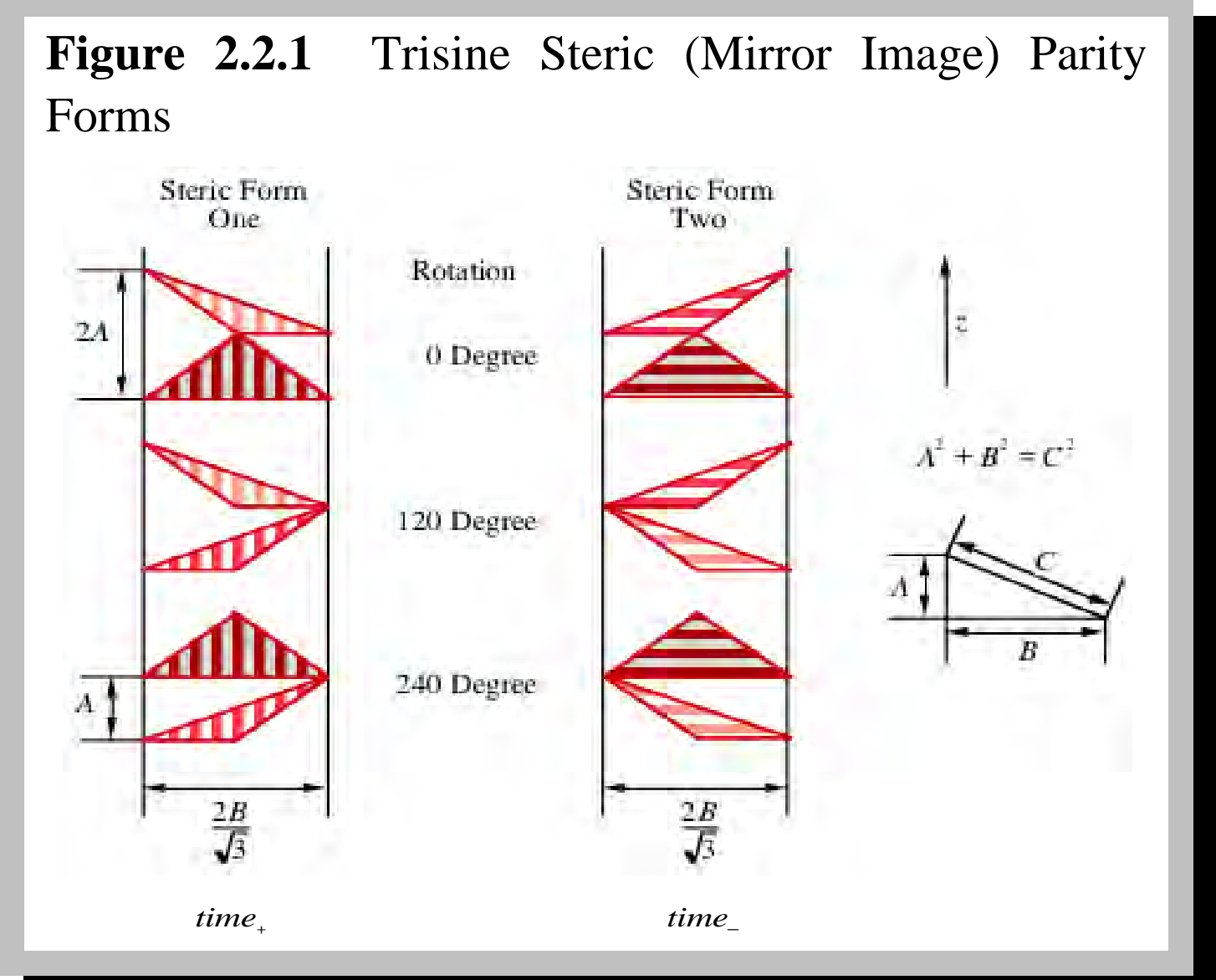

**Figure 2.2.1** Trisine Steric (Mirror Image) Parity Forms

Essentially, there are two mirror image *side*s to each *cavity*. At the nuclear scale (condition 1.2), such chiral geometry is consistent with the standard model (quarks, gluons etc.).

$$(2B)\times\left(\sqrt{A^2+B^2}\right) = 2B\sqrt{A^2+B^2}\,\sin(\theta) = 2AB \qquad (2.2.2)$$

$$(2B_r)\times\left(\sqrt{A_r^2+B_r^2}\right) = 2B\sqrt{A_r^2+B_r^2}\,\sin(\theta) = 2A_rB_r \qquad (2.2.2r)$$

$$cavity = 2\sqrt{3}AB^2 = \frac{4\pi Age_U m_T}{U} \qquad (2.2.3)$$

$$cavity_r = 2\sqrt{3}A_rB_r^2 = \frac{4\pi Age_{Ur} m_T}{U} \qquad (2.2.3r)$$

$$chain = \frac{2}{3}cavity \qquad (2.2.4)$$

$$chain_r = \frac{2}{3}cavity_r \qquad (2.2.4r)$$

$A$ and $B$ are proportional to the permeability $k_m$ and dielectric $\varepsilon$ as defined in section 2.5 as well as the trisine length $(l_T)$ and other constant factors.

$$A = \left(\frac{A}{B}\right)_T \pi l_T \frac{c}{v_{dx}} = \pi l_T\sqrt{k_{mx}\varepsilon_x} \qquad (2.2.5)$$

$$A_r = \left(\frac{A_r}{B_r}\right)_T \pi l_T \frac{c}{v_{dxr}} = \pi l_T\sqrt{k_{mxr}\varepsilon_{xr}} \qquad (2.2.5r)$$

$$B = \pi l_T \frac{c}{v_{dx}} = \pi l_T\sqrt{k_{mx}\varepsilon_x} \qquad (2.2.6)$$

$$B_r = \pi l_T \frac{c}{v_{dxr}} = \pi l_T\sqrt{k_{mxr}\varepsilon_{xr}} \qquad (2.2.6r)$$

Or in terms of the universal dimensional scale in the context of the trisine base length $(l_T)$.

$$B = \pi\hbar^{\frac{1}{3}}U^{\frac{-1}{3}}c^{\frac{-2}{3}}\left(\frac{c}{v_{dx}}\right) = \pi l_T\left(\frac{c}{v_{dx}}\right) \qquad (2.2.7)$$

$$B_r = \pi\hbar^{\frac{1}{3}}U^{\frac{-1}{3}}c^{\frac{-2}{3}}\left(\frac{c}{v_{dxr}}\right) = \pi l_T\left(\frac{c}{v_{dxr}}\right) \qquad (2.2.7r)$$

Or in terms of the Kolmogorov viscosity length scale[60]

$$B = \left(\frac{1}{\sqrt{2}}\right)\left(\left(\frac{B}{A}\right)_T^3 \frac{m_T\nu_U time_\pm}{k_bT_c}\right)^{\frac{1}{4}} \qquad (2.2.8)$$

$$B_r = \left(\frac{1}{\sqrt{2}}\right)\left(\left(\frac{B_r}{A_r}\right)_T^3 \frac{m_T\nu_U time_{r\pm}}{k_bT_{cr}}\right)^{\frac{1}{4}} \qquad (2.2.9r)$$

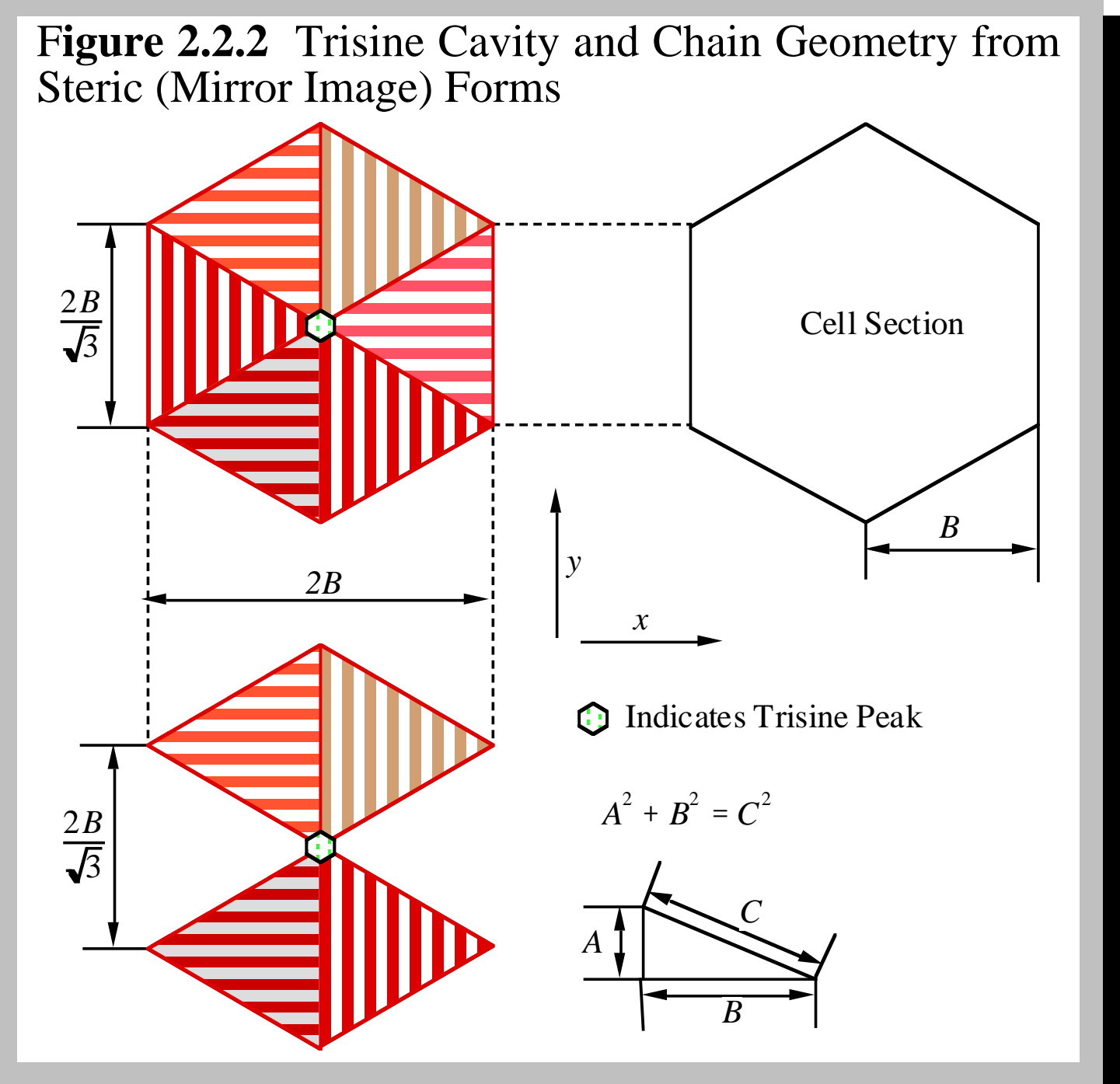

**Figure 2.2.2** Trisine Cavity and Chain Geometry from Steric (Mirror Image) Forms

The *section* is the trisine cell *cavity* projection on to the *x*, *y* plane as indicated in figure 2.2.2.

$$section = 2\sqrt{3}B^2 \qquad (2.2.10)$$

$$section_r = 2\sqrt{3}B_r^2 \qquad (2.2.10r)$$

The *approach* is the trisine cell *cavity* projection on to the *y*, *z* plane as indicated in figure 2.2.3.

$$approach = \frac{1}{2}\frac{3B}{\sqrt{3}}2A = \sqrt{3}AB \qquad (2.2.11)$$

$$approach_r = \frac{1}{2}\frac{3B_r}{\sqrt{3}}2A_r = \sqrt{3}A_rB_r \qquad (2.2.11r)$$

**Figure 2.2.3** Trisine *approach* from both *cavity* approaches

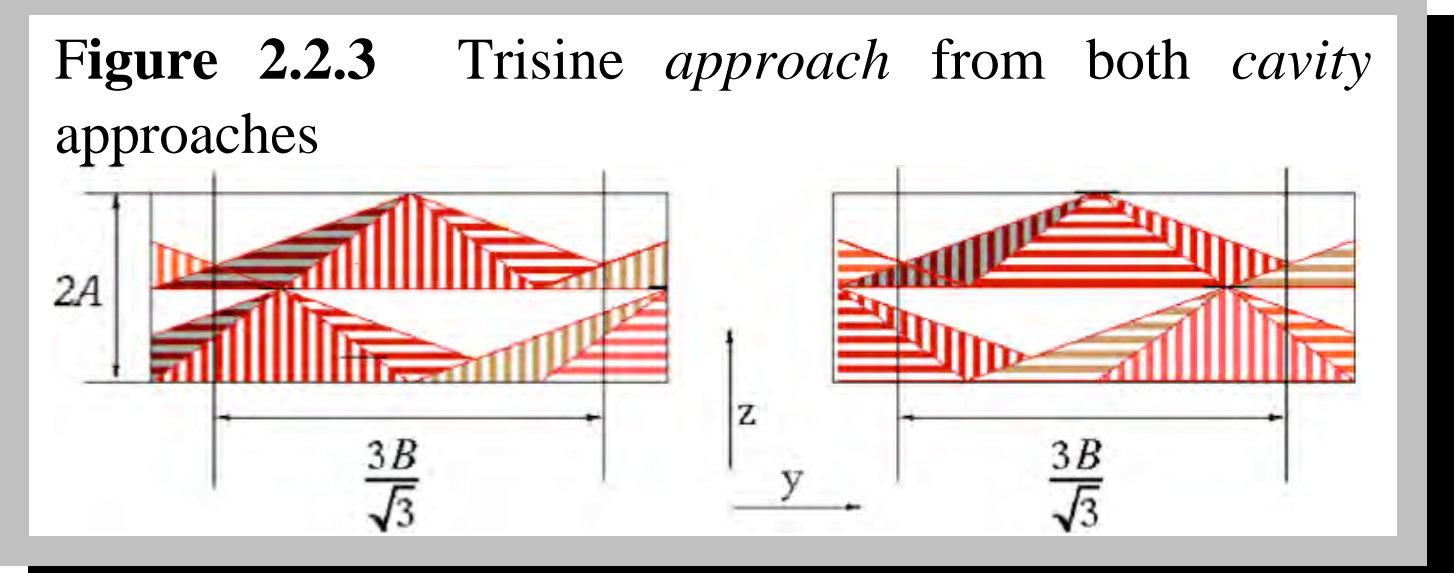

The *side* is the trisine cell *cavity* projection on to the *x*, *z* plane as indicated in figure 2.2.4.

$$side = \frac{1}{2}2A\ 2B = 2AB \qquad (2.2.12)$$

$$side_r = \frac{1}{2}2A_r\ 2B_r = 2A_rB_r \qquad (2.2.12r)$$

**Figure 2.2.4** Trisine *side* view from both *cavity* sides

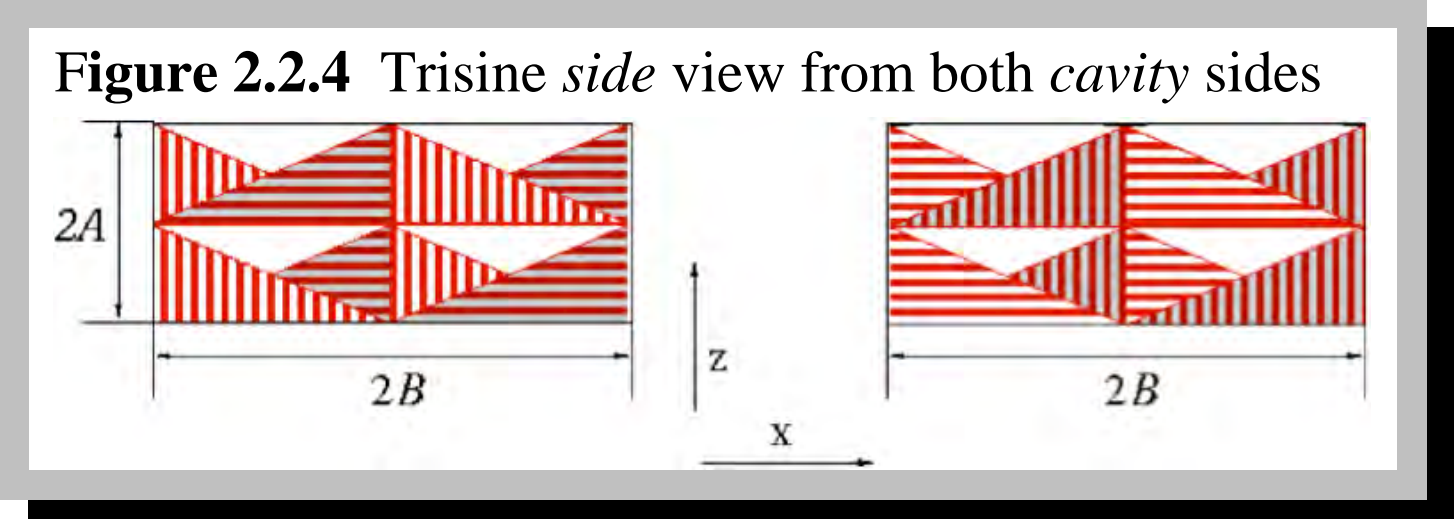

**Table 2.2.2** Trisine Cavity and Section with plot

| Condition | 1.2 | 1.3 | 1.6 | 1.9 | 1.10 | 1.11 |
|---|---|---|---|---|---|---|
| $T_a\left({}^0K\right)$ | 6.53E+11 | 6.53E+11 | 6.53E+11 | 6.53E+11 | 6.53E+11 | 6.53E+11 |
| $T_b\left({}^0K\right)$ | 9.05E+14 | 5.48E+13 | 9.22E+08 | 1.10E+07 | 2,868 | 2.723 |
| $T_{br}\left({}^0K\right)$ | 3.30E+12 | 5.45E+13 | 3.24E+18 | 2.72E+20 | 1.04E+24 | 1.10E+27 |
| $T_c\left({}^0K\right)$ | 8.96E+13 | 3.28E+11 | 93.0 | 0.0131 | 8.99E-10 | 8.10E-16 |
| $T_{cr}\left({}^0K\right)$ | 1.19E+09 | 3.25E+11 | 1.15E+21 | 8.11E+24 | 1.19E+32 | 1.32E+38 |
| $z_R+1$ | 3.67E+43 | 8.14E+39 | 3.89E+25 | 6.53E+19 | 1.17E+09 | 1.00E+00 |
| $z_B+1$ | 3.32E+14 | 2.01E+13 | 3.39E+08 | 4.03E+06 | 1.05E+03 | 1.00E+00 |
| $Age_{UG}\left(sec\right)$ | 7.14E-05 | 4.79E-03 | 6.94E+04 | 5.36E+07 | 1.27E+13 | 4.33E+17 |
| $Age_{UGv}\left(sec\right)$ | 4.59E-10 | 1.25E-07 | 4.42E+02 | 3.13E+06 | 1.22E+13 | 4.33E+17 |
| $Age_U\left(sec\right)$ | 1.18E-26 | 5.31E-23 | 1.11E-08 | 6.63E-03 | 3.70E+08 | 4.33E+17 |
| $cavity\left(cm^3\right)$ | 4.28E-40 | 1.93E-36 | 4.05E-22 | 2.41E-16 | 1.35E-05 | 15745 |
| $cavity_r\left(cm^3\right)$ | 8.85E-33 | 1.96E-36 | 9.36E-51 | 1.57E-56 | 2.81E-67 | 2.41E-76 |
| $chain\left(cm^3\right)$ | 2.86E-40 | 1.29E-36 | 2.70E-22 | 1.61E-16 | 8.98E-06 | 10497 |
| $chain_r\left(cm^3\right)$ | 5.90E-33 | 1.31E-36 | 6.24E-51 | 1.05E-56 | 1.88E-67 | 1.61E-76 |
| $section\left(cm^2\right)$ | 1.53E-26 | 4.19E-24 | 1.48E-14 | 1.05E-10 | 1.53E-03 | 1694 |
| $section_r\left(cm^2\right)$ | 1.15E-21 | 4.23E-24 | 1.20E-33 | 1.69E-37 | 1.16E-44 | 1.04E-50 |
| $approach\left(cm^2\right)$ | 3.22E-27 | 8.80E-25 | 3.10E-15 | 2.20E-11 | 3.21E-04 | 356 |
| $approach_r\left(cm^2\right)$ | 2.42E-22 | 8.88E-25 | 2.52E-34 | 3.56E-38 | 2.43E-45 | 2.19E-51 |
| $side\left(cm^2\right)$ | 3.72E-27 | 1.02E-24 | 3.58E-15 | 2.54E-11 | 3.71E-04 | 411 |
| $side_r\left(cm^2\right)$ | 2.80E-22 | 1.03E-24 | 2.91E-34 | 4.11E-38 | 2.81E-45 | 2.53E-51 |

**Figure 2.2.5** Trisine Geometry

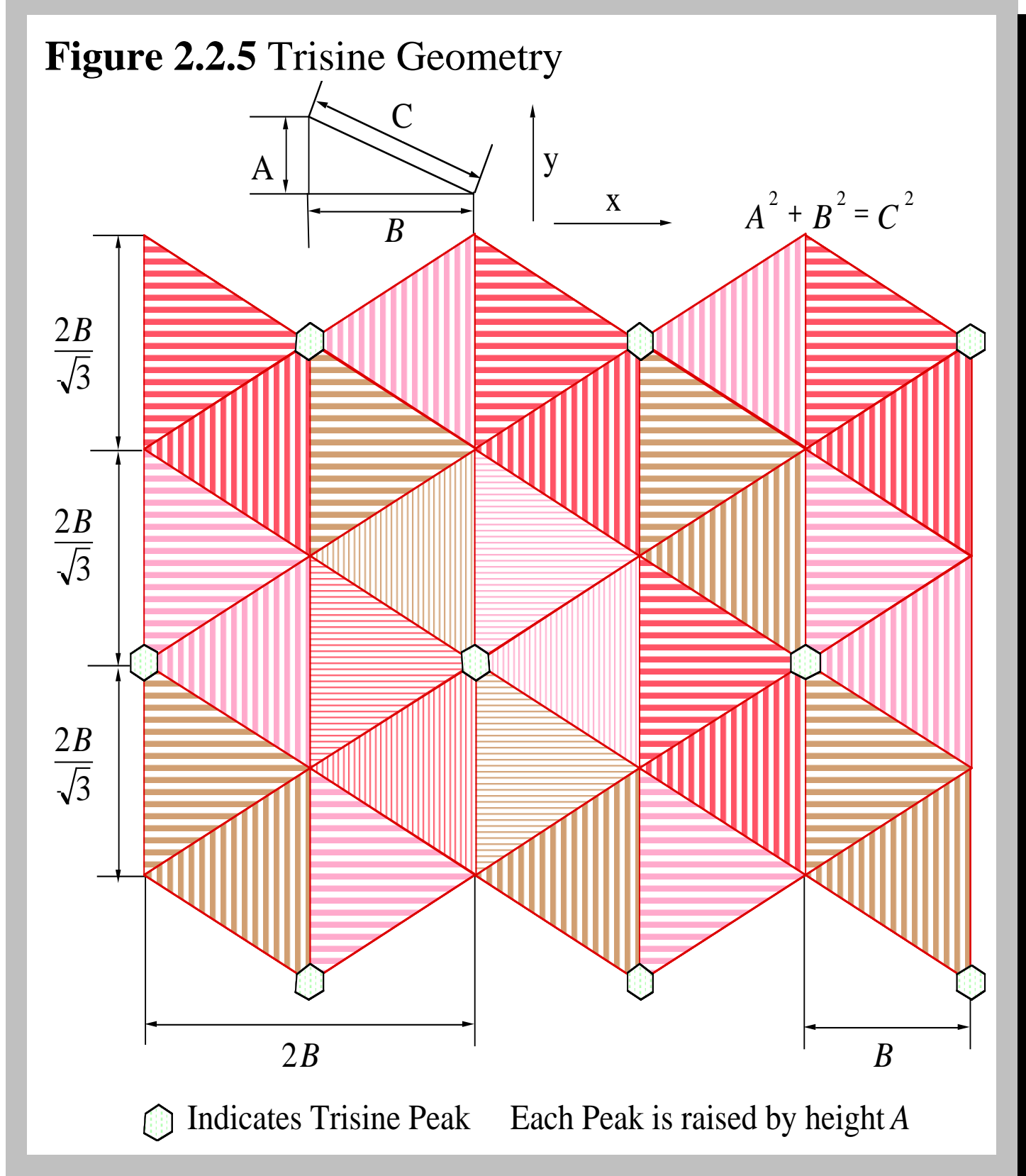

## 2.3 Trisine Characteristic Wave Vectors

Superconducting resonant model trisine characteristic wave vectors $K_{Dn}$, $K_{Ds}$ and $K_C$ are defined in equations 2.3.1 - 2.3.3 below where $K_{Dn}$ is the conventional Fermi surface vector.

$$K_{Dn} = \left(\frac{6\pi^2}{cavity}\right)^{\frac{1}{3}} = \left(\frac{3\pi^2\mathbb{C}}{cavity}\right)^{\frac{1}{3}} \qquad (2.3.1)$$

$$K_{Dnr} = \left(\frac{6\pi^2}{cavity_r}\right)^{\frac{1}{3}} = \left(\frac{3\pi^2\mathbb{C}}{cavity_r}\right)^{\frac{1}{3}} \qquad (2.3.1r)$$

The relationship in equation 2.3.1 represents the normal

Debye wave vector ($K_{Dn}$) assuming the *cavity* volume conforming to a sphere in $K$ space and also defined in terms of $K_A$, $K_P$ and $K_B$ as defined in equations 2.3.4, 2.3.5 and 2.3.6.

See Appendix B for the derivation of equation 2.3.1.

$$K_{Ds} = \left(\frac{8\pi^3}{cavity}\right)^{\frac{1}{3}} = \left(\frac{4\pi^3\mathbb{C}}{cavity}\right)^{\frac{1}{3}} = \left(K_A K_B K_P\right)^{\frac{1}{3}} \qquad (2.3.2)$$

$$K_{Ds} = \left(\frac{8\pi^3}{cavity}\right)^{\frac{1}{3}} = \left(\frac{4\pi^3\mathbb{C}}{cavity}\right)^{\frac{1}{3}} = \left(K_A K_B K_P\right)^{\frac{1}{3}} \qquad (2.3.2r)$$

The relationship in equation 2.3.2 represents the normal Debye wave vector($K_{Ds}$) assuming the *cavity* volume conforming to a characteristic trisine cell in $K$ space.

See Appendix B for the derivation of equation 2.3.2.

$$K_C = \frac{4\pi}{3\sqrt{3}A} \qquad (2.3.3)$$

$$K_{Cr} = \frac{4\pi}{3\sqrt{3}A_r} \qquad (2.3.3r)$$

The relationship in equation 2.3.3 represents the result of equating one-dimensional and trisine density of states. See Appendix C for the derivation of equation 2.3.3.

Equations 2.3.4, 2.3.5, and 2.3.6 translate the trisine wave vectors $K_C$, $K_{Ds}$, and $K_{Dn}$ ($K_{Cr}$, $K_{Dsr}$, and $K_{Dnr}$) into x, y, z Cartesian coordinates as represented by $K_B$, $K_P$ and $K_A$ ($K_{Br}$, $K_{Pr}$ and $K_{Ar}$ ) respectively. The superconducting current is in the x direction.

$$K_B = \frac{2\pi}{2B} \qquad (2.3.4)$$

$$K_{Br} = \frac{2\pi}{2B_r} \qquad (2.3.4r)$$

$$K_P = \frac{2\pi\sqrt{3}}{3B} \qquad (2.3.5)$$

$$K_{Pr} = \frac{2\pi\sqrt{3}}{3B_r} \qquad (2.3.5r)$$

$$K_A = \frac{2\pi}{A} \qquad (2.3.6)$$

$$K_{Ar} = \frac{2\pi}{A_r} \qquad (2.3.6r)$$

Note that the sorted energy relationships of wave vectors are as follows:

$$K_A^2 > K_C^2 > K_{Ds}^2 > K_P^2 > K_{Dn}^2 > K_B^2 \qquad (2.3.7)$$

$$K_{Ar}^2 > K_{Cr}^2 > K_{Dsr}^2 > K_{Pr}^2 > K_{Dnr}^2 > K_{Br}^2 \qquad (2.3.7r)$$

Equation 2.3.8 relates addition of wave function amplitudes ($B/\sqrt{3}$), ($P/2$), ($B/2$) in terms of superconducting resonate energy $\left(k_b T_c\right)$ congruent with Feynman[40].

$$k_b T_c = \frac{cavity}{chain}\,\frac{\hbar}{time_{\pm}}\begin{pmatrix} +\dfrac{K_B}{2B}\left(\dfrac{B}{\sqrt{3}}\right)^2 \\ +\dfrac{K_P}{2P}\left(\dfrac{P}{2}\right)^2 \\ +\dfrac{K_A}{2A}\left(\dfrac{A}{2}\right)^2 \end{pmatrix} \qquad (2.3.8)$$

$$k_b T_{cr} = \frac{cavity_r}{chain_r}\,\frac{\hbar}{time_{r\pm}}\begin{pmatrix} +\dfrac{K_{Br}}{2B}\left(\dfrac{B_r}{\sqrt{3}}\right)^2 \\ +\dfrac{K_{Pr}}{2P}\left(\dfrac{P_r}{2}\right)^2 \\ +\dfrac{K_{Ar}}{2A_r}\left(\dfrac{A_r}{2}\right)^2 \end{pmatrix} \qquad (2.3.8r)$$

The wave vector $\left(K_B\right)$ and $\left(K_{Br}\right)$, associated with the lowest energy $\left(\sim K_B^2\right)$ and $\left(\sim K_{Br}^2\right)$, is the carrier of the superconducting energy and in accordance with derivation in appendix A. All of the other wave vectors are contained within the cell *cavity* with many relationships with one indicated in 2.3.9 with the additional wave vector $\left(K_C\right)$ defined in 2.3.10

$$\frac{m_e m_T}{m_e + m_T}\frac{1}{K_B^2} + \frac{m_T m_p}{m_T + m_p}\frac{1}{K_A^2} = \frac{1}{6}g_s^2 m_t \frac{1}{K_C^2} \qquad (2.3.9)$$

$$K_C = \frac{1}{g_s}\frac{2\pi}{\sqrt{A^2 + B^2}} = \frac{4\pi}{3\sqrt{3}A} \qquad (2.3.10)$$

The wave vector $K_C$ is considered the hypotenuse vector because of its relationship to $K_A$ and $K_B$.

The conservation of momentum and energy relationships in 2.3.11, 2.3.12 and 2.1.22 result in a convergence to a trisine lattice $\left(B/A\right)_T$ or $\left(B_r/A_r\right)_T$ ratio of 2.379761271 and trisine mass $\left(m_T\right)$ 110.107178208 x electron mass $\left(m_e\right)$. Essentially a reversible process or reaction is defined recognizing that momentum is a vector and energy is a scalar.

Conservation Of Momentum

$$\sum_{n=B,C,Ds,Dn} \Delta p_n \Delta x_n = 0$$

$$g_s \left( \pm K_B \pm K_C \right) \rightleftarrows \pm K_{Ds} \pm K_{Dn} \qquad (2.3.11)$$

Conservation Of Momentum

$$\sum_{n=B_r,\, C_r,\, Dsr,\, Dnr} \Delta p_{rn} \Delta x_{rn} = 0$$

$$g_s \left( \pm K_{Br} \pm K_{Cr} \right) \rightleftarrows \pm K_{Dsr} \pm K_{Dnr} \qquad (2.3.11r)$$

Conservation Of Energy

$$\sum_{n=B,\, C,\, Ds,\, Dn} \Delta E_n \Delta t_n = 0$$

$$K_B^2 + K_C^2 = g_s \left( K_{Ds}^2 + K_{Dn}^2 \right) \qquad (2.3.12)$$

Conservation Of Energy

$$\sum_{n=B_r,\, C_r,\, Dsr,\, Dnr} \Delta E_{rn} \Delta t_{rn} = 0$$

$$K_{Br}^2 + K_{Cr}^2 = g_s \left( K_{Dsr}^2 + K_{Dnr}^2 \right) \qquad (2.3.12r)$$

The momentum and energy equations 2.3.11 and 2.3.12 solutions are exact and independent of *B* or *A* and totally dependent on the ratio *B/A*.

The momentum ratio is then expressed as:

$$\frac{g_s}{9} \pi \left( 9 + 4\sqrt{3} \left( \frac{B}{A} \right)_T \right) = \left( \frac{B}{A} \right)_T^{\frac{1}{3}} \left( 3 + 6^{\frac{2}{3}} \pi^{\frac{1}{3}} \right) \pi^{\frac{2}{3}} 3^{-\frac{5}{6}} \qquad (2.3.13)$$

$$\frac{g_s}{9} \pi \left( 9 + 4\sqrt{3} \left( \frac{B_r}{A_r} \right)_T \right) = \left( \frac{B_r}{A_r} \right)_T^{\frac{1}{3}} \left( 3 + 6^{\frac{2}{3}} \pi^{\frac{1}{3}} \right) \pi^{\frac{2}{3}} 3^{-\frac{5}{6}} \qquad (2.3.13r)$$

The energy is then expressed as:

$$\left( 27 + 16 \left( \frac{B}{A} \right)_T^2 \right) \pi^{\frac{2}{3}} = 9 \cdot 3^{\frac{1}{3}} \left( \frac{B}{A} \right)_T^{\frac{2}{3}} g_s \left( 3 + 2 \cdot 6^{\frac{1}{3}} \pi^{\frac{2}{3}} \right) \qquad (2.3.14)$$

$$\left( 27 + 16 \left( \frac{B_r}{A_r} \right)_T^2 \right) \pi^{\frac{2}{3}} = 9 \cdot 3^{\frac{1}{3}} \left( \frac{B_r}{A_r} \right)_T^{\frac{2}{3}} g_s \left( 3 + 2 \cdot 6^{\frac{1}{3}} \pi^{\frac{2}{3}} \right) \qquad (2.3.14r)$$

Representing an exact solution of two simultaneous equations with

Trisine size ratio $\left( B/A \right)_T$ or $\left( B_r/A_r \right)_T$

$= 2.379146658937169267\ldots$

Superconductorgyro-factor $g_s = 1.0098049781877999262\ldots$

Dirac number $g_d = 1 + \frac{1}{2\sqrt{3}} \left( \frac{A}{B} \right)_T \frac{m_e}{m_T} = 1.00110197701263$

Another perspective of this resonant energy and momentum convergence is contained in Appendix A.

Numerical values of wave vectors are listed in Table 2.3.1.

The visible reflective luminal resonant discontinuity ($v_{dx} \sim c$) occurs at the critical temperature $T_c \sim 1.3E12$ K, $K_B \sim 5.7E12$ $cm^{-1}$ and B ~ 5.5E-13 cm. This discontinuity forms potential barrier providing a probable mechanism for limiting nuclear size.

**Table 2.3.1** Trisine wave vectors

| | | | | | | |
|---|---|---|---|---|---|---|
| Condition | 1.2 | 1.3 | 1.6 | 1.9 | 1.10 | 1.11 |
| $T_a\left({}^0K\right)$ | 6.53E+11 | 6.53E+11 | 6.53E+11 | 6.53E+11 | 6.53E+11 | 6.53E+11 |
| $T_b\left({}^0K\right)$ | 9.05E+14 | 5.48E+13 | 9.22E+08 | 1.10E+07 | 2,868 | 2.723 |
| $T_{br}\left({}^0K\right)$ | 3.30E+12 | 5.45E+13 | 3.24E+18 | 2.72E+20 | 1.04E+24 | 1.10E+27 |
| $T_c\left({}^0K\right)$ | 8.96E+13 | 3.28E+11 | 93.0 | 0.0131 | 8.99E-10 | 8.10E-16 |
| $T_{cr}\left({}^0K\right)$ | 1.19E+09 | 3.25E+11 | 1.15E+21 | 8.11E+24 | 1.19E+32 | 1.32E+38 |
| $z_R + 1$ | 3.67E+43 | 8.14E+39 | 3.89E+25 | 6.53E+19 | 1.17E+09 | 1.00E+00 |
| $z_B + 1$ | 3.32E+14 | 2.01E+13 | 3.39E+08 | 4.03E+06 | 1.05E+03 | 1.00E+00 |
| $Age_{UG}\left(sec\right)$ | 7.14E-05 | 4.79E-03 | 6.94E+04 | 5.36E+07 | 1.27E+13 | 4.33E+17 |
| $Age_{UGv}\left(sec\right)$ | 4.59E-10 | 1.25E-07 | 4.42E+02 | 3.13E+06 | 1.22E+13 | 4.33E+17 |
| $Age_U\left(sec\right)$ | 1.18E-26 | 5.31E-23 | 1.11E-08 | 6.63E-03 | 3.70E+08 | 4.33E+17 |
| $K_B$ (/cm) | 4.72E+13 | 2.86E+12 | 4.81E+07 | 5.72E+05 | 1.50E+02 | 1.42E-01 |
| $K_{Br}$ (/cm) | 1.72E+11 | 2.84E+12 | 1.69E+17 | 1.42E+19 | 5.43E+22 | 5.72E+25 |
| $K_{Dn}$ (/cm) | 5.17E+13 | 3.13E+12 | 5.27E+07 | 6.26E+05 | 1.64E+02 | 1.56E-01 |
| $K_{Dnr}$ (/cm) | 1.88E+11 | 3.11E+12 | 1.85E+17 | 1.56E+19 | 5.95E+22 | 6.26E+25 |
| $K_{Ds}$ (/cm) | 8.33E+13 | 5.04E+12 | 8.49E+07 | 1.01E+06 | 2.64E+02 | 2.51E-01 |
| $K_{Dsr}$ (/cm) | 3.04E+11 | 5.02E+12 | 2.98E+17 | 2.51E+19 | 9.59E+22 | 1.01E+26 |
| $K_C$ (/cm) | 8.65E+13 | 5.23E+12 | 8.81E+07 | 1.05E+06 | 2.74E+02 | 2.60E-01 |
| $K_{Cr}$ (/cm) | 3.15E+11 | 5.21E+12 | 3.09E+17 | 2.60E+19 | 9.95E+22 | 1.05E+26 |
| $K_P$ (/cm) | 5.45E+13 | 3.30E+12 | 5.56E+07 | 6.60E+05 | 1.73E+02 | 1.64E-01 |
| $K_{Pr}$ (/cm) | 1.99E+11 | 3.28E+12 | 1.95E+17 | 1.64E+19 | 6.27E+22 | 6.61E+25 |
| $K_A$ (/cm) | 2.25E+14 | 1.36E+13 | 2.29E+08 | 2.72E+06 | 7.12E+02 | 6.76E-01 |
| $K_{Ar}$ (/cm) | 8.19E+11 | 1.35E+13 | 8.04E+17 | 6.76E+19 | 2.59E+23 | 2.72E+26 |

Negative and positive charge reversal as well as $time_\pm$ reversal takes place in mirror image parity pairs as is more visually seen in perspective presented in Figure 2.3.2.

It is important to reiterate that these triangular lattice representations are virtual and equivalent to a virtual reciprocal lattice but as geometric entities, do describe and correlate the physical data quite well.

The *cavity* arranged in a space filling lattice may be considered 3D tessellated throughout the universe and congruent with the 'dark energy' concept scaled back to Big Bang nucleosynthesis.

**Figure 2.3.1** Trisine Steric Charge Conjugate Pair Change Time Reversal CPT Geometry.

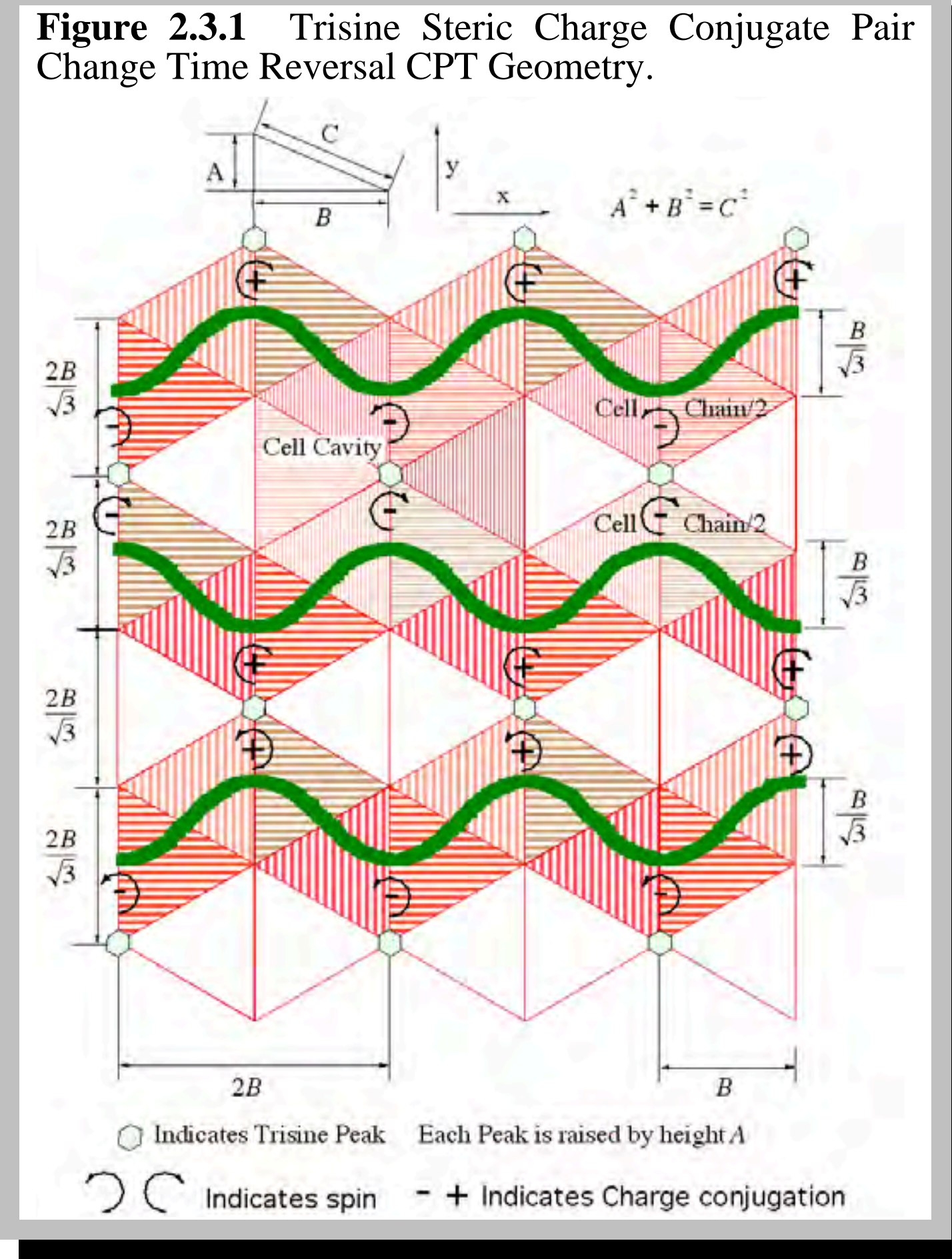

**Figure 2.3.2** Trisine Steric CPT Geometry in the Superconducting Mode showing the relationship between layered superconducting planes and wave vectors $K_B$. Layers are constructed in *x*, *y* and *z* directions making up an scaled lattice or Gaussian surface volume *xyz*.

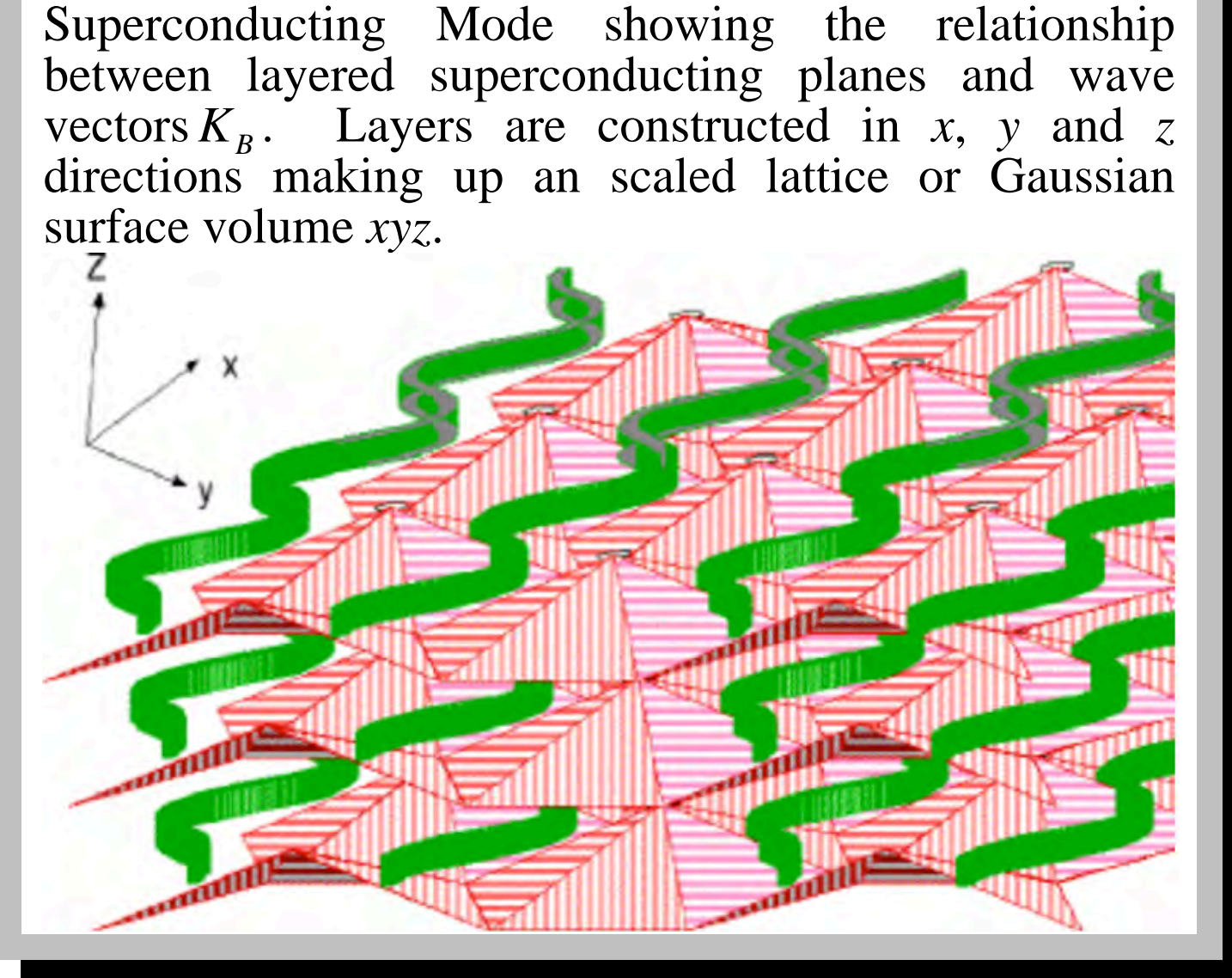

## 2.4 Trisine Characteristic De Broglie Velocities

The Cartesian de Broglie velocities $(v_{dx})$, $(v_{dy})$, and $(v_{dz})$ are computed with the trisine Cooper CPT Charge conjugated pair residence *resonant CPT* $time_\pm$ and characteristic angular frequency $(\omega)$ in phase with trisine superconducting dimension $(2B)$ in equations 2.4.1-2.4.5.

$$v_{dx} = \frac{\hbar K_B}{m_T} = \frac{2B}{time_\pm} = \frac{\omega}{2\pi}2B \tag{2.4.1}$$

$$v_{dxr} = \frac{\hbar K_{Br}}{m_T} = \frac{2B_r}{time_{r\pm}} = \frac{\omega_r}{2\pi}2B_r \tag{2.4.1r}$$

With $time_\pm$ and $B$ as a function of the universal constant(U) or more generally the trisine time $t_T$

$$time_\pm = 2\pi\hbar^{\frac{1}{3}}U^{\frac{-1}{3}}c^{\frac{-5}{3}}\left(\frac{c}{v_{dx}}\right)^2 = 2\pi t_T\left(\frac{c}{v_{dx}}\right)^2 \tag{2.4.2}$$

$$time_{r\pm} = 2\pi\hbar^{\frac{1}{3}}U^{\frac{-1}{3}}c^{\frac{-5}{3}}\left(\frac{c}{v_{dxr}}\right)^2 = 2\pi t_T\left(\frac{c}{v_{dxr}}\right)^2 \tag{2.4.2r}$$

In equation 2.4.1, note that $m_T v_{\varepsilon x}/eH_c$ is the particle spin axis precession rate in the critical magnetic field $(H_c)$. This electron spin rate is in tune with the electron moving with de Broglie velocity $(v_{dx})$ with a CPT residence $time_\pm$ in each *cavity*. In other words the Cooper CPT Charge conjugated pair flip spin twice per cavity and because each Cooper CPT Charge conjugated pair flips simultaneously, the quantum is an $2(1/2)$ or integer which corresponds to a boson.

$$time_\pm = \frac{g_s^3}{3}\frac{m_T v_{ex}}{eH_c} \tag{2.4.3}$$

$$time_{r\pm} = \frac{g_s^3}{3}\frac{m_T v_{exr}}{eH_{cr}} \tag{2.4.3r}$$

Equation 2.4.4 is a Higgs expression of superconductivity with the addition of the dielectric $(\varepsilon$ or $D/E)$ factor.

$$time_\pm = g_s^{-2}\frac{cavity}{chain}\left(\frac{\varepsilon m_T cavity}{4e^2}\right)^{\frac{1}{2}} \tag{2.4.4}$$

$$time_{r\pm} = g_s^{-2}\frac{cavity_r}{chain_r}\left(\frac{\varepsilon_r m_T cavity_r}{4e^2}\right)^{\frac{1}{2}} \tag{2.4.4r}$$

Equation 2.4.5 is an expression of universal oscillation.

$$time_{\pm} = \frac{1}{\sqrt{3}}\left(\frac{B}{A}\right)_T \frac{v_{dx}}{cH_U} \qquad (2.4.5)$$

$$time_{r\pm} = \frac{1}{\sqrt{3}}\left(\frac{B_r}{A_r}\right)_T \frac{v_{dxr}}{cH_{Ur}} \qquad (2.4.5r)$$

Or in terms of the Kolmogorov viscosity $(v_U)$ time scale[60]

$$time_{\pm} = \sqrt{\frac{B}{A}\frac{m_T v_U time_{\pm}}{k_b T_c}} \qquad (2.4.4)$$

$$time_{r\pm} = \sqrt{\frac{B_r}{A_r}\frac{m_T v_U time_{r\pm}}{k_b T_{cr}}} \qquad (2.4.4r)$$

Or in terms of universe mass and gravity dimensions

$$time_{\pm} = \frac{h}{6}\frac{G_U M_U}{m_T A c^4} \qquad (2.4.5)$$

$$time_{r\pm} = \frac{h}{6}\frac{G_{Ur} M_{Ur}}{m_T A_r c^4} \qquad (2.4.5r)$$

Then the de Broglie matter $(m_T)$ velocities are defined:

$$v_{dy} = \frac{\hbar K_P}{m_T} = \frac{2}{\sqrt{3}}\frac{2B}{time_{\pm}} \qquad (2.4.6)$$

$$v_{dyr} = \frac{\hbar K_{Pr}}{m_T} = \frac{2}{\sqrt{3}}\frac{2B_r}{time_{r\pm}} \qquad (2.4.6r)$$

$$v_{dz} = \frac{\hbar K_A}{m_T} = 2\left(\frac{B}{A}\right)_T^2 \frac{2A}{time_{\pm}} \qquad (2.4.7)$$

$$v_{dzr} = \frac{\hbar K_{Ar}}{m_T} = 2\left(\frac{B_r}{A_r}\right)_T^2 \frac{2A_r}{time_{r\pm}} \qquad (2.4.7r)$$

The vector sum of the $x$, $y$ and z de Broglie velocity components are used to compute a three dimensional helical or tangential de Broglie velocity $(v_{dT})$ as follows:

$$\begin{aligned} v_{dT} &= \sqrt{v_{dx}^2 + v_{dy}^2 + v_{dz}^2} \\ &= \frac{\hbar}{m_T}\sqrt{K_A^2 + K_B^2 + K_P^2} \end{aligned} \qquad (2.4.8)$$

$$\begin{aligned} v_{dTr} &= \sqrt{v_{dxr}^2 + v_{dyr}^2 + v_{dzr}^2} \\ &= \frac{\hbar}{m_T}\sqrt{K_{Ar}^2 + K_{Br}^2 + K_{Pr}^2} \end{aligned} \qquad (2.4.8r)$$

And in terms of reflective velocities:

$$v_{dxr} = \left(\frac{c}{v_{dx}}\right)c \qquad v_{dyr} = \frac{4}{3}\left(\frac{c}{v_{dy}}\right)c \qquad v_{dzr} = 4\left(\frac{B}{A}\right)_T^2\left(\frac{c}{v_{dz}}\right)c \qquad (2.4.9)$$

$$\frac{1}{v_{dTr}} = \sqrt{\frac{1}{v_{dxr}^2} + \frac{1}{v_{dyr}^2} + \frac{1}{v_{dzr}^2}} \qquad (2.4.10)$$

And the Majorana condition is fulfilled as follows resulting in a constant mass that can be considered a reciprocal of itself under the two time conditions $time_+$ and reversal $time_-$ ($time_{r+}$ and reversal $time_{r-}$).

*Schroedinger condition*

$$-i\hbar\frac{\partial}{\partial time_r}\Psi_+ + \frac{2\pi\hbar}{2time_r}\Psi_- = 0 \qquad (2.4.11)$$

where $\Psi_+ = e^{\frac{-im_t v_{dx}^2 time_+}{2\hbar}} \qquad \Psi_- = e^{\frac{-im_t v_{dx}^2 time_-}{2\hbar}}$

*Majorana condition*

$$-i\hbar\frac{\partial}{\partial time}\Psi_+ + m_t c^2 \Psi_- = 0 \qquad (2.4.12)$$

where $\Psi_+ = e^{\frac{-i\pi c^2 time_+^2}{(2B)^2}} \qquad \Psi_- = e^{\frac{-i\pi c^2 time_-^2}{(2B)^2}}$

and for both *Schroedinger* and *Majorana* conditions

$$\frac{(2B_r)^2}{time_{\pm}} = .06606194571\ \frac{cm^2}{\sec} \qquad (2.4.13)$$

and $m_T = \dfrac{\hbar 2\pi time_{\pm}}{(2B)^2} = 1.003008448E-25\ g$

*Schroedinger condition*

$$-i\hbar\frac{\partial}{\partial time_r}\Psi_{r+} + \frac{2\pi\hbar}{2time_r}\Psi_{r-} = 0 \qquad (2.4.11r)$$

where $\Psi_{r+} = e^{\frac{-im_t v_{dxr}^2 time_{r+}}{2\hbar}} \qquad \Psi_{r-} = e^{\frac{-im_t v_{dxr}^2 time_{r-}}{2\hbar}}$

*Majorana condition*

$$-i\hbar \frac{\partial}{\partial time_r} \Psi_{r+} + m_t c^2 \Psi_{r-} = 0 \qquad (2.4.12r)$$

where $\Psi_{r+} = e^{\frac{-i\pi c^2 time_{r+}^2}{(2B_r)^2}}$ $\quad \Psi_{-} = e^{\frac{-i\pi c^2 time_{r-}^2}{(2B_r)^2}}$

and for both *Schroedinger* and *Majorana* conditions

$$\frac{(2B_r)^2}{time_{r\pm}} = .06606194571 \ \frac{cm^2}{\sec} \qquad (2.4.13r)$$

and $m_T = \dfrac{\hbar 2\pi time_{r\pm}}{(2B_r)^2} = 1.003008448E-25 \ g$

**Table 2.4.1** Listing of de Broglie velocities, trisine cell pair residence resonant *CPT* $time_{\pm}$ and characteristic frequency $(\nu)$ as a function of selected critical temperature ($T_c$) and black body temperature ($T_b$) along with a $time_{\pm}$ plot.

| Condition | 1.2 | 1.3 | 1.6 | 1.9 | 1.10 | 1.11 |
|---|---|---|---|---|---|---|
| $T_a\left({}^0K\right)$ | 6.53E+11 | 6.53E+11 | 6.53E+11 | 6.53E+11 | 6.53E+11 | 6.53E+11 |
| $T_b\left({}^0K\right)$ | 9.05E+14 | 5.48E+13 | 9.22E+08 | 1.10E+07 | 2,868 | 2.723 |
| $T_{br}\left({}^0K\right)$ | 3.30E+12 | 5.45E+13 | 3.24E+18 | 2.72E+20 | 1.04E+24 | 1.10E+27 |
| $T_c\left({}^0K\right)$ | 8.96E+13 | 3.28E+11 | 93.0 | 0.0131 | 8.99E-10 | 8.10E-16 |
| $T_{cr}\left({}^0K\right)$ | 1.19E+09 | 3.25E+11 | 1.15E+21 | 8.11E+24 | 1.19E+32 | 1.32E+38 |
| $z_R + 1$ | 3.67E+43 | 8.14E+39 | 3.89E+25 | 6.53E+19 | 1.17E+09 | 1.00E+00 |
| $z_B + 1$ | 3.32E+14 | 2.01E+13 | 3.39E+08 | 4.03E+06 | 1.05E+03 | 1.00E+00 |
| $Age_{UG}\left(sec\right)$ | 7.14E-05 | 4.79E-03 | 6.94E+04 | 5.36E+07 | 1.27E+13 | 4.33E+17 |
| $Age_{UGv}\left(sec\right)$ | 4.59E-10 | 1.25E-07 | 4.42E+02 | 3.13E+06 | 1.22E+13 | 4.33E+17 |
| $Age_{U}\left(sec\right)$ | 1.18E-26 | 5.31E-23 | 1.11E-08 | 6.63E-03 | 3.70E+08 | 4.33E+17 |
| $v_{dx}$ (cm/sec) | 4.97E+11 | 3.00E+10 | 5.06E+05 | 6.01E+03 | 1.57E+00 | 1.49E-03 |
| $v_{dxr}$ (cm/sec | 1.81E+09 | 2.99E+10 | 1.78E+15 | 1.49E+17 | 5.71E+20 | 6.02E+23 |
| $v_{dy}$ (cm/sec) | 5.73E+11 | 3.47E+10 | 5.84E+05 | 6.94E+03 | 1.82E+00 | 1.72E-03 |
| $v_{dyr}$ (cm/sec) | 2.09E+09 | 3.45E+10 | 2.05E+15 | 1.73E+17 | 6.60E+20 | 6.95E+23 |
| $v_{dz}$ (cm/sec) | 2.36E+12 | 1.43E+11 | 2.41E+06 | 2.86E+04 | 7.49E+00 | 7.11E-03 |
| $v_{dzr}$ (cm/sec) | 8.61E+09 | 1.42E+11 | 8.45E+15 | 7.11E+17 | 2.72E+21 | 2.86E+24 |
| $v_{dTr}$ (cm/sec) | 2.48E+12 | 1.50E+11 | 2.53E+06 | 3.01E+04 | 7.86E+00 | 7.46E-03 |
| $v_{dTr}$ (cm/sec) | 9.04E+09 | 1.49E+11 | 8.88E+15 | 7.47E+17 | 2.85E+21 | 3.01E+24 |
| $time_{\pm}$ (sec) | 2.68E-25 | 7.32E-23 | 2.58E-13 | 1.83E-09 | 2.67E-02 | 2.96E+04 |
| $time_{r\pm}$ (sec) | 2.02E-20 | 7.39E-23 | 2.09E-32 | 2.96E-36 | 2.02E-43 | 1.82E-49 |
| $\nu$ (/sec) Hz | 3.73E+24 | 1.37E+22 | 3.88E+12 | 5.47E+08 | 3.75E+01 | 3.38E-05 |
| $\nu_r$ (/sec) Hz | 4.96E+19 | 1.35E+22 | 4.78E+31 | 3.38E+35 | 4.94E+42 | 5.48E+48 |

It is important to note that the one degree of freedom $time_{\pm}$ (2.96E04 sec or (164 x 3) minutes) in table 2.4.1 corresponds to a ubiquitous 160 minute resonant signal in the Universe.[68,69] and which is also consistent with the observed Active Galactic Nuclei (AGN) power spectrum. This would be an indication that the momentum and energy conserving (elastic) space lattice existing in universe space is dynamically in tune with stellar, galactic objects (Appendix H). This appears logical, based on the fact the major part of the universe mass consists of this space lattice which can be defined in other terms as dark energy and its associated matter.

Additionally, the earth's hum [87] with a harmonic of .09 Hz *($f_1$)* as measured by seismographs is correlated to universe frequency *(1/$time_{\pm}$)* 3.38E-5 Hz *($f_2$)* by the air/earth density unitless ratio (.0014/5.519) $\left(\rho_2/\rho_1 = E_2/E_1\right)$ by a factor of approximately 2/3, which is of course *chain/cavity*. This correlation is supported by telluric atmospheric sodium oscillations on the order of 3.38E-5 Hz[88]. This phenomenon may be explained in terms of De Broglie matter wave correlations:

$$E_1/f_1 = E_2/f_2 = h \quad \text{and} \quad \rho_2/\rho_1 = E_2/E_1 = f_2/f_1 \quad .$$

It is conceivable that these earth atmospheric resonance characteristics modulating the earth's center of mass (M) by resonant distance (*r*), may contribute to the flyby anomaly[86] by transferal of gravitational energy *GmM/(R+r)* to a spacecraft of mass (*m*) and radius (*R*) from the center of mass during flyby. The increase in spacecraft energy is reported in terms of ~1 mm/sec and oscillatory universe dark energy induced earth atmospheric tidal effects result in spacecraft velocity calculated deviations of this order.

Also, an observed base oscillatory frequency of 2.91E-5 Hz has been observed in the star $\beta$ Hydri that compares favorably (within 15%) to the anticipated universe frequency *(1/$time_{\pm}$)* 3.38E-5 Hz [102]. Harmonics of this base frequency are consistent with this base frequency.

This conceptual oscillatory view is supported by observed luminosity time periods for 2,327 RRc variable stars extracted from all constellations (The International Variable Star Index http://www.aavso.org/vsx/ with 182,631 variable stars) as compiled in Figure 2.4.1. The RRc variable stars are old with low mass (with perhaps low atmospheric density), with high absolute magnitude (40 or 50 times brighter than our Sun), with sinusoidal oscillation character and with metal-poor "Population II" star character. The RRc sinusoidal oscillation tidally (Appendix H) synchronizes with the universe trisine 'dark energy' oscillations. The average 2,327 RRc star time period is 0.316 day which differs from Table 2.4.1 prediction of 0.343 day (- 29,600 sec/(60x60x24) by 8.5% (which perhaps noncoincidentally $\sim 1/\cos(\theta)$ ).

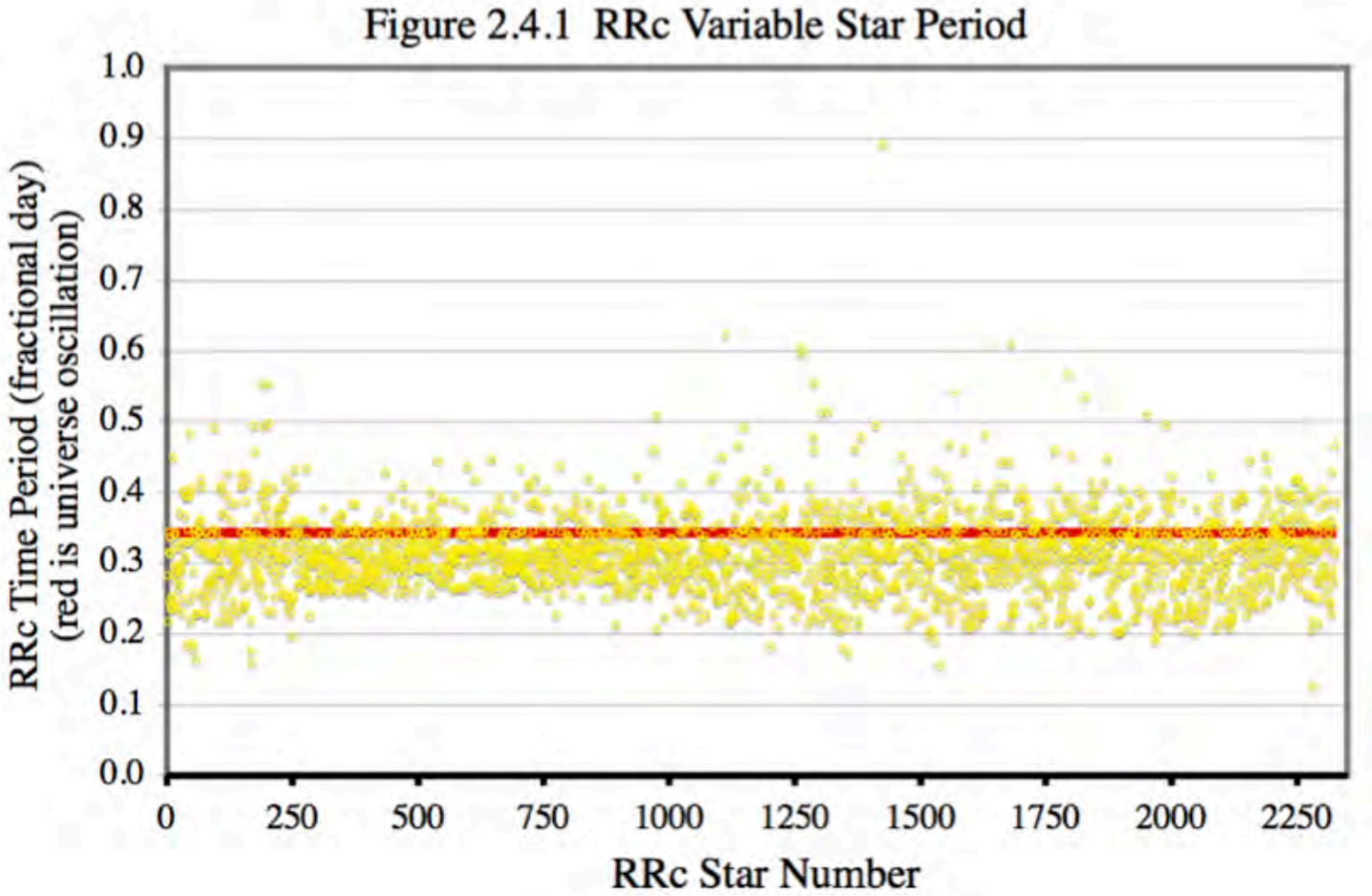

Figure 2.4.1 RRc Variable Star Period

*CPT time*$_{\pm}$ (9.23E-13 seconds) at $T_c = 26\ K$ is experimentally confirmed[132].

The trisine condition 1.11 (green) table 2.4.1 resonance *time* value of 29,600 seconds compares favorably with redshift adjusted Blazars[145 Figure 9] as presented in table 2.4.2.

| trisine condition 1.11 (green) table 2.4.1 | | |
|---|---|---|
| | 29,600 | Seconds |
| | .3426 | Days |
| | 18.34 | omega($2\pi$ /days) |
| | 5.45E-02 | 1/omega (days) |
| | -1.26 | $\text{Log}_{10}$(1/omega (days)) |

Table 2.4.2

| Blazars | $z_B$ | 1/omega $(1+z_B)^{-2}$ days | $\log_{10}$ 1/omega $(1+z_B)^{-2}$ days |
|---|---|---|---|
| B2 1633+38 | 1.814 | 6.89E-03 | -2.16 |
| PKS 1424-41 | 1.522 | 8.57E-03 | -2.07 |
| B2 1520+31 | 1.487 | 8.82E-03 | -2.05 |
| PKS 0454-234 | 1.003 | 1.36E-02 | -1.87 |
| 3C 454.3 | 0.859 | 1.58E-02 | -1.80 |
| 3C 279 | 0.536 | 2.31E-02 | -1.64 |
| PKS 1510-089 | 0.360 | 2.95E-02 | -1.53 |
| 3C 273 | 0.158 | 4.07E-02 | -1.39 |
| | 0 | 5.45E-02 | -1.26 |
| BL Lacs | | | |
| 3C 66A | 0.444 | 2.61E-02 | -1.58 |
| PKS 0716+714 | 0.3 | 3.23E-02 | -1.49 |
| PKS 2155-304 | 0.116 | 4.38E-02 | -1.36 |
| BL Lac | 0.069 | 4.77E-02 | -1.32 |
| Mkn 421 | 0.03 | 5.14E-02 | -1.29 |

The trisine condition 1.11 3426 day universal oscillation is also expressed in stellar core gravity wave "observed" period spacing patterns [154 -Figure 17).

This is evidence for a universal Machian oscillation presence.

## 2.5 Dielectric Permittivity And Magnetic Permeability

The development has origin in James Maxwell's establishment of the speed of light from previously experimentally observed constants by Michael Faraday

$$\begin{aligned}
&\frac{1}{\sqrt{\varepsilon_0 \mu_0}} = c\left(2.998E8\,\frac{m}{\text{sec}}\right) \\
&\varepsilon_0 = 8.854E-12\ \frac{farad}{m}\left(\frac{coulumb^2 \sec^2}{kg\ m^3}\right) \qquad (2.5.1) \\
&\mu_0 = 1.257E-6\ \frac{newton}{amp^2}\left(\frac{kg\ m}{coulumb^2}\right)
\end{aligned}$$

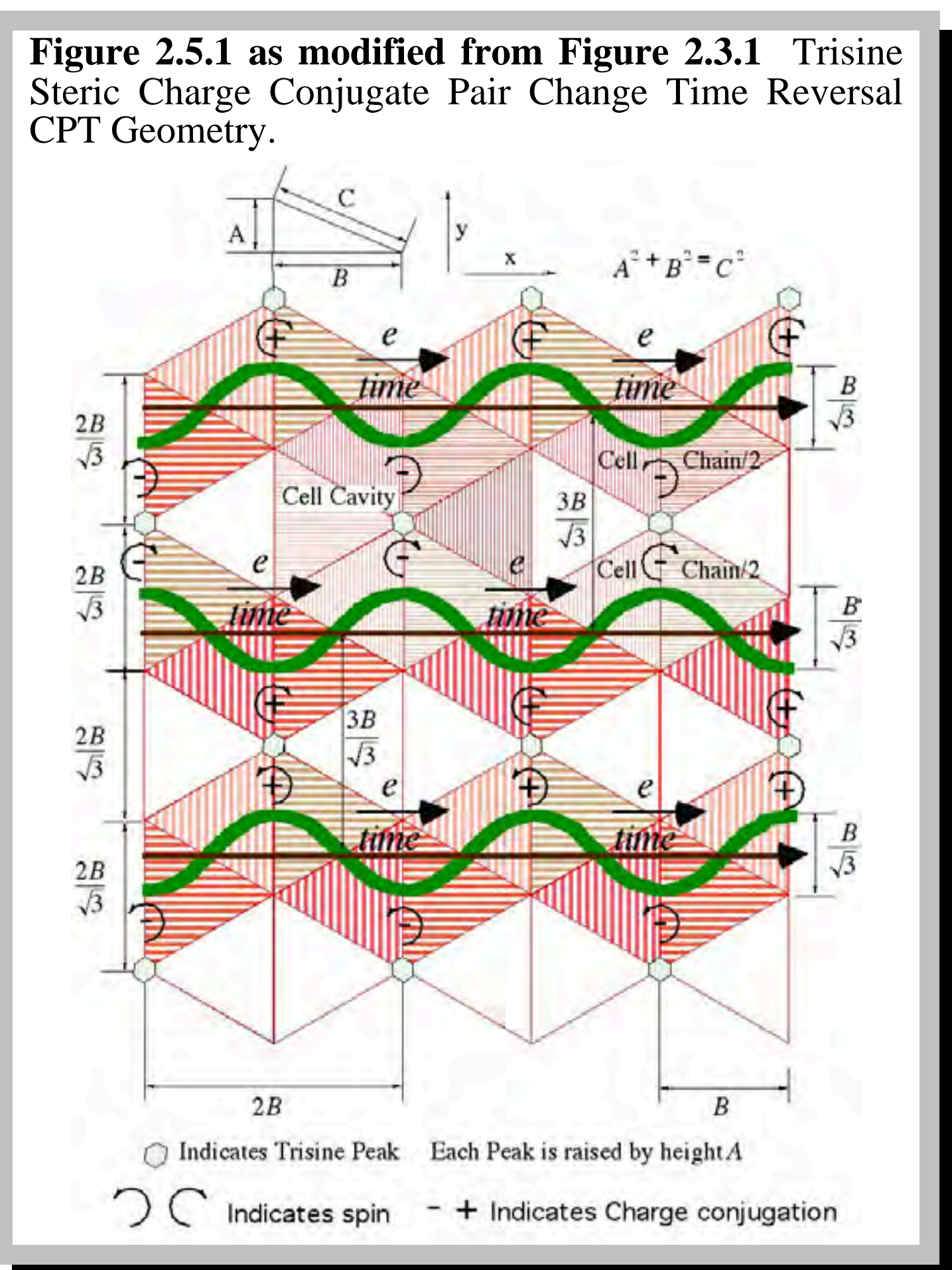

**Figure 2.5.1 as modified from Figure 2.3.1** Trisine Steric Charge Conjugate Pair Change Time Reversal CPT Geometry.

The material dielectric is computed by determining a displacement$(D)$ (equation 2.5.1) and electric field$(E)$ (equation 2.5.2) and which establishes a dielectric $(\varepsilon \text{ or } D/E)$ (equation 2.5.3) and modified speed of light $(v_{\varepsilon})$ (equations 2.5.4, 2.5.5). Then the superconducting fluxoid $(\Phi_{\varepsilon})$ (equation 2.5.6) can be calculated.

The electric field(E) is calculated by taking the translational energy $\left(m_T v_{dx}^2/2\right)$ (which is equivalent to $k_b T_c$) over a distance, which is the trisine cell volume to surface ratio. This concept of the surface to volume ratio is used extensively in fluid mechanics for computing energies of fluids passing through various geometric forms and is called the hydraulic ratio.

$$\left.\begin{matrix} E_{fx} \\ E_{fy} \\ E_{fz} \\ E_f \end{matrix}\right\} = \left\{\begin{matrix} \dfrac{k_b T_c}{e2B} \\ \dfrac{k_b T_c}{eP} \\ \dfrac{k_b T_c}{eA} \\ \dfrac{1}{\dfrac{1}{E_{fx}}+\dfrac{1}{E_{fy}}+\dfrac{1}{E_{fz}}} = \dfrac{k_b T_c}{\pi^2 eA} \end{matrix}\right. \qquad (2.5.2)$$

$$\left.\begin{matrix} E_{fxr} \\ E_{fyr} \\ E_{fzr} \\ E_{fr} \end{matrix}\right\} = \left\{\begin{matrix} \dfrac{k_b T_{cr}}{e2B_r} \\ \dfrac{k_b T_{cr}}{eP_r} \\ \dfrac{k_b T_{cr}}{eA_r} \\ \dfrac{1}{\dfrac{1}{E_{fxr}}+\dfrac{1}{E_{fyr}}+\dfrac{1}{E_{fzr}}} = \dfrac{k_b T_{cr}}{\pi^2 eA_r} \end{matrix}\right. \qquad (2.5.2r)$$

The trisine cell surface is (*section/cos*$(\theta)$) and the volume is expressed as *cavity*. Note that the trisine *cavity*, although bounded by (2*section/cos*$(\theta)$), still has the passageways for Cooper CPT Charge conjugated pairs $(\mathbb{C}e)$ to move from cell *cavity* to cell *cavity*.

The displacement$(D)$ as a measure of Gaussian surface containing free charges$(\mathbb{C}\, e_{\pm})$ is computed by taking the $(4\pi)$ solid angle divided by the characteristic trisine area $(section/\cos(\theta))$ with the two(2) charges $(\mathbb{C}\, e_{\pm})$ contained therein and in accordance with Gauss's law as expressed in equation 2.1.1.

$$\left.\begin{matrix} D_{fx} \\ D_{fy} \\ D_{fz} \\ D_f \end{matrix}\right\} = \left\{\begin{matrix} \dfrac{\pi}{g_s}\dfrac{4\pi\mathbb{C}e}{approach} \\ \dfrac{4\pi}{3g_s}\dfrac{4\pi\mathbb{C}e}{side} \\ \dfrac{\sqrt{3}}{2g_d^2}\dfrac{4\pi\mathbb{C}e}{section} \\ \dfrac{1}{\dfrac{1}{D_{fx}}+\dfrac{1}{D_{fy}}+\dfrac{1}{D_{fz}}} = \left(\dfrac{m_e}{m_t}\right)^{\frac{1}{2}}\dfrac{4\pi\mathbb{C}e}{g_s AB} \end{matrix}\right.$$

$$= \left\{\begin{matrix} \dfrac{\pi}{g_s}\dfrac{4\pi\mathbb{C}e}{\sqrt{3}AB} = \dfrac{\pi}{g_s}\dfrac{4\pi\mathbb{C}e}{\sqrt{3}\left(\dfrac{B}{A}\right)_t B^2} \\ \dfrac{4\pi}{3g_s}\dfrac{4\pi\mathbb{C}e}{2AB} = \dfrac{4\pi}{3g_s}\dfrac{4\pi\mathbb{C}e}{2\left(\dfrac{B}{A}\right)_t B^2} \\ \dfrac{\sqrt{3}}{2g_d^2}\dfrac{4\pi\mathbb{C}e}{2\sqrt{3}B^2} \\ \dfrac{1}{\dfrac{1}{D_{fx}}+\dfrac{1}{D_{fy}}+\dfrac{1}{D_{fz}}} = \left(\dfrac{m_e}{m_T}\right)^{\frac{1}{2}}\dfrac{4\pi\mathbb{C}e}{g_s\left(\dfrac{B}{A}\right)_t B^2} \end{matrix}\right. \qquad (2.5.3)$$

*where after cancelling* $B^2$

$$\frac{m_T}{m_e} = 110.107176860301$$

$$\left.\begin{array}{l} D_{fxr} \\ D_{fyr} \\ D_{fzr} \\ D_{fr} \end{array}\right\} = \left\{\begin{array}{c} \dfrac{\pi}{g_s}\dfrac{4\pi\mathbb{C}e}{approach_r} \\ \dfrac{4\pi}{3g_s}\dfrac{4\pi\mathbb{C}e}{side_r} \\ \dfrac{\sqrt{3}}{2g_s^2}\dfrac{4\pi\mathbb{C}e}{section_r} \\ \dfrac{1}{\dfrac{1}{D_{fxr}}+\dfrac{1}{D_{fyr}}+\dfrac{1}{D_{fzr}}} = \left(\dfrac{m_e}{m_t}\right)^{\frac{1}{2}}\dfrac{4\pi\mathbb{C}e}{g_sA_rB_r} \end{array}\right.$$

$$= \left\{\begin{array}{c} \dfrac{\pi}{g_s}\dfrac{4\pi\mathbb{C}e}{\sqrt{3}A_rB_r} = \dfrac{\pi}{g_s}\dfrac{4\pi\mathbb{C}e}{\sqrt{3}\left(\dfrac{B_r}{A_r}\right)_t B_r^2} \\ \dfrac{4\pi}{3g_s}\dfrac{4\pi\mathbb{C}e}{2A_rB_r} = \dfrac{4\pi}{3g_s}\dfrac{4\pi\mathbb{C}e}{2\left(\dfrac{B_r}{A_r}\right)_t B_r^2} \\ \dfrac{\sqrt{3}}{2g_d^2}\dfrac{4\pi\mathbb{C}e}{2\sqrt{3}B_r^2} \\ \dfrac{1}{\dfrac{1}{D_{fx}}+\dfrac{1}{D_{fy}}+\dfrac{1}{D_{fz}}} = \left(\dfrac{m_e}{m_T}\right)^{\frac{1}{2}}\dfrac{4\pi\mathbb{C}e}{g_s\left(\dfrac{B_r}{A_r}\right)_t B_r^2} \end{array}\right. \qquad (2.5.3r)$$

*where after cancelling* $B_r^2$

$$\frac{m_T}{m_e} = 110.107176860301$$

Now a dielectric coefficient $(\varepsilon)$ can be calculated from the electric field $(E)$ and displacement field $(D)$ [61]. Note that $\varepsilon < 1$ for group phase reflection velocities as justified by resonant condition as numerically indicated in table 2.5.1.

$$\left.\begin{array}{l} \varepsilon_x \\ \varepsilon_y \\ \varepsilon_z \\ \varepsilon \end{array}\right\} = \left\{\begin{array}{c} \dfrac{D_{fx}}{E_{fx}} \\ \dfrac{D_{fy}}{E_{fy}} \\ \dfrac{D_{fz}}{E_{fz}} \\ \dfrac{1}{9g_s}\dfrac{D_f}{E_f} = \dfrac{1}{\dfrac{1}{\varepsilon_x}+\dfrac{1}{\varepsilon_y}+\dfrac{1}{\varepsilon_z}} \end{array}\right. \qquad (2.5.4)$$

$$\left.\begin{array}{l} \varepsilon_{xr} \\ \varepsilon_{yr} \\ \varepsilon_{zr} \\ \varepsilon_r \end{array}\right\} = \left\{\begin{array}{c} \dfrac{D_{fxr}}{E_{fxr}} \\ \dfrac{D_{fyr}}{E_{fyr}} \\ \dfrac{D_{fzr}}{E_{fzr}} \\ \dfrac{1}{9g_s}\dfrac{D_{fr}}{E_{fr}} = \dfrac{1}{\dfrac{1}{\varepsilon_{xr}}+\dfrac{1}{\varepsilon_{yr}}+\dfrac{1}{\varepsilon_{zr}}} \end{array}\right. \qquad (2.5.4r)$$

Velocity of light dielectric $\varepsilon$ modified

$$\left.\begin{array}{l} v_{\varepsilon x} \\ v_{\varepsilon y} \\ v_{\varepsilon z} \\ v_{\varepsilon} \end{array}\right\} = \left\{\begin{array}{c} \dfrac{c}{\sqrt{\varepsilon_x}} \\ \dfrac{c}{\sqrt{\varepsilon_y}} \\ \dfrac{c}{\sqrt{\varepsilon_z}} \\ \dfrac{c}{\sqrt{\varepsilon_r}} \end{array}\right\} = \left\{\begin{array}{c} v_{dx}\sqrt{k_{mx}} \\ v_{dy}\sqrt{k_{my}} \\ v_{dz}\sqrt{k_{mz}} \\ 2v_{dT}\sqrt{k_m} \end{array}\right. \qquad (2.5.5)$$

$$\left.\begin{array}{l} v_{\varepsilon xr} \\ v_{\varepsilon yr} \\ v_{\varepsilon zr} \\ v_{\varepsilon r} \end{array}\right\} = \left\{\begin{array}{c} \dfrac{c}{\sqrt{\varepsilon_{xr}}} \\ \dfrac{c}{\sqrt{\varepsilon_{yr}}} \\ \dfrac{c}{\sqrt{\varepsilon_{zr}}} \\ \dfrac{c}{\sqrt{\varepsilon_r}} \end{array}\right\} = \left\{\begin{array}{c} v_{dxr}\sqrt{k_{mxr}} \\ v_{dyr}\sqrt{k_{myr}} \\ v_{dzr}\sqrt{k_{mzr}} \\ 2v_{dTr}\sqrt{k_{mr}} \end{array}\right. \qquad (2.5.5r)$$

Trisine incident/reflective angle (Figure 2.3.1) of 30 degrees is less than Brewster angle of $\tan^{-1}(\varepsilon/\varepsilon)^{1/2} = 45^0$ assuring total reflectivity as particles travel from trisine lattice cell to trisine lattice cell.

Now the fluxoid $(\Phi_\varepsilon)$ can be computed quantized according to $(\mathbb{C}\ e)$ as experimentally observed in superconductors.

$$\Phi_\varepsilon = \frac{2\pi\hbar v_\varepsilon}{\mathbb{C}\ e_\pm} \qquad (2.5.6)$$

What is generally called the Tao effect [65,66,67], wherein it is found that superconducting micro particles in the presence of an electrostatic field aggregate into balls of macroscopic dimensions, is modeled by the calculated values for E in table 2.5.1. using equation 2.5.2. When the electrostatic field exceeds a critical value coincident with the critical temperature $(T_b)$, the microscopic balls dissipate. This experimental evidence can be taken as another measure of superconductivity equivalent to the Meissner effect.

**Table 2.5.1** Displacement (D) and Electric (E) Fields

| Condition | 1.2 | 1.3 | 1.6 | 1.9 | 1.10 | 1.11 |
|---|---|---|---|---|---|---|
| $T_a\left({}^0K\right)$ | 6.53E+11 | 6.53E+11 | 6.53E+11 | 6.53E+11 | 6.53E+11 | 6.53E+11 |
| $T_b\left({}^0K\right)$ | 9.05E+14 | 5.48E+13 | 9.22E+08 | 1.10E+07 | 2,868 | 2.723 |
| $T_{br}\left({}^0K\right)$ | 3.30E+12 | 5.45E+13 | 3.24E+18 | 2.72E+20 | 1.04E+24 | 1.10E+27 |
| $T_c\left({}^0K\right)$ | 8.96E+13 | 3.28E+11 | 93.0 | 0.0131 | 8.99E-10 | 8.10E-16 |
| $T_{cr}\left({}^0K\right)$ | 1.19E+09 | 3.25E+11 | 1.15E+21 | 8.11E+24 | 1.19E+32 | 1.32E+38 |
| $z_R+1$ | 3.67E+43 | 8.14E+39 | 3.89E+25 | 6.53E+19 | 1.17E+09 | 1.00E+00 |
| $z_B+1$ | 3.32E+14 | 2.01E+13 | 3.39E+08 | 4.03E+06 | 1.05E+03 | 1.00E+00 |
| $Age_{UG}\left(sec\right)$ | 7.14E-05 | 4.79E-03 | 6.94E+04 | 5.36E+07 | 1.27E+13 | 4.33E+17 |
| $Age_{UGv}\left(sec\right)$ | 4.59E-10 | 1.25E-07 | 4.42E+02 | 3.13E+06 | 1.22E+13 | 4.33E+17 |
| $Age_U\left(sec\right)$ | 1.18E-26 | 5.31E-23 | 1.11E-08 | 6.63E-03 | 3.70E+08 | 4.33E+17 |
| $D_{fx}$ (erg/e/cm) | 1.17E+19 | 4.27E+16 | 1.21E+07 | 1.71E+03 | 1.17E-04 | 1.05E-10 |
| $D_{fy}$ (erg/e/cm) | 1.35E+19 | 4.93E+16 | 1.40E+07 | 1.97E+03 | 1.35E-04 | 1.22E-10 |
| $D_{fz}$ (erg/e/cm) | 6.81E+17 | 2.49E+15 | 7.07E+05 | 9.98E+01 | 6.83E-06 | 6.16E-12 |
| $D_f$ (erg/e/cm) | 6.14E+17 | 2.25E+15 | 6.37E+05 | 9.00E+01 | 6.16E-06 | 5.55E-12 |
| $D_{fxr}$ (erg/e/cm) | 1.55E+14 | 4.23E+16 | 1.49E+26 | 1.06E+30 | 1.54E+37 | 1.71E+43 |
| $D_{fyr}$ (erg/e/cm) | 1.79E+14 | 4.88E+16 | 1.72E+26 | 1.22E+30 | 1.78E+37 | 1.98E+43 |
| $D_{fzr}$ (erg/e/cm) | 9.04E+12 | 2.47E+15 | 8.71E+24 | 6.16E+28 | 9.01E+35 | 9.99E+41 |
| $D_{fr}$ (erg/e/cm) | 8.15E+12 | 2.23E+15 | 7.85E+24 | 5.56E+28 | 8.12E+35 | 9.01E+41 |
| $E_{fx}$ (erg/e/cm) | 1.94E+20 | 4.29E+16 | 2.05E+02 | 3.44E-04 | 6.15E-15 | 5.27E-24 |
| $E_{fy}$ (erg/e/cm) | 2.23E+20 | 4.95E+16 | 2.36E+02 | 3.97E-04 | 7.11E-15 | 6.08E-24 |
| $E_{fz}$ (erg/e/cm) | 9.21E+20 | 2.04E+17 | 9.74E+02 | 1.64E-03 | 2.93E-14 | 2.51E-23 |
| $E_f$ (erg/e/cm) | 9.32E+19 | 2.07E+16 | 9.86E+01 | 1.66E-04 | 2.96E-15 | 2.54E-24 |
| $E_f$ (volt/cm) | 2.79E+22 | 6.19E+18 | 2.96E+04 | 4.96E-02 | 8.89E-13 | 7.61E-22 |
| $E_{fxr}$ (erg/e/cm) | 9.37E+12 | 4.23E+16 | 8.86E+30 | 5.28E+36 | 2.95E+47 | 3.44E+56 |
| $E_{fyr}$ (erg/e/cm) | 1.08E+13 | 4.88E+16 | 1.02E+31 | 6.09E+36 | 3.40E+47 | 3.98E+56 |
| $E_{fzr}$ (erg/e/cm) | 4.46E+13 | 2.01E+17 | 4.21E+31 | 2.51E+37 | 1.40E+48 | 1.64E+57 |
| $E_{fr}$ (erg/e/cm) | 4.51E+12 | 2.04E+16 | 4.27E+30 | 2.54E+36 | 1.42E+47 | 1.66E+56 |
| $\varepsilon_x$ $D_{fx}/E_{fx}$ | 6.02E-02 | 9.96E-01 | 5.91E+04 | 4.97E+06 | 1.90E+10 | 2.00E+13 |
| $\varepsilon_y$ $D_{fy}/E_{fy}$ | 6.02E-02 | 9.96E-01 | 5.91E+04 | 4.97E+06 | 1.90E+10 | 2.00E+13 |
| $\varepsilon_z$ $D_{fz}/E_{fz}$ | 7.39E-04 | 1.22E-02 | 7.25E+02 | 6.10E+04 | 2.33E+08 | 2.46E+11 |
| $\varepsilon$ $D_f/E_f$ | 7.21E-04 | 1.19E-02 | 7.08E+02 | 5.96E+04 | 2.28E+08 | 2.40E+11 |
| $\varepsilon_{xr}$ $D_{fxr}/E_{fxr}$ | 1.65E+01 | 1.00E+00 | 1.68E-05 | 2.00E-07 | 5.24E-11 | 4.97E-14 |
| $\varepsilon_{yr}$ $D_{fyr}/E_{fyr}$ | 1.65E+01 | 1.00E+00 | 1.68E-05 | 2.00E-07 | 5.24E-11 | 4.97E-14 |
| $\varepsilon_{zr}$ $D_{fzr}/E_{fzr}$ | 2.03E-01 | 1.23E-02 | 2.07E-07 | 2.46E-09 | 6.42E-13 | 6.10E-16 |
| $\varepsilon_r$ $D_{fr}/E_{fr}$ | 1.98E-01 | 1.20E-02 | 2.02E-07 | 2.40E-09 | 6.27E-13 | 5.95E-16 |
| $v_{\varepsilon x}$ (cm/sec) | 1.22E+11 | 3.00E+10 | 1.23E+08 | 1.34E+07 | 2.17E+05 | 6.70E+03 |
| $v_{\varepsilon y}$ (cm/sec) | 1.22E+11 | 3.00E+10 | 1.23E+08 | 1.34E+07 | 2.17E+05 | 6.70E+03 |
| $v_{\varepsilon z}$ (cm/sec) | 1.10E+12 | 2.71E+11 | 1.11E+09 | 1.21E+08 | 1.96E+06 | 6.05E+04 |
| $v_{\varepsilon}$ (cm/sec) | 1.12E+12 | 2.75E+11 | 1.13E+09 | 1.23E+08 | 1.99E+06 | 6.12E+04 |
| $v_{\varepsilon xr}$ (cm/sec) | 7.36E+09 | 2.99E+10 | 7.29E+12 | 6.69E+13 | 4.13E+15 | 1.34E+17 |
| $v_{\varepsilon yr}$ (cm/sec) | 7.36E+09 | 2.99E+10 | 7.29E+12 | 6.69E+13 | 4.13E+15 | 1.34E+17 |
| $v_{\varepsilon zr}$ (cm/sec) | 6.66E+10 | 2.71E+11 | 6.60E+13 | 6.05E+14 | 3.74E+16 | 1.21E+18 |

A conversion between superconducting temperature and black body temperature is calculated in equation 2.5.9. Two primary energy related factors are involved, the first being the superconducting velocity $\left(v_{dx}^2\right)$ to light velocity $\left(v_\varepsilon^2\right)$ and the second is normal Debye wave vector $\left(K_{Dn}^2\right)$ to superconducting Debye wave vector $\left(K_{Ds}^2\right)$ all of this followed by a minor rotational factor for each factor involving $\left(m_e\right)$ and $\left(m_p\right)$. For reference, a value of 2.729 K is used for the universe black body temperature $\left(T_b\right)$ as indicated by experimentally observed microwave radiation by the Cosmic Background Explorer (COBE) and later satellites. Although the observed minor fluctuations (1 part in 100,000) in this universe background radiation indicative of clumps of matter forming shortly after the big bang, for the purposes of this report we will assume that the experimentally observed uniform radiation is indicative of present universe that is isotropic and homogeneous.

Now assume that the superconductor material magnetic permeability $\left(k_m\right)$ is defined as per equation 2.5.7.

$$\begin{Bmatrix} k_{mx} \\ k_{my} \\ k_{mz} \\ k_m \end{Bmatrix} = \begin{Bmatrix} \dfrac{D_{fx} v_{\varepsilon x}^2}{E_{fx}} & \dfrac{E_{fx}}{D_{fx} v_{dx}^2} \\ \dfrac{D_{fy} v_{\varepsilon y}^2}{E_{fy}} & \dfrac{E_{fy}}{D_{fy} v_{dy}^2} \\ \dfrac{D_{fz} v_{\varepsilon z}^2}{E_{fz}} & \dfrac{E_{fz}}{D_{fz} v_{dz}^2} \\ \dfrac{D_f v_\varepsilon^2}{E_f} & \dfrac{E_f}{D_f v_{dT}^2} \end{Bmatrix} = \begin{Bmatrix} \dfrac{v_{\varepsilon x}^2}{v_{dx}^2} \\ \dfrac{v_{\varepsilon y}^2}{v_{dy}^2} \\ \dfrac{v_{\varepsilon z}^2}{v_{dz}^2} \\ \dfrac{v_\varepsilon^2}{v_{dT}^2} \end{Bmatrix} \qquad (2.5.7)$$

$$\begin{Bmatrix} k_{mxr} \\ k_{myr} \\ k_{mzr} \\ k_{mr} \end{Bmatrix} = \begin{Bmatrix} \dfrac{D_{fxr} v_{\varepsilon xr}^2}{E_{fxr}} & \dfrac{E_{fxr}}{D_{fxr} v_{dxr}^2} \\ \dfrac{D_{fyr} v_{\varepsilon yr}^2}{E_{fyr}} & \dfrac{E_{fyr}}{D_{fyr} v_{dyr}^2} \\ \dfrac{D_{fzr} v_{\varepsilon zr}^2}{E_{fzr}} & \dfrac{E_{fzr}}{D_{fzr} v_{dzr}^2} \\ \dfrac{D_{fr} v_{\varepsilon r}^2}{E_{fr}} & \dfrac{E_{fr}}{D_{fr} v_{dTr}^2} \end{Bmatrix} = \begin{Bmatrix} \dfrac{v_{\varepsilon xr}^2}{v_{dxr}^2} \\ \dfrac{v_{\varepsilon yr}^2}{v_{dyr}^2} \\ \dfrac{v_{\varepsilon zr}^2}{v_{dzr}^2} \\ \dfrac{v_{\varepsilon r}^2}{v_{dTr}^2} \end{Bmatrix} \qquad (2.5.7r)$$

Then the de Broglie velocities ($v_{dx}$, $v_{dy}$ and $v_{dz}$) as per equations 2.4.1, 2.4.2, 2.4.3 can be considered a sub- and super- luminal speeds of light internal to the superconductor resonant medium as per equation 2.5.13. These sub- and super- luminal resonant de Broglie velocities as presented in equation 2.5.13 logically relate to wave vectors $K_A$, $K_B$ and $K_P$.

$$\begin{Bmatrix} v_{dx} \\ v_{dy} \\ v_{dz} \end{Bmatrix} = \begin{Bmatrix} \dfrac{c}{\sqrt{k_{mx}\varepsilon_x}} \\ \dfrac{c}{\sqrt{k_{my}\varepsilon_y}} \\ \dfrac{c}{\sqrt{k_{mz}\varepsilon_z}} \end{Bmatrix} = \begin{Bmatrix} \dfrac{v_{\varepsilon x}}{\sqrt{k_{mx}}} \\ \dfrac{v_{\varepsilon y}}{\sqrt{k_{my}}} \\ \dfrac{v_{\varepsilon z}}{\sqrt{k_{mz}}} \end{Bmatrix} \qquad (2.5.8)$$

$$\begin{Bmatrix} v_{dxr} \\ v_{dyr} \\ v_{dzr} \end{Bmatrix} = \begin{Bmatrix} \dfrac{c}{\sqrt{k_{mxr}\varepsilon_{xr}}} \\ \dfrac{c}{\sqrt{k_{myr}\varepsilon_{yr}}} \\ \dfrac{c}{\sqrt{k_{mzr}\varepsilon_{zr}}} \end{Bmatrix} = \begin{Bmatrix} \dfrac{v_{\varepsilon xr}}{\sqrt{k_{mxr}}} \\ \dfrac{v_{\varepsilon yr}}{\sqrt{k_{myr}}} \\ \dfrac{v_{\varepsilon zr}}{\sqrt{k_{mzr}}} \end{Bmatrix} \qquad (2.5.8r)$$

Combining equations 2.5.7 and 2.5.8, the Cartesian dielectric sub- and super- luminal velocities ($v_{\varepsilon x}$, $v_{\varepsilon y}$ & $v_{\varepsilon z}$) can be computed and are presented in Table 2.5.2.

**Table 2.5.2** Permittivities $(\varepsilon)$, Permeabilities $(k_m)$, Dielectric Velocities ($v_\varepsilon$)

| | | | | | | |
|---|---|---|---|---|---|---|
| Condition | 1.2 | 1.3 | 1.6 | 1.9 | 1.10 | 1.11 |
| $T_a\left({}^0K\right)$ | 6.53E+11 | 6.53E+11 | 6.53E+11 | 6.53E+11 | 6.53E+11 | 6.53E+11 |
| $T_b\left({}^0K\right)$ | 9.05E+14 | 5.48E+13 | 9.22E+08 | 1.10E+07 | 2,868 | 2.723 |
| $T_{br}\left({}^0K\right)$ | 3.30E+12 | 5.45E+13 | 3.24E+18 | 2.72E+20 | 1.04E+24 | 1.10E+27 |
| $T_c\left({}^0K\right)$ | 8.96E+13 | 3.28E+11 | 93.0 | 0.0131 | 8.99E-10 | 8.10E-16 |
| $T_{cr}\left({}^0K\right)$ | 1.19E+09 | 3.25E+11 | 1.15E+21 | 8.11E+24 | 1.19E+32 | 1.32E+38 |
| $z_R+1$ | 3.67E+43 | 8.14E+39 | 3.89E+25 | 6.53E+19 | 1.17E+09 | 1.00E+00 |
| $z_B+1$ | 3.32E+14 | 2.01E+13 | 3.39E+08 | 4.03E+06 | 1.05E+03 | 1.00E+00 |
| $Age_{UG}\,(sec)$ | 7.14E-05 | 4.79E-03 | 6.94E+04 | 5.36E+07 | 1.27E+13 | 4.33E+17 |
| $Age_{UGv}\,(sec)$ | 4.59E-10 | 1.25E-07 | 4.42E+02 | 3.13E+06 | 1.22E+13 | 4.33E+17 |
| $Age_U\,(sec)$ | 1.18E-26 | 5.31E-23 | 1.11E-08 | 6.63E-03 | 3.70E+08 | 4.33E+17 |
| $\varepsilon_x$ $D_{fx}/E_{fx}$ | 6.02E-02 | 9.96E-01 | 5.91E+04 | 4.97E+06 | 1.90E+10 | 2.00E+13 |
| $\varepsilon_y$ $D_{fy}/E_{fy}$ | 6.02E-02 | 9.96E-01 | 5.91E+04 | 4.97E+06 | 1.90E+10 | 2.00E+13 |
| $\varepsilon_z$ $D_{fz}/E_{fz}$ | 7.39E-04 | 1.22E-02 | 7.25E+02 | 6.10E+04 | 2.33E+08 | 2.46E+11 |
| $\varepsilon$ $D_f/E_f$ | 7.21E-04 | 1.19E-02 | 7.08E+02 | 5.96E+04 | 2.28E+08 | 2.40E+11 |
| $\varepsilon_{xr}$ $D_{fxr}/E_{fxr}$ | 1.65E+01 | 1.00E+00 | 1.68E-05 | 2.00E-07 | 5.24E-11 | 4.97E-14 |
| $\varepsilon_{yr}$ $D_{fyr}/E_{fyr}$ | 1.65E+01 | 1.00E+00 | 1.68E-05 | 2.00E-07 | 5.24E-11 | 4.97E-14 |
| $\varepsilon_{zr}$ $D_{fzr}/E_{fzr}$ | 2.03E-01 | 1.23E-02 | 2.07E-07 | 2.46E-09 | 6.42E-13 | 6.10E-16 |
| $\varepsilon_r$ $D_{fr}/E_{fr}$ | 1.98E-01 | 1.20E-02 | 2.02E-07 | 2.40E-09 | 6.27E-13 | 5.95E-16 |
| $k_{mx}$ | 6.05E-02 | 1.00E+00 | 5.94E+04 | 4.64E+06 | 1.83E+10 | 2.01E+13 |
| $k_{my}$ | 4.54E-02 | 7.50E-01 | 4.45E+04 | 3.48E+06 | 1.37E+10 | 1.51E+13 |
| $k_{mz}$ | 2.18E-01 | 3.60E+00 | 2.14E+05 | 1.67E+07 | 6.59E+10 | 7.24E+13 |
| $k_m$ | 2.02E-01 | 3.34E+00 | 1.99E+05 | 1.55E+07 | 6.12E+10 | 6.73E+13 |
| $k_{mxr}$ | 1.65E+01 | 1.00E+00 | 1.68E-05 | 2.15E-07 | 5.47E-11 | 4.97E-14 |
| $k_{myr}$ | 1.24E+01 | 7.50E-01 | 1.26E-05 | 1.61E-07 | 4.10E-11 | 3.73E-14 |
| $k_{mzr}$ | 5.98E+01 | 3.62E+00 | 6.09E-05 | 7.79E-07 | 1.98E-10 | 1.80E-13 |
| $k_{mr}$ | 5.55E+01 | 3.36E+00 | 5.65E-05 | 7.23E-07 | 1.84E-10 | 1.67E-13 |
| $v_{\varepsilon x}$ (cm/sec) | 1.22E+11 | 3.00E+10 | 1.23E+08 | 1.34E+07 | 2.17E+05 | 6.70E+03 |
| $v_{\varepsilon y}$ (cm/sec) | 1.22E+11 | 3.00E+10 | 1.23E+08 | 1.34E+07 | 2.17E+05 | 6.70E+03 |
| $v_{\varepsilon z}$ (cm/sec) | 1.10E+12 | 2.71E+11 | 1.11E+09 | 1.21E+08 | 1.96E+06 | 6.05E+04 |
| $v_\varepsilon$ (cm/sec) | 1.12E+12 | 2.75E+11 | 1.13E+09 | 1.23E+08 | 1.99E+06 | 6.12E+04 |
| $v_{\varepsilon xr}$ (cm/sec) | 7.36E+09 | 2.99E+10 | 7.29E+12 | 6.69E+13 | 4.13E+15 | 1.34E+17 |
| $v_{\varepsilon yr}$ (cm/sec) | 7.36E+09 | 2.99E+10 | 7.29E+12 | 6.69E+13 | 4.13E+15 | 1.34E+17 |
| $v_{\varepsilon zr}$ (cm/sec) | 6.66E+10 | 2.71E+11 | 6.60E+13 | 6.05E+14 | 3.74E+16 | 1.21E+18 |
| $v_{\varepsilon r}$ (cm/sec) | 6.74E+10 | 2.74E+11 | 6.68E+13 | 6.12E+14 | 3.79E+16 | 1.23E+18 |

Verification of equation 2.5.9 is indicated by the calculation that superconducting density $\left(m_T/cavity\right)$ (table 2.9.1) and present universe density (equation 2.11.21) are equal at this Debye black body temperature $\left(T_b\right)$ of $2.729\,{}^oK$.

$$T_b = \mathbb{C}\left(\frac{k_{mx}\varepsilon_x}{\varepsilon}\right)^2 T_c = \mathbb{C}\left(\frac{v_\varepsilon}{v_{dx}}\right)^2 T_c$$

$$= m_T v_\varepsilon^2 \frac{1}{k_b} \qquad (2.5.9)$$

*where*

$$m_T = \frac{1}{g_d^2 g_s}\frac{\mathbb{C}^2 e_\pm \hbar}{v_{\varepsilon z} H_{c1} cavity}$$

$$T_{br} = \mathbb{C}\left(\frac{k_{mxr}\varepsilon_{xr}}{\varepsilon_r}\right)^2 T_{cr} = \mathbb{C}\left(\frac{v_{\varepsilon r}}{v_{dxr}}\right)^2 T_{cr}$$

$$= m_t v_{\varepsilon r}^2 \frac{1}{k_b} \qquad (2.5.9r)$$

*where*

$$m_t = \frac{1}{g_d^2 g_s}\frac{\mathbb{C}^2 e_\pm \hbar}{v_{\varepsilon zr} H_{c1r} cavity_r}$$

Based on the Cartesian dielectric velocities ($v_{\varepsilon x}$, $v_{\varepsilon y}$ & $v_{\varepsilon z}$), corresponding Cartesian fluxoids can be computed as per equation 2.5.10.

$$\left\{\begin{array}{c} \Phi_{\varepsilon x} \\ \Phi_{\varepsilon y} \\ \Phi_{\varepsilon z} \\ \Phi_{\varepsilon} \end{array}\right\} = \left\{\begin{array}{c} \dfrac{2\pi\hbar v_{\varepsilon x}}{\mathbb{C}\ e_\pm} \\ \dfrac{2\pi\hbar v_{\varepsilon y}}{\mathbb{C}\ e_\pm} \\ \dfrac{2\pi\hbar v_{\varepsilon z}}{\mathbb{C}\ e_\pm} \\ \dfrac{2\pi\hbar v_{\varepsilon}}{\mathbb{C}\ e_\pm} \end{array}\right\} \qquad (2.5.10)$$

$$\left\{\begin{array}{c} \Phi_{\varepsilon xr} \\ \Phi_{\varepsilon yr} \\ \Phi_{\varepsilon zr} \\ \Phi_{\varepsilon r} \end{array}\right\} = \left\{\begin{array}{c} \dfrac{2\pi\hbar v_{\varepsilon xr}}{\mathbb{C}\ e_\pm} \\ \dfrac{2\pi\hbar v_{\varepsilon yr}}{\mathbb{C}\ e_\pm} \\ \dfrac{2\pi\hbar v_{\varepsilon zr}}{\mathbb{C}\ e_\pm} \\ \dfrac{2\pi\hbar v_{\varepsilon r}}{\mathbb{C}\ e_\pm} \end{array}\right\} \qquad (2.5.10r)$$

The trisine residence *resonant CPT* $time_\pm$ is confirmed in terms of conventional capacitance $(C)$ and inductance $(L)$ resonant circuit relationships in *x*, *y* and *z* as well as *trisine* dimensions.

$$time_\pm = \left\{\begin{array}{c} \sqrt{L_x}\sqrt{C_x} \\ \sqrt{L_y}\sqrt{C_y} \\ \sqrt{L_z}\sqrt{C_z} \\ \sqrt{L}\sqrt{C} \end{array}\right\} = \left\{\begin{array}{c} \sqrt{\dfrac{\Phi_{\varepsilon x}\ time_\pm}{4\pi\mathbb{C}\ e_\pm\ v_{\varepsilon x}}}\sqrt{\dfrac{approach}{2\pi}\dfrac{\varepsilon_x}{2B}} \\ \sqrt{\dfrac{\Phi_{\varepsilon y}\ time_\pm}{4\pi\mathbb{C}\ e_\pm\ v_{\varepsilon y}}}\sqrt{g_s\dfrac{\sqrt{3}}{2}\dfrac{side}{2\pi}\dfrac{\sqrt{3}\varepsilon_y}{3B}} \\ \sqrt{\dfrac{\Phi_{\varepsilon z}\ time_\pm}{4\pi\mathbb{C}\ e_\pm\ v_{\varepsilon z}}}\sqrt{\dfrac{2\sqrt{3}\pi}{3}\dfrac{section}{2\pi}\dfrac{\varepsilon_z}{A}} \\ \sqrt{\dfrac{\Phi_{\varepsilon}\ time_\pm}{4\pi\mathbb{C}\ e_\pm\ v_{\varepsilon}}}\sqrt{\dfrac{2\sqrt{3}\pi}{\cos(\theta)}\dfrac{section}{2\pi}\dfrac{\varepsilon}{A}} \end{array}\right\} \qquad (2.15.11)$$

$$time_{r\pm} = \left\{\begin{array}{c} \sqrt{L_{xr}}\sqrt{C_{xr}} \\ \sqrt{L_{yr}}\sqrt{C_{yr}} \\ \sqrt{L_{zr}}\sqrt{C_{zr}} \\ \sqrt{L_r}\sqrt{C_r} \end{array}\right\} = \left\{\begin{array}{c} \sqrt{\dfrac{\Phi_{\varepsilon xr}\ time_{r\pm}}{4\pi\mathbb{C}\ e_\pm\ v_{\varepsilon xr}}}\sqrt{\dfrac{approach_r}{2\pi}\dfrac{\varepsilon_{xr}}{2B_r}} \\ \sqrt{\dfrac{\Phi_{\varepsilon yr}\ time_{r\pm}}{4\pi\mathbb{C}\ e_\pm\ v_{\varepsilon yr}}}\sqrt{g_s\dfrac{\sqrt{3}}{2}\dfrac{side_r}{2\pi}\dfrac{\sqrt{3}\varepsilon_{yr}}{3B_r}} \\ \sqrt{\dfrac{\Phi_{\varepsilon z}\ time_{r\pm}}{4\pi\mathbb{C}\ e_\pm\ v_{\varepsilon zr}}}\sqrt{\dfrac{2\sqrt{3}\pi}{3}\dfrac{section_r}{2\pi}\dfrac{\varepsilon_{zr}}{A_r}} \\ \sqrt{\dfrac{\Phi_{\varepsilon}\ time_{r\pm}}{4\pi\mathbb{C}\ e_\pm\ v_{\varepsilon r}}}\sqrt{\dfrac{2\sqrt{3}\pi}{\cos(\theta)}\dfrac{section_r}{2\pi}\dfrac{\varepsilon}{A_r}} \end{array}\right\} \qquad (2.5.11r)$$

Where relationships between Cartesian and trisine capacitance and inductance is as follows:

$$C = C_x = C_y = C_z \qquad (2.5.12)$$

$$C_r = C_{xr} = C_{yr} = C_{zr} \qquad (2.5.12r)$$

$$L = L_x = L_y = L_z \qquad (2.5.13)$$

$$L_r = L_{xr} = L_{yr} = L_{zr} \qquad (2.5.13r)$$

**Table 2.5.3** Capacitance and Inductive Density

| | | | | | | |
|---|---|---|---|---|---|---|
| Condition | 1.2 | 1.3 | 1.6 | 1.9 | 1.10 | 1.11 |
| $T_a\left({}^0K\right)$ | 6.53E+11 | 6.53E+11 | 6.53E+11 | 6.53E+11 | 6.53E+11 | 6.53E+11 |
| $T_b\left({}^0K\right)$ | 9.05E+14 | 5.48E+13 | 9.22E+08 | 1.10E+07 | 2,868 | 2.723 |
| $T_{br}\left({}^0K\right)$ | 3.30E+12 | 5.45E+13 | 3.24E+18 | 2.72E+20 | 1.04E+24 | 1.10E+27 |
| $T_c\left({}^0K\right)$ | 8.96E+13 | 3.28E+11 | 93.0 | 0.0131 | 8.99E-10 | 8.10E-16 |
| $T_{cr}\left({}^0K\right)$ | 1.19E+09 | 3.25E+11 | 1.15E+21 | 8.11E+24 | 1.19E+32 | 1.32E+38 |
| $z_R+1$ | 3.67E+43 | 8.14E+39 | 3.89E+25 | 6.53E+19 | 1.17E+09 | 1.00E+00 |
| $z_B+1$ | 3.32E+14 | 2.01E+13 | 3.39E+08 | 4.03E+06 | 1.05E+03 | 1.00E+00 |
| $Age_{UG}\,(sec)$ | 7.14E-05 | 4.79E-03 | 6.94E+04 | 5.36E+07 | 1.27E+13 | 4.33E+17 |

| $Age_{UGv}\,(sec)$ | 4.59E-10 | 1.25E-07 | 4.42E+02 | 3.13E+06 | 1.22E+13 | 4.33E+17 |
|---|---|---|---|---|---|---|
| $Age_{U}\,(sec)$ | 1.18E-26 | 5.31E-23 | 1.11E-08 | 6.63E-03 | 3.70E+08 | 4.33E+17 |
| $C$ | 2.36E-16 | 6.46E-14 | 2.28E-04 | 1.61E+00 | 2.36E+07 | 2.61E+13 |
| $C$ | 1.78E-11 | 6.52E-14 | 1.85E-23 | 2.61E-27 | 1.79E-34 | 1.61E-40 |
| $C/\mathrm{v}$ ( /cm$^3$) | 5.52E+23 | 3.34E+22 | 5.62E+17 | 6.68E+15 | 1.75E+12 | 1.66E+09 |
| $C/\mathrm{v}$ ( /cm$^3$) | 2.01E+21 | 3.32E+22 | 1.97E+27 | 1.66E+29 | 6.35E+32 | 6.69E+35 |
| $C/\mathrm{v}$ (fd/cm$^3$) | 6.14E+11 | 3.72E+10 | 6.26E+05 | 7.44E+03 | 1.95E+00 | 1.85E-03 |
| $C/\mathrm{v}$ (fd/cm$^3$) | 2.24E+09 | 3.70E+10 | 2.20E+15 | 1.85E+17 | 7.06E+20 | 7.44E+23 |
| $L$ | 3.06E-34 | 8.36E-32 | 2.95E-22 | 2.09E-18 | 3.05E-11 | 3.38E-05 |
| $L$ | 2.30E-29 | 8.44E-32 | 2.39E-41 | 3.38E-45 | 2.31E-52 | 2.09E-58 |
| $L/\mathrm{v}$ ( /cm$^3$) | 7.14E+05 | 4.32E+04 | 7.28E-01 | 8.65E-03 | 2.26E-06 | 2.15E-09 |
| $L/\mathrm{v}$ ( /cm$^3$) | 2.60E+03 | 4.30E+04 | 2.56E+09 | 2.15E+11 | 8.22E+14 | 8.66E+17 |
| $L/\mathrm{v}$ (h/cm$^3$) | 6.42E+17 | 3.89E+16 | 6.54E+11 | 7.78E+09 | 2.03E+06 | 1.93E+03 |
| $L/\mathrm{v}$ ($h/cm^3$) | 2.34E+15 | 3.87E+16 | 2.30E+21 | 1.93E+23 | 7.39E+26 | 7.78E+29 |

In terms of an extended Thompson cross-section $(\sigma_T)$, it is noted that the $(1/R^2)$ factor [62] is analogous to a dielectric $(\varepsilon)$ equation 2.5.2. Dimensionally the dielectric $(\varepsilon)$ is proportional to trisine cell length dimension.

The extended Thompson scattering cross section $(\sigma_T)$ [43 equation 78.5, 62 equation 33] then becomes as in equation 2.5.14.

$$\sigma_T = \left(\frac{B}{A}\right)\left(\frac{\mathbb{C}^2 e_\pm^2}{\varepsilon m_t v_{dy}^2}\right)^2 \approx side \tag{2.5.14}$$

Force(F) relationships are maintained

$$\left.\begin{matrix} F_x \\ F_y \\ F_z \\ F \end{matrix}\right\} = \left\{\begin{matrix} \dfrac{\hbar K_B}{time_\pm} \\ \dfrac{\hbar K_P}{time_\pm} \\ \dfrac{\hbar K_A}{time_\pm} \\ \dfrac{\hbar\sqrt{K_B^2 + K_P^2 + K_A^2}}{time_\pm} \end{matrix}\right\} \tag{2.5.15}$$

$$\left.\begin{matrix} F_{xr} \\ F_{yr} \\ F_{zr} \\ F_r \end{matrix}\right\} = \left\{\begin{matrix} \dfrac{\hbar K_{Br}}{time_{r\pm}} \\ \dfrac{\hbar K_{Pr}}{time_{r\pm}} \\ \dfrac{\hbar K_{Ar}}{time_{r\pm}} \\ \dfrac{\hbar\sqrt{K_{Br}^2 + K_{Pr}^2 + K_{Ar}^2}}{time_{r\pm}} \end{matrix}\right\} \tag{2.5.15r}$$

$$\left.\begin{matrix} F_x \\ F_y \\ F_z \\ F \end{matrix}\right\} = \left\{\begin{matrix} \dfrac{m_T v_{dx}}{time_\pm} \\ \dfrac{m_T v_{dy}}{time_\pm} \\ \dfrac{m_T v_{dz}}{time_\pm} \\ \dfrac{m_T\sqrt{v_{dx}^2 + v_{dy}^2 + v_{dz}^2}}{time_\pm} \end{matrix}\right\} \tag{2.5.16}$$

$$\left.\begin{matrix} F_{xr} \\ F_{yr} \\ F_{zr} \\ F_r \end{matrix}\right\} = \left\{\begin{matrix} \dfrac{m_T v_{dxr}}{time_{r\pm}} \\ \dfrac{m_T v_{dyr}}{time_{r\pm}} \\ \dfrac{m_T v_{dyr}}{time_{r\pm}} \\ \dfrac{m_T\sqrt{v_{dxr}^2 + v_{dyr}^2 + v_{dzr}^2}}{time_{r\pm}} \end{matrix}\right\} \tag{2.5.16r}$$

$$\left.\begin{matrix} F_x \\ F_y \\ F_z \\ F \end{matrix}\right\} = \left\{\begin{matrix} \mathbb{C} e E_{fx} \\ \mathbb{C} e E_{fy} \\ \mathbb{C} e E_{fz} \\ \mathbb{C} e\sqrt{E_{fx}^2 + E_{fy}^2 + E_{fz}^2} \end{matrix}\right\} \tag{2.5.17}$$

$$\left\{\begin{matrix} F_{xr} \\ F_{yr} \\ F_{zr} \\ F_r \end{matrix}\right\} = \left\{\begin{matrix} \mathbb{C}eE_{fxr} \\ \mathbb{C}eE_{fyr} \\ \mathbb{C}eE_{fzr} \\ \mathbb{C}e\sqrt{E_{fxr}^2 + E_{fyr}^2 + E_{fzr}^2} \end{matrix}\right\} \tag{2.5.17r}$$

$$\left\{\begin{matrix} F_x \\ F_y \\ F_z \\ F \end{matrix}\right\} = \left\{\begin{matrix} 2\dfrac{1}{g_s^2}\left(\dfrac{m_T}{m_e}\right)\dfrac{(\mathbb{C}e)^2}{\varepsilon_x}\left(\dfrac{1}{2B}\right)^2 \\ \dfrac{3}{4}\left(\dfrac{m_T}{m_e}\right)\dfrac{(\mathbb{C}e)^2}{\varepsilon_y}\left(\dfrac{\sqrt{3}}{2B}\right)^2 \\ \dfrac{1}{g_s^2}\dfrac{1}{2\cos(\theta)}\dfrac{(\mathbb{C}e)^2}{\varepsilon_z}\left(\dfrac{1}{A}\right)^2 \\ \sqrt{F_x^2 + F_y^2 + F_z^2} \end{matrix}\right\} \tag{2.5.18}$$

$$\left\{\begin{matrix} F_{xr} \\ F_{yr} \\ F_{zr} \\ F_r \end{matrix}\right\} = \left\{\begin{matrix} 2\dfrac{1}{g_s^2}\left(\dfrac{m_T}{m_e}\right)\dfrac{(\mathbb{C}e)^2}{\varepsilon_{xr}}\left(\dfrac{1}{2B_r}\right)^2 \\ \dfrac{3}{4}\left(\dfrac{m_T}{m_e}\right)\dfrac{(\mathbb{C}e)^2}{\varepsilon_{yr}}\left(\dfrac{\sqrt{3}}{2B_r}\right)^2 \\ \dfrac{1}{g_s^2}\dfrac{1}{2\cos(\theta)}\dfrac{(\mathbb{C}e)^2}{\varepsilon_{zr}}\left(\dfrac{1}{A_r}\right)^2 \\ \sqrt{F_{xr}^2 + F_{yr}^2 + F_{zr}^2} \end{matrix}\right\} \tag{2.5.18r}$$

## 2.6 Fluxoid And Critical Fields

Based on the material fluxoid $(\Phi_\varepsilon)$ (equation 2.5.6), the critical fields $(H_{c1})$ (equation 2.6.1), $(H_{c2})$ (equation 2.6.2) & $(H_c)$ (equation 2.6.3) as well as penetration depth $(\lambda)$ (equation 2.6.5) and Ginzburg-Landau coherence length $(\ell)$ (equation 2.6.6) are computed. Also the critical field $(H_c)$ is alternately computed from a variation on the Biot-Savart law.

$$H_{c1} = \frac{\Phi_\varepsilon}{\pi\lambda^2} \tag{2.6.1}$$

$$H_{c1r} = \frac{\Phi_\varepsilon}{\pi\lambda^2} \tag{2.6.1r}$$

$$H_{c2} = \frac{4\pi\ \mathbb{C}}{section}\Phi_\varepsilon \tag{2.6.2}$$

$$H_{c2r} = \frac{4\pi\ \mathbb{C}}{section_r}\Phi_{\varepsilon r} \tag{2.6.2r}$$

$$H_c = \sqrt{H_{c1}H_{c2}} = g_s^2\frac{B}{A}\sqrt{\pi D_f E_f} = \frac{g_s^3 m_T v_{\varepsilon x}}{3e_\pm time_\pm} \tag{2.6.3}$$

$$H_{cr} = \sqrt{H_{c1r}H_{c2r}} = g_s^2\frac{B_r}{A_r}\sqrt{\pi D_f E_f} = \frac{g_s^3 m_T v_{\varepsilon xr}}{3e_\pm time_{r\pm}} \tag{2.6.3r}$$

$$n_c = \frac{\mathbb{C}}{cavity} \tag{2.6.4}$$

$$n_c = \frac{\mathbb{C}}{cavity} \tag{2.6.4r}$$

where $n_c$ = Ginzburg-Landau order parameter as indicated in equation D.1 in Appendix D. (2.6.4r)

And the fluxoid linked the resonant time as follows:

$$time_\pm = \frac{2eA}{\alpha^2 v_{ex}\Phi_\varepsilon} \tag{2.6.5}$$

$$time_{r\pm} = \frac{2eA_r}{\alpha^2 v_{exr}\Phi_{\varepsilon r}} \tag{2.6.5r}$$

Huxley [70] observed that an increasing magnetic field:

quenched the URhGe superconductor at 2 tesla (20,000 gauss) initiated the URhGe superconductor at 8 tesla (80,000 gauss) quenched the URhGe superconductor at 13 tesla (130,000 gauss)

The trisine model $(H_{c2})$ correlates well with the reported [70] 2 tesla (20,000 gauss) with a calculated modeled superconductor magnetic critical field value $(H_{c2})$ of 1.8 tesla (18,000 gauss) for critical temperature $(T_c)$ of .28 K.

At critical temperature $(T_c)$ of .4 kelvin, the trisine model predicts superconductor magnetic critical field $(H_{c2})$ of 2.9 tesla (29,000 gauss), which is substantially lower than the observed 13 tesla (130,000 gauss). This could be explained in terms of the ferromagnetic properties of URhGe, which essentially mask the superconductor critical magnetic field. But it still remarkable how well the model correlates to the observed superconducting magnetic critical field $(H_{c2})$ in this case.

In as much as the trisine lattice can be considered a plasma $\mathbb{C}\rho_U = \mathbb{C}m_T/cavity = \mathbb{C}M_U/V_U$, then trisine de Broglie velocities can be computed as plasma Alfvén speeds as per equation 2.6.6. Corresponding plasma Alfvén times would be congruent with $time_\pm = B/v_{dx}$.

$$\begin{Bmatrix} v_{dx} \\ v_{dy} \\ v_{dz} \end{Bmatrix} = \begin{Bmatrix} \frac{4}{3\sqrt{3}} \frac{A}{B} \frac{H_c}{\sqrt{\mathbb{C}\rho_U}} \\ \frac{2}{3} \frac{1}{\sqrt{3}} \frac{H_c}{\sqrt{\mathbb{C}\rho_U}} \\ \frac{2}{3} \frac{B}{A} \frac{H_c}{\sqrt{\mathbb{C}\rho_U}} \end{Bmatrix} \qquad (2.6.6)$$

$$\begin{Bmatrix} v_{dxr} \\ v_{dyr} \\ v_{dzr} \end{Bmatrix} = \begin{Bmatrix} \frac{4}{3\sqrt{3}} \frac{A_r}{B_r} \frac{H_{cr}}{\sqrt{\mathbb{C}\rho_{Ur}}} \\ \frac{2}{3} \frac{1}{\sqrt{3}} \frac{H_{cr}}{\sqrt{\mathbb{C}\rho_{Ur}}} \\ \frac{2}{3} \frac{B_r}{A_r} \frac{H_{cr}}{\sqrt{\mathbb{C}\rho_{Ur}}} \end{Bmatrix} \qquad (2.6.6r)$$

Also it interesting to note that the following relationship holds

$$2\pi \frac{e_{\pm}^2}{A} \frac{e_{\pm}^2}{B} = m_e v_{dx}^2 m_e c^2 \qquad (2.6.7)$$

$$2\pi \frac{e_{\pm}^2}{A_r} \frac{e_{\pm}^2}{B_r} = m_e v_{dxr}^2 m_e c^2 \qquad (2.6.7r)$$

$$\lambda^2 = m_T v_{\varepsilon}^2 \frac{1}{n_c} \frac{1}{e^2} \qquad (2.6.8)$$

$$\lambda_r^2 = m_T v_{\varepsilon r}^2 \frac{1}{n_{cr}} \frac{1}{e^2} \qquad (2.6.8r)$$

From equation 2.6.8, the trisine penetration depth $\lambda$ is calculated. This is plotted in the above figure along with Harshman[17] and Homes[51, 59] penetration data. The Homes data is multiplied by a factor of $(n_e \, chain/\mathbb{C})^{1/2}$ to get $n_c$ in equation 2.6.8. This is consistent with $\lambda^2 = 1/(\varepsilon n_c)$ in equation 2.6.8 and consistent with Uemura Plot $\lambda^2 = 1/n_c$ over short increments of penetration depth $\lambda$. It appears to be a good fit and better than the fit with data compiled by Harshman[17].

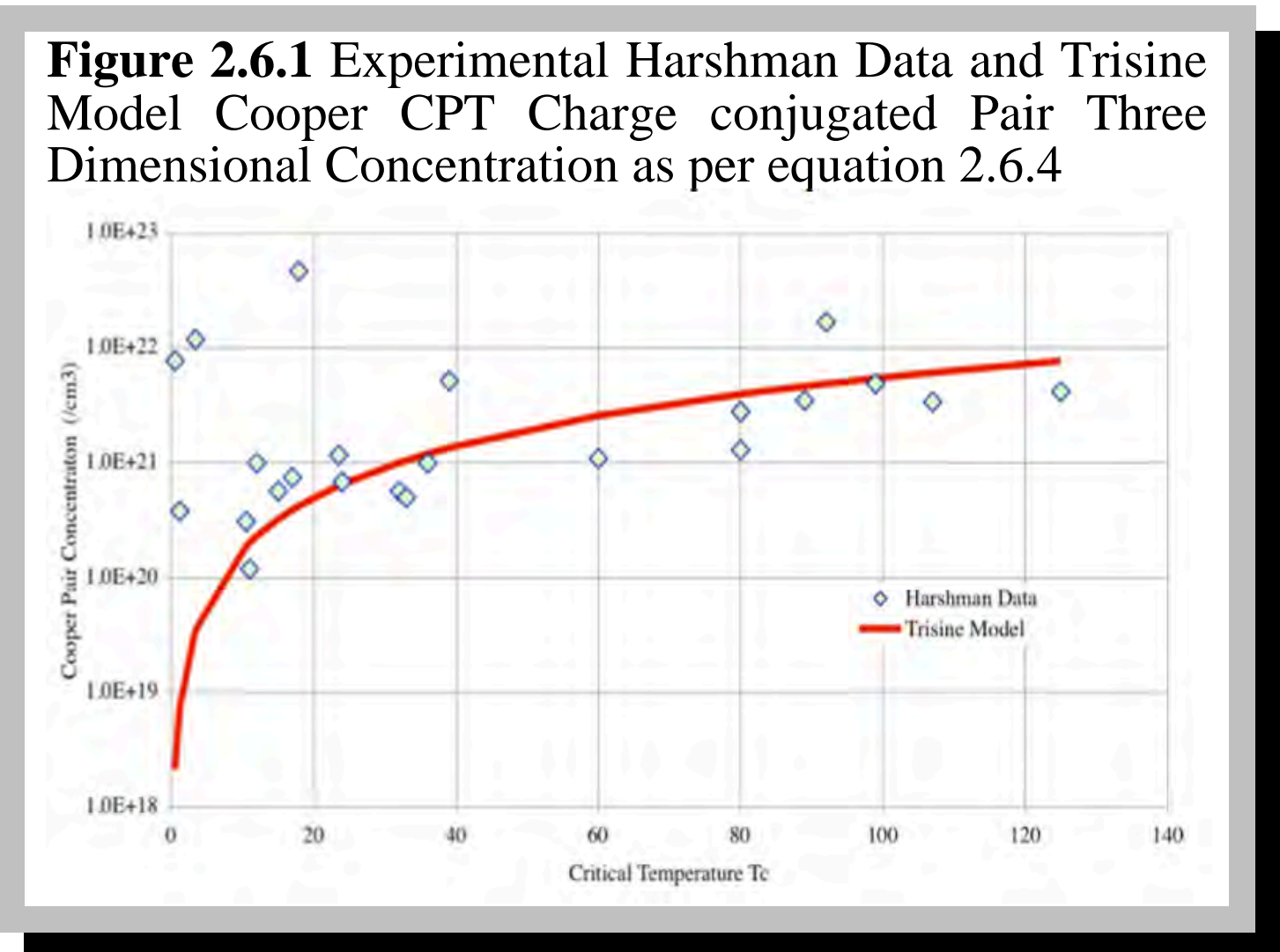

**Figure 2.6.1** Experimental Harshman Data and Trisine Model Cooper CPT Charge conjugated Pair Three Dimensional Concentration as per equation 2.6.4

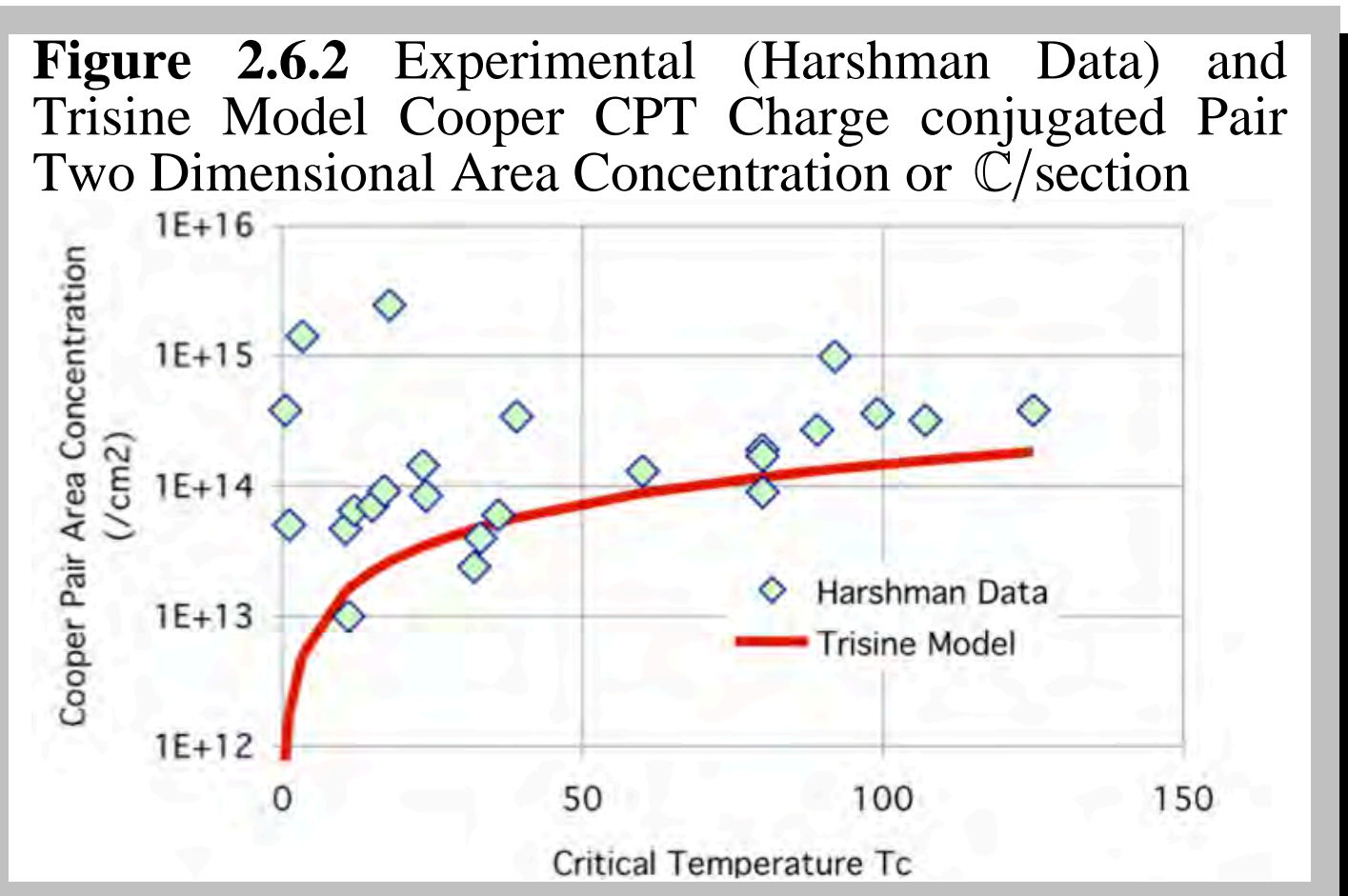

**Figure 2.6.2** Experimental (Harshman Data) and Trisine Model Cooper CPT Charge conjugated Pair Two Dimensional Area Concentration or $\mathbb{C}$/section

Penetration depth and gap data for magnesium diboride $MgB_2$ [55], which has a critical temperature $T_c$ of 39 K indicates a fit to the trisine model when a critical temperature of 39/3 or 13 K is assumed (see Table 2.6.1). This is interpreted as an indication that $MgB_2$ is a three dimensional superconductor.

**Table 2.6.1** Experimental ($MgB_2$ Data [55]) and the related Trisine Model Prediction

| | Trisine Data at $T_c$ 39/3 K or 13 K | Observed Data [55] |
|---|---|---|
| Penetration Depth (nm) | 276 | 260 ±20 |
| Gap (meV) | 1.12 | 3.3/3 or 1.1 ±.3 |

As indicated in equation 2.5.5, the material medium dielectric sub- and super- luminal speed of light $(v_\varepsilon)$ is a function of medium dielectric $(\varepsilon)$ constant, which is calculated in general, and specifically for trisine by $\varepsilon = D/E$ (equation 2.5.4). The fact that chain rather than cavity (related by 2/3 factor) must be used in obtaining a trisine model fit to Homes' data indicates that the stripe concept may be valid in as much as chains in the trisine lattice visually form a striped pattern as in Figure 2.3.1.

See derivation of Ginzburg-Landau coherence length $\ell$ in equation 2.6.9 in Appendix D. It is interesting to note that $B/\xi = 2.73 \approx \mathrm{e}$ for all $T_c$, which makes the trisine superconductor mode fit the conventional description of operating in the "dirty" limit. [54]

$$\xi = \sqrt{\frac{m_t}{m_e}\frac{1}{K_B^2}\frac{chain}{cavity}} \tag{2.6.9}$$

The geometry presented in Figure 2.6.4 is produced by three intersecting coherent polarized standing waves, which can be called the trisine wave function $\Psi$. This can be compared to figure 2.6.5, which presents the dimensional trisine geometry. Equation 2.6.10 represents the trisine wave function $\Psi$ that is analogous to Bloch sphere vector addition of three sin functions that are 120 degrees from each other in the x y plane and form an angle 22.8 degrees with this same x y plane. Also note that 22.8 degrees is related to the $(B/A)$ ratio by $\tan\left(90^0 - 22.8^0\right) = B/A$.

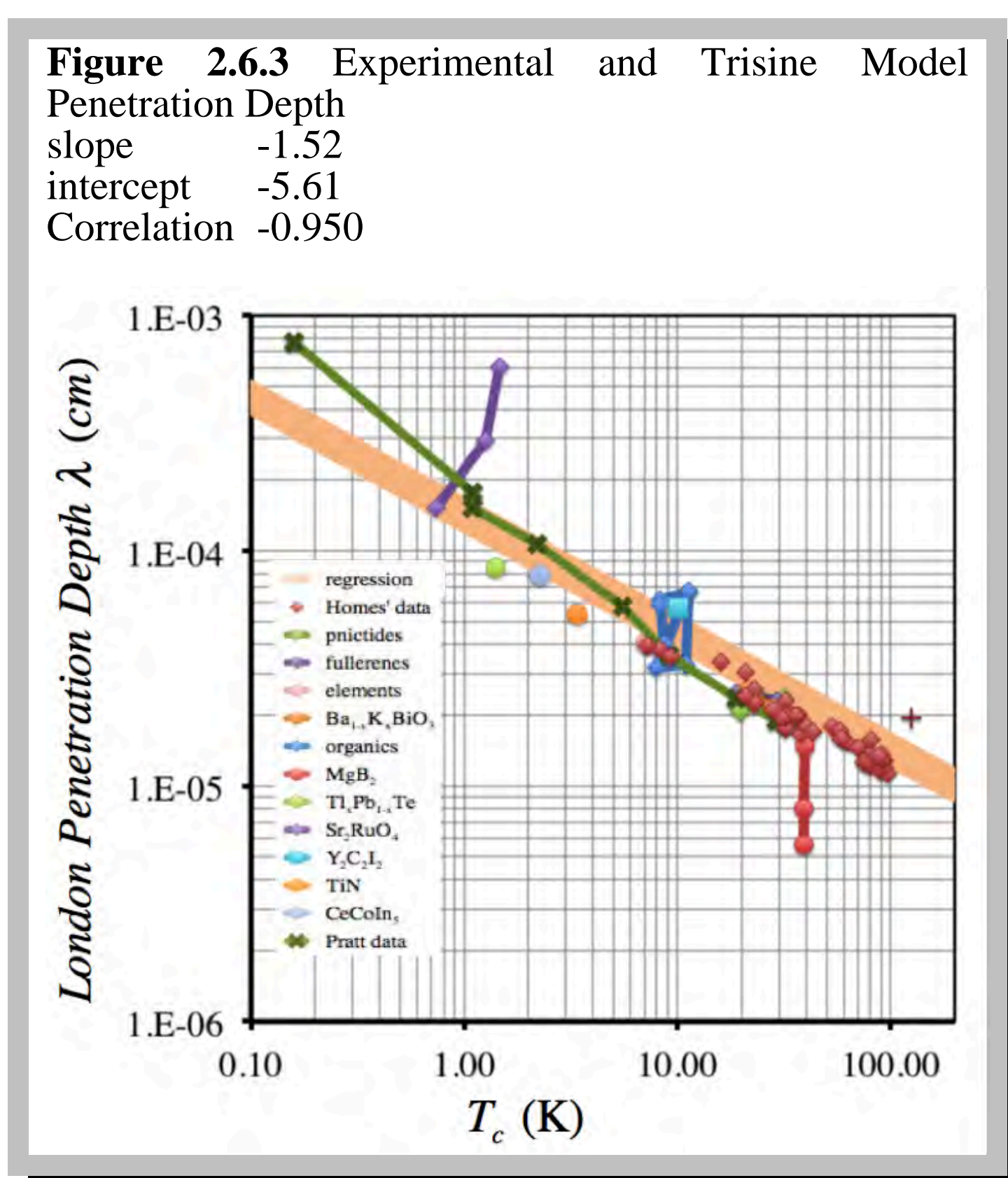

**Figure 2.6.3** Experimental and Trisine Model Penetration Depth
slope -1.52
intercept -5.61
Correlation -0.950

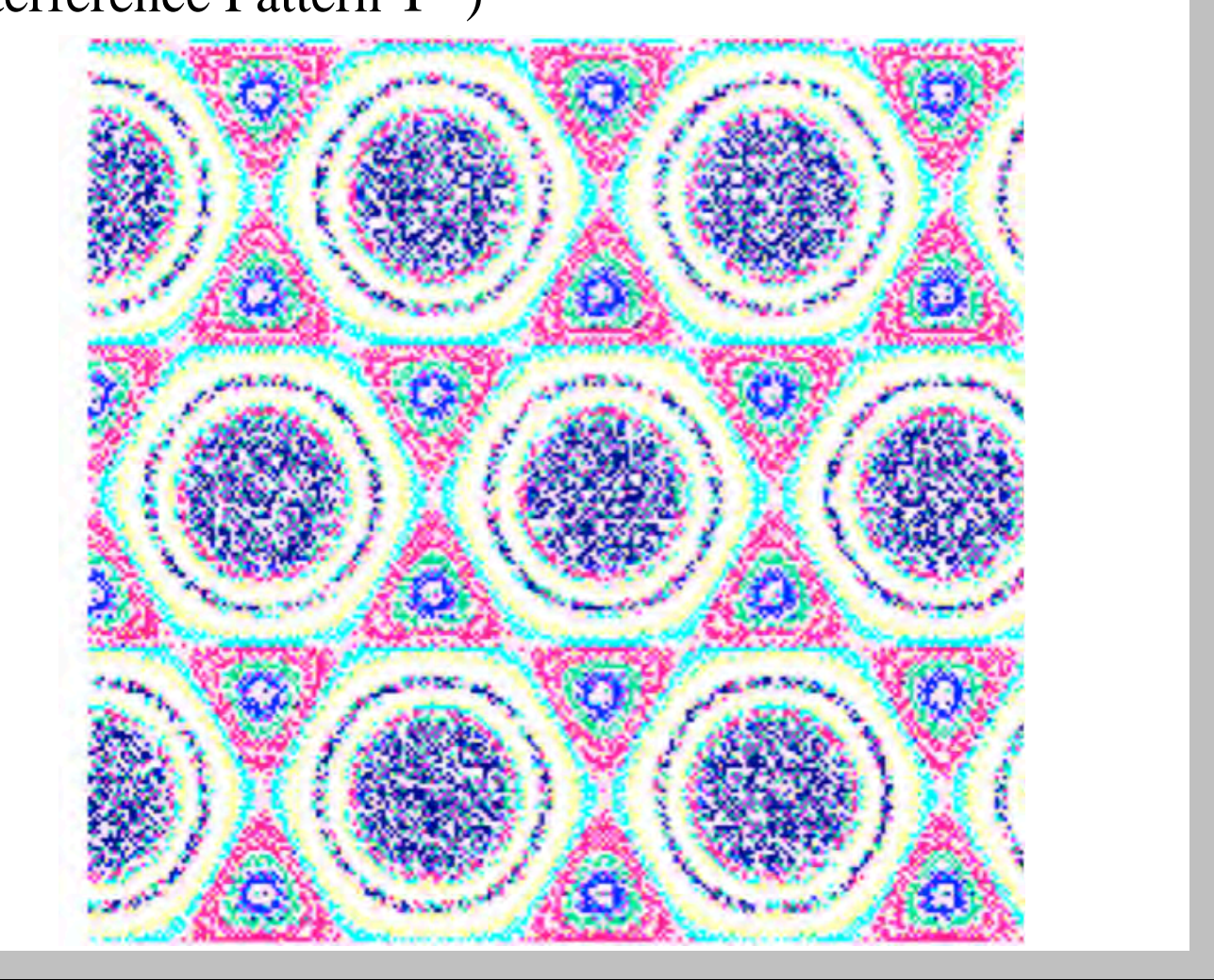

**Figure 2.6.4** Trisine Geometry in the Fluxoid Mode (Interference Pattern $\Psi^2$)

$$\begin{aligned}\Psi &= \exp\left(\sin\left(2\left(a_{x1}x_0 + b_{x1}y_0 + c_{x1}z_0\right)\right)\right) \\ &+ \exp\left(\sin\left(2\left(a_{x2}x_0 + b_{x2}y_0 + c_{x2}z_0\right)\right)\right) \\ &+ \exp\left(\sin\left(2\left(a_{x3}x_0 + b_{x3}y_0 + c_{x3}z_0\right)\right)\right)\end{aligned} \tag{2.6.10}$$

Where:

$a_{x1} = \cos\left(030^0\right)\cos\left(22.8^0\right) \quad b_{x1} = \sin\left(030^0\right)\cos\left(22.8^0\right)$

$c_{x1} = -\sin\left(030^0\right)$

$a_{x2} = \cos\left(150^0\right)\cos\left(22.8^0\right) \quad b_{x2} = \sin\left(150^0\right)\cos\left(22.8^0\right)$

$c_{x2} = -\sin\left(150^0\right)$

$a_{x3} = \cos\left(270^0\right)\cos\left(22.8^0\right) \quad b_{x3} = \sin\left(270^0\right)\cos\left(22.8^0\right)$

$c_{x3} = -\sin\left(270^0\right)$

Functionally, the interference of these planar intersecting waves would generate Laguerre-Gaussian optical vortices (fluxoid mode and its related superconducting mode). Equation 2.6.11 is an expression of reference [96] equation A36 in terms of trisine dimensionality.

$$\frac{E_z}{P} - 3\frac{E_y}{A} + \frac{H_{c2}}{v_\varepsilon time_\pm} \sim 0 \tag{2.6.11}$$

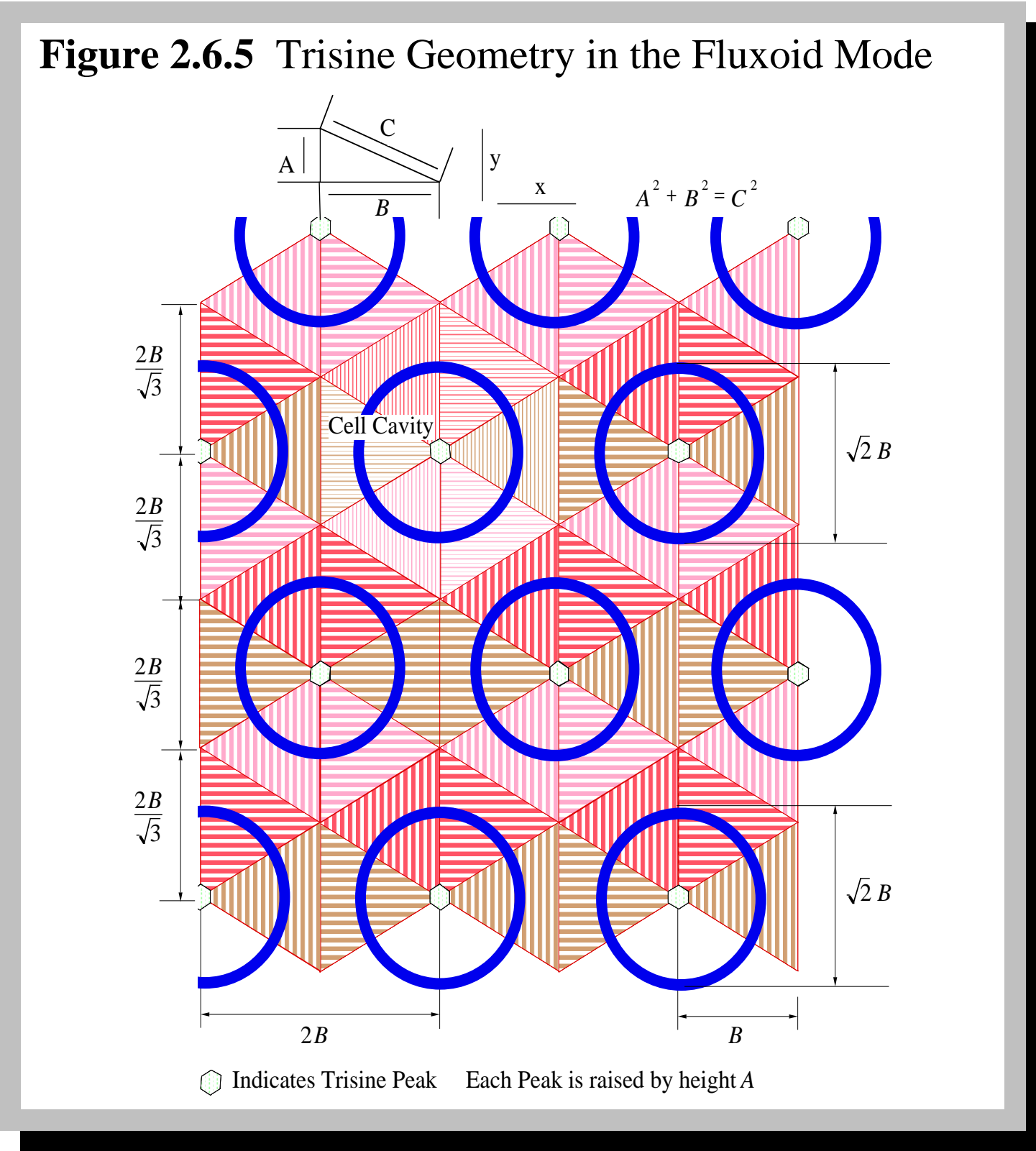

**Figure 2.6.5** Trisine Geometry in the Fluxoid Mode

**Table 2.6.2** Critical Fields and Lengths

| Condition | 1.2 | 1.3 | 1.6 | 1.9 | 1.10 | 1.11 |
|---|---|---|---|---|---|---|
| $T_a\left({}^0K\right)$ | 6.53E+11 | 6.53E+11 | 6.53E+11 | 6.53E+11 | 6.53E+11 | 6.53E+11 |
| $T_b\left({}^0K\right)$ | 9.05E+14 | 5.48E+13 | 9.22E+08 | 1.10E+07 | 2,868 | 2.723 |
| $T_{br}\left({}^0K\right)$ | 3.30E+12 | 5.45E+13 | 3.24E+18 | 2.72E+20 | 1.04E+24 | 1.10E+27 |
| $T_c\left({}^0K\right)$ | 8.96E+13 | 3.28E+11 | 93.0 | 0.0131 | 8.99E-10 | 8.10E-16 |
| $T_{cr}\left({}^0K\right)$ | 1.19E+09 | 3.25E+11 | 1.15E+21 | 8.11E+24 | 1.19E+32 | 1.32E+38 |
| $z_R + 1$ | 3.67E+43 | 8.14E+39 | 3.89E+25 | 6.53E+19 | 1.17E+09 | 1.00E+00 |
| $z_B + 1$ | 3.32E+14 | 2.01E+13 | 3.39E+08 | 4.03E+06 | 1.05E+03 | 1.00E+00 |
| $Age_{UG}\,(sec)$ | 7.14E-05 | 4.79E-03 | 6.94E+04 | 5.36E+07 | 1.27E+13 | 4.33E+17 |
| $Age_{UGv}\,(sec)$ | 4.59E-10 | 1.25E-07 | 4.42E+02 | 3.13E+06 | 1.22E+13 | 4.33E+17 |
| $Age_U\,(sec)$ | 1.18E-26 | 5.31E-23 | 1.11E-08 | 6.63E-03 | 3.70E+08 | 4.33E+17 |
| $H_c$ (gauss) | 3.27E+19 | 2.94E+16 | 3.42E+04 | 5.27E-01 | 5.83E-10 | 1.62E-17 |
| $H_{cr}$ (gauss) | 2.62E+13 | 2.91E+16 | 2.50E+28 | 1.62E+33 | 1.47E+42 | 5.28E+49 |
| $H_{c1}$ (gauss) | 4.22E+16 | 3.80E+13 | 4.43E+01 | 6.82E-04 | 7.55E-13 | 2.10E-20 |
| $H_{c1r}$ (gauss) | 3.39E+10 | 3.76E+13 | 3.23E+25 | 2.10E+30 | 1.90E+39 | 6.83E+46 |
| $H_c \frac{c}{v_{dx}} \frac{m_e}{m_t} \frac{2}{3}$ | 2.04E+20 | 1.83E+17 | 2.13E+05 | 3.29E+00 | 3.64E-09 | 1.01E-16 |
| $H_{cr} \frac{c}{v_{dxr}} \frac{m_e}{m_t} \frac{2}{3}$ | 2.63E+12 | 1.76E+14 | 2.55E+21 | 1.97E+24 | 4.66E+29 | 1.59E+34 |
| $H_{c2}$ (gauss) | 2.53E+22 | 2.27E+19 | 2.65E+07 | 4.08E+02 | 4.51E-07 | 1.25E-14 |
| $H_{c2r}$ (gauss) | 6.13E+13 | 6.81E+16 | 5.85E+28 | 3.80E+33 | 3.43E+42 | 1.24E+50 |
| $n_c$ (/cm$^3$) | 4.66E39 | 1.91E22 | 1.03E21 | 1.03E16 | 1.69E05 | 1.27E-04 |
| $n_{cr}$ (/cm$^3$) | 2.26E+32 | 1.02E+36 | 2.14E+50 | 1.27E+56 | 7.11E+66 | 8.30E+75 |
| $\lambda$ (cm) | 1.08E-11 | 1.78E-10 | 1.06E-05 | 8.90E-04 | 3.40E+00 | 3.58E+03 |
| $\lambda_r$ (cm) | 1.79E-11 | 1.08E-12 | 1.82E-17 | 2.17E-19 | 5.67E-23 | 5.38E-26 |
| $\xi$ (cm) | 1.81E-13 | 3.00E-12 | 1.78E-07 | 1.50E-05 | 5.73E-02 | 60.31 |
| $\xi_r$ (cm) | 4.98E-11 | 3.01E-12 | 5.07E-17 | 6.03E-19 | 1.58E-22 | 1.50E-25 |

In table 2.6.2, note that the ratio of London penetration length $(\lambda)$ to Ginzburg-Landau correlation length $(\ell)$ is the constant $(\kappa)$ of 58. The trisine values compares favorably with London penetration length $(\lambda)$ and constant $(\kappa)$ of 1155(3) Å and 68 respectively as reported in reference [11, 17] based on experimental data.

Measured intergalactic magnetic field (IGMF)[92,93] ~ 1E-15 Gauss is consistent with critical field $(H_c)$ value 1.64E-17 Gauss in Table 2.6.2 further verifying the trisine model at the universe 'dark energy' scale. Also, in this context, the quantity $H_c(c/v_{dx})(m_e/m_t)(2/3)$ is considered a magnetic B field that approximates Interstellar Magnetic Field (ISMF) at about 2E-6 Gauss[78,140].

**Figure 2.6.6** Critical Magnetic Field Data (gauss) compared to Trisine Model as a function of $T_c$

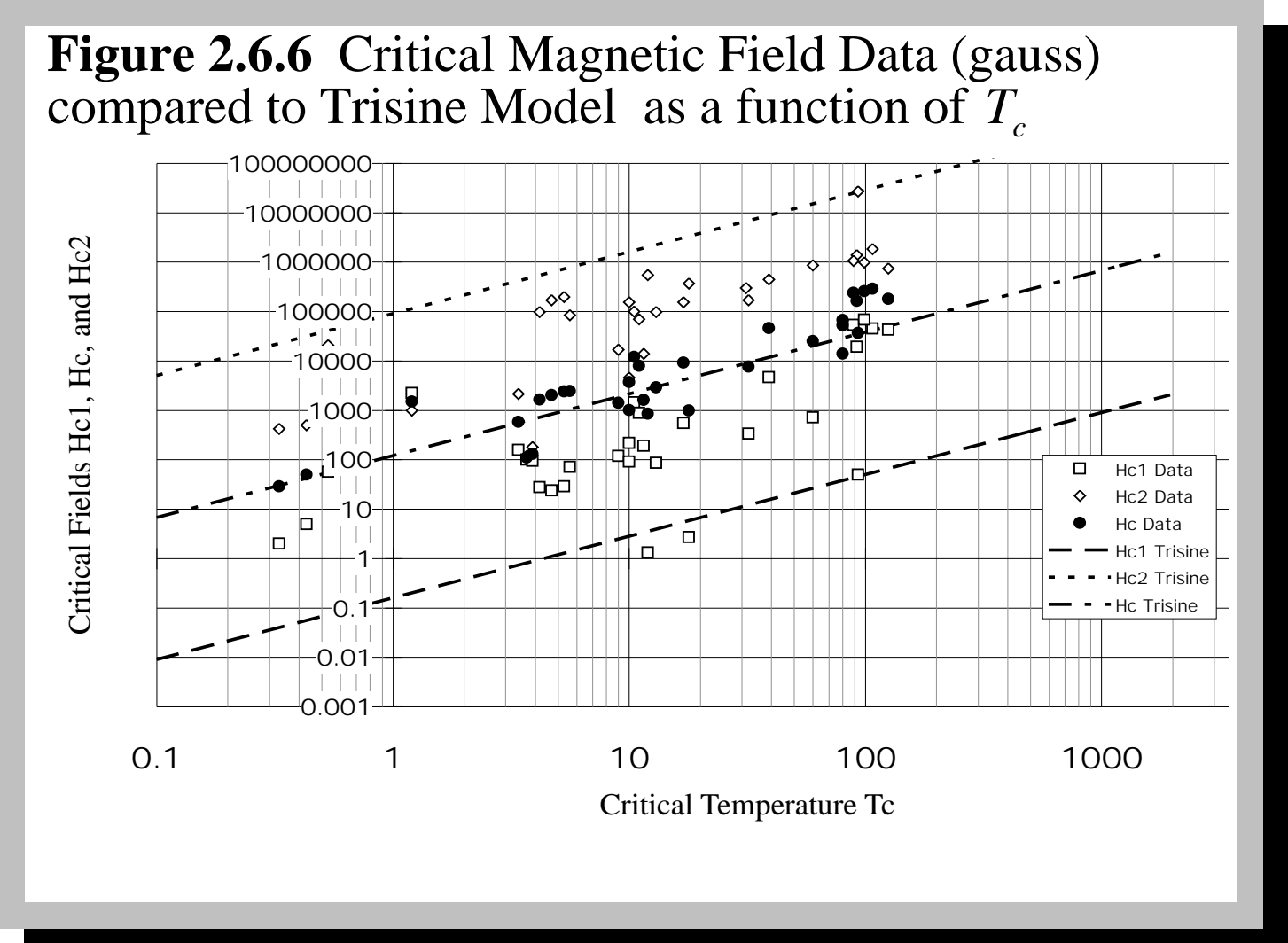

Figure 2.6.6 is graph of generally available critical field data as compile in Weast [16].

## 2.7 Superconductor Internal Pressure, Casimir Force And Deep Space Thrust Force (Pioneer 10 & 11)

The following pressure $(p_U)$ values in Table 2.7.1 as based on equations 2.7.1 indicate rather high values as superconducting $(T_c)$ increases. This is an indication of the forces that must be contained in these anticipated materials or lattices within potential barrier(c) restrictions expressed in 1.1 – 1.11. These forces are rather large for conventional materials especially at nuclear dimensions, but range to very small to the virtual lattice existing in outer space and are considered repulsive Casimir forces (eqn 2.7.1) on evaluation of eqn 7.44, Milonni[45] with $(\pm\varepsilon_1 = \pm\varepsilon_3 = \pm\varepsilon_2)$ and $(\pm K_1 = \mp K_3 = \pm K_2)$ with the polarities $(\pm, \mp)$ changing in the context of CPT.

$$Casimir\ Force(repulsive) = \frac{\hbar K_{B\mp}}{time_{\pm}} \tag{2.7.1}$$

$$p_U = \begin{Bmatrix} \dfrac{\hbar K_B}{time_{\pm}\ approach} \\ \dfrac{\hbar K_P}{time_{\pm}\ side} \\ \dfrac{\hbar K_A}{time_{\pm}\ section} \end{Bmatrix} = \begin{Bmatrix} \dfrac{F_x}{approach} \\ \dfrac{F_y}{side} \\ \dfrac{F_z}{section} \end{Bmatrix} = \frac{\mathbb{C} k T_c}{cavity} \tag{2.7.1a}$$

$$p_U = \frac{\sqrt{A^2 + B^2}}{B} \frac{cavity}{chain} \frac{\pi^2 \hbar\ v_{dx}}{240\ A^4} \quad \begin{pmatrix} \text{Casimir} \\ \text{pressure} \end{pmatrix} \tag{2.7.1b}$$

$$p_U = \begin{Bmatrix} \dfrac{m_T v_{dx}}{time_{\pm}\ approach} \\ \dfrac{m_T v_{dy}}{time_{\pm}\ side} \\ \dfrac{m_T v_{dz}}{time_{\pm}\ section} \end{Bmatrix} = \frac{\mathbb{C}\ k_b T_c}{cavity} \tag{2.7.1c}$$

These trisine cellular pressure and force relationship are universally correlated to the gravitational force as follows assuming that the Newtonian gravitational parameter is a variable such that $G_U = R_U c^2 / M_U$.

$$\begin{aligned} F_x &= 3\left(\frac{A}{B}\right)_T \frac{G_U M_U m_T}{R_U^2} \\ &= 3\frac{A}{B}\frac{R_U c^2}{M_U}\frac{M_U m_T}{R_U^2} = 3\left(\frac{A}{B}\right)_T \frac{m_T c^2}{R_U} \end{aligned} \tag{2.7.1d}$$

Cellular pressure $(p_U)$, as expressed in eqn 2.7.1c, is congruent with the universal pressure $(p_U)$ expressed as energy $(m_t c^2)$ spread over the volumes $(approach \cdot R_U)$, $(side \cdot R_U)$ and $(section \cdot R_U)$.

$$p_U = \begin{Bmatrix} \dfrac{F_x}{approach} \\ \dfrac{F_y}{side} \\ \dfrac{F_z}{section} \end{Bmatrix} = \begin{Bmatrix} 3\dfrac{A}{B}\dfrac{m_T c^2}{approach \cdot R_U} \\ \dfrac{6}{\sqrt{3}}\dfrac{A}{B}\dfrac{m_T c^2}{side \cdot R_U} \\ 6\dfrac{m_T c^2}{section \cdot R_U} \end{Bmatrix} = \frac{\mathbb{C}\ k_b T_c}{cavity} \tag{2.7.1e}$$

$$p_{Ur} = \begin{Bmatrix} \dfrac{F_x}{approach_r} \\ \dfrac{F_y}{side_r} \\ \dfrac{F_z}{section_r} \end{Bmatrix} = \begin{Bmatrix} 3\dfrac{A_r}{B_r}\dfrac{m_T c^2}{approach_r \cdot R_{Ur}} \\ \dfrac{6}{\sqrt{3}}\dfrac{A_r}{B_r}\dfrac{m_T c^2}{side_r \cdot R_{Ur}} \\ 6\dfrac{m_T c^2}{section_r \cdot R_{Ur}} \end{Bmatrix} = \frac{\mathbb{C}\ k_b T_{cr}}{cavity_r} \tag{2.7.1er}$$

To put the calculated pressures in perspective, the C-C bond has a reported energy of 88 kcal/mole and a bond length of 1.54 Å [24]. Given these parameters, the internal chemical bond pressure $(CBP)$ for this bond is estimated to be:

$$\begin{aligned} &88\frac{kcal}{mole} 4.18x10^{10}\frac{erg}{kcal}\frac{1}{6.02x10^{23}}\frac{mole}{bond}\frac{1}{\left(1.54x10^{-8}\right)^3}\frac{bond}{cm^3} \\ &= 1.67x10^{12}\frac{erg}{cm^3} \end{aligned}$$

Within the context of the superconductor internal pressure numbers in table 2.7.1, the internal pressure requirement for the superconductor is less than the C-C chemical bond pressure at the order of design $T_c$ of a few thousand degrees kelvin indicating that there is the possibility of using materials with the bond strength of carbon to chemically engineer high performance resonant superconducting materials.

In the context of interstellar space vacuum, the total pressure $m_t c^2 / cavity$ may be available (potential energy) and will be of such a magnitude as to provide for universe expansion as astronomically observed. (see equation 2.11.21). Indeed, there is evidence in the Cosmic Gamma Ray Background Radiation for energies on the order of $m_t c^2$ or 56 MeV or ($\log_{10}$(Hz) = 22.13) or ( 2.20E-12 cm) [73]. In the available spectrum [73], there is a hint of a black body peak at 56 MeV or ($\log_{10}$(Hz) = 22.13) or (2.20E-12 cm). This radiation has not had an identifiable source until now.

**Table 2.7.1** CPT Lattice Pressure

| | | | | | | |
|---|---|---|---|---|---|---|
| Condition | 1.2 | 1.3 | 1.6 | 1.9 | 1.10 | 1.11 |
| $T_a\left({}^0K\right)$ | 6.53E+11 | 6.53E+11 | 6.53E+11 | 6.53E+11 | 6.53E+11 | 6.53E+11 |
| $T_b\left({}^0K\right)$ | 9.05E+14 | 5.48E+13 | 9.22E+08 | 1.10E+07 | 2,868 | 2.723 |
| $T_{br}\left({}^0K\right)$ | 3.30E+12 | 5.45E+13 | 3.24E+18 | 2.72E+20 | 1.04E+24 | 1.10E+27 |
| $T_c\left({}^0K\right)$ | 8.96E+13 | 3.28E+11 | 93.0 | 0.0131 | 8.99E-10 | 8.10E-16 |
| $T_{cr}\left({}^0K\right)$ | 1.19E+09 | 3.25E+11 | 1.15E+21 | 8.11E+24 | 1.19E+32 | 1.32E+38 |
| $z_R + 1$ | 3.67E+43 | 8.14E+39 | 3.89E+25 | 6.53E+19 | 1.17E+09 | 1.00E+00 |

| $z_B+1$ | 3.32E+14 | 2.01E+13 | 3.39E+08 | 4.03E+06 | 1.05E+03 | 1.00E+00 |
|---|---|---|---|---|---|---|
| $Age_{UG}(sec)$ | 7.14E-05 | 4.79E-03 | 6.94E+04 | 5.36E+07 | 1.27E+13 | 4.33E+17 |
| $Age_{UGv}(sec)$ | 4.59E-10 | 1.25E-07 | 4.42E+02 | 3.13E+06 | 1.22E+13 | 4.33E+17 |
| $Age_U(sec)$ | 1.18E-26 | 5.31E-23 | 1.11E-08 | 6.63E-03 | 3.70E+08 | 4.33E+17 |
| $k_bT_c/cavity$ $erg/cm^3$ | 2.89E+37 | 2.34E+31 | 3.17E+07 | 7.52E-03 | 9.21E-21 | 7.11E-36 |
| $k_bT_{cr}/cavity_r$ $erg/cm^3$ | 1.86E+25 | 2.29E+31 | 1.69E+55 | 7.13E+64 | 5.81E+82 | 7.54E+97 |
| $m_tc^2/cavity$ $erg/cm^3$ | 2.10E+35 | 4.66E+31 | 2.23E+17 | 3.74E+11 | 6.69E+00 | 5.73E-09 |
| $m_Tc^2/cavity_r$ $erg/cm^3$ | 1.02E+28 | 4.60E+31 | 9.63E+45 | 5.73E+51 | 3.20E+62 | 3.74E+71 |

In comparison to CPT Lattice Pressure, Table 2.7.1a indicates Black Body Emission and Energy Density (Pressure). The rest mass CPT lattice $(T_c)$ energy density or pressure is greater than the corresponding Black Body $(T_b)$ energy density or pressure.

**Table 2.7.1a** Black Body $(T_b)$ Emission and Energy Density

| Condition | 1.2 | 1.3 | 1.6 | 1.9 | 1.10 | 1.11 |
|---|---|---|---|---|---|---|
| $T_a(^0K)$ | 6.53E+11 | 6.53E+11 | 6.53E+11 | 6.53E+11 | 6.53E+11 | 6.53E+11 |
| $T_b(^0K)$ | 9.05E+14 | 5.48E+13 | 9.22E+08 | 1.10E+07 | 2,868 | 2.723 |
| $T_{br}(^0K)$ | 3.30E+12 | 5.45E+13 | 3.24E+18 | 2.72E+20 | 1.04E+24 | 1.10E+27 |
| $T_c(^0K)$ | 8.96E+13 | 3.28E+11 | 93.0 | 0.0131 | 8.99E-10 | 8.10E-16 |
| $T_{cr}(^0K)$ | 1.19E+09 | 3.25E+11 | 1.15E+21 | 8.11E+24 | 1.19E+32 | 1.32E+38 |
| $z_R+1$ | 3.67E+43 | 8.14E+39 | 3.89E+25 | 6.53E+19 | 1.17E+09 | 1.00E+00 |
| $z_B+1$ | 3.32E+14 | 2.01E+13 | 3.39E+08 | 4.03E+06 | 1.05E+03 | 1.00E+00 |
| $Age_{UG}(sec)$ | 7.14E-05 | 4.79E-03 | 6.94E+04 | 5.36E+07 | 1.27E+13 | 4.33E+17 |
| $Age_{UGv}(sec)$ | 4.59E-10 | 1.25E-07 | 4.42E+02 | 3.13E+06 | 1.22E+13 | 4.33E+17 |
| $Age_U(sec)$ | 1.18E-26 | 5.31E-23 | 1.11E-08 | 6.63E-03 | 3.70E+08 | 4.33E+17 |
| $\sigma_sT_b^4$ $erg/cm^2/sec$ | 3.81E+55 | 5.11E+50 | 4.10E+31 | 8.19E+23 | 3.84E+09 | 3.12E-03 |
| $\sigma_sT_{br}^4$ $erg/cm^2/se$ | 6.72E+45 | 5.01E+50 | 6.23E+69 | 3.12E+77 | 6.67E+91 | 8.21E+103 |
| $\sigma_sT_b^4c/4$ $erg/cm^3$ | 5.08E+45 | 6.81E+40 | 5.48E+21 | 1.09E+14 | 5.12E-01 | 4.16E-13 |
| $\sigma_sT_{br}^4c/4$ $erg/cm^3$ | 8.96E+35 | 6.69E+40 | 8.32E+59 | 4.17E+67 | 8.90E+81 | 1.10E+94 |
| $photon/cm^3$ | 1.51E+46 | 3.33E+42 | 1.59E+28 | 2.67E+22 | 4.79E+11 | 4.10E+02 |
| $photon/cm^3$ | 7.29E+38 | 3.29E+42 | 6.89E+56 | 4.10E+62 | 2.29E+73 | 2.68E+82 |
| $wavelength$ $cm$ | 3.20E-16 | 5.29E-15 | 3.14E-10 | 2.64E-08 | 1.01E-04 | 1.06E-01 |
| $wavelength$ $cm$ | 8.78E-14 | 5.31E-15 | 8.95E-20 | 1.06E-21 | 2.78E-25 | 2.64E-28 |
| $Frequency$ $/sec$ | 5.32E+25 | 3.22E+24 | 5.42E+19 | 6.44E+17 | 1.69E+14 | 1.60E+11 |
| $Frequency$ $/sec$ | 1.94E+23 | 3.21E+24 | 1.90E+29 | 1.60E+31 | 6.12E+34 | 6.45E+37 |

The following figure indicates the Planck Black Body curve for CMBR at 2.729 K

The hydrogen BEC 1S-2S transition

| | |
|---|---|
| absorbed energy | 1.301E-12 erg/cm$^3$ |
| absorbed freq | 1.234E+15 /sec |
| absorbed wavelength | 2.430E-05 cm |
| band pass | ~1E-6 Hz  Ref: [90] |

is of such high frequency that CMBR intensity at 1.234E+15 /sec is negligible and band pass ~1E-6 Hz and will not degrade any intervening hydrogen dense (1 g/cc) BEC. The moles of photons(einsteins) at the tail of this curve (too far off the curve to be shown) are not enough to degrade a hydrogen dense (1 g/cc) BEC. This is further substantiate by Peeples[95] page 157 indicating electrons captured directly to the ground state emitted with not net neutral hydrogen change.

This is also the case for CMBR = 3000 K at recombination as indicated in the following figures. The moles of photons(einsteins) at the tail of this curve are not enough to degrade a hydrogen dense (1 g/cc) BEC. This is particularly true for the present universe condition where CMBR = 2.729 K.

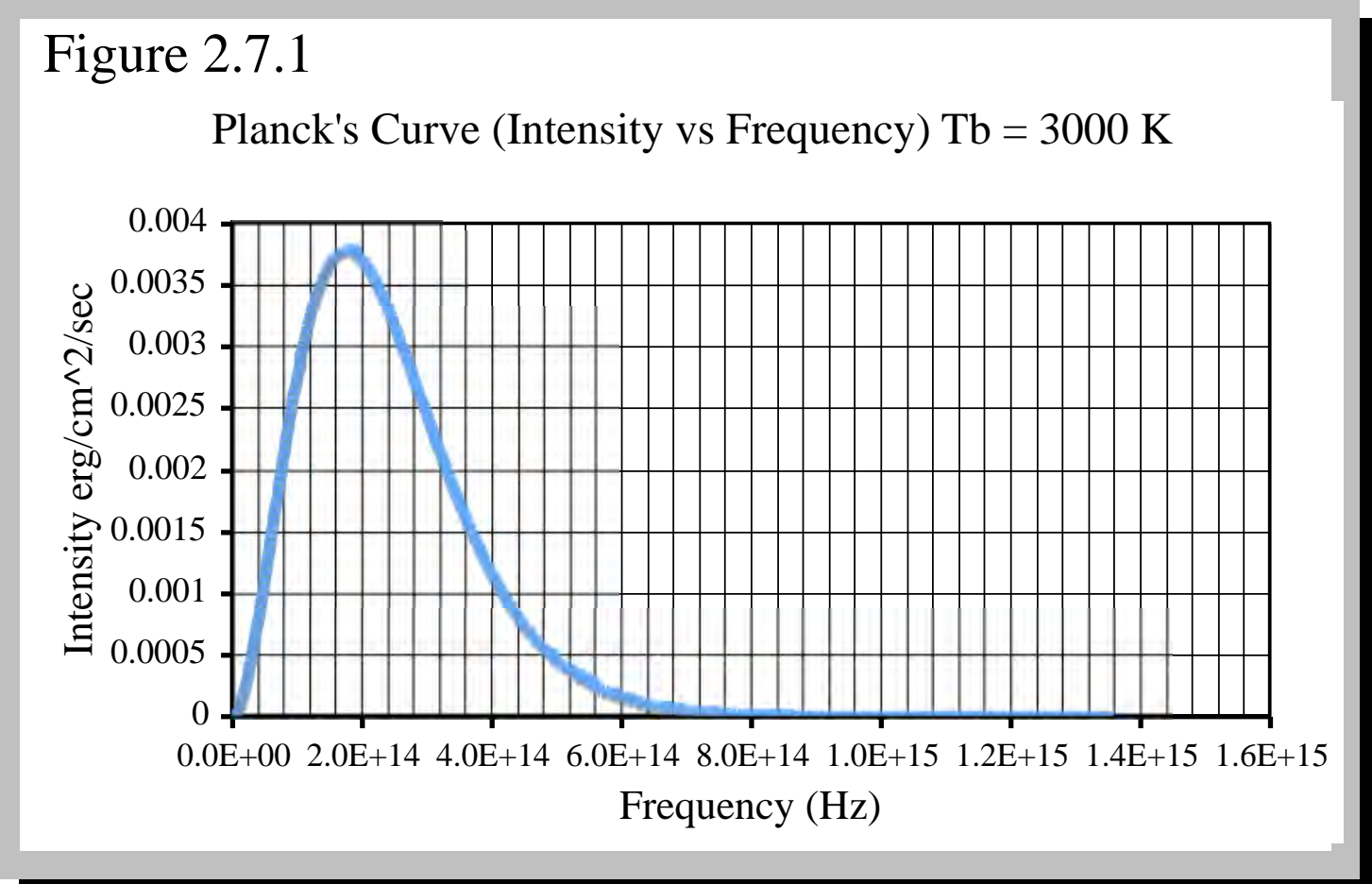

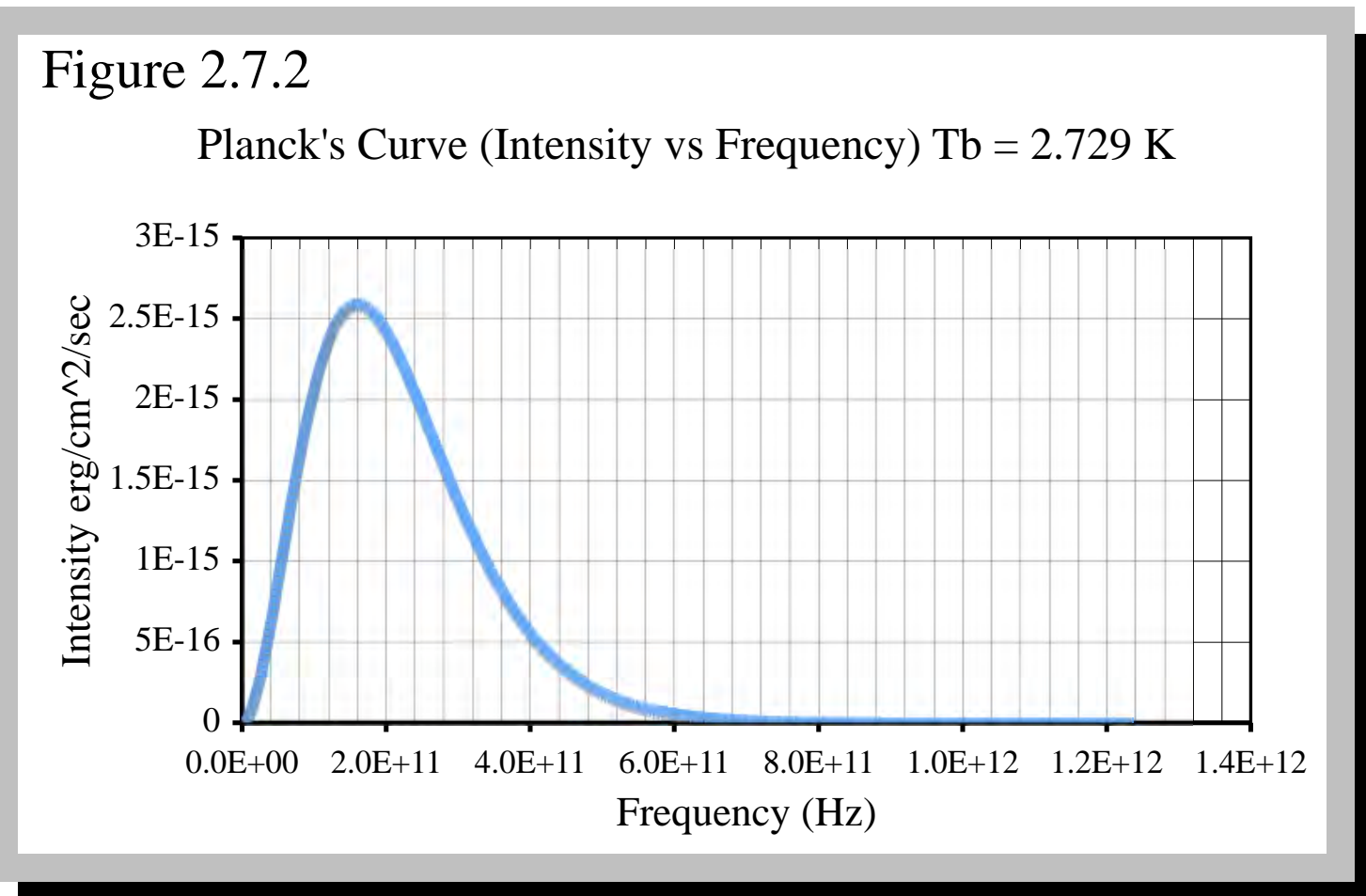

The red line indicates the hydrogen BEC absorption frequency of 1.234E+15 /sec in the context of universe expansion from the Big Bang nucleosynthetic event.

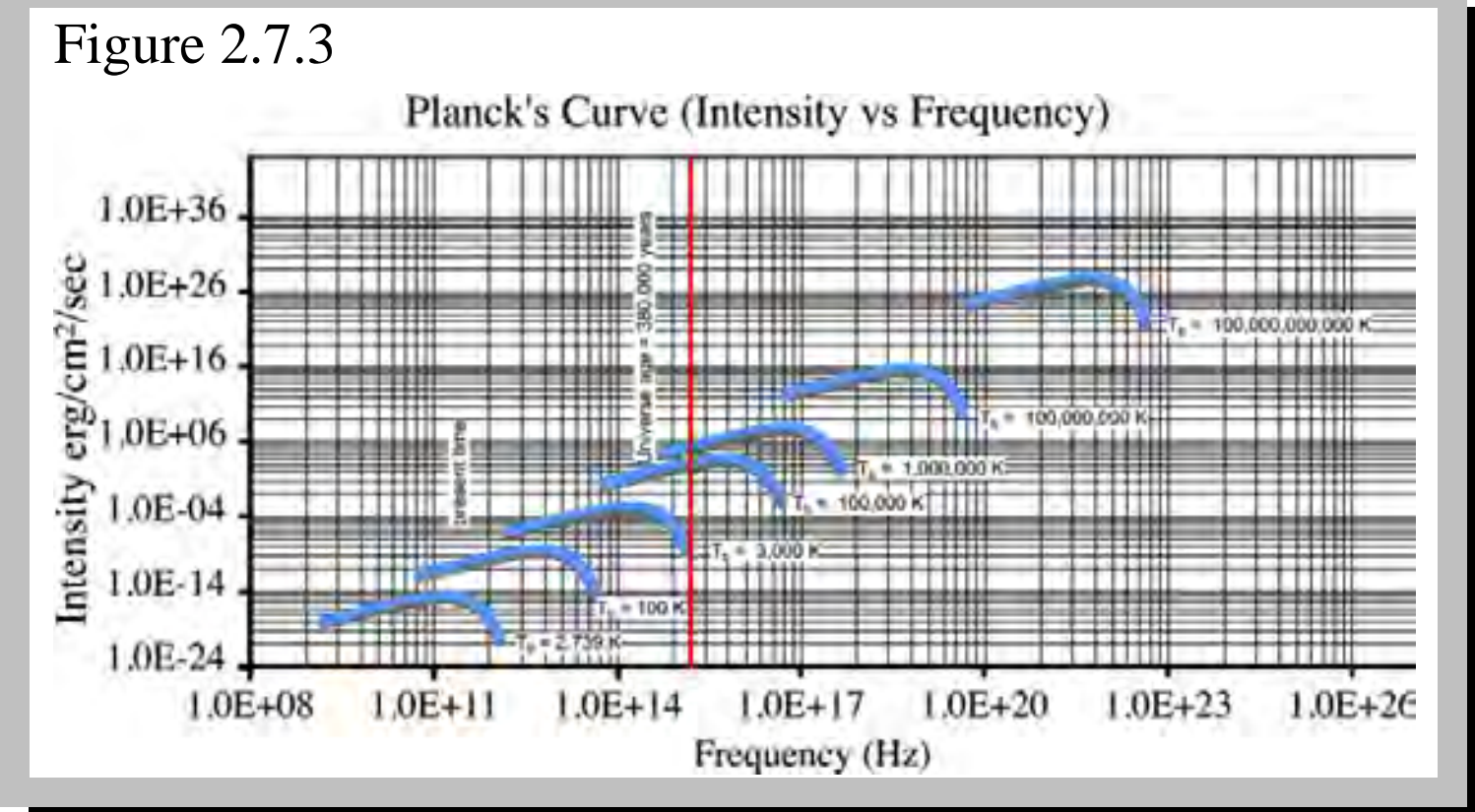

Figure 2.7.3

Assuming the superconducting current moves as a longitudinal wave with an adiabatic character, then equation 2.7.2 for current or de Broglie velocity $(v_{dx})$ holds where the ratio of heat capacity at constant pressure to the heat capacity at constant volume $(\delta)$ equals 1.

$$v_{dx} = \sqrt{\delta\ pressure \frac{cavity}{m_T}} \qquad (2.7.2)$$

The adiabatic nature of an expanding universe is indicated by the expansion relationships for $(T_c,\ T_b,\ T_a\ and\ T_p)$.

$$\begin{aligned}
&\left(T = T_c,\ \delta = 5/3\ \ \&\ \ p = p_U = \mathbb{C}k_bT_c / cavity\right) \text{ or} \\
&\left(T = T_b,\ \delta = 4/3\ \ \&\ \ p = \mathbb{C}k_bT_b / cavity\right) \text{ or} \\
&\left(T = T_a,\ \delta = 1\ \ \ \ \ \ \&\ \ p = \mathbb{C}k_bT_a / cavity\right) \text{ or} \\
&\left(T = T_p,\ \delta = 1/3\ \ \&\ \ p = \mathbb{C}k_bT_p / cavity\right)
\end{aligned} \qquad (2.7.2a)$$

$$p_1\ cavity_1^{\delta} = p_2\ cavity_2^{\delta} \qquad (2.7.3a)$$

$$T_1 cavity_1^{\delta-1} = T_2 cavity_2^{\delta-1} \qquad (2.7.3b)$$

$$T_1 p_1^{\frac{(1-\delta)}{\delta}} = T_2 p_2^{\frac{(1-\delta)}{\delta}} \qquad (2.7.3c)$$

$T_c$ degrees of freedom = 2/(5/3 – 1) = 3
$T_b$ degrees of freedom = 2/(4/3 – 1) = 6
$T_a$ degrees of freedom = 2/(1 – 1) = undefined
$T_p$ degrees of freedom = 2/(1/3 – 1) = -3

Equation 2.7.2a is particularly confirmed with the equation 2.7.3d where $(\Omega_b = .037)$ representing the universe luminescent fraction and

$T_{solar} = 5780K$
$\sigma_s T_{solar} = 1{,}367{,}000$ erg/(cm$^2$ sec AU$^2$ solar mass unit)

and Astronomical Unit $AU$ (distance between sun and earth)
$AU = 1.50E13\ cm$ and *solar mass* = 2.0E33 g

$$\sigma_s T_b^4 = \left(\frac{cavity}{cavity_{present}}\right)^{\frac{4}{3}} \Omega_b \frac{M_U}{solar\ mass} \sigma_s T_{solar}^4 \left(\frac{AU}{R_U}\right)^2 \qquad (2.7.3d)$$

It is particularly important to note this scaling relationship over universe time $(1/H_U)$. Equation 2.7.3d indicates a equi-partition of energy between CMBR and universe luminescent matter $(\Omega_b)$ (commonly called baryonic matter) going back to the Big Bang. (see Appendix J)

As an extension of this work, in a report[53] by Jet Propulsion Laboratory, an unmodeled deceleration was observed in regards to Pioneer 10 and 11 spacecraft as they exited the solar system. This deceleration can be related empirically to the following dimensional correct force expression (2.7.4) where $(m_t/cavity)$ is the mass equivalent energy density (pressure)

$$\left(m_t c^2 / cavity\right) \text{ value at } T_c = 8.11E-16\ {}^0K$$

representing conditions in space, which is 100 percent, transferred to an object passing through it.

$$Force = -AREA\frac{m_T}{cavity}c^2 = -AREA\left(\rho_U\right)c^2 \qquad (2.7.4)$$

Note that equation 2.7.4 is similar to the conventional drag equation as used in design of aircraft with $v^2$ replaced with $c^2$. The use of $c^2$ vs. $v^2$ means that the entire resonant energy content *($m_Tc^2$/cavity)* is swept out of volume as the spacecraft passes through it. Of course $Force = Ma$, so the Pioneer spacecraft $M/AREA$, is an important factor in establishing deceleration, which was observed to be on the order of 8.74E-8 cm/sec$^2$.

The force equation is repeated below as the conventional drag equation in terms of Pioneer spacecraft and the assumption that space density $(\rho_U)$ (2.11.16) is essentially what has been determined to be a potential candidate for dark energy:

$$F = Ma = C_d A_c \rho_U \frac{v^2}{2}$$

$$where: \ C_d = \frac{24}{R_e} + \frac{6}{1+\sqrt{R_e}} + .4 \qquad [60] \text{ empirical} \tag{2.7.4a}$$

and where:

$F$ = force $M$ = Pioneer 10 & 11 mass
$a$ = acceleration $A_c$ = Pioneer 10 & 11 cross section
$\rho_U$ = space fluid critical density (6.38E-30 g/cm$^3$)
$v$ = Pioneer 10 & 11 velocity
$d$ = Pioneer 10 & 11 diameter
$\mu_{Uc}$ = absolute vacuum viscosity (1.21E-16 g/(cm sec))
$\nu_{Uc}$ = kinematic viscosity $\mu_{Uc} / \rho_U$ (1.90E13 cm$^2$/sec)
$R_e$ = Reynolds' number = inertial force/viscous force
= $\rho_U v d / \mu_U$

When the Reynolds' number *($R_e$)* is low (laminar flow condition), then $C_d = (24/R_e)$ and the drag equation reduces to the Stokes' equation $F = 3\pi\mu_U vd$. [60]. When the Pioneer spacecraft data is fitted to this drag equation, the computed drag force is 6 orders of magnitude lower than observed values.

In other parts of this report, a model is developed as to what this dark energy actually is. The model dark energy is keyed to a 6.38E-30 g/cc value $(\rho_U)$ (2.11.16) in trisine model (NASA observed value of 6E-30 g/cc). The proposed dark energy consists of mass units of $m_T$ (110.107178208 x electron mass) per volume (Table 2.2.2 *cavity*- 15733 cm$^3$). These mass units are virtual particles that exist in a coordinated lattice at base energy $\hbar\omega/2$ under the condition that momentum and energy are conserved - in other words complete elastic character. This dark energy lattice would have internal pressure (table 2.7.1) to withstand collapse into gravitational clumps.

Under these circumstances, the traditional drag equation form is valid but velocity *(v)* should be replaced by speed of light (c). Conceptually, the equation 2.7.4a becomes a thrust equation rather than a drag equation. Correspondingly, the space viscosity $(\mu_{Uc})$ is computed as momentum/area or *($m_T c/(2\Delta x^2)$)* where $\Delta x$ is uncertainty dimension in table 2.5.1 at $T_c$ = 8.1E-16 K and as per equation 2.1.20. This is in general agreement with gaseous kinetic theory [19].

The general derivation is in one linear dimensional path (s) and in accordance with Newton's Second Law, Resonant Mass and Work-Energy Principles in conjunction with the concept of Momentum Space, which defines the trisine elastic space lattice. These considerations are developed and presented in Appendix F with path *'$ds/C_t$'* used here.

$$\int_B^0 F \frac{ds}{C_t} = m_t \int_{v_{dx}+vo}^{c} \frac{v}{\sqrt{1-\frac{v^2}{c^2}}} dv = m_t \int_{k_m\varepsilon+\frac{v_o^2}{c^2}}^{1} \frac{-1}{\alpha^2} \frac{c^2}{\sqrt{1-\frac{1}{\alpha}}} d\alpha \tag{2.7.5}$$

then

$$F \left. \frac{s}{C_t} \right|_B^0 = -m_t c^2 \sqrt{1-\frac{v^2}{c^2}} \Bigg|_{v_{dx}+vo}^{c} = -m_t c^2 \sqrt{1-\frac{1}{\alpha}} \Bigg|_{k_m\varepsilon+\frac{v_o^2}{c^2}}^{1}$$

$$\approx -m_t c^2 = -\frac{(m_t \text{c})^2}{m_t} \text{ at } \quad v_{dx} << c \quad \text{and} \quad \left(k_m\varepsilon + \frac{v_o^2}{c^2}\right) >> 1 \tag{2.7.5a}$$

$$= 0 \ \text{ at } \ v = c \ \text{ (photon has zero mass)}$$

Clearly, an adiabatic process is described wherein the change in internal energy of the system (universe) is equal in absolute magnitude to the work (related to objects moving through the universe) or Work = Energy. Momentum is transferred from the trisine elastic space lattice to the object moving through it.

The hypothesis can be stated as:

**<u>An object moving through momentum space will accelerate.</u>**

It is understood that the above development assumes that $dv/dt \neq 0$, an approach which is not covered in standard texts[61], but is assumed to be valid here because of it describes in part the actual physical phenomenon taking place when an object impacts or collides with an trisine elastic space lattice cell. The elastic space lattice cell essentially collapses $(B \to 0)$ and the entire cell momentum $(m_t c)$ is transferred to the object with cell permittivity $(\varepsilon)$ and permeability $(k_m)$ proceeding to a values of one(1).

In terms of an object passing through the trisine elastic space CPT lattice at some velocity $(v_o)$, the factor $(Fs/m_t)$ is no longer dependent of object velocity $(v_o)$ as defined in $(v_{dx} < v_o << c)$ but is a constant relative to $c^2$ which reflects the fact that the speed of light is a constant in the universe. Inspection would indicate that as $(v_o)$ is in range $(v_{dx} < v_o << c)$ then the observed object deceleration "a" is a constant (not tied to any particular reference frame either translational or rotational):

$$a = \frac{F}{M}$$

and specifically for object and trisine elastic space lattice relative velocities $(v_o)$ as follows:

The sun referenced Pioneer velocity of about 12 km/s
or earth (around sun) orbital velocity of 29.8 km/sec
or sun (circa milky way center) rotation velocity of 400 km/sec
or the CMBR dipole of 620 km/s
or any other velocity *'v'* magnitude *(0<v<<c)*

do not contribute to the observed Pioneer deceleration (as observed) and (noting that c ~ 300,000 km/sec).
and in consideration of the following established relationship:

$$F \approx -C_t \frac{m_t c^2}{s} = C_t A_c \rho_U c^2 \qquad (2.7.6)$$

where: $C_t = \frac{24}{R_{eo}} + \frac{6}{1+\sqrt{R_{eo}}} + .4$ and $R_{eo} = \frac{\rho_U c d}{\mu_{Uc}}$

Where constant $C_t$ is in the form of the standard fluid mechanical drag coefficient analogous to 2.7.4a, but is defined as a thrust coefficient.

This trisine model was used to analyze the Pioneer unmodeled deceleration data 8.74E-08 cm/sec$^2$ and there appeared to be a good fit with the observed space density $(\rho_U)$ 6E-30 g/cm$^3$ (trisine model $(\rho_U)$ 6.38E-30 g/cm$^3$). A thrust coefficient $(C_t)$ of 59.67 indicates a laminar flow condition. An absolute space viscosity seen $\left(\mu_{Uc} = m_t c / \left(2\Delta x^2\right)\right)$ of 1.21E-16 g/(cm sec) is established within trisine model and determines a space kinematic viscosity seen $(\mu_{Uc} / \rho_U)$ of $(1.21E-16/6.38E-30)$ or 1.90E13 cm$^2$/sec as indicated in 2.7.4a.

**Table 2.7.2** Pioneer (P) Translational Calculations

| | | |
|---|---|---|
| P mass *(M)* | 241,000 | gram |
| P diameter *(d)* | 274 | cm (effective) |
| P cross section *($A_c$)* | 58,965 | cm$^2$ (effective) |
| P area/mass *($A_c$/M)* | 0.24 | cm$^2$/g |
| P velocity | 1,117,600 | cm/sec |
| Pioneer Reynold's number *($R_{eo}$)* | 4.31E-01 | unitless |
| thrust coefficient *($C_t$)* | 59.67 | unitless |
| P deceleration *(a)* | 8.37E-08 | cm/sec$^2$ |
| one year thrust distance | 258.51 | mile |
| laminar thrust force | 9.40E-03 | dyne |
| time of object to stop | 1.34E13 | sec |

These parameters are in the context of radiation pressure

$$radiation\ pressure = \Upsilon \sigma_s T^4 \left(\frac{4}{3c}\right)$$

with a multiplier factor F as a measure of material emissivity or absorptivity.

Pioneer recoil deceleration(*a*) may be then calculated as:

*a = radiation pressure* $A_c$/M

This deceleration(*a*) is then proportional to a net system temperature(T)$^4$. This system temperature*(T)* scales with the RTG radioactive source half_life 87 years.

The Pioneer deceleration(a) half_life then varies with RTG half_life as $(1/2)^4 = 1/16$.

The Pioneer deceleration(aP) half_life should then be on the order of 87/16 or 5.4 years (not ~25 to ~40 years as indicated in ref 116a). This relatively short aP half_life (~5 years) is correctly modeled for both Pioneers with a combination exponential decay and constant deceleration with the constant *aP(infinity)* on the order of 8 E-8 cm/sec$^2$.

The half_lives associated with power consumption (heat, T) ref 116 Table I are shorter than 87 years making the above argument more compelling.

Pioneer internal temperature related radiation pressure decelerations ($\Upsilon$=1) are as follows:

| Temperature *T* (Kelvin) | Pioneer deceleration cm/sec$^2$ due to radiation pressure |
|---|---|
| 25 | 2.41E-10 |
| 50 | 3.86E-09 |
| 25 | 1.95E-08 |
| 100 | 6.17E-08 |
| 125 | 1.61E-07 |

It is difficult to escape the conclusion that the independently developed Pioneer thermal recoil model (based solely on the decay hypothesis $dx/dt = -kx$ is deficient in modeling the Pioneer anomaly.

The Jet Propulsion Laboratory (JPL), NASA raised a question concerning the universal application of the observed deceleration on Pioneer 10 & 11. In other words, why do not the planets and their satellites experience such a deceleration? The answer is in the $A_c/M$ ratio in the modified thrust relationship.

$$a \approx -C_t \frac{m_T}{M}\frac{c^2}{s} = -C_t \frac{A_c}{M}\rho_U c^2 = -C_t \frac{A_c}{M}\frac{m_T}{cavity} c^2 \qquad (2.7.6a)$$

Using the earth as an example, the earth differential movement per year due to modeled deceleration of 4.94E-19 $cm/\sec^2$ would be 0.000246 centimeters (calculated as $at^2/2$, an unobservable distance amount). Also assuming the trisine superconductor resonant symmetric model, the reported 6 nanotesla (nT) (6E-5 gauss) interplanetary magnetic field (IMF) in the vicinity of earth, (as compared to .2 nanotesla (nT) (2E-6 gauss [78,140]) interstellar magnetic field congruent with CMBR) will not allow the formation of a superconductor CPT lattice due to the well known properties of a superconductor in that magnetic fields above critical fields will destroy it. In this case, the space superconductor at $T_c$ = 8.1E-16 K will be destroyed by a magnetic field above

$$H_c\left(c/v_{dx}\right)\left(m_e/m_T\right)\left(chain/cavity\right) = 2.\text{E-6 gauss (2 nT)}$$

as in table 2.6.2. It is conceivable that the IMF decreases by some power law with distance from the sun. Perhaps at some distance from the sun, the IMF diminishes to an extent and defining an associated energy condition, wherein the space superconductor resonant CPT lattice is allowed to form. This IMF condition may explain the observed 'ramp', 'step' or "kick in" of the Pioneer spacecraft deceleration phenomenon at 10 Astronomical Units (AU) (Asteroid belt distance from the sun).

**Table 2.7.3** Earth Translational Calculations

| | | |
|---|---|---|
| Earth *(M)* | 5.98E27 | Gram |
| Earth *(d)* | 1.27E09 | cm |
| Earth cross section *($A_c$)* | 1.28E18 | cm$^2$ |
| Earth area/mass *($A_c$/M)* | 2.13E-10 | cm$^2$/g |
| Earth velocity | 2,980,010 | cm/sec |
| Earth Reynolds number *($R_{eo}$)* | 2.01E06 | unitless |
| thrust coefficient *($C_t$)* | 0.40 | unitless |
| thrust force *(F)* | 2.954E09 | dyne |
| Earth deceleration *(a)* | 4.94E-19 | cm/sec$^2$ |
| one year thrust distance | 0.000246 | cm |
| laminar thrust force | 4.61E-01 | dyne |
| time for Earth to stop rotating | 6.03E24 | sec |

Now to test the universal applicability of the unmodeled Pioneer 10 & 11 decelerations for other man made satellites, one can use the dimensions and mass of the Hubble space telescope. One arrives at a smaller deceleration of 6.79E-09 cm/sec$^2$ because of its smaller area/mass ratio. This low deceleration is swamped by other decelerations in the vicinity of earth, which would be multiples of those well detailed in reference 1, Table II Pioneer Deceleration Budget. Also the trisine elastic space CPT lattice probably does not exist in the immediate vicinity of the earth because of the earth's milligauss magnetic field, which would destroy the coherence of the trisine elastic space CPT lattice.

**Table 2.7.4** Hubble Satellite Translational Calculations

| | | |
|---|---|---|
| Hubble mass *(M)* | 1.11E07 | gram |
| Hubble diameter *(d)* | 7.45E02 | centimeter |
| Hubble cross section *($A_c$)* | 5.54E05 | cm$^2$ |
| Hubble area/mass *($A_c$/M)* | 0.0499 | cm$^2$/g |
| Hubble velocity | 790,183 | cm/sec |
| Hubble Reynolds' number *($R_{eo}$)* | 1.17E00 | unitless |
| thrust coefficient *($C_t$)* | 23.76 | |
| thrust force *(F)* | 7.549E-02 | dyne |
| Hubble deceleration *(a)* | 6.79E-09 | cm/sec$^2$ |
| One year thrust distance | 3,378,634 | cm |
| laminar thrust force | 7.15E-08 | dyne |
| Time for Hubble to stop rotating | 1.16E14 | sec |

The fact that the unmodeled Pioneer 10 & 11 deceleration data are statistically equal to each other and the two space craft exited the solar system essentially 190 degrees from each other implies that the supporting fluid space density through which they are traveling is co-moving with the solar system. But this may not be the case. Assuming that the supporting fluid density is actually related to the cosmic background microwave radiation (CMBR), it is known that the solar system is moving at 500 km/sec relative to CMBR. Equivalent decelerations in irrespective to spacecraft heading relative to the CMBR would further support the hypothesis that deceleration is independent of spacecraft velocity.

Reference 150 indicates dark matter cross-section and resultant drag characteristics for 72 studied galactic collisions .47 cm$^2$/g. The Pioneer 10 and 11 cross-sections are .24 cm$^2$/g. The proximity of these two values is an indication of a common drag mechanism?

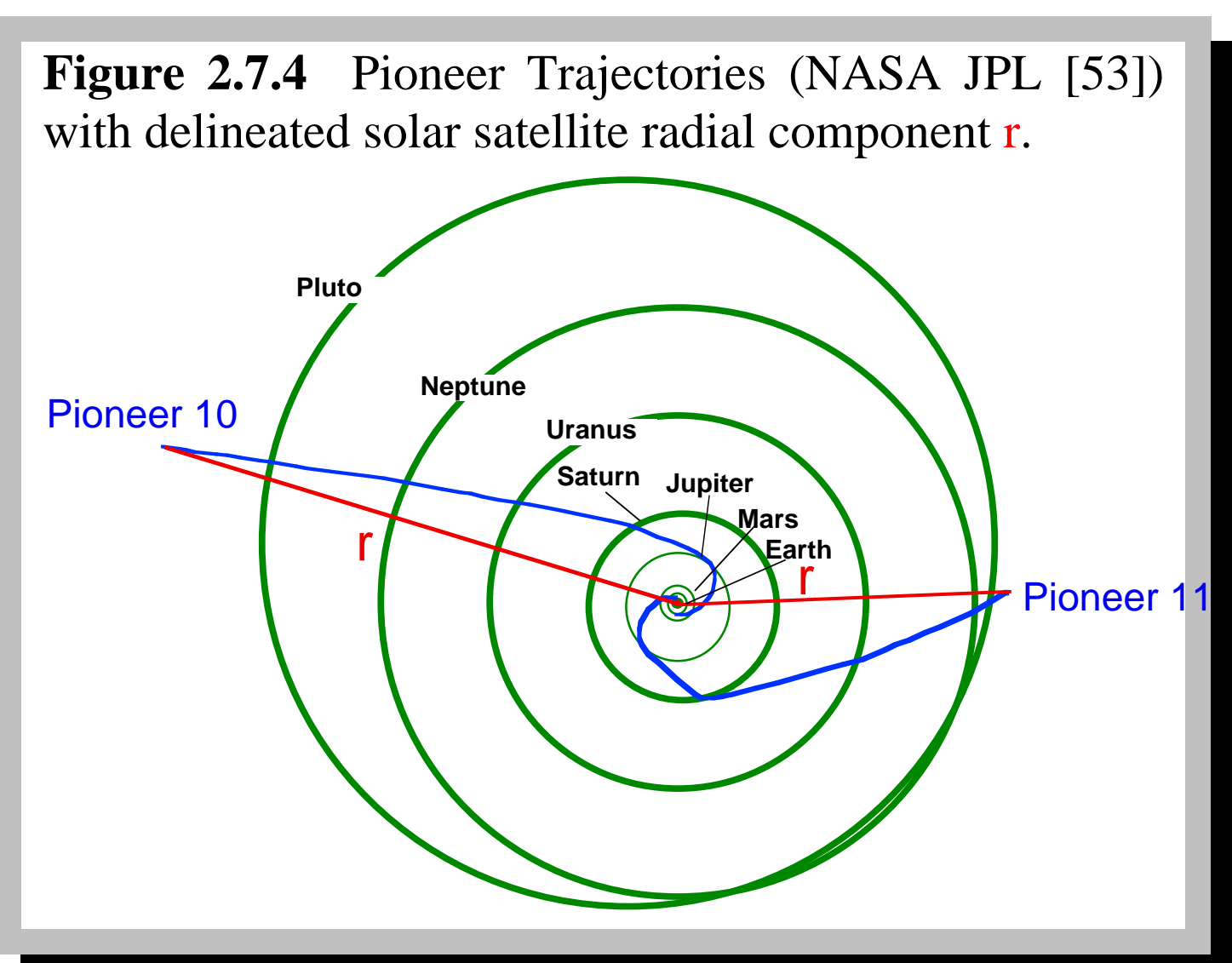

**Figure 2.7.4** Pioneer Trajectories (NASA JPL [53]) with delineated solar satellite radial component r.

Executing a web based [58] computerized radial rate dr/dt (where r is distance of spacecraft to sun) of Pioneer 10 & 11 Figure 2.7.1), it is apparent that the solar systems exiting velocities are not equal. Based on this graphical method, the spacecraft's velocities were in a range of about 9% (28,600 – 27,500 mph for Pioneer 10 "3Jan87 - 22July88" and 26,145 – 26,339 mph for Pioneer 11 "5Jan87 - 01Oct90"). This reaffirms the basic thrust equation used for computation in that, velocity magnitude of the spacecraft is not a factor, only its direction with the thrust force opposite to that direction. The observed equal deceleration at different velocities would appear to rule out deceleration due to Kuiper belt dust as a source of the deceleration for a variation at $1.09^2$ or 1.19 would be expected.

Also it is observed that the Pioneer spacecraft spin is slowing down during periods not affected by gaseous flows or other articulated spacecraft movements. JPL[57] suggests an unmodeled deceleration rate of .0067 rpm/year or 3.54E-12 rotations/sec$^2$ during these undisturbed periods.

Now using the standard torque formula:

$$\Gamma = I\dot{\omega} \tag{2.7.7}$$

The deceleration value of 3.54E-12 rotation/sec$^2$ can be replicated assuming a Pioneer Moment of Inertia about spin axis(I) of 5.88E09 g cm$^2$ and a 'paddle' cross section area of 3,874 cm$^2$ which is 6.5% spacecraft 'frontal" cross section which seems reasonable.

Also it is important to note that the angular deceleration rate for each spacecraft (Pioneer 10, 11) is the same even though they are spinning at the two angular rates 4 and 7 rpm respectfully. This would further confirm that velocity is not a factor in measuring the space mass or energy density. The calculations below are based on NASA JPL problem set values[57]. Also, the sun is moving through the CMBR at 600 km/sec. According to this theory, this velocity would not be a factor in the spacecraft deceleration.

**Table 2.7.5** Pioneer (P) Rotational Calculations

| | | |
|---|---|---|
| P mass *(M)* | 241,000 | g |
| P moment of inertia *(I)* | 5.88E09 | g cm$^2$ |
| P diameter *(d)* | 274 | cm |
| P translational cross section | 58,965 | cm$^2$ |
| P radius of gyration k | 99 | cm |
| P radius r | 137 | cm |
| P paddle cross section | 3,874 | cm$^2$ |
| P paddle area/mass | 0.02 | cm$^2$/g |
| P rotation speed at k | 4,517 | cm/sec |
| P rotation rate change | 0.0067 | rpm/year |
| P rotation rate change | 3.54E-12 | rotation/sec$^2$ |
| P rotation deceleration at k | 2.20E-09 | cm/sec$^2$ |
| P force slowing it down *(F)* | 1.32E-03 | dyne |
| P Reynolds' number *($R_{eo}$)* | 4.31E-01 | unitless |
| thrust coefficient *($C_t$)* | 59.67 | |
| thrust force *(F)* | 1.32E-03 | dyne |
| one year rotational distance | 1,093,344 | cm |
| P rotational laminar force | 7.32E-23 | dyne |
| time for P to stop rotating | 2.05E12 | sec |

These calculations suggest a future spacecraft with general design below to test this hypothesis. This design incorporates general features to measure translational and rotational thrust. A wide spectrum of variations on this theme is envisioned. General Space Craft dimensions should be on the order of Trisine geometry B at $T_b = 2.711\,K$ $T_c = 8.12E-16\,K$ or 22 cm (Table 2.2.1) or larger. The Space Craft Reynold's number would be nearly 1 at this dimension and the resulting thrust the greatest. The coefficient of thrust would decrease towards one with larger Space Craft dimensions. It is difficult to foresee what will happen with Space Craft dimensions below present dark energy dimension $B_{present}$ of 22 cm.

The spacecraft (Figure 2.7.5) would rotate on green bar axis while traveling in the direction of the green bar. The red paddles would interact with the space matter as $mc^2$ slowing the rotation with time in conjunction with the overall spacecraft slowing decelerating.

A basic experiment cries out here to be executed and that is to launch a series of spacecraft of varying $A_c/M$ ratios and exiting the solar system at various angles while measuring their rotational and translational decelerations. This would give a more precise measure of the spatial orientation of dark energy in the immediate vicinity of our solar system. An equal calculation of space energy density by translational and rotational observation would add credibility to accumulated data.

**Figure 2.7.5** Space craft general design to quantify translational and rotary thrust effect. Brown is parabolic antenna, green is translational capture element and blue is rotational capture element.

The assumption that the Pioneer spacecraft will continue on forever and eventually reach the next star constellation may be wrong. Calculations indicate the Pioneers will come to a stop relative to space fluid in about 400,000 years. We may be more isolated in our celestial position than previously thought.

When this thrust model is applied to a generalized design for a Solar Sail [63], a deceleration of 3.49E-05 cm/sec$^2$ is calculated which is 3 orders higher than experienced by Pioneer spacecraft. Consideration should be given to spin paddles for these solar sails in order to accumulate spin deceleration data also. In general, a solar sail with its characteristic high $A_c/M$ ratio such as this example should be very sensitive to the trisine elastic space CPT lattice. The integrity of the sail should be maintained beyond its projected solar wind usefulness as the space craft travels into space beyond Jupiter.

**Table 2.7.6** Sail Deceleration Calculations

| | | |
|---|---|---|
| Sail mass *(M)* | 300,000 | gram |
| Sail diameter *(d)* | 40,000 | centimeter |
| Sail cross section *($A_c$)* | 1,256,637,000 | cm$^2$ |
| Sail area/mass *($A_c$/M)* | 4,189 | cm$^2$/g |
| Sail velocity | 1,117,600 | cm/sec |
| Sail Reynold's number | 6.30E01 | unitless |
| thrust coefficient *($C_t$)* | 1.45 | |
| thrust force *(F)* | 10 | dyne |
| Sail acceleration *(a)* | 3.49E-05 | cm/sec$^2$ |
| one year thrust distance | 1.73E10 | Cm |
| laminar thrust force *(F)* | 1.37E00 | dyne |
| time of Sail to stop | 1015 | year |

The following calculation establishes a distance from the sun at which the elastic space CPT lattice kicks in. There is a energy density competition between the solar wind and the elastic space CPT lattice. Assuming the charged particle density at earth is 5 / cm$^3$ [57] and varying as 1/R$^2$ and comparing with elastic space density of $kT_c$*/cavity*, then an equivalent energy density is achieved at 3.64 AU. This of course is a dynamic situation with solar wind changing with time, which would result in the boundary of the elastic space CPT lattice changing with time. This is monitored by a spacecraft (Advanced Composition Explorer (ACE)) at the L1 position (1% distance to sun) and can be viewed at the website http://space.rice.edu/ISTP/dials.html. A more descriptive term would be making the elastic space CPT lattice decoherent or coherent. It is interesting to note that the calculated boundary is close to that occupied by the Astroids and is considered the Astroid Belt.

**Table 2.7.7** Space CPT lattice intersecting with solar wind

| | | |
|---|---|---|
| Mercury | 0.39 | |
| Venus | 0.72 | |
| Earth | 1.00 | |
| Mars | 1.53 | |
| Jupiter | 5.22 | |
| Saturn | 9.57 | |
| Uranus | 19.26 | |
| Neptune | 30.17 | |
| Pluto | 39.60 | |
| AU | 3.64 | |
| n at AU | 0.38 | cm$^{-3}$ |
| $r$ 1/n$^{1/3}$ | 1.38 | cm |
| coordination # | 6 | unitless |
| *volume* | 0.44 | cm$^3$ |
| $\frac{e^2}{r}\frac{1}{volume}\frac{T_c}{T_b}$ $\frac{energy}{volume}$ | 1.13E-34 | erg / cm$^3$ |
| Trisine $\frac{k_b T_c}{cavity}$ $\frac{energy}{volume}$ | 1.13E-34 | erg / cm$^3$ |

Now assuming the basic conservation of total kinetic and potential energy for an object in circular orbit of the sun with velocity $(v_0)$ and at distance $(R_0)$ equation 2.7.8 provides an indication of the time required to perturb the object out of the circular orbit in the direction of the sun.

$$v_0^2 + \frac{GM_\odot}{R_0} = v^2 + (v_0 - at)^2 \qquad (2.7.8a)$$

$$v_0^2 + \frac{GM_\odot}{R_0} = v^2 + \left(v_0 - C_t \frac{Area}{M}\rho c^2 t\right)^2 \qquad (2.7.8b)$$

$$where \quad v = \delta v_0$$

Assuming the sun orbital velocity perturbing factor $\delta$ to be .9 and the time (t) at the estimated age of the solar system (5,000,000,000 years), then the smallest size Dark Energy Influenced objects that could remain in sun orbit are listed below as a function of distance (AU) from the sun.

**Table 2.7.8a** Solar Space Dark Energy Influenced object upper size vs AU

| | AU | diameter (meter) |
|---|---|---|
| Mercury | 0.39 | 9.04 |
| Venus | 0.72 | 10.67 |
| Earth | 1.00 | 11.63 |
| Mars | 1.53 | 13.04 |
| Jupiter | 5.22 | 18.20 |
| Saturn | 9.57 | 21.50 |
| Uranus | 19.26 | 26.10 |
| Neptune | 30.17 | 29.60 |
| Pluto | 39.60 | 31.96 |
| Kuiper Belt | 50 | 35.95 |
| | 60 | 37.57 |
| Oort Cloud? | 80 | 40.36 |
| | 100 | 41.60 |
| | 200 | 50.82 |
| | 400 | 62.24 |

This calculation would indicate that there is not small particle dust remaining in the solar system, which would include the Kuiper Belt and hypothesized Oort Cloud. Also, the calculated small object size would seem to be consistent with Asteroid Belt objects, which exist at the coherence/decoherence boundary for the elastic space CPT lattice.

This momentum transfer between dark energy elastic space CPT lattice and impinging objects as represented by this equations 2.7.8a and 2.7.8b analysis has implications for large-scale galactic structures as well. Indeed, there is evidence in the Cosmic Gamma Ray Background Radiation for energies associated with this momentum transfer on the order of $m_T c^2$ or 56 MeV or ($\log_{10}$(Hz) = 22.13) or (2.20E-12 cm) [73]. In the available spectrum [73], there is a hint of a black body peak at 56 MeV or ($\log_{10}$(Hz) = 22.13) or ( 2.20E-12 cm). This radiation has not had an identifiable source until now.

Application of equations 2.7.8a and 2.7.8b at the Galactic scale replicate the observed constant rotation rate (220 km/sec) associated with the Milky Way Galaxy and others as well as indicated in Table 2.7.8b.

**Table 2.7.8b** Idealized flat rotation Milky Way Galactic Space object dark energy influenced upper size (after 13 billion years) vs galactic radius based on Milky Way mass of 2.15E11 solar masses contained within a galactic radius of 60,000 light years, galactic density $(\rho_g)$ of 6.43E-23 g/cm$^3$, object density $(\rho_o)$ of 1 g/cm$^3$ and galactic disc thickness of 12,000 light years

| Galactic Radius (lty) | Object Diameter (cm) $(d_o)$ | Object A/m (cm$^2$/g) | Object m/A (g/cm$^2$) | Object Velocity (km/sec) | Object M F P (lty) $(\rho_o d_o/\rho_g)$ | Optical Depth |
|---|---|---|---|---|---|---|
| 5,000 | 10.80 | 0.0463 | 21.6 | 220 | 3.55E+05 | 2.82E-02 |
| 7,500 | 11.96 | 0.0418 | 23.9 | 220 | 3.93E+05 | 3.82E-02 |
| 10,000 | 12.85 | 0.0389 | 25.7 | 220 | 4.23E+05 | 4.73E-02 |
| 12,500 | 13.59 | 0.0368 | 27.2 | 220 | 4.47E+05 | 5.60E-02 |
| 15,000 | 14.23 | 0.0351 | 28.5 | 220 | 4.68E+05 | 6.41E-02 |
| 17,500 | 14.79 | 0.0338 | 29.6 | 220 | 4.86E+05 | 7.20E-02 |
| 20,000 | 15.29 | 0.0327 | 30.6 | 220 | 5.03E+05 | 7.96E-02 |
| 22,500 | 15.75 | 0.0318 | 31.5 | 220 | 5.18E+05 | 8.69E-02 |
| 25,000 | 16.17 | 0.0309 | 32.3 | 220 | 5.32E+05 | 9.41E-02 |
| 27,500 | 16.56 | 0.0302 | 33.1 | 220 | 5.44E+05 | 1.01E-01 |
| 30,000 | 16.92 | 0.0295 | 33.8 | 220 | 5.56E+05 | 1.08E-01 |
| 32,500 | 17.27 | 0.0290 | 34.5 | 220 | 5.68E+05 | 1.14E-01 |
| 35,000 | 17.59 | 0.0284 | 35.2 | 220 | 5.78E+05 | 1.21E-01 |
| 37,500 | 17.90 | 0.0279 | 35.8 | 220 | 5.88E+05 | 1.27E-01 |
| 40,000 | 18.19 | 0.0275 | 36.4 | 220 | 5.98E+05 | 1.34E-01 |
| 42,500 | 18.47 | 0.0271 | 36.9 | 220 | 6.07E+05 | 1.40E-01 |
| 45,000 | 18.74 | 0.0267 | 37.5 | 220 | 6.16E+05 | 1.46E-01 |
| 47,500 | 18.99 | 0.0263 | 38.0 | 220 | 6.24E+05 | 1.52E-01 |
| 50,000 | 19.24 | 0.0260 | 38.5 | 220 | 6.32E+05 | 1.58E-01 |
| 52,500 | 19.47 | 0.0257 | 38.9 | 220 | 6.40E+05 | 1.64E-01 |
| 55,000 | 19.70 | 0.0254 | 39.4 | 220 | 6.48E+05 | 1.70E-01 |
| 57,500 | 19.92 | 0.0251 | 39.8 | 220 | 6.55E+05 | 1.76E-01 |
| 60,000 | 20.14 | 0.0248 | 40.3 | 220 | 6.62E+05 | 1.81E-01 |

The upper size objects vs distance from sun as calculated above and presented in Table 2.7.8a may provide a clue for constant velocity with distance observed in galactic rotation curves (in this case 220 km/sec). The calculated mean free path of such 'dark matter' objects indicates that they would be optically invisible as observed from outside the galaxy.

Dark energy and Dark matter may be very closely linked by momentum transfer forming Dark matter clumping. Reference [76] indicates non luminous but gravitational existence of dark matter clumping in the dynamics of galactic collisions and a thrust effect of the dark energy on galactic material. Reference [79] indicates the independence of dark and visible matter distribution but perhaps this independence is illusionary. According to the correlation herein, the observed dark matter distributions are composed of small dense (1 g/cc) hydrogen BEC objects (<1 meter) contributing to a space density of ~ 1.47E-23 g/cc with resulting optical path(=<1) dictating invisibility other than through the observed gravitational weak lensing effect.

**Table 2.7.8c** Idealized Dark Matter object dark energy influenced upper size (after 13 billion years) vs galactic radius based on Milky Way mass of 3.40E14 solar masses contained within a cluster radius of 240,000 light years, cluster density $(\rho_g)$ of 1.47E-23 g/cm$^3$, object density $(\rho_o)$ of 1 g/cm$^3$ and galactic disc thickness of 400,000 light years

| Galactic Radius (lty) | Object Diameter (cm) $(d_o)$ | Object A/m (cm$^2$/g) | Object m/A (g/cm$^2$) | Object Velocity (km/sec) | Object M F P (lty) $(\rho_o d_o/\rho_g)$ | Optical Depth |
|---|---|---|---|---|---|---|
| 20,000 | 4.60 | 0.1088 | 9.2 | 1211 | 3.31E+05 | 1.21E-01 |
| 30,000 | 5.09 | 0.0983 | 10.2 | 1211 | 3.67E+05 | 1.64E-01 |
| 40,000 | 5.47 | 0.0915 | 10.9 | 1211 | 3.94E+05 | 2.03E-01 |
| 50,000 | 5.78 | 0.0865 | 11.6 | 1211 | 4.17E+05 | 2.40E-01 |
| 60,000 | 6.05 | 0.0827 | 12.1 | 1211 | 4.36E+05 | 2.75E-01 |
| 70,000 | 6.29 | 0.0795 | 12.6 | 1211 | 4.53E+05 | 3.09E-01 |
| 80,000 | 6.50 | 0.0769 | 13.0 | 1211 | 4.69E+05 | 3.41E-01 |
| 90,000 | 6.70 | 0.0747 | 13.4 | 1211 | 4.83E+05 | 3.73E-01 |
| 100,000 | 6.87 | 0.0727 | 13.7 | 1211 | 4.96E+05 | 4.04E-01 |
| 110,000 | 7.04 | 0.0710 | 14.1 | 1211 | 5.07E+05 | 4.34E-01 |
| 120,000 | 7.20 | 0.0695 | 14.4 | 1211 | 5.19E+05 | 4.63E-01 |
| 130,000 | 7.34 | 0.0681 | 14.7 | 1211 | 5.29E+05 | 4.91E-01 |
| 140,000 | 7.48 | 0.0669 | 15.0 | 1211 | 5.39E+05 | 5.19E-01 |
| 150,000 | 7.61 | 0.0657 | 15.2 | 1211 | 5.48E+05 | 5.47E-01 |
| 160,000 | 7.73 | 0.0647 | 15.5 | 1211 | 5.57E+05 | 5.74E-01 |
| 170,000 | 7.85 | 0.0637 | 15.7 | 1211 | 5.66E+05 | 6.01E-01 |
| 180,000 | 7.96 | 0.0628 | 15.9 | 1211 | 5.74E+05 | 6.27E-01 |
| 190,000 | 8.07 | 0.0619 | 16.1 | 1211 | 5.82E+05 | 6.53E-01 |
| 200,000 | 8.18 | 0.0611 | 16.4 | 1211 | 5.89E+05 | 6.79E-01 |
| 210,000 | 8.28 | 0.0604 | 16.6 | 1211 | 5.97E+05 | 7.04E-01 |
| 220,000 | 8.37 | 0.0597 | 16.7 | 1211 | 6.04E+05 | 7.29E-01 |
| 230,000 | 8.47 | 0.0590 | 16.9 | 1211 | 6.10E+05 | 7.54E-01 |
| 240,000 | 8.56 | 0.0584 | 17.1 | 1211 | 6.17E+05 | 7.78E-01 |

Reference [64] presents an experimental protocol in which a comoving optical and sodium atom lattice is studied at nanokelvin a temperature. When the moving lattice is stopped (generating laser beams extinguished), the sodium atoms stop rather than proceeding inertially. This unexplained anomaly may be explained in a similar manner as Pioneer deceleration. The following calculation indicates a sodium atom deceleration large enough to indicate essentially an instantaneous stop.

**Table 2.7.**9 Sodium Atom Deceleration

| | | |
|---|---|---|
| sodium temperature | 1.00E-06 | kelvin |
| sodium mass | 3.82E-23 | gram |
| sodium van der Waals diameter | 4.54E-08 | centimeter |
| sodium cross-section | 1.62E-15 | $cm^2$ |
| sodium area/mass | 4.24E07 | $cm^2/g$ |
| optical lattice velocity | 3.0 | cm/sec |
| sodium Reynold's number | 7.15E-11 | unitless |
| thrust coefficient | 3.36E11 | unitless |
| thrust force (with $C_t$) | 3.116E-12 | dyne |
| sodium deceleration | 8.16E10 | $cm/sec^2$ |
| time to stop sodium atom | 3.68E-11 | sec |
| distance to stop sodium atom | 5.51E-11 | cm |

An experiment is envisioned whereby larger atomic clusters are decelerated in a more observable manner because of their increased mass. It assumed that such an anomalous deceleration will occur only at nanokelvin temperatures where the $k_bT$ energies are essentially zero.

It is understood that the New Horizons spacecraft launched on January 19, 2006 presents an opportunity to replicate the Pioneer 10 & 11 deceleration. The tables below provide a predictive estimate of this deceleration. Because of the New Horizons smaller area to mass ratio than Pioneer, it is anticipated that the New Horizons anomalous deceleration will be smaller than the Pioneer anomalous deceleration by a factor of about .4. Also, because the New Horizons Moment of Inertia about the spin axis is smaller than Pioneer, it is predicted that the rate of spin deceleration will be greater for the New Horizons than observed for the Pioneer space crafts.

**Table 2.7.9** New Horizons (NH) Translation Predictions

| | |
|---|---|
| New Horizons mass | 470,000 gram |
| New Horizons diameter | 210 cm (effective) |
| New Horizons cross section | 34,636(effective) $cm^2$ |
| New Horizons area/mass | 0.07 $cm^2/g$ |
| trisine area/mass | 5.58E30 $cm^2/g$ |
| New Horizons velocity | 1,622,755 cm/sec |
| New Horizons Reynold's number | 3.31E-01 unitless |
| drag coefficient | 76.82 |
| drag force (with drag coefficient) | 1.525E-02 dyne |
| New Horizons deceleration | 3.24E-08 $cm/sec^2$ |
| one year drag distance | 16,132,184 cm |
| laminar drag force | 7.21E-03 dyne |
| time of New Horizons to stop | 5.00E13 sec |

**Table 2.7.10** New Horizons (NH) Rotational Predictions

| | |
|---|---|
| NH mass | 470,000 g |
| NH moment of inertia NASA JPL | 4.01E09 g $cm^2$ |
| NH diameter | 315 cm |
| NH translational cross-section | 77,931 $cm^2$ |
| NH radius of gyration k | 131 cm |
| NH radius r | 158 cm |
| paddle cross-section | 60,210 $cm^2$ |
| paddle area/mass | 1.28E-01 $cm^2/g$ |
| NH rotation speed at k | 5,971 cm/sec |
| NH rotation rate change | 0.2000 rpm/year |
| NH rotation rate change | 1.06E-10 $rotation/sec^2$ |
| NH rotation deceleration at k | 8.68E-08 $cm/sec^2$ |
| NH force slowing it down | 2.04E-02 dyne |
| NH Reynold's number | 4.36E-01 unitless |
| thrust coefficient | 59.09 |
| thrust (with $C_t$) | 2.04E-02 dyne |
| one year rotational thrust distance | 43,139,338 cm |
| NH rotational laminar thrust force | 1.90E-02 dyne |
| time for NH to stop rotating | 6.88E10 sec |

This New Horizons deceleration characteristics will be monitored in the vicinity of Jupiter passage on February 28, 2007 and thereafter.

In addition to the Linear and Exponential Decay Pioneer deceleration $aP$ models presented in refs 116 & 116a, a combination Exponential Deceleration Decay To Constant Linear Deceleration approached at infinity is evaluated.

$$a_P(t) = a_P(t_0)\, e^{-\beta(t-t_0)\ln 2} + a_P(infinity)$$

For this evaluation, stochastic data is digitized from graphical presentation in refs: 116 and 116a. Initial conditions are at January 1, 1980.

For Pioneer 10 Model, given initial $a_P(t_0) = 9.82\ x\ 10^{-10}\ m/s^2$ at time 8.79 years from Table 2.7.11, then multivariable regression minimizing stochastic – model Root Mean Square(RMS) data from Table 2.7.13 yields the following modeled constants:

$$\beta = 1/5.00\ years$$

$$a_P(infinity) = 7.00\ x\ 10^{-10}\ m/s^2$$

$$RMS = .224\ x\ 10^{-10}\ m/s^2$$

This is compared to analysis with $a_P(infinity) = 0$

$$\beta = 1/61.0\ years$$

$$a_P(t_0) = 9.82\ x\ 10^{-10}\ m/s^2$$

$$RMS = .478\ x\ 10^{-10}\ m/s^2$$

For Pioneer 11 Model, given $a_P(t_0) = 9.274\ x\ 10^{-10}\ m/s^2$ at 10.632 years from Table 2.7.12, then multivariable regression minimizing stochastic – model Root Mean Square(RMS) data from Table 2.7.14 yields the following modeled constants:

$$\beta = 1/3.30\ years$$

$$a_P(infinity) = 8.20\ x\ 10^{-10}\ m/s^2$$

$$RMS = .160\ x\ 10^{-10}\ m/s^2$$

This is compared to analysis with $a_P(infinity) = 0$

$$\beta = 1/160\ years$$

$$a_P(t_0) = 9.274\ x\ 10^{-10}\ m/s^2$$

$$RMS = .273\ x\ 10^{-10}\ m/s^2$$

Table 2.7.11 Pioneer 10
Modeled Stochastic Deceleration Data

| Time($t$) (years) | Stochastic $a_P$ x $10^{-10}$ m/s$^2$ | Modeled $a_P$ x $10^{-10}$ m/s$^2$ |
|---|---|---|
| 8.79 | 9.82 | 9.91 |
| 10.78 | 9.33 | 9.20 |
| 12.79 | 8.78 | 8.65 |
| 14.82 | 8.21 | 8.24 |
| 16.81 | 8.21 | 7.94 |
| 18.80 | 7.39 | 7.71 |
| 20.81 | 7.34 | 7.53 |
| 22.82 | 7.23 | 7.40 |
| 24.85 | 7.72 | 7.30 |

Table 2.7.12 Pioneer 11
Modeled Stochastic Deceleration Data

| Time($t$) (years) | Stochastic $a_P$ x $10^{-10}$ m/s$^2$ | Modeled $a_P$ x $10^{-10}$ m/s$^2$ |
|---|---|---|
| 10.632 | 9.274 | 9.187 |
| 12.623 | 8.840 | 8.832 |
| 14.631 | 8.358 | 8.604 |
| 16.602 | 8.635 | 8.460 |

Table 2.7.13 Pioneer 10
Minimized Stochastic – Model (RMS x $10^{-10}$ m/s$^2$)

| $1/\beta$ years \ $a_P inf$ x $10^{-10}$ m/s$^2$ | 6.80 | 6.90 | 7.00 | 7.10 | 7.20 |
|---|---|---|---|---|---|
| 4.80 | .372 | .299 | .246 | .228 | .251 |
| 4.90 | .336 | .270 | .230 | .231 | .271 |
| 5.00 | .303 | .247 | **.224** | .244 | .298 |
| 5.10 | .274 | .232 | .229 | .266 | .330 |
| 5.40 | .252 | .226 | .242 | .294 | .336 |

Table 2.7.14 Pioneer 11
Minimized Stochastic – Model (RMS x $10^{-10}$ m/s$^2$)

| $1/\beta$ years \ $a_P inf$ x $10^{-10}$ m/s$^2$ | 8.00 | 8.10 | 8.20 | 8.30 | 8.40 |
|---|---|---|---|---|---|
| 3.10 | .325 | .244 | .183 | .167 | .205 |
| 3.20 | .284 | .210 | .165 | .175 | .232 |
| 3.30 | .246 | .182 | **.160** | .196 | .267 |
| 3.40 | .221 | .164 | .170 | .227 | .306 |
| 3.50 | .183 | .160 | .194 | .264 | .349 |

**Figure 2.7.6** Pioneer 10 Anomalous Deceleration $a_P$ x10$^{-8}$ *cm/sec*$^2$

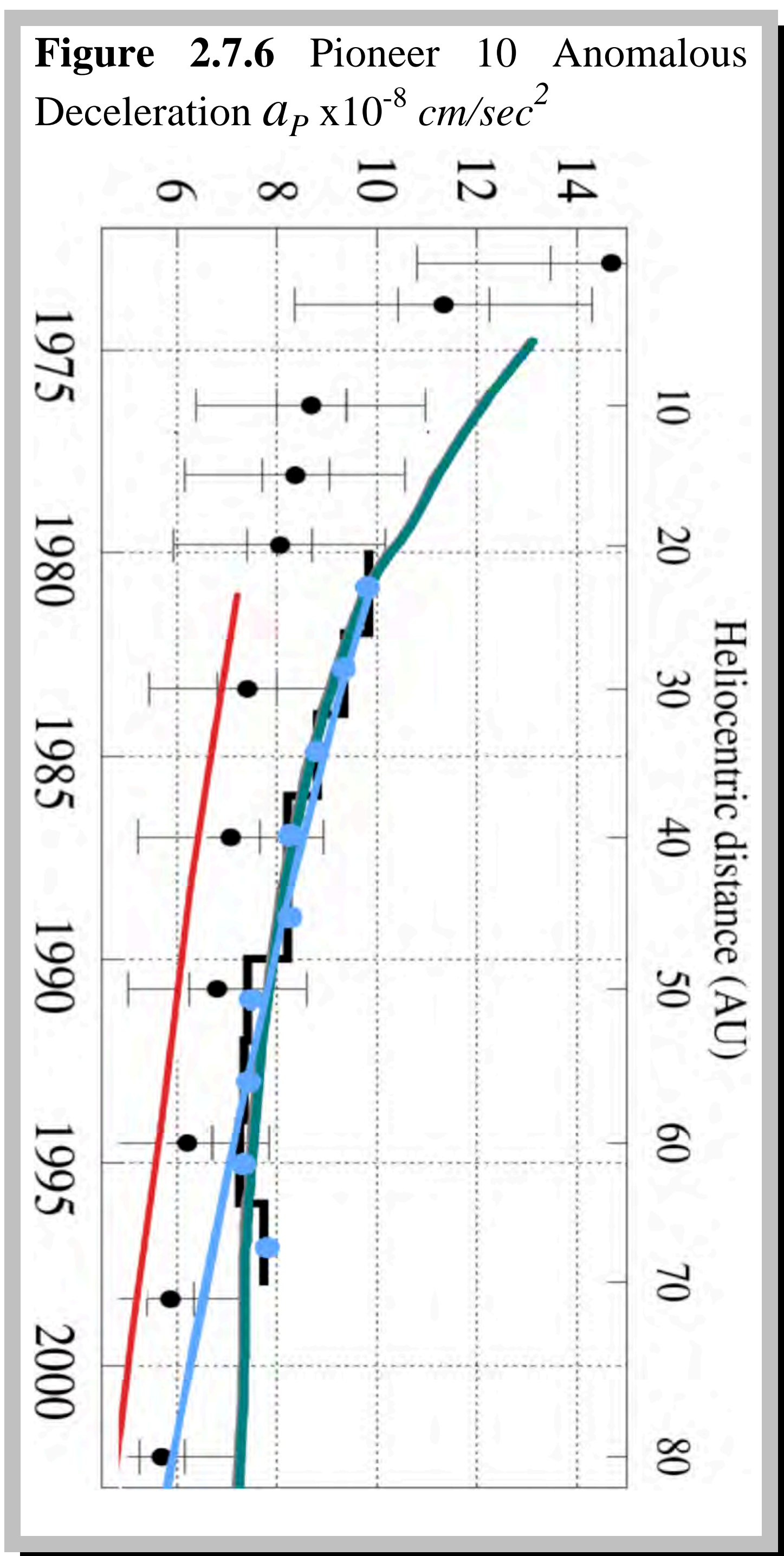

**Figure 2.7.7** Pioneer 11 Anomalous Deceleration $a_P$ x10$^{-8}$ *cm/sec*$^2$

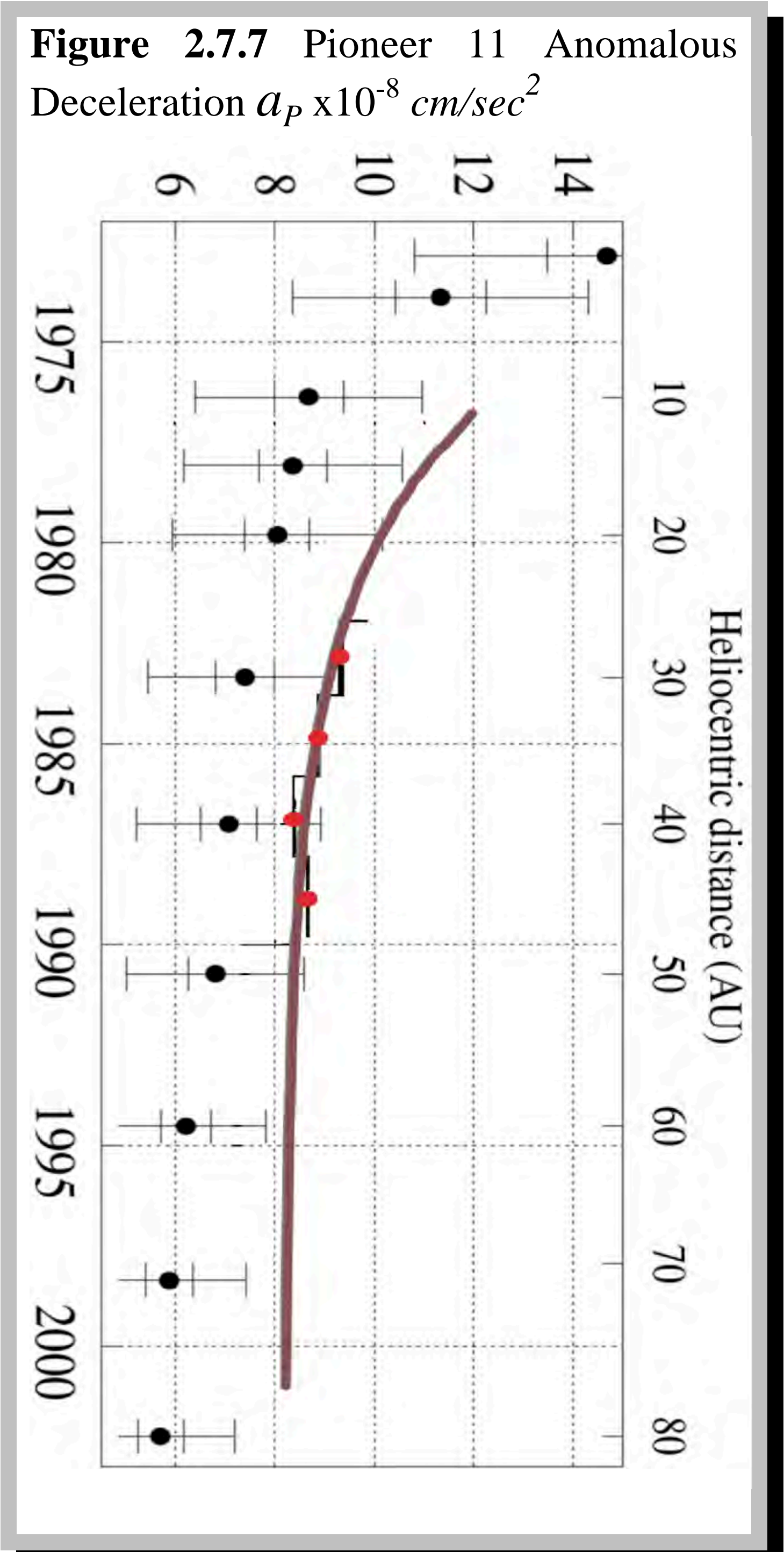

Figure 2.7.6 Pioneer 10 deceleration profile

Stochastic data is represented by black steps with blue midpoint circles ref 116

Blue line is Pioneer deceleration decay midpoint regression

$$(10.1\pm1.0)2^{-(t-t_0)/(28.8\pm2.0)} \quad \text{ref 116a} \qquad (2.7.9)$$

Red line is thermal deceleration decay midpoint regression

$$(7.4 \pm 2.5) 2^{-(t-t_0)/(36.9 \pm 6.7)} \text{ ref 116a} \qquad (2.7.10)$$

Green Line is Exponential Deceleration Decay To Constant Linear Deceleration

$$9.82 e^{-\frac{1}{5.00}(t-t_0)\ln 2} + 7.00 \text{ ref 116a} \qquad (2.7.11)$$

with extrapolation beyond data points

Figure 2.7.7 Pioneer 11 deceleration profile with thermal profile from Pioneer 10
Stochastic data is represented by black steps with red midpoint circles ref 116 with extrapolation beyond data points
Brown Line is Exponential Deceleration Decay To Constant Linear Deceleration

$$9.274 e^{-\frac{1}{3.30}(t-t_0)\ln 2} + 8.20 \qquad (2.7.12)$$

with extrapolation beyond data points.

Figures 2.7.6 and 2.7.7 indicate Pioneer thermal force has less contribution to Pioneer 10 deceleration with time approaching a constant Pioneer deceleration and not negating

$$a_P = (8.74 \pm 1.33) \text{ x } 10^{-10} \text{ m/s}^2$$

as originally suggested by John Anderson et al.[53].

Further analysis based on established theory indicates the Pioneer recoil deceleration may be calculated as:

$$pressure \text{ x } pioneer\ area / pioneer\ mass = a_P \qquad (2.7.13)$$

This deceleration $a_P$ is then proportional to a net system temperature$(T)^4$.

$$radiation\ pressure = F\sigma_s T^4 \left(\frac{4}{3c}\right) \qquad (2.7.14)$$

with a multiplier factor $F$ as a measure of material emissivity or absorptivity.

Pioneer recoil deceleration(aP) may be calculated as:

$$radiation\ pressure = F\sigma_s T^4 \left(\frac{4}{3c}\right) \qquad (2.7.15)$$

Then the pioneer deceleration $a_P$ may be represented as:

$$a_P = \frac{(radiation\ pressure)\ (pioneer\ area)}{pioneer\ mass} \qquad (2.7.16)$$

or

$$a_P = F\sigma_s T^4 \left(\frac{4}{3c}\right) \frac{pioneer\ area}{pioneer\ mass} \qquad (2.7.17)$$

A theoretically consistent model was used[116,116a]:

$$a_P = \frac{nQ}{\text{pioneer mass c}} \qquad (2.7.18)$$

where $Q$ is directed radiation power and $n$ represents empirically determined $F$ and pioneer asymmetric areas.

Deceleration ($a_P$) as well as radiation power half_life vary as 1/4 of RTG half_life.

Deceleration ($a_P$) as well as radiation power half_life should then be on the order of 87.7/4 or 21.9 years

These half_lives should be considered very accurate and this accuracy is far greater than earth monitors can observe through thermal telemetry and on-board radiation pressureshould follow the well established decay law $dx/dt = -kx$.

model one

$$\frac{da_P}{dt} = -ka_P = -\frac{\ln(2)}{87.2/4} a_P \qquad (2.7.19)$$

The stochastic data half life more accurately follows a modified version (not addressed in refs 116,116a).

model two

$$\frac{da_P}{dt} = -k\left(a_P - a_{P\infty}\right) \qquad (2.7.20)$$

indicating $a_P$ approaches a constant value ($a_{P\infty}$) with time($t$).

In summary, the references[116.116a] model the Pioneer deceleration $a_{P\infty}$ as the theoretically correct

$$a_P = \frac{\eta Q}{pioneer\ mass\ c} \tag{2.7.21}$$

with two components

$$a_P = \eta_{rtg}\frac{Q_{rtg}}{pioneer\ mass\ c} + \eta_{elec}\frac{Q_{elec}}{pioneer\ mass\ c} \tag{2.7.22}$$

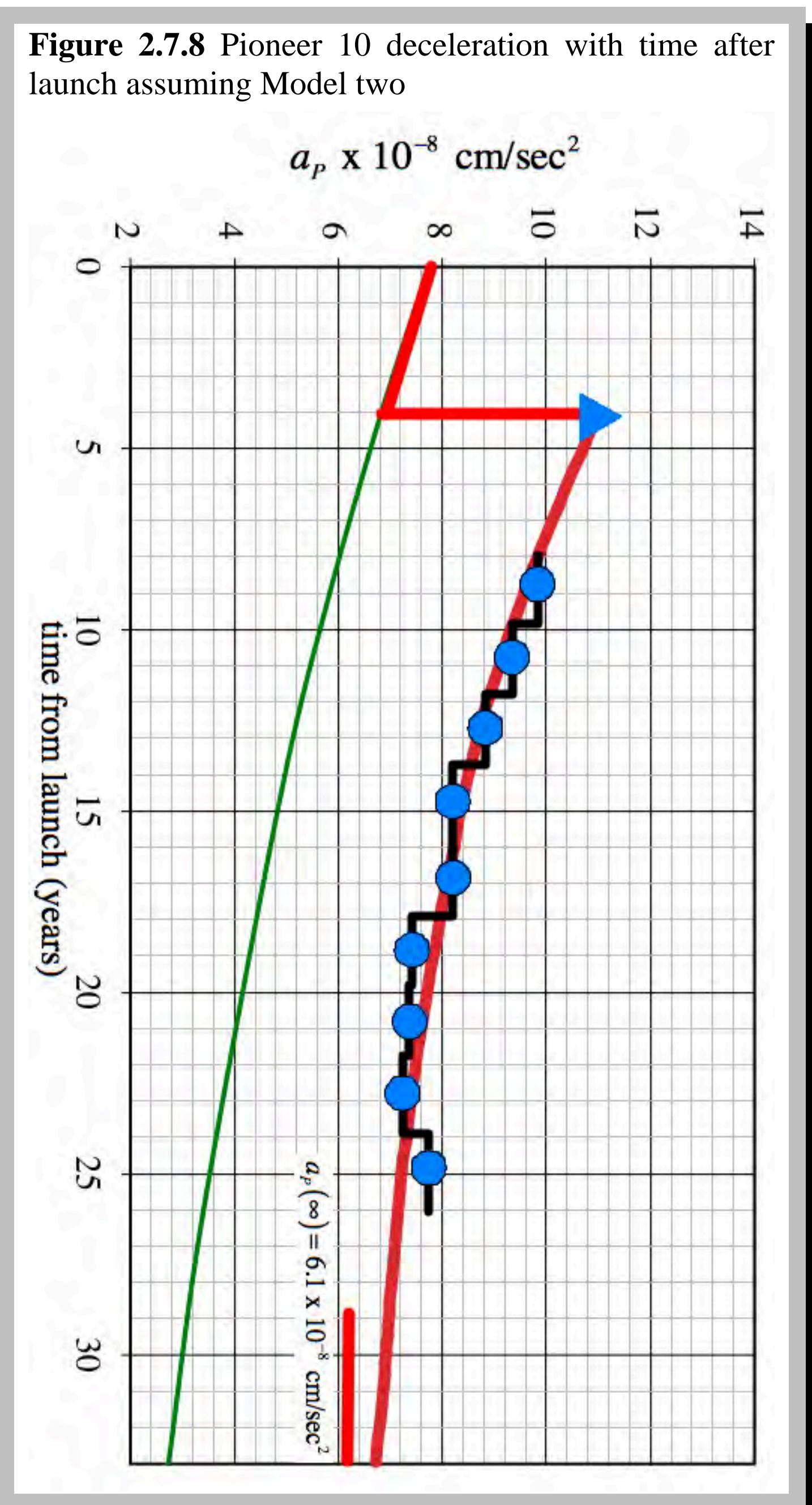

**Figure 2.7.8** Pioneer 10 deceleration with time after launch assuming Model two

The data for component:

$$a_P = \eta_{elec}\frac{Q_{elec}}{pioneer\ mass\ c} \tag{2.7.23}$$

follows the relationship linked to 1/4 of RTG half life (87.7 years)

$$\frac{da_P}{dt} = -\frac{\ln(2)}{87.2/4}a_P \tag{2.7.24}$$

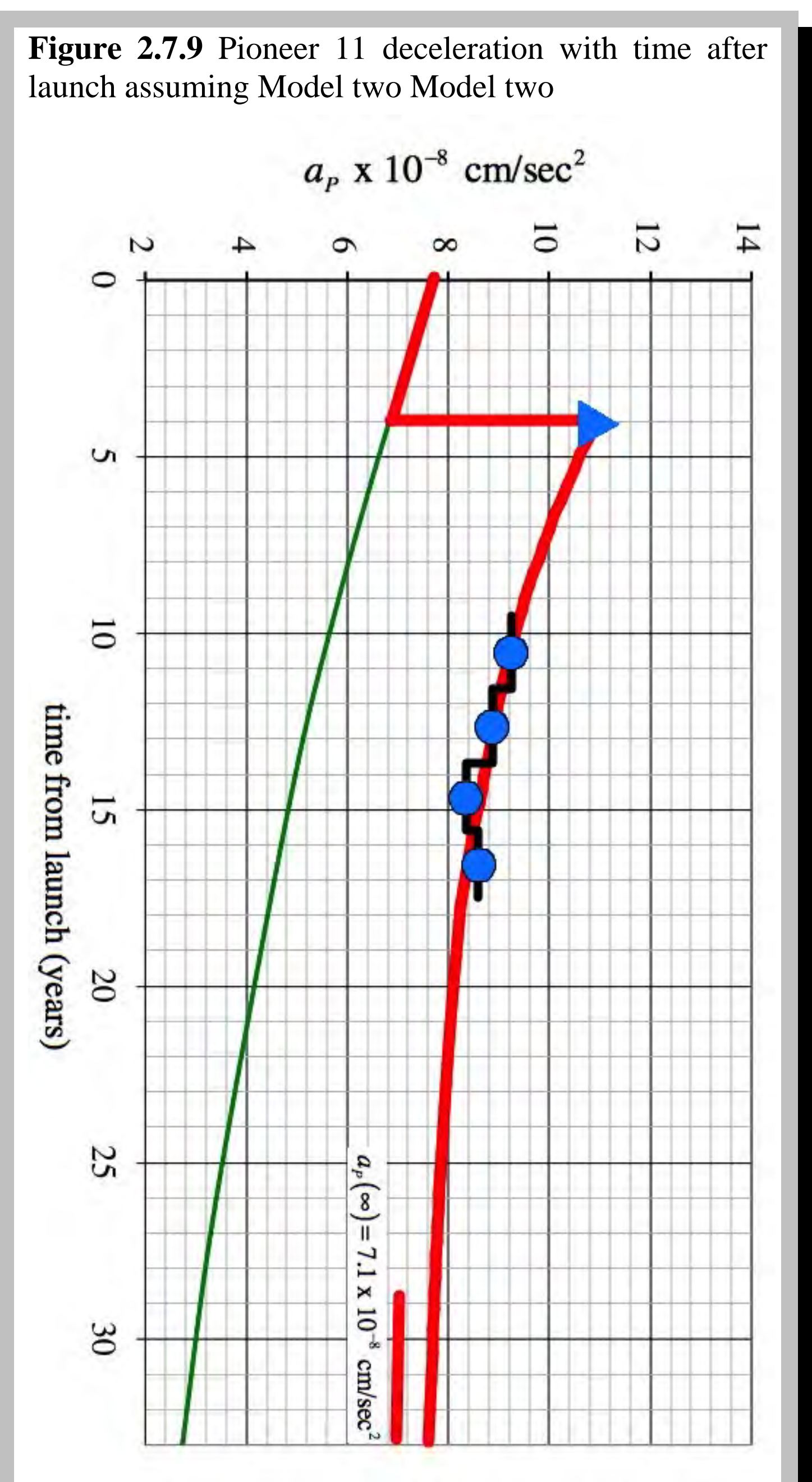

**Figure 2.7.9** Pioneer 11 deceleration with time after launch assuming Model two Model two

It is suggested that RTG deceleration component:

$$a_P = \eta_{rtg} \frac{Q_{rtg}}{pioneer\ mass\ c} \qquad (2.7.25)$$

is directly linked to RTG 87.7 year half life[116a].

$$\frac{da_P}{dt} = -\frac{\ln(2)}{87.2} a_P \qquad (2.7.26)$$

But this is theoretically not possible because of equations 2.7.16 and 2.7.17 where deceleration is related to $T^4$

Further, the RTG geometric design symmetry does not in principle allow for asymmetric radiation pressure contributing to $a_P$ .

Further, the theoretical $a_P$ identities (eqns 2.7.16 and 2.7.17) indicates $\eta$ represents empirically determined $F$ and pioneer asymmetric areas. Any anticipated extremely small RTG manufacturing area asymmetries would contribute in a congruent extremely small manner.

Further, the RTG component decay is only about 20 percent over 25 years.

It is suggested the RTG component is incorrectly identified in refs[116,116a] as contributing to $a_P$ . The actual contribution to model stochastic data is represented in eqn 2.7.23 plus an anomalous constant factor (aPinfinity) as represented in eqn 2.7.20 and $a_{P\infty}$ is testing the space vacuum[130] as per equation 2.11.9a.

Application of model two for stepped Pioneer 10 and 11 doppler data is represented in Figures 2.7.8 and 2.7.9.

The averaged per interval (black steps) Pioneer data are represented buy blue circles.

The green line represents the electric component 2.7.23 decaying in accordance with equation 2.7.24.

The blue triangle represents the beginning of a constant Pioneer deceleration estimation in the neighborhood of the asteroid belt.

The red line follow the Pioneer deceleration from a purely electrical (thermal) (eqn 2.7.19) contribution transitioning to the electrical plus constant following model two (equation 2.7.20). Model two fits Pioneer 10 Doppler data with RMS = .272x10$^{-8}$ cm/sec$^2$ and Pioneer 11 Doppler data with RMS = .180x10$^{-8}$ cm/sec$^2$. The Pioneer 10 $a_{P\infty}$ is 6.1x10$^{-8}$ cm/sec$^2$ and Pioneer 11 $a_{P\infty}$ is 7.1x10$^{-8}$ cm/sec$^2$ .

Pioneers could be probing the space vacuum viscosity in the same manner as the Voyagers test interstellar cosmic rays as they exist beyond the heliosphere (all initiated at JPL). Other object decelerations could then be dimensionally scaled to the Pioneer spacecraft. (planetary sized object deceleration would be negligible) The recently commissioned Dark Energy Survey http://www.darkenergysurvey.org/ detailing universe expansion may be correlated to this phenomenon if galactic components are primarily of Pioneer size or smaller.

****************************************

Aa. time (years after launch)(after AU to time conversion)
Bb. electric contribution to $a_{Pelec}$x10$^{-8}$ cm/sec$^2$
n = .406 half life 24.5 years regressing [116a]

$$a_{Pelec} = 7.75x10^{-8}\ e^{\frac{-\ln(2) time\ after\ launch}{24.5}}$$

Cc. RTG contribution to $a_{Prtg}$x10^-10 cm/sec$^2$
n = 0.0104 half life 87.72 years (Plutonium)

$$a_{Prtg} = 3.71x10^{-8}\ e^{\frac{-\ln(2) time\ after\ launch}{87.72}}$$

Dd. or Bb + Cc Total contribution to aPx10^-8 m/sec$^2$
Ee. Doppler measured $a_P$x10$^{-8}$ cm/sec$^2$

Scenario one

| Aa | Bb | Cc | Dd | Ee |
|---|---|---|---|---|
| 8.79 | 6.05 | 3.46 | 9.51 | 9.82 |
| 10.78 | 5.71 | 3.41 | 9.12 | 9.33 |
| 12.79 | 5.40 | 3.35 | 8.75 | 8.78 |
| 14.82 | 5.10 | 3.30 | 8.40 | 8.21 |
| 16.81 | 4.82 | 3.25 | 8.07 | 8.21 |
| 18.80 | 4.56 | 3.20 | 7.75 | 7.39 |
| 20.81 | 4.30 | 3.15 | 7.45 | 7.34 |
| 22.82 | 4.07 | 3.10 | 7.17 | 7.23 |
| 24.85 | 3.84 | 3.05 | 6.89 | 7.72 |

Dd-Ee has RMS = .34

****************************************

Statistical analysis of fin root temperatures[53a Figure 20] indicate temperature($T$) half life of 130 years. This translates into power half life ~emissivity x $T^4$ or 130/4 or 32.5 years. This is not reflected in reported RTG power performance. Radiation flux during Pioneer 10 Jupiter flyby (10,000 times that of Earth) did not cause differential RTG emissivity. Report [53 section C] does not provide any further mechanistic origin of Cc and none presented in [116a]. Assuming RTG contribution(Cc) is replaced by a constant(Ff) the electric(Bb) + constant(Ff) or Gg yields RMS = .33 (equal to scenario one .34) when compared to doppler(Ee)

Scenario two

| Aa | Bb | Ff | Gg | Ee |
|---|---|---|---|---|
| 8.79 | 6.05 | 3.33 | 9.38 | 9.82 |
| 10.78 | 5.71 | 3.33 | 9.04 | 9.33 |
| 12.79 | 5.40 | 3.33 | 8.73 | 8.78 |
| 14.82 | 5.10 | 3.33 | 8.43 | 8.21 |
| 16.81 | 4.82 | 3.33 | 8.15 | 8.21 |
| 18.80 | 4.56 | 3.33 | 7.89 | 7.39 |
| 20.81 | 4.30 | 3.33 | 7.63 | 7.34 |
| 22.82 | 4.07 | 3.33 | 7.40 | 7.23 |
| 24.85 | 3.84 | 3.33 | 7.17 | 7.72 |

Gg-Ee has RMS = .33

****************************************

A more detailed model Hh minimizing RMS at .27 (the best available fit) assumes aP approaching a constant 5.9

$$a_P = (12.587 - 5.9)e^{(-.068*time)} + 5.9$$

Scenario three

| Aa | Hh | Ee |
|---|---|---|
| 8.79 | 9.60 | 9.82 |
| 10.78 | 9.14 | 9.33 |
| 12.79 | 8.74 | 8.78 |
| 14.82 | 8.39 | 8.21 |
| 16.81 | 8.09 | 8.21 |
| 18.80 | 7.83 | 7.39 |
| 20.81 | 7.60 | 7.34 |
| 22.82 | 7.40 | 7.23 |
| 24.85 | 7.23 | 7.72 |

Hh-Ee has RMS = .27

A Scenario three analysis of more limited Pioneer 11 data:

$$a_P = (12.587 - 7.4)e^{(-.098*time)} + 7.4$$

RMS = .18

There is statistical logic within the context of [116a] and historical Pioneer reports for a constant $a_P$ contribution. The astrophysical importance of this constant $a_P$ contribution demands attention. [116a] conclusions should be executed:

First: Pioneer 11 data should be analyzed

Second: the anomalous spin-down of both spacecraft should be addressed considering the asymmetrically electrical (door) $a_P$ would have very little influence on the Pioneer moment of inertia $(MR^2)$ primarily represented by the 30 foot magnetometer boom. The spin deceleration may be another indication of constant $a_P$ linking Pioneer rotational angular momentum to translational momentum.

Third: Analysis early trajectory data for indication of $a_P$ onset comparable to a constant 5.9E-8 cm/sec$^2$

Fourth: Address the possibility of outgassing from surface materials, with potential observable contribution to the anomalous acceleration.

Fifth: Our understanding of systematics in the Doppler tracking data can be improved by a detailed auto correlation analysis.

Sixth: Measure RTG paint emissivity by a thermal vacuum chamber test of a hot RTG analogue. (Resolve scenarios one and two)

Seventh: Analyze other redundant data in the form of Deep Space Network signal strength measurements, which could be used to improve our understanding of the spacecraft's precise orientation.

## 2.8 Superconducting Resonant Current, Voltage And Conductance (and Homes' Law)

The superconducting electrical current $(I_e)$ is expressed as in terms of a volume *chain* containing a Cooper CPT Charge conjugated pair moving through the trisine CPT lattice at a de Broglie velocity $(v_{dx})$ and the same as Maxwell's Ampere Law.

$$J_x = \frac{I_e}{area} = \frac{D_{fx}}{time_\pm} \tag{2.8.1}$$

$$J_{xr} = \frac{I_e}{area_r} = \frac{D_{fxr}}{time_{r\pm}} \tag{2.8.1r}$$

The superconducting mass current $(I_m)$ is expressed as in terms of a trisine resonant transformed mass $(m_t)$ per *cavity* containing a Cooper CPT Charge conjugated pair and the *cavity* Cooper CPT Charge conjugated pair velocity $(v_{dx})$.

$$J_{mx} = \frac{I_m}{area} = \left(\frac{m_T}{cavity}\right)v_{dx} \tag{2.8.2}$$

$$J_{mxr} = \frac{I_m}{area_r} = \left(\frac{m_T}{cavity_r}\right)v_{dxr} \tag{2.8.2r}$$

It is recognized that under strict CPT symmetry, there would be zero current flow. It must be assumed that pinning of one side of CPT will result in observed current flow. Pinning (beyond that which naturally occurs) of course has been observed in real superconductors to enhance actual supercurrents. This is usually performed by inserting unlike atoms into the superconducting molecular lattice or mechanically distorting the lattice at localized points [72]. Also, there are reports that superconductor surface roughness enhances super currents by 30% [72]. This could be due to the inherent "roughness" of the CPT symmetry better expressing itself at a rough or corrugated surface rather than a smooth one.

Secondarily, this reasoning would provide reasoning why a net charge or current is not observed in a vacuum such as trisine model extended to critical space density in section 2.11.

A standard Ohm's law can be expressed as follows with the resistance expressed as the Hall quantum resistance with a value of 25,815.62 ohms (von Klitzing constant $2\pi\hbar/e_{\pm}^{2}$) in eqn 2.8.3.

$$voltage = \left( current \right)\left( resistance \right)$$

$$\frac{k_b T_c}{\mathbb{C}\ e_{\pm}} = \left( \frac{e_{\pm}}{\sqrt{3}\alpha\ approach\ time} \right)\left( \frac{2\pi\hbar}{\left(\mathbb{C}\ e_{\pm}\right)^2} \right) \qquad (2.8.3)$$

$$\frac{k_b T_c}{\mathbb{C}\ e_{\pm}} = \left( \frac{D_{fx}}{time} \right)\left( \frac{2\pi\hbar}{\left(\mathbb{C}\ e_{\pm}\right)^2} \right)$$

$$voltage = \left( current \right)\left( resistance \right)$$

$$\frac{k_b T_{cr}}{\mathbb{C}\ e_{\pm}} = \left( \frac{e_{\pm}}{\sqrt{3}\alpha\ approach_r\ time_r} \right)\left( \frac{2\pi\hbar}{\left(\mathbb{C}\ e_{\pm}\right)^2} \right) \qquad (2.8.3r)$$

$$\frac{k_b T_{cr}}{\mathbb{C}\ e_{\pm}} = \left( \frac{D_{fxr}}{time_r} \right)\left( \frac{2\pi\hbar}{\left(\mathbb{C}\ e_{\pm}\right)^2} \right)$$

**Table 2.8.1** with super current $(J_x)$ $\left(amp/cm^2\right)$ plot.

| Condition | 1.2 | 1.3 | 1.6 | 1.9 | 1.10 | 1.11 |
|---|---|---|---|---|---|---|
| $T_a\left({}^0K\right)$ | 6.53E+11 | 6.53E+11 | 6.53E+11 | 6.53E+11 | 6.53E+11 | 6.53E+11 |
| $T_b\left({}^0K\right)$ | 9.05E+14 | 5.48E+13 | 9.22E+08 | 1.10E+07 | 2,868 | 2.723 |
| $T_{br}\left({}^0K\right)$ | 3.30E+12 | 5.45E+13 | 3.24E+18 | 2.72E+20 | 1.04E+24 | 1.10E+27 |
| $T_c\left({}^0K\right)$ | 8.96E+13 | 3.28E+11 | 93.0 | 0.0131 | 8.99E-10 | 8.10E-16 |
| $T_{cr}\left({}^0K\right)$ | 1.19E+09 | 3.25E+11 | 1.15E+21 | 8.11E+24 | 1.19E+32 | 1.32E+38 |
| $z_R+1$ | 3.67E+43 | 8.14E+39 | 3.89E+25 | 6.53E+19 | 1.17E+09 | 1.00E+00 |
| $z_B+1$ | 3.32E+14 | 2.01E+13 | 3.39E+08 | 4.03E+06 | 1.05E+03 | 1.00E+00 |
| $Age_{UG}\left(sec\right)$ | 7.14E-05 | 4.79E-03 | 6.94E+04 | 5.36E+07 | 1.27E+13 | 4.33E+17 |
| $Age_{UGv}\left(sec\right)$ | 4.59E-10 | 1.25E-07 | 4.42E+02 | 3.13E+06 | 1.22E+13 | 4.33E+17 |
| $Age_{U}\left(sec\right)$ | 1.18E-26 | 5.31E-23 | 1.11E-08 | 6.63E-03 | 3.70E+08 | 4.33E+17 |
| $I\left(amp/cm^2\right)$ | 7.05E+24 | 9.45E+19 | 7.60E+00 | 1.52E-07 | 7.10E-22 | 5.77E-34 |
| $I\left(amp/cm^2\right)$ | 1.24E+15 | 9.28E+19 | 1.15E+39 | 5.78E+46 | 1.23E+61 | 1.52E+73 |
| $I\left(g/cm^2/\text{sec}\right)$ | 1.16E+26 | 1.56E+21 | 1.25E+02 | 2.50E-06 | 1.17E-20 | 9.52E-33 |
| $I\left(g/cm^2/\text{sec}\right)$ | 2.05E+16 | 1.53E+21 | 1.90E+40 | 9.54E+47 | 2.04E+62 | 2.51E+74 |

Equation 2.8.4 represents the Poynting vector relationship or in other words the power per area per resonant CPT $time_{\pm}$ being transmitted from and absorbed in each superconducting *cavity*:

$$k_b T_c = \frac{m_e}{m_T}\frac{1}{2}\ E_x\ H_c\left(2\ side\right)\ time_{\pm}\ v_{\varepsilon} \qquad (2.8.4)$$

Equation 2.8.4 can be rearranged in terms of a relationship between the de Broglie velocity $\left(v_{dx}\right)$ and modified speed of light velocity $\left(v_{\varepsilon x}\right)$ as expressed in equation 2.8.5.

$$v_{\varepsilon x}^2 = \frac{\sqrt{3}}{48\pi^2}\left(\frac{m_T}{m_e}\right)^2\frac{\left(\mathbb{C}\ e_{\pm}\right)^2}{\hbar}\left(\frac{B}{A}\right)_T v_{dx}$$

$$= 3^{\frac{1}{4}}\left(\frac{B}{A}\right)_T\cdot c\cdot v_{dx} = \pi\cdot c\cdot v_{dx} \qquad (2.8.5)$$

This leads to the group/phase velocity relationship in equation 2.8.6, which holds for the sub- and super- luminal nuclear conditions with the length dimensional transition point at the nuclear radius as established by dimensional analysis in Appendix I.

$$\left(v_{dx} < c,\ v_{\varepsilon x} < c\ \&\ v_{dxr} > c\right)\ \&\ \left(v_{dx} > c,\ v_{\varepsilon x} > c\ \&\ v_{dxr} < c\right)$$

$$(group\ velocity)\cdot(phase\ velocity(v_p)) = c^2$$

$$(v_{dx})\left(\frac{1}{\pi}\frac{v_{\varepsilon x}^2}{v_{dx}^2}\cdot c\right) \cong \left(\frac{1}{2}v_{dx}\right)\left(\frac{chain}{cavity}\frac{v_{\varepsilon x}^2}{v_{dx}^2}\cdot c\right) \qquad (2.8.6)$$

$$\cong (v_{dx})\left(\frac{c}{v_{dx}}\cdot c\right) \cong c^2$$

In the Nature reference [51] and in [59], Homes reports experimental data conforming to the empirical relationship Homes' Law relating superconducting number density $\rho_s$ at just above $T_c$, superconductor conductivity $\sigma_{dc}$ and $T_c$.

$$\rho_s \propto \sigma_{dc} T_c \qquad \text{Homes' Law} \quad \text{ref: 51} \qquad (2.8.7)$$

The trisine model is developed for $T_c$, but using the Tanner's law observation that $\mathrm{n_S} = \mathrm{n_N}/4$ (ref 52), we can assume that $\mathrm{E_{fN}} = \mathrm{E_{fS}}/4$. This is verified by equation 2.5.1. Now substituting $\mathrm{I_e}/\left(\mathrm{E_{fN}}\,area\right)$ for superconductor conductivity $\sigma_{dc}$, and using $\mathbb{C}/\text{cavity}$ for super current number density $\left(\rho_s\right)$, the following dimensional relationship is obtained and is given the name Homes' constant $\left(\mho\right)$ while recognizing mass density $\left(\rho_U\right) = m_t\rho_s$.

(Note: $\sigma_{dc}(cgs) = 8.99\text{E}11\ \sigma_{dc}(\text{Siemen/cm})$

which is dimensionally equivalent to:

$$\sigma_{dc}\left(e^2/\left(erg\ sec\ cm\right)\right) = 8.99\text{E}11\ \sigma_{dc}\left(coulumb^2/\left(joule\ sec\ cm\right)\right)$$

Then:

$$\sigma_{dc} = \frac{\mathrm{I_e}}{\mathrm{E_{fN}}\,area} = \frac{\dfrac{\mathbb{C}ev_{dx}}{chain}}{\dfrac{1}{2}m_t v_{dx}^2\dfrac{1}{\mathbb{C}eA}} = \frac{\sqrt{3}\mathbb{C}^2 e^2}{hB} \qquad (2.8.8)$$

$$\sigma_{dcr} = \frac{\mathrm{I}_{er}}{\mathrm{E}_{fNr} area_r} = \frac{\dfrac{\mathbb{C}ev_{dxr}}{chain_r}}{\dfrac{1}{2} m_t v_{dxr}^2 \dfrac{1}{\mathbb{C}eA_r}} = \frac{\sqrt{3}\mathbb{C}^2 e^2}{hB_r} \qquad (2.8.8r)$$

All of the dimensional resources are now available to define a numerical constant that called the Homes' constant $(\mho)$

$$\mho = \frac{\sigma_{dc} k_b T_c}{m_T \rho_s} = \frac{\dfrac{\mathbb{C}ev_{dx}}{chain}}{\dfrac{1}{2} m_T v_{dx}^2 \dfrac{1}{\mathbb{C}eA}} \frac{1}{2} m_T v_{dx}^2 \frac{cavity}{m_t}$$

$$= \frac{3\pi}{2} \frac{\mathbb{C}^2 e_\pm^2 \hbar}{m_T^2} \frac{A}{B} = 191{,}606 \frac{cm^5}{\sec^3} \qquad (2.8.9)$$

*or*

$$\rho_U = \rho_s m_T = \frac{\sigma_{dc} k_b T_c}{\mho} = \frac{m_t}{2} \frac{P}{k_b T_c} \quad \text{(cgs units)}$$

$$\mho_r = \frac{\sigma_{dcr} k_b T_{cr}}{m_T \rho_{sr}} = \frac{\dfrac{\mathbb{C}ev_{dxr}}{chain_r}}{\dfrac{1}{2} m_T v_{dxr}^2 \dfrac{1}{\mathbb{C}eA}} \frac{1}{2} m_T v_{dxr}^2 \frac{cavity_r}{m_t}$$

$$= \frac{3\pi}{2} \frac{\mathbb{C}^2 e_\pm^2 \hbar}{m_t^2} \frac{A_r}{B_r} = 191{,}606 \frac{cm^5}{\sec^3} \qquad (2.8.9r)$$

*or*

$$\rho_{Ur} = \rho_{sr} m_T = \frac{\sigma_{dcr} k_b T_{cr}}{\mho_r} = \frac{m_T}{2} \frac{P_r}{k_b T_{cr}} \quad \text{(cgs units)}$$

Equation 2.8.8 is consistent with equation 2.8.3.

The experimentally measured number density $(\rho_{sm})$ is calculated from plasma frequency relationship as follows:

$$\rho_{sm} = \frac{\omega_{\rho_{sm}}^2 c^2 m_T}{4\pi e^2} = \frac{\omega_m^2 m_T}{4\pi e^2} \qquad (2.8.10)$$

Where the measured plasma frequency $(\omega_m)$ is related to the penetration depth $(\lambda_m)$

$$\omega_m = \frac{1}{2\pi\lambda_m} \qquad (2.8.11)$$

The experimentally measured mass density $(\rho_{sm} m_e)$ compared to the trisine model $\rho_U = (m_t / cavity)$ is an indicator of sample purity or fractional material superconductor phase.

It is interesting to note that this mass density ratio is mostly less than one but on occasion is greater than one indicating superconductor lattice phase overlay as allowed by Bose-Einstein statistics.

$$\text{superconductor mass density ratio} = \frac{\rho_{sm} m_T}{\rho_U} \qquad (2.8.12)$$

This optical resistivity is further confirmed to be a representation of the BCS theory[138].

A modification of 2.8.9 and 2.8.12 by factor $m_e/m_T$ is consistent with equation 2.5.15c.

$$\frac{m_T}{m_e} = \frac{g_d^2}{g_s^2} \frac{2}{3} \frac{\varepsilon_x + \varepsilon_y + \varepsilon_z}{\varepsilon} \qquad (2.5.12)$$

It is important to note that the superconductivity $\sigma_{dc}$ in equation 2.8.9 is a function of redshift $(1 + z_B)$ as follows:

$$\rho_U (1 + z_B)^3 = \frac{m_T}{cavity (1 + z_B)^{-3}} = \frac{m_T}{A (1 + z_B)^{-1} section}$$

$$= \frac{\sigma_{dc} (1 + z_B)^1 k_b T_c (1 + z_B)^2}{\mho}$$

but in terms of area density becomes:

$$\frac{m_T}{section (1 + z_B)^{-2}} = \frac{\sigma_{dc} (1 + z_B)^1 A (1 + z_B)^{-1} k_b T_c (1 + z_B)^2}{\mho}$$

or

$$\frac{m_t}{section} = \frac{\sigma_{dc} A k_b T_c}{\mho}$$

or in other words area density is directly proportional to Tc and not volume density as a restatement of Homes' Law.

In Table 2.8.2, the trisine parameter $(time_\pm)$ is consistent with equation 2.1.27.

The measured dielectric $(\varepsilon_{dcm})$ is represented by:

$$\varepsilon_{dcm} = avg(a,b)^{-1/2} \frac{\sigma_{dc}(cgs)}{\omega_m}$$

The trisine dielectric $(\varepsilon)$ parameter is consistent with equations 2.5.4 and 2.6.5.

the number density $(\rho_{sm})$ is represented by:

$$\rho_{sm} = \frac{\omega_m m_T}{4\pi e^2}$$

The trisine mass density $(\rho_U)$ parameter is consistent with equations 2.9.1, 2.11.5, 2.11.16, 2.11.41, 2.11.42, 2.11.43, 2.11.44, 2.11.45 and 2.11.46.

The measured fractional superconductor components in the larger material matrix are represented by

$$a = \frac{\sigma_{dcm} k_b T_c}{\mho \rho_{sm} m_T} \frac{m_e}{m_T} \quad \text{and} \quad b = \frac{\rho_{sm} m_T}{\rho_U} \frac{m_e}{m_T}$$

**Table 2.8.2** Superconductor optical resistivity scaling

| ref | Superconducting Molecule | $Siemen/cm$ | $\sigma_{dcm}(cgs)$ | $\omega_{\rho_{sm}}(cm^{-1})$ | $\omega_m$ or $c\omega_{\rho_{sm}}(sec^{-1})$ | $1/time_{\pm}(sec^{-1})$ | $\varepsilon_{dcm}$ | $\varepsilon$ | $\rho_{sm}(\#/cm^3)$ | $\rho_U(g/cm^3)$ | $T_c(^oK)$ | a $(\rho_{sm} m_T/\rho_U)/(m_e/m_T)$ | b $(\rho_{sm} m_T/\rho_U)/(m_e/m_T)$ | $average$(a,b) |
|---|---|---|---|---|---|---|---|---|---|---|---|---|---|---|
| [51] | $YBa_2Cu_3O_{6.60}$ | 6,500 | 5.42E+15 | 6,400 | 1.92E+14 | 2.46E+12 | 261 | 228 | 1.27E+21 | 1.25E-04 | 59.00 | 9.27E-03 | 1.80E-02 | 1.37E-02 |
| [51] | $YBa_2Cu_3O_{6.95}$ | 10,500 | 2.94E+15 | 9,950 | 2.98E+14 | 3.88E+12 | 241 | 181 | 3.08E+21 | 2.49E-04 | 93.20 | 1.13E-02 | 2.32E-02 | 1.72E-02 |
| [51] | $Pr\text{-}YBa_2Cu_3O_{7\text{-}\delta}$ | 2,500 | 3.70E+15 | 3,700 | 1.11E+14 | 1.67E+12 | 243 | 277 | 4.26E+20 | 6.99E-05 | 40.00 | 5.55E-03 | 8.42E-03 | 6.99E-03 |
| [51] | $Pr\text{-}YBa_2Cu_3O_{7\text{-}\delta}$ | 4,900 | 2.42E+15 | 7,350 | 2.20E+14 | 3.13E+12 | 197 | 202 | 1.68E+21 | 1.79E-04 | 75.00 | 8.53E-03 | 1.21E-02 | 1.03E-02 |
| [51] | $YBa_2Cu_3O_{7\text{-}\delta}$ | 8,700 | 3.32E+15 | 9,200 | 2.76E+14 | 3.83E+12 | 235 | 183 | 2.63E+21 | 2.44E-04 | 92.00 | 9.83E-03 | 1.93E-02 | 1.46E-02 |
| [51] | $YBa_2Cu_4O_8$ | 6,000 | 3.68E+15 | 8,000 | 2.40E+14 | 3.33E+12 | 208 | 196 | 1.99E+21 | 1.98E-04 | 80.00 | 9.17E-03 | 1.43E-02 | 1.17E-02 |
| [51] | $Bi_2Ca_2SrCu_2O_{8+\delta}$ | 8,000 | 3.43E+15 | 8,800 | 2.64E+14 | 3.54E+12 | 228 | 190 | 2.41E+21 | 2.17E-04 | 85.00 | 1.01E-02 | 1.85E-02 | 1.43E-02 |
| [51] | $Bi_2Ca_2SrCu_2O_{8+\delta}$ | 9,800 | 3.53E+15 | 9,600 | 2.88E+14 | 3.79E+12 | 239 | 184 | 2.87E+21 | 2.40E-04 | 91.00 | 1.09E-02 | 2.19E-02 | 1.64E-02 |
| [51] | $Y/Pb - Bi_2Ca_2SrCu_2O_{8+\delta}$ | 2,500 | 3.66E+15 | 3,200 | 9.59E+13 | 1.46E+12 | 279 | 296 | 3.18E+20 | 5.72E-05 | 35.00 | 5.07E-03 | 9.01E-03 | 7.04E-03 |
| [51] | $Y - Bi_2Ca_2SrCu_2O_{8+\delta}$ | 2,600 | 2.27E+15 | 3,800 | 1.14E+14 | 1.67E+12 | 240 | 277 | 4.49E+20 | 6.99E-05 | 40.00 | 5.85E-03 | 8.76E-03 | 7.31E-03 |
| [51] | $Tl_2Ba_2SrCuO_{6+\delta}$ | 5,000 | 2.42E+15 | 6,630 | 1.99E+14 | 3.67E+12 | 247 | 187 | 1.37E+21 | 2.28E-04 | 88.00 | 5.46E-03 | 1.14E-02 | 8.41E-03 |
| [51] | $Nd_{1.85}Ce_{0.15}CuO_4$ | 28,000 | 3.59E+15 | 10,300 | 3.09E+14 | 9.58E+11 | 244 | 365 | 3.30E+21 | 3.05E-05 | 23.00 | 9.86E-02 | 1.24E-01 | 1.12E-01 |
| [51] | $La_{1.87}Sr_{0.13}CuO_4$ | 7,000 | 1.84E+15 | 5,600 | 1.68E+14 | 1.38E+12 | 256 | 305 | 9.75E+20 | 5.24E-05 | 33.00 | 1.70E-02 | 2.60E-02 | 2.15E-02 |
| [51] | $La_{1.86}Sr_{0.14}CuO_4$ | 9,000 | 2.20E+15 | 6,000 | 1.80E+14 | 1.54E+12 | 291 | 288 | 1.12E+21 | 6.22E-05 | 37.00 | 1.64E-02 | 3.15E-02 | 2.40E-02 |
| [51] | $Nb$ | 250,000 | 2.33E+15 | 17,600 | 5.28E+14 | 3.46E+11 | 338 | 608 | 9.63E+21 | 6.61E-06 | 8.30 | 1.33E+00 | 1.85E+00 | 1.59E+00 |
| [51] | $Nb$ | 850,000 | 1.10E+15 | 35,800 | 1.07E+15 | 3.88E+11 | 310 | 575 | 3.99E+22 | 7.84E-06 | 9.30 | 4.63E+00 | 5.94E+00 | 5.29E+00 |
| [51] | $Pb$ | 1,400,000 | 1.17E+15 | 41,000 | 1.23E+15 | 3.00E+11 | 324 | 653 | 5.23E+22 | 5.34E-06 | 7.20 | 8.92E+00 | 1.11E+01 | 1.00E+01 |
| [51] | $YBa_2Cu_3O_{6.50}$ | 9 | 1.03E+15 | 204 | 6.12E+12 | 2.21E+12 | 306 | 241 | 1.29E+18 | 1.07E-04 | 53.00 | 1.11E-05 | 2.63E-05 | 1.87E-05 |
| [51] | $YBa_2Cu_3O_{6.60}$ | 12 | 2.79E+15 | 244 | 7.31E+12 | 2.42E+12 | 303 | 230 | 1.85E+18 | 1.22E-04 | 58.00 | 1.38E-05 | 3.36E-05 | 2.37E-05 |
| [51] | $YBa_2Cu_3O_{6.70}$ | 14 | 2.92E+15 | 308 | 9.23E+12 | 2.63E+12 | 255 | 221 | 2.95E+18 | 1.38E-04 | 63.00 | 1.95E-05 | 3.76E-05 | 2.85E-05 |
| [51] | $YBa_2Cu_3O_{6.80}$ | 27 | 3.04E+15 | 465 | 1.39E+13 | 3.25E+12 | 250 | 198 | 6.72E+18 | 1.90E-04 | 78.00 | 3.22E-05 | 6.52E-05 | 4.87E-05 |
| [51] | $YBa_2Cu_3O_{6.85}$ | 47 | 3.38E+15 | 790 | 2.37E+13 | 3.71E+12 | 187 | 186 | 1.94E+19 | 2.32E-04 | 89.00 | 7.62E-05 | 1.06E-04 | 9.12E-05 |
| [51] | $YBa_2Cu_3O_{6.90}$ | 88 | 3.62E+15 | 1,003 | 3.01E+13 | 3.81E+12 | 210 | 183 | 3.13E+19 | 2.42E-04 | 91.50 | 1.18E-04 | 1.96E-04 | 1.57E-04 |
| [51] | $YBa_2Cu_3O_{6.95}$ | 220 | 3.67E+15 | 1,580 | 4.74E+13 | 3.88E+12 | 213 | 181 | 7.76E+19 | 2.49E-04 | 93.20 | 2.84E-04 | 4.86E-04 | 3.85E-04 |
| [51] | $YBa_2Cu_3O_{6.99}$ | 450 | 3.70E+15 | 2,070 | 6.21E+13 | 3.75E+12 | 236 | 185 | 1.33E+20 | 2.36E-04 | 90.00 | 5.15E-04 | 1.01E-03 | 7.63E-04 |
| [51] | $YBa_2Cu_4O_8$ | 320 | 3.64E+15 | 1,670 | 5.01E+13 | 3.33E+12 | 238 | 196 | 8.67E+19 | 1.98E-04 | 80.00 | 4.00E-04 | 7.62E-04 | 5.81E-04 |
| [51] | $Tl_2Ba_2CuO_{6+\delta}$ | 3 | 3.43E+15 | 130 | 3.90E+12 | 3.38E+12 | 318 | 195 | 5.26E+17 | 2.01E-04 | 81.00 | 2.38E-06 | 7.10E-06 | 4.74E-06 |
| [51] | $Hg_2Ba_2CuO_{4+\delta}$ | 6 | 3.45E+15 | 290 | 8.69E+12 | 4.04E+12 | 187 | 178 | 2.62E+18 | 2.64E-04 | 97.00 | 9.03E-06 | 1.30E-05 | 1.10E-05 |
| [51] | $Nd_{1.85}Ce_{0.15}CuO_4$ | 2 | 3.77E+15 | 60 | 1.80E+12 | 9.58E+11 | 335 | 365 | 1.12E+17 | 3.05E-05 | 23.00 | 3.35E-06 | 6.67E-06 | 5.01E-06 |
| [51] | $La_{1.92}Sr_{0.08}CuO_4$ | 1 | 1.84E+15 | 63 | 1.89E+12 | 1.17E+12 | 259 | 331 | 1.23E+17 | 4.09E-05 | 28.00 | 2.75E-06 | 4.03E-06 | 3.39E-06 |
| [51] | $La_{1.90}Sr_{0.10}CuO_4$ | 2 | 2.03E+15 | 108 | 3.24E+12 | 1.33E+12 | 193 | 310 | 3.63E+17 | 5.00E-05 | 32.00 | 6.61E-06 | 6.78E-06 | 6.69E-06 |
| [51] | $La_{1.88}Sr_{0.12}CuO_4$ | 4 | 2.17E+15 | 160 | 4.80E+12 | 1.33E+12 | 188 | 310 | 7.96E+17 | 5.00E-05 | 32.00 | 1.45E-05 | 1.43E-05 | 1.44E-05 |
| [51] | $La_{1.875}Sr_{0.125}CuO_4$ | 5 | 2.17E+15 | 153 | 4.59E+12 | 1.33E+12 | 245 | 310 | 7.28E+17 | 5.00E-05 | 32.00 | 1.33E-05 | 1.88E-05 | 1.60E-05 |
| [51] | $La_{1.875}Sr_{0.125}CuO_4$ | 7 | 2.17E+15 | 159 | 4.77E+12 | 1.33E+12 | 278 | 310 | 7.86E+17 | 5.00E-05 | 32.00 | 1.43E-05 | 2.45E-05 | 1.94E-05 |
| [51] | $La_{1.875}Sr_{0.15}CuO_4$ | 16 | 2.17E+15 | 344 | 1.03E+13 | 1.58E+12 | 191 | 284 | 3.68E+18 | 6.47E-05 | 38.00 | 5.18E-05 | 5.53E-05 | 5.36E-05 |
| [51] | $La_{1.83}Sr_{0.17}CuO_4$ | 32 | 2.36E+15 | 360 | 1.08E+13 | 1.50E+12 | 285 | 292 | 4.03E+18 | 5.97E-05 | 36.00 | 6.15E-05 | 1.14E-04 | 8.76E-05 |
| [51] | $La_{1.80}Sr_{0.20}CuO_4$ | 95 | 2.30E+15 | 515 | 1.54E+13 | 1.33E+12 | 347 | 310 | 8.25E+18 | 5.00E-05 | 32.00 | 1.50E-04 | 3.58E-04 | 2.54E-04 |
| [54] | $YBa_2Cu_3O_{6.95}$ | 4,400 | 2.17E+15 | 5,750 | 1.72E+14 | 2.92E+12 | 249 | 209 | 1.03E+21 | 1.62E-04 | 70.00 | 5.79E-03 | 1.12E-02 | 8.50E-03 |
| [54] | $YBa_2Cu_3O_{6.95}$ | 6,500 | 3.21E+15 | 8,840 | 2.65E+14 | 3.33E+12 | 191 | 196 | 2.43E+21 | 1.98E-04 | 80.00 | 1.12E-02 | 1.55E-02 | 1.33E-02 |
| [54] | $YBa_2Cu_3O_{6.95}$ | 9,200 | 3.43E+15 | 9,220 | 2.76E+14 | 3.54E+12 | 235 | 190 | 2.64E+21 | 2.17E-04 | 85.00 | 1.11E-02 | 2.13E-02 | 1.62E-02 |
| [54] | $YBa_2Cu_3O_{6.95}$ | 10,500 | 3.53E+15 | 11,050 | 3.31E+14 | 3.90E+12 | 210 | 181 | 3.80E+21 | 2.50E-04 | 93.50 | 1.38E-02 | 2.31E-02 | 1.85E-02 |
| [54] | $YBa_2Cu_3O_{6.60}$ | 6,500 | 3.71E+15 | 6,400 | 1.92E+14 | 2.46E+12 | 261 | 228 | 1.27E+21 | 1.25E-04 | 59.00 | 9.27E-03 | 1.80E-02 | 1.37E-02 |
| [54] | $YBa_2Cu_3O_{6.95}$ | 10,500 | 2.94E+15 | 9,950 | 2.98E+14 | 3.88E+12 | 241 | 181 | 3.08E+21 | 2.49E-04 | 93.20 | 1.13E-02 | 2.32E-02 | 1.72E-02 |
| [54] | $YBa_2Cu_3O_{6.95}$ | 8,700 | 3.70E+15 | 9,200 | 2.76E+14 | 3.83E+12 | 235 | 183 | 2.63E+21 | 2.44E-04 | 92.00 | 9.83E-03 | 1.93E-02 | 1.46E-02 |
| [54] | $YBa_2Cu_3O_{7-\delta}$ | 2,500 | 3.68E+15 | 3,700 | 1.11E+14 | 1.67E+12 | 243 | 277 | 4.26E+20 | 6.99E-05 | 40.00 | 5.55E-03 | 8.42E-03 | 6.99E-03 |
| [54] | $Pr-YBa_2Cu_3O_{7-\delta}$ | 4,900 | 2.42E+15 | 7,350 | 2.20E+14 | 3.13E+12 | 197 | 202 | 1.68E+21 | 1.79E-04 | 75.00 | 8.53E-03 | 1.21E-02 | 1.03E-02 |
| [54] | $Pr-YBa_2Cu_3O_{7-\delta}$ | 6,000 | 3.32E+15 | 8,000 | 2.40E+14 | 3.33E+12 | 208 | 196 | 1.99E+21 | 1.98E-04 | 80.00 | 9.17E-03 | 1.43E-02 | 1.17E-02 |
| [54] | $YBa_2Cu_3O_8$ | 11,500 | 3.43E+15 | 9,565 | 2.87E+14 | 3.75E+12 | 266 | 185 | 2.85E+21 | 2.36E-04 | 90.00 | 1.10E-02 | 2.58E-02 | 1.84E-02 |
| [54] | $Bi_2Ca_2SrCu_2O_{8+\delta}$ | 9,800 | 3.64E+15 | 9,600 | 2.88E+14 | 3.79E+12 | 239 | 184 | 2.87E+21 | 2.40E-04 | 91.00 | 1.09E-02 | 2.19E-02 | 1.64E-02 |
| [54] | $Bi_2Ca_2SrCu_2O_{8+\delta}$ | 8,500 | 3.66E+15 | 8,710 | 2.61E+14 | 3.54E+12 | 241 | 190 | 2.36E+21 | 2.17E-04 | 85.00 | 9.93E-03 | 1.96E-02 | 1.48E-02 |
| [54] | $Bi_2Ca_2SrCu_2O_{8+\delta}$ | 2,500 | 3.53E+15 | 3,200 | 9.59E+13 | 1.46E+12 | 279 | 296 | 3.18E+20 | 5.72E-05 | 35.00 | 5.07E-03 | 9.01E-03 | 7.04E-03 |
| [54] | $Y/Pb-Bi_2Ca_2SrCu_2O_{8+\delta}$ | 2,600 | 2.27E+15 | 3,800 | 1.14E+14 | 1.67E+12 | 240 | 277 | 4.49E+20 | 6.99E-05 | 40.00 | 5.85E-03 | 8.76E-03 | 7.31E-03 |
| [54] | $Y-Bi_2Ca_2SrCu_2O_{8+\delta}$ | 4,200 | 2.42E+15 | 5,140 | 1.54E+14 | 1.79E+12 | 227 | 267 | 8.22E+20 | 7.79E-05 | 43.00 | 9.61E-03 | 1.37E-02 | 1.16E-02 |
| [54] | $Tl_2Ba_2CuO_{6+\delta}$ | 5,000 | 2.51E+15 | 6,630 | 1.99E+14 | 3.67E+12 | 247 | 187 | 1.37E+21 | 2.28E-04 | 88.00 | 5.46E-03 | 1.14E-02 | 8.41E-03 |
| [54] | $Nd_{1.85}Ce_{0.15}CuO_4$ | 28,000 | 3.59E+15 | 10,300 | 3.09E+14 | 9.58E+11 | 244 | 365 | 3.30E+21 | 3.05E-05 | 23.00 | 9.86E-02 | 1.24E-01 | 1.12E-01 |
| [54] | $\mathrm{Pr}_{1.85}Ce_{0.15}CuO_4$ | 30,000 | 1.84E+15 | 10,300 | 3.09E+14 | 9.58E+11 | 257 | 365 | 3.30E+21 | 3.05E-05 | 23.00 | 9.86E-02 | 1.33E-01 | 1.16E-01 |
| [54] | $\mathrm{Pr}_{1.85}Ce_{0.15}CuO_4$ | 15,000 | 1.84E+15 | 4,820 | 1.44E+14 | 8.75E+11 | 429 | 382 | 7.22E+20 | 2.66E-05 | 21.00 | 2.48E-02 | 6.98E-02 | 4.73E-02 |
| [54] | $\mathrm{Pr}_{1.87}Ce_{0.15}CuO_4$ | 50,000 | 1.76E+15 | 7,960 | 2.39E+14 | 6.67E+11 | 439 | 438 | 1.97E+21 | 1.77E-05 | 16.00 | 1.02E-01 | 2.66E-01 | 1.84E-01 |
| [54] | $La_{1.87}Sr_{0.13}CuO_4$ | 7,000 | 1.53E+15 | 5,600 | 1.68E+14 | 1.33E+12 | 252 | 310 | 9.75E+20 | 5.00E-05 | 32.00 | 1.78E-02 | 2.64E-02 | 2.21E-02 |
| [54] | $La_{1.86}Sr_{0.14}CuO_4$ | 9,000 | 2.17E+15 | 6,000 | 1.80E+14 | 1.50E+12 | 287 | 292 | 1.12E+21 | 5.97E-05 | 36.00 | 1.71E-02 | 3.20E-02 | 2.45E-02 |
| [54] | $Ba_{0.62}K_{0.38}BiO_3$ | 3,800 | 2.30E+15 | 4,240 | 1.27E+14 | 1.29E+12 | 239 | 315 | 5.59E+20 | 4.77E-05 | 31.00 | 1.07E-02 | 1.45E-02 | 1.26E-02 |
| [54] | $Ba_{0.60}K_{0.40}BiO_3$ | 4,400 | 2.13E+15 | 4,240 | 1.27E+14 | 1.17E+12 | 254 | 331 | 5.59E+20 | 4.09E-05 | 28.00 | 1.24E-02 | 1.77E-02 | 1.51E-02 |
| [54] | $Ba_{0.54}K_{0.46}BiO_3$ | 6,000 | 2.03E+15 | 4,240 | 1.27E+14 | 8.75E+11 | 277 | 382 | 5.59E+20 | 2.66E-05 | 21.00 | 1.92E-02 | 2.79E-02 | 2.35E-02 |
| [54] | $Nb$ | 250,000 | 1.76E+15 | 17,600 | 5.28E+14 | 3.46E+11 | 338 | 608 | 9.63E+21 | 6.61E-06 | 8.30 | 1.33E+00 | 1.85E+00 | 1.59E+00 |
| [54] | $Nb$ | 850,000 | 1.10E+15 | 35,800 | 1.07E+15 | 3.88E+11 | 310 | 575 | 3.99E+22 | 7.84E-06 | 9.30 | 4.63E+00 | 5.94E+00 | 5.29E+00 |
| [54] | $Nb$ | 1,400,000 | 1.17E+15 | 41,000 | 1.23E+15 | 3.00E+11 | 324 | 653 | 5.23E+22 | 5.34E-06 | 7.20 | 8.92E+00 | 1.11E+01 | 1.00E+01 |
| [137] | organic | 3,800 | 1.03E+15 | 1,966 | 5.89E+13 | 3.58E+11 | 389 | 597 | 1.20E+20 | 6.97E-06 | 8.60 | 1.57E-02 | 2.76E-02 | 2.17E-02 |
| [137] | organic | 3,700 | 1.12E+15 | 1,098 | 3.29E+13 | 3.46E+11 | 765 | 608 | 3.75E+19 | 6.61E-06 | 8.30 | 5.17E-03 | 2.74E-02 | 1.63E-02 |
| [137] | organic | 4,000 | 1.10E+15 | 1,019 | 3.05E+13 | 4.71E+11 | 953 | 521 | 3.23E+19 | 1.05E-05 | 11.30 | 2.80E-03 | 2.54E-02 | 1.41E-02 |
| [137] | organic | 13,150 | 1.29E+15 | 4,749 | 1.42E+14 | 4.58E+11 | 294 | 528 | 7.01E+20 | 1.01E-05 | 11.00 | 6.34E-02 | 8.45E-02 | 7.39E-02 |
| [137] | organic | 25 | 1.27E+15 | 259 | 7.76E+12 | 3.33E+11 | 180 | 619 | 2.08E+18 | 6.25E-06 | 8.00 | 3.04E-04 | 1.88E-04 | 2.46E-04 |
| [137] | organic | 6 | 1.08E+15 | 40 | 1.21E+12 | 4.71E+11 | 967 | 521 | 5.07E+16 | 1.05E-05 | 11.30 | 4.40E-06 | 4.06E-05 | 2.25E-05 |
| [137] | organic | 4 | 1.29E+15 | 38 | 1.14E+12 | 3.48E+11 | 714 | 606 | 4.53E+16 | 6.67E-06 | 8.35 | 6.18E-06 | 2.95E-05 | 1.78E-05 |
| [137] | pnictides | 18 | 1.11E+15 | 215 | 6.43E+12 | 1.32E+12 | 339 | 311 | 1.43E+18 | 4.92E-05 | 31.66 | 2.65E-05 | 6.67E-05 | 4.66E-05 |
| [137] | pnictides | 5,744 | 2.16E+15 | 4,406 | 1.32E+14 | 1.06E+12 | 275 | 347 | 6.04E+20 | 3.56E-05 | 25.53 | 1.54E-02 | 2.42E-02 | 1.98E-02 |
| [137] | pnictides | 7,871 | 1.94E+15 | 5,024 | 1.51E+14 | 1.06E+12 | 277 | 347 | 7.85E+20 | 3.57E-05 | 25.56 | 2.00E-02 | 3.32E-02 | 2.66E-02 |
| [137] | pnictides | 4,443 | 1.94E+15 | 5,120 | 1.53E+14 | 8.20E+11 | 154 | 395 | 8.15E+20 | 2.41E-05 | 19.69 | 3.08E-02 | 2.13E-02 | 2.61E-02 |
| [137] | $K_3C_{60}$ | 1,200 | 1.70E+15 | 1,893 | 5.68E+13 | 7.92E+11 | 252 | 402 | 1.11E+20 | 2.29E-05 | 19.00 | 4.44E-03 | 5.87E-03 | 5.15E-03 |
| [137] | $Rb_3C_{60}$ | 1,300 | 1.67E+15 | 1,966 | 5.89E+13 | 1.21E+12 | 316 | 325 | 1.20E+20 | 4.31E-05 | 29.00 | 2.54E-03 | 5.14E-03 | 3.84E-03 |
| [137] | $CeCoIn_5$ | 251,064 | 2.06E+15 | 8,185 | 2.45E+14 | 9.39E+10 | 542 | 1167 | 2.08E+21 | 9.34E-07 | 2.25 | 2.03E+00 | 3.56E+00 | 2.80E+00 |
| [137] | elements | 244,565 | 5.75E+14 | 17,021 | 5.10E+14 | 3.51E+11 | 335 | 604 | 9.01E+21 | 6.76E-06 | 8.42 | 1.21E+00 | 1.80E+00 | 1.51E+00 |
| [137] | elements | 838,544 | 1.11E+15 | 34,734 | 1.04E+15 | 3.95E+11 | 306 | 569 | 3.75E+22 | 8.06E-06 | 9.48 | 4.24E+00 | 5.81E+00 | 5.02E+00 |
| [137] | elements | 1,403,835 | 1.18E+15 | 39,612 | 1.19E+15 | 2.97E+11 | 318 | 657 | 4.88E+22 | 5.25E-06 | 7.12 | 8.46E+00 | 1.12E+01 | 9.84E+00 |
| [137] | $MgB_2$ | 40,000 | 1.02E+15 | 7,178 | 2.15E+14 | 1.65E+12 | 579 | 278 | 1.60E+21 | 6.88E-05 | 39.60 | 2.12E-02 | 1.35E-01 | 7.83E-02 |
| [137] | $MgB_2$ | 137,000 | 2.41E+15 | 13,087 | 3.92E+14 | 1.63E+12 | 408 | 281 | 5.33E+21 | 6.73E-05 | 39.00 | 7.21E-02 | 4.68E-01 | 2.70E-01 |
| [137] | $MgB_2$ | 49,000 | 2.39E+15 | 18,696 | 5.60E+14 | 1.63E+12 | 275 | 281 | 1.09E+22 | 6.73E-05 | 39.00 | 1.47E-01 | 1.67E-01 | 1.57E-01 |
| [137] | $Tl_xPb_{1-x}Te$ | 1,111 | 2.39E+15 | 569 | 1.71E+13 | 5.88E+10 | 401 | 1475 | 1.01E+19 | 4.63E-07 | 1.41 | 1.98E-02 | 1.99E-02 | 1.99E-02 |
| [137] | $Ba_{1-x}K_xBiO_3$ | 2,891 | 4.55E+14 | 4,165 | 1.25E+14 | 1.34E+12 | 201 | 310 | 5.39E+20 | 5.01E-05 | 32.04 | 9.80E-03 | 1.09E-02 | 1.03E-02 |
| [137] | $Ba_{1-x}K_xBiO_3$ | 4,439 | 2.17E+15 | 4,165 | 1.25E+14 | 1.16E+12 | 256 | 332 | 5.39E+20 | 4.05E-05 | 27.81 | 1.21E-02 | 1.79E-02 | 1.50E-02 |
| [137] | $Ba_{1-x}K_xBiO_3$ | 6,818 | 2.02E+15 | 4,165 | 1.25E+14 | 8.96E+11 | 308 | 378 | 5.39E+20 | 2.76E-05 | 21.51 | 1.78E-02 | 3.13E-02 | 2.46E-02 |
| [137] | $Sr_2RuO_4$ | 4,760,000 | 1.78E+15 | 3,721 | 1.12E+14 | 6.13E+10 | 2209 | 1445 | 4.31E+20 | 4.92E-07 | 1.47 | 7.97E-01 | 8.37E+01 | 4.22E+01 |
| [137] | $Sr_2RuO_4$ | 2,060,000 | 4.65E+14 | 5,120 | 1.53E+14 | 5.24E+10 | 2551 | 1562 | 8.15E+20 | 3.90E-07 | 1.26 | 1.90E+00 | 3.91E+01 | 2.05E+01 |
| [137] | $Sr_2RuO_4$ | 1,380,000 | 4.30E+14 | 9,512 | 2.85E+14 | 3.08E+10 | 2160 | 2037 | 2.81E+21 | 1.76E-07 | 0.74 | 1.46E+01 | 3.42E+01 | 2.44E+01 |
| [137] | $Y_2C_2I_2$ | 400 | 3.30E+14 | 391 | 1.17E+13 | 4.22E+11 | 758 | 551 | 4.76E+18 | 8.89E-06 | 10.12 | 4.88E-04 | 2.68E-03 | 1.58E-03 |
| [137] | $TiN$ | 6,350 | 1.22E+15 | 2,159 | 6.47E+13 | 1.42E+11 | 319 | 950 | 1.45E+20 | 1.73E-06 | 3.40 | 7.62E-02 | 7.34E-02 | 7.48E-02 |
| [74] | $\kappa$-$BETS_2GaCl_4$ | 250,000 | 7.07E+14 | 684 | 2.05E+13 | 6.67E+09 | 4087 | 4380 | 1.45E+19 | 1.77E-08 | 0.16 | 7.49E-01 | 1.33E+01 | 7.03E+00 |
| [74] | $(TMTSF)_2ClO_4$ | 39,000 | 1.53E+14 | 1,238 | 3.71E+13 | 4.58E+10 | 1370 | 1671 | 4.77E+19 | 3.19E-07 | 1.10 | 1.36E-01 | 7.92E-01 | 4.64E-01 |
| [74] | a-$ET_2NH_4Hg(SCN)_4$ | 36,000 | 4.02E+14 | 1,429 | 4.29E+13 | 4.58E+10 | 1105 | 1671 | 6.35E+19 | 3.19E-07 | 1.10 | 1.82E-01 | 7.32E-01 | 4.57E-01 |
| [74] | $\beta$-$ET_2IBr_2$ | 26,000 | 4.02E+14 | 1,747 | 5.24E+13 | 9.17E+10 | 910 | 1181 | 9.49E+19 | 9.02E-07 | 2.20 | 9.59E-02 | 3.74E-01 | 2.35E-01 |
| [74] | $\lambda$-$BETS_2GaCl_4$ | 11,000 | 5.68E+14 | 2,184 | 6.55E+13 | 2.29E+11 | 569 | 747 | 1.48E+20 | 3.56E-06 | 5.50 | 3.79E-02 | 1.00E-01 | 6.89E-02 |
| [74] | $\kappa$-$ET_2Cu(NCS)_2$ | 6,000 | 8.99E+14 | 2,912 | 8.73E+13 | 3.83E+11 | 319 | 578 | 2.64E+20 | 7.71E-06 | 9.20 | 3.12E-02 | 4.22E-02 | 3.67E-02 |
| [74] | $K_3C_{60}$ | 2,900 | 1.16E+15 | 3,276 | 9.82E+13 | 7.88E+11 | 223 | 403 | 3.34E+20 | 2.27E-05 | 18.90 | 1.34E-02 | 1.42E-02 | 1.38E-02 |
| [74] | $Rb_3C_{60}$ | 2,500 | 1.67E+15 | 3,744 | 1.12E+14 | 1.22E+12 | 204 | 324 | 4.36E+20 | 4.38E-05 | 29.30 | 9.06E-03 | 9.84E-03 | 9.45E-03 |

The results for both c axis and a-b plane (ref 51) are graphically presented in the following figure 2.8.1. On average the Homes' constant calculated from data exceeds the Homes' constant derived from Trisine geometry by 1.72. There may be some degrees of freedom considerations $1\cdot k_b T_c/2$, $2\cdot k_b T_c/2$, $3\cdot k_b T_c/2$ that warrant looking into that may account for this divergence although conformity is noteworthy. Also 1.72 is close to sqrt(3), a value that is inherent to Trisine geometry. I will have to give this a closer look. In general, the conventional and high temperature superconductor data falls in the same pattern. I would conclude that Homes' Law can be extrapolated to any critical temperature $T_c$ as a universal phenomenon.

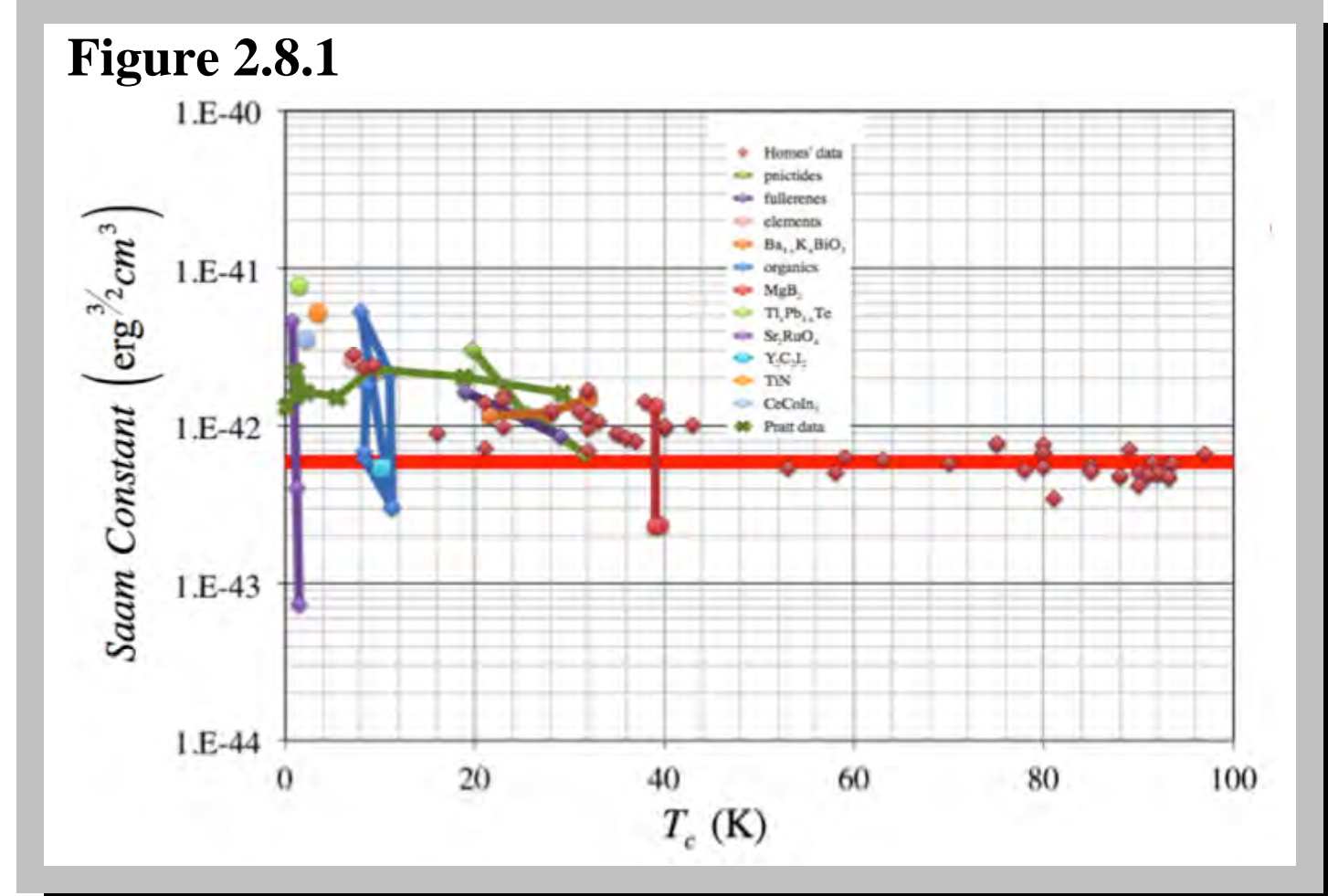

**Figure 2.8.1**

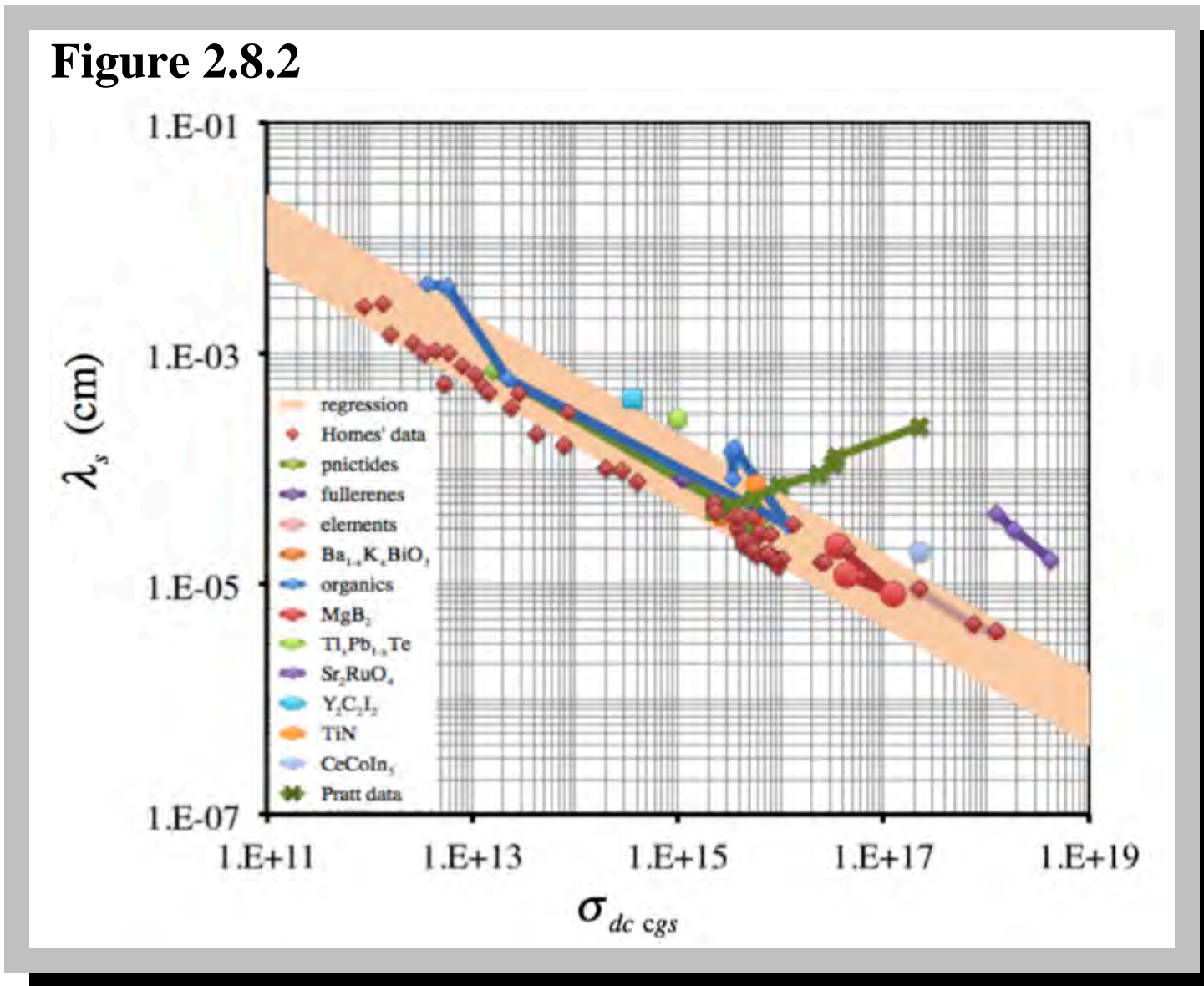

**Figure 2.8.2**

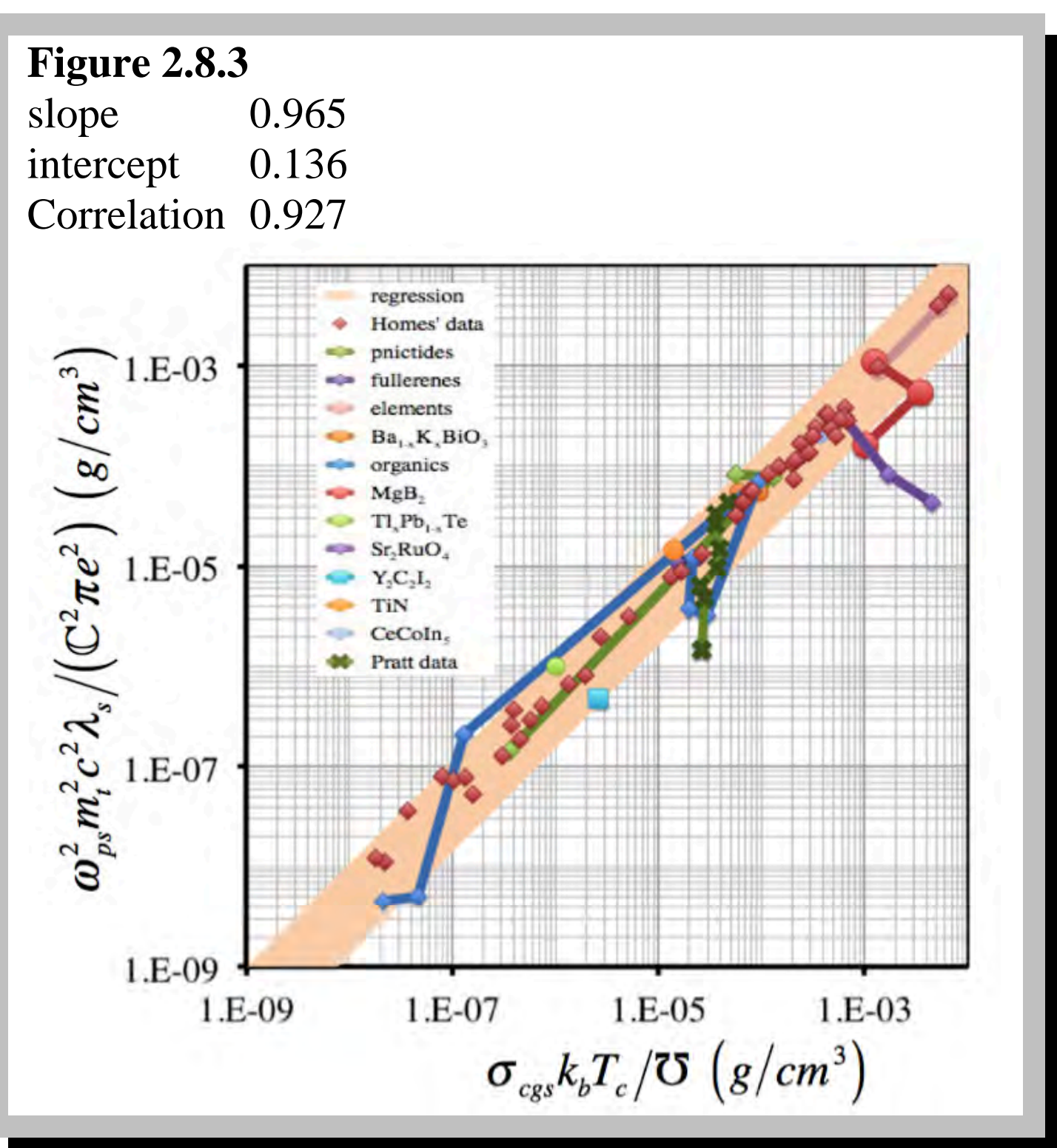

**Figure 2.8.3**

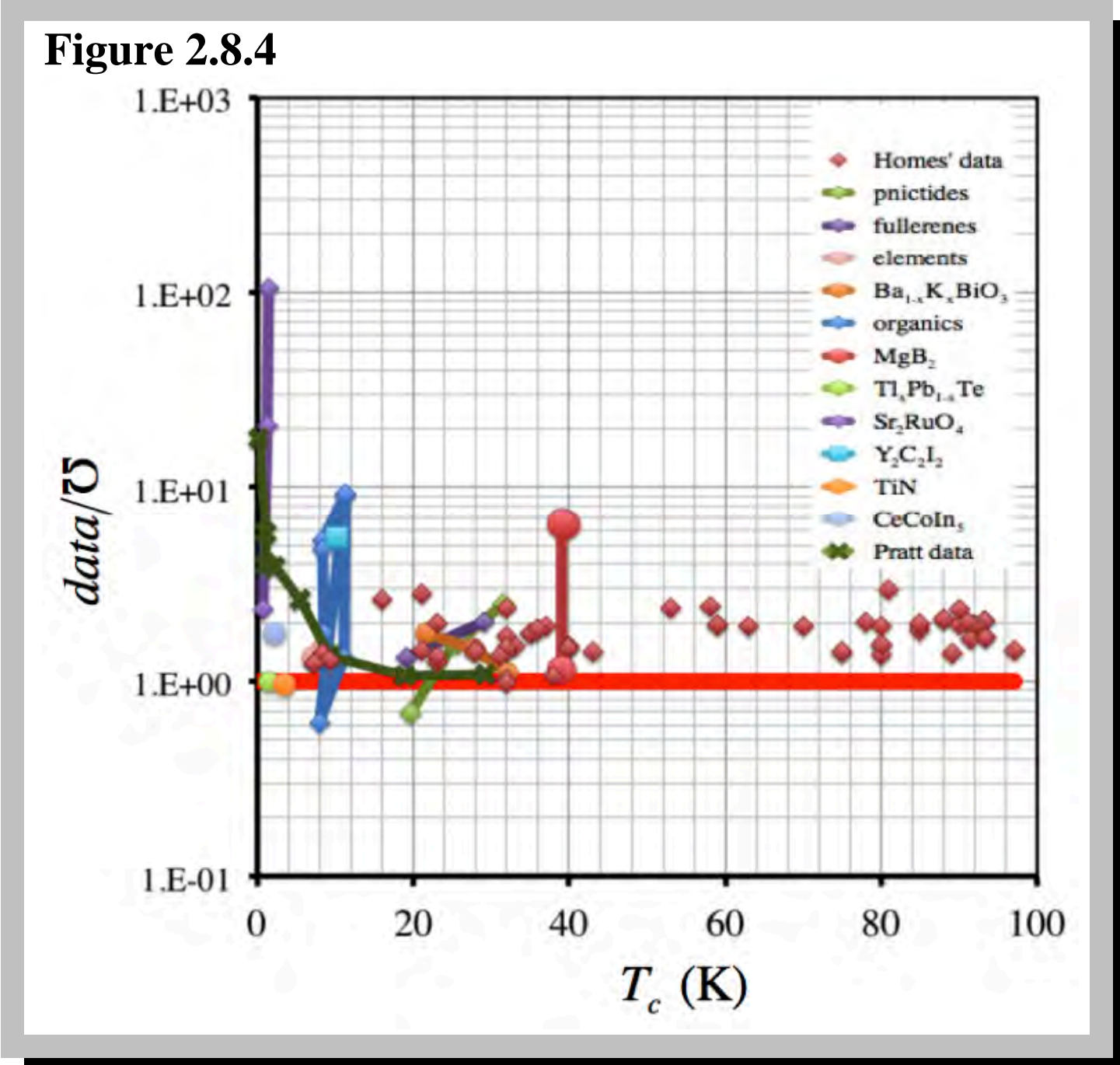

**Figure 2.8.4**

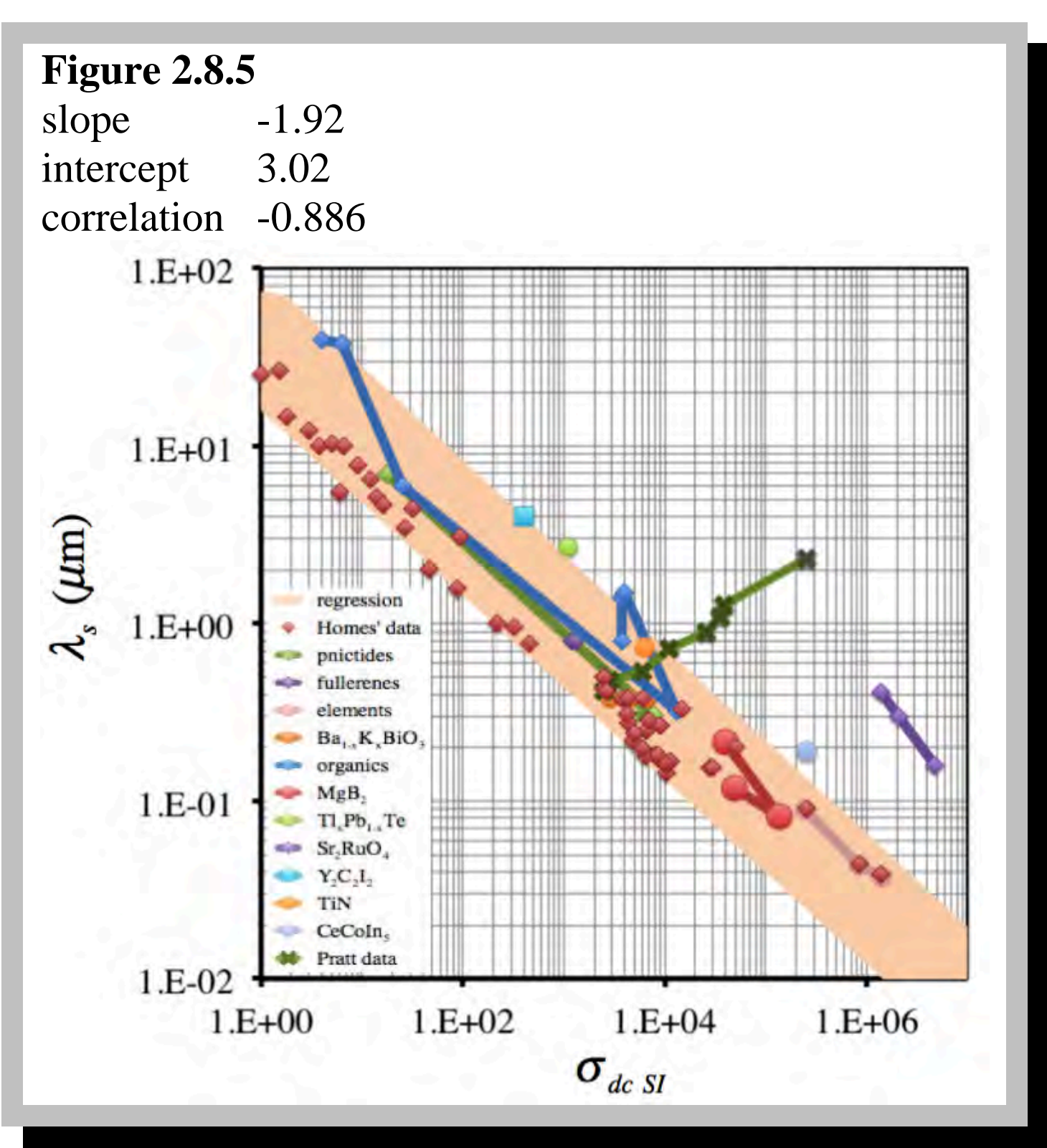

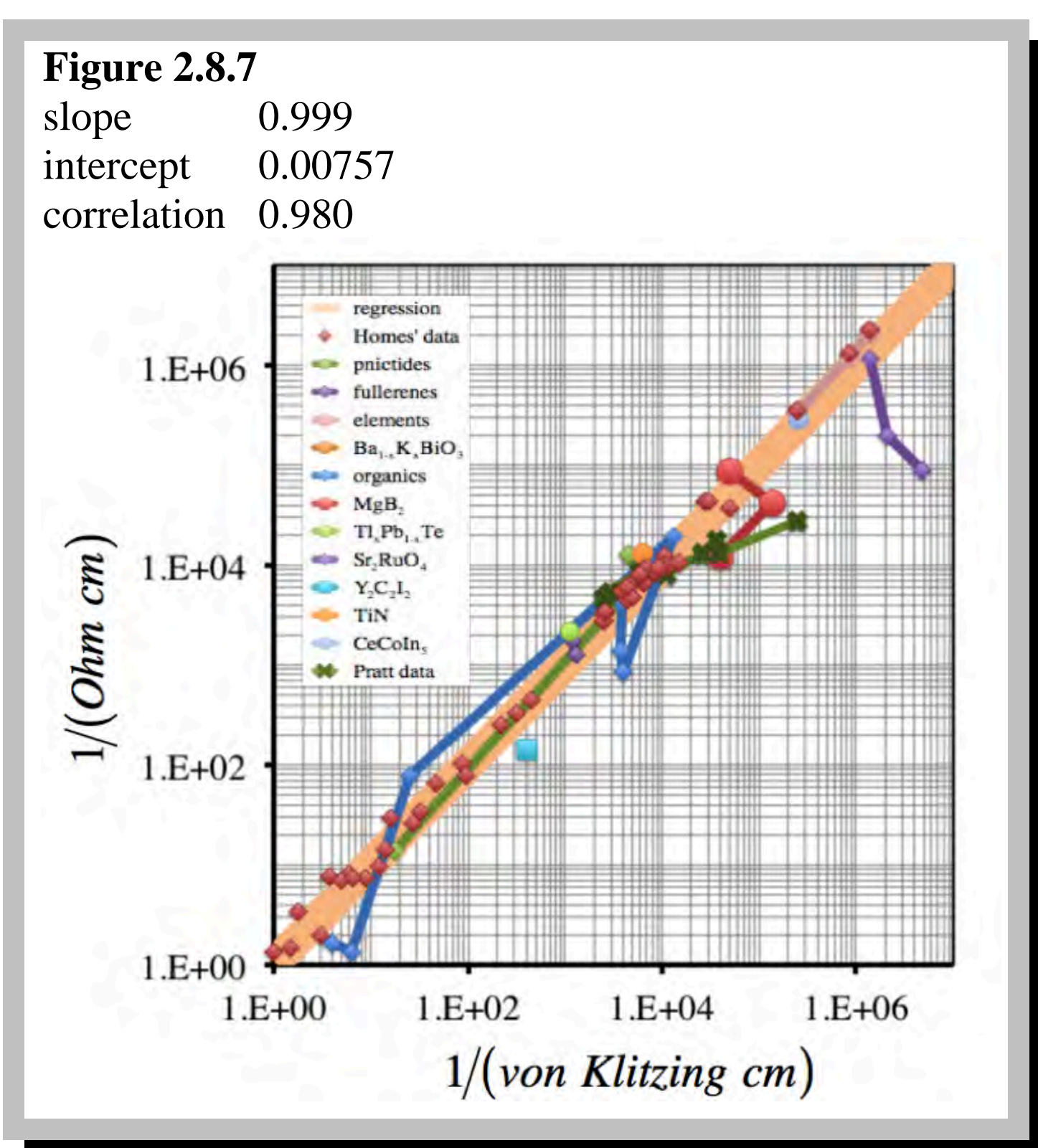

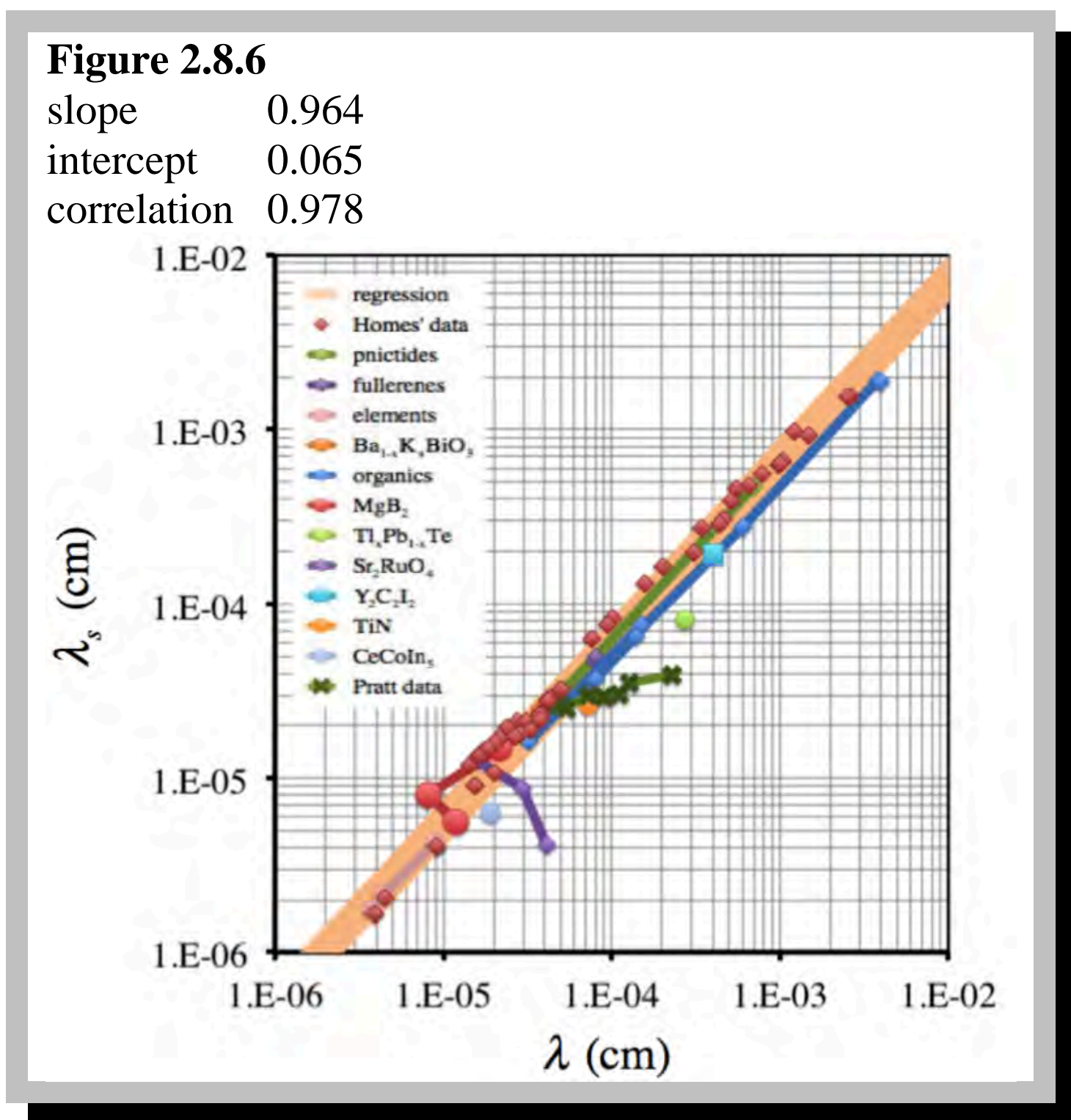

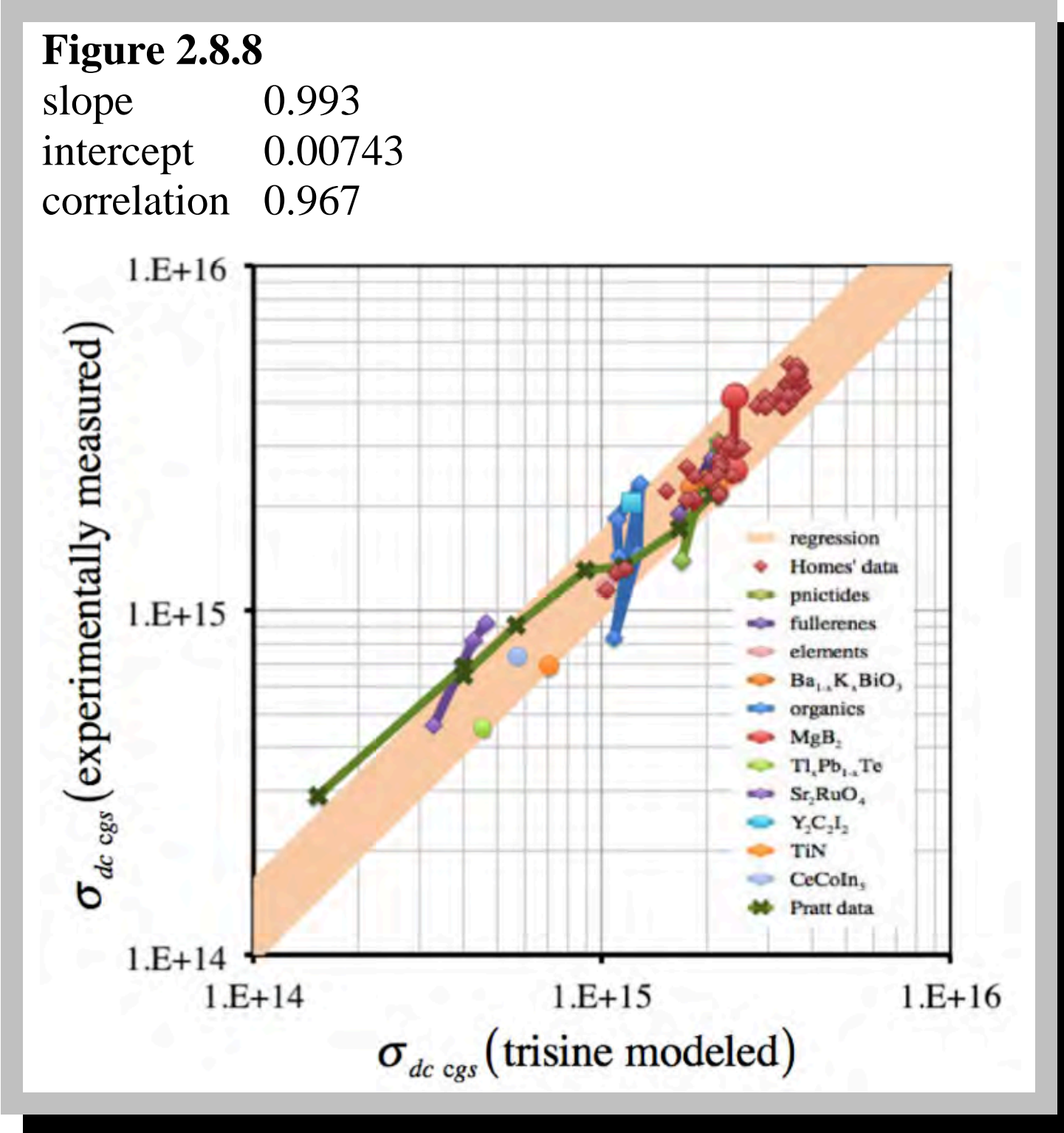

## 2.9 Superconductor Apparent Weight Reduction In A Gravitational Field

Based on the superconducting helical or tangential velocity $(v_{dT})$, the superconducting gravitational shielding effect $(\Theta)$ is computed based on Newton's gravitational law and the geometry as presented in Figure 2.9.1.

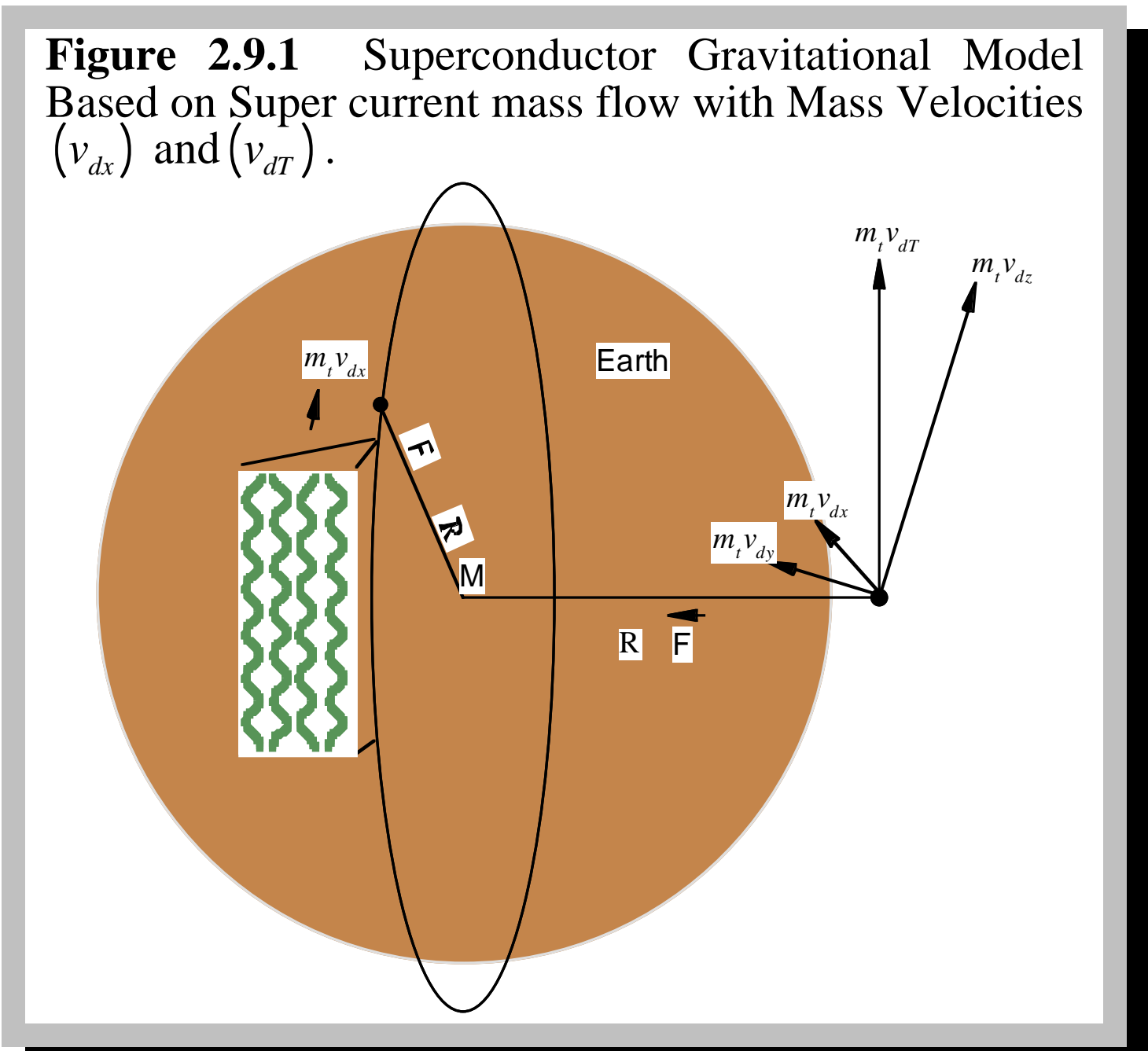

**Figure 2.9.1** Superconductor Gravitational Model Based on Super current mass flow with Mass Velocities $(v_{dx})$ and $(v_{dT})$.

With a material density $(\rho)$ of 6.39 g/cc [17] and the $YBa_2Cu_3O_{7-x}$ critical temperature of 93K, the gravitational apparent shielding effect $(\Theta_{dT})(\rho_c/\rho)$ is .05% as observed by Podkletnov and Nieminen (as per following equations or extrapolating on Table 2.9.1).

$$\rho_U = \frac{m_T}{cavity} = \frac{n\ m_T}{2} \tag{2.9.1}$$

$$F_{\uparrow} = m_T \frac{v_{dx}^2}{R} = \frac{2k_b T_c}{R} \qquad F_{\downarrow} = \frac{G m_T M}{R^2} \tag{2.9.2}$$

$$\Theta_{dx} = \frac{Superconducting\ Force\ \left(F_{\uparrow dx}\right)}{Earth's\ Gravitational\ Force\ \left(F_{\downarrow}\right)} = \frac{Earth\ Radius}{Earth\ Mass}\frac{v_{dx}^2}{G} \tag{2.9.3}$$

$$\Theta_{dT} = \frac{Superconducting\ Force\ \left(F_{\uparrow dT}\right)}{Earth's\ Gravitational\ Force\ \left(F_{\downarrow}\right)} = \frac{Earth\ Radius}{Earth\ Mass}\frac{v_{dT}^2}{G} \tag{2.9.4}$$

**Table 2.9.1** Cooper Pair Concentration and Shielding $(\Theta)$ with Plot of Cooper $(\mathbb{C})$ Pair density

| Condition | 1.2 | 1.3 | 1.6 | 1.9 | 1.10 | 1.11 |
|---|---|---|---|---|---|---|
| $T_a\left({}^0K\right)$ | 6.53E+11 | 6.53E+11 | 6.53E+11 | 6.53E+11 | 6.53E+11 | 6.53E+11 |
| $T_b\left({}^0K\right)$ | 9.05E+14 | 5.48E+13 | 9.22E+08 | 1.10E+07 | 2,868 | 2.723 |
| $T_{br}\left({}^0K\right)$ | 3.30E+12 | 5.45E+13 | 3.24E+18 | 2.72E+20 | 1.04E+24 | 1.10E+27 |
| $T_c\left({}^0K\right)$ | 8.96E+13 | 3.28E+11 | 93.0 | 0.0131 | 8.99E-10 | 8.10E-16 |
| $T_{cr}\left({}^0K\right)$ | 1.19E+09 | 3.25E+11 | 1.15E+21 | 8.11E+24 | 1.19E+32 | 1.32E+38 |
| $z_R+1$ | 3.67E+43 | 8.14E+39 | 3.89E+25 | 6.53E+19 | 1.17E+09 | 1.00E+00 |
| $z_B+1$ | 3.32E+14 | 2.01E+13 | 3.39E+08 | 4.03E+06 | 1.05E+03 | 1.00E+00 |
| $Age_{UG}(sec)$ | 7.14E-05 | 4.79E-03 | 6.94E+04 | 5.36E+07 | 1.27E+13 | 4.33E+17 |
| $Age_{UGv}(sec$ | 4.59E-10 | 1.25E-07 | 4.42E+02 | 3.13E+06 | 1.22E+13 | 4.33E+17 |
| $Age_U(sec)$ | 1.18E-26 | 5.31E-23 | 1.11E-08 | 6.63E-03 | 3.70E+08 | 4.33E+17 |
| $\rho_U\ (g/cm^3)$ | 2.34E+14 | 5.19E+10 | 2.48E-04 | 4.16E-10 | 7.44E-21 | 6.37E-30 |
| $\Theta_{dx}$ | 3.94E+11 | 1.44E+09 | 4.09E-01 | 5.77E-05 | 3.95E-12 | 3.56E-18 |
| $\Theta_{dT}$ | 9.83E+12 | 3.60E+10 | 1.02E+01 | 1.44E-03 | 9.87E-11 | 8.90E-17 |

Experiments have been conducted to see if a rotating body's weight is changed from that of a non-rotating body [21, 22, 23]. The results of these experiments were negative as they clearly should be because the center of rotation equaled the center mass. A variety of gyroscopes turned at various angular velocities were tested. Typically a gyroscope with effective radius 1.5 cm was turned at a maximum rotational frequency of 22,000 rpm. This equates to a tangential velocity of 3,502 cm/sec. At the surface of the earth, the tangential velocity is calculated to be:

$$\sqrt{gR_E} \text{ or } 791{,}310 \text{ cm/sec.}$$

The expected weight reduction if center of mass was different then center of rotation $(r)$ according to equations 2.9.2 and 2.9.3 would be:

$$\left(3{,}502/791{,}310\right)^2 \text{ or } 1.964E\text{-}5$$

of the gyroscope weight. The experiments were done with gyroscopes with weights of about 142 grams. This would indicate a required balance sensitivity of 1.964E-5 x 142 g or 2.8 mg, which approaches the sensitivity of laboratory balances. Both positive[21] and negative results [22, 23] are experimentally observed in this area although it has been generally accepted that the positive results were spurious.

Under the geometric scenario for superconducting materials presented herein, the Cooper CPT Charge conjugated pair centers of mass and centers of rotation $(r)$ are different making the numerical results in table 2.9.1 valid. Under this scenario, it is the superconducting material that looses weight due to transverse Cooper pair movement relative to a center of gravity. The experimentally observed superconducting shielding effect cannot be explained in these terms.

## 2.10 BCS Verifying Constants and Maxwell's Equations

Equations 2.9.1 - 2.9.8 represent constants as derived from BCS theory [2] and considered by Little [5] to be primary constraints on any model depicting the superconducting phenomenon. It can be seen that the trisine model constants compare very favorably with BCS constants in brackets { }.

$$C_D = 10.2 D\left(\in_T\right) k_b^2 T_c \frac{\mathbb{C}}{cavity} = 10.2 k_b \frac{\mathbb{C}}{cavity} \tag{2.10.1}$$

$$\gamma = \frac{2}{3}\pi^2 D\left(\varepsilon_T\right) k_b^2 \frac{\mathbb{C}}{cavity} = \frac{2}{3}\frac{\pi^2 k_b}{T_c}\frac{\mathbb{C}}{cavity} \tag{2.10.2}$$

$$\frac{1}{\mathbb{C}}\frac{\gamma T_c^2}{H_c^2} = .178\left\{\frac{\pi}{18} or .170\right\} \tag{2.10.3}$$

$$\frac{C_D}{\gamma T_c} = \frac{3 \cdot jump}{2\pi^2}\{1.52\} = 1.55\{1.52\} \tag{2.10.4}$$

$$\frac{\hbar^2\left(\frac{K_A^2}{\mathbb{C}} - K_{Dn}^2\right)}{2 m_T k_b T_c} = jump\{10.2\} = 10.12\{10.2\} \tag{2.10.5}$$

$$\left(k_b T_c\right)^{3/2} cavity = \frac{3\pi^3\hbar^3}{6^{1/2} m_T^{3/2}\left(\frac{B}{A}\right)_t} = 5.890351329\text{E-}43$$

$$\approx \frac{3\cos(30^0)}{2.6124}\left(\frac{2\pi}{m_T}\right)^{\frac{3}{2}}\hbar^3 = \frac{3\cos(30^0)}{\zeta\left(3/2\right)}\left(\frac{2\pi}{m_T}\right)^{\frac{3}{2}}\hbar^3 \tag{2.10.6}$$

$$\frac{2e^{Euler}\left(K_C^2 + K_{Ds}^2\right)}{\pi K_B^2 \sinh(trisine)} = \frac{\left(K_C^2 + K_{Ds}^2\right)}{K_B^2}\frac{2}{\mathrm{e}^{trisine} - 1} = \mathbb{C} \tag{2.10.6a}$$

Equation 2.10.6 basically assumes the ideal gas law in the adiabatic thermodynamic condition for superconducting charge carriers and Bose Einstein Condensate.

When the $1/\rho_s$ data from reference [51, 59] is used as a measure of cavity volume in the equation 2.10.6 , the following plot (figure 2.10.1) is achieved. It is a good fit for some of the data and not for others. One can only conclude that the number of charge carriers in the cavity volume varies considerably from one superconductor to another with perhaps only the Cooper CPT Charge conjugated pair being a small fraction of the total. Also, degrees of freedom may be a factor as in the kinetic theory of gas or other words – is a particular superconductor one-, two- or three- dimensional $1 \cdot kT_c/2$, $2 \cdot kT_c/2$, $3 \cdot kT_c/2$ as Table 2.6.1 data for $MgB_2$ would imply? Also it is important to remember that Homes' data is from just above $T_c$ and not at or below $T_c$.

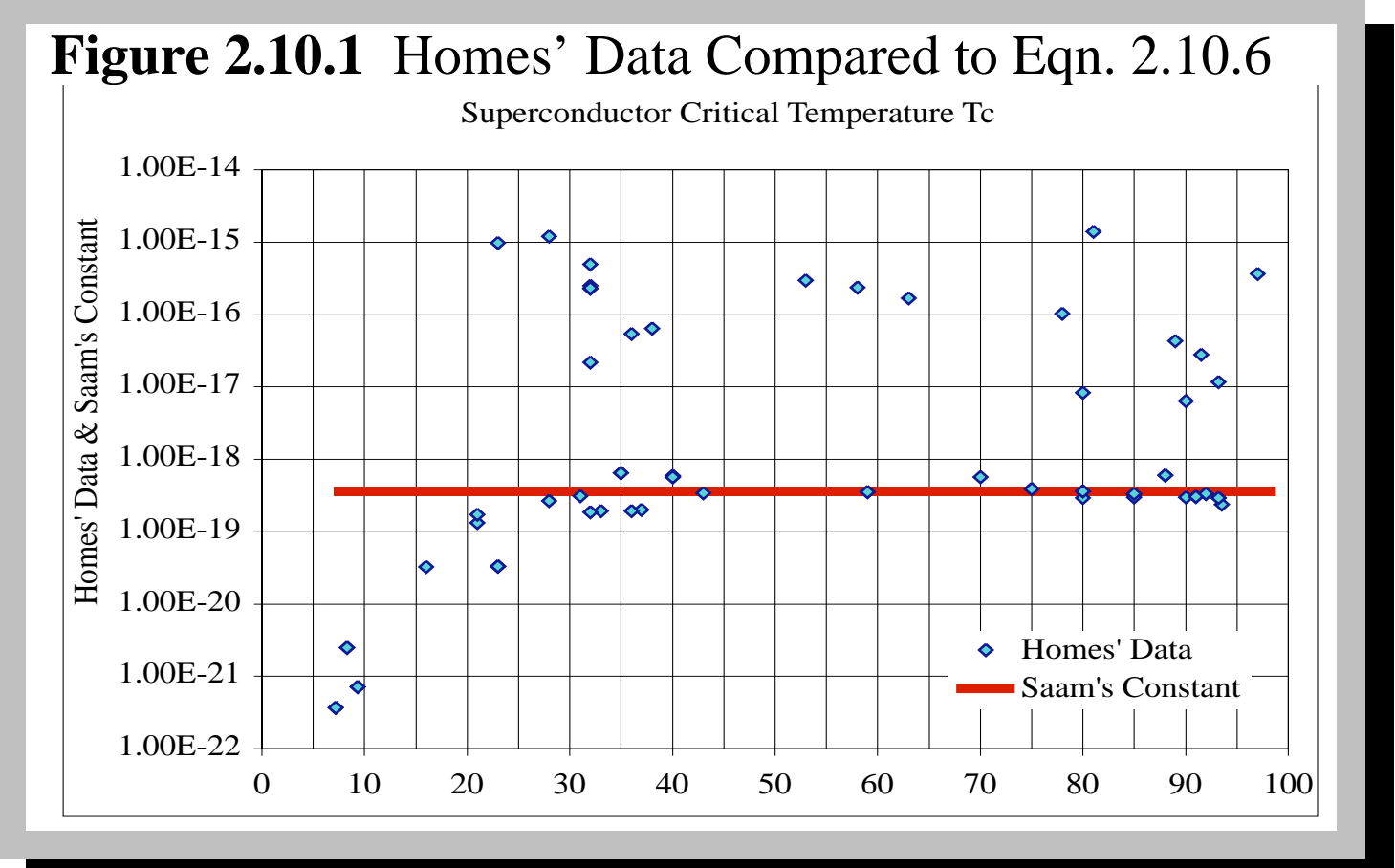

**Figure 2.10.1** Homes' Data Compared to Eqn. 2.10.6

$$\frac{\gamma T_c}{k_b}\frac{cavity}{\mathbb{C}} = \left\{\frac{2\pi^2}{3}\right\} \tag{2.10.7}$$

$$\frac{2\Delta_o^{BCS}}{k_b T_c} = 2\pi e^{-Euler}\{3.527754\} \tag{2.10.8}$$

Within the context of the trisine model, it is important to note that the Bohr radius is related to the nuclear radius $\left(B_{nucleus}\right)$ by this BCS gap as defined in 2.10.8

$$\frac{cavity}{\Delta x \Delta y \Delta z}\frac{m_T}{m_e}\frac{e^{Euler}}{2\pi} = \frac{Bohr\ radius}{B_{nucleus}} \tag{2.10.9}$$

Also the significant Euler constant is related to other universal constants

$$Euler \sim \frac{F_P}{M_U\ H_U c} = \frac{c^3}{M_U\ G_U H_U} = \frac{1}{3^{1/2}} \tag{2.10.9a}$$

**Table 2.10.1** Electronic Heat Jump $k_b\mathbb{C}/cavity$ at $T_c$

| Condition | 1.2 | 1.3 | 1.6 | 1.9 | 1.10 | 1.11 |
|---|---|---|---|---|---|---|
| $T_a\left({}^0K\right)$ | 6.53E+11 | 6.53E+11 | 6.53E+11 | 6.53E+11 | 6.53E+11 | 6.53E+11 |
| $T_b\left({}^0K\right)$ | 9.05E+14 | 5.48E+13 | 9.22E+08 | 1.10E+07 | 2,868 | 2.723 |
| $T_{br}\left({}^0K\right)$ | 3.30E+12 | 5.45E+13 | 3.24E+18 | 2.72E+20 | 1.04E+24 | 1.10E+27 |
| $T_c\left({}^0K\right)$ | 8.96E+13 | 3.28E+11 | 93.0 | 0.0131 | 8.99E-10 | 8.10E-16 |
| $T_{cr}\left({}^0K\right)$ | 1.19E+09 | 3.25E+11 | 1.15E+21 | 8.11E+24 | 1.19E+32 | 1.32E+38 |
| $z_R + 1$ | 3.67E+43 | 8.14E+39 | 3.89E+25 | 6.53E+19 | 1.17E+09 | 1.00E+00 |
| $z_B + 1$ | 3.32E+14 | 2.01E+13 | 3.39E+08 | 4.03E+06 | 1.05E+03 | 1.00E+00 |

| $Age_{UG}\left(sec\right)$ | 7.14E-05 | 4.79E-03 | 6.94E+04 | 5.36E+07 | 1.27E+13 | 4.33E+17 |
|---|---|---|---|---|---|---|
| $Age_{UGv}\left(sec\right.$ | 4.59E-10 | 1.25E-07 | 4.42E+02 | 3.13E+06 | 1.22E+13 | 4.33E+17 |
| $Age_{U}\left(sec\right)$ | 1.18E-26 | 5.31E-23 | 1.11E-08 | 6.63E-03 | 3.70E+08 | 4.33E+17 |
| $\frac{10.2k_b\mathbb{C}}{cavity}$ $erg/cm^3$ | 6.52E+24 | 1.45E+21 | 6.90E+06 | 1.16E+01 | 2.07E-10 | 1.78E-19 |

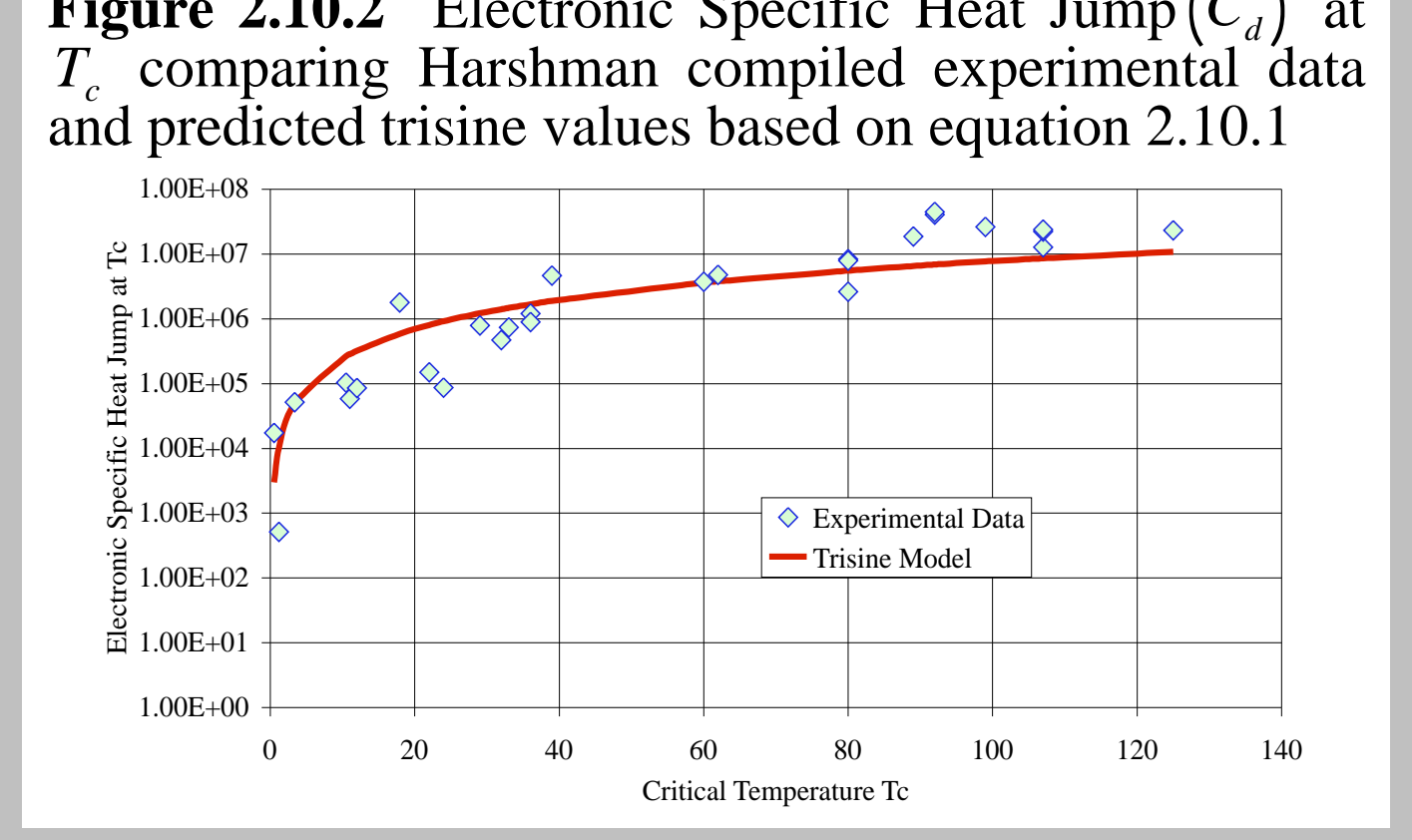

**Figure 2.10.2** Electronic Specific Heat Jump $\left(C_d\right)$ at $T_c$ comparing Harshman compiled experimental data and predicted trisine values based on equation 2.10.1

Harshman [17] has compiled and reported an extended listing of electronic specific heat jump at $T_c$ data for a number of superconducting materials. This data is reported in units of mJ/mole $K^2$. With the volume per formula weight data for the superconductors also reported by Harshman and after multiplying by $T_c^2$, the specific heat jump at $T_c$ is expressed in terms of erg/cm$^3$ and graphed in figure 2.10.2.

In addition it is important to note a universal resonant superconductor diffusion coefficient $D_c$ based on Einstein's 1905 publication [75] which adds to the universality of the trisine model.

$$D_c = \frac{chain}{cavity}\frac{\sqrt{3}k_bT_c}{\mu_U B} = \frac{(2B)^2}{time_\pm} = 2v_{dx}B = \frac{2\pi\hbar}{m_t}$$
$$= 2\pi c^{\frac{1}{3}}\hbar^{\frac{1}{3}}U^{-\frac{1}{3}}\frac{1}{\cos\left(\theta\right)} = 6.60526079\times 10^{-2} \tag{2.10.10}$$

The Diffusion Coefficient is integral to the Higgs mass as demonstrated by complementary analysis to Jay Wacker Quora Post Nov 29, 2011

The Higgs field, Hf , can take on different values, how much energy it costs to take on different values is characterized by the Higgs Potential Energy(V). The Higgs potential can be written as

$$V(H_f) = \lambda(|H_f|^2 - v^2/2)^2 \tag{2.10.11}$$

The minimum of the potential energy is not at $|H_f| = 0$ , but instead at $|H_f| = v/\sqrt{2}$ and where v is the vacuum expectation value. It uses the lowest energy configuration of the Higgs field.

The Higgs field takes on a uniform value over all space and time equal to the vacuum expectation value. To understand what things look like, we should expand our fields around the minimum by a Taylor expansion about the minimum:

$$H_f = \frac{v+B}{\sqrt{2}} \tag{2.10.12}$$

where B is the fluctuation about the miniumum representing the Trisine configuration as per the rest of this report. This resolves into:

$$V\left(B\right) = \lambda\left(\sqrt{2}vB + \frac{B^2}{2}\right)^2 \tag{2.10.13}$$

The quadratic, cubic and quartic Taylor expansion terms are $2\lambda v^2,\ \sqrt{2}\lambda v\ and\ \lambda/4$

The quadratic term is related to the Higgs mass $2\lambda v^2 = m_H^2/2$ which means $m_H = 2\lambda^{1/2}v$

Now assume the real and reciprocal Trisine lattice energy assumption:

$$\frac{1}{2}m_H\left(\frac{2B}{time}\right)^2 = \frac{\hbar^2}{2m_H}\left(\frac{\pi}{B}\right)^2 \tag{2.10.14}$$

A solution of equations 2.10.13 and 2.10.14 defines a Higgs mass $m_H$ within the Trisine

$$m_H = 2\pi\hbar\frac{time}{\left(2B\right)^2} = \frac{2\pi\hbar}{D_c} = \frac{m_t}{2} \tag{2.10.15}$$

**Figure 2.10.3** Higgs particles (including charged particles) are not at Higgs Field = zero but are offset(Equation 2.10.11) in a manner similar to Trisine cellular configuration as presented in Figure 2.5.1 after going through a charge conjugation transition state (lightly shaded).

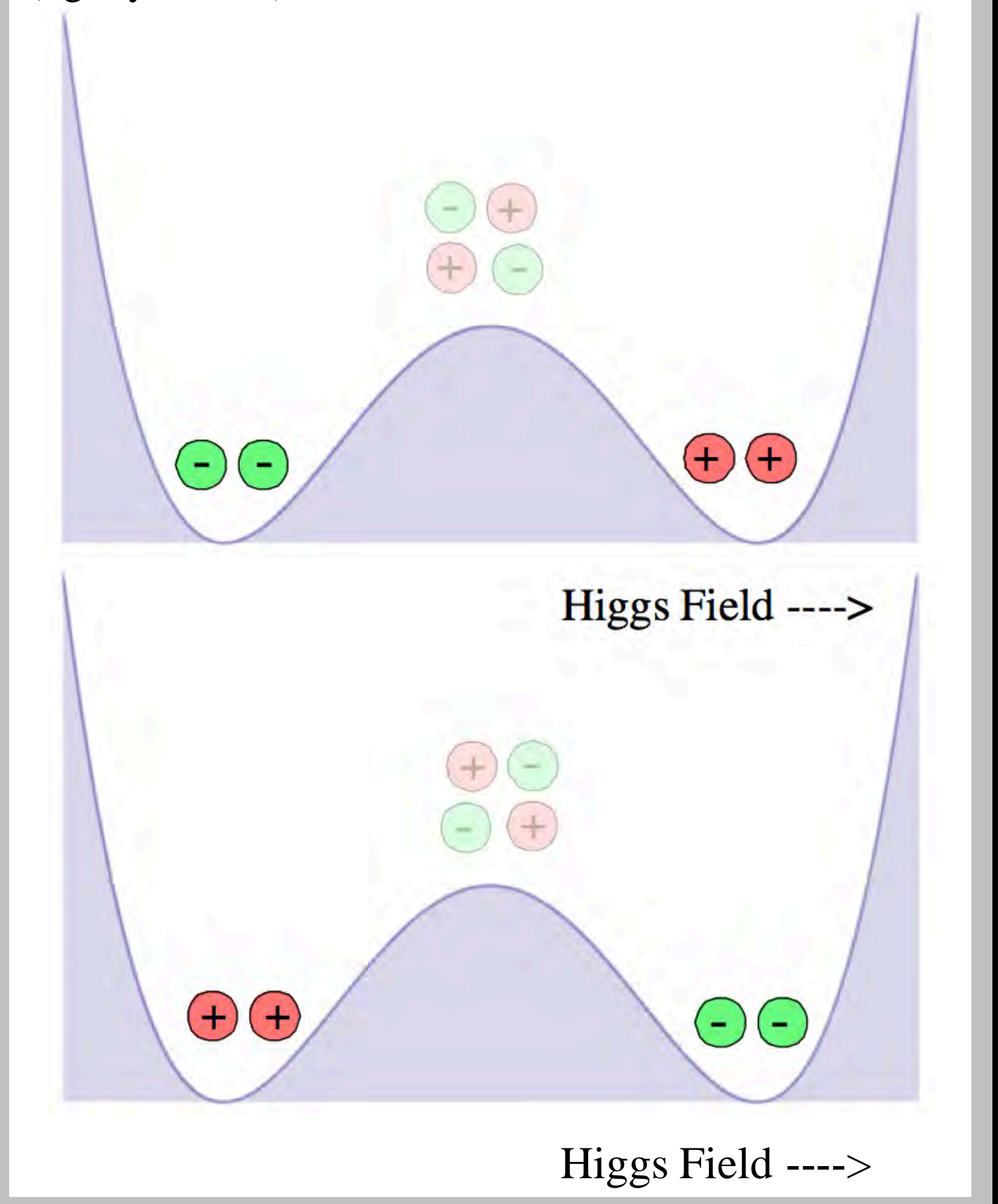

And also in terms of the Lamb shift frequency $\left(\nu_{Lamb\ shift}\right)$ indicative of the anomalous energy difference between $\left({}^2S_{1/2} \text{ and } {}^2P_{1/2}\right)$ orbitals that differ in their spatial orientation. Energetics associated with spatial orientation is consistent with the trisine geometric space concept.

$$h\nu_{Lamb\ Shift} = \frac{3}{5}\frac{(\mathbb{C}e)^2}{k_{mx}\varepsilon_x B_{Bohr\ Radius}}\ln\left(2\frac{B}{A}\right) \tag{2.10.16}$$

at $T_c = 14150.44\ K^o$ $\quad \nu_{Lamb\ Shift} = 1.06E+09\ Hz$

The trisine vacuum *cavity* ratio 2*B/A* represents the *cavity* minimum/maximum frequencies generally used in Lamb Shift derivations[137].

Maxwell's equations may be expressed in trisine geometrical dimensions:

*Gauss's Law (e)*

$$\frac{1}{2\pi}D_{ftrisine}\frac{section}{\cos(\theta)} = \mathbb{C}e$$

$$\frac{1}{2g_d}\left(-D_{fx}approach + D_{fy}side - D_{fz}side\right)\cos(\theta) \tag{2.10.17}$$

*Gauss's Law (H)*

$$H\ section - H\ section = 0 \tag{2.10.18}$$

*Faraday's Law*

$$voltage = EMF = \frac{\hbar}{time}\frac{\pi}{\mathbb{C}e} \tag{2.10.19}$$

*Ampere's Law*

$$\frac{chain}{cavity}H_c\frac{v_{\varepsilon x}}{K_B} = \mathbb{C}e\frac{2}{time} + D\frac{2section}{\cos(\theta)}\frac{2}{time} \tag{2.10.20}$$

## 2.11 Cosmological Parameters

The following relations of *Hubble radius* $\left(R_U\right)$, *mass* $\left(M_U\right)$, *density* $\left(\rho_U\right)$ and *time* $\left(T_U\right)$ in terms of the *cosmological parameter* $\left(\Lambda\ or\ 6C_U^2/\cos(\theta)\right)$ are in general agreement with known data. This could be significant. The cosmological parameter $\left(\Lambda\ or\ 6C_U^2/\cos(\theta)\right)$ as presented in equation 2.11.1 is based on energy/area or *trisine surface tension* $\left(\sigma_T\right)$, as well as energy/volume or *universe pressure* $\left(p_U\right)$. The *trisine surface tension* $\left(\sigma_T\right)$ is a function of $\left(T_c\right)$ while the *universe pressure* $\left(p_U\right)$ is the value at current CMBR temperature of $2.729\,^oK$ as presented in table 2.7.1 and observed by COBE [26] and WMAP. The *cosmological parameter* $\left(\Lambda\ or\ 6C_U^2/\cos(\theta)\right)$ also reduces to an expression relating trisine resonant velocity transformed mass $\left(m_t\right)$, $G_U$ and $\hbar$. Also a reasonable value of 71.2 km/sec-million parsec for the Hubble constant $\left(H_U\right)$ from equation 2.11.14 is achieved based on a *universe absolute viscosity* $\left(\mu_U\right)$ and *energy dissipation rate* $\left(P_U\right)$.

$$C_U = \frac{16\pi \mathrm{G_U}}{v_{dx}^4}\sigma_T = \frac{H_U}{\sqrt{3}c} = \frac{4}{\sqrt{3}\pi}\frac{m_t^3 G_U}{\hbar^2} = \frac{1}{R_U}$$

$$\Lambda_U = \frac{2H_U^2}{c^2\cos(\theta)} \qquad \frac{H_U}{G_U} = U = 3.46\text{E-11 g sec/cm}^3 \tag{2.11.1}$$

$$C_{Ur} = \frac{16\pi \mathrm{G_{Ur}}}{v_{dxr}^4}\sigma_{Tr} = \frac{H_{Ur}}{\sqrt{3}c} = \frac{4}{\sqrt{3}\pi}\frac{m_t^3 G_{Ur}}{\hbar^2} = \frac{1}{R_{Ur}}$$

$$\Lambda_{Ur} = \frac{2H_{Ur}^2}{c^2\cos(\theta)} \qquad \frac{H_{Ur}}{G_{Ur}} = U = 3.46\text{E-11 g sec/cm}^3 \tag{2.11.1r}$$

$$\Lambda_U = 1.29\text{E-56}\ cm^{-2} \qquad \text{(present)}$$

$$\sigma_T = \frac{k_b T_c}{section} \quad (2.11.2)$$

$$\sigma_{Tr} = \frac{k_b T_{cr}}{section_r} \quad (2.11.2r)$$

$\sigma_T = 5.32\text{E-}08\ dyne/cm^2$ (present)

$$R_U = \frac{1}{C_U} = \frac{\sqrt{3}c}{\sqrt{4\pi G_U \rho_U}} \quad \left(\sqrt{3} \cdot \text{Einstein radius}\right) \quad (2.11.3)$$

$$R_{Ur} = \frac{1}{C_{Ur}} = \frac{\sqrt{3}c}{\sqrt{4\pi G_{Ur} \rho_{Ur}}} \quad \left(\sqrt{3} \cdot \text{Einstein radius}\right) \quad (2.11.3r)$$

The Einstein radius is sometimes called the Jeans' length [101]. The factor $\sqrt{3}$ represent an internal deviation with the trisine structure that adds an inflation factor to the static state represented by the Einstein Radius.

$$R_U = \frac{\sqrt{3}}{g_s^3} \frac{cavity}{chain} k_{mx} \varepsilon_x B \quad (2.11.4)$$

$$R_{Ur} = \frac{\sqrt{3}}{g_s^3} \frac{cavity_r}{chain_r} k_{mxr} \varepsilon_{xr} B_r \quad (2.11.4r)$$

**Figure 2.11.1** Observed Universe with Mass $M_U$ and Hubble radius of curvature $R_U$ and smooth trisine lattice expansion (three dimensional tessellation of space) all as a function of universe age($Age_U$)

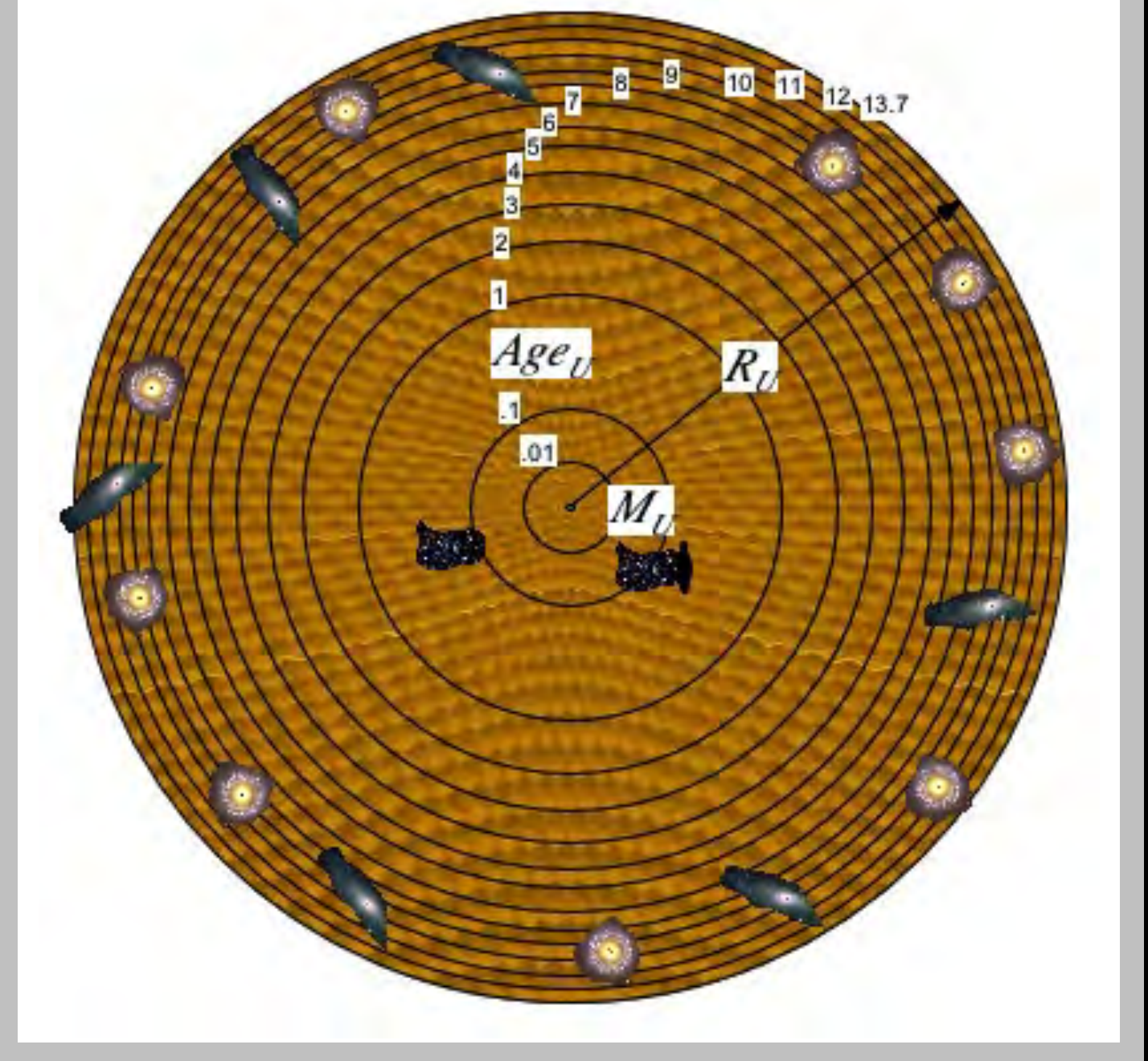

Under conditions expressed in equation 2.11.42, $R_U$ is equal to the Schwarz gravitational viscous scale.

$$R_U = \frac{g_s}{3} \frac{m_T c}{\hbar 2\pi} \sqrt{\frac{3}{2} \frac{\nu_U}{G_U \rho_U}} \quad (2.11.5)$$

$$R_{Ur} = \frac{g_s}{3} \frac{m_T c}{\hbar 2\pi} \sqrt{\frac{3}{2} \frac{\nu_{Ur}}{G_{Ur} \rho_{Ur}}} \quad (2.11.5r)$$

where $\nu_U = \nu_{Ur}$

$R_U = 2.25E28$ cm ($\sqrt{3}$ c$\cdot$13.71 Gyr)

**Figure 2.11.2** The universe expands with $B$ and $R_U$.

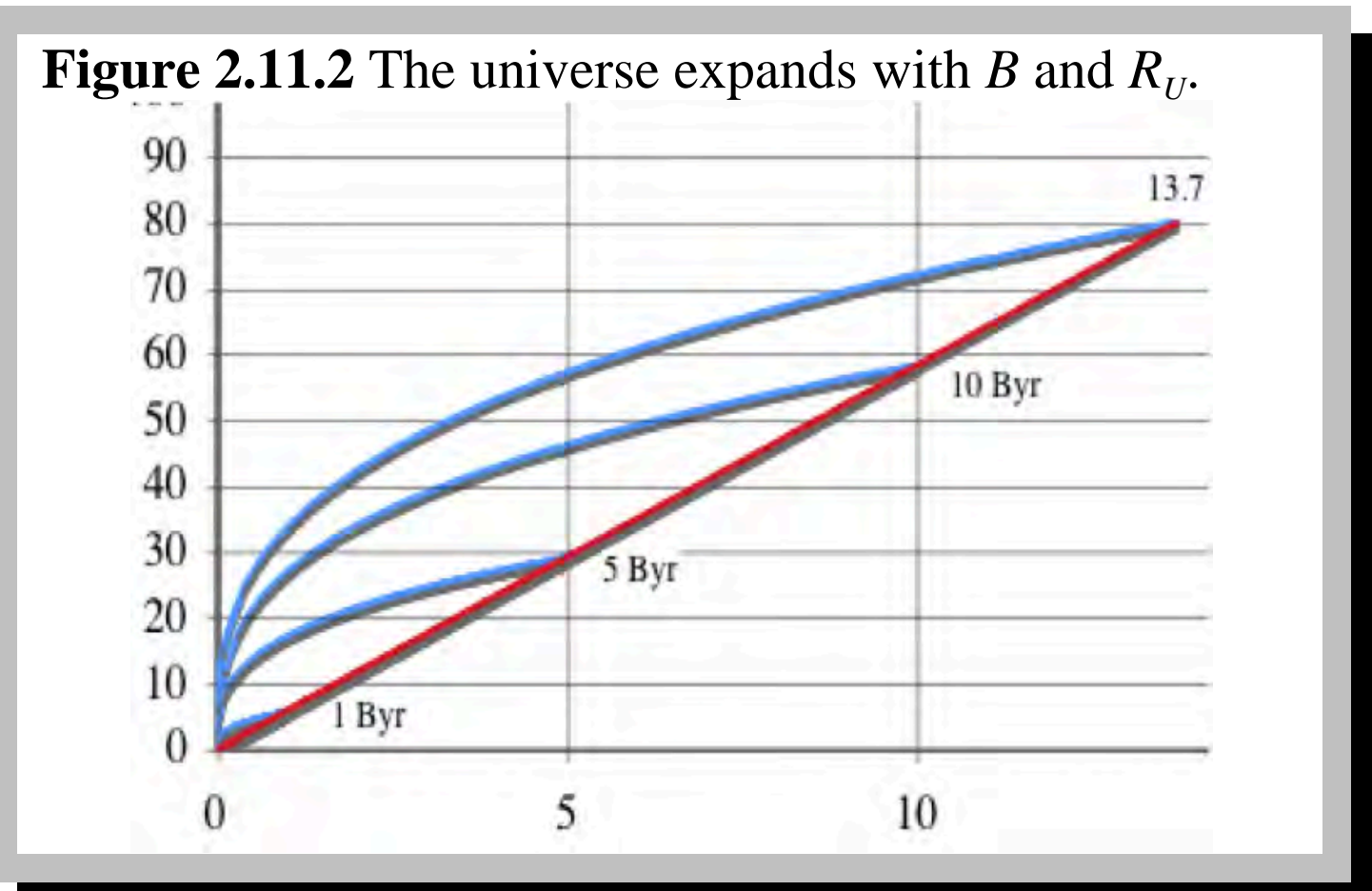

The universe $R_U$ and Trisine Lattice as measured by cellular dimension $B$ are equal at any proper time in Figure 2.11.2.

The mass of the universe $\left(M_U\right)$ is established by assuming a spherical isotropic universe (center of mass in the universe center) of Hubble radius of curvature $\left(R_U\right)$ such that a particle with the speed of light cannot escape from it as expressed in equation 2.11.6 and graphically described in figure 2.11.1.

$$M_U = R_U \cdot \frac{c^2}{G_U} = \frac{\sqrt{3}\pi}{4} \frac{\hbar^2 c^2}{m_T^3 G_U^2} = \frac{F_P}{Euler\ a_U} \quad (2.11.6)$$

$$M_{Ur} = R_{Ur} \cdot \frac{c^2}{G_{Ur}} = \frac{\sqrt{3}\pi}{4} \frac{\hbar^2 c^2}{m_T^3 G_{Ur}^2} = \frac{F_{Pr}}{Euler\ a_{Ur}} \quad (2.11.6r)$$

$M_U = 3.02E56\ g\ = 3.01E61\ m_t\ = 3.32E83\ m_e$

$$\rho_U = \frac{M_U}{V_U} = M_U \cdot \frac{3}{4\pi} \frac{1}{R_U^{\ 3}} = \frac{4}{\pi^3} \frac{m_T^6 c^2 G_U}{\hbar^4} = \frac{m_T}{cavity} \quad (2.11.7)$$

$$\rho_{Ur} = \frac{M_{Ur}}{V_{Ur}} = M_{Ur} \cdot \frac{3}{4\pi} \frac{1}{R_{Ur}^{\ 3}} = \frac{4}{\pi^3} \frac{m_T^6 c^2 G_{Ur}}{\hbar^4} = \frac{m_T}{cavity_r} \quad (2.11.7r)$$

$\rho_U = 6.38E-30$ g/cm$^3$

This value for Universe density is much greater than the density of the universe calculated from observed celestial objects. As reported in reference [27], standard cosmology model allows accurate determination of the universe baryon density of between 1.7E-31 to 4.1E-31 g/cm$^3$. These values are 2.7 - 6.4% of the universe density reported herein based the background universe as superconductor.

$$Universe\ Age\left(Age_U\right) = \frac{R_U}{\sqrt{3}c} = \frac{\pi}{4}\frac{\hbar^2}{m_T^3 c\ G_U} \qquad (2.11.6)$$

$$= 4.32E17 \text{ sec } (13.71 \text{ Gyr})$$

Equation 2.11.7 may be rearranged in terms of a universal surface constant.

$$\frac{M_U}{R_U^2} = \frac{M_{Ur}}{R_{Ur}^2} = 0.599704 \qquad (2.11.6a)$$

Which compares favorability to 13.75 ± 0.13 Gyr [20]. This is an indication that the universe is expanding at the speed of light ($c$) but not accelerating.

As illustrated in Figure 2.11.3 equation 2.11.14 for the Hubble constant $\left(H_U\right)$ is derived by equating the forces acting on a cube of fluid in shear in one direction to those in the opposite direction, or

$$p_U\ \Delta y\ \Delta z + \left[\tau + \frac{d\tau}{dz}\Delta z\right]\Delta x\ \Delta y = \tau\ \Delta x\ \Delta y + \left[p_U + \frac{dp_U}{dx}\Delta x\right]\Delta y\ \Delta z \qquad (2.11.7)$$

$$p_{Ur}\ \Delta y_r\ \Delta z_r + \left[\tau_r + \frac{d\tau_r}{dz_r}\Delta z_r\right]\Delta x_r\ \Delta y_r = \tau_r\ \Delta x_r\ \Delta y_r + \left[p_{U_r} + \frac{dp_{Ur}}{dx_r}\Delta x_r\right]\Delta y_r\ \Delta z_r \qquad (2.11.7r)$$

**Figure 2.11.3** Shear forces along parallel planes of an elemental volume of fluid

Here $p_U$ is the universe pressure intensity $\left(\mathbb{C}k_B T_c/cavity\right)$, $\tau$ the shear intensity, and $\Delta x$, $\Delta y$, and $\Delta z$ are the dimensions of the cube. It follows that:

$$\frac{d\tau}{dz} = \frac{dp_U}{dx} \qquad (2.11.8)$$

$$\frac{d\tau_r}{dz_r} = \frac{dp_{Ur}}{dx_r} \qquad (2.11.8r)$$

The power expended, or the rate at which the couple does work $\left(\tau\ \Delta x\ \Delta y\right)$, equals the torque $\left(\tau\ \Delta x\ \Delta y\right)\Delta z$ times the angular velocity $dv/dz$. Hence the universe power $\left(P_U\right)$ consumption per universe volume $\left(V_U\right)$ of fluid is

$P_U/V_U = \left[\left(\tau\ \Delta x\ \Delta y\right)\Delta z\ \ dv/dz\right]/\left(\Delta x\ \Delta y\ \Delta z\right) = \tau\ \left(\text{dv/dz}\right)$.

Defining $\quad \tau = \mu_U\ \ dv/dz \quad$ and $\quad H_U = dv/dz$

then $\quad P_U/V_U = \mu_U\left(dv/dz\right)^2 = \mu_U H_U^2$

or $\quad H_U^2 = P_U/\left(\mu_U V_U\right)$

where $\quad p_{U_1} cavity_{U_1}^{5/3} = p_{U_2} cavity_{U_2}^{5/3}$

with the exponent factor '5/3' being representative of a monatomic ideal gas adiabatic expansion and with inflation number $\#_U$ representing the number of cavities in the universe

and $\quad V_U = \left(M_U/m_t\right)cavity = \#_U\ cavity$

The *universe absolute viscosity* $\left(\mu_U\right)$ is based on momentum transferred per surface area as an outcome of classical kinetic theory of gas viscosity [19]. In terms of our development, the momentum changes over the *section* in each *cavity* such that:

$$\mu_U = \frac{m_T v_{dx}}{section} = \frac{1}{2}\left(\frac{A}{B}\right)_T \frac{h}{cavity} \qquad (2.11.9)$$

$$\mu_{Ur} = \frac{m_T v_{dxr}}{section_r} = \frac{1}{2}\left(\frac{A_r}{B_r}\right)_T \frac{h}{cavity_r} \qquad (2.11.9r)$$

Equation 2.11.10 is equivalent to entropy density derivation[130].

$$\frac{\eta}{s} = \mu_U cavity\left(\frac{B}{A}\right)_T \frac{1}{4\pi^2} = \frac{\hbar}{4\pi} \qquad (2.11.10)$$

$$\frac{\eta}{s} = \mu_{Ur} cavity_r\left(\frac{B}{A}\right)_T \frac{1}{4\pi^2} = \frac{\hbar}{4\pi} \qquad (2.11.10r)$$

Derivation of the kinematic viscosity $\left(\upsilon_U\right)$ in terms of absolute viscosity as typically presented in fluid mechanics [18], a constant universe kinematic viscosity $\left(\upsilon_U\right)$ is achieved in terms of Planck's constant $(\hbar)$ and Cooper CPT Charge conjugated pair resonant transformed trisine mass $(m_T)$.

$$\upsilon_U = \frac{cavity}{m_T}\mu_U = \frac{\mu_U}{\rho_U} = \pi\left(\frac{A}{B}\right)_T\frac{l_T^2}{t_T} = \pi\frac{\hbar}{m_T}\left(\frac{A}{B}\right)_T = \frac{4\pi^2}{m_T}\frac{\eta}{s}\left(\frac{A}{B}\right)_T = 1.39E-02\ \frac{cm^2}{sec} \qquad (2.11.11)$$

$$\upsilon_U = \frac{cavity_r}{m_T}\mu_{Ur} = \frac{\mu_{Ur}}{\rho_{Ur}} = \pi\left(\frac{A_r}{B_r}\right)_T\frac{l_T^2}{t_T} = \pi\frac{\hbar}{m_T}\left(\frac{A_r}{B_r}\right)_T = \frac{4\pi^2}{m_T}\frac{\eta}{s}\left(\frac{A_r}{B_r}\right)_T = 1.39E-02\ \frac{cm^2}{sec} \qquad (2.11.11r)$$

Derivation of the Universe Energy Dissipation Rate $(P_U)$ assumes that all of the universe mass $(M_U)$ in a universe volume $(V_U)$ is in the superconducting state. In other words, our visible universe contributes an insignificant amount of mass to the universe mass $(M_U)$. This assumption appears to be valid due to the subsequent calculation of the Hubble constant $(H_U)$ of 71.23 km/sec-million parsec (equation 2.11.14) which falls within the experimentally observed WMAP value[20] of 71.0 ± 2.5 km/sec-million parsec.

$$P_U = \frac{A}{B}\frac{4\cos(\theta)}{Euler}\frac{m_Tc^2}{t_U} = \frac{M_U v_{dx}^2}{2}\frac{v_{dx}}{R_U} = \mu_U H_U^2 V_U \qquad (2.11.12)$$

$$P_{Ur} = \frac{A_r}{B_r}\frac{4\cos(\theta)}{Euler}\frac{m_Tc^2}{t_U} = \frac{M_{U_r} v_{dxr}^2}{2}\frac{v_{dxr}}{R_{Ur}} = \mu_{Ur} H_{Ur}^2 V_{Ur} \qquad (2.11.12r)$$

$$P_U = P_{Ur} = 2.24E+19\ erg/sec$$

$$V_U = \frac{4}{3}\pi R_U^3 = \frac{\sqrt{3}\pi^4\hbar^6}{16 m_T^9 G_U^3} \qquad (2.11.13)$$

$$V_{Ur} = \frac{4}{3}\pi R_{Ur}^3 = \frac{\sqrt{3}\pi^4\hbar^6}{16 m_T^9 G_{Ur}^3} \qquad (2.11.13r)$$

$$= 4.74E85\ cm^3 = 3.01E81\ cavitys$$

$$H_U = \sqrt{\frac{P_U}{\mu_U V_U}} = \frac{4\ m_t^3 G_U c}{\pi\hbar^2} = \frac{F_P}{Euler\ M_U c} = \sqrt{3}\frac{c}{R_U} = \frac{1}{\sqrt{3}}\frac{B}{A}\frac{v_{dx}}{c}\frac{1}{time_\pm} \qquad (2.11.14)$$

$$H_{Ur} = \sqrt{\frac{P_U}{\mu_{Ur} V_{Ur}}} = \frac{4\ m_t^3 G_{Ur} c}{\pi\hbar^2} = \frac{F_{Pr}}{Euler\ M_{Ur} c} = \sqrt{3}\frac{c}{R_{Ur}} = \frac{1}{\sqrt{3}}\frac{B_r}{A_r}\frac{v_{dxr}}{c}\frac{1}{time_{r\pm}} \qquad (2.11.14r)$$

With the universal constant mass $(m_t)$ (analogous to the Weinberg mass[99]) established as:

$$m_T^3 = \frac{m_T\rho_U k_b T_c section}{\sqrt{3}H_U c} = \frac{\pi^2\rho_U\hbar^2}{H_U c} = \frac{U\hbar^2\cos^3(\theta)}{c} \qquad (2.11.15)$$

$$m_T^3 = \frac{m_T\rho_{Ur} k_b T_{cr} section_r}{\sqrt{3}H_{Ur} c} = \frac{\pi^2\rho_{Ur}\hbar^2}{H_{Ur} c} = \frac{U\hbar^2\cos^3(\theta)}{c} \qquad (2.11.15r)$$

With proximity to Heisenberg Uncertainty condition $\hbar/2$:

$$2\frac{\hbar}{2} = m_T c\frac{4}{\pi}\frac{m_T^2}{U\hbar} = m_T c\frac{4}{\pi}\frac{G_U m_T^2}{H_U\hbar} = m_T c l_T = m_T c^2 t_T \qquad (2.11.16)$$

$$2\frac{\hbar}{2} = m_T c\frac{4}{\pi}\frac{m_T^2}{U\hbar} = m_T c\frac{4}{\pi}\frac{G_{Ur} m_T^2}{H_{Ur}\hbar} = m_T c l_T = m_T c^2 t_T \qquad (2.11.16r)$$

$$H_U = 2.31E-18/\sec \qquad \left(71.23\frac{\text{km}}{\text{sec}}\frac{1}{\text{million parsec}}\right)$$

$$x_U = 1.75E-13\ cm \qquad \text{(nuclear dimension)}$$

Now from equation 2.11.1 and 2.11.14 the *cosmological parameter* $(\Lambda_U\ or\ 6C_U^2/\cos(\theta))$ is defined in terms of the Hubble constant $(H_U)$ in equation 2.11.17. The *cosmological parameter* $\Lambda_U$ is considered a scalable multiple and not addend factor

$$C_U = \left(\frac{\Lambda_U\cos(\theta)}{6}\right)^{\frac{1}{2}} = \frac{H_U}{\sqrt{3}c} = \frac{1}{R_U} = \frac{c^2}{M_U G_U} \qquad (2.11.17)$$

$$C_{Ur} = \left(\frac{\Lambda_{Ur}\cos(\theta)}{6}\right)^{\frac{1}{2}} = \frac{H_{Ur}}{\sqrt{3}c} = \frac{1}{R_{Ur}} = \frac{c^2}{M_{Ur} G_{Ur}} \qquad (2.11.17r)$$

$$= 4.45E-29\ cm^{-1}$$

And the Universe Escape Velocity $(v_U)$ equivalent to the speed of light *(c)*.

$$v_U = c = \frac{H_U R_U}{\sqrt{3}} = \sqrt{\frac{G_U M_U}{R_U}} \qquad (2.11.18)$$

$$v_U = c = \frac{H_{Ur} R_{Ur}}{\sqrt{3}} = \sqrt{\frac{G_{Ur} M_{Ur}}{R_{Ur}}} \qquad (2.11.18r)$$

Equation (2.11.19) implies a constant gravitational energy for any test mass($m$) at any universe time epoch.

$$mv_U^2 = mc^2 = \frac{mH_U^2 R_U^2}{3} = \frac{G_U m M_U}{R_U}$$
$$\text{where} \quad G_U R_U = 1.49867\text{E+21} \tag{2.11.19}$$
$$\text{and} \quad \frac{M_U}{R_U^2} = .599704$$

$$mv_U^2 = mc^2 = \frac{mH_{Ur}^2 R_{Ur}^2}{3} = \frac{G_{Ur} m M_{Ur}}{R_{Ur}}$$
$$\text{where} \quad G_{Ur} R_{Ur} = 1.49867\text{E+21} \tag{2.11.19r}$$
$$\text{and} \quad \frac{M_{Ur}}{R_{Ur}^2} = .599704$$

Now from equation 2.11.20 and 2.11.15 the *universe mass* $(M_U)$ is defined in terms of the Hubble constant $(H_U)$ in equation 2.11.14.

$$M_U = \frac{m_t}{cavity} V_U = \rho_U \frac{4\pi}{3} R_U^3$$
$$= \frac{\sqrt{3}c^3}{H_U G_U} = \frac{\sqrt{3}\pi\hbar^2 c^2}{4 m_T^3 G_U^2} = \frac{R_U c^2}{G_U} \tag{2.11.20}$$

$$M_{Ur} = \frac{m_T}{cavity_r} V_{Ur} = \rho_{Ur} \frac{4\pi}{3} R_{Ur}^3$$
$$= \frac{\sqrt{3}c^3}{H_{Ur} G_{Ur}} = \frac{\sqrt{3}\pi\hbar^2 c^2}{4 m_T^3 G_{Ur}^2} = \frac{R_{Ur} c^2}{G_{Ur}} \tag{2.11.20r}$$

$$M_{Upresent} = \frac{P_U}{c^2} \int_{v_{dx}=c}^{v_{dx}\text{ at present}} \left( \frac{Age_{Upresent}}{Age_U} \right)^2 dAge_U$$
$$= -\frac{P_U Age_{Upresent}^2}{c^2} \left( \frac{1}{Age_{Uatpresent}} - \frac{1}{Age_{U\ at\ vdx=c}} \right)$$
$$= 3.02E56\ g \quad \text{(conforming to condition 1.11)}$$
$$\text{Number of particles} = \frac{M_{Upresent}}{m_T} = \frac{V_U}{cavity} = 3.01E81$$

**Figure 2.11.4** Trisine Steric Charge Conjugate Pair Change Time Reversal CPT Geometry as depicted in Figure 2.3.1 with Relativistic Pressure '*-p*', Quantum Mechanical Pressure '*p*' and Density '$\rho_U$'.

**Figure 2.11.5** Graphical presentation of universe mass

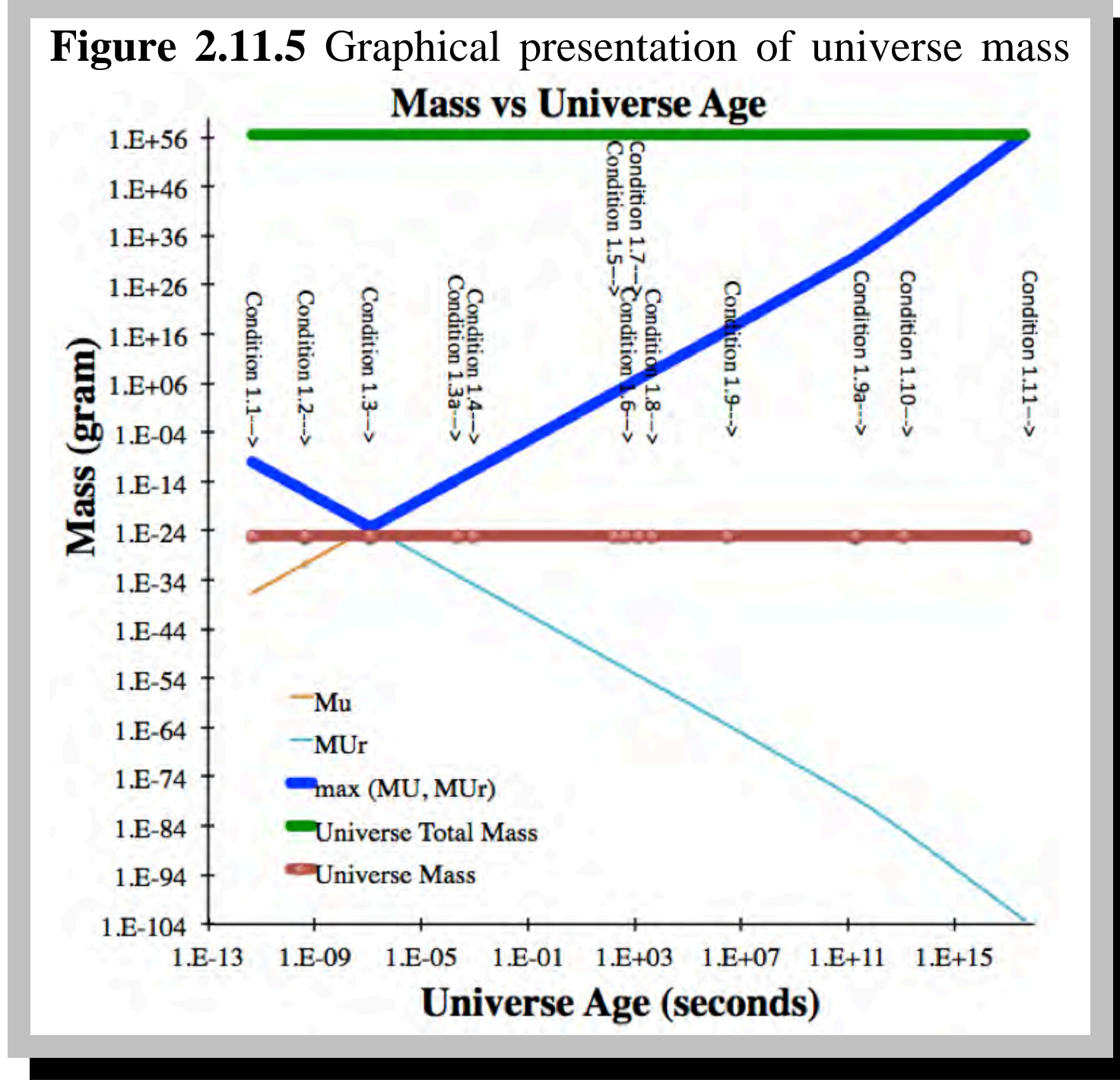

The trisine model conforms to Einstein Energy Tensor equations at all lattice similarities assuming the universe mass $M_U$ increases with time.

**Figure 2.11.5a** Graphical presentation of universe length

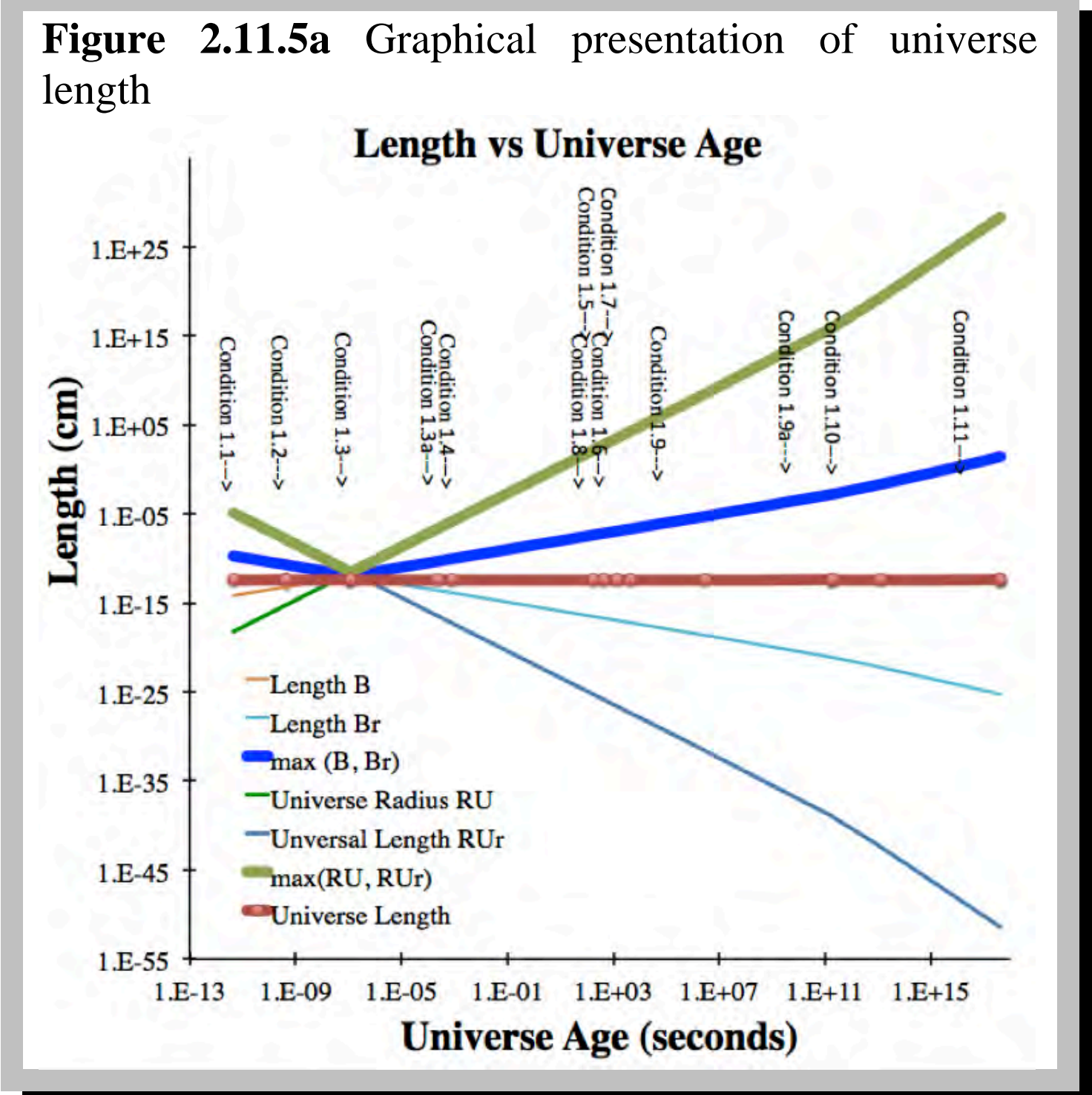

This is consistent with conventional Friedmann-Lemaitre theory for universe constant mass under assumption that universe began with extremely large number of mass points ( $M_{Upresent}/M_U$ ~1E90) at trisine condition 1.1 each quantum mechanically in phase with each other in as much as they form a time fleeting colligative Bose Einstein Condensate aggregation adiabatically expanding to the present. These initial ~1E90 in phase points converging in to the present one universe explains the present expanded one universe homogeneity without that group phase reflection inflation step.

Now from equation 2.11.14 and 2.11.5 the *universe density* $(\rho_U)$ is defined in terms of the Hubble constant $(H_U)$ in equation 2.11.21. This differs from the universe critical density equation in reference [27] by a factor of as a consequence of Robertson Walker equation 2.11.34 reduction, which indicates an expanding but not necessarily inflationary universe.

$$\rho_U = \frac{2}{8\pi}\frac{H_U^2}{G_U} = \frac{3}{8\pi}\frac{chain}{cavity}\frac{H_U^2}{G_U} = \frac{m_T}{cavity} = \frac{1}{8\pi}\frac{c^2}{G_U}\Lambda\cos(\theta) \qquad (2.11.21)$$

$$\Lambda_U = \frac{2H_U^2}{c^2\cos(\theta)} = \frac{8\pi G_U\rho_U}{c^2\cos(\theta)}$$

$$\rho_{Ur} = \frac{2}{8\pi}\frac{H_{Ur}^2}{G_{Ur}} = \frac{3}{8\pi}\frac{chain}{cavity}\frac{H_{Ur}^2}{G_{Ur}} = \frac{m_T}{cavity_r} = \frac{1}{8\pi}\frac{c^2}{G_{Ur}}\Lambda_r\cos(\theta) \qquad (2.11.21r)$$

$$\Lambda_{Ur} = \frac{2H_{Ur}^2}{c^2\cos(\theta)} = \frac{8\pi G_{Ur}\rho_{Ur}}{c^2\cos(\theta)}$$

$$\rho_U = 6.38E-30\ \frac{g}{cm^3}$$

$$\Omega_{\Lambda_U} = \frac{1}{8\pi}\frac{c^2}{G_U}\Lambda_U\left(\frac{3}{8\pi}\frac{H_U^2}{G_U}\right)^{-1} = \frac{\Lambda_U c^2}{3H_U^2} = 0.72315 \approx \frac{2}{3\cos(\theta)} \qquad (2.11.22)$$

$$\Omega_{\Lambda_{Ur}} = \frac{1}{8\pi}\frac{c^2}{G_{Ur}}\Lambda_U\left(\frac{3}{8\pi}\frac{H_{Ur}^2}{G_{Ur}}\right)^{-1} = \frac{\Lambda_{Ur} c^2}{3H_{Ur}^2} = 0.72315 \approx \frac{2}{3\cos(\theta)} \qquad (2.11.22r)$$

Now correlating equation 2.11.15 and 2.11.7, the *universe age* $(Age_U)$ is defined in terms of the Hubble constant $(H_U)$ in equation 2.11.23.

$$Age_U = \frac{1}{H_U} \qquad (2.11.23)$$

$$= 4.32E{+}17 \text{ or } 1.37E10 \text{ years}$$

$$\textit{Universe mass creation } (M_{UC}) \textit{ at } T_c \qquad (2.11.24)$$

$$Age_{UG} = Age_U\sqrt{G_U/G} = Age_U\sqrt{R_U c^2/(GM_U)} \qquad (2.11.25)$$

$$Age_{UGv} = Age_U\sqrt{R_U c^2/\left(G\left(M_U + M_v\right)\right)} \qquad (2.11.26)$$

$Age_{UGv} = Age_{UG} = Age_U$ = 13.7E+9 years at the present time

**Table 2.11.1** Vacuum Space parameters

| Condition | 1.2 | 1.3 | 1.6 | 1.9 | 1.10 | 1.11 |
|---|---|---|---|---|---|---|
| $T_a\left({}^0K\right)$ | 6.53E+11 | 6.53E+11 | 6.53E+11 | 6.53E+11 | 6.53E+11 | 6.53E+11 |
| $T_b\left({}^0K\right)$ | 9.05E+14 | 5.48E+13 | 9.22E+08 | 1.10E+07 | 2,868 | 2.723 |
| $T_{br}\left({}^0K\right)$ | 3.30E+12 | 5.45E+13 | 3.24E+18 | 2.72E+20 | 1.04E+24 | 1.10E+27 |
| $T_c\left({}^0K\right)$ | 8.96E+13 | 3.28E+11 | 93.0 | 0.0131 | 8.99E-10 | 8.10E-16 |
| $T_{cr}\left({}^0K\right)$ | 1.19E+09 | 3.25E+11 | 1.15E+21 | 8.11E+24 | 1.19E+32 | 1.32E+38 |
| $z_R + 1$ | 3.67E+43 | 8.14E+39 | 3.89E+25 | 6.53E+19 | 1.17E+09 | 1.00E+00 |
| $z_B + 1$ | 3.32E+14 | 2.01E+13 | 3.39E+08 | 4.03E+06 | 1.05E+03 | 1.00E+00 |
| $Age_{UG}\left(sec\right)$ | 7.14E-05 | 4.79E-03 | 6.94E+04 | 5.36E+07 | 1.27E+13 | 4.33E+17 |
| $Age_{UGv}\left(sec\right)$ | 4.59E-10 | 1.25E-07 | 4.42E+02 | 3.13E+06 | 1.22E+13 | 4.33E+17 |
| $Age_U\left(sec\right)$ | 1.18E-26 | 5.31E-23 | 1.11E-08 | 6.63E-03 | 3.70E+08 | 4.33E+17 |
| $\mu_U$ (g/cm/sec) | 3.25E+12 | 7.20E+08 | 3.44E-06 | 5.77E-12 | 1.03E-22 | 8.84E-32 |
| $\mu_{Ur}$ (g/cm/sec) | 1.57E+05 | 7.10E+08 | 1.49E+23 | 8.86E+28 | 4.95E+39 | 5.78E+48 |
| $\sigma_U$ (erg/cm$^2$) | 8.07E+23 | 1.08E+19 | 8.70E-01 | 1.74E-08 | 8.13E-23 | 6.61E-35 |
| $\sigma_{Ur}$ (erg/cm$^2$) | 1.42E+14 | 1.06E+19 | 1.32E+38 | 6.62E+45 | 1.41E+60 | 1.74E+72 |
| $P_U$ (erg/sec) | 2.24E+19 | 2.24E+19 | 2.24E+19 | 2.24E+19 | 2.24E+19 | 2.24E+19 |
| $P_{Ur}$ (erg/sec) | 2.24E+19 | 2.24E+19 | 2.24E+19 | 2.24E+19 | 2.24E+19 | 2.24E+19 |
| $M_{UC}$ (g) | 3.97E+59 | 8.80E+55 | 4.20E+41 | 7.05E+35 | 1.26E+25 | 0.00E+00 |

This development would allow the galactic collision rate calculation in terms of the Hubble constant $\left(H_U\right)$ (which is considered a shear rate in our fluid mechanical concept of the universe) in accordance with the Smolukowski relationship [12, 13, 14] where $J_U$ represents the collision rate of galaxies of two number concentrations $\left(n_i, n_j\right)$ and diameters $\left(d_i, d_j\right)$:

$$J_U = \frac{1}{6} n_i n_j H_U \left(d_i + d_j\right)^3 \tag{2.11.27}$$

$$d = \frac{24}{\pi}\frac{C\sigma}{\mu_1 H} \quad \text{After Appendix L2.12} \tag{2.11.28}$$

The superconducting Reynolds' number $\left(R_e\right)$, which is defined as the ratio of inertial forces to viscous forces is presented in equation 2.11.20 and has the value of one(1). In terms of conventional fluid mechanics this would indicate a condition of laminar flow [18].

$$\begin{aligned} R_e &= \frac{v_{dx}}{\nu_U}\frac{cavity}{section} = \frac{v_{dx}}{\nu_U} B = \frac{v_{dx}}{\nu_U} 4R_h = \frac{v_{dx}}{\nu_U} 4\frac{B}{4} \\ &= \frac{m_t}{cavity}\frac{v_{dx}}{\mu}\frac{cavity}{section} = \frac{m_T v_{dx}}{\mu\ section} = 1 \end{aligned} \tag{2.11.29}$$

$$\begin{aligned} a_U &= \pi e^{-Euler}\left(\frac{v_{dy}^2}{2P}\right)\Bigg|_{\substack{T_c=8.11E-16 \\ T_b=2.729}} \\ &= cH_U = \frac{R_U H_U^2}{\sqrt{3}} = \sqrt{\frac{P_U c^2}{\mu_U V_U}} = 6.93x10^{-8}\frac{cm}{sec^2} \end{aligned} \tag{2.11.30}$$

The numerical value of the universal acceleration $\left(a_U\right)$ expressed in equation 2.11.30 is nearly equal to that deceleration *(a)* observed with Pioneer 10 & 11 [53], but this near equality is viewed as consequence of equations 2.7.6a and 2.11.31-2.11.33 for the particular dimensional characteristics (cross section $\left(A_c\right)$, Mass $\left(M\right)$ and thrust coefficient $\left(C_t\right)$) of the Pioneer space crafts (Table 2.7.2) with appropriate negative sign indicating deceleration. Free body physical analysis of such a fact (in this case the positive acceleration $cH_U$) would indicate the possibility of opposing component forces expressed as negative acceleration. Pioneer or any other object traveling through this universe vacuum viscosity[130] would have a negative acceleration due to their momentum transfer with the space vacuum and relative to their area/mass and may in aggregate match the positive universal $cH_U$

$$a_U = cH_U = \frac{H_U^2 R_U}{\sqrt{3}} = 6.93x10^{-8}\frac{cm}{sec^2} \tag{2.11.31}$$

$$a = -\frac{A_c}{M}\frac{1}{G_U}\left(cH_U\right)^2 = -7.20x10^{-8}\frac{cm}{sec^2} \tag{2.11.32}$$

$$a = -C_t\frac{A_c}{M}\rho_U c^2 = -8.37x10^{-8}\frac{cm}{sec^2} \tag{2.11.33}$$

Analogously, the observed type Ia supernovae acceleration interpretation may actually be due to a deceleration of our local baryonic matter (star systems, galaxies, etc.) related to momentum transfer boundary between the elastic space CPT lattice and baryonic matter (star systems, galaxies, etc.) and under the universal gravity and quantum mechanical energy relationships in equations 2.11.32 and 2.11.33.

$$m_T c^2 = G_U \frac{m_t M_U}{R_U} \qquad (2.11.34)$$

$$= \frac{\sqrt{3}}{8\pi}\frac{M_U}{m_T}\frac{h^2}{2m_t}\frac{1}{R_U^2} \quad where \quad \frac{M_U}{R_U^2} = 0.599704 \ \text{g/cm}^2$$

$$= \frac{2}{3}\left(\frac{\sqrt{3}}{8\pi}\right)^2 \frac{h^2}{2m_T}\frac{1}{B_{nucleus}^2} = 3\pi k_m \varepsilon g_s^2 \frac{h^2}{2m_T}\frac{1}{A^2}$$

Eqn 2.11.34 fits well into the relativistic derivations in Appendix F and also the energy ($pc$) relationship:

$$\frac{1}{2}hH_U = \frac{1}{2}h\frac{\sqrt{3}c}{R_U} = \frac{3}{4}G_U\frac{m_T^2}{B_{nucleus}} = 8\pi G_U m_T^3 \frac{c}{h} \qquad (2.11.35)$$

The Milky Way galactic rotational energy $\left(I\omega^2\right)$ is nearly equal to relativistic energy $\left(Mc^2\right)$ by the equipartition of energy principle where the accepted dimensions are as follows with the assumption that the Milky Way is composed of 5.8E11 solar masses which rotated 20 times during the last 4.5 billion years(the age of the earth):

**Table 2.11.2** Galactic Equipartition of Energy

| | | | | |
|---|---|---|---|---|
| Diameter | 9.46E+22 | *Cm* | 100,000 | *light years* |
| Radius | 4.73E+22 | *Cm* | 50,000 | *light years* |
| Width | 2.37E+22 | *Cm* | 25,000 | *light years* |
| Volume | 1.66E+68 | *cm*$^3$ | 1.96E+14 | *light years*$^3$ |
| Density | 6.38E-30 | *g/cm*$^3$ | | |
| dark energy mass | 1.06E+39 | *G* | *M* | |
| dark energy | 9.53E+59 | *Erg* | $Mc^2$ | |
| Density | 6.94E-24 | *g/cm*$^3$ | | |
| Mass | 1.15E+45 | *G* | | |
| moment of inertia (I) | 1.29E+90 | *g cm*$^2$ | *I* | |
| rotation $\omega$ | 8.86E-16 | *radian/sec* | 1.41E-16 | *rev/sec* |
| Energy | 1.01E+60 | *erg* | $I\omega^2$ | |

These ideas are consistent with an equilibrium droplet size as derived in Appendix L2.12 and shear stated in terms of $H_U$ verses the more conventional fluid dynamic notation $G_I$. The Jeans' length relationship scales properly by constants (.546, .546, .772).

$$Universe\ Diameter = \frac{24}{\pi}\frac{C\sigma}{\mu H_U} = \frac{24}{\pi}\frac{C}{H_U}c$$

$$= .546\sqrt{\frac{15k_bT_a}{4\pi G_U m_T \rho_U}} \sim R_U \qquad (2.11.36a)$$

$$Galaxy\ Diameter = \frac{24}{\pi}\frac{C\sigma}{\mu H_U} = \frac{24}{\pi}\frac{C}{H_U}v_\varepsilon$$

$$= .546\sqrt{\frac{15k_bT_b}{4\pi G_U m_T \rho_U}} \qquad (2.11.36b)$$

$$Solar\ System\ Diameter = \frac{24}{\pi}\frac{C\sigma}{\mu H_U} = \frac{24}{\pi}\frac{C}{H_U}v_{dx}$$

$$= .772\sqrt{\frac{15k_bT_c}{4\pi G_U m_T \rho_U}} \qquad (2.11.36c)$$

Calculations are presented in Table 2.11.3 for Dark Matter as Hydrogent Bose Einstein Condensate Concentration(C) $1-\left(\Omega_{\Lambda_U}\right) = 1-.72315 = .27685$.

Universe material arranges itself in accordance with

$$c,\ v_\varepsilon,\ v_{dx} \text{ or } c,\ \left(\frac{c}{\sqrt{\varepsilon}}\right),\ \left(\frac{c}{\sqrt{k_{mx}\varepsilon}}\right)$$

**Table 2.11.3** Universe, Galaxy, Solar Systems Size in terms of universe expansion

| Condition | 1.2 | 1.3 | 1.6 | 1.9 | 1.10 | 1.11 |
|---|---|---|---|---|---|---|
| $T_a\left(^0K\right)$ | 6.53E+11 | 6.53E+11 | 6.53E+11 | 6.53E+11 | 6.53E+11 | 6.53E+11 |
| $T_b\left(^0K\right)$ | 9.05E+14 | 5.48E+13 | 9.22E+08 | 1.10E+07 | 2,868 | 2.723 |
| $T_{br}\left(^0K\right)$ | 3.30E+12 | 5.45E+13 | 3.24E+18 | 2.72E+20 | 1.04E+24 | 1.10E+27 |
| $T_c\left(^0K\right)$ | 8.96E+13 | 3.28E+11 | 93.0 | 0.0131 | 8.99E-10 | 8.10E-16 |
| $T_{cr}\left(^0K\right)$ | 1.19E+09 | 3.25E+11 | 1.15E+21 | 8.11E+24 | 1.19E+32 | 1.32E+38 |
| $z_R+1$ | 3.67E+43 | 8.14E+39 | 3.89E+25 | 6.53E+19 | 1.17E+09 | 1.00E+00 |
| $z_B+1$ | 3.32E+14 | 2.01E+13 | 3.39E+08 | 4.03E+06 | 1.05E+03 | 1.00E+00 |
| $Age_{UG}\left(sec\right)$ | 7.14E-05 | 4.79E-03 | 6.94E+04 | 5.36E+07 | 1.27E+13 | 4.33E+17 |
| $Age_{UGv}\left(sec\right)$ | 4.59E-10 | 1.25E-07 | 4.42E+02 | 3.13E+06 | 1.22E+13 | 4.33E+17 |
| $Age_U\left(sec\right)$ | 1.18E-26 | 5.31E-23 | 1.11E-08 | 6.63E-03 | 3.70E+08 | 4.33E+17 |
| Universe *Diameter*$\left(cm\right)$ | 7.46E-16 | 3.37E-12 | 7.06E+02 | 4.20E+08 | 2.35E+19 | 2.74E+28 |
| Universe *Diameter*$\left(lyr\right)$ | 7.89E-34 | 3.56E-30 | 7.46E-16 | 4.44E-10 | 2.48E+01 | 2.90E+10 |
| Galaxy *Diameter*$\left(cm\right)$ | 2.78E-14 | 3.09E-11 | 2.65E+01 | 1.72E+06 | 1.56E+15 | 5.60E+22 |
| Galaxy *Diameter*$\left(lyr\right)$ | 2.94E-32 | 3.26E-29 | 2.80E-17 | 1.82E-12 | 1.64E-03 | 5.92E+04 |
| Solar System | 1.24E-14 | 3.38E-12 | 1.19E-02 | 8.43E+01 | 1.23E+09 | 1.37E+15 |
| *Diameter*(*cm*) | | | | | | |
| Solar System *Diameter*(*lyr*) | 1.31E-32 | 3.57E-30 | 1.26E-20 | 8.91E-17 | 1.30E-09 | 1.44E-03 |

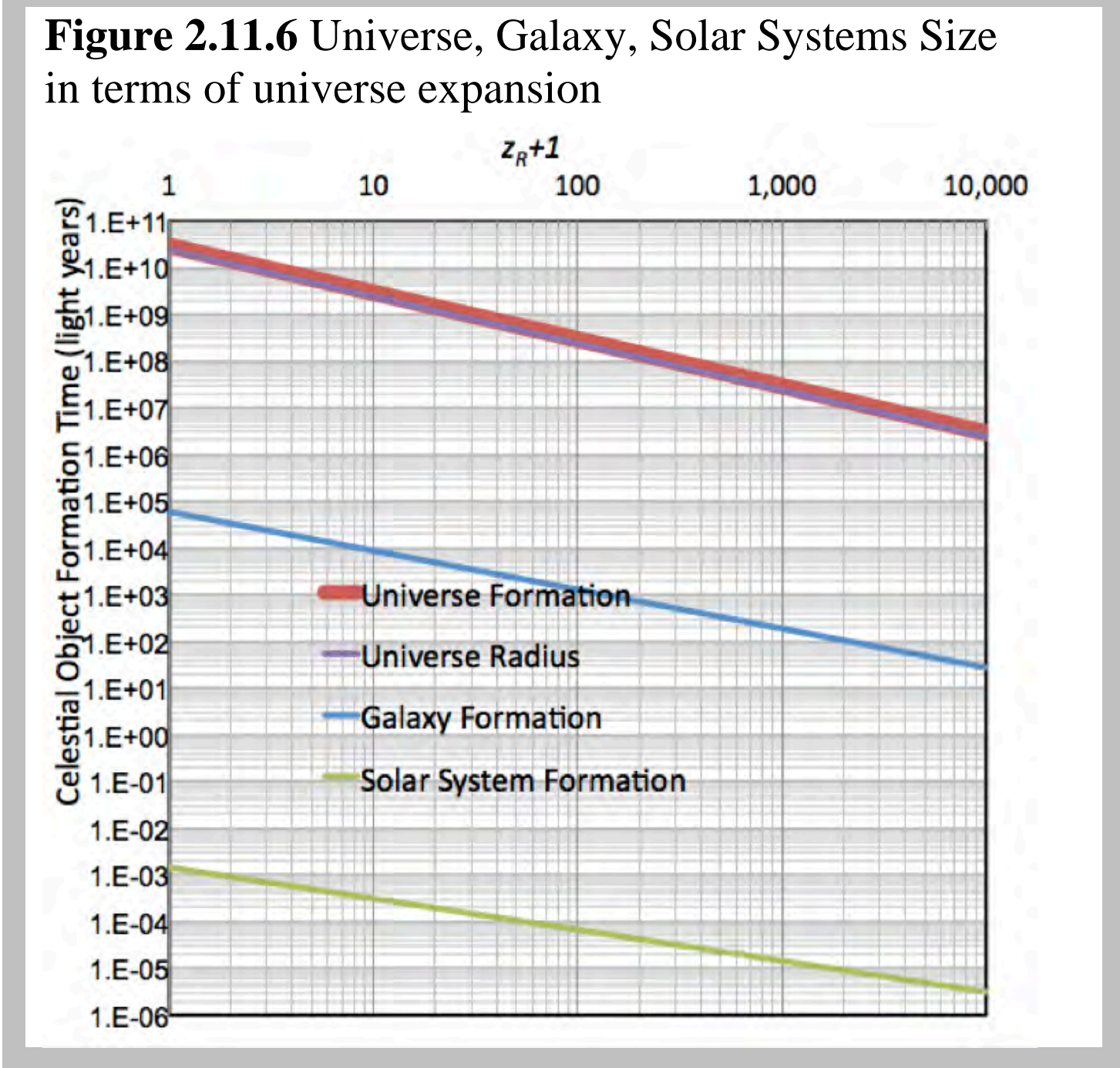

**Figure 2.11.6** Universe, Galaxy, Solar Systems Size in terms of universe expansion

These Table 2.11.3 Galactic Formation Diameter calculations indicate present galaxies with diameters typically on the order of 2.9E-3 light years(lty) were first formed at the first light event($T_b$ = 3,000K) (possibly before) and continued to increase in size (Figure 2.11.6) to present on average 1.17E+05 *light years* at $Age_U$ and $Age_{UG}$ of 4.32E+17sec (13.7 billion years). Galactic surveys should be able to verify this relationship. Dark energy shear expressed as a changing Hubble parameter $H_U$ was the principle mechanism. The gravitational mechanism was secondary.

In summary and referencing Stabell and Refsdal[128] and Robertson[129] the Robertson and Walker homogeneous and isotropic universe metric is:

$$ds^2 = c^2dt^2 - R^2(t)\left[\frac{dr^2}{1-kr^2} + r^2\left(d\theta^2 + \sin^2\theta\, d\phi^2\right)\right] \qquad (2.11.37)$$

where k is equal to +1, =1 or 0, and $R$ is a scale-factor with dimension length $R$ not necessarily related to $R_U$ in this paper and resulting in the following two non-trivial equations:

$$8\pi G\rho = \frac{3}{R^2}\left(kc^2 + \dot{R}^2\right) - \Lambda \qquad (2.11.38)$$

$$\frac{8\pi G}{c^2}p = -\frac{2\ddot{R}}{R} - \frac{\dot{R}^2}{R^2} - \frac{kc^2}{R^2} + \Lambda \qquad (2.11.39)$$

and as per Harrison[122] & D'Inverno[32], the equation 2.11.39 is integrated with constant of integration provided by equation 2.11.38.

$$\dot{R}^2 = \frac{8\pi G\rho R^2}{3} + \frac{\Lambda R^2}{3} - kc^2 \qquad (2.11.40)$$

The Einstein-de Sitter universe has $k = 0,\ \Lambda = 0$ and $\Omega = 1$.

$$\dot{R}^2 = \frac{8\pi G\rho R^2}{3} \qquad (2.11.41)$$

or

$$\frac{\dot{R}^2}{R^2} = \left(\frac{\dot{R}}{R}\right)^2 = \frac{8\pi G\rho}{3} \qquad (2.11.42)$$

That is by definition

$$H^2 = \frac{8\pi G\rho}{3} \qquad (2.11.43)$$

and the universe density $\rho_U$

$$\rho = \frac{3}{8\pi}\frac{H^2}{G} \qquad (2.11.44)$$

Given that:

$$\frac{H_{UG}}{H_U} = \left(\frac{G}{G_U}\right)^{\frac{1}{2}} \qquad (2.11.45)$$

then equations 2.11.38 and 2.11.39 dimensionally are identified with the following equations when $k = 0$, $\Lambda = 0$ and trisine volume factor $cavity/chain = 3/2$.

$$8\pi G\frac{3}{2}\rho_U = \frac{3}{B^2}\left(kc^2 + \left(H_{UG}B\right)^2\right) - \Lambda \qquad (2.11.46)$$

$$\frac{8\pi G}{c^2}\frac{3}{2}\frac{m_T c^2}{cavity} = -\frac{2\left(BH_{UG}^2\right)}{B} - \frac{\left(BH_{UG}\right)^2}{B^2} - \frac{kc^2}{B^2} + \Lambda \qquad (2.11.47)$$

And also equivalent to:

$$8\pi G_U\frac{3}{2}\rho_U = \frac{3}{B^2}\left(kc^2 + \left(H_U B\right)^2\right) - \Lambda \qquad (2.11.48)$$

and

$$8\pi G_U\frac{3}{2}\rho_U = 3H_U^2 \qquad (2.11.49)$$

$$\frac{8\pi G_U}{c^2}\frac{3}{2}\frac{m_T c^2}{cavity} = -\frac{2\left(BH_U^2\right)}{B} - \frac{\left(BH_U\right)^2}{B^2} - \frac{kc^2}{B^2} + \Lambda \qquad (2.11.50)$$

and

$$\frac{8\pi G_U}{c^2}\frac{3}{2}\frac{m_T c^2}{cavity} = -3H_U^2 \qquad (2.11.51)$$

but what is the $G$ and $H$ and $\rho_U$ proper scaling? Equation 2.11.37 and 2.11.38 reduces to several expressions of this $\rho_U$ scaling within the context of redshift '$z$' with a special note to the 2/8 factor versus 3/8 factor indicating a result applicable to only 2/3 (or *chain/cavity* volume ratio) of tessellated 3D space assumed to be dark energy. The remaining 1/3 of tessellated space is assumed to be related to dark matter.

$$\rho_U\left(1+z\right)^3 = \frac{2}{8\pi}\frac{\left(H_U\left(1+z\right)^3\right)^2}{G_U\left(1+z\right)^3} \qquad (2.11.52)$$

$$\rho_U\left(1+z\right)^3 = \frac{2}{8\pi}\frac{\left(H_{UG}\left(1+z\right)^{\frac{3}{2}}\right)^2}{G\left(1+z\right)^0} \qquad (2.11.53)$$

$$\rho_U\left(1+z\right)^3 = \frac{m_t\left(1+z\right)^0}{cavity\left(1+z\right)^{-3}} \qquad (2.11.54)$$

$$\rho_U\left(1+z\right)^3 = \frac{3}{4\pi}\frac{M_U\left(1+z\right)^{-6}}{\left(R_U\left(1+z\right)^{-3}\right)^3} = \frac{M_U\left(1+z\right)^{-6}}{V_U\left(1+z\right)^{-9}} \qquad (2.11.55)$$

$$\rho_U\left(1+z\right)^3 = \frac{c}{\pi^2\left(\hbar\left(1+z\right)^0\right)^2}H_U\left(1+z\right)^3\left(m_t\left(1+z\right)^0\right)^3 \qquad (2.11.56)$$

As the '$z$+1' redshift factors indicate with their zero summation exponents, $G_U$ approaches $G$ at the present time from larger values at the Big Bang and $H_{UG}$ approaches $H_U$ at the present time from smaller values at the Big Bang and $1/H_{UG} = 1/H_U$ = 13.7 billion years at the present time and $m_T$ represents a constant mass 1.003E-25 gram over all time contained in each cavity (~ 15,000 $cm^3$ at present) tessellating volume $R^3$ in accordance with the CPT theory and each *cavity* having a conservation of energy and momentum 'elastic' property and the universe mass $M_U$ increasing from the Big Bang and the universe density $\rho_U$ is equated to these changing conditions.

Within these scaling conditions, the university density $\rho_U$ scaling condition 2.11.52 and 2.11.45.

$$\rho_U\left(1+z\right)^3 = \frac{2}{8\pi}\frac{\left(H_{UG}\left(1+z\right)^{\frac{3}{2}}\right)^2}{G\left(1+z\right)^0} \qquad (2.11.57)$$

represents Big Bang modeling with constant Newtonian G as conventionally modeled represents the universe 'dark energy'. This is confirmed by the conventional CMBR temperature $T_b$ universe first light representation at $z$ = 1100:

$$T_b\left(1+z\right) = 3000K \text{ at } \frac{1}{H_{UG}\left(1+z\right)^{\frac{3}{2}}} = 376{,}000 \text{ years}$$

The universe expands $R_U$ at the speed of light $c$ while tessellating with *cavities* from within.

$$R_U\left(1+z\right)^{-3} = \frac{\sqrt{3}c}{H_U\left(1+z\right)^3} \qquad (2.11.58)$$

$$cavity\left(1+z\right)^{-3} = \frac{4\pi}{3}\frac{m_T\left(1+z\right)^0\left(R_U\left(1+z\right)^{-3}\right)^3}{M_U\left(1+z\right)^{-6}} \qquad (2.11.59)$$

with a universal Machian mass $M_U$ and Schwarzschild radius $R_U$ type gravitational energy with respect to all test masses $m$.

$$mc^2 = G_U\left(1+z\right)^3\frac{mM_U\left(1+z\right)^{-6}}{R_U\left(1+z\right)^{-3}} \qquad (2.11.60)$$

with energy $hH_U$ related to nuclear dimensions.

$$h\left(1+z\right)^0 H_U\left(1+z\right)^3 = G_U\left(1+z\right)^3\frac{\left(m_T\left(1+z\right)^0\right)^2}{nuclear\ radius\left(1+z\right)^0} \qquad (2.11.61)$$

with the constant mass $m_T$ (considered to be the Weinberg mass[99]) related to $\rho_U$ and $H_U$ (density condition 2.11.44)

$$\left(m_T\left(1+z\right)^0\right)^3 = \frac{\pi^2\left(\hbar\left(1+z\right)^0\right)^2}{c}\frac{\rho_U\left(1+z\right)^3}{H_U\left(1+z\right)^3} \qquad (2.11.62)$$

and $$\frac{M_U\left(1+z\right)^{-6}}{\left(R_U\left(1+z\right)^{-3}\right)^2} = \text{constant}(.599704\ \text{g/cm}^3) \qquad (2.11.64)$$

indicating a continued universe synthesis. (no inflationary period required) and a decelerating expanding universe from *cavity* reference. But horizon $R_U$ appears to be accelerating away from an observer immersed in the *cavity* tessellation.

$$\left(cavity\left(1+z\right)^{-3}\right)^{\frac{1}{3}} H_U\left(1+z\right)^3 = v\left(1+z\right)^2 \qquad (2.11.65)$$

and all consistent with Einstein stress energy tensor :

$$G_{\mu\nu} = \frac{8\pi G_U}{c^4} T_{\mu\nu} \qquad (2.11.66)$$

## 2.12 Superconducting Resonant Variance At T/Tc

Establish Gap and Critical Fields as a Function of $T/T_c$ using Statistical Mechanics [19]

$$z_s = e^{\frac{-\hbar^2 K_B^2}{2m_e k_b T_c}\frac{chain}{cavity}} = e^{-\frac{2}{3}} \qquad (2.12.1)$$

$$cavity = \frac{3\pi^3\hbar^3}{6^{\frac{1}{2}} m_T^{\frac{3}{2}}\left(\frac{B}{A}\right)_T k_b^{\frac{3}{2}} T_c^{\frac{3}{2}}} \qquad (2.12.2)$$

$$z_t = \frac{\left(2\pi m_T k_b T_c\right)^{\frac{3}{2}} cavity}{8\pi^3\hbar^3} \overset{1.3\%}{\approx} \left(\frac{T}{T_c}\right)^{\frac{3}{2}} \qquad (2.12.3)$$

$$k_t = \frac{z_t}{z_s} e^{\frac{k_b T_c}{k_b T}} = \left(\frac{T}{T_c}\right)^{\frac{3}{2}} e^{\frac{2}{3}} e^{-\frac{T_c}{T}} = \left(\frac{T}{T_c}\right)^{\frac{3}{2}} e^{\left(\frac{2}{3}-\frac{T_c}{T}\right)} \qquad (2.12.4)$$

$$k_s = \frac{2}{3} - k_t = \frac{chain}{cavity} - k_t \qquad (2.12.5)$$

$$\Delta_T = 1 - \frac{1}{3}\frac{T}{T_c}\frac{1}{\ln\left(\frac{1}{k_t}\right)} = 1 + \frac{1}{3}\frac{T}{T_c}\frac{1}{\ln\left(k_t\right)} \qquad (2.12.6)$$

$$H_{T1}^2 = H_o^2\left[1 - \frac{2}{3}\pi^2\left(\frac{T}{\pi e^{-Euler} T_c}\right)^2\right] \quad \text{at} \quad \frac{T}{T_c} \sim 0 \qquad (2.12.7)$$

$$H_{T2}^2 = H_o^2 k_s \quad \text{at} \quad \frac{T}{T_c}\ \text{near}\ 1 \qquad (2.12.8)$$

$$AvgH_{cT}^2 = \left[\text{if } H_{cT1}^2 > H_{cT2}^2 \text{ then } H_{cT1}^2 \text{ else } H_{cT2}^2\right] \qquad (2.12.9)$$

## 2.13 Superconductor Resonant Energy Content

Table 2.13.1 presents predictions of superconductivity energy content in various units that may be useful in anticipating the uses of these materials.

**Table 2.13.1** Trisine Resonant Energy/Volume $\left(m_t c^2 / cavity\right)$

| Condition | 1.2 | 1.3 | 1.6 | 1.9 | 1.10 | 1.11 |
| --- | --- | --- | --- | --- | --- | --- |
| $T_a\left({}^0K\right)$ | 6.53E+11 | 6.53E+11 | 6.53E+11 | 6.53E+11 | 6.53E+11 | 6.53E+11 |
| $T_b\left({}^0K\right)$ | 9.05E+14 | 5.48E+13 | 9.22E+08 | 1.10E+07 | 2,868 | 2.723 |
| $T_{br}\left({}^0K\right)$ | 3.30E+12 | 5.45E+13 | 3.24E+18 | 2.72E+20 | 1.04E+24 | 1.10E+27 |
| $T_c\left({}^0K\right)$ | 8.96E+13 | 3.28E+11 | 93.0 | 0.0131 | 8.99E-10 | 8.10E-16 |
| $T_{cr}\left({}^0K\right)$ | 1.19E+09 | 3.25E+11 | 1.15E+21 | 8.11E+24 | 1.19E+32 | 1.32E+38 |
| $z_R+1$ | 3.67E+43 | 8.14E+39 | 3.89E+25 | 6.53E+19 | 1.17E+09 | 1.00E+00 |
| $z_B+1$ | 3.32E+14 | 2.01E+13 | 3.39E+08 | 4.03E+06 | 1.05E+03 | 1.00E+00 |
| $Age_{UG}\left(sec\right)$ | 7.14E-05 | 4.79E-03 | 6.94E+04 | 5.36E+07 | 1.27E+13 | 4.33E+17 |
| $Age_{UGv}\left(sec\right)$ | 4.59E-10 | 1.25E-07 | 4.42E+02 | 3.13E+06 | 1.22E+13 | 4.33E+17 |
| $Age_U\left(sec\right)$ | 1.18E-26 | 5.31E-23 | 1.11E-08 | 6.63E-03 | 3.70E+08 | 4.33E+17 |
| erg/cm$^3$ | 2.10E+35 | 4.66E+31 | 2.23E+17 | 3.74E+11 | 6.69E+00 | 5.73E-09 |
| joule/liter | 2.10E+31 | 4.66E+27 | 2.23E+13 | 3.74E+07 | 6.69E-04 | 5.73E-13 |
| BTU/ft$^3$ | 5.65E+29 | 1.25E+26 | 5.97E+11 | 1.00E+06 | 1.80E-05 | 1.54E-14 |
| kwhr/ft$^3$ | 1.65E+26 | 3.67E+22 | 1.75E+08 | 2.94E+02 | 5.26E-09 | 4.50E-18 |
| hp-hr/gallon | 2.97E+25 | 6.57E+21 | 3.14E+07 | 5.27E+01 | 9.43E-10 | 8.07E-19 |
| gasoline equivalent | 5.66E+23 | 1.25E+20 | 5.98E+05 | 1.00E+00 | 1.80E-11 | 1.54E-20 |

Table 2.13.2 represents energy level of trisine energy states and would be key indicators material superconductor characteristics at a particular critical temperature. The BCS gap is presented as a reference energy.

**Table 2.13.2** The energies (erg) and momenta (g cm/sec) associated with trisine wave vectors are presented. These energies are computed by the equation $\left(\hbar^2/2m_t\right)K^2$ and and momentum ($\left(\hbar K\right)$. The BCS gap (erg) is computed by the equation $\left(\pi/\exp\left(Euler\right)\right)\left(\hbar^2/2m_T\right)K_B^2$.

| Condition | 1.2 | 1.3 | 1.6 | 1.9 | 1.10 | 1.11 |
| --- | --- | --- | --- | --- | --- | --- |
| $T_a\left({}^0K\right)$ | 6.53E+11 | 6.53E+11 | 6.53E+11 | 6.53E+11 | 6.53E+11 | 6.53E+11 |
| $T_b\left({}^0K\right)$ | 9.05E+14 | 5.48E+13 | 9.22E+08 | 1.10E+07 | 2,868 | 2.723 |
| $T_{br}\left({}^0K\right)$ | 3.30E+12 | 5.45E+13 | 3.24E+18 | 2.72E+20 | 1.04E+24 | 1.10E+27 |
| $T_c\left({}^0K\right)$ | 8.96E+13 | 3.28E+11 | 93.0 | 0.0131 | 8.99E-10 | 8.10E-16 |
| $T_{cr}\left({}^0K\right)$ | 1.19E+09 | 3.25E+11 | 1.15E+21 | 8.11E+24 | 1.19E+32 | 1.32E+38 |
| $z_R+1$ | 3.67E+43 | 8.14E+39 | 3.89E+25 | 6.53E+19 | 1.17E+09 | 1.00E+00 |
| $z_B+1$ | 3.32E+14 | 2.01E+13 | 3.39E+08 | 4.03E+06 | 1.05E+03 | 1.00E+00 |

| $Age_{UG}(sec)$ | 7.14E-05 | 4.79E-03 | 6.94E+04 | 5.36E+07 | 1.27E+13 | 4.33E+17 |
|---|---|---|---|---|---|---|
| $Age_{UGv}(sec)$ | 4.59E-10 | 1.25E-07 | 4.42E+02 | 3.13E+06 | 1.22E+13 | 4.33E+17 |
| $Age_{U}(sec)$ | 1.18E-26 | 5.31E-23 | 1.11E-08 | 6.63E-03 | 3.70E+08 | 4.33E+17 |
| $K_B^2$ | 1.24E-02 | 4.53E-05 | 1.28E-14 | 1.81E-18 | 1.24E-25 | 1.12E-31 |
| $K_{Dn}^2$ | 1.48E-02 | 5.43E-05 | 1.54E-14 | 2.17E-18 | 1.49E-25 | 1.34E-31 |
| $K_P^2$ | 1.65E-02 | 6.04E-05 | 1.71E-14 | 2.42E-18 | 1.66E-25 | 1.49E-31 |
| $K_{BCS}^2$ | 2.18E-02 | 7.99E-05 | 2.26E-14 | 3.20E-18 | 2.19E-25 | 1.97E-31 |
| $K_{Ds}^2 - K_{Dn}^2$ | 2.37E-02 | 8.67E-05 | 2.46E-14 | 3.47E-18 | 2.38E-25 | 2.14E-31 |
| $K_C^2 - K_B^2$ | 2.91E-02 | 1.07E-04 | 3.02E-14 | 4.27E-18 | 2.92E-25 | 2.63E-31 |
| $K_{Dn}^2 + K_P^2$ | 3.13E-02 | 1.15E-04 | 3.25E-14 | 4.59E-18 | 3.14E-25 | 2.83E-31 |
| $K_{Ds}^2$ | 3.85E-02 | 1.41E-04 | 4.00E-14 | 5.65E-18 | 3.87E-25 | 3.48E-31 |
| $K_C^2$ | 4.15E-02 | 1.52E-04 | 4.31E-14 | 6.08E-18 | 4.16E-25 | 3.75E-31 |
| $K_{Ds}^2 + K_B^2$ | 5.09E-02 | 1.86E-04 | 5.28E-14 | 7.46E-18 | 5.11E-25 | 4.60E-31 |
| $K_{Ds}^2 + K_{Dn}^2$ | 5.33E-02 | 1.95E-04 | 5.54E-14 | 7.82E-18 | 5.35E-25 | 4.82E-31 |
| $K_B^2 + K_C^2$ | 5.39E-02 | 1.97E-04 | 5.59E-14 | 7.90E-18 | 5.41E-25 | 4.87E-31 |
| $K_{Ds}^2 + K_P^2$ | 5.50E-02 | 2.01E-04 | 5.71E-14 | 8.07E-18 | 5.52E-25 | 4.98E-31 |
| $K_{Dn}^2 + K_C^2$ | 5.63E-02 | 2.06E-04 | 5.85E-14 | 8.26E-18 | 5.65E-25 | 5.09E-31 |
| $K_{Ds}^2 + K_C^2$ | 8.00E-02 | 2.93E-04 | 8.30E-14 | 1.17E-17 | 8.03E-25 | 7.24E-31 |
| $K_A^2$ | 2.80E-01 | 1.03E-03 | 2.91E-13 | 4.11E-17 | 2.81E-24 | 2.53E-30 |
| $K_B$ | 4.98E-14 | 3.01E-15 | 5.08E-20 | 6.03E-22 | 1.58E-25 | 1.50E-28 |
| $K_{Dn}$ | 5.45E-14 | 3.30E-15 | 5.56E-20 | 6.60E-22 | 1.73E-25 | 1.64E-28 |
| $K_P$ | 5.75E-14 | 3.48E-15 | 5.86E-20 | 6.96E-22 | 1.82E-25 | 1.73E-28 |
| $K_{BCS}$ | 6.62E-14 | 4.00E-15 | 6.74E-20 | 8.01E-22 | 2.10E-25 | 1.99E-28 |
| $K_{Ds} - K_{Dn}$ | 3.34E-14 | 2.02E-15 | 3.40E-20 | 4.04E-22 | 1.06E-25 | 1.00E-28 |
| $K_C - K_B$ | 4.14E-14 | 2.51E-15 | 4.22E-20 | 5.02E-22 | 1.31E-25 | 1.25E-28 |
| $K_{Dn} + K_P$ | 1.12E-13 | 6.78E-15 | 1.14E-19 | 1.36E-21 | 3.55E-25 | 3.37E-28 |
| $K_{Ds}$ | 8.79E-14 | 5.32E-15 | 8.96E-20 | 1.06E-21 | 2.78E-25 | 2.64E-28 |
| $K_C$ | 9.12E-14 | 5.52E-15 | 9.30E-20 | 1.10E-21 | 2.89E-25 | 2.74E-28 |
| $K_{Ds} + K_B$ | 1.38E-13 | 8.33E-15 | 1.40E-19 | 1.67E-21 | 4.36E-25 | 4.14E-28 |
| $K_{Ds} + K_{Dn}$ | 1.42E-13 | 8.62E-15 | 1.45E-19 | 1.72E-21 | 4.51E-25 | 4.28E-28 |
| $K_B + K_C$ | 1.41E-13 | 8.53E-15 | 1.44E-19 | 1.71E-21 | 4.47E-25 | 4.24E-28 |
| $K_{Ds} + K_P$ | 1.45E-13 | 8.80E-15 | 1.48E-19 | 1.76E-21 | 4.61E-25 | 4.37E-28 |
| $K_{Dn} + K_C$ | 1.46E-13 | 8.82E-15 | 1.49E-19 | 1.76E-21 | 4.62E-25 | 4.38E-28 |
| $K_{Ds} + K_C$ | 1.79E-13 | 1.08E-14 | 1.83E-19 | 2.17E-21 | 5.67E-25 | 5.39E-28 |
| $K_A$ | 2.37E-13 | 1.43E-14 | 2.41E-19 | 2.87E-21 | 7.51E-25 | 7.13E-28 |
| $K_{Br}^2$ | 1.64E-07 | 4.49E-05 | 1.58E+05 | 1.12E+09 | 1.64E+16 | 1.82E+22 |
| $K_{Dnr}^2$ | 1.97E-07 | 5.38E-05 | 1.90E+05 | 1.34E+09 | 1.96E+16 | 2.18E+22 |
| $K_{Pr}^2$ | 2.19E-07 | 5.98E-05 | 2.11E+05 | 1.49E+09 | 2.18E+16 | 2.42E+22 |
| $K_{BCSr}^2$ | 2.90E-07 | 7.91E-05 | 2.79E+05 | 1.98E+09 | 2.89E+16 | 3.20E+22 |
| $K_{Dsr}^2 - K_{Dnr}^2$ | 3.15E-07 | 8.59E-05 | 3.03E+05 | 2.15E+09 | 3.13E+16 | 3.48E+22 |
| $K_{Cr}^2 - K_{Br}^2$ | 3.87E-07 | 1.06E-04 | 3.72E+05 | 2.64E+09 | 3.85E+16 | 4.27E+22 |
| $K_{Dnr}^2 + K_{Pr}^2$ | 4.16E-07 | 1.14E-04 | 4.01E+05 | 2.84E+09 | 4.14E+16 | 4.60E+22 |
| $K_{Dsr}^2$ | 5.11E-07 | 1.40E-04 | 4.93E+05 | 3.49E+09 | 5.10E+16 | 5.65E+22 |
| $K_{Cr}^2$ | 5.51E-07 | 1.50E-04 | 5.31E+05 | 3.76E+09 | 5.49E+16 | 6.09E+22 |
| $K_{Dsr}^2 + K_{Br}^2$ | 6.76E-07 | 1.85E-04 | 6.51E+05 | 4.61E+09 | 6.73E+16 | 7.47E+22 |
| $K_{Dsr}^2 + K_{Dnr}^2$ | 7.08E-07 | 1.93E-04 | 6.82E+05 | 4.83E+09 | 7.06E+16 | 7.83E+22 |
| $K_{Br}^2 + K_{Cr}^2$ | 7.15E-07 | 1.95E-04 | 6.89E+05 | 4.88E+09 | 7.13E+16 | 7.91E+22 |
| $K_{Dsr}^2 + K_{Pr}^2$ | 7.31E-07 | 2.00E-04 | 7.04E+05 | 4.98E+09 | 7.28E+16 | 8.07E+22 |
| $K_{Dnr}^2 + K_{Cr}^2$ | 7.48E-07 | 2.04E-04 | 7.20E+05 | 5.10E+09 | 7.45E+16 | 8.27E+22 |
| $K_{Dsr}^2 + K_{Cr}^2$ | 1.06E-06 | 2.90E-04 | 1.02E+06 | 7.25E+09 | 1.06E+17 | 1.17E+23 |
| $K_{Ar}^2$ | 3.72E-06 | 1.02E-03 | 3.58E+06 | 2.54E+10 | 3.71E+17 | 4.11E+23 |
| $K_{Br}$ | 1.82E-16 | 3.00E-15 | 1.78E-10 | 1.50E-08 | 5.73E-05 | 6.04E-02 |
| $K_{Dnr}$ | 1.99E-16 | 3.28E-15 | 1.95E-10 | 1.64E-08 | 6.27E-05 | 6.61E-02 |
| $K_{Pr}$ | 2.10E-16 | 3.46E-15 | 2.06E-10 | 1.73E-08 | 6.62E-05 | 6.97E-02 |
| $K_{BCSr}$ | 2.41E-16 | 3.98E-15 | 2.37E-10 | 1.99E-08 | 7.61E-05 | 8.02E-02 |
| $K_{Dsr} - K_{Dnr}$ | 1.22E-16 | 2.01E-15 | 1.19E-10 | 1.00E-08 | 3.84E-05 | 4.04E-02 |
| $K_{Cr} - K_{Br}$ | 1.51E-16 | 2.49E-15 | 1.48E-10 | 1.25E-08 | 4.76E-05 | 5.02E-02 |
| $K_{Dnr} + K_{Pr}$ | 4.08E-16 | 6.75E-15 | 4.01E-10 | 3.37E-08 | 1.29E-04 | 1.36E-01 |
| $K_{Dsr}$ | 3.20E-16 | 5.29E-15 | 3.14E-10 | 2.65E-08 | 1.01E-04 | 1.06E-01 |
| $K_{Cr}$ | 3.32E-16 | 5.49E-15 | 3.26E-10 | 2.75E-08 | 1.05E-04 | 1.11E-01 |
| $K_{Dsr} + K_{Br}$ | 5.02E-16 | 8.29E-15 | 4.93E-10 | 4.14E-08 | 1.58E-04 | 1.67E-01 |
| $K_{Dsr} + K_{Dnr}$ | 5.19E-16 | 8.58E-15 | 5.09E-10 | 4.29E-08 | 1.64E-04 | 1.73E-01 |
| $K_{Br} + K_{Cr}$ | 5.14E-16 | 8.49E-15 | 5.04E-10 | 4.24E-08 | 1.62E-04 | 1.71E-01 |
| $K_{Dsr} + K_{Pr}$ | 5.30E-16 | 8.76E-15 | 5.20E-10 | 4.38E-08 | 1.67E-04 | 1.76E-01 |
| $K_{Dnr} + K_{Cr}$ | 5.31E-16 | 8.78E-15 | 5.21E-10 | 4.39E-08 | 1.68E-04 | 1.77E-01 |
| $K_{Dsr} + K_{Cr}$ | 6.53E-16 | 1.08E-14 | 6.41E-10 | 5.39E-08 | 2.06E-04 | 2.17E-01 |
| $K_{Ar}$ | 8.64E-16 | 1.43E-14 | 8.48E-10 | 7.13E-08 | 2.73E-04 | 2.87E-01 |

## 2.14 Superconductor Resonant Gravitational Energy

As per reference[25], a spinning nonaxisymmetric object rotating about its minor axis with angular velocity $\omega$ will radiate gravitational energy in accordance with equation 2.14.1

$$\dot{E}_g = \frac{32G}{5c^5} I_3^2 \varsigma^2 \omega^6 \qquad (2.14.1)$$

This expression has been derived based on weak field theory assuming that the object has three principal moments of inertia $I_1$, $I_2$, $I_3$, respectively, about three principal axes $a > b > c$ and $\varsigma$ is the ellipticity in the equatorial plane.

$$\varsigma = \frac{a-b}{\sqrt{ab}} \tag{2.14.2}$$

Equations 2.14.1 with 2.14.2 were developed from first principles (based on a quadrapole configuration) for predicting gravitational energy emitted from pulsars [25], but we use them to calculate the gravitational energy emitted from rotating Cooper CPT Charge conjugated pairs with resonant trisine mass ($m_T$) in the trisine resonant superconducting mode.

Assuming the three primary trisine axes *C>B>A* corresponding to a>b>c in reference [25] as presented above, then the trisine ellipticity (or measure of trisine moment of inertia) would be:

$$\varsigma = \frac{C-B}{\sqrt{CB}} = \frac{chain}{cavity}\frac{A}{B} = .2795 \tag{2.14.3}$$

Assuming a mass $(m_t)$ rotating around a nonaxisymmetric axis of resonant radius $B$ to establish moment of inertia $(I_3)$ then equation 2.14.1 becomes a resonant gravitational radiation emitter with no net energy emission outside of resonant *cavity*:

$$\frac{E_g}{time_\pm} = \frac{32G_U}{5}\frac{(\varsigma)^2\left(m_T B^2\right)^2}{v_{dx}^5}\left(\frac{2}{time_\pm}\right)^6 \tag{2.14.4}$$

where angular velocity $\omega = 2/time_\pm$, and replacing the speed of light $(c)$ with resonant consequential sub- and super-luminal de Broglie speed of light $(v_{dx})$, then equation 2.14.4 in terms of dimension *'B'* becomes the Newtonian gravitational energy relationship in equation 2.14.5:

$$E_g = -G_U\frac{m_T^2}{B} = -G_U\frac{m_T^2}{3\sqrt{3}B_{proton}k_m g_s^2} \tag{2.14.5}$$

Note the general characteristic length B related to the extended proton length. All this is consistent with the Virial Theorem.

Gravitational Potential Energy/2 = Kinetic Energy

$$\frac{1}{2}G_U m_T^2\frac{3}{\Delta x+\Delta y+\Delta z} = \frac{1}{2}m_r v_{dx}^2 \tag{2.14.6}$$

and

$$hH_U = \frac{cavity}{chain}G_U\frac{m_t^2}{B_p} \tag{2.14.7}$$

Equation 2.14.7 provides a mechanism explaining the CERN (July 4, 2012) observed Higgs particle) at 126.5 GeV as well as the astrophysical Fermi ~130 GeV [127]. The astrophysical ~130 GeV energy is sourced to a Trisine Condition 1.2 nucleosynthetic event with a fundamental $G_U$ gravity component.

$$g_s\frac{3}{4}\cos(\theta)G_U m_T^2\frac{1}{proton\ radius} = 125.27\ GeV \tag{2.14.8}$$

$$g_s\frac{3}{4}\cos(\theta)G_U m_T^2\frac{2\sin(\theta)}{B_p} = 125.27\ GeV \tag{2.14.9}$$

$$g_s\frac{3}{2}\sin(\theta)\cos(\theta)G_U m_t^2\frac{1}{B_p} = 125.27\ GeV$$

falling within the measured range

$$125.36 \pm 0.37\,(statistical\ uncertainty) \pm 0.18\,(systematic\ uncertainty) \tag{2.14.10}$$

$125.36 \pm 0.41\ GeV$ [149]

The Equation 2.14.7 is of the same form as 2.14.5 and in conjunction with equation 2.11.15 implies the following nuclear condition:

$$\hbar K_B\Big|_{\substack{T_c=8.95E13\\T_b=9.07E14}} = 8\pi\frac{chain}{cavity}m_t c \tag{2.14.11}$$

also:

$$\left\{\begin{matrix}\dfrac{1}{2\varepsilon_x k_m}m_T v_{dx}^2\\[2ex]\dfrac{chain}{cavity}\dfrac{1}{2\varepsilon_x k_m}\dfrac{1}{\varepsilon}\dfrac{e^2}{B_{CMBR}}\end{matrix}\right\}_{\substack{T_c=8.11E-16\\T_b=2.729}} = \frac{1}{2}m_r v_{dx}^2\Big|_{\substack{T_c=8.11E-16\\T_b=2.729}}$$

$$where:\quad 2\varepsilon_x k_m = \frac{2c^2}{v_{dx}^2}\Big|_{\substack{T_c=8.11E-16\\T_b=2.729}} \tag{2.14.12}$$

Where $m_T$ and $m_r$ are related as in Equation 2.1.25 and $\Delta x$, $\Delta y, \Delta z$ are Heisenberg Uncertainties as expressed in equation 2.1.20.

In the case of a superconductor, the radius R is the radius of curvature for pseudo particles described in this report as the variable B. All of the superconductor pseudo particles move in asymmetric unison, making this virial equation applicable to this particular engineered superconducting situation. This is consistent with Misner Gravitation [35] page 978 where it is indicated that gravitational power output is related to "power flowing from one side of a system to the other" carefully incorporating "only those internal power flows with a time-changing quadrupole moment". The assumption in this report is that a super current flowing through a trisine superconducting CPT lattice has the appropriate quadrupole moment at the dimensional level of *B*. The key maybe a linear back and forth resonant trisine system. Circular superconducting systems such as represented by Gravity Probe B gyroscopic elements indicate the non gravitational radiation characteristic of such circular systems reinforcing the null results of mechanical gyroscopes [21, 22, 23]. The density $\left(m_T/cavity\right)$ (7.96E-06 g/cm$^3$) associated with niobium super mass currents at $T_c$ = 9.4 K are subject to tidal oscillations(Appendix H) imposed by the universe dark energy at $T_c$ = 8.11E-16 K with its characteristic frequency of 3.38E-5 Hz (table 2.4.1) may fundamentally and harmonically contribute to observed gyro asymptotic polhode frequency anomalies associated with Gravity Probe B (see condition 1.8) (Alex Silbergleit, John Conklin, Stanford University[94]).

**Table 2.14.1** Gravity Probe B Asymptotic Gyros' Polhode Frequencies (as harmonics of present universe dark energy)

| | Asymptotic Polhode Period (hours) | Asymptotic Polhode Frequency (Hz) | Asymptotic Polhode Harmonic | Asymptotic Polhode Harmonic Deviation |
|---|---|---|---|---|
| Gyro 1 | 0.867 | 3.20E-04 | 9 | 1.05 |
| Gryo 3 | 1.529 | 1.82E-04 | 5 | 1.08 |
| Gyro 2 | 2.581 | 1.08E-04 | 3 | 1.06 |
| Gyro 4 | 4.137 | 6.71E-05 | 2 | 0.99 |
| Universe | 8.22 | 3.38E-05 | 1 | 1.00 |

Based on March 2017 Stefano Vitale's Twitter graph(extended mission data):

https://twitter.com/VitaleTrident

the 1, 1.5, 2, 2.5 and 3d harmonics 33.8 microHz clearly seen in the differentiated representation of the March 2017 extended mission data.

The 33.8 $\mu Hz$ value is dimensionally predicted in Condition 1.11 green as an indicator of present universe mass($M_U$) (3.03E56 g at critical density $\left(m_T/cavity\right)$ 6.37E30 g/cm$^3$) coherent oscillation

($1/time$ =1/29,600 sec = 3.38E10-5 Hz)

defining a corresponding present universe zero point temperature ($\pi\hbar v/k_b$ = 8.10E-16 K) as a dark vacuum energy heat sink representing 73% of the universe.

The dark matter (23% of the universe probably as baryonic meter sized chunks) is in near thermal equilibrium with this dark energy heat sink and is necessarily electromagnetically undetectable because of its extremely cold state that is quantum mechanically distinct and a necessary complement to the Cosmic Microwave Background Radiation( CMBR) at 2.7 K.

This model extends back to highly energetic Big Bang nucleosynthesis requiring new theories for the Big Bang nucleosynthesis that may be necessary in light of Large Hadron Collider(LHC) primordial quark soup studies.

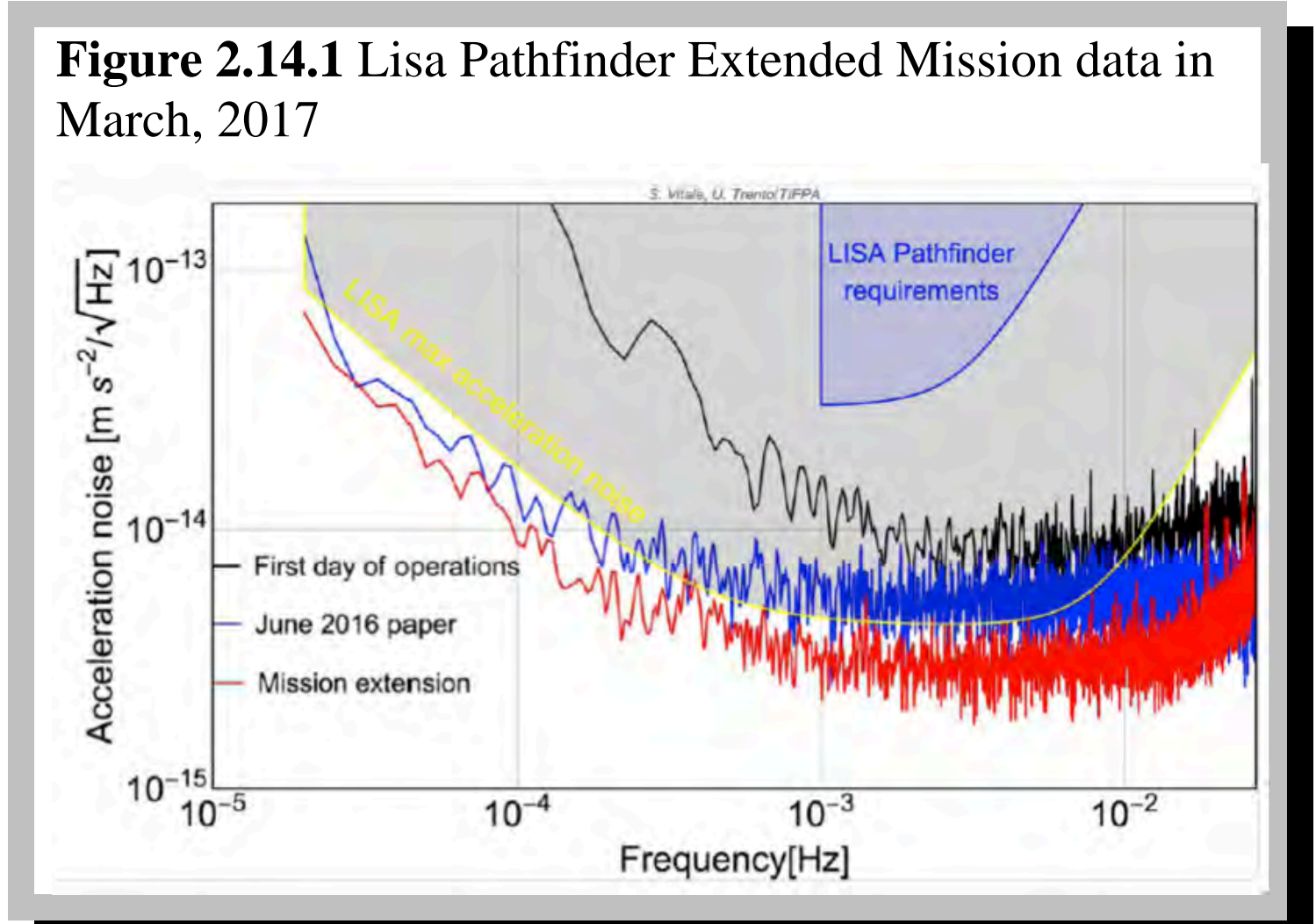

**Figure 2.14.1** Lisa Pathfinder Extended Mission data in March, 2017

These following ten(10) Mach universe representations are consistent with Trisine geometry.

Mach0: The universe, as represented by the average motion of distant galaxies, does not appear to rotate relative to local inertial frames.

Mach1: Newton's gravitational constant G is a dynamical field. (It changes with redshift z)

Mach2: An isolated body in otherwise empty space has no inertia.

Mach3: Local inertial frames are affected by the cosmic motion and distribution of matter.

Mach4: The universe is spatially closed.

Mach5: The total energy, angular and linear momentum of the universe are zero.

Mach6: Inertial mass is affected by the global distribution of matter.

Mach7: If you take away all matter, there is no more space.

Mach8:

$1 = 4\pi \dfrac{G_U T^2}{\rho_U}$ where T is the Hubble time($1/H_U$) and consistent with the Trisine formulation.

$$\rho_U\left(1+z\right)^3 = \frac{2}{8\pi}\frac{\left(H_U\left(1+z\right)^3\right)^2}{G_U\left(1+z\right)^3}$$

Mach9: The theory contains no absolute elements.

Mach10: Overall rigid rotations and translations of a system are unobservable.

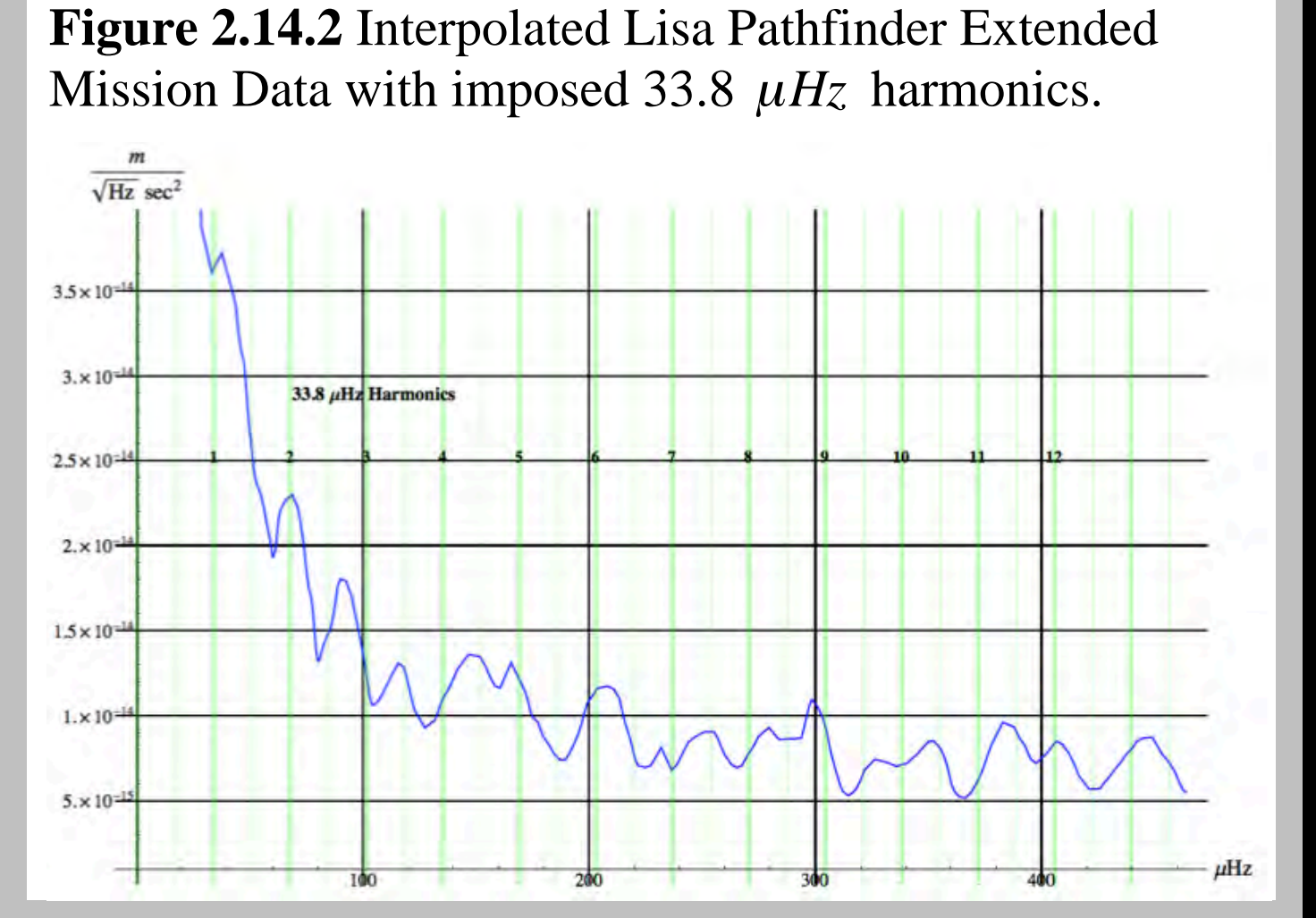

**Figure 2.14.2** Interpolated Lisa Pathfinder Extended Mission Data with imposed 33.8 $\mu Hz$ harmonics.

Differentiation more clearly exposes the 33.8 $\mu Hz$ 1, 1.5, 2, 2.5 and 3d harmonics.

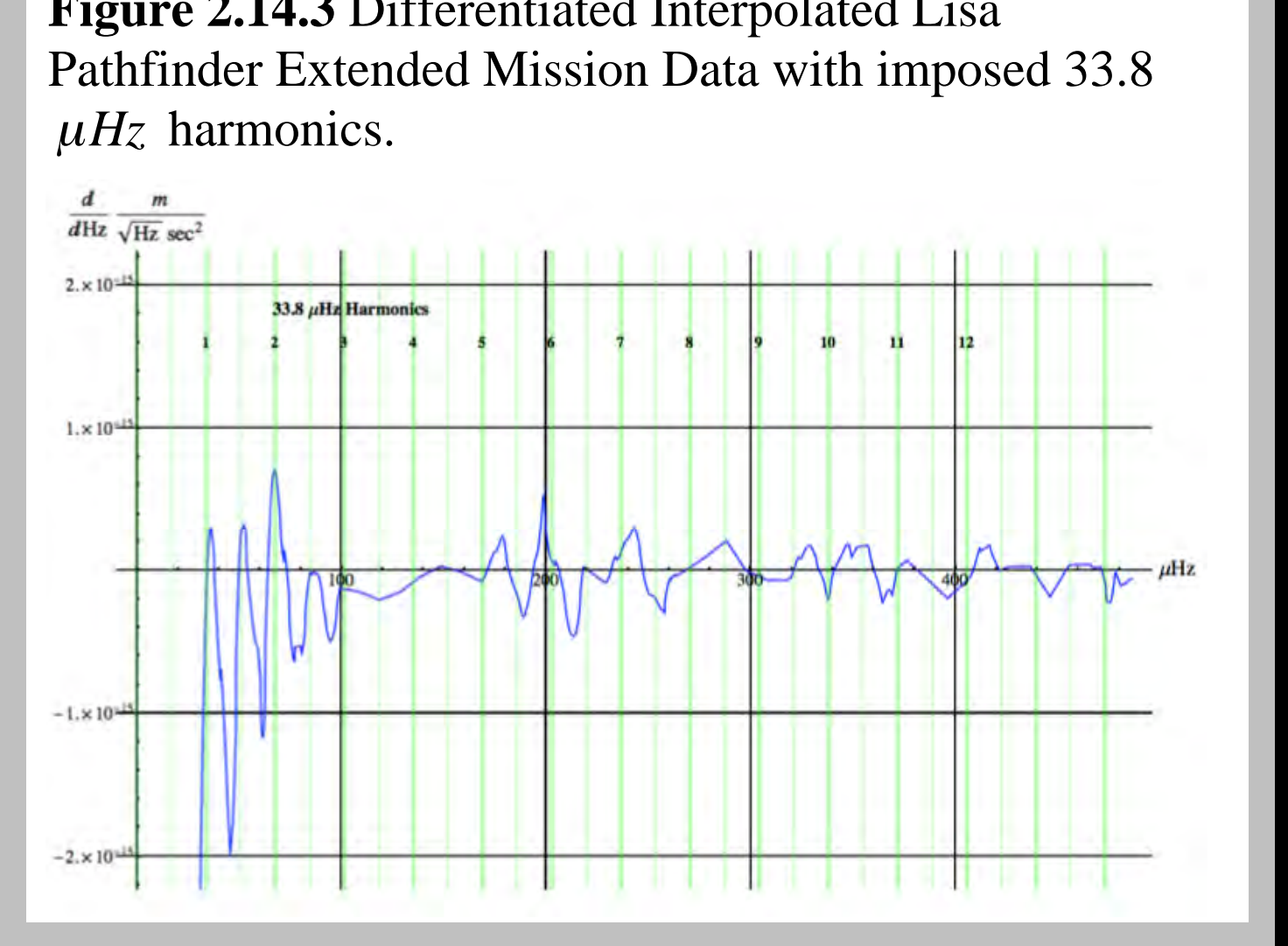

**Figure 2.14.3** Differentiated Interpolated Lisa Pathfinder Extended Mission Data with imposed 33.8 $\mu Hz$ harmonics.

The LISA Pathfinder expressed purpose[Ref 1] was to experimentally provide a platform to study gravitational waves in anticipation of a comprehensive LISA to be launched in approximately 10 years. The LISA Pathfinder performed that mission extremely well[Ref 2]. In addition to this central mission, the LISA Pathfinder in its near perfect Free-Fall state, was an excellent platform to observe other universal mass oscillations and in particular that associated with a 33.8 $\mu Hz$ oscillation and related frequencies $\left(\sim 1/29{,}600\ s \sim 1/8.22\ hr \sim 1/.342\ day\right)$ (Condition 1.11).

Ref 152, Figure 1 red oscillation data was digitized(Digitzeit) to visual limits, abscissa made linear, ordinate made logarithmic and then graphically(Figure 1) expressed in blue with overlaid 33.8 $\mu Hz$ harmonics and systematic brown baseline .
(All data analysis was executed with Stephen Wolfram's Mathematica version 11.3)

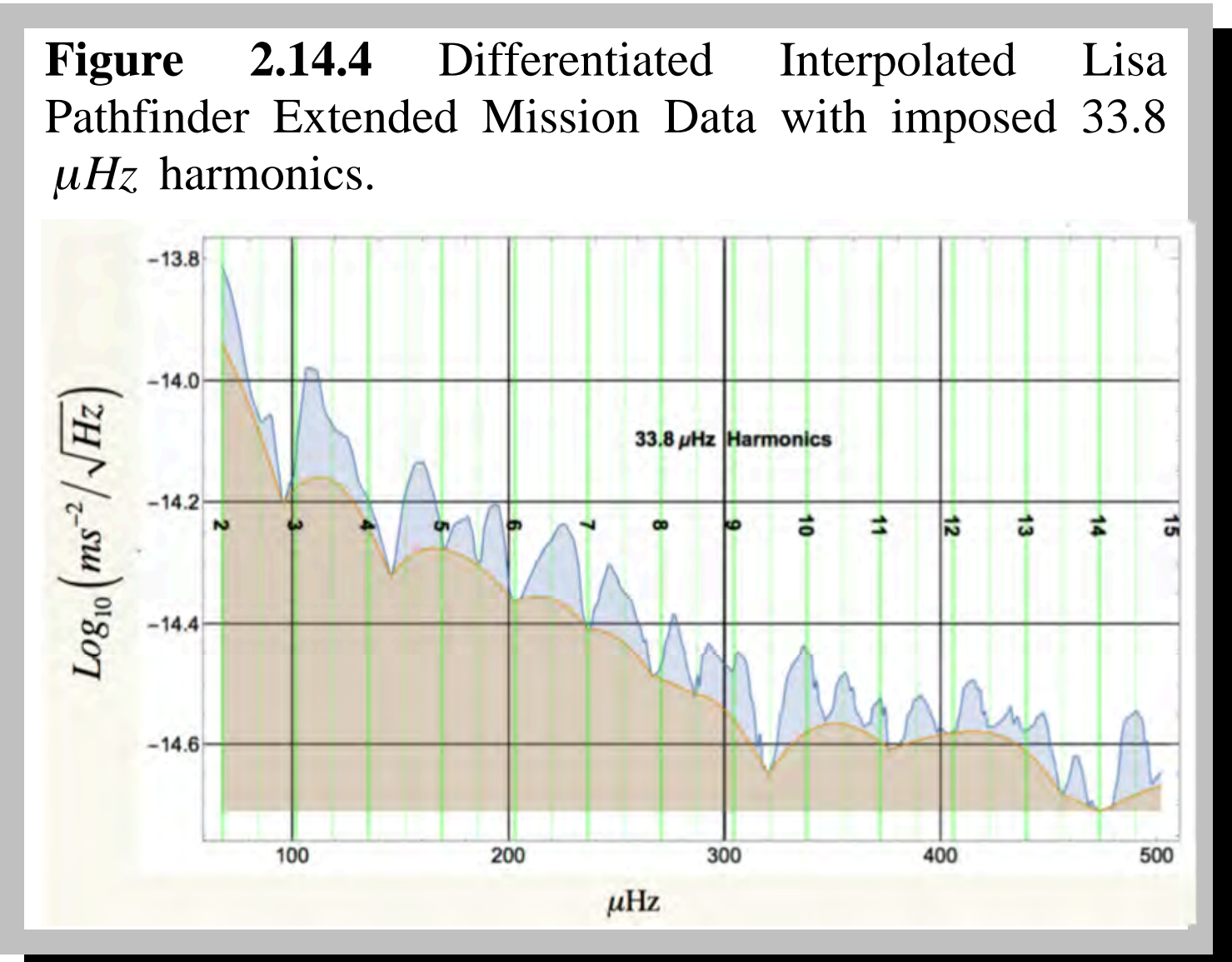

**Figure 2.14.4** Differentiated Interpolated Lisa Pathfinder Extended Mission Data with imposed 33.8 $\mu Hz$ harmonics.

The LISA Pathfinder systematic brown baseline is then subtracted from the oscillation data as graphically follows in Figure 2.14.5. An intuitive 33.8 $\mu Hz$ fit is observed necessitating a more quantitative Fourier Analysis.

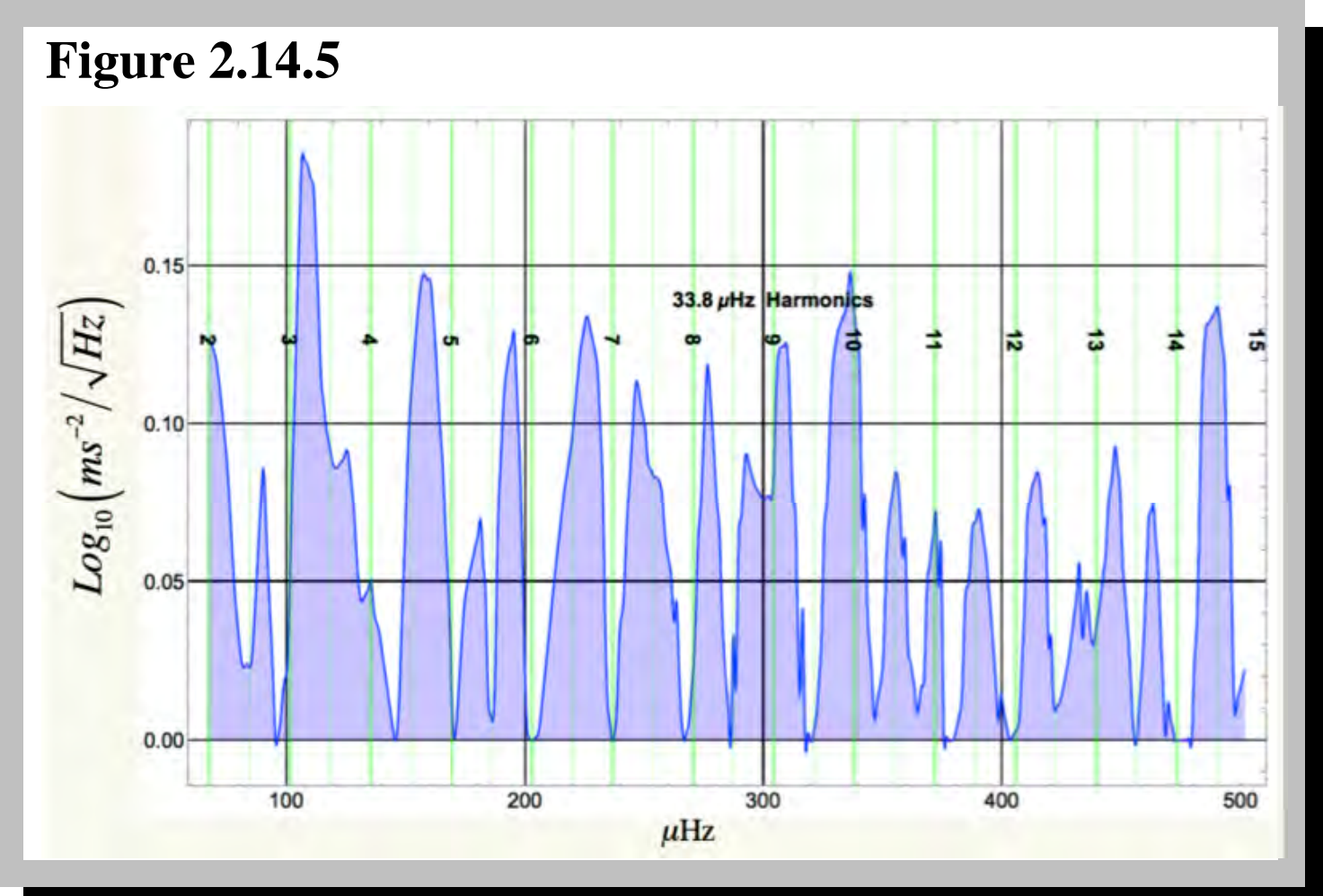

**Figure 2.14.5**

Notice the near uniform nature of these oscillations with the removed LISA Pathfinder systematic baseline. A further Fourier Discrete Sine Wave analysis indicates a predominate first 33.8 $\mu Hz$ Harmonic in Figure 2.14.5.

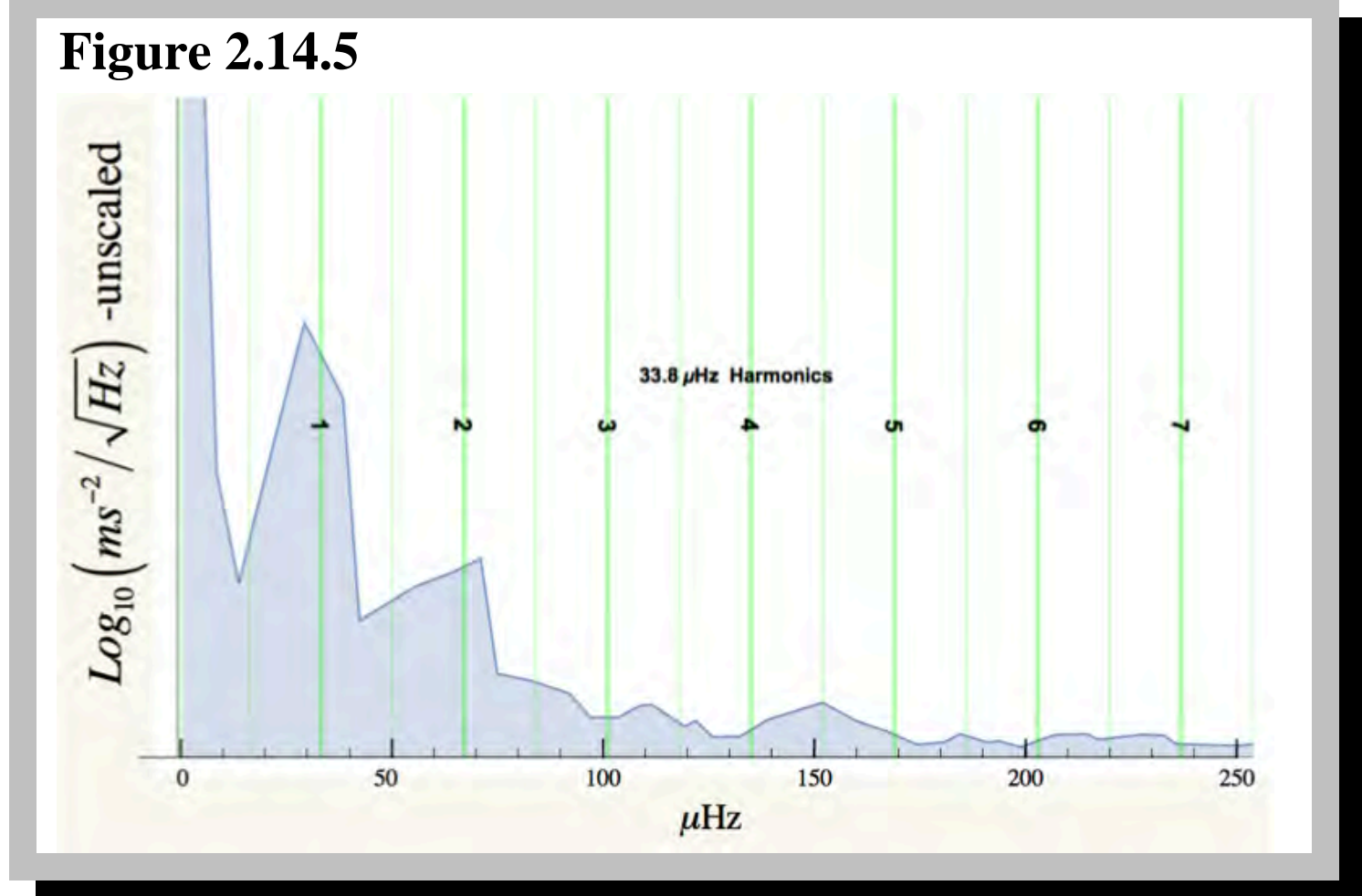

**Figure 2.14.5**

Conclusion:
In conjunction with The LISA Pathfinder testing the possibility of detecting gravitational waves from a space platform, it also demonstrated the capability for detected extremely low frequency 33.8 $\mu\text{Hz}\left(\sim 1/29,600\ s \sim 1/8.22\ hr \sim 1/.342\ day\right)$ oscillations indicative of a Machian universal source as evidenced in other observations such as:

1. Kuiper Belt object rotations -Vatican Advanced Technology Telescope
2. Asteroid rotations (not as good a correlation due to collisions
   possibly modifying their original 8 hr periods)
3. Lisa Pathfinder accelerometer oscillations (1/8 hrs)
   (A Machian oscillating universe)
4. Gravity Probe B anomalous gyro oscillations (1/8 hrs)
5. Solar oscillations (1/8 hrs)
6. Distant stellar oscillation observed by the Planck space craft
7. Several thousand RRC stars with 8 hr oscillation periods
8. 'Oumuamua has a rotation period of 8.1 hours
   https://arxiv.org/abs/1711.04927
9. The 8 hr period is dimensionally linked to a trinitarian space time
   at the quantum level.

Such a coherent space oscillation or frequency property $\left(frequency \sim 33.8\mu\text{Hz} \sim 1/29,600\ s \sim 1/8.22\ hr \sim 1/.342\ day\right)$ would necessarily define an extremely low temperature(T).

$$T = h\ frequency / k\ \sim 10^{-16} K$$

This extremely low temperature(T) would provide a 'dark energy' milieu in thermal equilibrium with dark matter that is quantum mechanically distinct from the Cosmic Microwave Background Radiation at 2.7 K (Condition 1.11),

## 2.15 Black Body Relationships[18]

There are many unique characteristics if the Planck Black Body curve as it was introduced over 100 years ago that have been used in Universe evolutionary processes.

The number '*N*' of possible frequencies in box of edge '*l*' in the range $\nu$ to $\nu + d\nu$ (The '2' accounts for both oscillating electric and magnetic fields) [19 page 468]:

$$2\,dN = \frac{2 \cdot 4\pi l^3 \nu^2}{c^3} d\nu \qquad (2.15.1)$$

Therefore:

$$2\,N = \frac{2 \cdot 4}{3} \frac{\pi \nu^3}{c^3} l^3 \qquad (2.15.2)$$

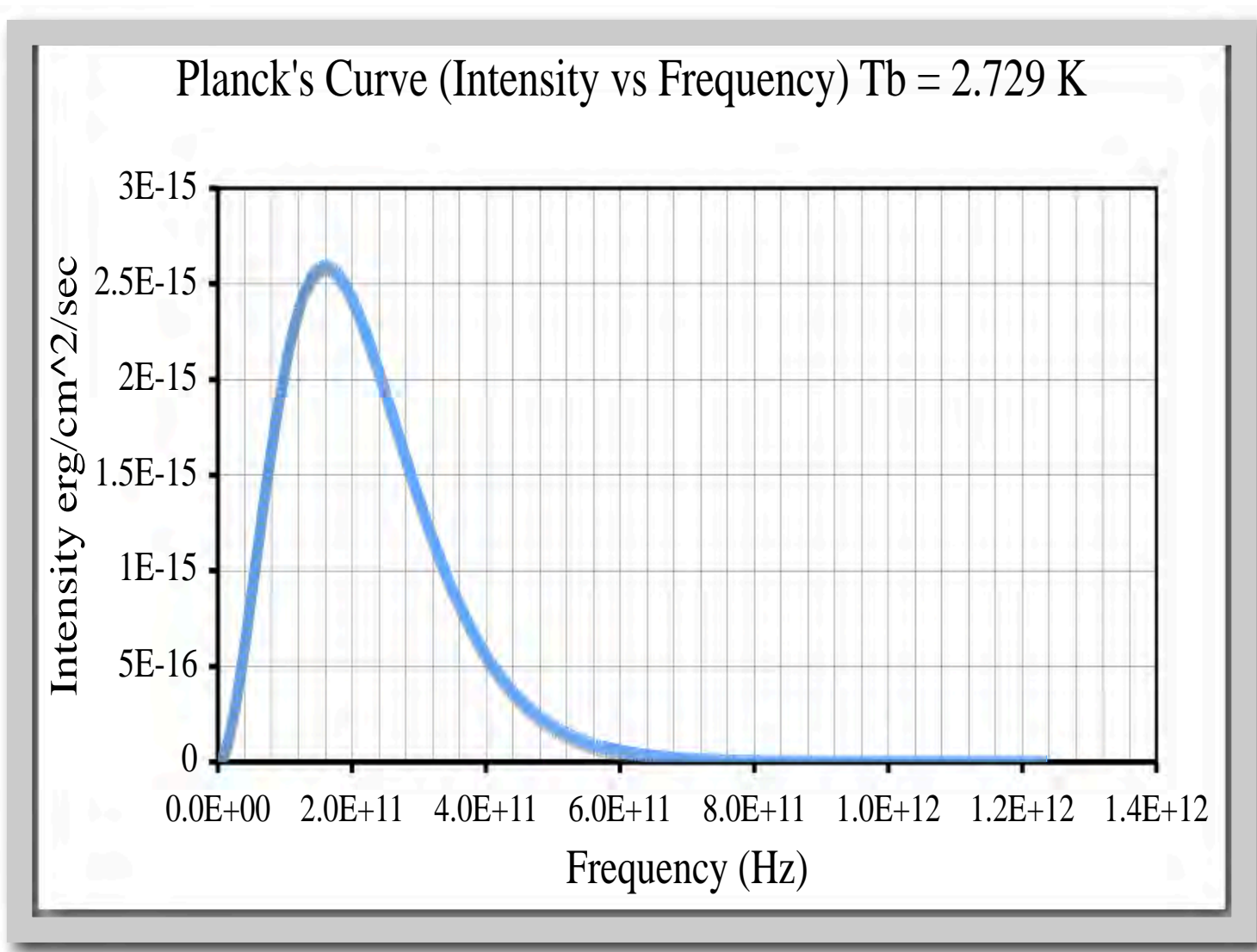

The energy $(2U)$ of a system consisting of N oscillators of frequency $(\nu)$:

$$2U = \frac{2h\nu}{e^{h\nu/k_b T_b} - 1} \qquad (2.15.3)$$

The energy $(U)$ of a system consisting of N oscillators of frequency $(\nu)$:

$$\begin{aligned} 2UdN &= \frac{2h\nu 2dN}{e^{h\nu/k_b T_b} - 1} \\ &= \frac{2h\nu}{e^{h\nu/k_b T_b} - 1} \frac{2 \cdot 4\pi l^3 \nu^2}{c^3} d\nu \\ &= \frac{2h\nu^3}{e^{h\nu/k_b T_b} - 1} \frac{2 \cdot 4\pi l^3}{c^3} d\nu \end{aligned} \qquad (2.15.4)$$

Or by rearranging in terms of energy density $(\rho_v)$

$$2\rho_v dv = \frac{2hv^3}{e^{hv/k_bT_b}-1}\frac{2\cdot 4\pi}{c^3}dv \qquad (2.15.5)$$

The energy volume density $(\rho_v)$ related to the surface emissivity $(K_v)$ which is the ratio of radiation emission $(E_v)$ over radiation absorption coefficient $(A_v)$ (The absorption coefficient $(A_v)$ is assumed to be 1).

$$2\rho_v = \frac{2\cdot 8\pi K_v}{c} = \frac{2\cdot 8\pi E_v}{c} \qquad (2.15.6)$$

$$\begin{gathered}\frac{2\cdot 8\pi K_v}{c}dv = \frac{2hv^3}{e^{hv/k_bT_b}-1}\frac{2\cdot 4\pi}{c^3}dv \\ or \\ \frac{2\cdot 8\pi E_v}{c}dv = \frac{2hv^3}{e^{hv/k_bT_b}-1}\frac{2\cdot 4\pi}{c^3}dv\end{gathered} \qquad (2.15.7)$$

And by rearranging:

$$2E_v dv = \frac{2}{c^2}\frac{hv^3}{e^{hv/k_bT_b}-1}dv \qquad (2.15.8)$$

The total radiation $(2E_v)$ is obtained by integrating over all frequencies $(v)$

$$2\int_0^\infty E_v\,dv = \frac{2h}{c^2}\int_0^\infty \frac{v^3}{e^{hv/k_bT_b}-1}dv \qquad (2.15.9)$$

The black body energy emission ( $2E_v$ or energy/area/time) is:

$$\begin{aligned}2E_v &= \frac{2\pi^4 k_b^4}{15c^2h^3}T_b^4 \\ &= \sigma_S T_b^4\end{aligned} \qquad (2.15.10)$$

$$\begin{aligned}2E_{vr} &= \frac{2\pi^4 k_b^4}{15c^2h^3}T_{br}^4 \\ &= \sigma_S T_{br}^4\end{aligned} \qquad (2.15.10r)$$

This black body radiation emission is through a solid angle and conversion to the total energy radiated from one side of a plane then $\sigma_S = \pi\sigma_S^{'}$.

The black body energy density ( $\rho_{BBenergy}$ or energy/volume) is:

$$\rho_{BBenergy} = \sigma_S T_b^4 \frac{4}{c} \qquad (2.15.11)$$

$$\rho_{BBrenergy} = \sigma_S T_{br}^4 \frac{4}{c} \qquad (2.15.11r)$$

The black body mass density ( $\rho_{BBmass}$ or mass/volume) is:

$$\rho_{BBmass} = \sigma_S T_b^4 \frac{4}{c^3} \qquad (2.15.12)$$

$$\rho_{BBrmass} = \sigma_S T_{br}^4 \frac{4}{c^3} \qquad (2.15.12r)$$

The black body maximum wave length $(\lambda_{BB\max})$ is:

$$E_v dv = E_{\lambda_b} d\lambda_b$$

$$|dv| = \frac{c}{\lambda_b^2}|d\lambda_b|$$

$$\begin{gathered}2E_{\lambda_b} d\lambda_b = \frac{2hc^2}{\lambda_b^5}\frac{d\lambda_b}{e^{\frac{hc}{k_bT_b\lambda}}-1} \\ x = \frac{ch}{\lambda_b k_b T_b} \qquad \lambda_b = \frac{ch}{x k_b T_b}\end{gathered} \qquad (2.15.13)$$

Finding the differential peak:

$$\frac{dE_{\lambda_b}}{d\lambda_b} = \frac{10hc^2}{\lambda_b^6\left(e^{\frac{hc}{k_bT_b\lambda}}-1\right)} - \frac{2h^2c^3e^{\frac{hc}{k_bT_b\lambda_b}}}{T_b\lambda_b^7k_b\left(e^{\frac{hc}{k_bT_b\lambda_b}}-1\right)^2} = 0 \qquad (2.15.14)$$

By reduction then

$$e^x(x-5)+5=0 \qquad x = 4.965364560 \qquad (2.15.15)$$

and

$$\lambda_{BB\max} = \frac{hc}{xk_bT_b} = \frac{hc}{4.96536456k_bT_b} \qquad (2.15.16)$$

$$\lambda_{BBr\max} = \frac{hc}{xk_bT_{br}} = \frac{hc}{4.96536456k_bT_{br}} \qquad (2.15.16r)$$

BlackBody most probable energy factor based on maximum(peak) wavelength $hc/\lambda_{BB\max} = 4.965364560\ k_b T_b$ and Wein displacement constant $= hc/\left(4.965364560\ k_b\right)$

$= 0.28977683\ cm\ K$

Differentiating 2.15.8 yields energy and frequency $(\nu)$ relationship.

$$\frac{dE_\nu}{d\nu} = \frac{6h\nu^2}{c^2\left(e^{\frac{h\nu}{k_b T_b}} - 1\right)} - \frac{2h^2\nu^3 e^{\frac{h\nu}{k_b T_b}}}{T_b c^2 k_b \left(e^{\frac{h\nu}{k_b T_b}} - 1\right)^2} = 0 \qquad (2.15.17)$$

$$x = \frac{h\nu}{k_b T_b} \qquad \nu = \frac{x k_b T_b}{h}$$

By reduction then:

$$e^x\left(x-3\right) + 3 = 0 \qquad x = 2.821439372 \qquad (2.15.18)$$

Then

$$\nu_{BB\max} = \frac{2.821439372 k_b T_b}{h} \qquad (2.15.19)$$

$$\nu_{BBr\max} = \frac{2.821439372 k_b T_{br}}{h} \qquad (2.15.19r)$$

BlackBody most probable energy factor based on maximum(peak) frequency $h\nu_{BB\max} = 2.821439372\ k_b T_b$

The velocity computed by $\nu_{BB\max}\lambda_{BB\max} = 1.70E+10$ cm/sec is less than the speed of light (.568224012c). This is important consideration in performing black body energy momentum calculations.

Now for average photon number density ($n$)

$$n d\nu = \frac{8\pi\nu^2 d\nu}{c^3\left(e^{\frac{h\nu}{k_b T_b}} - 1\right)} \qquad (2.15.20)$$

$$x = \frac{h\nu}{k_b T_b} \qquad \nu = \frac{k_b T_b x}{h} \qquad d\nu = \frac{k_b T_b dx}{h}$$

By change of variable and reduction then:

$$n d\nu = \frac{8\pi k_b^3 T_b^3 x^2 dx}{c^3 h^3\left(e^{\frac{h\nu}{k_b T_b}} - 1\right)} \qquad (2.15.21)$$

$$\int_0^\infty \frac{x^2}{\left(e^x - 1\right)} = 2.404113806$$

Then

$$\int_0^{2.356763057} \frac{x^2}{\left(e^x - 1\right)} = \int_{2.356763057}^{\infty} \frac{x^2}{\left(e^x - 1\right)} = \frac{2.404113806}{2} = 1.202056903 \qquad (2.15.22)$$

or

$$\rho_{BBphoton} = n = 8\pi 2.404113806\left(\frac{k_b T_b}{hc}\right)^3$$

$$\rho_{BBrphoton} = n = 8\pi 2.404113806\left(\frac{k_b T_{br}}{hc}\right)^3 \qquad (2.15.22r)$$

This means that half of photons (median photon density) are below x = 2.356763057 and half above. By similar analysis, one quarter(1/4) of photons (median photon density) are below x = 1.40001979 and three quarters(3/4) above and three quarters(3/4) of photons (median photon density) are below x = 3.63027042 and one quarter(1/4) above.

$$\nu_{BBphoton} = \frac{2.356763057 k_b T_b}{h} = \frac{\zeta(3) k_b T_b}{h} \qquad (2.15.23)$$

$$\nu_{BBrphoton} = \frac{2.356763057 k_b T_{br}}{h} = \frac{\zeta(3) k_b T_{br}}{h} \qquad (2.15.23r)$$

Now for average photon energy density:

$$\rho_\nu d\nu = \frac{8\pi\nu^3 d\nu}{c^3\left(e^{\frac{h\nu}{k_b T_b}} - 1\right)} \qquad (2.15.25)$$

$$x = \frac{h\nu}{k_b T_b} \qquad \nu = \frac{k_b T_b x}{h} \qquad d\nu = \frac{k_b T_b dx}{h}$$

By change of variable and reduction then:

$$\rho_\nu d\nu = \frac{8\pi k_b^4 T_b^4 x^3 dx}{c^3 h^3\left(e^x - 1\right)}$$

$$\int_0^\infty \frac{x^3 dx}{\left(e^x - 1\right)} = 6.493939402 = \frac{\pi^4}{15} \qquad (2.15.26)$$

Then

$$\int_{0}^{3.503018826} \frac{x^3 dx}{\left(e^x - 1\right)} = \int_{3.503018826}^{\infty} \frac{x^3 dx}{\left(e^x - 1\right)} \qquad (2.15.27)$$

$$= \frac{6.493939402}{2} = 3.246969701$$

This means that half of photons (average photon energy density) are below x = 3.503018826 and half above designating a characteristic black body frequency $\left(\nu_{BBphoton} = c/\lambda_{BBphoton}\right)$.

$$\nu_{BBphoton} = \frac{c}{\lambda_{BBphoton}} = \frac{3.503018826\, k_b T_b}{h} \qquad (2.15.28)$$

Then combining equations 2.15.21 and 2.15.24 results in a photon mean energy density.

$$\frac{\rho_\nu}{n} = \frac{\dfrac{8\pi k_b^4 T_b^4}{c^3 h^3} 6.493939402}{\dfrac{8\pi k_b^3 T_b^3}{c^3 h^3} 2.404113806} \qquad (2.15.29)$$

$$= 2.70117803\, k_b T_b = \frac{\pi^4}{30\zeta(3)} k_b T_b$$

$\rho_U (1+z)^3 = \rho_\nu (1+z)^4 = 1.62\text{E-}17 \text{ g/cm}^3$

$T_b = 37{,}200$ K $\quad z_B = 13{,}600 \quad Age_{UG} = 8{,}600$ years

The black body photons momentum $\left(p_{BBmomentum}\right)$ is:

$$p_{BBmomentum} = \frac{h}{\lambda_{BB\max}} \qquad (2.15.30)$$

The black body photons in the universe $\left(U\#\, photons\right)$ is:

$$U\#\, photons = \frac{\rho_{BBenergy}}{h f_{BB\max}} V_U \qquad (2.15.31)$$

$$U_r\#\, photons = \frac{\rho_{BBrenergy}}{h f_{BBr\max}} V_U \qquad (2.15.31r)$$

The universe viscosity due to black body photons $\left(\mu_{BB}\right)$ is:

$$\mu_{BB} = \sigma_S T_b^4 \frac{4}{c^2} \lambda_{BB\max} \qquad (2.15.32)$$

$$\mu_{BBr} = \sigma_S T_{br}^4 \frac{4}{c^2} \lambda_{BBr\max} \qquad (2.15.32r)$$

**Table 2.15.1** Universe Black Body parameters

| Condition | 1.2 | 1.3 | 1.6 | 1.9 | 1.10 | 1.11 |
|---|---|---|---|---|---|---|
| $T_a\left({}^0K\right)$ | 6.53E+11 | 6.53E+11 | 6.53E+11 | 6.53E+11 | 6.53E+11 | 6.53E+11 |
| $T_b\left({}^0K\right)$ | 9.05E+14 | 5.48E+13 | 9.22E+08 | 1.10E+07 | 2,868 | 2.723 |
| $T_{br}\left({}^0K\right)$ | 3.30E+12 | 5.45E+13 | 3.24E+18 | 2.72E+20 | 1.04E+24 | 1.10E+27 |
| $T_c\left({}^0K\right)$ | 8.96E+13 | 3.28E+11 | 93.0 | 0.0131 | 8.99E-10 | 8.10E-16 |
| $T_{cr}\left({}^0K\right)$ | 1.19E+09 | 3.25E+11 | 1.15E+21 | 8.11E+24 | 1.19E+32 | 1.32E+38 |
| $z_R + 1$ | 3.67E+43 | 8.14E+39 | 3.89E+25 | 6.53E+19 | 1.17E+09 | 1.00E+00 |
| $z_B + 1$ | 3.32E+14 | 2.01E+13 | 3.39E+08 | 4.03E+06 | 1.05E+03 | 1.00E+00 |
| $Age_{UG}\,(sec)$ | 7.14E-05 | 4.79E-03 | 6.94E+04 | 5.36E+07 | 1.27E+13 | 4.33E+17 |
| $Age_{UG\nu}\,(sec)$ | 4.59E-10 | 1.25E-07 | 4.42E+02 | 3.13E+06 | 1.22E+13 | 4.33E+17 |
| $Age_U\,(sec)$ | 1.18E-26 | 5.31E-23 | 1.11E-08 | 6.63E-03 | 3.70E+08 | 4.33E+17 |
| $2E_\nu$ | 3.84E+55 | 5.15E+50 | 4.14E+31 | 8.26E+23 | 3.87E+09 | 3.14E-03 |
| $\rho_{BBenergy}$ | 5.08E+45 | 6.81E+40 | 5.48E+21 | 1.09E+14 | 5.12E-01 | 4.16E-13 |
| $\rho_{BBmass}$ | 5.65E+24 | 7.58E+19 | 6.09E+00 | 1.22E-07 | 5.70E-22 | 4.63E-34 |
| $\lambda_{BB\max}$ | 3.20E-16 | 5.29E-15 | 3.14E-10 | 2.64E-08 | 1.01E-04 | 1.06E-01 |
| $\nu_{BB\max}$ | 5.32E+25 | 3.22E+24 | 5.42E+19 | 6.44E+17 | 1.69E+14 | 1.60E+11 |
| $\rho_{BBphoton}$ | 1.51E+46 | 3.33E+42 | 1.59E+28 | 2.67E+22 | 4.79E+11 | 4.10E+02 |
| $p_{BB}$ | 2.07E-11 | 1.25E-12 | 2.11E-17 | 2.51E-19 | 6.56E-23 | 6.23E-26 |
| $\mu_{BB}$ | 5.42E+19 | 1.20E+16 | 5.74E+01 | 9.63E-05 | 1.73E-15 | 1.48E-24 |
| $U\#\, photons$ | 1.44E+01 | 2.93E+08 | 1.29E+37 | 4.57E+48 | 1.42E+70 | 1.95E+88 |
| $2E_\nu$ | 6.72E+45 | 5.01E+50 | 6.23E+69 | 3.12E+77 | 6.67E+91 | 8.21E103 |
| $\rho_{BBrenergy}$ | 8.96E+35 | 6.69E+40 | 8.32E+59 | 4.17E+67 | 8.90E+81 | 1.10E+94 |
| $\rho_{BBrmass}$ | 9.97E+14 | 7.44E+19 | 9.25E+38 | 4.64E+46 | 9.90E+60 | 1.22E+73 |
| $\lambda_{BBr\max}$ | 8.78E-14 | 5.31E-15 | 8.95E-20 | 1.06E-21 | 2.78E-25 | 2.64E-28 |
| $\nu_{BBr\max}$ | 1.94E+23 | 3.21E+24 | 1.90E+29 | 1.60E+31 | 6.12E+34 | 6.45E+37 |
| $\rho_{BBrphoton}$ | 7.29E+38 | 3.29E+42 | 6.89E+56 | 4.10E+62 | 2.29E+73 | 2.68E+82 |
| $p_{BBr}$ | 7.54E-14 | 1.25E-12 | 7.40E-08 | 6.23E-06 | 2.38E-02 | 2.51E+01 |
| $\mu_{BBr}$ | 2.63E+12 | 1.19E+16 | 2.48E+30 | 1.48E+36 | 8.26E+46 | 9.65E+55 |
| $U_r\#\, photons$ | 6.97E-07 | 2.89E+08 | 5.57E+65 | 7.01E+88 | 6.82E+131 | 1.27E168 |

Black body viscosity 1.49E-24 g/(cm sec) is much lower than trisine viscosity $m_t c/\left(2\Delta x^2\right)$ or 1.21E-16 g/(cm sec) at present universe temperature of 8.11E-16 K. This indicates negligible drag of universe objects as measured in the Pioneer 10 and 11 anomaly and dark matter impacted galactic rotation curves.

## 2.16 Critical Optical Volume Energy (COVE) and Schwinger Pair Production.

This section considers the concept of critical volume, which would be analogous to critical mass in atomic fission and in the context of the Lawson Criterion. The critical volume would establish a parameter beyond which the output energy would exceed the input energy. It is anticipated that this energy would have a benign character and not be usable in explosive applications. The basic principle is that a coherent beam length is not related to input power creating that beam while virtual particle generation (in accordance with the CPT theorem and Schwinger limit) is a function of length. Given that volume is length cubed, then there must be a critical volume at which the energy of created virtual particles exceeds the input power to the volume. The following is a derivation of this concept.

$$\begin{aligned} Poytning\ Vector(S_{Input}) &= c\left(\frac{1}{2}\right)\left(\frac{\hbar\omega}{cavity}\right) \\ &= c\left(\frac{1}{2}\right)\left(\frac{h}{cavity}\frac{1}{time_{\pm}}\right) \\ &= c\left(\frac{1}{2}\right)\left(\frac{h}{cavity}\frac{v_{dx}}{2B}\right) \end{aligned} \tag{2.16.1}$$

Now consider the potential out put power per trisine cell which is consistent with Larmor radiation of accelerated virtual particles and associated Schwinger virtual pair production within that cell and its resonant $time_{\pm}$.

$$Poytning\ Vector(S_{Output}) = v_{dx}\left(\frac{1}{2}\right)\left(\frac{m_T v_{dx}^2}{cavity}\right) \tag{2.16.2}$$

Now calculate trisine length for output power to balance input power.

$$Length = 2B\frac{S_{Input}}{S_{Output}} = 2B\frac{c}{v_{dx}} = 2B^2\frac{m_T c}{\hbar\pi} = c\ time_{\pm} \tag{2.16.3}$$

The critical volume is then:

$$Critical\ Volume = Length^3 = c^3 time_{\pm}^3 \tag{2.16.4}$$

The critical area is then:

$$Critical\ Area = Length^2 \tag{2.16.5}$$

Then the power requirement at criticality is:

$$\begin{aligned} Power &= S_{Input} Length^2 = S_{Output}\frac{Length}{2B}Length^2 \\ &= \frac{1}{2}h\frac{time_{\pm}}{cavity}c^3 \end{aligned} \tag{2.16.6}$$

The Optical Volume incident wavelength, frequency and trisine $B$ are:

$$\begin{aligned} &Incident\ wavelength = 2c\ time \\ &Incident\ frequency = \frac{1}{2\ time} \\ &B = \left(\frac{\hbar\pi\ time}{2m_T}\right)^{1/2} \end{aligned} \tag{2.16.6a}$$

It is envisioned that this length would be a measure of trisine lattice size necessary for criticality and in particular a defined Critical Optical Volume defined at the of intersection of three lengths. To retrieve the power, one side (+ or -) of the virtual pair would have to be pinned (perhaps optically). In this case power could be retrieved by electrically coupling lattice to a pick up coil or arranging to intercept 56 MeV radiation.

This all appears consistent with Schwinger pair production [83] per volume and area (as well as length $B$) and frequency (1/ time) as expressed as follows as derived and verified with equations 2.16.7 - 2.16.11 with particular note to correlation with Stefan's law as express in equation 2.16.9 with a special note that $\hbar\left(h/2\pi\right)$ or $m_t(2B)^2/\left(2\pi\ time_{\pm}\right)$:

$$\frac{chain\ Ee^2}{cavity\ DB} = \frac{chain}{cavity}\frac{\sqrt{3}BeE}{\mathbb{C}\pi\cos(\theta)} = \frac{\pi g_s\hbar}{time_{\pm}} \tag{2.16.7}$$

$$\begin{aligned} \Gamma_{BP} &= \frac{1\ \text{pair}\left(\mathbb{C}\right)}{B\ time_{\pm}} \\ &= \frac{4\sqrt{3}}{3}\left(\frac{eE}{g_s\mathbb{C}2\pi^2\hbar\cos\left(\theta\right)}\right)\gamma \end{aligned} \tag{2.16.8}$$

$$\begin{aligned} \Gamma_{AP} &= \frac{1\ \text{pair}\left(\mathbb{C}\right)}{section\ time_{\pm}} \\ &= \frac{4\sqrt{2}\,3^{3/4}}{9}\left(\frac{eE}{g_s\mathbb{C}2\pi^2\hbar\cos\left(\theta\right)}\right)^{\frac{3}{2}}\frac{\gamma}{v_{dx}^{1/2}} \end{aligned} \tag{2.16.9a}$$

$$\Gamma_{AP} = \frac{1}{2}\frac{1}{(B/A)_t}\frac{1}{k_b T_c} Poynting\ Vector$$
$$= \frac{1}{\cos(\theta)}\frac{1}{(B/A)_t}\frac{c^2}{m_t v_{dx}^4}\sigma_S T_c^4 \quad (2.16.9b)$$

$$\Gamma_{VP} = \frac{1\ \text{pair}(\mathbb{C})}{cavity\ time_{\pm}}$$
$$= \frac{16\sqrt{3}}{9}\left(\frac{B}{A}\right)_t \left(\frac{eE}{g_s \mathbb{C} 2\pi^2 \hbar \cos(\theta)}\right)^2 \frac{\gamma}{v_{dx}} \quad (2.16.10)$$

And where summation is cutoff at $n = 1$ for volume $(1/n^2)$, area $(1/n^{3/2})$ and linear $(1/n)$:

$$\phi = \frac{1}{n^2}\ \exp\left(\frac{-n\pi m_e^2 v_{dx}^3}{eE\hbar}\right)$$

$$\phi = \frac{1}{n^{3/2}}\ \exp\left(\frac{-n\pi m_e^2 v_{dx}^3}{eE\hbar}\right) \quad (2.16.11)$$

$$\phi = \frac{1}{n}\ \exp\left(\frac{-n\pi m_e^2 v_{dx}^3}{eE\hbar}\right)$$

$$\gamma = \sum_{n=1} \phi = 0.9997 \quad \gamma = \sum_{n=1}^{\infty} \phi = 2.5493$$

**Table 2.16.1** Schwinger Virtual Pair Production per Trisine unit Length (*B*), per Area (*section*) and per Volume (*cavity*)

| Condition | 1.2 | 1.3 | 1.6 | 1.9 | 1.10 | 1.11 |
|---|---|---|---|---|---|---|
| $T_a\left({}^0K\right)$ | 6.53E+11 | 6.53E+11 | 6.53E+11 | 6.53E+11 | 6.53E+11 | 6.53E+11 |
| $T_b\left({}^0K\right)$ | 9.05E+14 | 5.48E+13 | 9.22E+08 | 1.10E+07 | 2,868 | 2.723 |
| $T_{br}\left({}^0K\right)$ | 3.30E+12 | 5.45E+13 | 3.24E+18 | 2.72E+20 | 1.04E+24 | 1.10E+27 |
| $T_c\left({}^0K\right)$ | 8.96E+13 | 3.28E+11 | 93.0 | 0.0131 | 8.99E-10 | 8.10E-16 |
| $T_{cr}\left({}^0K\right)$ | 1.19E+09 | 3.25E+11 | 1.15E+21 | 8.11E+24 | 1.19E+32 | 1.32E+38 |
| $z_R + 1$ | 3.67E+43 | 8.14E+39 | 3.89E+25 | 6.53E+19 | 1.17E+09 | 1.00E+00 |
| $z_B + 1$ | 3.32E+14 | 2.01E+13 | 3.39E+08 | 4.03E+06 | 1.05E+03 | 1.00E+00 |
| $Age_{UG}(sec)$ | 7.14E-05 | 4.79E-03 | 6.94E+04 | 5.36E+07 | 1.27E+13 | 4.33E+17 |
| $Age_{UGv}(sec)$ | 4.59E-10 | 1.25E-07 | 4.42E+02 | 3.13E+06 | 1.22E+13 | 4.33E+17 |
| $Age_U(sec)$ | 1.18E-26 | 5.31E-23 | 1.11E-08 | 6.63E-03 | 3.70E+08 | 4.33E+17 |
| E Field (V/cm) | 2.79E+22 | 6.19E+18 | 2.96E+04 | 4.96E-02 | 8.89E-13 | 7.61E-22 |
| $\Gamma_{BP}$ (cm$^{-1}$sec$^{-1}$) | 5.61E+37 | 1.24E+34 | 5.93E+19 | 9.96E+13 | 1.78E+03 | 1.53E-06 |
| $\Gamma_{AP}$ (cm$^{-2}$sec$^{-1}$) | 2.43E+50 | 3.26E+45 | 2.62E+26 | 5.23E+18 | 2.45E+04 | 1.99E-08 |
| $\Gamma_{VP}$ (cm$^{-3}$sec$^{-1}$) | 8.70E+63 | 7.06E+57 | 9.55E+33 | 2.27E+24 | 2.78E+06 | 2.14E-09 |
| critical volume | 5.18E-43 | 1.06E-35 | 4.63E-07 | 1.64E+05 | 5.12E+26 | 6.99E+44 |

It is important to note that under unperturbed resonant conditions, the rate of Schwinger pair (with energy $k_b T_c$) production and pair (with energy $k_b T_c$) absorption are equal.

## 3. Discussion

This effort began as an attempt towards formulating a numerical and dimensional frame work towards explaining the Podkletnov's and Nieminen's [7] results but in the process, relationships between electron/proton mass, charge and gravitational forces were established which appear to be unified under this correlation with links to the cosmological parameter, the dark energy of the universe, and the universe black body radiation as a continuation of its big bang origins and nuclear forces. Essentially, this effort is an attempt to explain universe stability. Most (99.99% ) of the universe exists in an elastic state and does not change state at the atomic level for extremely large times. So rather that explain universe dynamics (human activity, planetary tectonics, industrial mechanics, etc), I start from the base that the universe essentially runs in place and is generally in conformance with the conservation of energy and momentum (elastic).

An importantly, the universe appears to be constructed of a visible and invisible or real and reflected nature that transforms between two at the nuclear dimension as governed by the speed of light(c) that is expressed conventionally under the concept of phase and group velocity.

Solid-state physics provides an intuitive understanding of these correlations. In solid state semiconductors, the roles of charged particles and photons in field theory are replaced by electrons in the conduction band, holes in the valence band, and phonons or vibrations of the crystal lattice.

The overall correlation is an extension to what Sir Arthur Eddington proposed during the early part of the 20th century. He attempted to relate proton and electron mass to the number of particles in the universe and the universe radius. Central to his approach was the fine structure constant:

$$1/\alpha = \hbar c / e^2$$

which he knew to be approximately equal to 137. This relationship is also found as a consequence of our development in equation 2.1.48, the difference being that a material dielectric $(\varepsilon)$ modified velocity of light $(v_\varepsilon)$ is used. To determine this dielectric $(v_\varepsilon)$, displacement $(D)$ and electric fields $(E)$ are determined for Cooper CPT Charge conjugated pairs moving through the trisine CPT lattice under conditions of the trisine CPT lattice diamagnetism equal to $-1/4\pi$ (the Meissner effect). Standard principles involving Gaussian surfaces are employed and in terms of this model development the following relationship is deduced from equation 2.1.48 repeated here

$$\frac{2}{1}\frac{\mathbb{C}e_{\pm}\hbar}{2m_e v_{\varepsilon y}} = chain\ H_c$$

As related to cosmological variables

$$\rho_U = \frac{m_T}{cavity} = \frac{2}{8\pi}\frac{H_U^2}{G_U} = \frac{M_U}{V_U} = \frac{3M_U}{4\pi R_U^3} \qquad (3.1)$$

$$\rho_{Ur} = \frac{m_T}{cavity_r} = \frac{2}{8\pi}\frac{H_{Ur}^2}{G_{Ur}} = \frac{M_{Ur}}{V_{Ur}} = \frac{3M_{Ur}}{4\pi R_{Ur}^3} \qquad (3.1r)$$

and

$$\frac{G_U M_U H_U}{c^3} = 3^{1/2} \qquad (3.2)$$

$$\frac{G_{Ur} M_{Ur} H_{Ur}}{c^3} = 3^{1/2} \qquad (3.2r)$$

The extension to traditional superconductivity discussion is postulated wherein the underlying principle is resonance expressed as a conservation of energy and momentum (elastic character). The mechanics of this approach are in terms of a cell (*cavity*) defined by

momentum states: $K_1,\ K_2,\ K_3\ and\ K_4$

and corresponding energy states: $K_1^2,\ K_2^2,\ K_3^2\ and\ K_4^2$

by a required new mass called trisine mass $(m_t)$, which is intermediate to the electron and proton masses. The cell dimensions can be scaled from nuclear standard model particles to universe dimensions (invariant with length '*B*' *~1/K* and related cell volumes '*cavity*' and '*chain*') with the trisine mass $(m_t)$ remaining constant as per the following relationships expressed in terms of the superconductor critical temperature $(T_c)$.

$$k_b T_c = h(1/time_{\pm})/2 = (m_T v_{dx}/time_{\pm})B \qquad (3.3)$$

$$k_b T_{cr} = h(1/time_{r\pm})/2 = (m_T v_{dxr}/time_{r\pm})B_r \qquad (3.3r)$$

$$k_b T_c = \hbar\pi/time_{\pm} \quad \text{Aharonov – Bohm effect} \qquad (3.4)$$

$$k_b T_c = (m_T v_{dx}/time_{\pm})B \quad Torque \qquad (3.5)$$

$$k_b T_{cr} = (m_T v_{dxr}/time_{r\pm})B_r \quad Torque \qquad (3.5r)$$

$$k_b T_c = (h^2/(2m_T))(1/(2B)^2) \qquad (3.6)$$

$$k_b T_{cr} = (h^2/(2m_T))(1/(2B_r)^2) \qquad (3.6r)$$

$$k_b T_c = (1/2)m_T v_{dx}^2 \qquad (3.7)$$

$$k_b T_{cr} = (1/2)m_T v_{dxr}^2 \qquad (3.7r)$$

$$k_b T_c = 3\sqrt{3}e_{\pm}^2/(g_s\varepsilon 2B) \qquad (3.8)$$

$$k_b T_{cr} = 3\sqrt{3}e_{\pm}^2/(g_s\varepsilon_r B_r) \qquad (3.8r)$$

$$k_b T_c = (2/3)\left(1/\sqrt{3}\right)(A/B)_T\ (1/(4\pi)\ cavity\ D_x E_x \qquad (3.9)$$

$$k_b T_{cr} = (2/3)\left(1/\sqrt{3}\right)(A_r/B_r)_T(1/(4\pi)\ cavity\ D_{xr}E_{xr} \qquad (3.9r)$$

$$k_b T_c = (2/3)(1/\sqrt{3})(A/B)_T(1/(4\pi)cavity E_x^2\varepsilon_x \qquad (3.10)$$

$$k_b T_{cr} = (2/3)(1/\sqrt{3})(A_r/B_r)_T(1/(4\pi)cavity E_{xr}^2\varepsilon_{xr} \qquad (3.10r)$$

$$k_b T_c = (2/3)(1/\sqrt{3})(A/B)_T(1/(4\pi)cavity D_x^2/\varepsilon_x \qquad (3.11)$$

$$k_b T_{cr} = (2/3)(1/\sqrt{3})(A_r/B_r)(1/(4\pi)cavity D_{xr}^2/\varepsilon_{xr} \qquad (3.11r)$$

$$k_b T_c = (chain)H_c^2/(8\pi) \qquad (3.12)$$

$$k_b T_{cr} = (chain_r)H_{cr}^2/(8\pi) \qquad (3.12r)$$

$$k_b T_c = (pressure)(cavity) \qquad (3.13)$$

$$k_b T_{cr} = (chain)H_{cr}^2/(8\pi) \qquad (3.13r)$$

$$k_b T_c = m_t c^2/(2k_{mx}\varepsilon_x) \qquad (3.14)$$

$$k_b T_{cr} = m_t c^2/(2k_{mxr}\varepsilon_{xr}) \qquad (3.14r)$$

$$k_b T_c = m_r c^2 \qquad (3.15)$$

$$k_b T_{cr} = m_{rr} c^2 \qquad (3.15r)$$

$$k_b T_c = g_d \frac{6}{\pi}\left(\frac{1}{\varsigma(3/2)}\right)^{\frac{2}{3}}\frac{2\pi\hbar^2}{m_T}\left(\frac{1}{cavity}\right)^{\frac{2}{3}} \qquad (3.16)$$

$$k_b T_{cr} = g_d \frac{6}{\pi}\left(\frac{1}{\varsigma(3/2)}\right)^{\frac{2}{3}}\frac{2\pi\hbar^2}{m_T}\left(\frac{1}{cavity_r}\right)^{\frac{2}{3}} \qquad (3.16r)$$

$$k_b T_c = \frac{h^2}{2\pi m_T}\left(\frac{\mathbb{C}}{\varsigma(3/2)chain}\right)^{\frac{2}{3}}\cos(\theta) \qquad (3.17)$$

$$k_b T_{cr} = \frac{h^2}{2\pi m_T}\left(\frac{\mathbb{C}}{\varsigma(3/2)chain_r}\right)^{\frac{2}{3}}\cos(\theta) \qquad (3.17r)$$

$$k_b T_c = G_U m_r M_U/R_U \qquad (3.18)$$

$$k_b T_{cr} = G_{Ur} m_{rr} M_{Ur}/R_{Ur} \qquad (3.18r)$$

$$k_b T_c = 4\hbar v_{xr}/R_U \qquad (3.19)$$

$$k_b T_c = h^2 R_U c^2/(8B^2 G_U M_U m_T) \quad \text{Hawking Radiation} \qquad (3.20)$$

$$k_b T_{cr} = h^2 R_{Ur} c^2/(8B^2 G_{Ur} M_{Ur} m_T) \quad \text{Hawking Radiation} \qquad (3.20r)$$

$$k_b T_b = m_T c^2/(4\varepsilon) \qquad (3.21)$$

$$k_b T_{br} = m_T c^2/\varepsilon_r \qquad (3.21r)$$

$$k_b T_a = m_T c^2 \qquad (3.22)$$

$$k_b T_a = (2\pi)^{2/3} h^{4/3} c^{1/3}/(8U^{1/3}B^2\cos(\theta)) \qquad (3.23)$$

These correlations then can be placed in the form of the

Einstein tensor $G_{\mu\nu}$ and stress energy tensor $T_{\mu\nu}$ relationship:

$$G_{\mu\nu} = \frac{8\pi G_U}{c^4} T_{\mu\nu}$$

$$G_{\mu\nu} = \frac{8\pi G\ sherd^{1/2}}{c^4} T_{\mu\nu} \qquad (3.24)$$

And reflective tensor condition

$$G_{\mu\nu r} = \frac{8\pi G_{Ur}}{c^4} T_{\mu\nu r}$$

$$G_{\mu\nu r} = \frac{8\pi G\ sherd_r^{1/2}}{c^4} T_{\mu\nu r} \qquad (3.24r)$$

where the Newton gravitational parameter $\left(G_U \text{ or } G\ sherd^{1/2}\right)$ varies with universe age $(Age_U)$ and $\left(G_{Ur} \text{ or } G\ sherd_r^{1/2}\right)$ varies with universe age $(Age_{Ur})$ or more specifically within the context of stress energy tensor components:

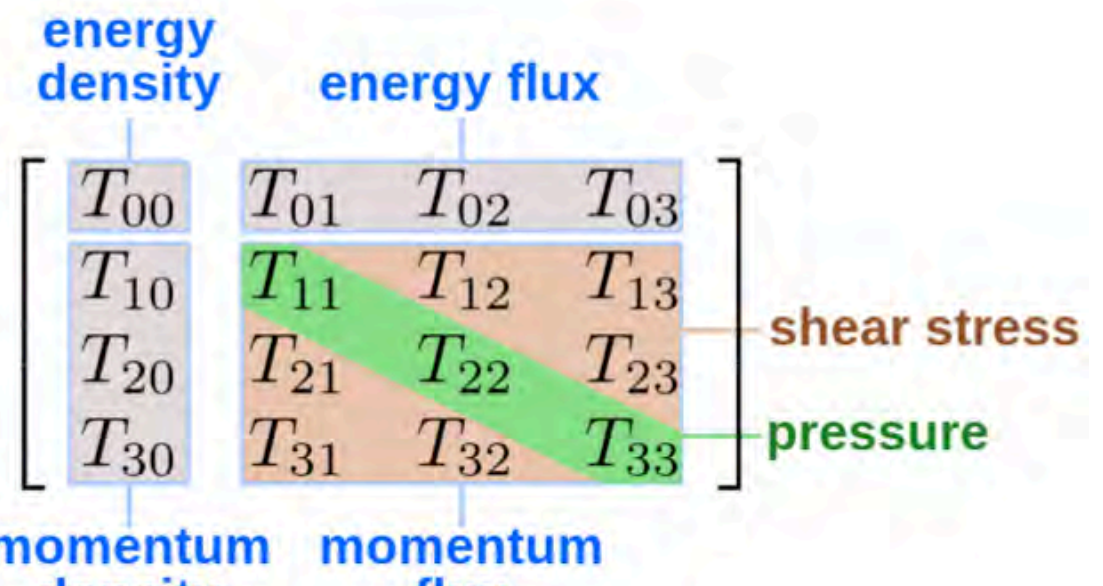

and more definitively:

$$\begin{vmatrix} \frac{k_b T_c}{cavity} & \frac{1}{\sqrt{3}} \frac{c\sqrt{2m_T k_b T_c}}{approach\ time_\pm} & \frac{2}{\sqrt{3}} \frac{Euler\ c\sqrt{2m_T k_b T_c}}{side\ time_\pm} & \frac{2}{\sqrt{3}} \frac{B}{A} \frac{c\sqrt{2m_T k_b T_c}}{section\ time_\pm} \\ \frac{m_r c}{cavity} & \frac{1}{2} \frac{\sqrt{2m_T k_b T_c}}{approach\ time_\pm} & \frac{2}{\sqrt{3}} \frac{A}{B} \frac{m_T c}{approach} & \frac{2}{\sqrt{3}} \frac{A}{B} \frac{m_T c}{side} \\ \frac{m_r c}{cavity} & \frac{1}{\sqrt{3}} \frac{\sqrt{2m_T k_b T_c}}{approach\ time_\pm} & \frac{Euler\sqrt{2m_t k_b T_c}}{side\ time_\pm} & \frac{2m_T c}{section} \\ \frac{m_r c}{cavity} & \frac{2}{\sqrt{3}} \frac{Euler\sqrt{2m_T k_b T_c}}{side\ time_\pm} & \frac{2}{\sqrt{3}} \frac{B}{A} \frac{\sqrt{2m_T k_b T_c}}{section\ time_\pm} & \frac{B}{A} \frac{\sqrt{2m_T k_b T_c}}{section\ time_\pm} \end{vmatrix}$$

$$= \frac{8\pi G_U}{c^4} \left( \frac{A}{B} \frac{c^4}{Euler} \frac{m_T^3}{h} \sqrt{\frac{1}{2m_T k_b T_c}} \right) \begin{vmatrix} \frac{M_U c^2}{V_U} & \frac{M_U c^2}{A_U Age_U} & \frac{M_U c^2}{A_U Age_U} & \frac{M_U c^2}{A_U Age_U} \\ \frac{M_U c}{V_U} & \frac{M_U c^2}{V_U} & \frac{M_U c}{A_U} & \frac{M_U c}{A_U} \\ \frac{M_U c}{V_U} & \frac{M_U c}{A_U Age_U} & \frac{M_U c^2}{V_U} & \frac{M_U c}{A_U} \\ \frac{M_U c}{V_U} & \frac{M_U c}{A_U Age_U} & \frac{M_U c}{A_U Age_U} & \frac{M_U c^2}{V_U} \end{vmatrix}$$

and reflective tensor condition

$$\begin{vmatrix} \frac{k_b T_{cr}}{cavity_r} & \frac{1}{\sqrt{3}} \frac{c\sqrt{2m_T k_b T_{cr}}}{approach_r\ time_{r\pm}} & \frac{2}{\sqrt{3}} \frac{Euler\ c\sqrt{2m_T k_b T_{cr}}}{side_r\ time_{r\pm}} & \frac{2}{\sqrt{3}} \frac{B}{A} \frac{c\sqrt{2m_T k_b T_{cr}}}{section_r\ time_{r\pm}} \\ \frac{m_{rr} c}{cavity_r} & \frac{1}{2} \frac{\sqrt{2m_T k_b T_{cr}}}{approach_r\ time_{r\pm}} & \frac{2}{\sqrt{3}} \frac{A_r}{B_r} \frac{m_T c}{approach_r} & \frac{2}{\sqrt{3}} \frac{A}{B} \frac{m_T c}{side_r} \\ \frac{m_{rr} c}{cavity_r} & \frac{1}{\sqrt{3}} \frac{\sqrt{2m_T k_b T_{cr}}}{approach_r\ time_{r\pm}} & \frac{Euler\sqrt{2m_T k_b T_{cr}}}{side_r\ time_\pm} & \frac{2m_T c}{section_r} \\ \frac{m_{rr} c}{cavity_r} & \frac{2}{\sqrt{3}} \frac{Euler\sqrt{2m_T k_b T_{cr}}}{side_r\ time_{r\pm}} & \frac{2}{\sqrt{3}} \frac{B_r}{A_r} \frac{\sqrt{2m_T k_b T_{cr}}}{section_r\ time_{r\pm}} & \frac{B}{A} \frac{\sqrt{2m_T k_b T_{cr}}}{section_r\ time_{r\pm}} \end{vmatrix}$$

$$= \frac{8\pi G_{Ur}}{c^4} \left( \frac{A_r}{B_r} \frac{c^4}{Euler} \frac{m_T^3}{h} \sqrt{\frac{1}{2m_t k_b T_{cr}}} \right) \begin{vmatrix} \frac{M_{Ur} c^2}{V_{Ur}} & \frac{M_{Ur} c^2}{A_{Ur} Age_{Ur}} & \frac{M_{Ur} c^2}{A_U Age_{Ur}} & \frac{M_{Ur} c^2}{A_{Ur} Age_{Ur}} \\ \frac{M_{Ur} c}{V_{Ur}} & \frac{M_U c^2}{V_{Ur}} & \frac{M_{Ur} c}{A_{Ur}} & \frac{M_{Ur} c}{A_{Ur}} \\ \frac{M_{Ur} c}{V_{Ur}} & \frac{M_{Ur} c}{A_{Ur} Age_{Ur}} & \frac{M_{Ur} c^2}{V_{Ur}} & \frac{M_{Ur} c}{A_{Ur}} \\ \frac{M_{Ur} c}{V_{Ur}} & \frac{M_{Ur} c}{A_{Ur} Age_{Ur}} & \frac{M_{Ur} c}{A_{Ur} Age_{Ur}} & \frac{M_{Ur} c^2}{V_{Ur}} \end{vmatrix}$$

Note that $8\pi G_U / c^4$ has '1/force' dimensionality as does $h\sqrt{m_t k_b T_c} / \left(c^4 m_t^3\right)$ making their ratio dimensionless but defining the quantum Einstein tensor $G_{uv}$ cellular 'curvature' as

$$G_{\mu\nu} = \frac{8\pi G_U}{c^4} T_{\mu\nu} \text{ (assuming } 1 \Big/ \left( \frac{A}{B} \frac{c^4}{Euler} \frac{m_t^3}{h} \sqrt{\frac{1}{2m_T k_b T_c}} \right)$$

is multiplied into '$uv$' elements of $G_{uv}$ ). These $G_{uv}$ curvatures are on a cosmic scale ~1E28 cm. (essentially a flat universe)

The above approach is backwards to defining Einstein tensor $G_{uv}$ in terms of the Ricci, metric tensors and scalar curvature.

$$G_{uv} = R_{uv} - (1/2) g_{uv} R + g_{uv} \Lambda$$

Note how the quantum *(h)* as well as charge *(e)* and all other parameters as reflected by correlations 3.2 – 3.17 are naturally expressed within these Einstein field equations or perhaps more correctly, solutions to the Einstein field equations with a special note that the Weinberg mass[99] is inherent to this formulation here repeated from equations 2.11.15 and 2.11.15r.

$$m_T^3 = \frac{m_t \rho_U k_b T_c section}{\sqrt{3} H_U c} = \frac{\pi^2 \rho_U \hbar^2}{H_U c} = \frac{U \hbar^2 \cos^3(\theta)}{c}$$

$$m_T^3 = \frac{m_t \rho_{Ur} k_b T_{cr} section_r}{\sqrt{3} H_{Ur} c} = \frac{\pi^2 \rho_{Ur} \hbar^2}{H_{Ur} c} = \frac{U \hbar^2 \cos^3(\theta)}{c}$$

The triangular hexagonal cell character is generally accepted by conventional solid-state mechanics to be equivalent in momentum and real space making the energy $(\sim K^2)$ and momentum $(\sim K)$ cell (*cavity*) identity possible. Also, the hexagonal or triangulation energy $(\sim K^2)$ and momentum $(\sim K)$ have a 1/3, 2/3 character inherent to the standard model 1/3, 2/3 quark definitions.

$$\begin{vmatrix} \frac{\sqrt{3}Euler\ h^4}{8cavity^2 m_T^4 c^4} & \frac{Euler\ h^4}{4cavity^2 m_T^4 c^3} & \frac{Euler\ h^4}{4cavity^2 m_T^4 c^3} & \frac{Euler\ h^4}{4cavity^2 m_T^4 c^3} \\ \frac{\sqrt{3}Euler\ h^4}{8cavity^2 m_T^4 c^5} & \frac{\sqrt{3}Euler\ h^4}{8cavity^2 m_T^4 c^4} & \frac{3^{\frac{1}{2}} Euler^2\ h^2}{cavity\ m_T^2 c^3} & \frac{3^{\frac{1}{2}} Euler^2\ h^2}{cavity\ m_T^2 c^3} \\ \frac{\sqrt{3}Euler\ h^4}{8cavity^2 m_T^4 c^5} & \frac{\sqrt{3}Euler^2\ h^4}{4cavity^2 m_T^4 c^5} & \frac{\sqrt{3}Euler\ h^4}{8cavity^2 m_T^4 c^4} & \frac{3^{\frac{1}{2}} Euler^2\ h^2}{cavity\ m_T^2 c^3} \\ \frac{\sqrt{3}Euler\ h^4}{8cavity^2 m_T^4 c^5} & \frac{\sqrt{3}Euler^2\ h^4}{4cavity^2 m_T^4 c^5} & \frac{\sqrt{3}Euler^2\ h^4}{4cavity^2 m_T^4 c^5} & \frac{\sqrt{3}Euler\ h^4}{8cavity^2 m_T^4 c^4} \end{vmatrix}$$

$$= \frac{8\pi G_U}{c^4} \begin{vmatrix} \frac{M_U c^2}{V_U} & \frac{M_U c^2}{A_U Age_U} & \frac{M_U c^2}{A_U Age_U} & \frac{M_U c^2}{A_U Age_U} \\ \frac{M_U c}{V_U} & \frac{M_U c^2}{V_U} & \frac{M_U c}{A_U} & \frac{M_U c}{A_U} \\ \frac{M_U c}{V_U} & \frac{M_U c}{A_U Age_U} & \frac{M_U c^2}{V_U} & \frac{M_U c}{A_U} \\ \frac{M_U c}{V_U} & \frac{M_U c}{A_U Age_U} & \frac{M_U c}{A_U Age_U} & \frac{M_U c^2}{V_U} \end{vmatrix}$$

and reflective tensor condition

$$\begin{vmatrix} \frac{\sqrt{3}Euler\ h^4}{8cavity_r^2 m_T^4 c^4} & \frac{Euler\ h^4}{4cavity_r^2 m_T^4 c^3} & \frac{Euler\ h^4}{4cavity_r^2 m_T^4 c^3} & \frac{Euler\ h^4}{4cavity_r^2 m_T^4 c^3} \\ \frac{\sqrt{3}Euler\ h^4}{8cavity_r^2 m_T^4 c^5} & \frac{\sqrt{3}Euler\ h^4}{8cavity_r^2 m_T^4 c^4} & \frac{3^{\frac{1}{2}} Euler^2\ h^2}{cavity_r\ m_T^2 c^3} & \frac{3^{\frac{1}{2}} Euler^2\ h^2}{cavity_r\ m_T^2 c^3} \\ \frac{\sqrt{3}Euler\ h^4}{8cavity_r^2 m_T^4 c^5} & \frac{\sqrt{3}Euler^2\ h^4}{4cavity_r^2 m_T^4 c^5} & \frac{\sqrt{3}Euler\ h^4}{8cavity_r^2 m_T^4 c^4} & \frac{3^{\frac{1}{2}} Euler^2\ h^2}{cavity_r\ m_T^2 c^3} \\ \frac{\sqrt{3}Euler\ h^4}{8cavity_r^2 m_T^4 c^5} & \frac{\sqrt{3}Euler^2\ h^4}{4cavity_r^2 m_T^4 c^5} & \frac{\sqrt{3}Euler^2\ h^4}{4cavity_r^2 m_T^4 c^5} & \frac{\sqrt{3}Euler\ h^4}{8cavity_r^2 m_T^4 c^4} \end{vmatrix}$$

$$= \frac{8\pi G_{Ur}}{c^4} \begin{vmatrix} \frac{M_{Ur} c^2}{V_{Ur}} & \frac{M_{Ur} c^2}{A_{Ur} Age_{Ur}} & \frac{M_{Ur} c^2}{A_{Ur} Age_{Ur}} & \frac{M_{Ur} c^2}{A_{Ur} Age_{Ur}} \\ \frac{M_{Ur} c}{V_{Ur}} & \frac{M_{Ur} c^2}{V_{Ur}} & \frac{M_{Ur} c}{A_{Ur}} & \frac{M_{Ur} c}{A_{Ur}} \\ \frac{M_{Ur} c}{V_{Ur}} & \frac{M_{Ur} c}{A_{Ur} Age_{Ur}} & \frac{M_{Ur} c^2}{V_{Ur}} & \frac{M_{Ur} c}{A_{Ur}} \\ \frac{M_{Ur} c}{V_{Ur}} & \frac{M_{Ur} c}{A_{Ur} Age_{Ur}} & \frac{M_{Ur} c}{A_{Ur} Age_{Ur}} & \frac{M_{Ur} c^2}{V_{Ur}} \end{vmatrix}$$

and then assuming the Planck constant dimensional definition $\hbar\left(h/2\pi\right)$ or $m_t(2B)^2/\left(2\pi\ time_\pm\right)$ and $cavity = 2\sqrt{3}\left(A/B\right)B^3$ resulting in the trisine cellular curvature accelerated expansion $\left(B/time_\pm^2\right)$ and $\left(B/time_\pm^2\right)^2$.

$$\begin{vmatrix} \left(\frac{B}{A}\right)^2 \frac{8Euler\ B^2}{\sqrt{3time_\pm^4 c^4}} & \left(\frac{B}{A}\right)^2 \frac{16Euler\ B^2}{3time_\pm^4 c^3} & \left(\frac{B}{A}\right)^2 \frac{16Euler\ B^2}{3time_\pm^4 c^3} & \left(\frac{B}{A}\right)^2 \frac{16Euler\ B^2}{3time_\pm^4 c^3} \\ \left(\frac{B}{A}\right)^2 \frac{4Euler\ B^2}{\sqrt{3time_\pm^4 c^5}} & \left(\frac{B}{A}\right)^2 \frac{8Euler\ B^2}{\sqrt{3time_\pm^4 c^4}} & \left(\frac{B}{A}\right) \frac{8Euler^2\ B}{time_\pm^2 c^3} & \left(\frac{B}{A}\right) \frac{8Euler^2\ B}{time_\pm^2 c^3} \\ \left(\frac{B}{A}\right)^2 \frac{4Euler\ B^2}{\sqrt{3time_\pm^4 c^5}} & \left(\frac{B}{A}\right)^2 \frac{16Euler^2\ B^2}{\sqrt{3time_\pm^4 c^4}} & \left(\frac{B}{A}\right)^2 \frac{8Euler\ B^2}{\sqrt{3time_\pm^4 c^4}} & \left(\frac{B}{A}\right) \frac{8Euler^2\ B}{time_\pm^2 c^3} \\ \left(\frac{B}{A}\right)^2 \frac{4Euler\ B^2}{\sqrt{3time_\pm^4 c^5}} & \left(\frac{B}{A}\right)^2 \frac{16Euler^2\ B^2}{\sqrt{3time_\pm^4 c^4}} & \left(\frac{B}{A}\right)^2 \frac{16Euler^2\ B^2}{\sqrt{3time_\pm^4 c^4}} & \left(\frac{B}{A}\right)^2 \frac{8Euler\ B^2}{\sqrt{3time_\pm^4 c^4}} \end{vmatrix}$$

$$= \frac{8\pi G_U}{c^4} \begin{vmatrix} \frac{M_U c^2}{V_U} & \frac{M_U c^2}{A_U Age_U} & \frac{M_U c^2}{A_U Age_U} & \frac{M_U c^2}{A_U Age_U} \\ \frac{M_U c}{V_U} & \frac{M_U c^2}{V_U} & \frac{M_U c}{A_U} & \frac{M_U c}{A_U} \\ \frac{M_U c}{V_U} & \frac{M_U c}{A_U Age_U} & \frac{M_U c^2}{V_U} & \frac{M_U c}{A_U} \\ \frac{M_U c}{V_U} & \frac{M_U c}{A_U Age_U} & \frac{M_U c}{A_U Age_U} & \frac{M_U c^2}{V_U} \end{vmatrix}$$

and reflective tensor condition

$$\begin{vmatrix} \left(\frac{B_r}{A_r}\right)^2 \frac{8Euler\ B_r^2}{\sqrt{3time_{r\pm}^4 c^4}} & \left(\frac{B_r}{A_r}\right)^2 \frac{16Euler\ B_r^2}{3time_{r\pm}^4 c^3} & \left(\frac{B_r}{A_r}\right)^2 \frac{16Euler\ B_r^2}{3time_{r\pm}^4 c^3} & \left(\frac{B_r}{A_r}\right)^2 \frac{16Euler\ B_r^2}{3time_{r\pm}^4 c^3} \\ \left(\frac{B_r}{A_r}\right)^2 \frac{4Euler\ B_r^2}{\sqrt{3time_{r\pm}^4 c^5}} & \left(\frac{B_r}{A_r}\right)^2 \frac{8Euler\ B_r^2}{\sqrt{3time_{r\pm}^4 c^4}} & \left(\frac{B_r}{A_r}\right) \frac{8Euler^2\ B_r}{time_{r\pm}^2 c^3} & \left(\frac{B_r}{A_r}\right) \frac{8Euler^2\ B_r}{time_{r\pm}^2 c^3} \\ \left(\frac{B_r}{A_r}\right)^2 \frac{4Euler\ B_r^2}{\sqrt{3time_{r\pm}^4 c^5}} & \left(\frac{B_r}{A_r}\right)^2 \frac{16Euler^2\ B_r^2}{\sqrt{3time_{r\pm}^4 c^4}} & \left(\frac{B_r}{A_r}\right)^2 \frac{8Euler\ B_r^2}{\sqrt{3time_{r\pm}^4 c^4}} & \left(\frac{B_r}{A_r}\right) \frac{8Euler^2\ B_r}{time_{r\pm}^2 c^3} \\ \left(\frac{B_r}{A_r}\right)^2 \frac{4Euler\ B_r^2}{\sqrt{3time_{r\pm}^4 c^5}} & \left(\frac{B_r}{A_r}\right)^2 \frac{16Euler^2\ B_r^2}{\sqrt{3time_{r\pm}^4 c^4}} & \left(\frac{B_r}{A_r}\right)^2 \frac{16Euler^2\ B_r^2}{\sqrt{3time_{r\pm}^4 c^4}} & \left(\frac{B_r}{A_r}\right)^2 \frac{8Euler\ B_r^2}{\sqrt{3time_{r\pm}^4 c^4}} \end{vmatrix}$$

$$= \frac{8\pi G_{Ur}}{c^4} \begin{vmatrix} \frac{M_{Ur}c^2}{V_{Ur}} & \frac{M_{Ur}c^2}{A_{Ur}Age_{Ur}} & \frac{M_{Ur}c^2}{A_{Ur}Age_{Ur}} & \frac{M_{Ur}c^2}{A_{Ur}Age_{Ur}} \\ \frac{M_{Ur}c}{V_{Ur}} & \frac{M_{Ur}c^2}{V_{Ur}} & \frac{M_{Ur}c}{A_{Ur}} & \frac{M_{Ur}c}{A_{Ur}} \\ \frac{M_{Ur}c}{V_{Ur}} & \frac{M_{Ur}c}{A_{Ur}Age_{Ur}} & \frac{M_{Ur}c^2}{V_{Ur}} & \frac{M_{Ur}c}{A_{Ur}} \\ \frac{M_{Ur}c}{V_{Ur}} & \frac{M_{Ur}c}{A_{Ur}Age_{Ur}} & \frac{M_{Ur}c}{A_{Ur}Age_{Ur}} & \frac{M_{Ur}c^2}{V_{Ur}} \end{vmatrix}$$

And as a general expression of these matrices, note that

$$8\pi G_U/c^4 = 8\pi 3m_T time_{\pm} A/\left(M_U \hbar\pi\right)$$

which is conceptually an expression of the impulse momentum theorem or Newton's second law at the universal level:

$$\begin{aligned} dp/dt = c^4/\left(8\pi G_U\right) &= M_U v_{dz}/\left(8\pi 6 time_{\pm}\right) \\ &= M_U \hbar\pi/\left(8\pi 6 m_T time_{\pm} A\right) \end{aligned}$$

These relationships expressed purely at the universal level are as follows:

$$\begin{vmatrix} \frac{12\pi}{A_U} & \frac{4}{R_U Age_U} & \frac{4}{R_U Age_U} & \frac{4}{R_U Age_U} \\ \frac{8\pi\sqrt{3}Age_U}{V_U} & \frac{12\pi}{A_U} & \frac{8\pi\sqrt{3}Age_U}{A_U} & \frac{8\pi\sqrt{3}Age_U}{A_U} \\ \frac{8\pi\sqrt{3}Age_U}{V_U} & \frac{8\pi\sqrt{3}}{A_U} & \frac{12\pi}{A_U} & \frac{8\pi\sqrt{3}Age_U}{A_U} \\ \frac{8\pi\sqrt{3}Age_U}{V_U} & \frac{8\pi\sqrt{3}}{A_U} & \frac{8\pi\sqrt{3}}{A_U} & \frac{12\pi}{A_U} \end{vmatrix}$$

$$= \frac{8\pi G_U}{c^4} \begin{vmatrix} \frac{M_U c^2}{V_U} & \frac{M_U c^2}{A_U Age_U} & \frac{M_U c^2}{A_U Age_U} & \frac{M_U c^2}{A_U Age_U} \\ \frac{M_U c}{V_U} & \frac{M_U c^2}{V_U} & \frac{M_U c}{A_U} & \frac{M_U c}{A_U} \\ \frac{M_U c}{V_U} & \frac{M_U c}{A_U Age_U} & \frac{M_U c^2}{V_U} & \frac{M_U c}{A_U} \\ \frac{M_U c}{V_U} & \frac{M_U c}{A_U Age_U} & \frac{M_U c}{A_U Age_U} & \frac{M_U c^2}{V_U} \end{vmatrix}$$

and reflective tensor condition

$$\begin{vmatrix} \frac{12\pi}{A_{Ur}} & \frac{4}{R_{Ur} Age_{Ur}} & \frac{4}{R_{Ur} Age_{Ur}} & \frac{4}{R_{Ur} Age_{Ur}} \\ \frac{8\pi\sqrt{3}Age_{Ur}}{V_{Ur}} & \frac{12\pi}{A_{Ur}} & \frac{8\pi\sqrt{3}Age_U}{A_{Ur}} & \frac{8\pi\sqrt{3}Age_U}{A_{Ur}} \\ \frac{8\pi\sqrt{3}Age_{Ur}}{V_{Ur}} & \frac{8\pi\sqrt{3}}{A_{Ur}} & \frac{12\pi}{A_{Ur}} & \frac{8\pi\sqrt{3}Age_{Ur}}{A_{Ur}} \\ \frac{8\pi\sqrt{3}Age_{Ur}}{V_{Ur}} & \frac{8\pi\sqrt{3}}{A_{Ur}} & \frac{8\pi\sqrt{3}}{A_{Ur}} & \frac{12\pi}{A_{Ur}} \end{vmatrix}$$

$$= \frac{8\pi G_{Ur}}{c^4} \begin{vmatrix} \frac{M_{Ur}c^2}{V_{Ur}} & \frac{M_{Ur}c^2}{A_{Ur}Age_{Ur}} & \frac{M_{Ur}c^2}{A_{Ur}Age_{Ur}} & \frac{M_{Ur}c^2}{A_{Ur}Age_{Ur}} \\ \frac{M_{Ur}c}{V_{Ur}} & \frac{M_{Ur}c^2}{V_{Ur}} & \frac{M_{Ur}c}{A_{Ur}} & \frac{M_{Ur}c}{A_{Ur}} \\ \frac{M_{Ur}c}{V_{Ur}} & \frac{M_{Ur}c}{A_{Ur}Age_{Ur}} & \frac{M_{Ur}c^2}{V_{Ur}} & \frac{M_{Ur}c}{A_{Ur}} \\ \frac{M_{Ur}c}{V_{Ur}} & \frac{M_{Ur}c}{A_{Ur}Age_{Ur}} & \frac{M_{Ur}c}{A_{Ur}Age_{Ur}} & \frac{M_{Ur}c^2}{V_{Ur}} \end{vmatrix}$$

Observation of the universe from the smallest (nuclear particles) to the largest (galaxies) indicates a particularly static condition making our elastic assumption reasonable and applicable at all dimensional scales. Indeed, this is the case. This simple cellular model is congruent with proton density and radius as well as interstellar space density and energy density.

In order to scale this cellular model to these dimensions varying by 15 orders of magnitude, special relativity Lorentz transforms are justifiably suspended due to resonant cancellation in the context of the de Broglie hypothesis. This defines a sub- super- luminal resonant discontinuity ($v_{dx} \sim c$) at a critical temperature $T_c \sim$ 1.3E12 K, $K_B \sim$ 5.7E12 $cm^{-1}$ and B ~ 5.5E-13 cm. This discontinuity forms potential barrier providing a probable mechanism for limiting nuclear size.

Most importantly, the gravity force is linked to the Hubble constant through Planck's constant. This indicates that gravity is linked to other forces at the proton scale and not the Planck scale. Indeed, when the model trisine model is scaled to the Planck scale (B= 1.62E-33 cm), a mass of 1.31E40 $GeV/c^2$ (2.33E16 g) is obtained. This value would appear arbitrary. Indeed the model can be scaled to the Universe mass *($M_U$)* of 3.02E56 grams at a corresponding dimension (*B*) of 1.42E-53 cm.

In a particular Schwinger mechanism associated with graphene structure [83] which is similar to the trisine structure projected in 2D. This is substantiated by inserting the bond length value for graphene (*B* = 1.22E-8 cm) the trisine value for potential field *(E)* eqn 2.5.1 and velocity ($v_{dx}$) eqn 2.4.1, ref 83 eqn 3 (repeated paragraph 2.16) scales nicely (within 5 parts/thousand) with the potential trisine pair production rate per area:

$$1 \; pair/(section \; time_{\pm}) = 2.06\text{E}29 \;\; cm^{-2} \; sec^{-1}$$

for the graphene dimension (f=1 and 2.612 in denominator). And then it is noted that ref 83 eqn 1 equates (within 9 parts/thousand) trisine pair production rate per volume

$$1 \; pair/(cavity \; time_{\pm}) = 3.92\text{E}37 \; cm^{-3} \; sec^{-1}$$

And then it is noted that by logic there exists a trisine pair production rate per length (B)

$$1 \; pair/(2B \; time_{\pm}) = 8.78\text{E}21 \; cm^{-1} \; sec^{-1}$$

and this potential graphene pair creation taking place at a frequency($1/time_{\pm}$) of 1.08E14 /sec.

It is known that paper [83] assumes production of pairs out of a static classical electric field of infinite extent whereas this paper assumes a resonant condition of resonant period *'$time_{\pm}$'* not of infinite extent but represented by condition n=1 for each cell within lattice of ~infinite extent. The correlation can be made that a static field exists for a resonant period *'$time_{\pm}$'* and then it flips to another 'static' condition for another resonant period *'$time_{\pm}$'*.

A correlation between trisine geometry as related to Dark Energy and DAMA is noted. The DAMA experiment is located in the Gran Sasso National Laboratory, Italy at 3500 mwe. The DAMA observed annual oscillation [84], dictates a question in terms of the following logic:

The Milky Way is moving at approximately 627 km/s (at 276 degree galactic longitude, 30 degree galactic latitude) with respect to the photons of the Cosmic Microwave Background (CMB). This has been observed by satellites such as COBE and WMAP as a dipole contribution to the CMB, as photons in equilibrium at the CMB frame get blue-shifted in the direction of the motion and red-shifted in the opposite direction.

and also: the solar system is orbiting galactic center at about 217 km/sec which is essentially additive (217 + 627) or 844 km/sec relative to CMB dipole rest frame.

and also: the earth has a velocity of 30 km/sec around the sun.

and also: refs: [84] Fig 10, harmonic DAMA/NaI experimental data and [111] Fig 10, harmonic CoGeNT experimental data indicates an annual oscillation of detected radiation flux detected on earth.

Is there a correlation of annual DAMA/ CoGeNT experimental data on earth due to earth_solar annual differential velocity projection to CMB velocity vector?

In answer to this question, the CMB 'hot spot' or 'wind origin' (+3.5 mK above the present CMBR 2.729 K) is in constellation Leo at essentially the celestial equator: http://aether.lbl.gov/www/projects/u2/

Given that the earth orbits the sun at 30 km/sec, the hypothesis would be that average flux would be when earth - sun line would be parallel to earth - Leo line which occurs September 1 (sun in Leo) and March 1 with maximum flux (earth against the CMB dipole wind) on December 1 and minimum (CMB dipole wind to earth's back) on June 1. This CMB dipole wind and earth solar rotation are then approximately in phase with DAMA NAI [84] Fig 10 where $t_0$ = 152.5 is ~June 1 (minimum flux). This phase relationship is congruent and also supported with the energy ratios (temperature ratio vs velocity$^2$ ratio) identity $.0035/2.732 = 30^2/(627+217)^2$ within 1 percent.

In contrast, the model presented on the DAMA web site http://people.roma2.infn.it/~dama/web/nai_dmp.html with its earth velocity 30 km/sec relative to galactic rotation indicates a maximum at this date ~June 1. Also note that the CMB 'hot spot' is essentially at the celestial equator (declination = zero). A detector optimized at pointing in this direction relative to that detectors earth latitude could conceivably have more chance of particle detection.

Also it is noted that the Pioneers 10 & 11 had annual deceleration peaks [57] that are generally in phase (maximum on December 1 and minimum on June 1) with the

DAMA/CoGeNT annual flux peaks making both effects having possible correlation with dark energy

The trisine model can be adiabatically correlated to any time in the universe age. There is a correlation of X rays produced at trisine condition 1.3 with X-rays observed by DAMA/CoGeNT [148] implying a congruency between these observations and the early universe expressed in trisine condition 1.3 all within the context of the Heisenberg Uncertainty constraint. The model predicts an oscillation at ~20 particle counts per 30 days at ~1 keVee that is consistent with DAMA/CoGeNT observations and originating at Trisine Condition 1.3 and congruent with the CMBR annual oscillations due to earth rotation around sun and its relationship to CMBR flux from above .0035/2.732 = $30^2/(627+217)^2$.

Particle production per area and time is computed with the following equation times the CoGeNT crystal frontal area based on diameter of 2.2 cm.

$$\sigma T_{cr}^4 \xi \left(k_b T_r\right)^{-\frac{8}{8}} = \left(\frac{B}{A}\right)^{\frac{12}{8}} \frac{\pi^{\frac{8}{8}} G^{\frac{8}{8}} m_T^{\frac{47}{8}} c^{\frac{66}{8}}}{2^{\frac{67}{8}} 3^{\frac{10}{8}} 4^{\frac{8}{8}} 15^{\frac{8}{8}} \hbar^{\frac{28}{8}}} \left(\frac{B_{Present}}{\mathrm{R}_{U_{Present}}}\right)^{\frac{6}{8}} \frac{1}{2\pi} \left(k_b T_r\right)^{-\frac{15}{8}}$$

This directional congruency apparently exists from the time of the Big Bang nucleosynthesis.

Now wind this universe age ($Age_U$) further back to at time in which the universe was at condition 1.2 where proton density (2.34E14 g/cc), background milieu temperature *($T_c$)* 8.96E13 K, CMBR temperature *($T_b$)* 9.07E14K. In the extant universe, both this CMBR *($T_b$)* and background milieu temperature *($T_c$)* are unobserved except for current nuclear observations (trisine condition 1.2) and cosmic rays evident in conclusion 4.33.

It is important to note that at CMBR temperature *($T_b$)* 9.07E14K, the radiation is not longer microwave as CMBR implies, but the term is used for consistency.

At this proton density point, background milieu temperature > CMBR (because of luminal - group phase reflection considerations). With adiabatic expansion, an extremely cold medium at 9.8E-10K. At this temperature, dense (~1 g/cm$^3$) Bose Einstein Condensates (BEC) form from hydrogen helium created by nucleosynthesis through adiabatic expansion with milieu temperature *($T_c$)* going to 9.8E-10K and congruent with extant CMBR *($T_b$)* 3,000 K. or dimensionally

($T_c$ after expansion)/ ($T_c$ before expansion)

= (densityafterExpansion/densitybeforeExpansion)$^{2/3}$

(9.8E-10 /8.96E13) = (9.01E-21/2.34E14)$^{2/3}$

The hypothesis would be that there was a time in which the entire universe was at $T_c$ = 9.8E-10 K and the extant CMBR $T_b$ = 3000 K due to adiabatic expansion after the Big Bang. All the atomic entities (primarily hydrogen) would be in the form of Bose Einstein Condensates BEC's at 9.8E-10 K. This extremely low temperature would allow for dense BEC formation where the fundamental BEC formation formula:

$$k_b T_{BEC} = k_b T_c = \left(n/\zeta\left(3/2\right)\right)^{2/3}\left(h^2/2\pi m\right)$$

is modified in terms of density (*$am_p$/volume* or $\rho$ replacing *n* or */volume*) and multiple proton masses *( $am_p$ )*:

$$k_b T_{BEC} = k_b T_c = \left(\rho/\zeta\left(3/2\right)\right)^{2/3}\left(h^2\Big/2\pi\left(am_p\right)^{5/3}\right)$$

$$\text{where } \zeta\left(3/2\right) = 2.6124$$

BEC's with density (1 g/cm$^3$) could be formed at $T_{BEC} = T_c =$ 9.8E-10 K with *'a'* ~ 1E6 proton masses $m_p$.

This condition (1.10) would not last very long as these dense (~1 g/cc) BEC's (those greater than ~1E6 cm) gravitationally collapsed resulting in heating that initiated stellar nuclear reactions distributed in galaxies with aggregate density ~1E-24 g/cc aided perhaps by the extant CMBR 3000 K at that time but leaving the residual hydrogen BECs as ~ meter objects in thermal equilibrium with all encompassing and ubiquitous $T_b$ ($T_b$ decreasing from 1E-9K to ~1E-16K due to further universe expansion) over the next 4.32E17 sec (13.7 billion years) gravitationally observed today as dark matter.

This extant hydrogen BEC would not be observable by the Beers Lambert Law. As generally described in elementary texts, the optical path established by this law is derived assuming that that particles may be described as having an *area* perpendicular to the path of light through a solution (or space media), such that a photon of light is absorbed if it strikes the particle, and is transmitted if it does not. Expressing the number of photons absorbed by the concentration C in direction *z* in the absorbing volume as $dI_z$, and the total number of photons incident on the absorbing volume as $I_z$, the fraction of photons absorbed by the absorbing volume is given by:

*$dI_z/I_z$ = -area c dz.* There is a large difference (~6E7) in *area* for one mole of individual hydrogen atoms and that same mole in one BEC at 1 g/cc (the concentration c is the same for both cases) *area* of one mole of individual hydrogen atoms

6E23 x 1E-16 cm$^2$/hydrogen atom ~ 6E7 cm$^2$ *area* of one mole of hydrogen BEC with density 1 g/cc ~1 cm$^2$

There is an argument that the Intergalactic Medium (IGM) composed of 'hot' atoms would 'melt' any BECs that were there at the initial Big Bang Nucleosynthesis. In order to address this question, assume the IGM is totally ionized at the given universe critical density of ~1 proton mass per meter$^3$. (It is very clear that this ionized IGM was not there in the beginning but ejected into space from stars etc.) Now take a 1 meter$^3$ hydrogen BEC. It would have $100^3$(*Avogadro's number*) hydrogen atoms in it.

If there was no relative motion between the hydrogen dense (1 g/cc) BEC and the hot atoms, the hydrogen dense (1 g/cc) BEC would not degrade. If there was relative motion between the hydrogen dense (1 g/cc) BEC and the hot atoms, the hydrogen dense (1 g/cc) BEC could travel $100^3$(*Avogadro's number*) meters or 6.023E29 meters through space before 'melting'. The length 6.023E29 meters is larger than the size of the visual universe (2.25E26 meters). There could have been multi kilometer sized hydrogen BEC's at the adiabatic expansion still remaining at nearly that size. Also, as a BEC, it would not have a vapor pressure. It would not sublime. The hydrogen dense (1 g/cc) BEC s are still with us.

As a followup to this reasoning, is `Oumuamua a BEC? Upper limits on its flux' indicate a non standard(comet) object[157]. `Oumuamua non emissivity may indicate `Oumuamua as a Bose Einstein Condensate(BEC). This is in the context of the observed acceleration [155] .

$a_0 = (4.92 \pm 0.16) \times 10^{-4}$ cm/sec$^2$

with modified density (much lower than 1-3 g/cm$^3$) conforming to the standard Bose Einstein Condensate relationship from above, but consider that the reflectivity Coefficient(CR) should be 2 vs 1 expressing near 100% BEC radiation reflectivity and particle mass is the proton mass(hydrogen) then CR=2 and (m/A) =.2 g/cm$^2$

| width(cm) | density (g/cm$^3$) | hBEC Temp(K) | |
|---|---|---|---|
| (cm) | (g/cm$^3$) | (K) | |
| 5.35E+01 | 3.74E-03 | 2.73E+00 | $T_b$ (present) |
| 3.98E+04 | 5.03E-06 | 3.32E-02 | $T_i$ (present) |
| 1.00E+28 | 1.92E-26 | 8.11E-16 | $T_c$ (present) |

The energy associated with the present Temp($T_i$) is measured at wavelengths 9.29 - 44.23 cm encompassing the 21 cm line[158]

The energy associated with present critical Temp($T_c$) is related to a univesal Machian time 33.8 microHz ~1/(8.22 hr) near the `Oumuamua rotaion rate

Note that

$(T_b \text{ density}/T_i \text{ density})^{1/3}$=(3.74E-03/5.03E-06)$^{1/3}$=9.1/1

which is near the 10/1 length to width ratio `Oumuamua dimensional characteristic as optically observed ref1.

Where did `Oumuamua come from[159]? Maybe `Oumuamua came from the interstellar space out to the galactic halo(R). Given a typical halo density ~1E-26 g/cc, one can assume that the same aggregate density may be composed of individual BECs(d), perhaps shaped like `Oumuamua, that cannot be seen with our present instrumentation due to a halo mean free path limitation ~ hydrogen_BEC size(d) > R. Calculations indicate these particles with size on the order of `Oumuamua would be ~1,000,000 kilometers apart, with not too much chance to interact. This would be congruent with `Oumuamua originated from a very special frame of reference, the so-called Local Standard of Rest (LSR)[159].

The BEC stability may be due to its very discrete absorption/emission characteristics (1S-2S characteristic hydrogen_BEC hydrogen transition 243 nm (1.23E+15 Hz) absorption with a band width of ~1E6 Hz and a natural line width of 1.3 Hz) calculated herein as BECST.

As a BEC, `Oumuamua would reflect nearly 100 percent of EM radiation except for extremely absorbed ~1E6 Hz.

There is experimental and theoretical work on BEC reflectivity[160].

The actual BEC may be a mixture of elements other than hydrogen but not negating the above generalizations.

The Bottom line is: If you were Captain Kirk of the Star Ship Enterprise would you go full speed ahead into inter- intra-galactic space analogous to Captain Smith of the Titanic? I certainly would not.

## 4. Conclusion

A momentum and energy conserving (elastic) CPT lattice called trisine and associated superconducting theory is postulated whereby electromagnetic and gravitational forces are mediated by a particle of resonant transformed mass (110.107178208 x electron mass or 56.26465274 MeV/c$^2$) such that the established electron/proton mass is maintained, electron and proton charge is maintained and the universe radius as computed from Einstein's cosmological parameter $(\Lambda)$ is 1.29E-56 cm$^{-2}$, the universe mass is 3.0E56 gram, the universe density is 6.38E-30 g/cm$^3$ and the universe age($Age_U$) is 1.37E10 years.

The cosmological parameter is based on a universe surface tension directly computed from a superconducting resonant energy over surface area.

The calculated universe mass and density are based on an isotropic homogeneous media filling the vacuum of space analogous to the 'aether' referred to in the 19$^{th}$ century (but still in conformance with Einstein's relativity theory) and could be considered a candidate for the 'cold dark matter' (or more properly 'cold dark energy') in present universe theories and as developed by Primack [80]. Universe density is 2/3 of conventionally calculated critical density. This is in conformance with currently accepted inflationary state of the universe.

The 'cold dark energy', a postulated herein, is in equilibrium with and congruent to a CMBR of 2.729 $^oK$ as has been observed by COBE, WMAP and later satellites although its temperature in terms of superconducting critical temperature is an extremely cold 8.11E-16$^oK$.

Also, the reported results by Podkletnov and Nieminen wherein the weight of an object is lessened by .05% are theoretically confirmed where the object is the superconductor, although the gravitational shielding phenomenon of $YBa_2Cu_3O_{7-x}$ is yet to be explained. It is understood that the Podkletnov and Nieminen experiments have not been replicated at this time. It could be that the very high super currents that are required and theoretically possible have not be replicated from original experiment.

The trisine model provides a basis for considering the phenomenon of superconductivity in terms of an ideal gas with 1, 2 or 3 dimensions (degrees of freedom). The trisine model was developed primarily in terms of 1 dimension, but experimental data from $MgB_2$ would indicate a direct translation to 3 dimensions.

These trisine structures have an analog in various resonant structures postulated in the field of chemistry to explain properties of delocalized $\pi$ electrons in benzene, ozone and a myriad of other molecules with conservation of energy and momentum conditions defining a superconductive (elastic) condition.

Other verifying evidence is the trisine model correlation to the ubiquitous 160-minute (~third $T_c$ harmonic - 29,600/(60x3) resonant oscillation [68,69] in the universe and also the model's explanation of the Tao effect [65, 66, 67].

Also dimensional guidelines are provided for design of room temperature superconductors and beyond. These dimensional scaling guidelines have been verified by Homes' Law and generally fit within the conventionally described superconductor operating in the "dirty" limit. [54]

The deceleration observed by Pioneer 10 & 11 as they exited the solar system into deep space appear to verify the existence of an space energy density consistent with the amount theoretically presented in this report. This translational deceleration is independent of the spacecraft velocity. Also, the same space density explains the Pioneer spacecraft rotational deceleration. This is remarkable in that translational and rotational velocities differ by three (3) orders of magnitude.

4.1. Universe Dark Energy is represented by a trisine elastic space CPT lattice (observed as $\Omega_\Lambda$)which is composed of virtual particles within the Heisenberg uncertainty without recourse to net matter-antimatter annihilation concepts.

4.2. Dark Energy is related to Cosmic Microwave Background Radiation (CMBR) through very large space dielectric $(\varepsilon)$ (permittivity) and permeability $(k)$ constants.

4.3. A very large trisine permittivity (dielectric) constant (6.16E10), permeability (6.26E13) and CPT equivalence makes the Dark Energy invisible to electromagnetic radiation.

4.4. A velocity independent thrust force acts on all objects passing through the trisine space CPT lattice and models the translational and rotational deceleration of Pioneer 10 & 11 spacecraft as well as dust clumping at initial stages of stellar galactic formation and deceleration of local galactic baryonic matter relative to Type 1A supernovae (an observation more conventionally reported as universe expansion acceleration).

4.5. The Asteroid Belt marks the interface between solar wind and trisine, the radial extent an indication of the dynamics of the energetic interplay of these two fields

4.6. The trisine model predicts no 'dust' in the Kuiper Belt. The Kuiper Belt is made of large Asteroid like objects.

4.7. The observed Matter Clumping in Galaxies is due to magnetic or electrical energetics in these particular areas at such a level as to destroy the coherent trisine CPT lattice.

4.8. The apparent approximately constant star rotation (~ 220 km/sec for Milky Way galaxy) with radius ($R$) from galactic center in galaxies is due to preponderance of galactic mass ($M_d$) represented by small objects (~10 - 20 cm) which rotate around the galactic center in a Keplerian manner with orbital velocity ($v_d$) relationship:

$$v_d^2 = \frac{GM_d}{R}$$

Stars with their apparent mass ($M_x$) (gravity coupled to this dark matter) appear to rotate about the galactic center at distance R with orbital velocity ($v_s$) relationship:

$$v_s^2 = \frac{GM_s}{R} \quad \text{where} \quad \frac{M_s}{R} \sim \text{constant}$$

This is a residual effect left over from the time when galaxies were primarily dust or small particles when the formula:

$$deceleration = C_d A \rho_U c^2$$

$\rho_U$ = conventionally accepted space vacuum density

$$\rho_U = \frac{2}{8\pi}\frac{H_U^2}{G} \approx \frac{1}{8\pi}\frac{c^2}{G}\Lambda\cos(\theta)$$

A = dark object cross sectional area

$C_t$ = thrust coefficient

was primarily applicable and is still applicable in as much as galaxies still consist of small particles (< a few meters) gravitationally coupled to large objects such as stars. These small particles have mean free paths ~1,000 times greater than the galactic size which indicates that they are an optically invisible form outside the galaxy and can truly be called 'dark but real matter'. The same principle could be applied to Mira's tail formation although these tail components are visible due to some type of emission and/or absorption mechanism.

4.9. The observed Universe (unaccelerated) Expansion is due to trisine space CPT lattice $(\Omega_\Lambda)$ internal pressure and is at rate defined by the speed of light ($c$) escape velocity. Baryonic mass within that trisine space CPT lattice $(1-\Omega_\Lambda)$ 'B' frame is slowing down or decelerating relative to the universe '$R_U$' frame giving the appearance of expansion with distance in that $R_U$ frame.

$$Age_U \sim B^3 \qquad Age_U \sim R_U$$

$$dAge_U \sim B^2 dB \qquad dAge_U \sim R_U^0 dR_U$$

$$d^2Age_U \sim B^1 d^2 B \qquad d^2Age_U \sim R_U^{-1} d^2 R_U$$

$$\frac{d^2 B}{dAge_U^2} \sim B^{-1} \qquad \frac{d^2 R_U}{dAge_U^2} \sim R_U^1$$

4.10. When the trisine is scaled to molecular dimensions, superconductor resonant parameters such as critical fields, penetration depths, coherence lengths, Cooper CPT Charge conjugated pair densities are modeled.

4.11. A superconductor is consider electrically neutral with balanced conjugated charge as per CPT theorem. Observed current flow is due to one CPT time frame being pinned relative to observer.

4.12. The nuclear Delta particles $\Delta^{++}$, $\Delta^{+}$, $\Delta^{0}$, $\Delta^{-}$ with corresponding quark structure uuu, uud, udd, ddd, with masses of 1,232 MeV/c² are generally congruent with the trisine prediction of $(4/3)\hbar K_B c$ or 1,242 MeV/c² (compared to $\hbar K_B c$ or 932 MeV/c² for the proton 'ud' mass) where $B$ reflects the nuclear radius $3\hbar/(16 m_t c)$ *or* 6.65E-14 *cm.* Also, the experimentally determined delta particle (6E-24 seconds) mean lifetime is consistent and understandably longer than the proton resonant $time_\pm$ (2.68E-25 second from Table 2.4.1).

4.13. It is noted that the difference in neutron ($m_n$) and proton ($m_p$) masses about two electron ($m_e$) masses ($m_n$-$m_p$~2$m_e$).

4.14. The nuclear weak force boson particles $(W^-W^+)$ are associated with energies

$$\frac{B}{A}\frac{\hbar^2\left(K_B^2+K_C^2\right)}{2m_T} = 79.97\ GeV$$

$$\frac{B}{A}\frac{\hbar^2\left(K_{Ds}^2+K_{Dn}^2\right)}{2m_T}\cdot g_s = 79.97\ GeV$$

$$4\pi\frac{B}{A}\hbar\left(K_B+K_C\right)c = 79.67\ GeV$$

$$4\pi\frac{B}{A}\hbar\left(K_{Ds}+K_{Dn}\right)c\cdot g_s = 79.67\ GeV\left(80.385\pm.015\right)$$

The Z boson is associated with energy

$$\mathrm{e}\frac{\hbar^2\left(K_B^2+K_C^2\right)}{2m_T} = 91.37\ GeV$$

$$\mathrm{e}\frac{\hbar^2\left(K_{Ds}^2+K_{Dn}^2\right)}{2m_T}\cdot g_s = 91.37\ GeV$$

$$4\pi\cdot\mathrm{e}\cdot\hbar\left(K_B+K_C\right)c = 91.03\ GeV$$

$$4\pi\cdot\mathrm{e}\cdot\hbar\left(K_{Ds}+K_{Dn}\right)c\cdot g_s = 91.03\ GeV\left(91.1876\pm.0021\right)$$

e is natural log base 2.7182818284590

4.15. The top quark is associated with energy [156]

$$\frac{\hbar^2 K_A^2}{2m_T} = 174.798\ GeV\ \left(174.98\pm.76\right)$$

4.16. The up (u) quark is associated with

$$\frac{1}{3}\frac{m_e}{m_T}\hbar K_B\ c = 2.82\ MeV\left(1.7-3.1\right)$$

The down (d) quark is associated with

$$\frac{2}{3}\frac{m_e}{m_T}\hbar K_B\ c = 5.64\ MeV\left(4.1-5.7\right)$$

The charm (c) quark is associated with

$$\frac{1}{3}\hbar K_A\ c = 1.48\ GeV\left(1.29\ +.50\ -.11\right)$$

4.17. The strange quark (s) is associated with

$$\frac{1}{3}\frac{m_e}{m_T}\frac{\hbar^2}{2m_T}\left(K_{Ds}^2 + K_B^2\right) = 96.13\ MeV\ \left(95 \pm 5\right).$$

4.18. The bottom (b) quark is associated with

$$\hbar K_A\ c = 4,435\ MeV\left(4,190\ +180\ -60\right)$$

verified [147]

4.19. The Gluon ratio is 60/25 = 2.4 ≅ $\pi^0/m_t$ ≅ *B/A* =2.379761271.

4.20. Active Galactic Nuclei (AGN) power spectrum as typically presented in reference 89 is indicative of dark energy (related to $\Omega_\Lambda$) presence as described herein and coincident with the black body nature of trisine dark energy at 8.11E-16 K at a frequency of $1/time_\pm$ or 3.38E-5 Hz.

4.20a. Anomalous Cosmic Rays (ACRs) at 56 MeV are detected by the NASA Voyager I spacecraft prior to passage, during passage and transit beyond our solar system terminal shock. The solar system terminal shock is not the source of Anomalous Cosmic Rays (ACRs). The universe matter is presently 70% composed of a cellular fluid having density 6.38E-30 g/cc as indicated by astrophysical relationship Eqns 2.11.7 and 2.11.22 and each cell cavity (15,733 $cm^3$) composed of an ubiquitous unit mass ($m_T$ = 56 MeV/$c^2$) having that universe density $m_T/cavity$ = 6.38E-30 g/cc.

(This universal density can be considered 'dark energy' $m_T$ is of the same order as the Weinberg mass, but it is not the same form as represented in eqn 2.11.15.

The cellular cavity nature can best described within the:

Charge ($e_\pm$) - conjugation

Parity (*trisine mirror images* )- change

Time ($time_\pm$) - reversal (CPT) theorem

In this context, when any particle or object intersects these cavities, momentum is transferred to the particle or object resulting in a electromagnetic gamma ray release of ~56 MeV (the particle slows down) (This is the mechanism for Pioneer 10 and 11 deceleration as described in chapter 2.) After their release, the 56 MeV gamma ray can interact with all particles, atoms, molecules quantum mechanically changing their energy states on the order of MeV resulting in what has come to be called:

Anomalous Cosmic Rays (ACRs)

Voyager 1 and Voyager 2 will monitor these ACR's as they proceed beyond the terminal shock as long as particles, atoms, molecules exist within interstellar and intergalactic space.

Such an hypothesis is subject to simple 'back of the envelop analysis) or more extensive computer analysis with appropriate algorithms. Essentially Voyager 1 and Voyager 2 are indirectly measuring the 'dark energy'. The presently operating FERMI satellite may directly measure the ubiquitous 56 Mev (in direct 'dark energy' measurement) in the Extra Galactic Background Gamma Ray Emission as was indicated by the previous EGRET satellite.

FERMI instrument systematics are not favorable for 56 MeV detection but such results are being processed by the FERMI team.

(FERMI 56 MeV results are eagerly awaited in conjunction with Voyager1 and Voyager2 ACR's)

4.21. The numerical and dimensional correlation between trisine lattice supercurrent and Maxwell's Ampere's Law is indicated with experimental verification by magnesium boride $MgB_2$ supercurrent of 1E7 amp/cm$^2$ at $T_c$ of 39/3 K [82].

4.22. The cosmological parameter $(\Lambda)$ and Hubble Constant $(H_U)$ are dimensionally related by the speed of light $(c)$ and the trisine angle $(\theta)$ of 22.8$^o$ :

$$\Lambda_U = 2H_U^2/\left(c^2\cos(\theta)\right)$$

and also

$$\Lambda_U = \cos(\theta)^{2/3}\left(m_e/\left(\alpha l_P^{1/3} m_P\right)\right)^6$$

4.23. The cosmological constant simple dimensionless relationship $\left(\Omega_{\Lambda_U} = 2/(3\cos(\theta)\right)$ may be a direct measure of trisine space lattice angle $(\theta)$.

4.24. The present $\left(T_b = 2.729K\right)$ CMBR flux $=\sigma_S T_b^4$

=

$$2cos(\theta)\left(\frac{m_p}{m_T}\right)\left(\frac{m_T}{m_e}\right)\left(\frac{m_p}{m_e}\right)\left(\frac{2.821439372\ k_B T_B}{cavity}\right)$$

= 3.14E-03 (erg cm$^{-2}$ sec$^{-1}$)

Also the CMBR energy density is:

$$\sigma_S T_b^4 \left(4/c\right) = 4.20\text{E-}13 \; erg/cm^3$$

Which in terms of mass density is:

$$\sigma_S T_b^4 \left(4/c^3\right) = 4.67\text{E-}34 \; g/cm^3$$

or 7.32E-05 of universe critical density $\left(\rho_U\right)$ 6.38E-30 *g/cm³*
Consequently, the universe critical density is a very large cold reservoir capable of maintaining hydrogen dense (1 g/cc) BEC's at the critical temperature $T_c$ = 8.11E-16 K.

4.25. The Euler constant in terms of fundamental constants may be the Rosetta stone in correlating the superconducting phenomenon and in particular these two correlations.

$$Euler \sim 1/sqrt(3) = F_P / \left(M_U \; H_U c\right) \; = c^3 / \left(M_U \; GH_U\right)$$

$$Euler \sim 1/sqrt(3) = -\ln\left(2\Delta_o^{BCS} / \left(k_b T_c\right)\right) / \left(2\pi\right)$$

4.26. Definition of electron mass *($m_e$)*

$$m_e = \alpha m_P \left(\Lambda_U / \cos(\theta)^{2/3}\right)^{1/6} l_P^{1/3} = \alpha \hbar^{2/3} G_U^{-1/3} \left(\Lambda_U / \cos(\theta)^{2/3}\right)^{1/6}$$

and proton mass *($m_p$)*

$$m_p = h^2 \left(\Lambda_U \cos(\theta)/2\right)^{1/2} / \left(3 G_U m_t^2\right)$$

provide a universal cosmological continuity with generalized superconducting equation 2.1.13b repeated here:

$$k_b T_c = \frac{\hbar^2}{\dfrac{m_e m_T}{m_e + m_T}\left(2B\right)^2 + \dfrac{m_T m_p}{m_T + m_p}\left(A\right)^2}$$

This correlation would be applicable to any superconductory equation with the proton mass *($m_p$)* or electron mass *($m_e$)* as a component.

4.27. The increase in spacecraft flyby energy over classical Newtonian calculation is correlated to the universe resonance with earth's atmosphere and solid earth via De Broglie matter waves and its gravitational energy transfer to the spacecraft in unsymmetrical flyby earth or other planetary passage.

4.28. The inherent Trisine B/A ratio may explain the universe alignment of galactic rotation and CMBR anomalies according to what has become called 'The Axis of Evil'

4.29. A correlation would be that there was a time in which the entire universe was at $T_b$ ~1E-9 K and the CMBR $T_c$ was at ~3,000 K. The correlation follows:

$$(1\text{E-}9/1\text{E}13) \sim (1\text{E-}20/2\text{E}14)^{2/3}$$

due to adiabatic expansion after the Big Bang. All the atomic entities created during the nucleosynthesis event (primarily hydrogen and helium) would be in the form of BECs at ~1E-9 K. This condition would not last very long as these clouds of BEC's gravitationally condensed resulting in heating that initiated stellar nuclear reactions distributed in galaxies with aggregate density ~1E-24 g/cc aided perhaps by the extant CMBR 3,000 K at that time but leaving the residual hydrogen BECs in thermal equilibrium with all encompassing and ubiquitous $T_c$ decreasing from 9.80E-10K to 8.11E-16K due to further universe expansion) over the next 13.7 billion years gravitationally observed today as dark matter. This scenario would also resolve the issue of the observed hydrogen helium ratios in proto galaxies similar to nucleosynthesis values. The dark matter was simply hydrogen and helium produced by nucleosynthetic adiabatic expansion frozen out (as dense BEC's) of the rest leaving the star hydrogen and helium dynamics the same.
These hydrogen dense (1 g/cc) BEC's would be largely unnoticed by 21 cm line measurements. Assume that these primeval (through adiabatic Big Bang expansion) hydrogen dense (1 g/cc) BEC 's were in rather large chunks, say 10's of kilometers in diameter. The parallel to antiparallel hydrogen (proton electron spin) configuration resulting in 21 cm line would occur randomly in the BEC chunk hydrogens at the known 2.9E-15 /sec rate. Transmittance accordance with Beer's law would indicate that the 22 cm radiation from the BEC center would be attenuated more than that from the BEC near surface resulting in a lower than reality hydrogen spatial density reading obtained by a distant observer. Also, the momentum of a hydrogen nanokelvin BEC at density of 1 g/cc would interact by momentum transfer with a photon of wavelength ~140 cm which is much lower than CMBR (.1 cm) or the 21 cm line.

4.30. The critical energy $k_b T_{BEC}$ (at the zero point) associated with BEC formation is related to $kT_c$ through characteristic angle $\theta$ (22.8°) by $k_b T_c = k_b T_{BEC} \cos\left(\theta\right)$ in essence defining a BCS – BEC crossover.

4.31. A residual delta frequency (~3 MHz) is imposed on HI 1420.40575177 MHz (21 cm neutral hydrogen line) assuming to originate at the recombination event [98]. The residual delta frequency is imposed by continuum foreground sources.

It is suggested that the residual delta frequency is a direct indication of 'dark energy' as generally described in this paper as trisine geometry and can be quantified by the equation:

$$\frac{m_e}{m_t}\frac{c}{2B} = \text{6E6 Hz (6 MHz)}$$

Essentially the HI oscillatory frequency results from HI 1420.40575177 MHz electromagnetic radiation coupling with the 'dark energy' lattice with trisine geometry.

This ubiquitous trisine lattice is extremely tenuous, having a density of 6.38E-30 g/cm$^3$ $\left(m_t/cavity \text{ or } M_U/V_U\right)$ and critical temperature $T_c$ = 8.11E-16 K, and this lattice making up ~72.3 percent of the universe back to nucleosynthetic Big Bang epoch that has come to be known as 'dark energy'.

4.32 Mantz[107] provides galaxy cluster energy as a function of redshift '$z$' and analysis thereof. This cluster data is replicated in Table 4.32 without the error bars. Mantz[107] indicates a statistical relationship between E($z$) M2500 ($10^{14}$ solar mass) and $k_bT$ of log log slope 1.91±0.19. Within this context, the trisine model predicts a relationship between universe density $\rho_U$ and $Age_{UG}$ of log log slope 2 as indicated on Figure 4.32 blue line and this relationship expressed as

$$\rho_U = \left(constant/Age_{UG}\right)^2$$

where $Age_{UG} = Age_U \left(G_U/G\right)^{1/2} = \left(1/H_U\right)\left(G_U/G\right)^{1/2}$.

In the Mantz[107] analysis, $E(z) \sim \left(\rho(z)\right)^{1/2}$.

This Mantz[107] relationship assumes a Newtonian universally fixed gravity (G) in

$$\rho \sim H_U^2/G \text{ and } E(z) \sim H_U$$

and is not dimensionally consistent with universe dimensional expansion. In the trisine hypothesis, the universally constant ratio $H_U/G_U$ eliminates the square root and establishes the relationship

$$E(z) \sim \rho_U(z) \sim H_U^2/G_U .$$

In the Mantz[107] presentation, there is no attempt to dimensionally relate E($z$) and $k_bT$. This may be done in the future, but the slope = 2 related to $1/Age_{UG}$ or

$$\rho_U = \left(\text{constant}/Age_{UG}\right)^2$$

correlates the Mantz[107] analysis to the trisine model where $h/Age_{UG}$ is energy dimensionally and related to change in galaxy cluster plasma energy $k_bT$ with $z$. It is concluded that Newtonian gravity changes $\left(G_U\right)$ with $z$ as reflected in cluster spatial and plasma dynamics.

Table 4.32 Galactic Cluster Data

| Cluster | z | E(z) | M2500 (10^14 solarmass) | $k_bT$ (keV) | E(z) M2500 (10^14 solar mass) |
|---|---|---|---|---|---|
| 3C295 | 0.46 | 1.28 | 1.71 | 5.27 | 2.19 |
| Abell-1795 | 0.06 | 1.03 | 2.79 | 6.14 | 2.87 |
| Abell-1835 | 0.25 | 1.14 | 5.87 | 9.11 | 6.66 |
| Abell-2029 | 0.08 | 1.04 | 3.54 | 8.22 | 3.67 |
| Abell-2204 | 0.15 | 1.08 | 4.09 | 8.22 | 4.40 |
| Abell-2390 | 0.23 | 1.12 | 5.19 | 10.20 | 5.82 |
| Abell-2537 | 0.30 | 1.16 | 2.67 | 8.03 | 3.11 |
| Abell-383 | 0.19 | 1.10 | 2.16 | 5.36 | 2.37 |
| Abell-478 | 0.09 | 1.04 | 4.11 | 7.96 | 4.28 |
| Abell-611 | 0.29 | 1.16 | 2.65 | 6.85 | 3.07 |
| Abell-963 | 0.21 | 1.11 | 2.74 | 6.64 | 3.03 |
| CLJ1226.9+3332 | 0.89 | 1.65 | 5.46 | 11.90 | 9.03 |
| MACSJ0011.7−1523 | 0.38 | 1.22 | 2.59 | 6.42 | 3.16 |
| MACSJ0159.8−0849 | 0.40 | 1.24 | 4.63 | 9.08 | 5.73 |
| MACSJ0242.6−2132 | 0.31 | 1.18 | 2.14 | 6.10 | 2.51 |
| MACSJ0329.7−0212 | 0.45 | 1.27 | 2.56 | 6.34 | 3.25 |
| MACSJ0429.6−0253 | 0.40 | 1.23 | 1.83 | 6.61 | 2.26 |
| MACSJ0744.9+3927 | 0.69 | 1.46 | 3.07 | 8.08 | 4.49 |
| MACSJ0947.2+7623 | 0.35 | 1.20 | 4.26 | 8.80 | 5.09 |
| MACSJ1115.8+0129 | 0.36 | 1.20 | 6.02 | 8.08 | 7.24 |
| MACSJ1311.0−0311 | 0.49 | 1.30 | 2.37 | 6.00 | 3.09 |
| MACSJ1359.2−1929 | 0.45 | 1.27 | 2.20 | 6.27 | 2.79 |
| MACSJ1423.8+2404 | 0.54 | 1.34 | 2.59 | 6.92 | 3.47 |
| MACSJ1427.3+4408 | 0.49 | 1.30 | 1.88 | 8.19 | 2.44 |
| MACSJ1427.6−2521 | 0.32 | 1.18 | 1.37 | 4.61 | 1.61 |
| MACSJ1532.9+3021 | 0.36 | 1.21 | 3.32 | 6.86 | 4.01 |
| MACSJ1621.6+3810 | 0.46 | 1.28 | 2.83 | 6.93 | 3.62 |
| MACSJ1720.3+3536 | 0.39 | 1.23 | 3.02 | 7.08 | 3.71 |
| MACSJ1931.8−2635 | 0.35 | 1.20 | 4.02 | 8.05 | 4.83 |
| MACSJ2046.0−3430 | 0.42 | 1.25 | 1.57 | 4.78 | 1.96 |
| MACSJ2229.8−2756 | 0.32 | 1.18 | 1.41 | 5.33 | 1.67 |
| MS2137.3−2353 | 0.31 | 1.17 | 2.15 | 5.56 | 2.52 |
| RXJ0439.0+0521 | 0.21 | 1.11 | 1.63 | 5.19 | 1.81 |
| RXJ1347.5−1144 | 0.45 | 1.27 | 10.80 | 13.80 | 13.73 |
| RXJ1504.1−0248 | 0.22 | 1.11 | 5.33 | 8.19 | 5.93 |
| RXJ2129.6+0005 | 0.24 | 1.13 | 2.34 | 6.24 | 2.63 |
| Zwicky=3146 | 0.29 | 1.16 | 5.99 | 7.54 | 6.92 |

Figure 4.32

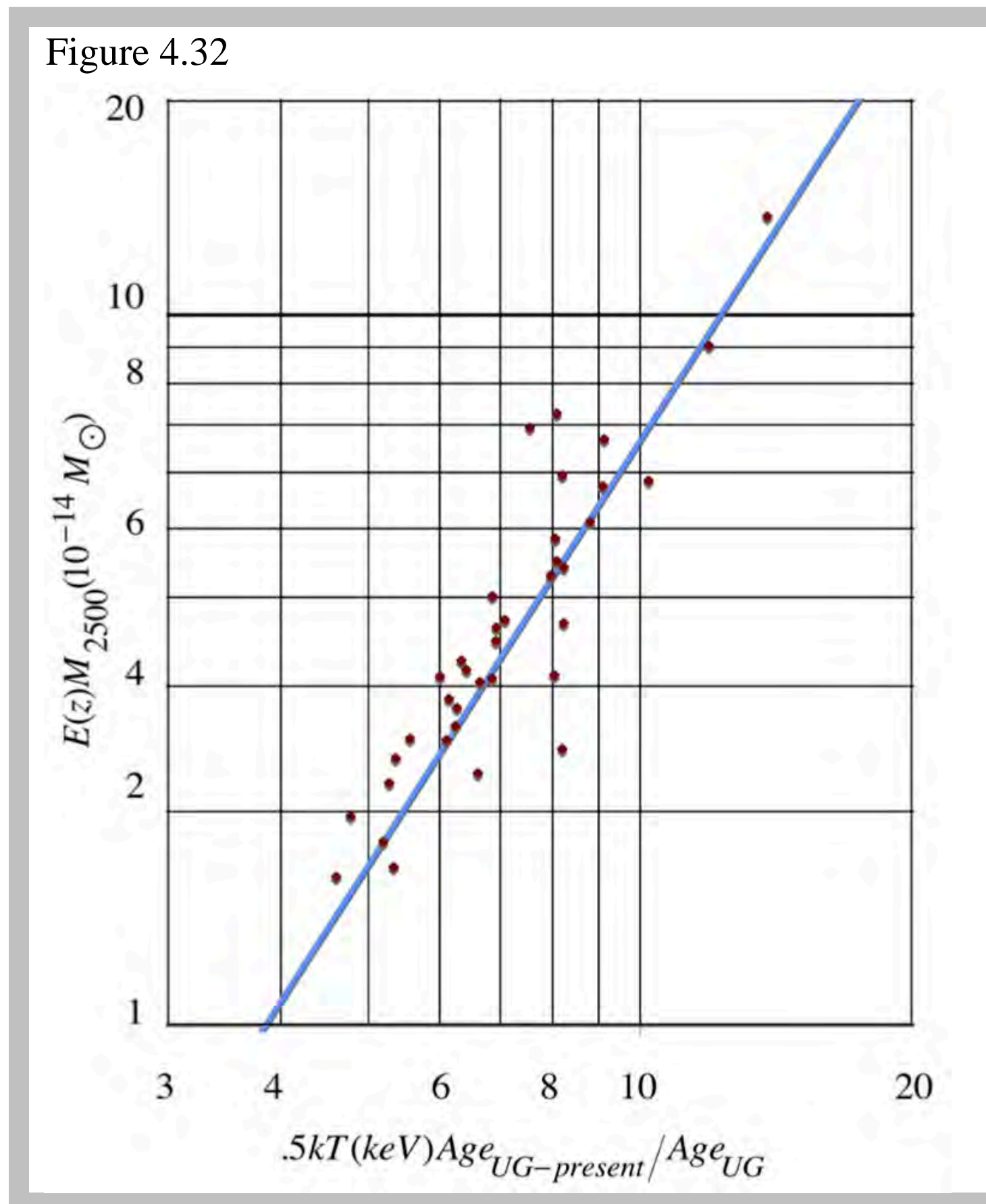

4.33. A model is developed for expressing this data using the Trisine parameters developed herein within the context of Black Body theory having the basic dimensional units of energy/area/time and then scaled to universe time and volume development:

$$T_{cr} = (1/2)\, m_T v_{dxr}^2 / k_b = m_T c^2 / (2 k_{mxr} \varepsilon_{xr} k_b)$$

$$\sigma T_{cr}^4 \xi = \sigma T_{cr}^4 \frac{cavity}{V_U} \frac{c}{4} \frac{Age_U}{Age_{UG}} \left( \frac{B_{Present}}{\mathrm{R}_{U_{Present}}} \right)^{\frac{6}{8}} \frac{1}{2\pi}$$

$$= \left( \frac{B}{A} \right)^{\frac{12}{8}} \frac{\pi^{\frac{8}{8}} G^{\frac{8}{8}} m_t^{\frac{47}{8}} c^{\frac{66}{8}}}{2^{\frac{67}{8}} 3^{\frac{10}{8}} 4^{\frac{8}{8}} 15^{\frac{8}{8}} \hbar^{\frac{28}{8}}} \left( \frac{B_{Present}}{\mathrm{R}_{U_{Present}}} \right)^{\frac{6}{8}} \frac{1}{2\pi} (k_b T_j)^{-\frac{7}{8}}$$

or in the numerical format in which this cosmological data is traditionally presented

$$\sigma T_{cr}^4 \xi (k_b T_j)^{-\frac{16}{8}} = \sigma T_{cr}^4 \frac{cavity}{V_U} \frac{c}{4} \frac{Age_U}{Age_{UG}} \left( \frac{B_{Present}}{\mathrm{R}_{U_{Present}}} \right)^{\frac{6}{8}} \frac{1}{2\pi} (k_b T_j)^{-\frac{16}{8}}$$

$$= \left( \frac{B}{A} \right)^{\frac{12}{8}} \frac{\pi^{\frac{8}{8}} G^{\frac{8}{8}} m_T^{\frac{47}{8}} c^{\frac{66}{8}}}{2^{\frac{67}{8}} 3^{\frac{10}{8}} 4^{\frac{8}{8}} 15^{\frac{8}{8}} \hbar^{\frac{28}{8}}} \left( \frac{B_{Present}}{\mathrm{R}_{U_{Present}}} \right)^{\frac{6}{8}} \frac{1}{2\pi} (k_b T_j)^{-\frac{7}{8}} (k_b T_j)^{-\frac{16}{8}}$$

$$= \left( \frac{B}{A} \right)^{\frac{12}{8}} \frac{\pi^{\frac{8}{8}} G^{\frac{8}{8}} m_T^{\frac{47}{8}} c^{\frac{66}{8}}}{2^{\frac{67}{8}} 3^{\frac{10}{8}} 4^{\frac{8}{8}} 15^{\frac{8}{8}} \hbar^{\frac{28}{8}}} \left( \frac{B_{Present}}{\mathrm{R}_{U_{Present}}} \right)^{\frac{6}{8}} \frac{1}{2\pi} (k_b T_j)^{-\frac{23}{8}}$$

where the resonant mass temperature $(T_j)$ is defined as:

$$T_j = (m_T v_{dT}^4 / 4)(1/c^2)/k_b$$
$$= (m_T (v_{dx}^2 + v_{dy}^2 + v_{dz}^2)^2 / 4)(1/c^2)/k_b$$
$$= \frac{1}{4} \frac{m_T c^2}{(k_m \varepsilon)^2} \frac{1}{k_b}$$

and resonant mass $(m_r)$

$$m_r = m_T v_{dx}^2 / c^2$$

and permiability $(\varepsilon)$ and permittivity $(\varepsilon)$ relationships

$$k_m = \frac{1}{\varepsilon} \frac{1}{\frac{1}{\varepsilon_x k_{mx}} + \frac{1}{\varepsilon_y k_{my}} + \frac{1}{\varepsilon_z k_{mz}}} \qquad \varepsilon = \frac{1}{\frac{1}{\varepsilon_x} + \frac{1}{\varepsilon_y} + \frac{1}{\varepsilon_z}}$$

$$T_j = \frac{1}{4} \frac{m_T c^2}{(k_m \varepsilon)^2} \frac{1}{k_b} \qquad T_{jr} = \frac{1}{4} \frac{m_T c^2}{(k_{mr} \varepsilon_r)^2} \frac{1}{k_b}$$

The following condition represents particle production per area and time.

$$\sigma T_{cr}^4 \xi (k_b T_j)^{-\frac{8}{8}} = \left( \frac{B}{A} \right)^{\frac{12}{8}} \frac{\pi^{\frac{8}{8}} G^{\frac{8}{8}} m_T^{\frac{47}{8}} c^{\frac{66}{8}}}{2^{\frac{67}{8}} 3^{\frac{10}{8}} 4^{\frac{8}{8}} 15^{\frac{8}{8}} \hbar^{\frac{28}{8}}} \left( \frac{B_{Present}}{\mathrm{R}_{U_{Present}}} \right)^{\frac{6}{8}} \frac{1}{2\pi} (k_b T_j)^{-\frac{15}{8}}$$

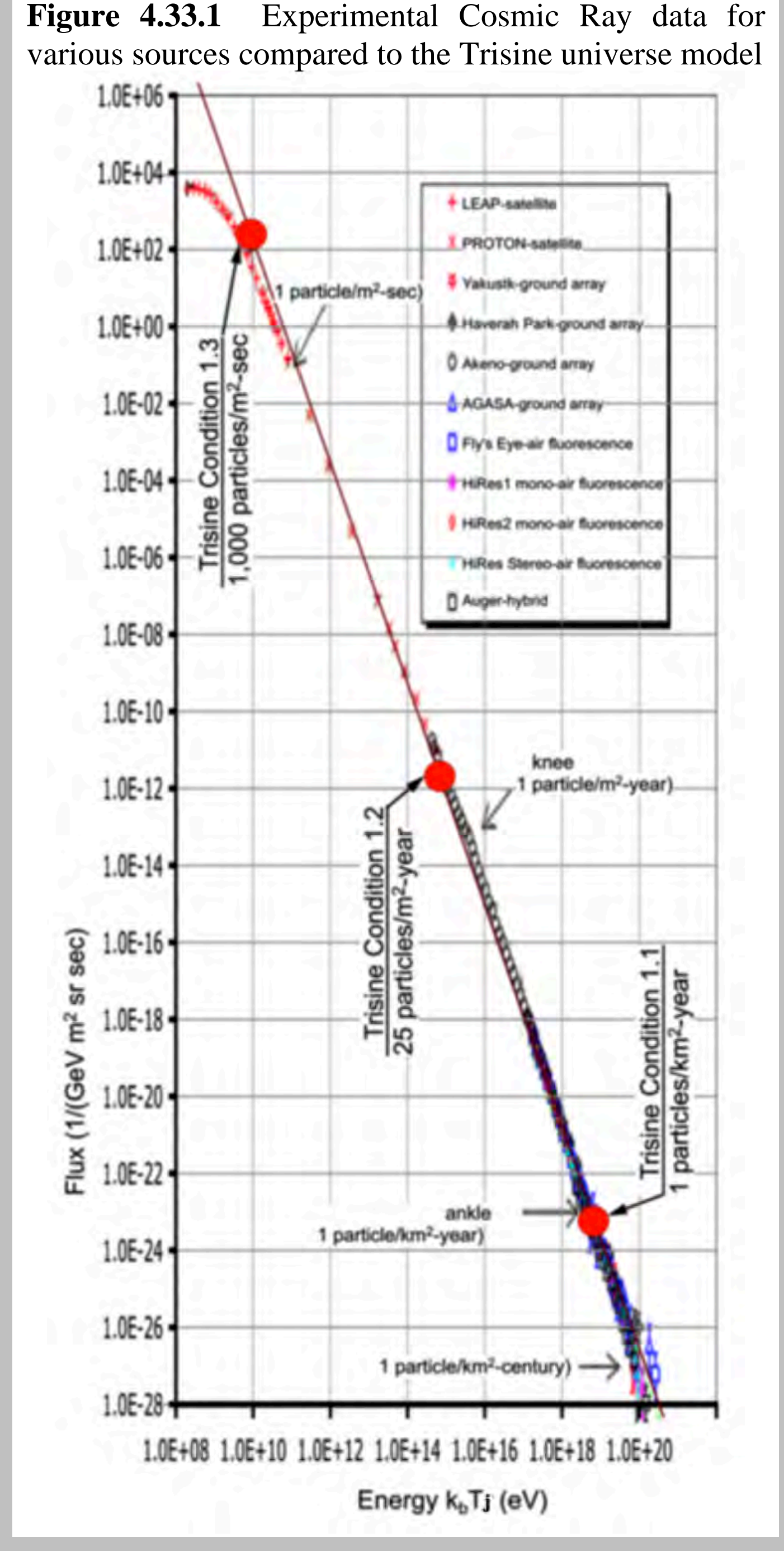

**Figure 4.33.1** Experimental Cosmic Ray data for various sources compared to the Trisine universe model

Conditions 1.1, 1.2 and 1.3 are clearly defined in Figure 4.33.1 as representing group phase reflection nucleosynthetic conditions originating the observed cosmic ray particles by various ground and satellite platforms including the recently launched Alpha Magnetic Spectrometer-02 (AMS-02) (repeated here from the introduction) and IceTop stations [119] in the knee proximity.

| | | | | | |
|---|---|---|---|---|---|
| 1.1 | $z_R$ | 3.82E+46 | $z_R$ | 3.82E+46 | *unitless* |
| | $z_B$ | 3.37E+15 | $z_B$ | 3.37E+15 | *unitless* |
| | $T_a$ | 6.53E+11 | $T_a$ | 6.53E+11 | $^oK$ |
| | $T_b$ | 9.19E+15 | $T_{br}$ | 9.19E+15 | $^oK$ |
| | $T_c$ | 9.19E+15 | $T_{cr}$ | 9.19E+15 | $^oK$ |
| | $T_d$ | 1.53E+14 | $T_{dr}$ | 1.53E+14 | $^oK$ |
| | $T_i$ | 1.10E+14 | $T_{ir}$ | 1.10E+14 | $^oK$ |
| | $T_r$ | 8.07E+22 | $T_{rr}$ | 8.07E+22 | $^oK$ |

This energy conjunction at $T_c = T_b$ marks the beginning of the universe where time and space are not defined in a group phase reflection condition where $v_d \sim v_\varepsilon \gg c$. The universe mass ($M_U$) is nominally at 2.08E-37 gram the radiation mass $\left(M_\nu\right) \sim \sqrt{3} m_t$. This nearly corresponds to the graphically displayed ankle position.

| | | | | | |
|---|---|---|---|---|---|
| 1.2 | $z_R$ | 3.67E+43 | $z_R$ | 3.67E+43 | *unitless* |
| | $z_B$ | 3.32E+14 | $z_B$ | 3.32E+14 | *unitless* |
| | $T_a$ | 6.53E+11 | $T_a$ | 6.53E+11 | $^oK$ |
| | $T_b$ | 9.07E+14 | $T_{br}$ | 6.05E+07 | $^oK$ |
| | $T_c$ | 8.96E+13 | $T_{cr}$ | 1.19E+09 | $^oK$ |
| | $T_d$ | 4.80E+13 | $T_{dr}$ | 8.89E+09 | $^oK$ |
| | $T_i$ | 1.08E+13 | $T_{ir}$ | 3.94E+10 | $^oK$ |
| | $T_r$ | 7.67E+18 | $T_{rr}$ | 6.98E+06 | $^oK$ |

This $T_c$ defines superconductive phenomenon as dictated by the proton density 2.34E14 g/cm$^3$ and radius *B* of 6.65E-14 cm (8.750E-14 cm NIST reported charge radius value) which is defined at a critical temperature $T_c$ of

$$T_c = T_{å} = g_d m_T \hbar c^3 / \left(e^2 k_b\right) = g_d m_t c^2 / \left(\alpha k_b\right)$$

that is a consequence of a group phase reflection condition:

$$v_{dx} = \left(\frac{2}{\alpha}\right)^{\frac{1}{2}} c$$

that is justifiable in elastic resonant condition (see equations 2.1.5 and 2.1.6 and follow-on explanation) and

$$\frac{\hbar c}{e^2} = \frac{m_r}{m_T} = \frac{v_{dx}^2}{2c^2} = \text{fine structure constant} = \frac{1}{\alpha}$$

This nearly corresponds to the graphically displayed knee position.

| | | | | | |
|---|---|---|---|---|---|
| 1.3 | $z_R$ | 8.14E+39 | $z_R$ | 8.14E+39 | *unitless* |
| | $z_B$ | 2.01E+13 | $z_B$ | 2.01E+13 | *unitless* |
| | $T_a$ | 6.53E+11 | $T_a$ | 6.53E+11 | $^oK$ |
| | $T_b$ | 5.49E+13 | $T_{br}$ | 1.00E+09 | $^oK$ |
| | $T_c$ | 3.28E+11 | $T_{cr}$ | 3.25E+11 | $^oK$ |
| | $T_d$ | 1.18E+13 | $T_{dr}$ | 3.61E+10 | $^oK$ |
| | $T_i$ | 6.54E+11 | $T_{ir}$ | 6.51E+11 | $^oK$ |
| | $T_r$ | 1.03E+14 | $T_{rr}$ | 5.21E+11 | $^oK$ |
| | This temperature condition represents the undefined but potentially high energy barrier luminal state where $v_{dx} = v_{dxr} = c$ and $v_{\varepsilon x} = v_{\varepsilon xr} > c$ , $T_b = T_{br}$ , $T_c = T_{cr}$ , $T_d = T_{dr}$ , $T_i = T_{ir} = T_a$ , $T_j = T_{jr}$ and trisine length $B = 4\pi\ nuclear\ radius$ which defines a condition for the size of the proton as it 'freezes' out of early expanding universe nucleosynthesis | | | | |

$$\sigma T_{cr}^4 \xi \left(k_b T_j\right)^{-\frac{8}{8}} = \left(\frac{B}{A}\right)^{\frac{12}{8}} \frac{\pi^{\frac{8}{8}} G^{\frac{8}{8}} m_T^{\frac{47}{8}} c^{\frac{66}{8}}}{2^{\frac{67}{8}} 3^{\frac{10}{8}} 4^{\frac{8}{8}} 15^{\frac{8}{8}} \hbar^{\frac{28}{8}}} \left(\frac{B_{Present}}{R_{U_{Present}}}\right)^{\frac{6}{8}} \frac{1}{2\pi} \left(k_b T_j\right)^{-\frac{15}{8}}$$

Energy per nucleon Voyager 1 data is presented in Figure 4.33.2 with the intention of matching to reported cosmic ray data as presented in Figure 4.33.1.

**Figure 4.33.2** Energy per nucleon Spectra for Voyager 1 for the first 295 days of 2012 Data from University of Maryland Website http://sd-www.jhuapl.edu/VOYAGER/

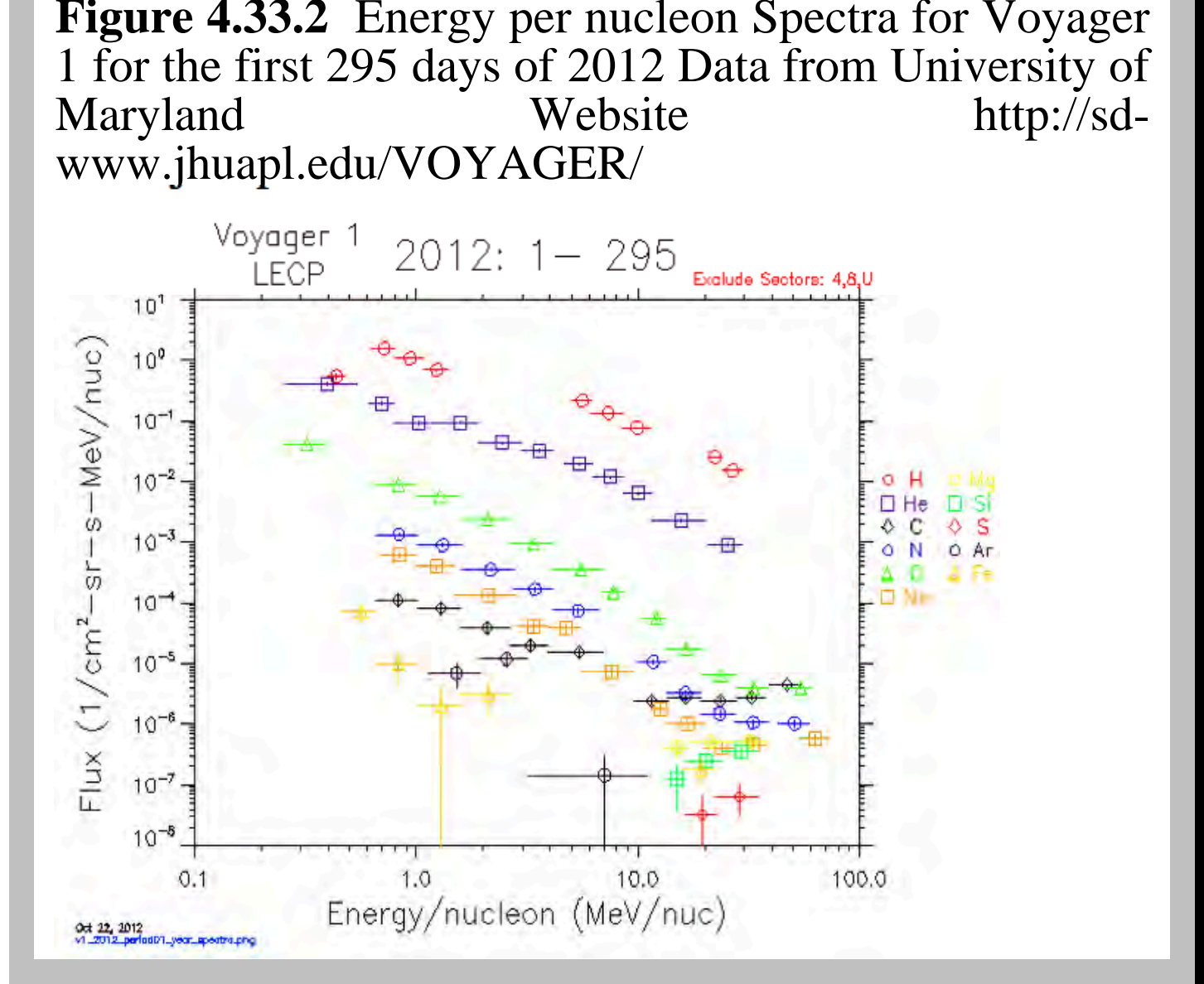

**Figure 4.33.3** Dimensionally changed total energy spectra for Voyager 1 for the first 295 days of 2012 as it proceeds beyond the helio shock wave into interstellar space as an extension of Figure 4.33.1. A discontinuity is noted between Alpha Magnetic Spectrometer and Voyager 1 data and most probably defined by the isotopic binding energy.

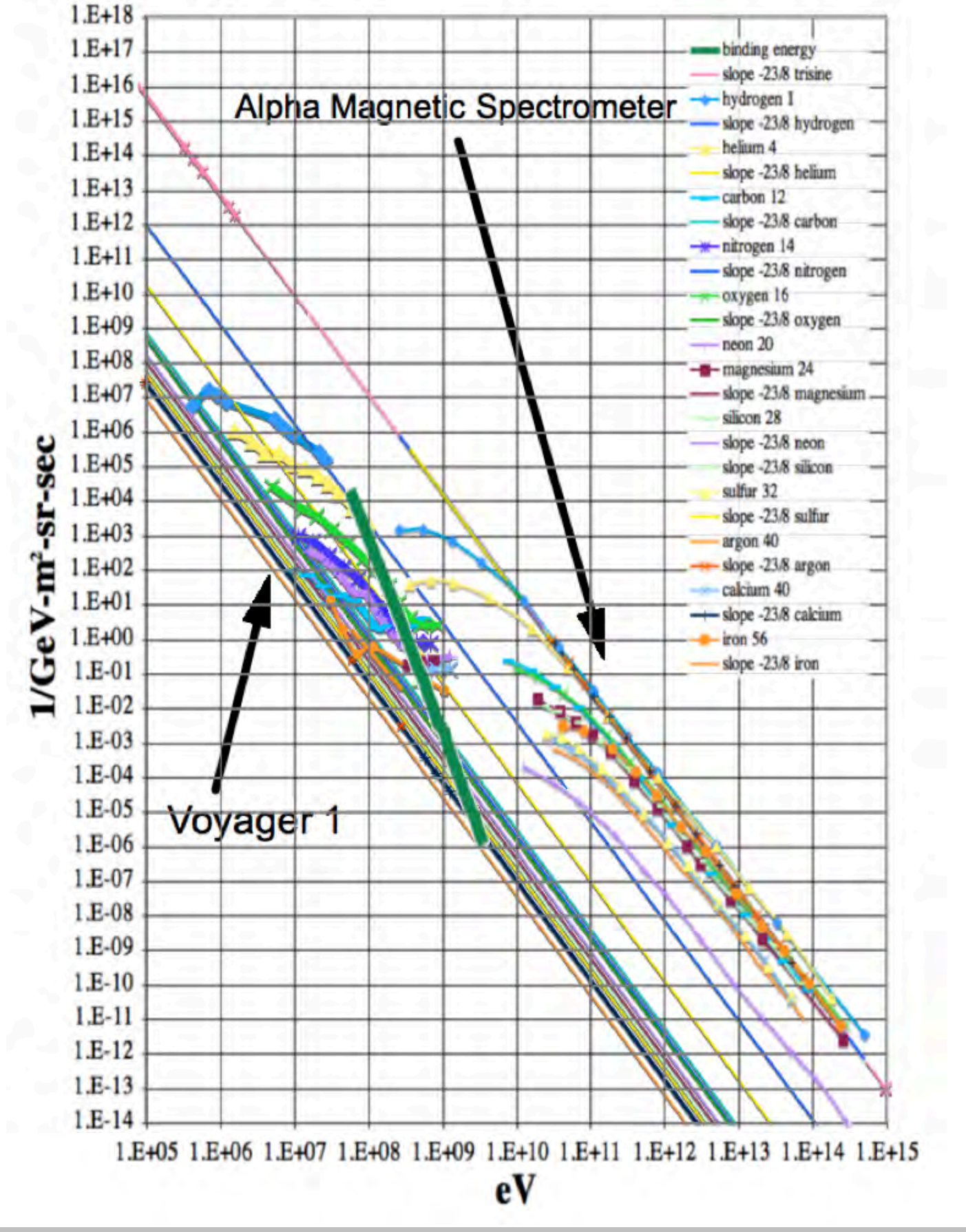

Linear intensities have log log -23/8 slopes as per the following

$$Intensity\left(\frac{1}{MeV\text{-sec-}sr\text{-}m^2}\right) \sim \left(Atomic\ Number\ \text{x}\ m_t c^2\right)^{-\frac{23}{8}}$$

Reported isotopic binding energies including rest mass energy

$$\left(Isotopic\ Binding\ Energy + Atomic\ Number\ \text{x}\ m_t c^2\right)^{-\frac{23}{8}}$$

(green line) are provided as a reference beyond which there may be a tendency for isotopic kinetic instability.

Selected IceCube[125 -table 3] data at the knee continues to confirm these relationships as per Figure 4.33.3.

Evidence for solar wind source and not extra helio source of tens-of-keV/n neutral ions [143] is concluded in that their log-log slope does not conform to interstellar-extragalactic log-log slope -23/8 as presented in Figure 4.33.1 of this report.

**Figure 4.33.4** Experimental IceCubeCosmic Ray proton data[125, -table 3] (red squares without error bars) at the knee with Trisine Condition 1.2 (green dot) on Trisine model near universe -23/8 log log linear slope nucleosynthetic condition (red line).

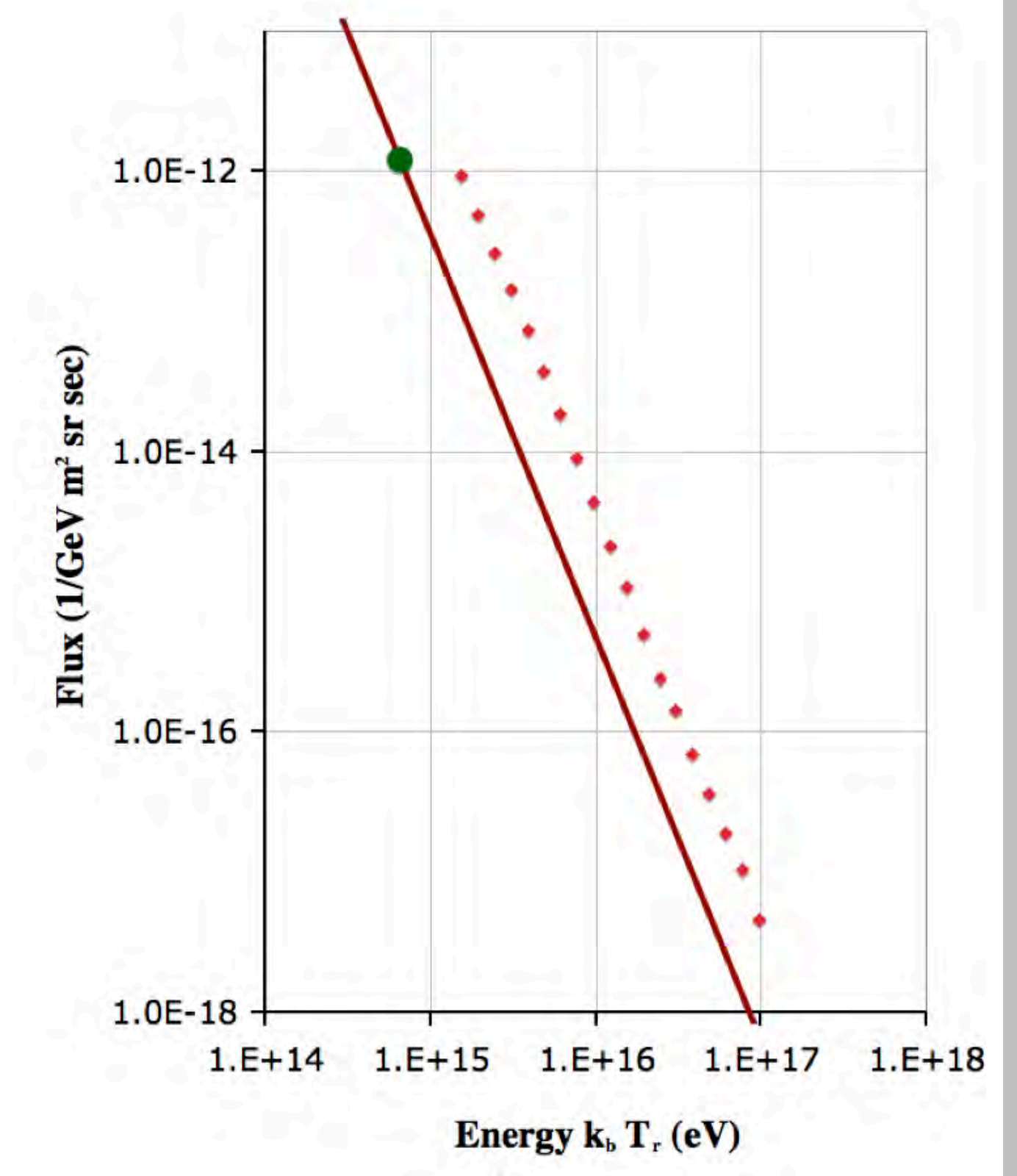

Electron and positron flux as measured by FERMI [136 Figure 3] (with energy and nucleon conversion) is consistent with -23/8 log log slope as presented in Figure 4.33.1 although reported as -3 log log slope.

4.34. Also, there is an indication of a 56 MeV peak in black body curve from the Energetic Gamma Ray Experiment Telescope (EGRET) instrument aboard The Compton Gamma Ray Observatory Mission (Comptel) spacecraft, which generally observed the extra galactic diffuse background x-ray radiation. The data graphic is replicated below in Figure 4.2 by author permission [73] with the 56 MeV resonant energy superimposed. In addition the following black body equation models the log linear characteristic of the energy data with $T_p$ conforming to the phase velocity being less then the speed of light $v_{cr} < v_{\varepsilon r} < c < v_\varepsilon < v_c$ and approaching $v_d \sim c$.

$$\sigma T_{cr}^4 \xi = \sigma T_{cr}^4 \frac{cavity}{V_U} \frac{c}{4} \left( \frac{1}{R_{U_{Present}}} \right)^{\frac{6}{3}} \frac{1}{2\pi} time_{\pm} H_U$$

$$= \frac{2 m_T^{\frac{13}{3}} c^{\frac{23}{3}}}{3^{\frac{13}{6}} 4^{\frac{6}{3}} 15^{\frac{3}{3}} \pi^{\frac{8}{3}} \hbar^{\frac{9}{3}}} \left( \frac{1}{R_{U_{Present}}} \right)^{\frac{6}{3}} \frac{1}{2\pi} \left( \frac{1}{2} \hbar H_U \right)^{-\frac{1}{3}}$$

*or*

$$= \left( \frac{B}{A} \right)^{\frac{10}{3}} \frac{\pi^{\frac{2}{3}} m_T^{\frac{13}{3}} c^{\frac{23}{3}}}{2^{\frac{6}{3}} 3^{\frac{13}{6}} 4^{\frac{16}{3}} 15^{\frac{3}{3}} \hbar^{\frac{9}{3}}} \left( \frac{1}{R_{U_{Present}}} \right)^{\frac{6}{3}} \frac{1}{2\pi} \left( \frac{1}{2} \hbar H_U \right)^{-\frac{1}{3}}$$

This equation is dimensionally correct in the CGS system with energy expressed graphically(green line) as MeV in Figures 4.34.1 to 4.34.3. Note that the abscissa is divided by an instrument $2\pi$ factor (to be checked). This graphic expression is dimensionally consistent with EGRET and FERMI featureless data and as log log linear with negative 1/3 slope slightly different than the observed FERMI negative 1/2.41 slope[91]. The phase velocity $v_p < c < v_\varepsilon < v_d$ (and approaching $v_d \sim c$) implies that the Figures 4.34.1 to 4.34.3 data signatures reflect universe conditions ($z_B$ = 2.01E+13) in the early stages of nucleosynthesis (conditions 1.1 to 1.3 as indicated in the introduction plus an interim condition 1.3a as defined below). These relic expressions as a presently observable phenomenon look back in time further than the Cosmic Microwave Background Radiation (CMBR) providing further insight into the universe evolution. Further analysis of FERMI data may provide descriptive features indicating nucleosynthetic processes

Table 4.34.1

| $T_c$ (°K) | 1.43E+12 | 2.10E+12 | 9.73E+12 | 4.52E+13 | 2.10E+14 |
|---|---|---|---|---|---|
| $T_b$ (°K)) | 1.15E+14 | 1.39E+14 | 2.99E+14 | 6.45E+14 | 1.39E+15 |
| $T_{cr}$ (°K)) | 7.35E+10 | 5.01E+10 | 1.08E+10 | 2.32E+09 | 5.00E+08 |
| $\sigma T_p^4 \xi$ | 2.03E-03 | 1.67E-03 | 7.77E-04 | 3.60E-04 | 1.67E-04 |
| $(1/2)\hbar H_U$ $(MeV)$ | 56.27 | 1.00E+02 | 1.00E+03 | 1.00E+04 | 1.00E+05 |
| $R_U$ (cm) | 3.04E-13 | 1.71E-13 | 1.71E-14 | 1.70E-15 | 1.70E-16 |
| $H_U$ (sec$^{-1}$) | 1.71E+23 | 3.04E+23 | 3.04E+24 | 3.05E+25 | 3.05E+26 |
| $G_U$ (cm) | 4.93E+33 | 8.77E+33 | 8.77E+34 | 8.79E+35 | 8.80E+36 |
| $Age_U$ (sec) | 5.85E-24 | 3.29E-24 | 3.29E-25 | 3.28E-26 | 3.28E-27 |
| $Age_{UG}$ (sec) | 1.59E-03 | 1.19E-03 | 3.77E-04 | 1.19E-04 | 3.77E-05 |

There seems to be two models tested by isotropic and galactic diffuse models that are provided by the NASA-FSSC[104].

1. Diffuse emission is from unresolved point sources that will be resolved with further analysis.

2. <u>Diffuse emission is from unresolved sources that will never be resolved as point sources with further analysis</u>.

Model one(1) would appear to be the easy approach. Under model two(2), one is left with the vexing question as to what is the source. Logically it could be another isotropic source analogous to the Cosmic Microwave Background Radiation (CMBR) as proposed herein.

**Figure 4.34.1** Diffuse X-ray Background radiation including that from EGRET and FERMI [73, 91]with a $m_T c^2$ = 56 MeV reference cutoff energy (vertical green line) and sloping (-1/3) log log green line imposed as a remnant from Big Bang nucleosynthesis with group phase reflection Condition 1.2 as green circle. The brown line is from NASA -FSSC Pass 7 (V6) Clean (front+back) P7CLEAN_V6 iso_p7v6clean.txt converted by x MeV$^2$ for dimensional comparison.

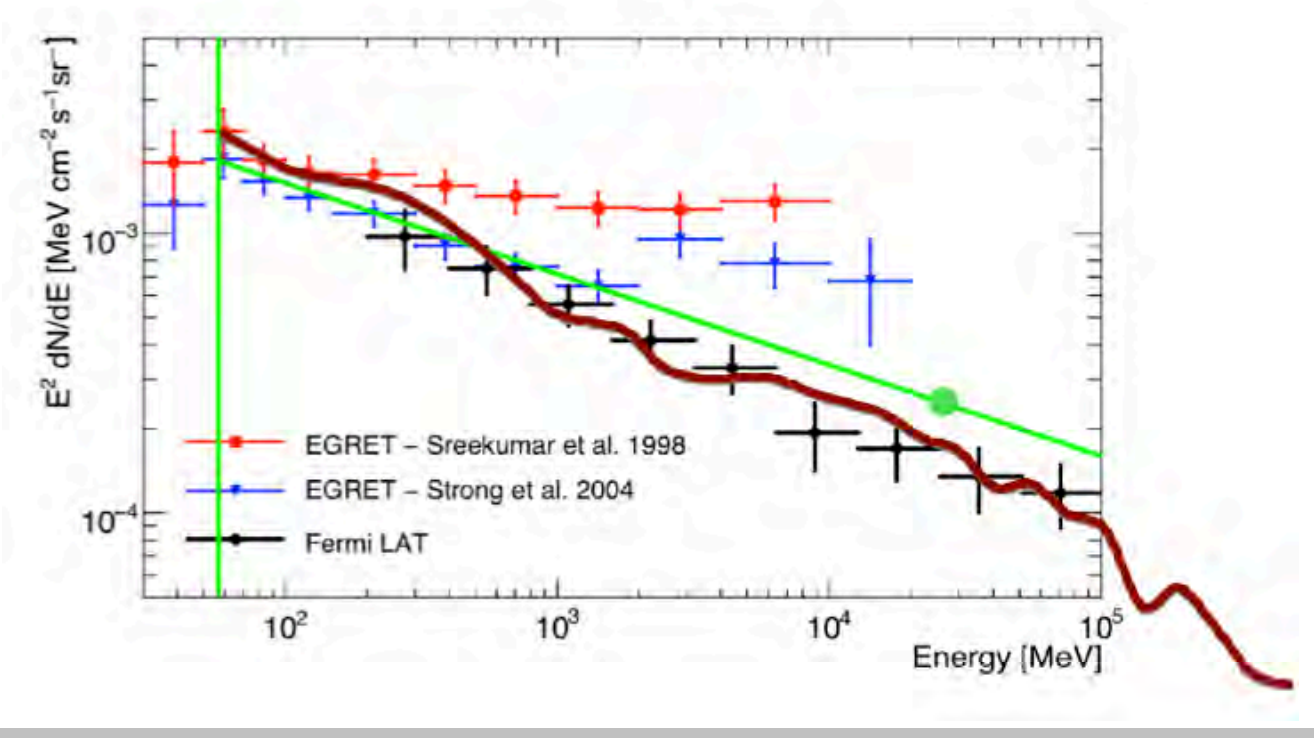

Inclusive in these observations would be the detected radiation from the Crab Nebula possibly correlated to its expansion shockwave. Verification is anticipated with the (FERMI) platform providing a much more detailed delineation of the extra galactic diffuse X-ray background radiation and potentially the trisine space lattice 56 MeV/c$^2$ dark energy component evidenced by interaction with known celestial bodies such as comets, planets, asteroids, stars, galaxies, intergalactic dust etcetera. Such data is presented in

The log log -1/3 slope describing Isotropic Diffuse Extragalactic Gamma radiation may be further described as having origins within group phase reflection Big Bang nucleosynthesis processes a phenomenon replicated in present standard model nuclear Asymtotic Freedom observations Figure 4.36.1

This Isotropic Diffuse Extragalactic Gamma radiation universe origin appears to be verified by the isotropic spectra[115] Fig. 29. The spectra of the isotropic component overlays with good visual agreement (log log -1/3 slope) on Figure 4.34.1 and displayed on Figure 4.34.2 appears to support hypothesis 2 from above.

The higher than isotropic FERMI energies[115] are a result of inverse Compton scattering of Big Bang istropic energies subjected to inverse Compton scattering while transiting galactic medium such as Black Holes and energetic gases or plasmas a similar conceptual process as Sunyaev-Zeldovich effect on Cosmic Microwave Background Radiation(CMBR).

**Figure 4.34.2** Diffuse X-ray Background radiation as indicated in Figure 4.34.1 and including that from FERMI [115] with a $m_T c^2$ = 56 MeV reference cutoff energy (vertical green line) and sloping (-1/3) log log green line imposed as a remnant from Big Bang nucleosynthesis with group phase reflection Condition 1.2 (green circle) and with Clean data removed and grey area correlating [115] with Fermi LAT data with uncertainty expressed in quadrature.

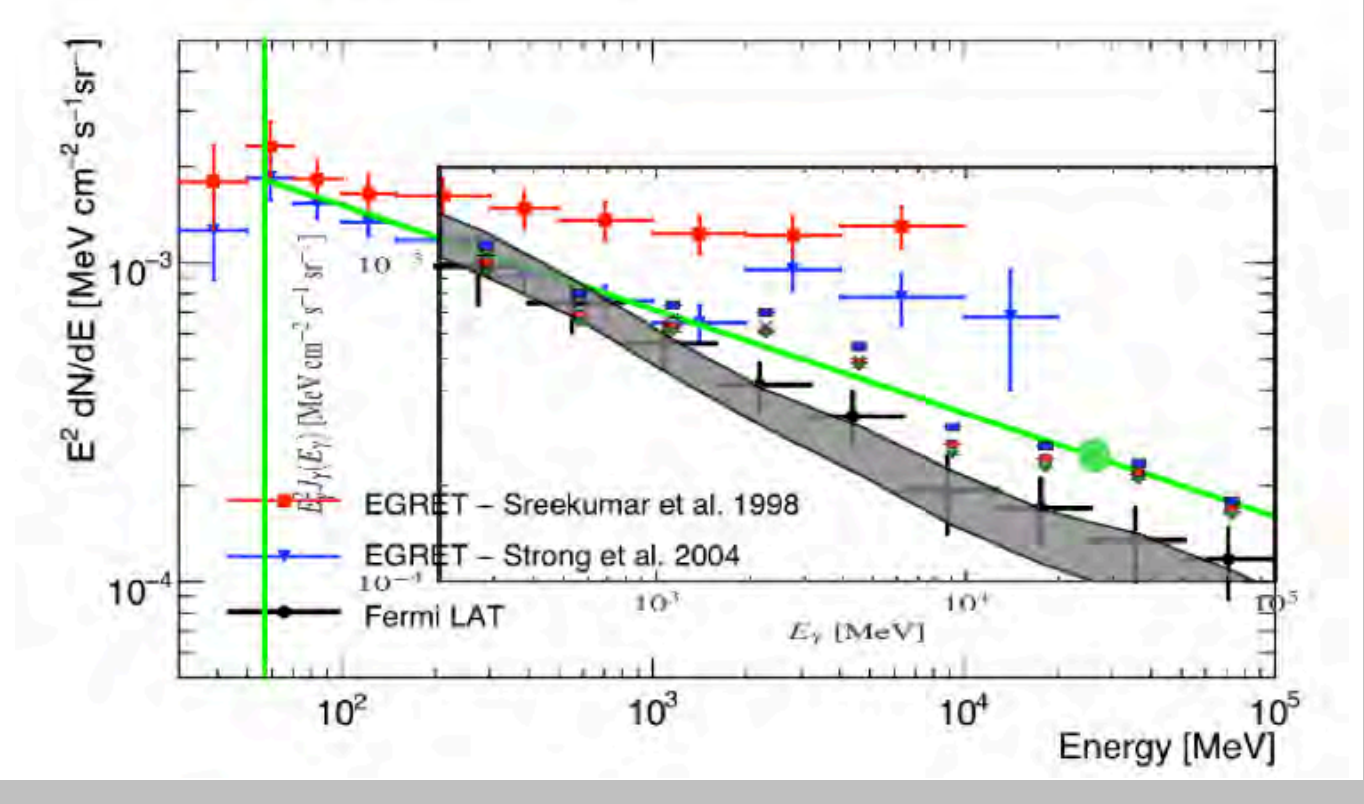

Fermi LAT Data from Table I [123] is replicated below and modified by Trisine Cosmic Ray photons achieving the Trisine slope of -1/3 vs the reported -1/2.438 indicating the possibility of Cosmic Ray(CR) contamination of Fermi LAT data. The following previously derived equation generates the Trisine Cosmic Ray particles:

$$\sigma T_{cr}^4 \xi \left(k_b T_{cr}\right)^{-\frac{8}{8}} = \left(\frac{B}{A}\right)^{\frac{12}{8}} \frac{\pi^{\frac{8}{8}} G^{\frac{8}{8}} m_T^{\frac{47}{8}} c^{\frac{66}{8}}}{2^{\frac{67}{8}} 3^{\frac{10}{8}} 4^{\frac{8}{8}} 15^{\frac{8}{8}} \hbar^{\frac{28}{8}}} \left(\frac{B_{Present}}{R_{U_{Present}}}\right)^{\frac{6}{8}} \frac{1}{2\pi} \left(k_b T_j\right)^{-\frac{15}{8}}$$

A 11,000 factor modifies the Fermi gamma ray $n$ by:

$n = n$ - 11,000 $x$ *Trisine CR*

Table 4.34.2

| FERMI Energy Range corrected | | Trisine(CR) Particles | | | Energy | | n flux | n flux |
|---|---|---|---|---|---|---|---|---|
| GeV - | GeV | n | # $cm^{-2}$ $sec^{-1}$ $sr^{-1}$ | n corrected | $GeV^{-1}$ $cm^{-2}$ $sec^{-1}$ $sr^{-1}$ | MeV | MeV $cm^{-2}$ $sec^{-1}$ $sr^{-1}$ | MeV $cm^{-2}$ $sec^{-1}$ $sr^{-1}$ |
| 4.8 - | 5.1 | 7,156 | 3.23E-01 | 3,604 | 6.29E-08 | 4,950 | 1.54E-03 | 7.76E-04 |
| 5.1 - | 5.3 | 6,638 | 2.94E-01 | 3,400 | 5.54E-08 | 5,200 | 1.50E-03 | 7.67E-04 |
| 5.3 - | 5.6 | 6,295 | 2.70E-01 | 3,329 | 4.95E-08 | 5,450 | 1.47E-03 | 7.78E-04 |
| 5.6 - | 5.9 | 5,684 | 2.44E-01 | 3,002 | 4.21E-08 | 5,750 | 1.39E-03 | 7.35E-04 |
| 5.9 - | 6.2 | 5,477 | 2.22E-01 | 3,039 | 3.83E-08 | 6,050 | 1.40E-03 | 7.78E-04 |
| 6.2 - | 6.5 | 5,162 | 2.02E-01 | 2,935 | 3.40E-08 | 6,350 | 1.37E-03 | 7.80E-04 |
| 6.5 - | 6.8 | 4,744 | 1.86E-01 | 2,702 | 2.94E-08 | 6,650 | 1.30E-03 | 7.41E-04 |
| 6.8 - | 7.1 | 4,317 | 1.71E-01 | 2,437 | 2.52E-08 | 6,950 | 1.22E-03 | 6.87E-04 |
| 7.1 - | 7.5 | 3,944 | 1.56E-01 | 2,230 | 2.17E-08 | 7,300 | 1.16E-03 | 6.54E-04 |
| 7.5 - | 7.9 | 3,744 | 1.41E-01 | 2,193 | 1.94E-08 | 7,700 | 1.15E-03 | 6.74E-04 |
| 7.9 - | 8.3 | 3,416 | 1.28E-01 | 2,005 | 1.67E-08 | 8,100 | 1.10E-03 | 6.43E-04 |
| 8.3 - | 8.7 | 3,012 | 1.17E-01 | 1,723 | 1.40E-08 | 8,500 | 1.01E-03 | 5.79E-04 |
| 8.7 - | 9.1 | 2,763 | 1.07E-01 | 1,581 | 1.21E-08 | 8,900 | 9.58E-04 | 5.48E-04 |
| 9.1 - | 9.6 | 2,630 | 9.80E-02 | 1,552 | 1.09E-08 | 9,350 | 9.53E-04 | 5.62E-04 |
| 9.6 - | 10 | 2,372 | 8.97E-02 | 1,385 | 9.30E-09 | 9,800 | 8.93E-04 | 5.22E-04 |
| 10 - | 10.5 | 2,301 | 8.25E-02 | 1,394 | 8.50E-09 | 10,250 | 8.93E-04 | 5.41E-04 |
| 10.5 - | 11.1 | 2,071 | 7.48E-02 | 1,248 | 7.20E-09 | 10,800 | 8.40E-04 | 5.06E-04 |
| 11.1 - | 11.6 | 1,994 | 6.81E-02 | 1,245 | 6.55E-09 | 11,350 | 8.44E-04 | 5.27E-04 |
| 11.6 - | 12.2 | 1,936 | 6.23E-02 | 1,250 | 6.00E-09 | 11,900 | 8.50E-04 | 5.49E-04 |
| 12.2 - | 12.8 | 1,833 | 5.69E-02 | 1,208 | 5.36E-09 | 12,500 | 8.38E-04 | 5.52E-04 |
| 12.8 - | 13.4 | 1,760 | 5.21E-02 | 1,187 | 4.87E-09 | 13,100 | 8.36E-04 | 5.64E-04 |
| 13.4 - | 14.1 | 1,625 | 4.75E-02 | 1,102 | 4.24E-09 | 13,750 | 8.02E-04 | 5.44E-04 |
| 14.1 - | 14.8 | 1,519 | 4.33E-02 | 1,042 | 3.75E-09 | 14,450 | 7.83E-04 | 5.37E-04 |
| 14.8 - | 15.6 | 1,418 | 3.94E-02 | 985 | 3.33E-09 | 15,200 | 7.69E-04 | 5.34E-04 |
| 15.6 - | 16.3 | 1,396 | 3.60E-02 | 1,000 | 3.11E-09 | 15,950 | 7.91E-04 | 5.67E-04 |
| 16.3 - | 17.2 | 1,337 | 3.28E-02 | 976 | 2.83E-09 | 16,750 | 7.94E-04 | 5.79E-04 |
| 17.2 - | 18 | 1,183 | 2.99E-02 | 854 | 2.37E-09 | 17,600 | 7.34E-04 | 5.30E-04 |
| 18 - | 18.9 | 1,122 | 2.74E-02 | 821 | 2.13E-09 | 18,450 | 7.25E-04 | 5.30E-04 |
| 18.9 - | 19.9 | 1,048 | 2.49E-02 | 774 | 1.88E-09 | 19,400 | 7.08E-04 | 5.22E-04 |
| 19.9 - | 20.9 | 1,012 | 2.27E-02 | 762 | 1.72E-09 | 20,400 | 7.16E-04 | 5.39E-04 |
| 20.9 - | 21.9 | 942 | 2.07E-02 | 714 | 1.52E-09 | 21,400 | 6.96E-04 | 5.27E-04 |
| 21.9 - | 23 | 901 | 1.90E-02 | 692 | 1.37E-09 | 22,450 | 6.90E-04 | 5.31E-04 |
| 23 - | 24.1 | 825 | 1.73E-02 | 634 | 1.19E-09 | 23,550 | 6.60E-04 | 5.07E-04 |
| 24.1 - | 25.4 | 791 | 1.58E-02 | 617 | 1.08E-09 | 24,750 | 6.62E-04 | 5.16E-04 |
| 25.4 - | 26.6 | 694 | 1.44E-02 | 536 | 8.90E-10 | 26,000 | 6.02E-04 | 4.64E-04 |
| 26.6 - | 27.9 | 690 | 1.32E-02 | 545 | 8.40E-10 | 27,250 | 6.24E-04 | 4.93E-04 |
| 27.9 - | 29.3 | 653 | 1.20E-02 | 521 | 7.60E-10 | 28,600 | 6.22E-04 | 4.96E-04 |
| 29.3 - | 30.8 | 644 | 1.10E-02 | 523 | 7.10E-10 | 30,050 | 6.41E-04 | 5.21E-04 |
| 30.8 - | 32.4 | 537 | 9.99E-03 | 427 | 5.60E-10 | 31,600 | 5.59E-04 | 4.45E-04 |
| 32.4 - | 34 | 538 | 9.11E-03 | 438 | 5.32E-10 | 33,200 | 5.86E-04 | 4.77E-04 |
| 34 - | 35.7 | 490 | 8.31E-03 | 399 | 4.60E-10 | 34,850 | 5.59E-04 | 4.54E-04 |
| 35.7 - | 37.5 | 484 | 7.58E-03 | 401 | 4.31E-10 | 36,600 | 5.77E-04 | 4.78E-04 |
| 37.5 - | 39.3 | 429 | 6.93E-03 | 353 | 3.64E-10 | 38,400 | 5.37E-04 | 4.41E-04 |
| 39.3 - | 41.3 | 437 | 6.33E-03 | 367 | 3.54E-10 | 40,300 | 5.75E-04 | 4.83E-04 |
| 41.3 - | 43.4 | 386 | 5.77E-03 | 323 | 2.98E-10 | 42,350 | 5.34E-04 | 4.47E-04 |
| 43.4 - | 45.5 | 360 | 5.27E-03 | 302 | 2.65E-10 | 44,450 | 5.24E-04 | 4.39E-04 |
| 45.5 - | 47.8 | 323 | 4.81E-03 | 270 | 2.26E-10 | 46,650 | 4.92E-04 | 4.11E-04 |
| 47.8 - | 50.2 | 320 | 4.39E-03 | 272 | 2.14E-10 | 49,000 | 5.14E-04 | 4.36E-04 |
| 50.2 - | 52.7 | 289 | 4.00E-03 | 245 | 1.85E-10 | 51,450 | 4.90E-04 | 4.15E-04 |
| 52.7 - | 55.3 | 242 | 3.66E-03 | 202 | 1.48E-10 | 54,000 | 4.32E-04 | 3.60E-04 |
| 55.3 - | 58.1 | 253 | 3.34E-03 | 216 | 1.48E-10 | 56,700 | 4.76E-04 | 4.07E-04 |
| 58.1 - | 61 | 208 | 3.04E-03 | 175 | 1.16E-10 | 59,550 | 4.11E-04 | 3.45E-04 |
| 61 - | 64.1 | 208 | 2.78E-03 | 177 | 1.11E-10 | 62,550 | 4.34E-04 | 3.71E-04 |
| 64.1 - | 67.3 | 189 | 2.53E-03 | 161 | 9.60E-11 | 65,700 | 4.14E-04 | 3.53E-04 |
| 67.3 - | 70.6 | 181 | 2.31E-03 | 156 | 8.80E-11 | 68,950 | 4.18E-04 | 3.60E-04 |
| 70.6 - | 74.2 | 148 | 2.11E-03 | 125 | 6.90E-11 | 72,400 | 3.62E-04 | 3.05E-04 |
| 74.2 - | 77.9 | 172 | 1.92E-03 | 151 | 7.70E-11 | 76,050 | 4.45E-04 | 3.91E-04 |
| 77.9 - | 81.8 | 128 | 1.76E-03 | 109 | 5.46E-11 | 79,850 | 3.48E-04 | 2.96E-04 |
| 81.8 - | 85.8 | 119 | 1.60E-03 | 101 | 4.85E-11 | 83,800 | 3.41E-04 | 2.90E-04 |
| 85.8 - | 90.1 | 129 | 1.47E-03 | 113 | 5.03E-11 | 87,950 | 3.89E-04 | 3.40E-04 |
| 90.1 - | 94.6 | 119 | 1.34E-03 | 104 | 4.45E-11 | 92,350 | 3.80E-04 | 3.33E-04 |
| 94.6 - | 99.4 | 99 | 1.22E-03 | 86 | 3.54E-11 | 97,000 | 3.33E-04 | 2.88E-04 |
| 99.4 - | 104.3 | 83 | 1.11E-03 | 71 | 2.85E-11 | 101,850 | 2.96E-04 | 2.52E-04 |
| 104 - | 110 | 101 | 1.01E-03 | 90 | 3.32E-11 | 107,000 | 3.80E-04 | 3.38E-04 |
| 110 - | 115 | 77 | 9.24E-04 | 67 | 2.42E-11 | 112,500 | 3.06E-04 | 2.66E-04 |
| 115 - | 121 | 82 | 8.45E-04 | 73 | 2.47E-11 | 118,000 | 3.44E-04 | 3.05E-04 |
| 121 - | 127 | 67 | 7.70E-04 | 59 | 1.94E-11 | 124,000 | 2.98E-04 | 2.61E-04 |
| 127 - | 133 | 70 | 7.04E-04 | 62 | 1.94E-11 | 130,000 | 3.28E-04 | 2.92E-04 |
| 133 - | 140 | 55 | 6.43E-04 | 48 | 1.46E-11 | 136,500 | 2.72E-04 | 2.37E-04 |
| 140 - | 147 | 53 | 5.85E-04 | 47 | 1.35E-11 | 143,500 | 2.78E-04 | 2.44E-04 |
| 147 - | 154 | 33 | 5.35E-04 | 27 | 8.10E-12 | 150,500 | 1.83E-04 | 1.51E-04 |
| 154 - | 162 | 60 | 4.89E-04 | 55 | 1.41E-11 | 158,000 | 3.52E-04 | 3.20E-04 |
| 162 - | 170 | 30 | 4.45E-04 | 25 | 6.77E-12 | 166,000 | 1.87E-04 | 1.56E-04 |
| 170 - | 178 | 26 | 4.08E-04 | 22 | 5.64E-12 | 174,000 | 1.71E-04 | 1.41E-04 |
| 178 - | 187 | 45 | 3.73E-04 | 41 | 9.40E-12 | 182,500 | 3.13E-04 | 2.85E-04 |
| 187 - | 197 | 35 | 3.39E-04 | 31 | 7.00E-12 | 192,000 | 2.58E-04 | 2.31E-04 |
| 197 - | 207 | 27 | 3.08E-04 | 24 | 5.20E-12 | 202,000 | 2.12E-04 | 1.86E-04 |
| 207 - | 217 | 30 | 2.82E-04 | 27 | 5.55E-12 | 212,000 | 2.49E-04 | 2.24E-04 |
| 217 - | 228 | 21 | 2.57E-04 | 18 | 3.73E-12 | 222,500 | 1.85E-04 | 1.60E-04 |
| 228 - | 239 | 18 | 2.35E-04 | 15 | 3.07E-12 | 233,500 | 1.67E-04 | 1.43E-04 |
| 239 - | 251 | 24 | 2.15E-04 | 22 | 3.94E-12 | 245,000 | 2.36E-04 | 2.13E-04 |
| 251 - | 264 | 18 | 1.96E-04 | 16 | 2.84E-12 | 257,500 | 1.88E-04 | 1.66E-04 |

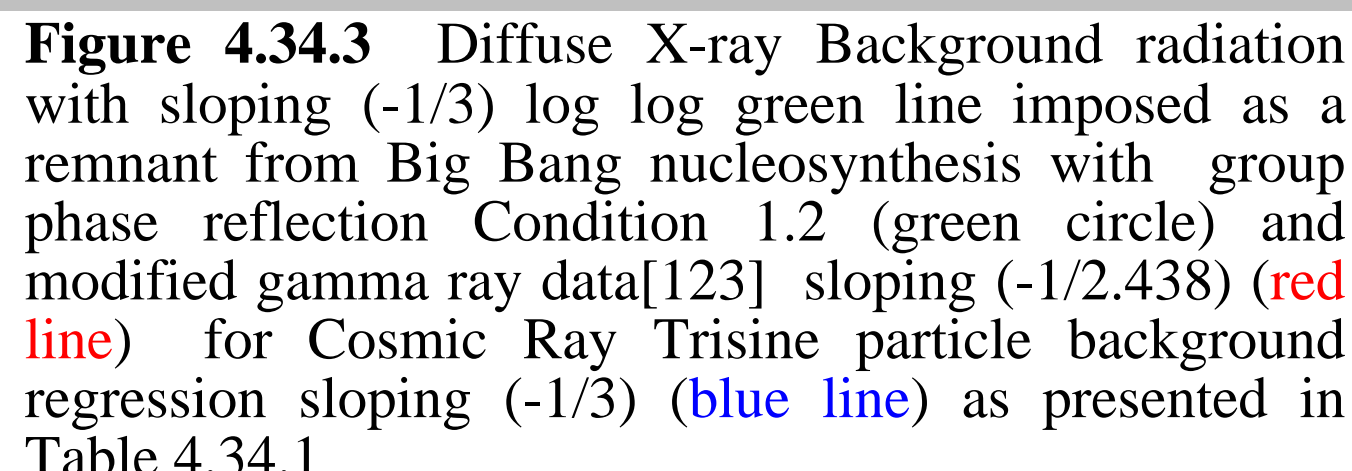

**Figure 4.34.3** Diffuse X-ray Background radiation with sloping (-1/3) log log green line imposed as a remnant from Big Bang nucleosynthesis with group phase reflection Condition 1.2 (green circle) and modified gamma ray data[123] sloping (-1/2.438) (red line) for Cosmic Ray Trisine particle background regression sloping (-1/3) (blue line) as presented in Table 4.34.1

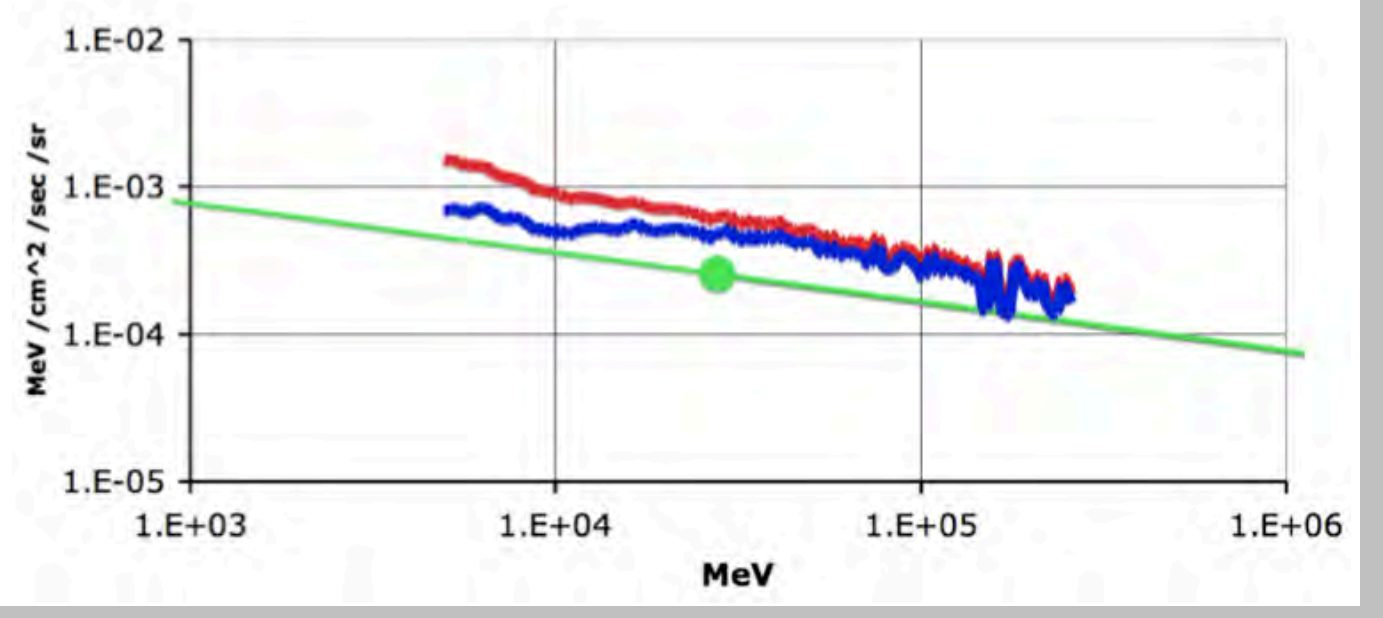

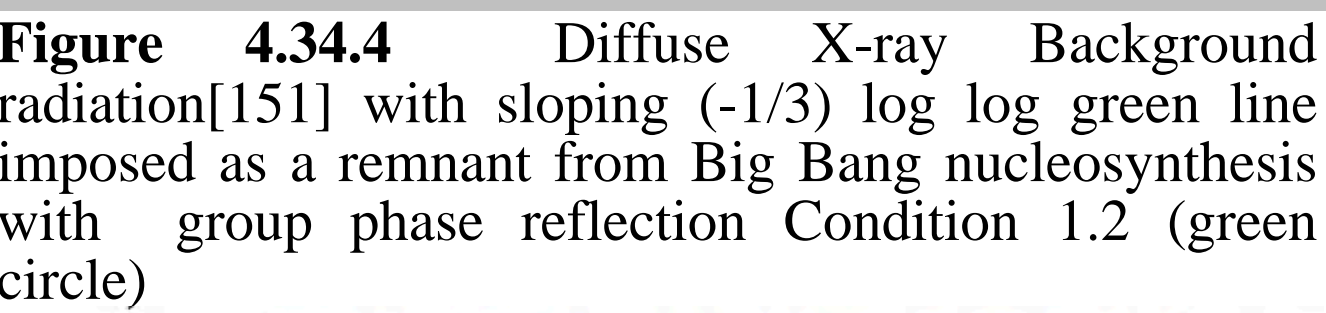

**Figure 4.34.4** Diffuse X-ray Background radiation[151] with sloping (-1/3) log log green line imposed as a remnant from Big Bang nucleosynthesis with group phase reflection Condition 1.2 (green circle)

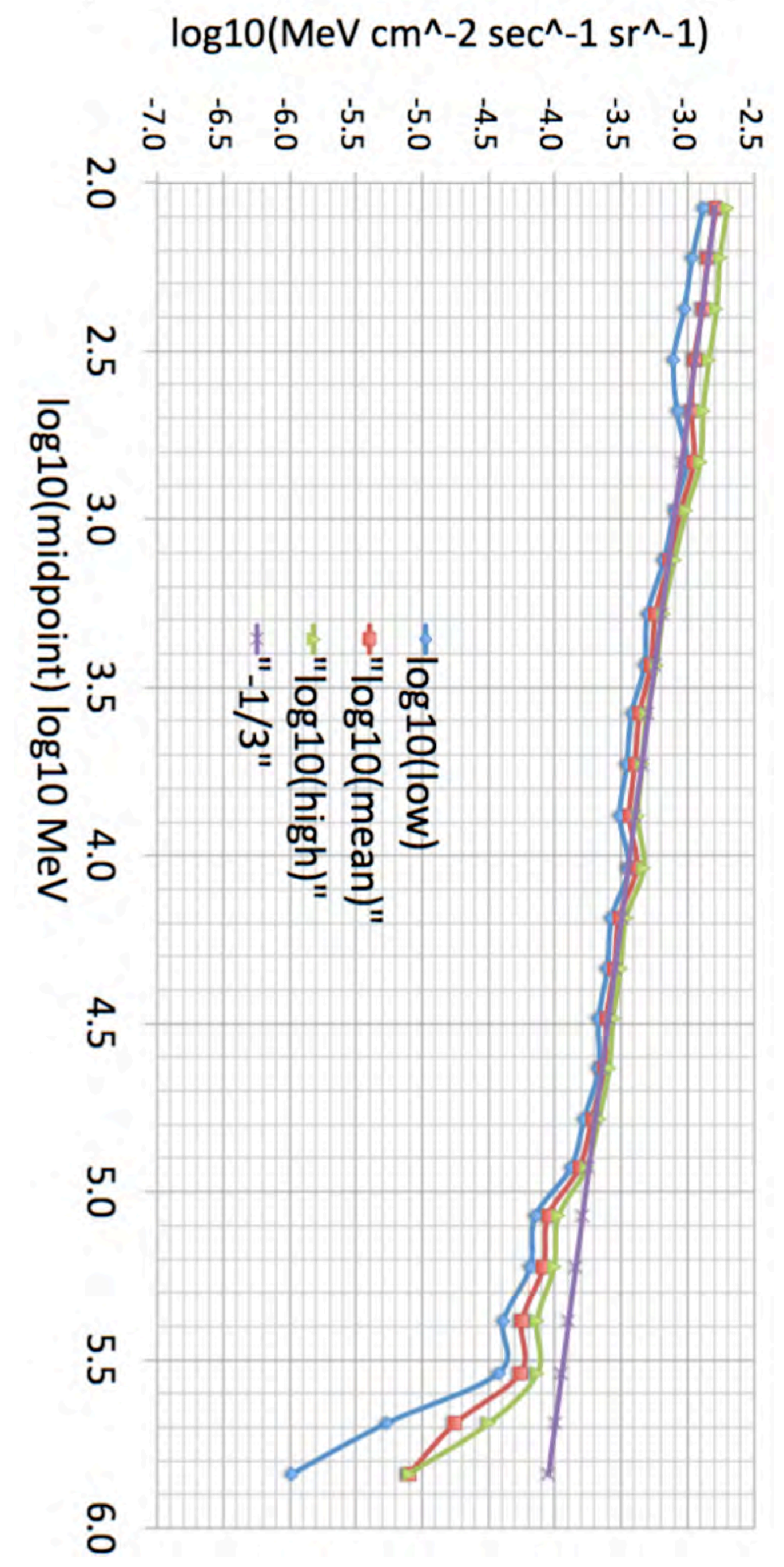

In summary, dimensional logic leads to the conclusion that empirically observed astrophysical X-Ray and Gamma Ray were created in the first second of the universe under group phase reflection Trisine Conditions 1.1 to 1.3a as presented as green circles in Figure 4.34.3. These green circles are not to any particular scale but their size is indicative becoming larger with universe time $Age_{UG}$ and universe size $R_U$ while fitting the nodes of X-Ray and Gamma Ray spectra. The additional 1.3a to introductory Trisine Conditions is as follows which appears appropriate for building up atomic structures in the Big Bang event at trisine condititon 1.3a:

| 1.3a | $z_R$ | 9.72E+34 | $z_R$ | 9.72E+34 | *unitless* |
|---|---|---|---|---|---|
| | $z_B$ | 4.60E+11 | $z_B$ | 4.60E+11 | *unitless* |
| | $T_a$ | 6.53E+11 | $T_a$ | 6.53E+11 | $^oK$ |
| | $T_b$ | 1.25E+12 | $T_{br}$ | 2.39E+15 | $^oK$ |
| | $T_c$ | 1.71E+08 | $T_{cr}$ | 6.22E+14 | $^oK$ |
| | $T_d$ | 9.04E+11 | $T_{dr}$ | 3.95E+13 | $^oK$ |
| | $T_i$ | 1.50E+10 | $T_{ir}$ | 2.85E+13 | $^oK$ |
| | $T_j$ | 2.80E+07 | $T_{jr}$ | 3.70E+20 | $^oK$ |
| | This temperature condition defined by $B = \frac{m_e}{m_T} Bohr\ radius = \frac{m_e}{m_T}\frac{\hbar}{\alpha m_e c} = \frac{\hbar}{\alpha m_T c}$ | | | | |

Equation 2.14.7a repeated here

$$hH_U = \frac{cavity}{chain} G_U \frac{m_T^2}{B_p} \tag{2.14.7a}$$

provides a mechanism explaining the CERN (July 4, 2012) observed particle at 126.5 GeV as well as the astrophysical Fermi ~130 GeV [127]. The astrophysical ~130 GeV energy is sourced to a Trisine Condition 1.2 nucleosynthetic event with a fundamental $G_U$ gravity component.

$$\begin{aligned} & g_s^2 \frac{1}{2} hH_U \cos(\theta) \\ & = g_s^2 \frac{1}{2}\frac{cavity}{chain} G_U \frac{m_T^2}{proton\ radius}\cos(\theta) \\ & = 126.5\ GeV \end{aligned}$$

Also the energy MeV scaling in Figures 4.34.1, 4.34.2, 4.34.3 & 4.34.5 may be a consequence of and directly related to $G_U$ gravity scaling.

$$H_U = \frac{1}{h}\frac{cavity}{chain} G_U \frac{m_T^2}{proton\ radius}$$

Where $proton\ radius = B_p / \left(2\sin(\theta)\right)$ or $x_U / \left(8\sin(\theta)\right)$

= 8.59E-14 (8E-14)

The value $B_p$ represents trisine cell dimension $B$ at Trisine Condition 1.2

**Figure 4.34.5** Diffuse broader band X-ray Background radiation including that from FERMI [115, 151] with a $m_T c^2$ = 56 MeV reference cutoff energy and sloping (-1/3) log log green line imposed as a remnant from Big Bang nucleosynthesis- group phase reflection Conditions 1.1 to 1.3a expressed as green circles.

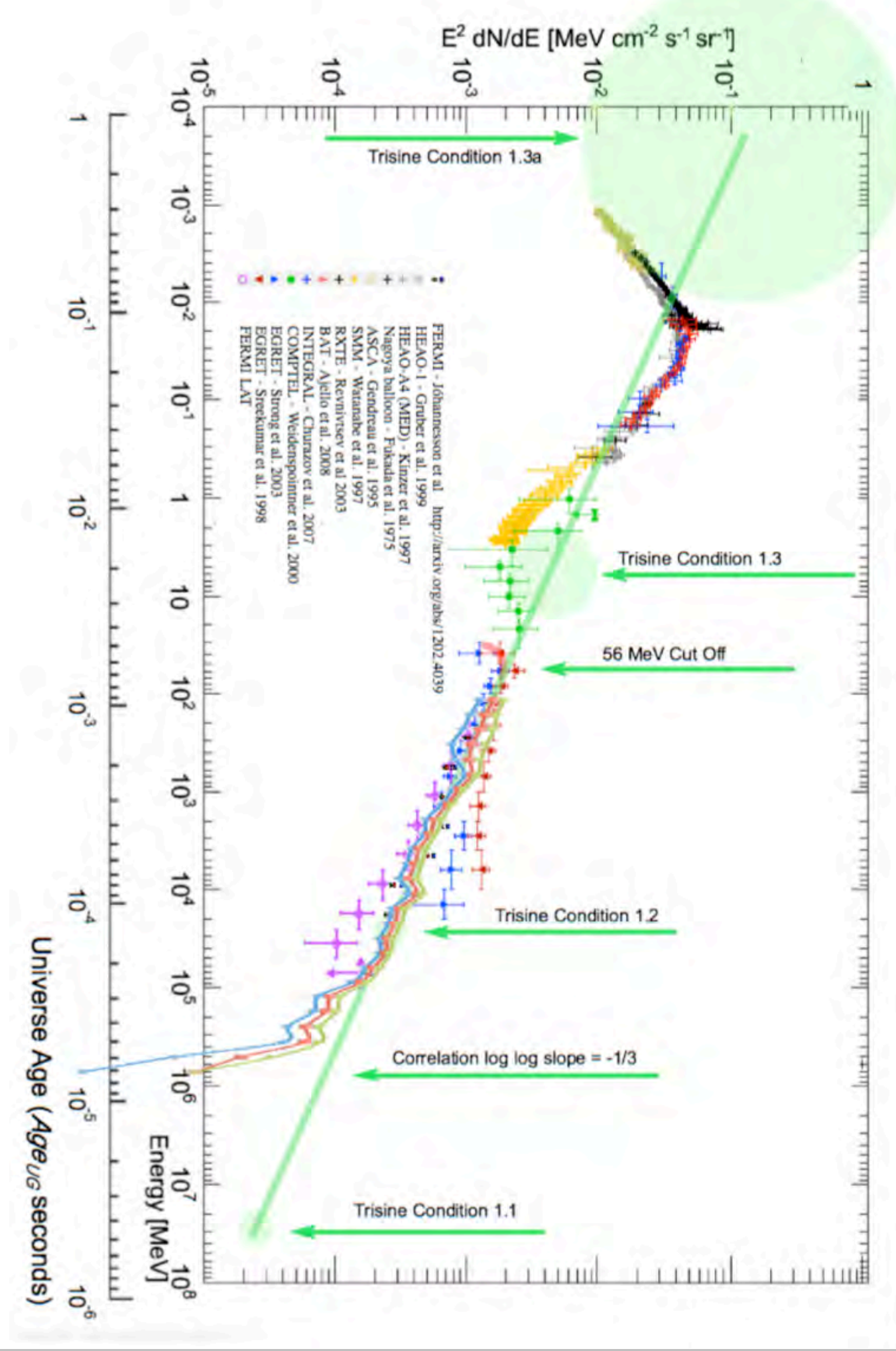

Reference [76] indicates the very existence of dark matter (observational by galactic collisions) and just such clumping thereof as well as a thrust effect of the dark energy on galactic material. Reference [79] indicates the independence of dark and visible matter distribution but perhaps this independence is illusionary. According to this correlation, the observed dark matter distributions are composed of small hydrogen dense (1 g/cc) BEC objects (<1 meter) contributing to a space density of < 1E-23 g/cm$^3$ with resulting optical mean free path dictating invisibility other than through the observed weak lensing effect.

4.36. The Absolute Radiometer for Cosmology, Astrophysics and Diffuse Emission(ARCADE 2) instrumented balloon was launched reporting radio measurements (3 to 90 GHz) from sky 20-30 degree radius[106]. An anomalous temperature($T$) signal was measured deviant from expected the well established present universe black body temperature of 2.729 K. This anomalous signal is well fitted by the trisine model component.

$$T = T_{cr} \cdot k_m \cdot \varepsilon \cdot H_{U_{present}} / H_U$$

where $H_{U_{present}}$ = 2.31E-18 /sec

and

$$T = T_{cr} \cdot k_m \cdot \varepsilon \cdot Age_U / Age_{U_{present}}$$

where $Age_{U_{present}}$ = 4.32E+17 *sec* (13.7 billion years)

**Figure 4.36.1** ARCADE 2 Data[106] (black dots with error bars) compared with the present universe black body spectrum at 2.729K (dotted line) and trisine model red line as a function of frequency /sec as indicated by the Hubble parameter.

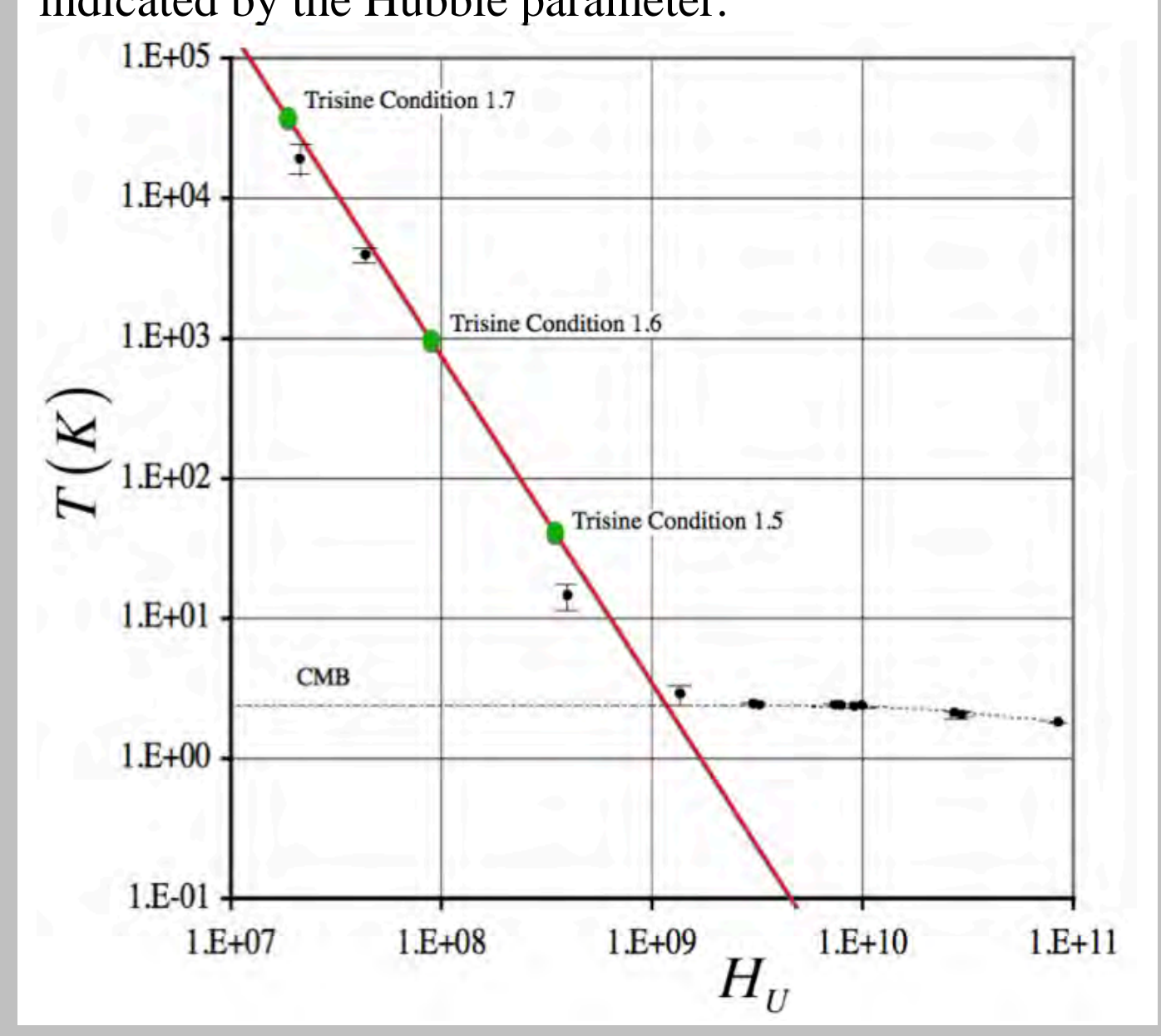

This data conforms to universe introductory Trisine conditions 1.5 to 1.7 or the universe at a very early stage $Age_{UG} \sim 3.5E4$ *sec* or .41 *days*

To $Age_{UG} \sim 1.5E5$ *sec* or 1.7 *days*

where

$$Age_{UG} = Age_U \sqrt{G_U / G}$$

and $k_m$ is the universe time variant permeability and $\varepsilon$ is the universe time variant permittivity .

Also the energy $H_U$ scaling in Figures 4.36.1 may be a consequence of and directly related to $G_U$ gravity scaling.

$$H_U = \frac{1}{h}\frac{cavity}{chain} G_U \frac{m_T^2}{proton\ radius}$$

Where *proton radius* = $B_p/\left(2\sin(\theta)\right)$ or $x_U/\left(8\sin(\theta)\right)$
= 8.59E-14 (8E-14)

4.37. The David Gross, Frank Wilczek, and David Politzer concept of Asymtotic Freedom is naturally linked to Trisine context in the same manner as the diffuse gamma ray background as observed by EGRET and FERMI and graphically presented in Figures 4.34.1, 4.34.2 & 4.34.3 and Asymtotic data as presented by Bethke [120]. This relationship is in the context of Big Bang nucleosynthesis and Trisine Conditions 1.1 and 1.2.

$$\sigma T_{cr}^4 \xi = \sigma T_{cr}^4 \frac{cavity}{V_U}\frac{c}{4}\left(\frac{1}{5.9\text{E-}19}\right)^3 \frac{1}{2\pi} time_{\pm} H_U$$

$$= \frac{2 m_T^{\frac{13}{3}} c^{\frac{23}{3}}}{3^{\frac{13}{6}} 4^{\frac{6}{3}} 15^{\frac{3}{3}} \pi^{\frac{8}{3}} \hbar^{\frac{9}{3}}}\left(\frac{1}{5.9\text{E-}19}\right)^3 \frac{1}{2\pi}\left(\frac{1}{2}\hbar H_U\right)^{-\frac{1}{3}}$$

*or*

$$= \left(\frac{B}{A}\right)^{\frac{10}{3}} \frac{\pi^{\frac{2}{3}} m_T^{\frac{13}{3}} c^{\frac{23}{3}}}{2^{\frac{6}{3}} 3^{\frac{13}{6}} 4^{\frac{16}{3}} 15^{\frac{3}{3}} \hbar^{\frac{9}{3}}}\left(\frac{1}{5.9\text{E-}19}\right)^3 \frac{1}{2\pi}\left(\frac{1}{2}\hbar H_U\right)^{-\frac{1}{3}}$$

These two concepts are compared in Figure 4.37.1 indicating a commonality within the group phase reflection nuclear dimension. There is a normalization length dimension 5.9E-19 cm. This length dimension equals the universe Hubble radius of curvature $R_U$ at Trisine Condition 1.1 indicative of the universe beginning. The Trisine model indicates length termination at Trisine Condition 1.1 which is far from the Planck dimensional length scale of 1.6E-33 cm.

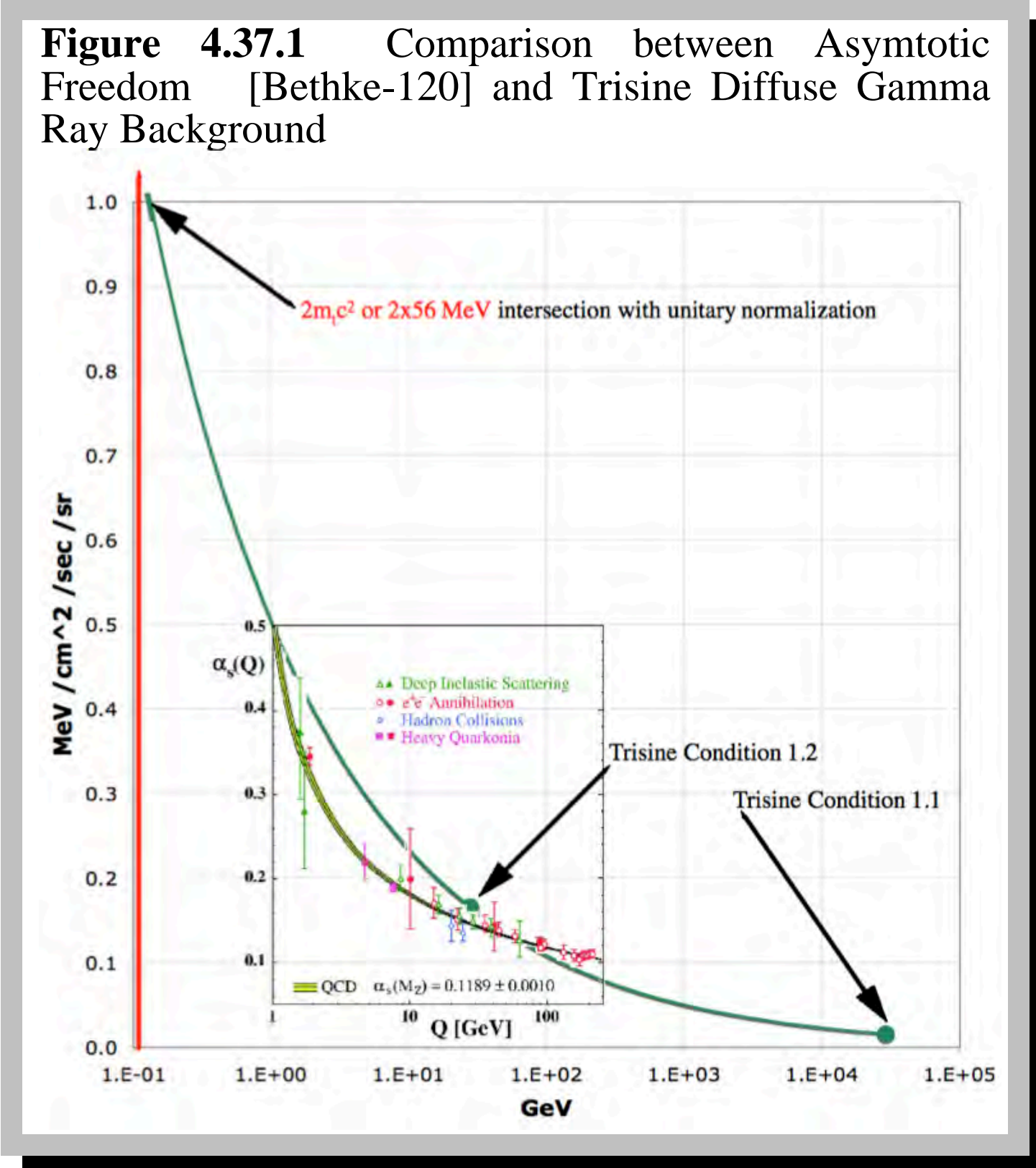

**Figure 4.37.1** Comparison between Asymtotic Freedom [Bethke-120] and Trisine Diffuse Gamma Ray Background

4.38. Within the Trisine context, the concept of entanglement or action at a distance is naturally resolved with the defined phase and group velocity definitions

$$(group\ velocity)\cdot(phase\ velocity) = c^2$$

$$v_{dx} v_{dxr} = v_{dx}\frac{c^2}{v_{dx}} = c^2 \qquad (2.1.60)$$

Table 2.1.3 indicates present universe phase velocity at 6.02E23 cm/sec. The average distance from the center of the Earth to the center of the Moon is 384,403 km (~3.84E10 cm). Earth to moon phase velocity $(v_{dxr})$ travel time would be 3.84E10/$v_{dxr}$ 3.84E10/5.97E23 = 6.43216E-24 sec. This extremely small number probably is unmeasurable at this time but instructive of the experimental confusion regarding entanglement or action at a distance discussions and dialogue. Indeed, entanglement and the related phase velocity $(v_{dxr})$ is more appropriately compared to universe dimensions $v_{dxr} = (1/3)(B/A)(R_U/time_{\pm})$. It is like casting your eyes from across the distant universe and this rate of view is comparable to the present universe phase velocity $(v_{dxr})$.

4.39. The OPERA experimental results[108] indicate neutrino group phase reflection velocity which was falsified by ICARUS detector at the CNGS beam and a subsequent FermiLab MINOS June 8, 2012 reported investigation. A group phase reflection result would be congruent with Trisine theory and in particular nuclear neutrino generating conditions as expressed in introductory trisine conditions 1.1 to 1.3. The nuclear generated superlumal neutrinos would transit the nuclear diameter and maintain the group phase reflection character as they pass into the surrounding media. Further nuclear neutrino generating experiments for all neutrino flavors (electron, muon, and tau and their antiparticles) are required to fully study this nuclear group phase reflection concept.

Also, observed group phase reflection motion in astrophysical radio sources may provide the basis for further confirmation of this concept in as much as the source of the these jets is nuclear in origin.

Also, there evidence of group phase reflection velocities as conducted in National Institute of Standards and Technology (NIST) experiments[118] that may feed into a more generalized view of this phenomenon.

4.40. Unidentified Infrared objects [112, 113] with observed wavelengths 3.3, 6.2, 7.7, 8.6, 11.3, 12.6 and 16.4 $\mu m$ may be associated with present day trisine dark energy condition 1.11 expressing itself within galactic space resulting from dark energy and particle collisions, all within the context of the Heisenberg uncertainty constraint with Interstellar Medium proton mass $\left(m_p\right)$ interaction with trisine mass $\left(m_t\right)$ with trisine (not CMBR) distribution governing. This all may be consistent with bow shock micrometer emissions from bow shock the Cygnus Loop Supernova Remnant [145] resulting from momentum transfer with the ubiquitous trisine lattice.

Reported unidentified AKARI [112,113] Near- to Mid-Infrared Unidentified Emission Bands in the Interstellar Medium of the Large Magellanic Cloud as microwaves in micrometers $\mu m$.

| 3.3 | 6.2 | 7.7 | 8.6 | 11.3 | 12.6 | 16.4 |
|---|---|---|---|---|---|---|

And these wavelengths converted to gigaHz (GHz).

| 90,846 | 48,354 | 38,934 | 34,860 | 26,530 | 23,793 | 18,280 |
|---|---|---|---|---|---|---|

The following are corresponding condition 1.11 trisine harmonics corresponding to unidentified microwaves [112,113] in micrometer units $\mu m$.

$$hc\Big/\left(\frac{\hbar^2}{2m_T}K_B^2\frac{m_T}{m_e}2\frac{v_\varepsilon^2}{v_{dx}^2}\right)=\lambda_{K_B^2}$$

$$hc\Big/\left(\frac{\hbar^2}{2m_T}K_{Dn}^2\frac{m_T}{m_e}2\frac{v_\varepsilon^2}{v_{dx}^2}\right)=\lambda_{K_{Dn}^2}$$

$$hc\Big/\left(\frac{\hbar^2}{2m_T}K_P^2\frac{m_T}{m_e}2\frac{v_\varepsilon^2}{v_{dx}^2}\right)=\lambda_{K_P^2}$$

$$hc\Big/\left(\frac{\hbar^2}{2m_T}K_{BCS}^2\frac{m_T}{m_e}2\frac{v_\varepsilon^2}{v_{dx}^2}\right)=\lambda_{K_{BCS}^2}$$

$$hc\Big/\left(\frac{\hbar^2}{2m_T}\left(K_{Ds}^2-K_{Dn}^2\right)\frac{m_T}{m_e}2\frac{v_\varepsilon^2}{v_{dx}^2}\right)=\lambda_{K_{Ds}^2-K_{Dn}^2}$$

$$hc\Big/\left(\frac{\hbar^2}{2m_T}\left(K_C^2-K_B^2\right)\frac{m_T}{m_e}2\frac{v_\varepsilon^2}{v_{dx}^2}\right)=\lambda_{K_C^2-K_B^2}$$

$$hc\Big/\left(\frac{\hbar^2}{2m_T}\left(K_{Dn}^2+K_P^2\right)\frac{m_T}{m_e}2\frac{v_\varepsilon^2}{v_{dx}^2}\right)=\lambda_{K_{Dn}^2+K_P^2}$$

$$hc\Big/\left(\frac{\hbar^2}{2m_T}K_{Ds}^2\frac{m_T}{m_e}2\frac{v_\varepsilon^2}{v_{dx}^2}\right)=\lambda_{K_{Ds}^2}$$

$$hc\Big/\left(\frac{\hbar^2}{2m_T}K_C^2\frac{m_T}{m_e}2\frac{v_\varepsilon^2}{v_{dx}^2}\right)=\lambda_{K_C^2}$$

$$hc\Big/\left(\frac{\hbar^2}{2m_T}\left(K_{Ds}^2+K_B^2\right)\frac{m_T}{m_e}2\frac{v_\varepsilon^2}{v_{dx}^2}\right)=\lambda_{K_{Ds}^2+K_B^2}$$

$$hc\Big/\left(\frac{\hbar^2}{2m_T}\left(K_{Ds}^2+K_{Dn}^2\right)\frac{m_T}{m_e}2\frac{v_\varepsilon^2}{v_{dx}^2}\right)=\lambda_{K_{Ds}^2+K_{Dn}^2}$$

$$hc\Big/\left(\frac{\hbar^2}{2m_T}\left(K_B^2+K_C^2\right)\frac{m_T}{m_e}2\frac{v_\varepsilon^2}{v_{dx}^2}\right)=\lambda_{K_B^2+K_C^2}$$

$$hc\Big/\left(\frac{\hbar^2}{2m_T}\left(K_{Ds}^2+K_P^2\right)\frac{m_T}{m_e}2\frac{v_\varepsilon^2}{v_{dx}^2}\right)=\lambda_{K_{Ds}^2+K_P^2}$$

$$hc\Big/\left(\frac{\hbar^2}{2m_T}\left(K_{Dn}^2+K_C^2\right)\frac{m_T}{m_e}2\frac{v_\varepsilon^2}{v_{dx}^2}\right)=\lambda_{K_{Dn}^2+K_C^2}$$

$$hc\Big/\left(\frac{\hbar^2}{2m_T}\left(K_{Ds}^2+K_C^2\right)\frac{m_T}{m_e}2\frac{v_\varepsilon^2}{v_{dx}^2}\right)=\lambda_{K_{Ds}^2+K_C^2}$$

$$hc\Big/\left(\frac{\hbar^2}{2m_T}K_A^2\frac{m_T}{m_e}2\frac{v_\varepsilon^2}{v_{dx}^2}\right)=\lambda_{K_A^2}$$

The AKARI spacecraft frequency spectrum is:

117,566 22,373 GHz

2.55 13.40 $\mu m$

| 1/Harmonic | 1 | 2 | 3 | 4 | 5 | 6 |
|---|---|---|---|---|---|---|
| Harmonic | 1/1 | 1/2 | 1/3 | 1/4 | 1/5 | 1/6 |
| $\lambda_{K_B^2}$ | 47.99 | 23.99 | 16.00 | 12.00 | 9.60 | 8.00 |
| $\lambda_{K_{Dn}^2}$ | 40.05 | 20.02 | 13.35 | 10.01 | 8.01 | 6.67 |
| $\lambda_{K_P^2}$ | 35.99 | 18.00 | 12.00 | 9.00 | 7.20 | 6.00 |
| $\lambda_{K_{BCS}^2}$ | 27.21 | 13.60 | 9.07 | 6.80 | 5.44 | 4.53 |
| $\lambda_{K_{Ds}^2-K_{Dn}^2}$ | 25.05 | 12.53 | 8.35 | 6.26 | 5.01 | 4.18 |
| $\lambda_{K_C^2-K_B^2}$ | 20.38 | 10.19 | 6.79 | 5.10 | 4.08 | 3.40 |
| $\lambda_{K_{Dn}^2+K_P^2}$ | 18.96 | 9.48 | 6.32 | 4.74 | 3.79 | 3.16 |
| $\lambda_{K_{Ds}^2}$ | 15.41 | 7.71 | 5.14 | 3.85 | 3.08 | 2.57 |
| $\lambda_{K_C^2}$ | 14.31 | 7.15 | 4.77 | 3.58 | 2.86 | 2.38 |
| $\lambda_{K_{Ds}^2+K_B^2}$ | 11.67 | 5.83 | 3.89 | 2.92 | 2.33 | 1.94 |
| $\lambda_{K_{Ds}^2+K_{Dn}^2}$ | 11.13 | 5.56 | 3.71 | 2.78 | 2.23 | 1.85 |
| $\lambda_{K_B^2+K_C^2}$ | 11.02 | 5.51 | 3.67 | 2.76 | 2.20 | 1.84 |
| $\lambda_{K_{Ds}^2+K_P^2}$ | 10.79 | 5.40 | 3.60 | 2.70 | 2.16 | 1.80 |
| $\lambda_{K_{Dn}^2+K_C^2}$ | 10.54 | 5.27 | 3.51 | 2.64 | 2.11 | 1.76 |
| $\lambda_{K_{Ds}^2+K_C^2}$ | 7.42 | 3.71 | 2.47 | 1.85 | 1.48 | 1.24 |
| $\lambda_{K_A^2}$ | 2.12 | 1.06 | 0.71 | 0.53 | 0.42 | 0.35 |

This trisine and stellar emission correlation represents a Compton like mechanism.

4.41. Recently quantified galactic haze may be associated with present day trisine dark energy condition 1.11 expressing itself within galactic space resulting from dark energy and particle collisions, all within the context of the Heisenberg uncertainty constraint with Interstellar Medium electron mass($m_e$) interaction with trisine mass($m_t$) with trisine (not CMBR) distribution governing.

The following are corresponding trisine condition 1.11 harmonics corresponding to galactic haze as frequency $(GHz)$.

$$\left(\frac{\hbar^2}{2m_T}K_B^2\frac{m_T}{m_p}2\frac{v_\varepsilon^2}{v_{dx}^2}\right)\Big/h=\nu_{K_B^2}$$

$$\left(\frac{\hbar^2}{2m_t}K_{Dn}^2\frac{m_t}{m_p}2\frac{v_\varepsilon^2}{v_{dx}^2}\right)\Big/h=\nu_{K_{Dn}^2}$$

$$\left(\frac{\hbar^2}{2m_t}K_P^2\frac{m_t}{m_p}2\frac{v_\varepsilon^2}{v_{dx}^2}\right)\Big/h=\nu_{K_P^2}$$

$$\left(\frac{\hbar^2}{2m_t}K_{BCS}^2\frac{m_t}{m_p}2\frac{v_\varepsilon^2}{v_{dx}^2}\right)\Big/h=\nu_{K_{BCS}^2}$$

$$\left(\frac{\hbar^2}{2m_t}\left(K_{Ds}^2-K_{Dn}^2\right)\frac{m_t}{m_p}2\frac{v_\varepsilon^2}{v_{dx}^2}\right)\Big/h=\nu_{K_{Ds}^2-K_{Dn}^2}$$

$$\left(\frac{\hbar^2}{2m_t}\left(K_C^2-K_B^2\right)\frac{m_t}{m_p}2\frac{v_\varepsilon^2}{v_{dx}^2}\right)\Big/h=\nu_{K_C^2-K_B^2}$$

$$\left(\frac{\hbar^2}{2m_T}\left(K_{Dn}^2+K_P^2\right)\frac{m_T}{m_p}2\frac{v_\varepsilon^2}{v_{dx}^2}\right)\Big/h=\nu_{K_{Dn}^2+K_P^2}$$

$$\left(\frac{\hbar^2}{2m_T}K_{Ds}^2\frac{m_T}{m_p}2\frac{v_\varepsilon^2}{v_{dx}^2}\right)\Big/h=\nu_{K_{Ds}^2}$$

$$\left(\frac{\hbar^2}{2m_T}K_C^2\frac{m_T}{m_p}2\frac{v_\varepsilon^2}{v_{dx}^2}\right)\Big/h=\nu_{K_C^2}$$

$$\left(\frac{\hbar^2}{2m_T}\left(K_{Ds}^2+K_B^2\right)\frac{m_T}{m_p}2\frac{v_\varepsilon^2}{v_{dx}^2}\right)\Big/h=\nu_{K_{Ds}^2+K_B^2}$$

$$\left(\frac{\hbar^2}{2m_T}\left(K_{Ds}^2+K_{Dn}^2\right)\frac{m_T}{m_p}2\frac{v_\varepsilon^2}{v_{dx}^2}\right)\Big/h=\nu_{K_{Ds}^2+K_{Dn}^2}$$

$$\left(\frac{\hbar^2}{2m_T}\left(K_B^2+K_C^2\right)\frac{m_T}{m_p}2\frac{v_\varepsilon^2}{v_{dx}^2}\right)\Big/h=\nu_{K_B^2+K_C^2}$$

$$\left(\frac{\hbar^2}{2m_T}\left(K_{Ds}^2+K_P^2\right)\frac{m_T}{m_p}2\frac{v_\varepsilon^2}{v_{dx}^2}\right)\Big/h=\nu_{K_{Ds}^2+K_P^2}$$

$$\left(\frac{\hbar^2}{2m_T}\left(K_{Dn}^2+K_C^2\right)\frac{m_T}{m_p}2\frac{v_\varepsilon^2}{v_{dx}^2}\right)\Big/h=\nu_{K_{Dn}^2+K_C^2}$$

$$\left(\frac{\hbar^2}{2m_T}\left(K_{Ds}^2+K_C^2\right)\frac{m_T}{m_p}2\frac{v_\varepsilon^2}{v_{dx}^2}\right)\Big/h=\nu_{K_{Ds}^2+K_C^2}$$

$$\left(\frac{\hbar^2}{2m_T}K_A^2\frac{m_T}{m_p}2\frac{v_\varepsilon^2}{v_{dx}^2}\right)\Big/h=\nu_{K_A^2}$$

The 30 – 44 GHz Milky Way haze spectrum within Planck spacecraft frequency observational capability confirming WMAP haze detection at 23 GHz[114] with proximate trisine resonant modes shaded:

| | | |
|---|---|---|
| 857 | 30 | GHz |
| 350 | 10,000 | $\mu m$ |

| Harmonic | 1 | 2 | 3 | 4 | 5 | 6 |
|---|---|---|---|---|---|---|
| 1/Harmonic | 1/1 | 1/2 | 1/3 | 1/4 | 1/5 | 1/6 |
| Frequency GHz | | | | | | |
| $\nu_{K_B^2}$ | 3.40 | 6.80 | 10.21 | 13.61 | 17.01 | 20.41 |
| $\nu_{K_{D_n}^2}$ | 4.08 | 8.15 | 12.23 | 16.31 | 20.38 | 24.46 |
| $\nu_{K_P}$ | 4.54 | 9.07 | 13.61 | 18.15 | 22.68 | 27.22 |
| $\nu_{K_{BCS}^2}$ | 6.00 | 12.00 | 18.00 | 24.00 | 30.01 | 36.01 |
| $\nu_{K_{Ds}^2-K_{Dn}^2}$ | 6.52 | 13.03 | 19.55 | 26.07 | 32.58 | 39.10 |
| $\nu_{K_C^2-K_B^2}$ | 8.01 | 16.02 | 24.03 | 32.04 | 40.05 | 48.06 |
| $\nu_{K_{Dn}^2+K_P^2}$ | 8.61 | 17.23 | 25.84 | 34.45 | 43.07 | 51.68 |

| $\nu_{K_{Ds}^2}$ | 10.59 | 21.19 | 31.78 | 42.37 | 52.97 | 63.56 |
|---|---|---|---|---|---|---|
| $\nu_{K_C^2}$ | 11.41 | 22.82 | 34.24 | 45.65 | 57.06 | 68.47 |
| $\nu_{K_{Ds}^2+K_B^2}$ | 14.00 | 27.99 | 41.99 | 55.98 | 69.98 | 83.98 |
| $\nu_{K_{Ds}^2+K_{Dn}^2}$ | 14.67 | 29.34 | 44.01 | 58.68 | 73.35 | 88.02 |
| $\nu_{K_B^2+K_C^2}$ | 14.81 | 29.63 | 44.44 | 59.26 | 74.07 | 88.89 |
| $\nu_{K_{Ds}^2+K_P^2}$ | 15.13 | 30.26 | 45.39 | 60.52 | 75.65 | 90.78 |
| $\nu_{K_{Dn}^2+K_C^2}$ | 15.49 | 30.98 | 46.47 | 61.96 | 77.44 | 92.93 |
| $\nu_{K_{Ds}^2+K_C^2}$ | 22.01 | 44.01 | 66.02 | 88.02 | 110.0 | 132.0 |
| $\nu_{K_A^2}$ | 77.03 | 154.1 | 231.1 | 308.1 | 385.2 | 462.2 |

This trisine and galactic haze correlation represents a Compton like mechanism.

4.42. The approach in estimating Pioneer 10 and 11 deceleration (section 2) can estimate the universe dark energy anomalous acceleration but in terms of an actual deceleration of the local observer to the supernovae type 1A candle explosion events. The earth its satellite detector within the local galactic environment would be considered the observer slowing down between the supernovae type 1A event and the detection event. Equation 2.7.6 (repeated here) indicates this observer slowing or deceleration would be isotropic in as much as $c^2$ is constant. The parameters $C_t$ and $A_c$, are characteristic of the local galactic environment and not the exploding supernovae type 1A environment.

$$F \approx -C_t \frac{m_t c^2}{s} = C_t A_c \rho_U c^2 \qquad (2.7.6)$$

$$\text{where: } C_t = \frac{24}{R_{eo}} + \frac{6}{1+\sqrt{R_{eo}}} + .4 \text{ and } R_{eo} = \frac{\rho_U c d}{\mu_{Uc}}$$

Where constant $C_t$ is in the form of the standard fluid mechanical drag coefficient analogous to 2.7.4a, but is defined as a thrust coefficient.

This trisine model was used to analyze the Pioneer unmodeled deceleration data 8.74E-08 cm/sec$^2$ and there appeared to be a good fit with the observed space density $(\rho_U)$ 6E-30 g/cm$^3$ (trisine model $(\rho_U)$ 6.38E-30 g/cm$^3$). A thrust coefficient $(C_t)$ of 59.67 indicates a laminar flow condition. An absolute space viscosity seen $\left(\mu_{Uc} = m_T c/\left(2\Delta x^2\right)\right)$ of 1.21E-16 g/(cm sec) is established within trisine model and determines a space kinematic viscosity seen $(\mu_{Uc}/\rho_U)$ of $(1.21\text{E}-16/6.38\text{E}-30)$ or 1.90E13 cm$^2$/sec as indicated in 2.7.4a. NASA, JPL raised a question concerning the universal application of the observed deceleration on Pioneer 10 & 11. In other words, why do not the planets and their satellites experience such a deceleration?

The answer is in the $A_c/M$ ratio in the modified thrust relationship.

$$a \approx -C_t \frac{m_t}{M}\frac{c^2}{s} = -C_t \frac{A_c}{M}\rho_U c^2 = -C_t \frac{A_c}{M}\frac{m_t}{cavity}c^2 \qquad (2.7.6a)$$

Using the earth as an example, the earth differential movement

$$L_1 = \frac{1}{2}a\left(Age_{UGpresent}^2 - Age_{UG}^2\right)$$

The $z$ factors $(1+z_R)^a$ or $(1+z_B)^b$ are presented for extrapolation of present values into the observational past consistent with tabular values in this paper. The two references $(1+z_R)^a$ or $(1+z_B)^b$ are presented representing both the universe Hubble radius of curvature $R_U$ and trisine cellular $B$ perspectives where $a/b = 1/3$.

$L_2$ scales as trisine local dimension $L_2 \sim B_{present}(1+z_R)^{-1/3}$

With relationship to universe Hubble volume $V_U$ dimension and Hubble radius of curvature dimension $R_U$ as follows:

$$\#_{Upresent} = \frac{m_t}{m_e}\frac{V_{Upresent}}{cavity_{present}}$$

$$where \quad V_{Upresent} = \frac{4\pi}{3}R_{Upresent}^3$$

$$and \quad cavity_{present} = 2\sqrt{3}AB^2 = 2\sqrt{3}\left(\frac{A}{B}\right)_t B^3$$

A

| $n$ | | $\sum_1^n\left((1+z_{Rn})^w\right)$ | $\sum_1^n\left((1+z_{Rn})^w\right)$ | $\sum_1^n\left((1+z_{Rn})^w\right)$ | $\sum_1^n\left((1+z_{Rn})^w\right)$ |
|---|---|---|---|---|---|
| $\sum_1^n\left((1+z_{Rn})^w\right)$ | $\sum_1^n\left((1+z_{Rn})^w\right)$ | $\sum_1^n\left((1+z_{Rn})^w\right)$ | $\sum_1^n\left((1+z_{Rn})^w\right)$ | $\sum_1^n\left((1+z_{Rn})^w\right)$ | |
| $\sum_1^n\left((1+z_{Rn})^w\right)$ | $\sum_1^n\left((1+z_{Rn})^w\right)$ | $\sum_1^n\left((1+z_{Rn})^w\right)$ | $\sum_1^n\left((1+z_{Rn})^w\right)$ | $\sum_1^n\left((1+z_{Rn})^w\right)$ | |
| $\sum_1^n\left((1+z_{Rn})^w\right)$ | $\sum_1^n\left((1+z_{Rn})^w\right)$ | $\sum_1^n\left((1+z_{Rn})^w\right)$ | $\sum_1^n\left((1+z_{Rn})^w\right)$ | $\sum_1^n\left((1+z_{Rn})^w\right)$ | |
| $\sum_1^n\left((1+z_{Rn})^w\right)$ | $\sum_1^n\left((1+z_{Rn})^w\right)$ | $\sum_1^n\left((1+z_{Rn})^w\right)$ | $\sum_1^n\left((1+z_{Rn})^w\right)$ | $\sum_1^n\left((1+z_{Rn})^w\right)$ | |

| b | Inverse(A) x b |
|---|---|
| $\sum_1^n C_{t(1+z_{Rn})}\left(\left(1+z_{Rn}\right)^w\right)^0$ | $a_{0R}$ |
| $\sum_1^n C_{t(1+z_{Rn})}\left(\left(1+z_{Rn}\right)^w\right)^1$ | $a_{1R}$ |
| $\sum_1^n C_{t(1+z_{Rn})}\left(\left(1+z_{Rn}\right)^w\right)^2$ | $a_{2R}$ |
| $\sum_1^n C_{t(1+z_{Rn})}\left(\left(1+z_{Rn}\right)^w\right)^3$ | $a_{3R}$ |
| $\sum_1^n C_{t(1+z_{Rn})}\left(\left(1+z_{Rn}\right)^w\right)^4$ | $a_{4R}$ |

$$C_{t(1+z_R)} = a_{0R}\left(\left(1+z_R\right)^w\right)^0 + a_{1R}\left(\left(1+z_R\right)^w\right)^1 + a_{2R}\left(\left(1+z_R\right)^w\right)^2 + a_{3R}\left(\left(1+z_R\right)^w\right)^3 + a_{4R}\left(\left(1+z_R\right)^w\right)^4$$

and combining exponents by multiplication

$$C_{t(1+z_R)} = a_{0R} + a_{1R}\left(1+z_R\right)^{1w} + a_{2R}\left(1+z_R\right)^{2w} + a_{3R}\left(1+z_R\right)^{3w} + a_{4R}\left(1+z_R\right)^{4w}$$

and similarly for B scaling:

A

$$\begin{array}{ccccc}
n & \sum_1^n\left(\left(1+z_{Bn}\right)^w\right. & \sum_1^n\left(\left(1+z_{Bn}\right)^w\right. & \sum_1^n\left(\left(1+z_{Bn}\right)^w\right. & \sum_1^n\left(\left(1+z_{Bn}\right)^w\right. \\
\sum_1^n\left(\left(1+z_{Bn}\right)^w\right. & \sum_1^n\left(\left(1+z_{Bn}\right)^w\right. & \sum_1^n\left(\left(1+z_{Bn}\right)^w\right. & \sum_1^n\left(\left(1+z_{Bn}\right)^w\right. & \sum_1^n\left(\left(1+z_{Bn}\right)^{-\frac{1}{3}}\right. \\
\sum_1^n\left(\left(1+z_{Bn}\right)^w\right. & \sum_1^n\left(\left(1+z_{Bn}\right)^w\right. & \sum_1^n\left(\left(1+z_{Bn}\right)^w\right. & \sum_1^n\left(\left(1+z_{Bn}\right)^{-\frac{1}{3}}\right. & \sum_1^n\left(\left(1+z_{Bn}\right)^w\right. \\
\sum_1^n\left(\left(1+z_{Bn}\right)^w\right. & \sum_1^n\left(\left(1+z_{Bn}\right)^w\right. & \sum_1^n\left(\left(1+z_{Bn}\right)^{-\frac{1}{3}}\right. & \sum_1^n\left(\left(1+z_{Bn}\right)^w\right. & \sum_1^n\left(\left(1+z_{Bn}\right)^w\right. \\
\sum_1^n\left(\left(1+z_{Bn}\right)^w\right. & \sum_1^n\left(\left(1+z_{Bn}\right)^{-\frac{1}{3}}\right. & \sum_1^n\left(\left(1+z_{Bn}\right)^w\right. & \sum_1^n\left(\left(1+z_{Bn}\right)^w\right. & \sum_1^n\left(\left(1+z_{Bn}\right)^w\right.
\end{array}$$

| b | Inverse(A) x b |
|---|---|
| $\sum_1^n C_{t(1+z_{Bn})}\left(\left(1+z_{Bn}\right)^w\right)^0$ | $a_{0B}$ |
| $\sum_1^n C_{t(1+z_{Bn})}\left(\left(1+z_{Bn}\right)^w\right)^1$ | $a_{1B}$ |
| $\sum_1^n C_{t(1+z_{Bn})}\left(\left(1+z_{Bn}\right)^w\right)^2$ | $a_{2B}$ |
| $\sum_1^n C_{t(1+z_{Bn})}\left(\left(1+z_{Bn}\right)^w\right)^3$ | $a_{3B}$ |
| $\sum_1^n C_{t(1+z_{Bn})}\left(\left(1+z_{Bn}\right)^w\right)^4$ | $a_{4B}$ |

$$C_{t(1+z_B)} = a_{0B}\left(\left(1+z_B\right)^w\right)^0 + a_{1B}\left(\left(1+z_B\right)^w\right)^1 + a_{2B}\left(\left(1+z_B\right)^w\right)^2 + a_{3B}\left(\left(1+z_B\right)^w\right)^3 + a_{4B}\left(\left(1+z_B\right)^w\right)^4$$

and combining exponents by multiplication

$$C_{t(1+z_B)} = a_{0B} + a_{1B}\left(1+z_B\right)^{1w} + a_{2B}\left(1+z_R\right)^{2w} + a_{3B}\left(1+z_B\right)^{3w} + a_{4B}\left(1+z_B\right)^{4w}$$

$$\left(z_B+1\right)^3 = \left(z_R+1\right)^1$$

$$z_B = \left(z_R+1\right)^{1/3} - 1 \qquad z_R = \left(z_B+1\right)^3 - 1$$

The universe expands at the speed of light $c$.

$$R_U = \frac{\sqrt{3}c}{H_U}$$

$$\begin{aligned}
L_{zR} &= \frac{1}{3\sqrt{3}} R_{Upresent} \int_0^{z_R} \frac{dz}{\left(1+z\right)^{1/3}} \\
&= \frac{1}{3\sqrt{3}} R_{Upresent} \left(\left(\frac{3}{2}\right)\left(z_R+1\right)^{2/3} - \frac{3}{2}\right) \\
&= \frac{1}{3\sqrt{3}} R_{Upresent} \int_0^{\left(1+z_B\right)^3-1} \frac{dz}{\left(1+z\right)^{1/3}} \\
&= \frac{1}{3\sqrt{3}} R_{Upresent} \left(\left(\frac{3}{2}\right)\left(z_B^3+3z_B^2+3z_B+1\right)^{2/3} - \frac{3}{2}\right)
\end{aligned}$$

And scaling as conventionally used [122 -Harrison]

$$L_{z_B} = \frac{1}{3^2} R_{Upresent} \int_0^{z_B} \frac{dz}{(1+z)}$$

$$= \frac{1}{3^2} R_{Upresent} \ln(1+z_B)$$

$$= \frac{1}{3^2} R_{Upresent} \int_0^{(z_R+1)^{1/3}-1} \frac{dz}{(1+z)}$$

$$= \frac{1}{3^3} R_{Upresent} \ln(1+z_R)$$

$$L_1 = \int_0^{L_1} 1 dL_1 = \int_0^{z_B} \frac{1}{2} a t^2 dz_B$$

$$= \int_0^{z_B} \frac{1}{2} C_{t(1+z_B)} \frac{A_c}{M} (1+z_B)^3 \rho_{present} c^2 \left( (1+z_B)^{-3/2} Age_{UGpresent} \right)^2 dz_B$$

$$= \frac{A_c}{M} \rho_{present} c^2 Age^2_{UGpresent} \begin{pmatrix} a_{0B} z_B / 2 \\ + \dfrac{a_{1B}\left(z_B (z_B+1)^w + (z_B+1)^w - 1\right)}{2w+2} \\ + \dfrac{a_{2B}\left(z_B (z_B+1)^{2w} + (z_B+1)^{2w} - 1\right)}{4w+2} \\ + \dfrac{a_{3B}\left(z_B (z_B+1)^{3w} + (z_B+1)^{3w} - 1\right)}{6w+2} \\ + \dfrac{a_{4B}\left(z_B (z_B+1)^{4w} + (z_B+1)^{4w} - 1\right)}{8w+2} \end{pmatrix}$$

$$= \frac{A_c}{M} factor$$

Then normalize the object area to mass ratio $A_c/M$ to $L_{z_B}$:

$$\frac{A_c}{M} = \frac{L_{z_B}}{factor}$$

Monte Carlo analysis indicates $w$ = -1/3.334 with reasonable correlation only for $z_B$ and not $z_R$ and a clear indication of laminar fluid condition as indicated in $C_t$ vs $R_{e0}$.

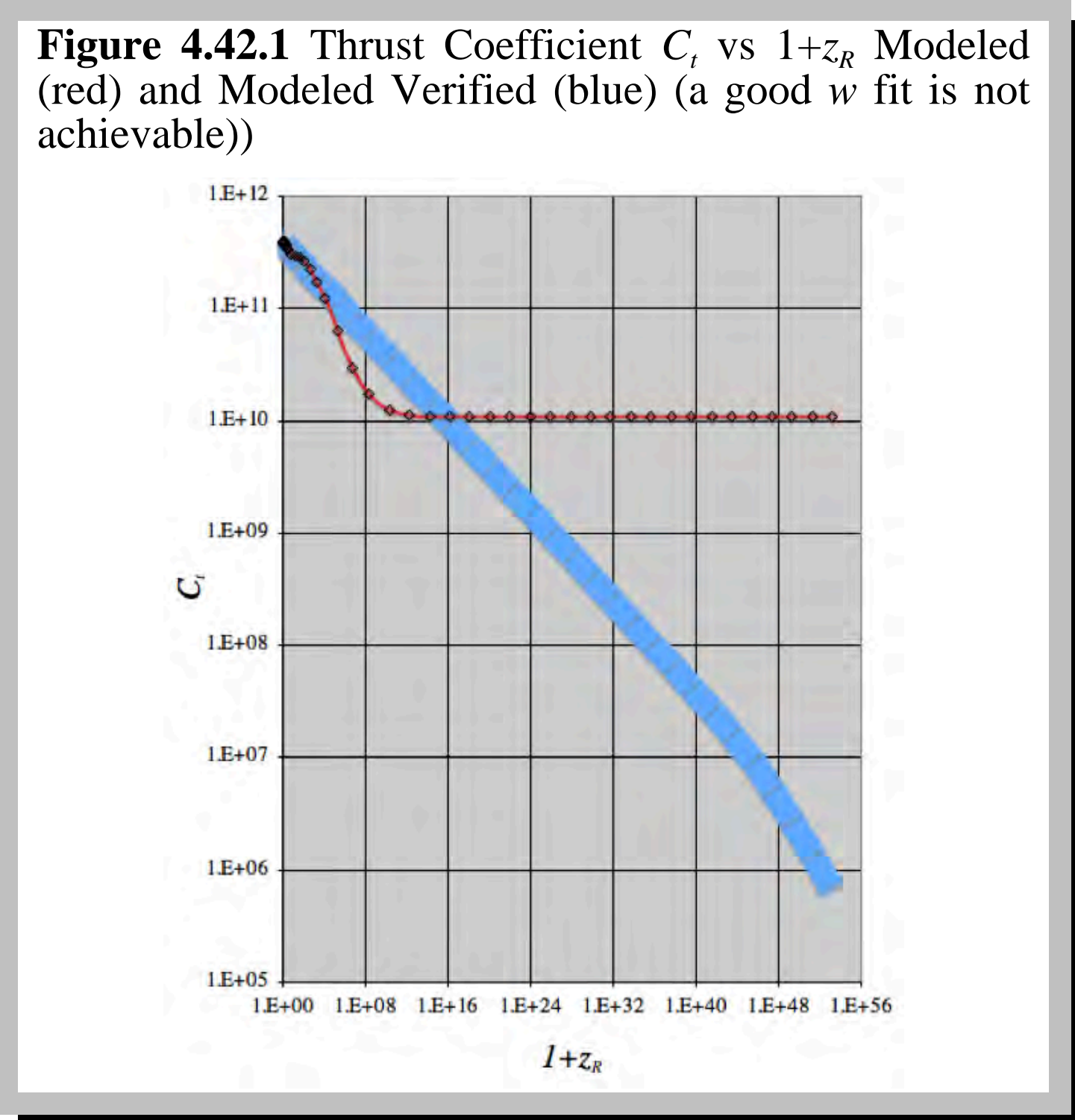

**Figure 4.42.1** Thrust Coefficient $C_t$ vs 1+$z_R$ Modeled (red) and Modeled Verified (blue) (a good $w$ fit is not achievable))

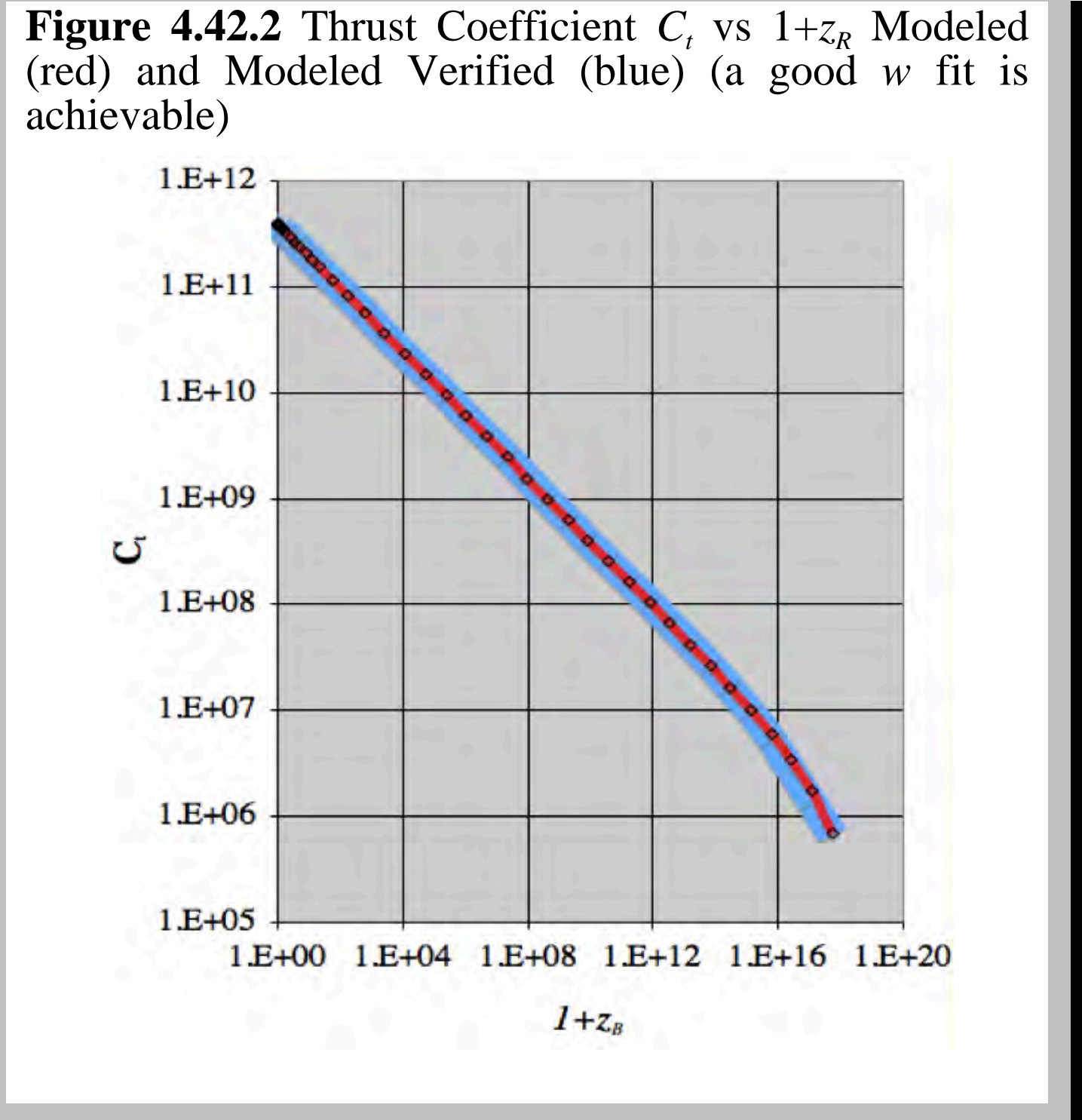

**Figure 4.42.2** Thrust Coefficient $C_t$ vs 1+$z_R$ Modeled (red) and Modeled Verified (blue) (a good $w$ fit is achievable)

For all energies represented by velocities ($v$) including $c$.

By definition the absolute magnitude(M) is the apparent magnitude(m) at a distance of 10 parsec that provides a distance modulus $(m-M)_z$.

$$(m-M)_{z_R} = 5\log_{10}\left(L_{z_R}\right) - 5$$

$$(m-M)_{z_B} = 5\log_{10}\left(L_{z_B}\right) - 5$$

$$(m-M)_{z_B} = 5\log_{10}\left(L_1\right) - 5$$

$$5\log_{10}\left(L_{z_B}\right) - 5 = 5\log_{10}\left(L_1\right) - 5$$

Here $L_1,\ L_{z_B} and\ L_{z_R}$

have been dimensionally changed to *parsecs*

Distance modulus conversion between $z_R$ and $z_B$ universe scaling is as follows:

$$(m-M)_{z_R} = 5\log_{10}\left(\sqrt{3}\ln\left(z_B+1\right)/\left(\frac{3}{2}\left(z_R+1\right)^{2/3}-\frac{3}{2}\right)10^{\left(\left((m-M)_{z_B}+5\right)/5\right)}\right) - 5$$

The curvature in Figure 4.42.2 is not due to accelerated expansion but dual linear expansion of universe radius *R* and contained lattice cellular structure with unit length *B*.

http://supernova.lbl.gov/Union/figures/SCPUnion2.1_mu_vs_z.txt

| $Uage_{GUzR}$ | $z_R$ | $(m-M)_{z_R}$ | $Uage_{GUzB}$ | $z_B$ | $(m-M)_{z_B}$ |
|---|---|---|---|---|---|
| 4.15E+17 | 0.088 | 34.092 | 4.15E+17 | 0.028 | 35.347 |
| 3.47E+17 | 0.158 | 35.383 | 4.02E+17 | 0.050 | 36.682 |
| 3.43E+17 | 0.167 | 35.512 | 4.00E+17 | 0.053 | 36.818 |
| 3.19E+17 | 0.225 | 36.105 | 3.91E+17 | 0.070 | 37.447 |
| 3.29E+17 | 0.200 | 36.157 | 3.95E+17 | 0.063 | 37.483 |
| 2.96E+17 | 0.286 | 36.851 | 3.81E+17 | 0.088 | 38.229 |
| 3.08E+17 | 0.255 | 36.129 | 3.86E+17 | 0.079 | 37.488 |
| 4.00E+17 | 0.053 | 33.424 | 4.21E+17 | 0.017 | 34.654 |
| 3.59E+17 | 0.132 | 35.053 | 4.06E+17 | 0.042 | 36.336 |
| 3.54E+17 | 0.142 | 35.351 | 4.05E+17 | 0.045 | 36.640 |
| 3.68E+17 | 0.113 | 34.634 | 4.10E+17 | 0.036 | 35.905 |
| 3.96E+17 | 0.060 | 33.350 | 4.20E+17 | 0.020 | 34.585 |
| 2.81E+17 | 0.334 | 37.052 | 3.74E+17 | 0.101 | 38.457 |
| 3.83E+17 | 0.084 | 33.834 | 4.15E+17 | 0.027 | 35.086 |
| 3.13E+17 | 0.241 | 36.237 | 3.88E+17 | 0.075 | 37.588 |
| 3.84E+17 | 0.082 | 34.231 | 4.16E+17 | 0.026 | 35.481 |
| 3.47E+17 | 0.157 | 35.268 | 4.02E+17 | 0.050 | 36.567 |
| 3.78E+17 | 0.095 | 34.292 | 4.13E+17 | 0.031 | 35.550 |
| 4.02E+17 | 0.050 | 32.816 | 4.22E+17 | 0.016 | 34.044 |
| 4.04E+17 | 0.047 | 32.715 | 4.23E+17 | 0.015 | 33.941 |
| 3.78E+17 | 0.094 | 34.341 | 4.13E+17 | 0.031 | 35.599 |
| 3.88E+17 | 0.075 | 33.813 | 4.17E+17 | 0.025 | 35.058 |
| 3.89E+17 | 0.074 | 33.724 | 4.17E+17 | 0.024 | 34.969 |
| 3.85E+17 | 0.080 | 34.118 | 4.16E+17 | 0.026 | 35.367 |
| 3.49E+17 | 0.154 | 35.434 | 4.03E+17 | 0.049 | 36.732 |
| 3.88E+17 | 0.075 | 33.864 | 4.17E+17 | 0.024 | 35.109 |
| 4.04E+17 | 0.046 | 32.876 | 4.23E+17 | 0.015 | 34.102 |
| 3.69E+17 | 0.111 | 34.691 | 4.10E+17 | 0.036 | 35.961 |
| 3.49E+17 | 0.154 | 35.085 | 4.03E+17 | 0.049 | 36.382 |
| 3.92E+17 | 0.067 | 33.613 | 4.19E+17 | 0.022 | 34.853 |
| 2.55E+17 | 0.422 | 37.592 | 3.63E+17 | 0.124 | 39.045 |
| 3.69E+17 | 0.112 | 34.551 | 4.10E+17 | 0.036 | 35.821 |
| 4.02E+17 | 0.050 | 32.789 | 4.22E+17 | 0.016 | 34.017 |
| 4.01E+17 | 0.051 | 32.997 | 4.22E+17 | 0.017 | 34.226 |
| 3.83E+17 | 0.085 | 34.398 | 4.15E+17 | 0.028 | 35.650 |
| 3.92E+17 | 0.067 | 33.734 | 4.19E+17 | 0.022 | 34.974 |
| 4.02E+17 | 0.050 | 32.953 | 4.22E+17 | 0.016 | 34.181 |
| 3.90E+17 | 0.071 | 33.843 | 4.18E+17 | 0.023 | 35.086 |
| 3.68E+17 | 0.113 | 34.863 | 4.10E+17 | 0.036 | 36.134 |
| 3.97E+17 | 0.059 | 33.718 | 4.20E+17 | 0.019 | 34.953 |
| 4.00E+17 | 0.054 | 33.113 | 4.21E+17 | 0.018 | 34.344 |
| 3.76E+17 | 0.098 | 34.468 | 4.13E+17 | 0.032 | 35.729 |
| 3.89E+17 | 0.072 | 33.926 | 4.18E+17 | 0.024 | 35.170 |
| 4.01E+17 | 0.051 | 32.774 | 4.22E+17 | 0.017 | 34.003 |
| 3.42E+17 | 0.170 | 35.169 | 4.00E+17 | 0.054 | 36.476 |
| 4.01E+17 | 0.052 | 33.149 | 4.22E+17 | 0.017 | 34.379 |
| 3.82E+17 | 0.086 | 33.841 | 4.15E+17 | 0.028 | 35.093 |
| 4.01E+17 | 0.052 | 33.031 | 4.22E+17 | 0.017 | 34.261 |
| 3.79E+17 | 0.093 | 34.715 | 4.14E+17 | 0.030 | 35.972 |
| 4.02E+17 | 0.051 | 33.115 | 4.22E+17 | 0.017 | 34.344 |
| 4.04E+17 | 0.046 | 32.938 | 4.23E+17 | 0.015 | 34.164 |
| 3.41E+17 | 0.172 | 35.646 | 3.99E+17 | 0.054 | 36.954 |
| 2.25E+17 | 0.545 | 37.714 | 3.48E+17 | 0.156 | 39.229 |
| 3.64E+17 | 0.123 | 35.057 | 4.08E+17 | 0.039 | 36.334 |
| 2.55E+17 | 0.420 | 37.370 | 3.63E+17 | 0.124 | 38.822 |
| 2.36E+17 | 0.498 | 37.344 | 3.53E+17 | 0.144 | 38.836 |
| 2.50E+17 | 0.443 | 37.516 | 3.60E+17 | 0.130 | 38.979 |
| 3.08E+17 | 0.254 | 36.323 | 3.86E+17 | 0.078 | 37.682 |
| 3.35E+17 | 0.185 | 35.716 | 3.97E+17 | 0.058 | 37.033 |
| 3.77E+17 | 0.096 | 34.670 | 4.13E+17 | 0.031 | 35.929 |
| 3.62E+17 | 0.127 | 35.086 | 4.07E+17 | 0.041 | 36.366 |
| 4.04E+17 | 0.046 | 32.791 | 4.23E+17 | 0.015 | 34.017 |
| 3.91E+17 | 0.069 | 33.706 | 4.18E+17 | 0.022 | 34.947 |
| 4.03E+17 | 0.049 | 32.947 | 4.22E+17 | 0.016 | 34.174 |
| 3.68E+17 | 0.113 | 34.716 | 4.10E+17 | 0.036 | 35.987 |
| 4.00E+17 | 0.053 | 33.020 | 4.21E+17 | 0.017 | 34.250 |
| 3.77E+17 | 0.097 | 34.367 | 4.13E+17 | 0.031 | 35.627 |
| 3.92E+17 | 0.068 | 33.671 | 4.18E+17 | 0.022 | 34.912 |
| 4.03E+17 | 0.049 | 32.597 | 4.22E+17 | 0.016 | 33.825 |
| 3.87E+17 | 0.077 | 33.557 | 4.17E+17 | 0.025 | 34.804 |
| 3.78E+17 | 0.094 | 34.370 | 4.14E+17 | 0.030 | 35.628 |
| 3.81E+17 | 0.087 | 34.267 | 4.15E+17 | 0.028 | 35.520 |
| 4.04E+17 | 0.046 | 33.033 | 4.23E+17 | 0.015 | 34.258 |
| 3.71E+17 | 0.107 | 34.711 | 4.11E+17 | 0.035 | 35.978 |
| 3.69E+17 | 0.112 | 34.409 | 4.10E+17 | 0.036 | 35.679 |
| 3.87E+17 | 0.076 | 34.010 | 4.17E+17 | 0.025 | 35.256 |
| 3.80E+17 | 0.090 | 34.737 | 4.14E+17 | 0.029 | 35.993 |
| 4.02E+17 | 0.050 | 33.225 | 4.22E+17 | 0.016 | 34.453 |
| 3.98E+17 | 0.057 | 33.815 | 4.21E+17 | 0.019 | 35.048 |
| 3.96E+17 | 0.060 | 33.522 | 4.20E+17 | 0.020 | 34.757 |
| 3.86E+17 | 0.079 | 34.437 | 4.16E+17 | 0.026 | 35.685 |
| 3.73E+17 | 0.105 | 34.579 | 4.11E+17 | 0.034 | 35.844 |
| 3.40E+17 | 0.173 | 35.300 | 3.99E+17 | 0.055 | 36.610 |
| 3.89E+17 | 0.074 | 33.932 | 4.17E+17 | 0.024 | 35.176 |
| 3.73E+17 | 0.104 | 34.740 | 4.12E+17 | 0.034 | 36.005 |
| 3.72E+17 | 0.106 | 34.576 | 4.11E+17 | 0.034 | 35.842 |
| 3.85E+17 | 0.080 | 34.111 | 4.16E+17 | 0.026 | 35.360 |
| 3.94E+17 | 0.065 | 33.422 | 4.19E+17 | 0.021 | 34.661 |
| 3.75E+17 | 0.099 | 34.634 | 4.12E+17 | 0.032 | 35.896 |
| 3.92E+17 | 0.068 | 33.682 | 4.18E+17 | 0.022 | 34.922 |
| 3.73E+17 | 0.104 | 34.615 | 4.12E+17 | 0.033 | 35.880 |
| 3.72E+17 | 0.106 | 34.677 | 4.11E+17 | 0.034 | 35.943 |
| 3.59E+17 | 0.132 | 35.117 | 4.06E+17 | 0.042 | 36.400 |
| 3.36E+17 | 0.183 | 35.765 | 3.98E+17 | 0.058 | 37.080 |
| 3.95E+17 | 0.063 | 33.380 | 4.19E+17 | 0.021 | 34.617 |

| | | | | | | | | | | | |
|---|---|---|---|---|---|---|---|---|---|---|---|
| 3.62E+17 | 0.126 | 35.096 | 4.08E+17 | 0.040 | 36.375 | 3.55E+17 | 0.141 | 35.261 | 4.05E+17 | 0.045 | 36.550 |
| 3.85E+17 | 0.080 | 34.133 | 4.16E+17 | 0.026 | 35.381 | 3.74E+17 | 0.102 | 34.704 | 4.12E+17 | 0.033 | 35.968 |
| 3.85E+17 | 0.080 | 34.167 | 4.16E+17 | 0.026 | 35.416 | 3.12E+17 | 0.244 | 36.228 | 3.88E+17 | 0.075 | 37.580 |
| 3.89E+17 | 0.073 | 33.791 | 4.17E+17 | 0.024 | 35.035 | 3.95E+17 | 0.062 | 33.420 | 4.20E+17 | 0.020 | 34.656 |
| 3.20E+17 | 0.222 | 36.227 | 3.91E+17 | 0.069 | 37.566 | 3.90E+17 | 0.071 | 33.892 | 4.18E+17 | 0.023 | 35.134 |
| 3.26E+17 | 0.208 | 35.975 | 3.93E+17 | 0.065 | 37.307 | 3.84E+17 | 0.083 | 34.010 | 4.16E+17 | 0.027 | 35.260 |
| 3.91E+17 | 0.070 | 33.955 | 4.18E+17 | 0.023 | 35.197 | 3.99E+17 | 0.055 | 33.118 | 4.21E+17 | 0.018 | 34.350 |
| 3.76E+17 | 0.098 | 34.391 | 4.13E+17 | 0.032 | 35.651 | 3.50E+17 | 0.152 | 35.390 | 4.03E+17 | 0.048 | 36.686 |
| 3.93E+17 | 0.066 | 33.694 | 4.19E+17 | 0.022 | 34.933 | 3.37E+17 | 0.180 | 35.650 | 3.98E+17 | 0.057 | 36.964 |
| 3.86E+17 | 0.078 | 34.472 | 4.16E+17 | 0.026 | 35.720 | 3.27E+17 | 0.204 | 35.987 | 3.94E+17 | 0.064 | 37.316 |
| 3.74E+17 | 0.101 | 34.550 | 4.12E+17 | 0.033 | 35.813 | 2.34E+17 | 0.506 | 38.063 | 3.52E+17 | 0.146 | 39.559 |
| 3.00E+17 | 0.275 | 36.681 | 3.83E+17 | 0.084 | 38.052 | 2.50E+17 | 0.440 | 37.456 | 3.60E+17 | 0.129 | 38.918 |
| 3.77E+17 | 0.095 | 34.370 | 4.13E+17 | 0.031 | 35.629 | 2.79E+17 | 0.341 | 37.079 | 3.73E+17 | 0.103 | 38.487 |
| 3.74E+17 | 0.101 | 34.831 | 4.12E+17 | 0.033 | 36.094 | 1.63E+17 | 0.918 | 38.449 | 3.12E+17 | 0.243 | 40.130 |
| 3.59E+17 | 0.132 | 35.109 | 4.06E+17 | 0.042 | 36.393 | 1.34E+17 | 1.189 | 39.278 | 2.92E+17 | 0.298 | 41.062 |
| 3.21E+17 | 0.220 | 36.393 | 3.92E+17 | 0.068 | 37.731 | 3.57E+17 | 0.137 | 35.100 | 4.06E+17 | 0.044 | 36.387 |
| 4.04E+17 | 0.047 | 33.481 | 4.23E+17 | 0.015 | 34.707 | 2.67E+17 | 0.379 | 37.122 | 3.68E+17 | 0.113 | 38.551 |
| 3.90E+17 | 0.072 | 33.639 | 4.18E+17 | 0.023 | 34.882 | 1.55E+17 | 0.984 | 38.903 | 3.07E+17 | 0.256 | 40.611 |
| 3.49E+17 | 0.155 | 35.432 | 4.02E+17 | 0.049 | 36.730 | 1.35E+17 | 1.175 | 39.358 | 2.93E+17 | 0.296 | 41.137 |
| 3.59E+17 | 0.133 | 34.644 | 4.06E+17 | 0.043 | 35.928 | 1.01E+17 | 1.630 | 39.727 | 2.67E+17 | 0.380 | 41.657 |
| 3.97E+17 | 0.059 | 33.502 | 4.20E+17 | 0.019 | 34.737 | 2.35E+17 | 0.504 | 37.562 | 3.53E+17 | 0.146 | 39.057 |
| 3.77E+17 | 0.095 | 34.520 | 4.13E+17 | 0.031 | 35.779 | 1.46E+17 | 1.065 | 38.988 | 3.01E+17 | 0.273 | 40.727 |
| 3.93E+17 | 0.065 | 33.609 | 4.19E+17 | 0.021 | 34.847 | 1.34E+17 | 1.184 | 38.989 | 2.93E+17 | 0.298 | 40.772 |
| 3.82E+17 | 0.085 | 34.453 | 4.15E+17 | 0.028 | 35.705 | 1.02E+17 | 1.622 | 39.654 | 2.67E+17 | 0.379 | 41.581 |
| 3.73E+17 | 0.104 | 34.709 | 4.12E+17 | 0.034 | 35.974 | 1.01E+17 | 1.630 | 39.342 | 2.67E+17 | 0.380 | 41.272 |
| 3.94E+17 | 0.064 | 33.558 | 4.19E+17 | 0.021 | 34.795 | 1.32E+17 | 1.206 | 39.703 | 2.91E+17 | 0.302 | 41.494 |
| 4.00E+17 | 0.053 | 33.000 | 4.21E+17 | 0.017 | 34.230 | 1.13E+17 | 1.451 | 39.434 | 2.76E+17 | 0.348 | 41.308 |
| 3.69E+17 | 0.112 | 34.876 | 4.10E+17 | 0.036 | 36.146 | 2.99E+17 | 0.280 | 36.624 | 3.82E+17 | 0.086 | 37.997 |
| 3.90E+17 | 0.072 | 33.955 | 4.18E+17 | 0.023 | 35.198 | 1.53E+17 | 1.003 | 38.808 | 3.06E+17 | 0.261 | 40.523 |
| 3.34E+17 | 0.187 | 35.793 | 3.97E+17 | 0.059 | 37.112 | 1.80E+17 | 0.796 | 38.666 | 3.23E+17 | 0.216 | 40.297 |
| 3.35E+17 | 0.185 | 35.741 | 3.97E+17 | 0.058 | 37.058 | 2.62E+17 | 0.396 | 37.141 | 3.66E+17 | 0.118 | 38.579 |
| 3.21E+17 | 0.221 | 36.148 | 3.91E+17 | 0.069 | 37.487 | 2.04E+17 | 0.652 | 38.027 | 3.36E+17 | 0.182 | 39.594 |
| 3.75E+17 | 0.099 | 34.386 | 4.12E+17 | 0.032 | 35.648 | 1.09E+17 | 1.502 | 39.439 | 2.73E+17 | 0.358 | 41.329 |
| 3.44E+17 | 0.165 | 35.370 | 4.01E+17 | 0.052 | 36.674 | 2.38E+17 | 0.489 | 37.787 | 3.54E+17 | 0.142 | 39.274 |
| 3.77E+17 | 0.095 | 34.334 | 4.13E+17 | 0.031 | 35.594 | 1.53E+17 | 1.003 | 39.093 | 3.06E+17 | 0.261 | 40.808 |
| 3.74E+17 | 0.102 | 34.678 | 4.12E+17 | 0.033 | 35.942 | 1.69E+17 | 0.874 | 38.541 | 3.16E+17 | 0.233 | 40.205 |
| 3.90E+17 | 0.071 | 33.828 | 4.18E+17 | 0.023 | 35.071 | 2.29E+17 | 0.528 | 37.649 | 3.50E+17 | 0.152 | 39.156 |
| 4.04E+17 | 0.046 | 33.155 | 4.23E+17 | 0.015 | 34.380 | 2.89E+17 | 0.309 | 36.782 | 3.78E+17 | 0.094 | 38.173 |
| 3.75E+17 | 0.099 | 34.609 | 4.12E+17 | 0.032 | 35.870 | 1.39E+17 | 1.130 | 39.275 | 2.96E+17 | 0.287 | 41.038 |
| 3.27E+17 | 0.206 | 35.846 | 3.94E+17 | 0.064 | 37.176 | 1.94E+17 | 0.704 | 38.372 | 3.31E+17 | 0.194 | 39.962 |
| 3.75E+17 | 0.099 | 34.565 | 4.12E+17 | 0.032 | 35.827 | 2.33E+17 | 0.509 | 37.806 | 3.52E+17 | 0.147 | 39.304 |
| 3.94E+17 | 0.064 | 33.465 | 4.19E+17 | 0.021 | 34.703 | 1.82E+17 | 0.779 | 38.937 | 3.24E+17 | 0.212 | 40.560 |
| 3.92E+17 | 0.067 | 33.609 | 4.19E+17 | 0.022 | 34.849 | 2.05E+17 | 0.644 | 38.076 | 3.37E+17 | 0.180 | 39.639 |
| 3.75E+17 | 0.099 | 34.327 | 4.12E+17 | 0.032 | 35.588 | 1.51E+17 | 1.017 | 39.047 | 3.04E+17 | 0.263 | 40.767 |
| 4.04E+17 | 0.046 | 33.300 | 4.23E+17 | 0.015 | 34.526 | 1.96E+17 | 0.694 | 38.454 | 3.32E+17 | 0.192 | 40.039 |
| 3.97E+17 | 0.059 | 33.257 | 4.20E+17 | 0.019 | 34.492 | 1.16E+17 | 1.400 | 39.448 | 2.79E+17 | 0.339 | 41.305 |
| 3.84E+17 | 0.082 | 34.073 | 4.16E+17 | 0.027 | 35.323 | 2.63E+17 | 0.395 | 37.308 | 3.66E+17 | 0.117 | 38.746 |
| 3.66E+17 | 0.117 | 34.522 | 4.09E+17 | 0.038 | 35.795 | 2.38E+17 | 0.491 | 37.628 | 3.54E+17 | 0.142 | 39.116 |
| 3.87E+17 | 0.076 | 33.668 | 4.17E+17 | 0.025 | 34.914 | 2.21E+17 | 0.564 | 37.794 | 3.46E+17 | 0.161 | 39.319 |
| 3.88E+17 | 0.074 | 33.944 | 4.17E+17 | 0.024 | 35.189 | 1.38E+17 | 1.139 | 39.077 | 2.96E+17 | 0.288 | 40.843 |
| 3.68E+17 | 0.114 | 34.700 | 4.10E+17 | 0.037 | 35.971 | 2.57E+17 | 0.416 | 37.351 | 3.63E+17 | 0.123 | 38.800 |
| 3.91E+17 | 0.070 | 33.896 | 4.18E+17 | 0.023 | 35.139 | 1.51E+17 | 1.018 | 38.827 | 3.04E+17 | 0.264 | 40.548 |
| 3.77E+17 | 0.097 | 34.623 | 4.13E+17 | 0.031 | 35.883 | 2.53E+17 | 0.429 | 37.254 | 3.62E+17 | 0.126 | 38.710 |
| 4.04E+17 | 0.046 | 32.886 | 4.23E+17 | 0.015 | 34.111 | 2.11E+17 | 0.613 | 37.955 | 3.41E+17 | 0.173 | 39.503 |
| 3.72E+17 | 0.106 | 34.504 | 4.11E+17 | 0.034 | 35.770 | 2.19E+17 | 0.576 | 37.864 | 3.44E+17 | 0.164 | 39.395 |
| 3.87E+17 | 0.077 | 33.701 | 4.17E+17 | 0.025 | 34.948 | 1.59E+17 | 0.951 | 39.086 | 3.10E+17 | 0.250 | 40.781 |
| 3.97E+17 | 0.058 | 33.141 | 4.20E+17 | 0.019 | 34.375 | 1.54E+17 | 0.990 | 38.942 | 3.07E+17 | 0.258 | 40.651 |
| 3.79E+17 | 0.092 | 34.214 | 4.14E+17 | 0.030 | 35.471 | 2.74E+17 | 0.356 | 37.220 | 3.71E+17 | 0.107 | 38.637 |
| 3.75E+17 | 0.100 | 34.116 | 4.12E+17 | 0.032 | 35.378 | 2.22E+17 | 0.560 | 37.828 | 3.46E+17 | 0.160 | 39.351 |
| 3.83E+17 | 0.085 | 34.223 | 4.15E+17 | 0.028 | 35.475 | 1.87E+17 | 0.750 | 38.414 | 3.27E+17 | 0.205 | 40.025 |
| 3.52E+17 | 0.148 | 35.201 | 4.04E+17 | 0.047 | 36.494 | 1.62E+17 | 0.927 | 38.517 | 3.12E+17 | 0.244 | 40.202 |
| 3.99E+17 | 0.056 | 33.139 | 4.21E+17 | 0.018 | 34.372 | 1.59E+17 | 0.946 | 38.580 | 3.10E+17 | 0.249 | 40.272 |
| 3.06E+17 | 0.260 | 36.324 | 3.85E+17 | 0.080 | 37.686 | 1.71E+17 | 0.854 | 38.599 | 3.18E+17 | 0.229 | 40.254 |
| 3.88E+17 | 0.074 | 33.807 | 4.17E+17 | 0.024 | 35.052 | 2.99E+17 | 0.280 | 36.578 | 3.82E+17 | 0.086 | 37.952 |
| 4.04E+17 | 0.046 | 32.725 | 4.23E+17 | 0.015 | 33.950 | 3.30E+17 | 0.197 | 35.807 | 3.95E+17 | 0.062 | 37.131 |
| 3.81E+17 | 0.088 | 34.294 | 4.15E+17 | 0.028 | 35.548 | 1.43E+17 | 1.087 | 39.091 | 2.99E+17 | 0.278 | 40.838 |
| 1.45E+17 | 1.075 | 39.001 | 3.00E+17 | 0.275 | 40.743 | 2.36E+17 | 0.496 | 37.695 | 3.54E+17 | 0.144 | 39.186 |
| 2.26E+17 | 0.542 | 37.804 | 3.48E+17 | 0.155 | 39.317 | 1.57E+17 | 0.965 | 38.818 | 3.09E+17 | 0.252 | 40.518 |
| 1.20E+17 | 1.355 | 39.386 | 2.82E+17 | 0.331 | 41.228 | 9.91E+16 | 1.670 | 39.950 | 2.65E+17 | 0.387 | 41.892 |
| 1.08E+17 | 1.526 | 39.486 | 2.72E+17 | 0.362 | 41.384 | 1.13E+17 | 1.453 | 39.426 | 2.76E+17 | 0.349 | 41.301 |
| 1.20E+17 | 1.356 | 39.210 | 2.82E+17 | 0.331 | 41.052 | 1.55E+17 | 0.979 | 38.780 | 3.07E+17 | 0.255 | 40.486 |
| 2.35E+17 | 0.500 | 37.798 | 3.53E+17 | 0.145 | 39.291 | 1.79E+17 | 0.801 | 38.728 | 3.22E+17 | 0.217 | 40.360 |
| 9.85E+16 | 1.681 | 39.657 | 2.64E+17 | 0.389 | 41.603 | 8.65E+16 | 1.924 | 39.303 | 2.53E+17 | 0.430 | 41.319 |
| 2.10E+17 | 0.618 | 37.982 | 3.40E+17 | 0.174 | 39.532 | 4.93E+16 | 3.252 | 40.904 | 2.10E+17 | 0.620 | 43.228 |
| 1.33E+17 | 1.199 | 39.059 | 2.92E+17 | 0.300 | 40.847 | 5.68E+16 | 2.870 | 40.244 | 2.20E+17 | 0.570 | 42.489 |
| 2.64E+17 | 0.391 | 37.310 | 3.67E+17 | 0.116 | 38.746 | 1.33E+17 | 1.197 | 39.175 | 2.92E+17 | 0.300 | 40.963 |
| 1.78E+17 | 0.810 | 38.563 | 3.21E+17 | 0.219 | 40.199 | 1.02E+17 | 1.628 | 40.134 | 2.67E+17 | 0.380 | 42.063 |
| 2.60E+17 | 0.404 | 37.313 | 3.65E+17 | 0.120 | 38.756 | 8.65E+16 | 1.924 | 40.383 | 2.53E+17 | 0.430 | 42.399 |
| 2.26E+17 | 0.539 | 37.811 | 3.49E+17 | 0.155 | 39.324 | 1.64E+17 | 0.907 | 39.066 | 3.13E+17 | 0.240 | 40.742 |
| 9.26E+16 | 1.793 | 39.862 | 2.59E+17 | 0.408 | 41.840 | 8.38E+16 | 1.986 | 40.017 | 2.50E+17 | 0.440 | 42.050 |
| 2.07E+17 | 0.633 | 38.513 | 3.38E+17 | 0.178 | 40.071 | 6.97E+16 | 2.375 | 40.231 | 2.35E+17 | 0.500 | 42.363 |
| 1.58E+17 | 0.956 | 39.049 | 3.09E+17 | 0.251 | 40.746 | 2.05E+16 | 6.645 | 39.992 | 1.56E+17 | 0.970 | 42.821 |
| 1.57E+17 | 0.961 | 38.843 | 3.09E+17 | 0.252 | 40.542 | 7.43E+16 | 2.235 | 40.254 | 2.40E+17 | 0.479 | 42.352 |
| 9.77E+16 | 1.695 | 39.559 | 2.63E+17 | 0.392 | 41.509 | 2.85E+16 | 5.128 | 40.907 | 1.75E+17 | 0.830 | 43.542 |
| 2.51E+17 | 0.438 | 37.403 | 3.61E+17 | 0.129 | 38.864 | 9.04E+16 | 1.839 | 40.432 | 2.57E+17 | 0.416 | 42.424 |
| 9.53E+16 | 1.742 | 39.765 | 2.61E+17 | 0.400 | 41.729 | 5.50E+16 | 2.952 | 39.801 | 2.18E+17 | 0.581 | 42.063 |
| 1.29E+17 | 1.245 | 39.214 | 2.89E+17 | 0.309 | 41.018 | 8.12E+16 | 2.049 | 39.783 | 2.48E+17 | 0.450 | 41.832 |
| 2.13E+17 | 0.604 | 37.931 | 3.41E+17 | 0.171 | 39.475 | 5.54E+16 | 2.937 | 40.927 | 2.18E+17 | 0.579 | 43.186 |
| 1.43E+17 | 1.092 | 39.019 | 2.99E+17 | 0.279 | 40.768 | 1.24E+17 | 1.300 | 39.421 | 2.85E+17 | 0.320 | 41.245 |
| 1.29E+17 | 1.241 | 39.105 | 2.89E+17 | 0.309 | 40.908 | 4.46E+16 | 3.550 | 40.594 | 2.03E+17 | 0.657 | 42.975 |
| 2.01E+17 | 0.667 | 38.251 | 3.35E+17 | 0.186 | 39.824 | 8.65E+16 | 1.924 | 39.751 | 2.53E+17 | 0.430 | 41.766 |
| 3.24E+17 | 0.213 | 36.038 | 3.93E+17 | 0.066 | 37.372 | 7.59E+16 | 2.190 | 39.883 | 2.42E+17 | 0.472 | 41.969 |
| 1.25E+17 | 1.281 | 39.578 | 2.86E+17 | 0.316 | 41.395 | 1.04E+17 | 1.594 | 41.266 | 2.68E+17 | 0.374 | 43.185 |
| 1.54E+17 | 0.988 | 38.885 | 3.07E+17 | 0.257 | 40.594 | 2.05E+17 | 0.643 | 38.627 | 3.37E+17 | 0.180 | 40.190 |
| 1.33E+17 | 1.191 | 39.244 | 2.92E+17 | 0.299 | 41.029 | 6.02E+16 | 2.724 | 42.130 | 2.24E+17 | 0.550 | 44.343 |
| 1.47E+17 | 1.050 | 38.936 | 3.02E+17 | 0.270 | 40.669 | 5.34E+16 | 3.035 | 41.872 | 2.15E+17 | 0.592 | 44.152 |
| 1.43E+17 | 1.094 | 38.751 | 2.99E+17 | 0.279 | 40.501 | 2.12E+17 | 0.610 | 37.756 | 3.41E+17 | 0.172 | 39.302 |
| 1.36E+17 | 1.159 | 39.145 | 2.94E+17 | 0.292 | 40.919 | 6.46E+16 | 2.554 | 39.781 | 2.29E+17 | 0.526 | 41.956 |
| 1.99E+17 | 0.680 | 38.213 | 3.34E+17 | 0.189 | 39.792 | 3.37E+16 | 4.480 | 41.933 | 1.85E+17 | 0.763 | 44.472 |
| 1.50E+17 | 1.028 | 38.774 | 3.04E+17 | 0.266 | 40.498 | 5.52E+16 | 2.944 | 41.044 | 2.18E+17 | 0.580 | 43.305 |
| 2.55E+17 | 0.421 | 37.268 | 3.63E+17 | 0.124 | 38.720 | 8.65E+16 | 1.924 | 39.786 | 2.53E+17 | 0.430 | 41.802 |
| 2.03E+17 | 0.654 | 38.217 | 3.36E+17 | 0.183 | 39.784 | 8.12E+16 | 2.049 | 40.222 | 2.48E+17 | 0.450 | 42.271 |
| 1.27E+17 | 1.263 | 39.204 | 2.87E+17 | 0.313 | 41.014 | 4.47E+16 | 3.541 | 40.765 | 2.03E+17 | 0.656 | 43.145 |
| 1.28E+17 | 1.246 | 39.388 | 2.89E+17 | 0.310 | 41.193 | 7.08E+16 | 2.341 | 39.995 | 2.37E+17 | 0.495 | 42.119 |
| 2.95E+17 | 0.292 | 36.448 | 3.81E+17 | 0.089 | 37.828 | 7.19E+16 | 2.308 | 39.671 | 2.38E+17 | 0.490 | 41.787 |
| 9.44E+16 | 1.758 | 39.943 | 2.60E+17 | 0.402 | 41.911 | 5.68E+16 | 2.870 | 40.426 | 2.20E+17 | 0.570 | 42.671 |
| 1.90E+17 | 0.731 | 38.250 | 3.29E+17 | 0.201 | 39.852 | 9.89E+16 | 1.674 | 40.264 | 2.64E+17 | 0.388 | 42.208 |
| 1.21E+17 | 1.334 | 39.137 | 2.83E+17 | 0.326 | 40.972 | 8.12E+16 | 2.049 | 40.353 | 2.48E+17 | 0.450 | 42.402 |
| 1.82E+17 | 0.779 | 38.357 | 3.24E+17 | 0.212 | 39.980 | 7.41E+16 | 2.242 | 40.060 | 2.40E+17 | 0.480 | 42.160 |
| 2.06E+17 | 0.642 | 38.159 | 3.38E+17 | 0.180 | 39.721 | 5.00E+16 | 3.212 | 40.226 | 2.11E+17 | 0.615 | 42.543 |
| 2.90E+17 | 0.306 | 36.892 | 3.78E+17 | 0.093 | 38.281 | 9.51E+16 | 1.744 | 40.349 | 2.61E+17 | 0.400 | 42.313 |
| 1.32E+17 | 1.209 | 39.241 | 2.91E+17 | 0.302 | 41.033 | 4.48E+16 | 3.533 | 39.937 | 2.03E+17 | 0.655 | 42.315 |
| 2.72E+17 | 0.363 | 37.230 | 3.70E+17 | 0.109 | 38.651 | 7.02E+16 | 2.362 | 40.855 | 2.36E+17 | 0.498 | 42.984 |
| 2.99E+17 | 0.280 | 36.620 | 3.82E+17 | 0.086 | 37.994 | 7.76E+16 | 2.144 | 39.751 | 2.44E+17 | 0.465 | 41.825 |
| 1.89E+17 | 0.739 | 38.362 | 3.28E+17 | 0.203 | 39.968 | 8.05E+16 | 2.068 | 40.767 | 2.47E+17 | 0.453 | 42.821 |
| 1.93E+17 | 0.714 | 38.330 | 3.30E+17 | 0.197 | 39.924 | 8.79E+16 | 1.894 | 39.195 | 2.54E+17 | 0.425 | 41.202 |
| 2.53E+17 | 0.430 | 37.458 | 3.62E+17 | 0.127 | 38.915 | 6.69E+16 | 2.470 | 40.631 | 2.32E+17 | 0.514 | 42.786 |
| 1.80E+17 | 0.792 | 38.573 | 3.23E+17 | 0.215 | 40.202 | 8.84E+16 | 1.881 | 39.562 | 2.55E+17 | 0.423 | 41.566 |
| 1.24E+17 | 1.302 | 39.278 | 2.85E+17 | 0.320 | 41.103 | 2.66E+16 | 5.424 | 41.416 | 1.71E+17 | 0.859 | 44.093 |
| 1.78E+17 | 0.808 | 38.585 | 3.22E+17 | 0.218 | 40.221 | 2.21E+16 | 6.256 | 40.524 | 1.61E+17 | 0.936 | 43.307 |
| 1.98E+17 | 0.684 | 38.359 | 3.33E+17 | 0.190 | 39.940 | 6.42E+16 | 2.568 | 40.276 | 2.29E+17 | 0.528 | 42.454 |
| 2.90E+17 | 0.306 | 36.757 | 3.78E+17 | 0.093 | 38.146 | 2.01E+16 | 6.739 | 40.660 | 1.55E+17 | 0.978 | 43.500 |
| 1.02E+17 | 1.626 | 39.504 | 2.67E+17 | 0.380 | 41.433 | 2.49E+16 | 5.698 | 41.471 | 1.67E+17 | 0.885 | 44.184 |
| 1.83E+17 | 0.776 | 38.362 | 3.25E+17 | 0.211 | 39.984 | 2.96E+16 | 4.979 | 41.460 | 1.77E+17 | 0.815 | 44.074 |
| 1.06E+17 | 1.552 | 39.954 | 2.71E+17 | 0.367 | 41.860 | 3.99E+16 | 3.896 | 41.326 | 1.95E+17 | 0.698 | 43.769 |
| 8.90E+16 | 1.869 | 40.133 | 2.55E+17 | 0.421 | 42.134 | 5.71E+16 | 2.855 | 40.467 | 2.20E+17 | 0.568 | 42.709 |
| 1.54E+17 | 0.991 | 38.934 | 3.06E+17 | 0.258 | 40.645 | 3.86E+16 | 4.009 | 41.104 | 1.93E+17 | 0.711 | 43.567 |
| 2.03E+17 | 0.658 | 38.142 | 3.36E+17 | 0.184 | 39.711 | 1.16E+17 | 1.404 | 39.225 | 2.79E+17 | 0.340 | 41.084 |
| 1.82E+17 | 0.783 | 38.463 | 3.24E+17 | 0.213 | 40.088 | 9.62E+16 | 1.723 | 39.530 | 2.62E+17 | 0.397 | 41.488 |
| 1.08E+17 | 1.516 | 39.263 | 2.73E+17 | 0.360 | 41.157 | 2.98E+16 | 4.949 | 41.044 | 1.77E+17 | 0.812 | 43.654 |
| 9.70E+16 | 1.709 | 39.827 | 2.63E+17 | 0.394 | 41.781 | 3.08E+16 | 4.822 | 40.787 | 1.79E+17 | 0.799 | 43.379 |
| 2.65E+17 | 0.385 | 37.236 | 3.67E+17 | 0.115 | 38.669 | 2.51E+16 | 5.666 | 40.666 | 1.67E+17 | 0.882 | 43.375 |

| | | | | | |
|---|---|---|---|---|---|
| 2.83E+16 | 5.159 | 41.048 | 1.74E+17 | 0.833 | 43.688 |
| 2.56E+16 | 5.581 | 40.598 | 1.69E+17 | 0.874 | 43.295 |
| 3.29E+16 | 4.564 | 40.958 | 1.83E+17 | 0.772 | 43.510 |
| 2.07E+17 | 0.635 | 37.897 | 3.38E+17 | 0.178 | 39.455 |
| 1.53E+17 | 1.000 | 39.109 | 3.06E+17 | 0.260 | 40.823 |
| 2.01E+17 | 0.668 | 38.138 | 3.35E+17 | 0.186 | 39.712 |
| 1.48E+17 | 1.044 | 39.046 | 3.03E+17 | 0.269 | 40.776 |
| 1.80E+17 | 0.794 | 38.749 | 3.23E+17 | 0.215 | 40.378 |
| 6.14E+16 | 2.674 | 40.277 | 2.26E+17 | 0.543 | 42.479 |
| 3.49E+16 | 4.359 | 40.723 | 1.87E+17 | 0.750 | 43.243 |
| 4.67E+16 | 3.411 | 40.409 | 2.06E+17 | 0.640 | 42.764 |
| 8.65E+16 | 1.924 | 40.169 | 2.53E+17 | 0.430 | 42.184 |
| 4.67E+16 | 3.411 | 40.810 | 2.06E+17 | 0.640 | 43.164 |
| 7.04E+16 | 2.355 | 40.197 | 2.36E+17 | 0.497 | 42.325 |
| 8.38E+16 | 1.986 | 39.978 | 2.50E+17 | 0.440 | 42.011 |
| 1.10E+17 | 1.488 | 39.457 | 2.74E+17 | 0.355 | 41.342 |
| 3.23E+16 | 4.640 | 41.037 | 1.82E+17 | 0.780 | 43.601 |
| 6.20E+16 | 2.652 | 40.227 | 2.26E+17 | 0.540 | 42.425 |
| 2.65E+16 | 5.435 | 41.244 | 1.70E+17 | 0.860 | 43.922 |
| 7.69E+16 | 2.164 | 40.469 | 2.43E+17 | 0.468 | 42.549 |
| 2.78E+16 | 5.230 | 41.223 | 1.73E+17 | 0.840 | 43.873 |
| 2.09E+16 | 6.530 | 40.798 | 1.58E+17 | 0.960 | 43.613 |
| 2.91E+16 | 5.046 | 41.191 | 1.76E+17 | 0.822 | 43.815 |
| 2.24E+16 | 6.189 | 40.778 | 1.61E+17 | 0.930 | 43.552 |
| 8.10E+16 | 2.055 | 39.746 | 2.47E+17 | 0.451 | 41.797 |
| 5.07E+16 | 3.173 | 40.597 | 2.12E+17 | 0.610 | 42.905 |
| 2.85E+16 | 5.128 | 41.415 | 1.75E+17 | 0.830 | 44.051 |
| 3.90E+16 | 3.974 | 40.826 | 1.94E+17 | 0.707 | 43.282 |
| 9.07E+16 | 1.833 | 39.887 | 2.57E+17 | 0.415 | 41.877 |
| 5.90E+16 | 2.775 | 40.332 | 2.23E+17 | 0.557 | 42.557 |
| 3.14E+16 | 4.745 | 40.997 | 1.80E+17 | 0.791 | 43.577 |
| 4.02E+16 | 3.870 | 40.773 | 1.96E+17 | 0.695 | 43.212 |
| 4.76E+16 | 3.355 | 40.706 | 2.07E+17 | 0.633 | 43.050 |
| 1.59E+17 | 0.947 | 38.918 | 3.10E+17 | 0.249 | 40.611 |
| 6.34E+16 | 2.596 | 40.381 | 2.28E+17 | 0.532 | 42.566 |
| 1.19E+17 | 1.358 | 39.235 | 2.82E+17 | 0.331 | 41.078 |
| 1.14E+17 | 1.439 | 39.495 | 2.77E+17 | 0.346 | 41.365 |
| 2.09E+16 | 6.541 | 41.447 | 1.57E+17 | 0.961 | 44.264 |
| 5.03E+16 | 3.197 | 40.681 | 2.11E+17 | 0.613 | 42.994 |
| 1.16E+17 | 1.407 | 39.468 | 2.79E+17 | 0.340 | 41.328 |
| 1.99E+16 | 6.798 | 41.311 | 1.55E+17 | 0.983 | 44.157 |
| 3.87E+16 | 4.000 | 40.561 | 1.93E+17 | 0.710 | 43.022 |
| 3.67E+16 | 4.178 | 40.786 | 1.90E+17 | 0.730 | 43.277 |
| 7.64E+16 | 2.177 | 40.052 | 2.43E+17 | 0.470 | 42.135 |
| 4.93E+16 | 3.252 | 40.685 | 2.10E+17 | 0.620 | 43.009 |
| 6.55E+16 | 2.519 | 40.013 | 2.31E+17 | 0.521 | 42.180 |
| 1.05E+17 | 1.566 | 39.724 | 2.70E+17 | 0.369 | 41.634 |
| 5.66E+16 | 2.877 | 40.152 | 2.20E+17 | 0.571 | 42.399 |
| 5.16E+16 | 3.127 | 40.228 | 2.13E+17 | 0.604 | 42.527 |
| 2.26E+16 | 6.157 | 41.170 | 1.62E+17 | 0.927 | 43.940 |
| 1.40E+17 | 1.122 | 39.094 | 2.97E+17 | 0.285 | 40.854 |
| 1.37E+17 | 1.153 | 39.068 | 2.95E+17 | 0.291 | 40.839 |
| 6.05E+16 | 2.709 | 40.086 | 2.25E+17 | 0.548 | 42.296 |
| 2.60E+16 | 5.518 | 40.804 | 1.69E+17 | 0.868 | 43.493 |
| 7.06E+16 | 2.348 | 40.089 | 2.36E+17 | 0.496 | 42.215 |
| 2.99E+16 | 4.940 | 40.796 | 1.77E+17 | 0.811 | 43.405 |
| 3.43E+16 | 4.415 | 41.285 | 1.86E+17 | 0.756 | 43.814 |
| 2.94E+16 | 4.999 | 41.034 | 1.77E+17 | 0.817 | 43.652 |
| 3.47E+16 | 4.378 | 40.779 | 1.86E+17 | 0.752 | 43.302 |
| 5.99E+16 | 2.735 | 40.087 | 2.24E+17 | 0.552 | 42.303 |
| 1.09E+17 | 1.503 | 39.544 | 2.73E+17 | 0.358 | 41.435 |
| 1.87E+16 | 7.121 | 41.130 | 1.52E+17 | 1.010 | 44.012 |
| 3.57E+16 | 4.277 | 41.210 | 1.88E+17 | 0.741 | 43.717 |
| 8.65E+16 | 1.924 | 39.795 | 2.53E+17 | 0.430 | 41.811 |
| 6.46E+16 | 2.554 | 40.225 | 2.29E+17 | 0.526 | 42.400 |
| 5.34E+16 | 3.035 | 40.297 | 2.15E+17 | 0.592 | 42.577 |
| 2.38E+16 | 5.913 | 40.898 | 1.64E+17 | 0.905 | 43.639 |
| 2.15E+16 | 6.403 | 40.644 | 1.59E+17 | 0.949 | 43.445 |
| 7.86E+16 | 2.117 | 40.004 | 2.45E+17 | 0.461 | 42.072 |
| 1.05E+17 | 1.576 | 39.755 | 2.69E+17 | 0.371 | 41.669 |
| 3.07E+16 | 4.832 | 41.121 | 1.79E+17 | 0.800 | 43.714 |
| 4.20E+16 | 3.733 | 41.043 | 1.99E+17 | 0.679 | 43.457 |
| 5.49E+16 | 2.957 | 40.408 | 2.17E+17 | 0.582 | 42.672 |
| 6.02E+16 | 2.724 | 40.062 | 2.24E+17 | 0.550 | 42.276 |
| 2.99E+16 | 4.930 | 40.763 | 1.78E+17 | 0.810 | 43.370 |
| 2.14E+16 | 6.415 | 41.175 | 1.59E+17 | 0.950 | 43.977 |
| 1.17E+17 | 1.392 | 39.440 | 2.80E+17 | 0.337 | 41.294 |
| 2.35E+16 | 5.968 | 41.558 | 1.64E+17 | 0.910 | 44.306 |
| 1.51E+17 | 1.015 | 38.915 | 3.05E+17 | 0.263 | 40.635 |
| 4.63E+16 | 3.435 | 40.652 | 2.05E+17 | 0.643 | 43.011 |
| 4.07E+16 | 3.835 | 40.656 | 1.97E+17 | 0.691 | 43.089 |
| 1.09E+17 | 1.499 | 39.536 | 2.74E+17 | 0.357 | 41.426 |
| 3.76E+16 | 4.097 | 40.698 | 1.92E+17 | 0.721 | 43.176 |
| 5.50E+16 | 2.952 | 40.479 | 2.18E+17 | 0.581 | 42.742 |
| 4.84E+16 | 3.305 | 40.424 | 2.08E+17 | 0.627 | 42.759 |
| 2.94E+16 | 5.009 | 40.774 | 1.76E+17 | 0.818 | 43.393 |
| 7.81E+16 | 2.129 | 39.972 | 2.44E+17 | 0.463 | 42.042 |
| 8.15E+16 | 2.042 | 39.975 | 2.48E+17 | 0.449 | 42.023 |
| 4.10E+16 | 3.810 | 40.626 | 1.97E+17 | 0.688 | 43.054 |
| 2.59E+16 | 5.539 | 41.539 | 1.69E+17 | 0.870 | 44.231 |
| 6.89E+16 | 2.404 | 40.199 | 2.34E+17 | 0.504 | 42.338 |
| 5.35E+16 | 3.027 | 40.930 | 2.15E+17 | 0.591 | 43.209 |
| 8.76E+16 | 1.900 | 39.753 | 2.54E+17 | 0.426 | 41.762 |
| 1.20E+17 | 1.347 | 39.501 | 2.82E+17 | 0.329 | 41.340 |
| 5.47E+16 | 2.967 | 40.135 | 2.17E+17 | 0.583 | 42.401 |
| 6.59E+16 | 2.505 | 40.908 | 2.31E+17 | 0.519 | 43.072 |
| 9.48E+16 | 1.750 | 39.729 | 2.61E+17 | 0.401 | 41.695 |
| 1.87E+17 | 0.750 | 38.293 | 3.27E+17 | 0.205 | 39.904 |
| 1.16E+17 | 1.406 | 39.354 | 2.79E+17 | 0.340 | 41.213 |
| 8.49E+16 | 1.961 | 39.854 | 2.51E+17 | 0.436 | 41.880 |
| 1.07E+17 | 1.532 | 39.652 | 2.72E+17 | 0.363 | 41.552 |
| 8.49E+16 | 1.961 | 39.900 | 2.51E+17 | 0.436 | 41.926 |
| 1.29E+17 | 1.243 | 39.370 | 2.89E+17 | 0.309 | 41.174 |
| 1.15E+17 | 1.417 | 39.522 | 2.78E+17 | 0.342 | 41.385 |
| 2.23E+17 | 0.557 | 37.895 | 3.47E+17 | 0.159 | 39.416 |
| 1.19E+17 | 1.363 | 39.410 | 2.81E+17 | 0.332 | 41.255 |
| 7.66E+16 | 2.170 | 40.273 | 2.43E+17 | 0.469 | 42.354 |
| 1.65E+17 | 0.902 | 38.584 | 3.14E+17 | 0.239 | 40.259 |
| 1.11E+17 | 1.471 | 39.543 | 2.75E+17 | 0.352 | 41.424 |
| 5.04E+16 | 3.189 | 40.502 | 2.11E+17 | 0.612 | 42.814 |
| 4.78E+16 | 3.339 | 40.036 | 2.08E+17 | 0.631 | 42.377 |
| 4.60E+16 | 3.451 | 40.454 | 2.05E+17 | 0.645 | 42.817 |
| 8.68E+16 | 1.918 | 39.876 | 2.53E+17 | 0.429 | 41.890 |
| 7.04E+16 | 2.355 | 39.948 | 2.36E+17 | 0.497 | 42.076 |
| 6.21E+16 | 2.645 | 40.031 | 2.26E+17 | 0.539 | 42.227 |
| 5.83E+16 | 2.804 | 40.641 | 2.22E+17 | 0.561 | 42.872 |
| 9.21E+16 | 1.803 | 39.366 | 2.58E+17 | 0.410 | 41.347 |
| 9.16E+16 | 1.815 | 39.439 | 2.58E+17 | 0.412 | 41.424 |
| 5.23E+16 | 3.088 | 40.452 | 2.14E+17 | 0.599 | 42.743 |
| 4.95E+16 | 3.244 | 40.734 | 2.10E+17 | 0.619 | 43.056 |
| 8.87E+16 | 1.875 | 39.726 | 2.55E+17 | 0.422 | 41.728 |
| 6.20E+16 | 2.652 | 40.314 | 2.26E+17 | 0.540 | 42.511 |
| 9.48E+16 | 1.750 | 40.589 | 2.61E+17 | 0.401 | 42.555 |
| 1.78E+17 | 0.807 | 38.440 | 3.22E+17 | 0.218 | 40.075 |
| 4.76E+16 | 3.355 | 39.858 | 2.07E+17 | 0.633 | 42.202 |
| 1.01E+17 | 1.645 | 39.707 | 2.66E+17 | 0.383 | 41.642 |
| 1.32E+17 | 1.207 | 39.520 | 2.91E+17 | 0.302 | 41.311 |
| 1.16E+17 | 1.406 | 39.221 | 2.79E+17 | 0.340 | 41.080 |
| 6.77E+16 | 2.443 | 39.732 | 2.33E+17 | 0.510 | 41.881 |
| 8.90E+16 | 1.869 | 40.136 | 2.55E+17 | 0.421 | 42.136 |
| 9.54E+16 | 1.738 | 39.526 | 2.61E+17 | 0.399 | 41.488 |
| 7.12E+16 | 2.328 | 40.026 | 2.37E+17 | 0.493 | 42.147 |
| 4.11E+16 | 3.801 | 40.570 | 1.97E+17 | 0.687 | 42.996 |
| 4.11E+16 | 3.801 | 40.408 | 1.97E+17 | 0.687 | 42.835 |
| 7.08E+16 | 2.341 | 40.124 | 2.37E+17 | 0.495 | 42.249 |
| 5.17E+16 | 3.119 | 40.349 | 2.13E+17 | 0.603 | 42.646 |
| 8.90E+16 | 1.869 | 40.179 | 2.55E+17 | 0.421 | 42.180 |
| 1.13E+17 | 1.449 | 39.716 | 2.76E+17 | 0.348 | 41.590 |
| 1.81E+17 | 0.785 | 38.483 | 3.24E+17 | 0.213 | 40.109 |
| 1.14E+17 | 1.428 | 39.307 | 2.78E+17 | 0.344 | 41.173 |
| 1.47E+17 | 1.053 | 38.798 | 3.02E+17 | 0.271 | 40.532 |
| 5.78E+16 | 2.826 | 40.137 | 2.21E+17 | 0.564 | 42.373 |
| 1.45E+17 | 1.068 | 38.985 | 3.01E+17 | 0.274 | 40.725 |
| 2.05E+17 | 0.647 | 38.119 | 3.37E+17 | 0.181 | 39.683 |
| 5.49E+16 | 2.959 | 40.899 | 2.17E+17 | 0.582 | 43.164 |
| 4.19E+16 | 3.742 | 40.488 | 1.99E+17 | 0.680 | 42.904 |
| 9.48E+16 | 1.750 | 39.970 | 2.61E+17 | 0.401 | 41.936 |
| 9.04E+16 | 1.839 | 39.566 | 2.57E+17 | 0.416 | 41.558 |
| 1.39E+17 | 1.127 | 39.449 | 2.97E+17 | 0.286 | 41.211 |
| 5.81E+16 | 2.811 | 40.818 | 2.22E+17 | 0.562 | 43.051 |
| 1.50E+17 | 1.029 | 38.666 | 3.04E+17 | 0.266 | 40.391 |
| 1.27E+17 | 1.269 | 39.419 | 2.87E+17 | 0.314 | 41.232 |
| 5.50E+16 | 2.952 | 41.406 | 2.18E+17 | 0.581 | 43.669 |
| 7.80E+16 | 2.131 | 39.877 | 2.44E+17 | 0.463 | 41.948 |
| 1.15E+17 | 1.411 | 39.130 | 2.78E+17 | 0.341 | 40.992 |
| 4.78E+16 | 3.339 | 40.541 | 2.08E+17 | 0.631 | 42.882 |
| 6.53E+16 | 2.526 | 40.509 | 2.30E+17 | 0.522 | 42.677 |
| 1.06E+17 | 1.560 | 39.559 | 2.70E+17 | 0.368 | 41.468 |
| 1.29E+17 | 1.243 | 39.057 | 2.89E+17 | 0.309 | 40.861 |
| 6.42E+16 | 2.568 | 40.192 | 2.29E+17 | 0.528 | 42.370 |
| 1.79E+17 | 0.798 | 38.774 | 3.22E+17 | 0.216 | 40.405 |
| 1.40E+17 | 1.117 | 39.077 | 2.97E+17 | 0.284 | 40.835 |
| 6.81E+16 | 2.429 | 40.049 | 2.34E+17 | 0.508 | 42.194 |
| 3.22E+16 | 4.649 | 40.872 | 1.82E+17 | 0.781 | 43.437 |
| 5.03E+16 | 3.197 | 40.304 | 2.11E+17 | 0.613 | 42.617 |
| 1.43E+17 | 1.087 | 38.819 | 2.99E+17 | 0.278 | 40.566 |
| 7.48E+16 | 2.222 | 39.960 | 2.41E+17 | 0.477 | 42.054 |
| 2.14E+16 | 6.415 | 40.836 | 1.59E+17 | 0.950 | 43.638 |
| 1.68E+16 | 7.704 | 41.207 | 1.47E+17 | 1.057 | 44.151 |
| 2.95E+16 | 4.989 | 41.075 | 1.77E+17 | 0.816 | 43.691 |
| 8.00E+16 | 2.080 | 40.266 | 2.46E+17 | 0.455 | 42.324 |
| 1.83E+16 | 7.242 | 41.467 | 1.51E+17 | 1.020 | 44.363 |
| 1.41E+16 | 8.800 | 41.273 | 1.38E+17 | 1.140 | 44.323 |
| 2.69E+16 | 5.373 | 40.939 | 1.71E+17 | 0.854 | 43.608 |
| 8.90E+15 | 12.312 | 41.720 | 1.19E+17 | 1.370 | 45.050 |
| 2.02E+16 | 6.704 | 41.498 | 1.56E+17 | 0.975 | 44.334 |
| 2.05E+16 | 6.645 | 41.634 | 1.56E+17 | 0.970 | 44.463 |
| 3.58E+16 | 4.268 | 40.782 | 1.88E+17 | 0.740 | 43.288 |
| 8.57E+15 | 12.652 | 41.523 | 1.17E+17 | 1.390 | 44.876 |
| 7.88E+16 | 2.112 | 40.082 | 2.45E+17 | 0.460 | 42.148 |
| 1.83E+16 | 7.242 | 41.268 | 1.51E+17 | 1.020 | 44.164 |
| 1.47E+16 | 8.528 | 41.489 | 1.40E+17 | 1.120 | 44.514 |
| 1.17E+16 | 10.090 | 41.858 | 1.30E+17 | 1.230 | 45.021 |
| 1.27E+16 | 9.503 | 41.250 | 1.33E+17 | 1.190 | 44.363 |
| 2.79E+16 | 5.219 | 40.750 | 1.73E+17 | 0.839 | 43.398 |
| 1.87E+16 | 7.121 | 42.030 | 1.52E+17 | 1.010 | 44.912 |
| 6.55E+16 | 2.519 | 40.212 | 2.31E+17 | 0.521 | 42.378 |
| 7.52E+16 | 2.209 | 40.014 | 2.41E+17 | 0.475 | 42.105 |
| 2.14E+16 | 6.415 | 41.082 | 1.59E+17 | 0.950 | 43.884 |
| 1.02E+16 | 11.167 | 41.769 | 1.24E+17 | 1.300 | 45.016 |
| 1.01E+16 | 11.247 | 41.487 | 1.24E+17 | 1.305 | 44.740 |
| 1.79E+17 | 0.798 | 38.925 | 3.22E+17 | 0.216 | 40.556 |
| 3.62E+16 | 4.223 | 40.594 | 1.89E+17 | 0.735 | 43.092 |
| 1.41E+16 | 8.800 | 41.146 | 1.38E+17 | 1.140 | 44.197 |
| 1.01E+16 | 11.278 | 42.155 | 1.23E+17 | 1.307 | 45.411 |
| 1.09E+16 | 10.620 | 41.739 | 1.27E+17 | 1.265 | 44.944 |
| 4.30E+16 | 3.657 | 40.742 | 2.00E+17 | 0.670 | 43.143 |
| 4.67E+16 | 3.411 | 40.569 | 2.06E+17 | 0.640 | 42.924 |
| 9.43E+15 | 11.813 | 41.773 | 1.21E+17 | 1.340 | 45.068 |
| 2.78E+16 | 5.230 | 40.864 | 1.73E+17 | 0.840 | 43.514 |
| 2.22E+16 | 6.245 | 40.759 | 1.61E+17 | 0.935 | 43.540 |
| 2.13E+16 | 6.449 | 41.468 | 1.58E+17 | 0.953 | 44.274 |
| 1.46E+16 | 8.582 | 41.537 | 1.40E+17 | 1.124 | 44.568 |
| 5.98E+16 | 2.738 | 40.294 | 2.24E+17 | 0.552 | 42.511 |
| 4.29E+16 | 3.666 | 40.580 | 2.00E+17 | 0.671 | 42.982 |
| 6.75E+16 | 2.450 | 40.223 | 2.33E+17 | 0.511 | 42.374 |
| 1.79E+16 | 7.365 | 41.331 | 1.50E+17 | 1.030 | 44.240 |
| 1.27E+16 | 9.532 | 41.343 | 1.33E+17 | 1.192 | 44.459 |
| 1.56E+16 | 8.156 | 41.019 | 1.43E+17 | 1.092 | 44.008 |
| 2.03E+16 | 6.692 | 41.000 | 1.56E+17 | 0.974 | 43.834 |
| 1.50E+16 | 8.394 | 41.613 | 1.41E+17 | 1.110 | 44.625 |
| 9.25E+15 | 11.978 | 41.521 | 1.20E+17 | 1.350 | 44.827 |
| 2.71E+16 | 5.332 | 40.830 | 1.72E+17 | 0.850 | 43.494 |
| 1.15E+16 | 10.254 | 41.406 | 1.29E+17 | 1.241 | 44.582 |
| 8.20E+15 | 13.067 | 41.423 | 1.15E+17 | 1.414 | 44.804 |
| 1.28E+16 | 9.475 | 41.497 | 1.34E+17 | 1.188 | 44.608 |
| 1.84E+16 | 7.206 | 41.402 | 1.51E+17 | 1.017 | 44.294 |
| 9.90E+15 | 11.407 | 41.707 | 1.23E+17 | 1.315 | 44.971 |
| 2.91E+16 | 5.039 | 41.018 | 1.76E+17 | 0.821 | 43.641 |
| 1.21E+16 | 9.867 | 42.103 | 1.31E+17 | 1.215 | 45.247 |
| 4.89E+16 | 3.275 | 40.186 | 2.09E+17 | 0.623 | 42.515 |

$$k_b T_{BEC} = \left( \frac{n}{\varsigma\left(3/2\right)} \right)^{\frac{2}{3}} \frac{h^2}{2\pi m}$$

where

$$m = am_p \qquad \rho = \frac{m}{Vol}$$

$$n = \frac{m}{am_p Vol} = \frac{\rho}{am_p}$$

then

$$k_b T_{BEC} = \frac{h^2}{2\pi am_p} \left( \frac{\rho}{\varsigma\left(3/2\right)} \right)^{\frac{2}{3}} \left( \frac{1}{am_p} \right)^{\frac{2}{3}} = \frac{h^2}{2\pi} \left( \frac{\rho}{\varsigma\left(3/2\right)} \right)^{\frac{2}{3}} \left( \frac{1}{am_p} \right)^{\frac{5}{3}}$$

$$\rho = \varsigma\left(3/2\right) \left( \frac{2\pi k_b T_c \left(am_p\right)^{\frac{5}{3}}}{h^2} \right)^{\frac{2}{3}}$$

$$\rho = \varsigma\left(3/2\right) \left( \frac{2\pi k_b T_b \left(bm_p\right)^{\frac{5}{3}}}{h^2} \right)^{\frac{2}{3}}$$

$\rho = \rho$ therefore $T_c \left(am_p\right)^{\frac{5}{3}} = T_b \left(bm_p\right)^{\frac{5}{3}}$

and

$$a=\frac{\left(\frac{T_b\left(bm_p\right)^{\frac{5}{3}}}{T_c}\right)^{\frac{3}{5}}}{m_p}$$

Object Spherical Perimeter $P=\pi d = A_c/d$

Object Hydraulic Radius $d$ to calculate Universe Reynolds Number Object(seen) $A_c/P=\left(\left(6/\pi\right)Vol\right)^{\frac{1}{3}} = d$

Object frontal area ($A$) to thrust cm$^2$

Object Spherical Diameter (d) SQRT(4 x A/pi)

Object number ($a$) of $m_p$

Object number ($b$ = 5.996E+07) of $m_p$

Object number $(a+b)$ of $m_p$

Object Mass $(a+b)$ $m_p$

Object Area to Mass ratio $A_c/m = A_c/\left(\left(a+b\right)m_p\right)$

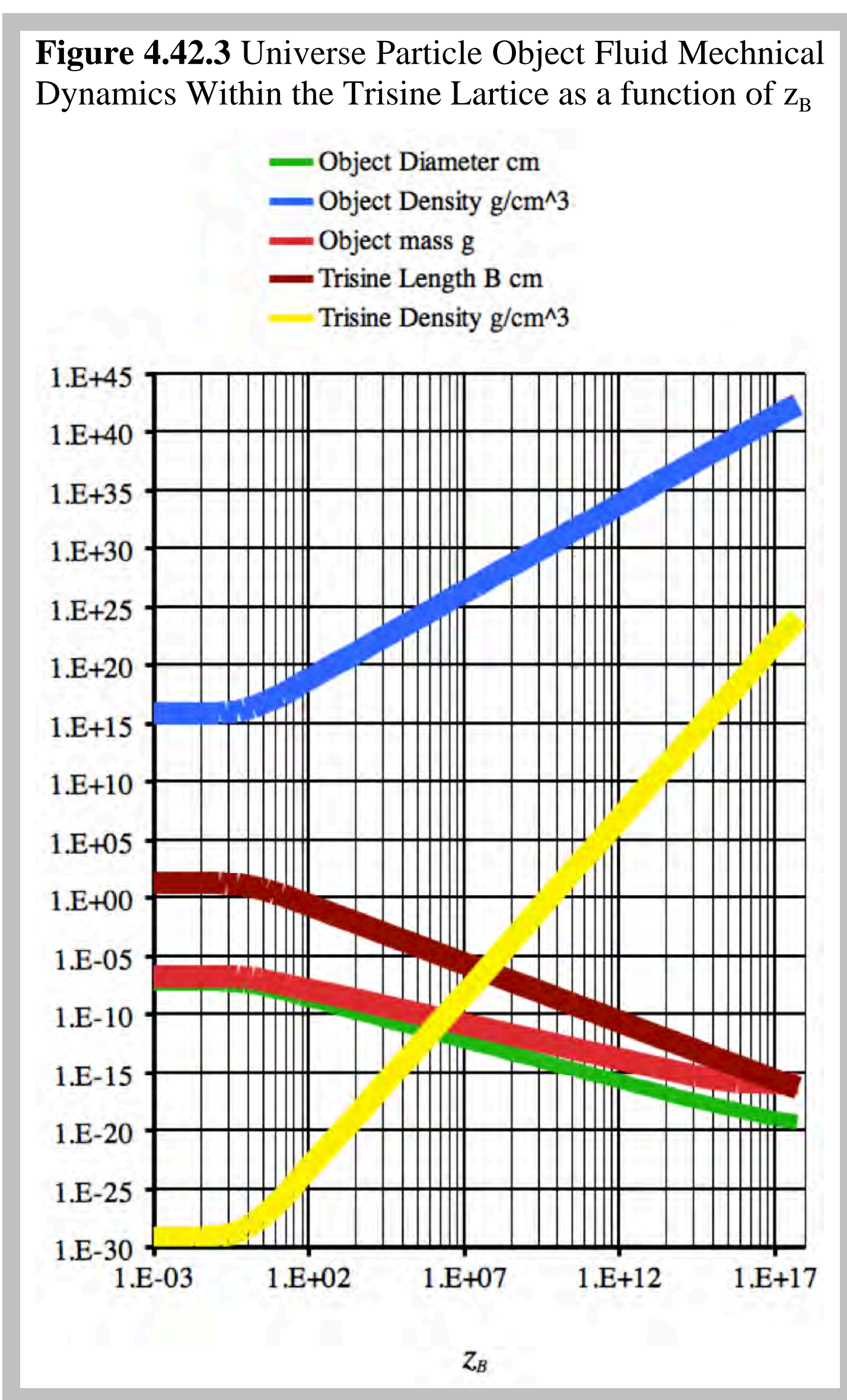

**Figure 4.42.3** Universe Particle Object Fluid Mechnical Dynamics Within the Trisine Lartice as a function of $z_B$

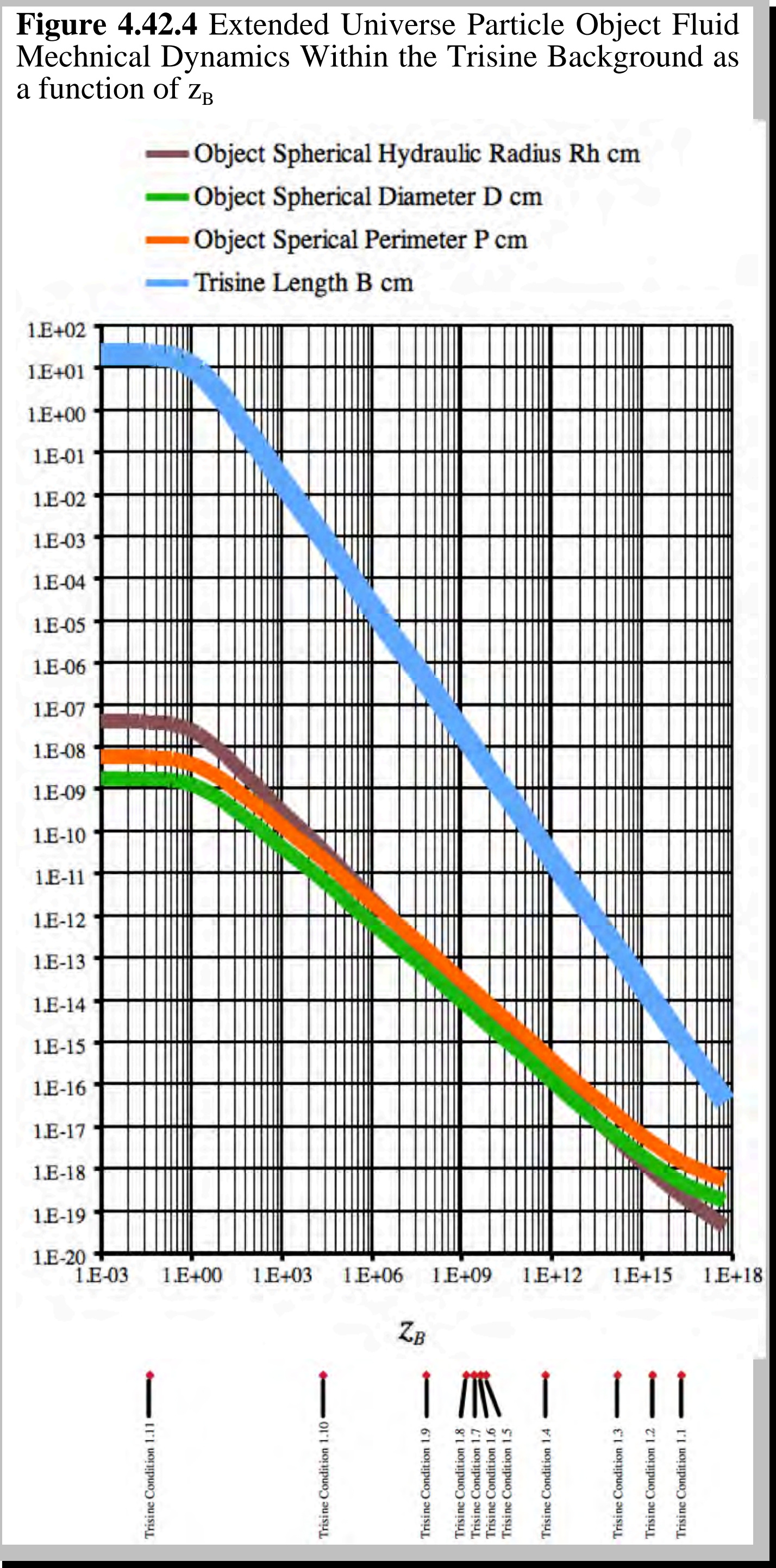

**Figure 4.42.4** Extended Universe Particle Object Fluid Mechnical Dynamics Within the Trisine Background as a function of $z_B$

The generalized fluid model (Appendix L) is used for this universe thrust model as illustrated in Figures 4.42.3 - 4.42. 9 The thrust model extends over the many orders of magnitude representing the universe at its beginning to present. The dimensional relationship between Trisine temperature $T_c$ (dark energy) and Cosmic Microwave Background Radiation (CMBR) temperature $T_b$ dictates a third phase (dark matter).

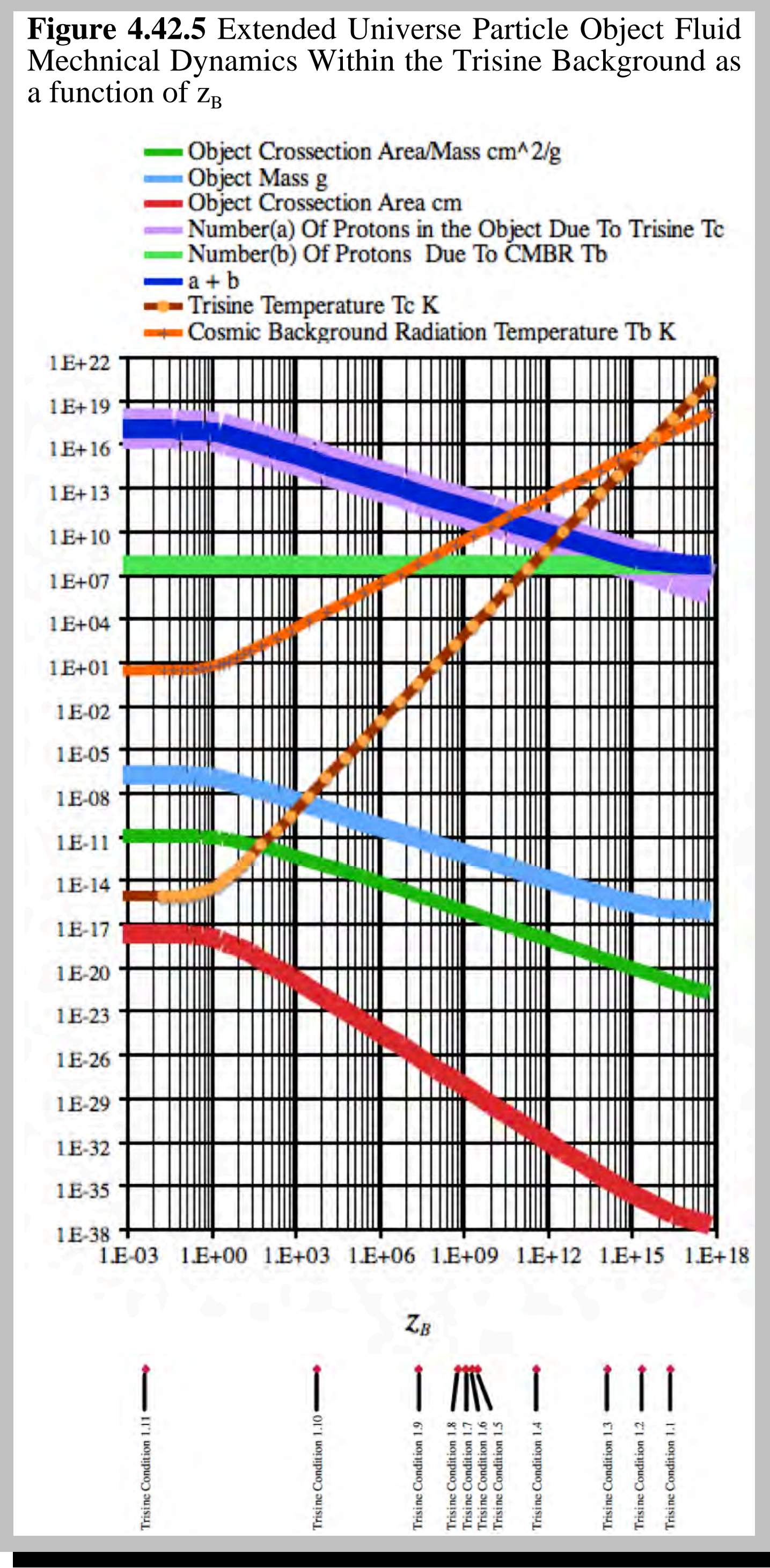

**Figure 4.42.5** Extended Universe Particle Object Fluid Mechnical Dynamics Within the Trisine Background as a function of $z_B$

The universe is shown to be 'thrust' into existence much like an explosion. In this context, dark matter is essentially another baryonic phase (like ice and water) with distinct density and size characteristics that are not observed by Baryonic Acoustic Oscillations (at $z_B$ ~ 3,000) which is oriented towards a continuous homogeneous gas phase.

The phase density Figures 4.42.3 (Blue line) has a slope correlation to Appendix K Figure K2. Magnitude difference can be explained in terms of particle coalesence dynamics as presented in Appendix L and in particular equation L2.14.

This thrust model assumes the unit proton mass as $m_p$. Other unit mass assumptions would not change the general developed relationship and may indeed provide further insights into early universe nucleosynthetic dynamics.

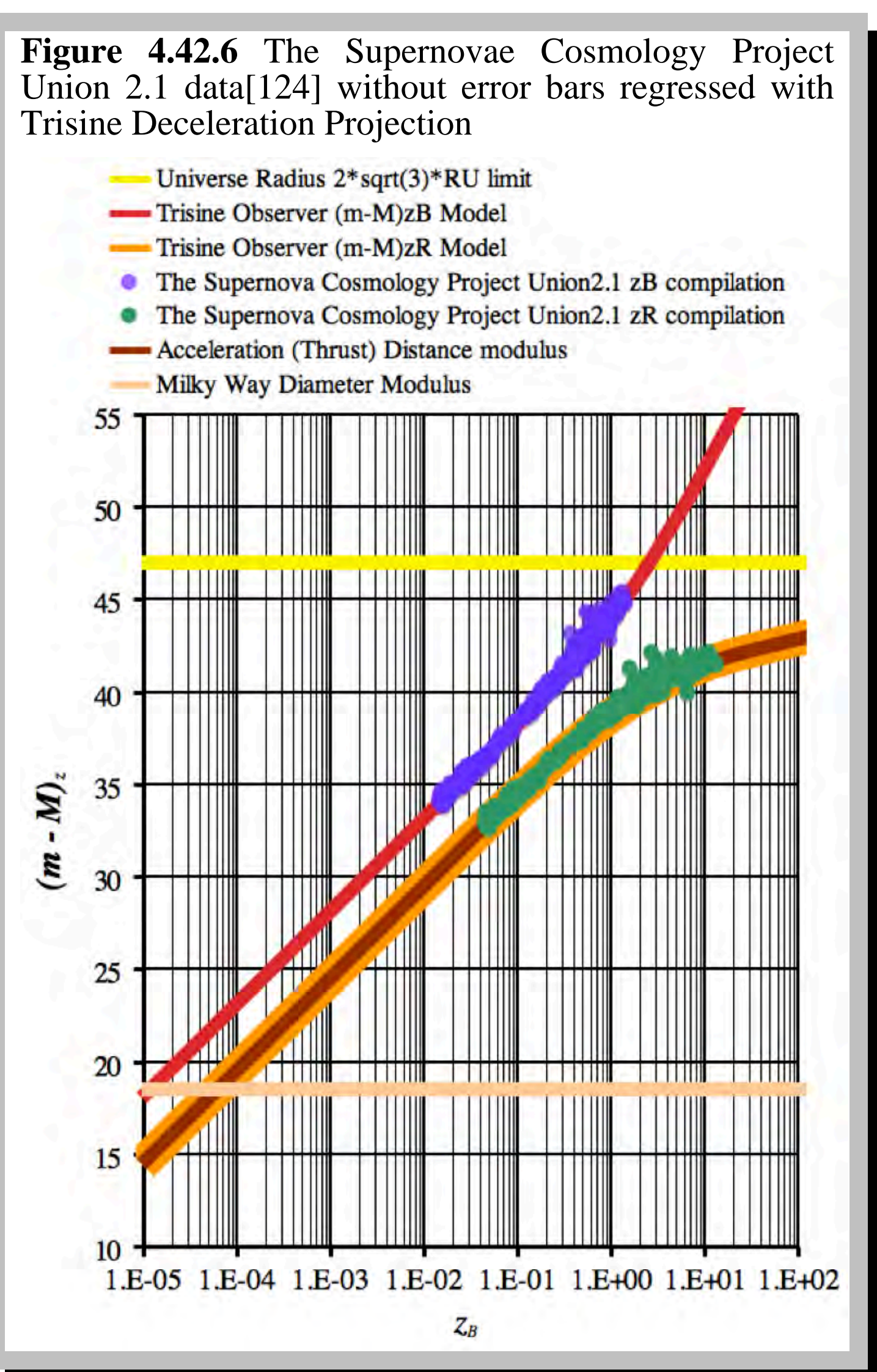

**Figure 4.42.6** The Supernovae Cosmology Project Union 2.1 data[124] without error bars regressed with Trisine Deceleration Projection

The Type 1A supernovae when plotted in the $z_B$ dimension (rather than $z_R$ ) further demonstrates the universe beginning expanding to the present at the speed of light $c$.

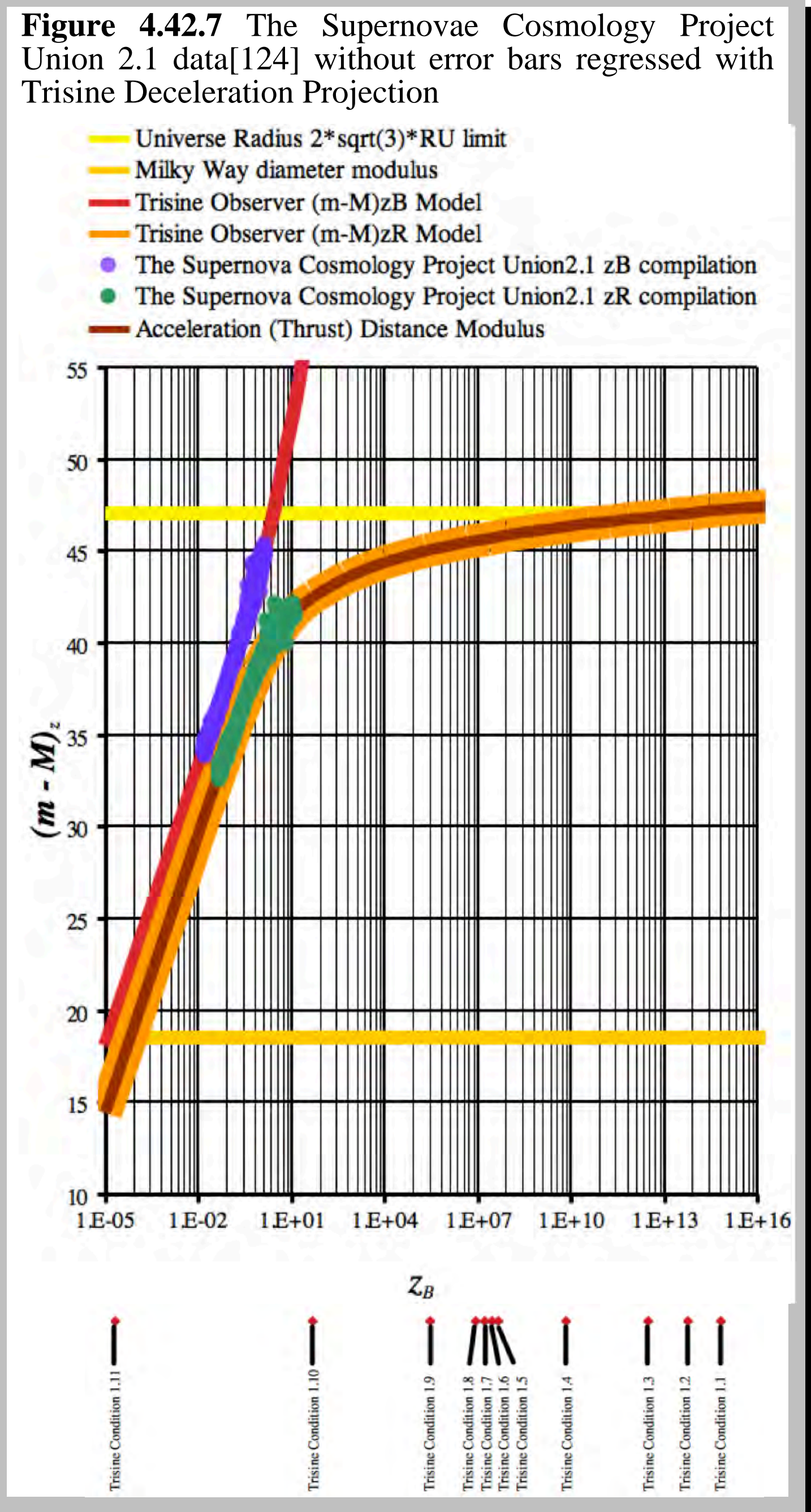

**Figure 4.42.7** The Supernovae Cosmology Project Union 2.1 data[124] without error bars regressed with Trisine Deceleration Projection

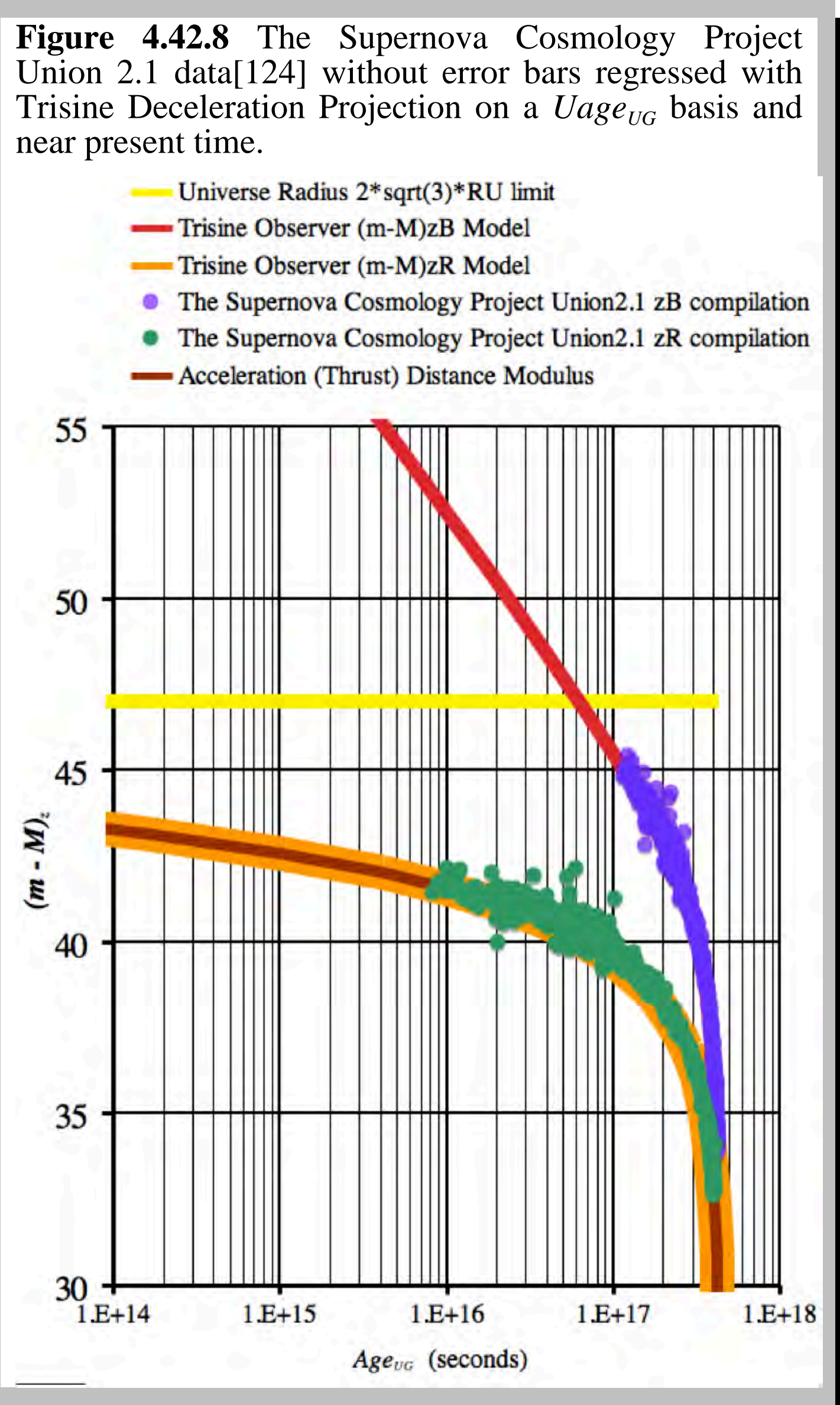

**Figure 4.42.8** The Supernova Cosmology Project Union 2.1 data[124] without error bars regressed with Trisine Deceleration Projection on a $Uage_{UG}$ basis and near present time.

The Type 1A Supernovae when plotted in the $Age_{UG}$ dimension further demonstrates the universe beginning expanding to the present at the speed of light *c*. This Type 1A supernovae day can be extrapolated or modeled back to the universe beginning in terms of this thrust model.

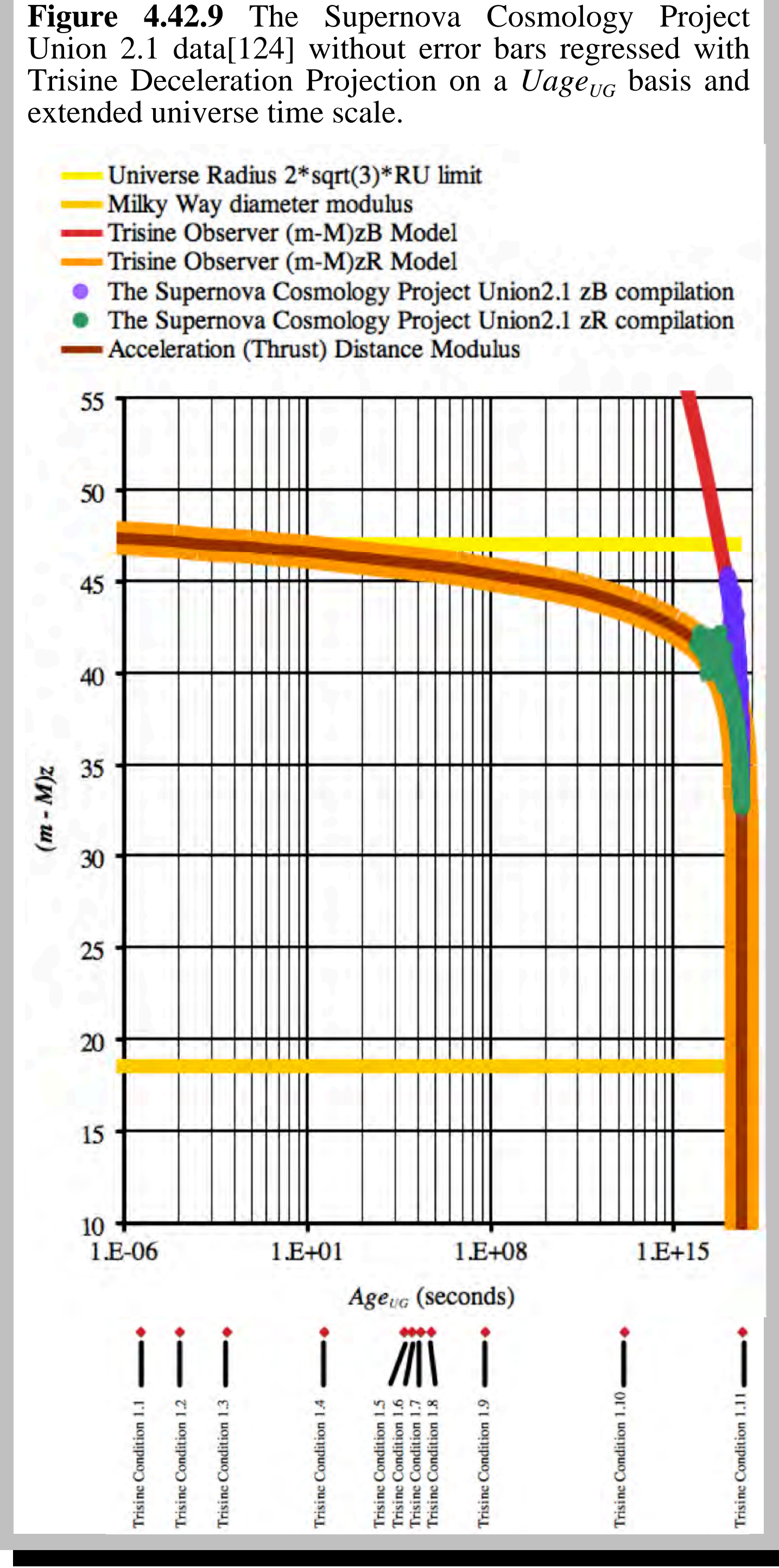

**Figure 4.42.9** The Supernova Cosmology Project Union 2.1 data[124] without error bars regressed with Trisine Deceleration Projection on a $Uage_{UG}$ basis and extended universe time scale.

4.43. It is noted that time picosecond(ps) fluctuations [132] Figure 4.c are of the same order as de Broglie matter wave cellular times as a function of $T_c$ numerically and graphically expressed in Table and Figure 2.4.

It is further noted that these De Broglie waves are an expression of the Charge conjugation, Parity change, Time reversal, CPT theorem in the space filling mode.

Anticipated space filling CPT times at $T_c$ 's, your charge density wave times and their differences are as follows:

Table 4.43.1

| $T_c$ | CPT time (ps) | CDW (ps) | CDW - CPT (ps) |
|---|---|---|---|
| 5.14 | 2.01 | 4.67 | -2.66 |
| 15.05 | 1.79 | 1.59 | 0.20 |
| 24.95 | 1.48 | 0.976 | 0.50 |
| 35.05 | 1.33 | 0.685 | 0.64 |
| 44.95 | 1.04 | 0.534 | 0.50 |
| 55.05 | 0.76 | 0.436 | 0.33 |
| 75.81 | 0.62 | 0.317 | 0.30 |
| 100.00 | 0.39 | 0.24 | 0.15 |

Numbers are graphically presented Figure 4.43.1.

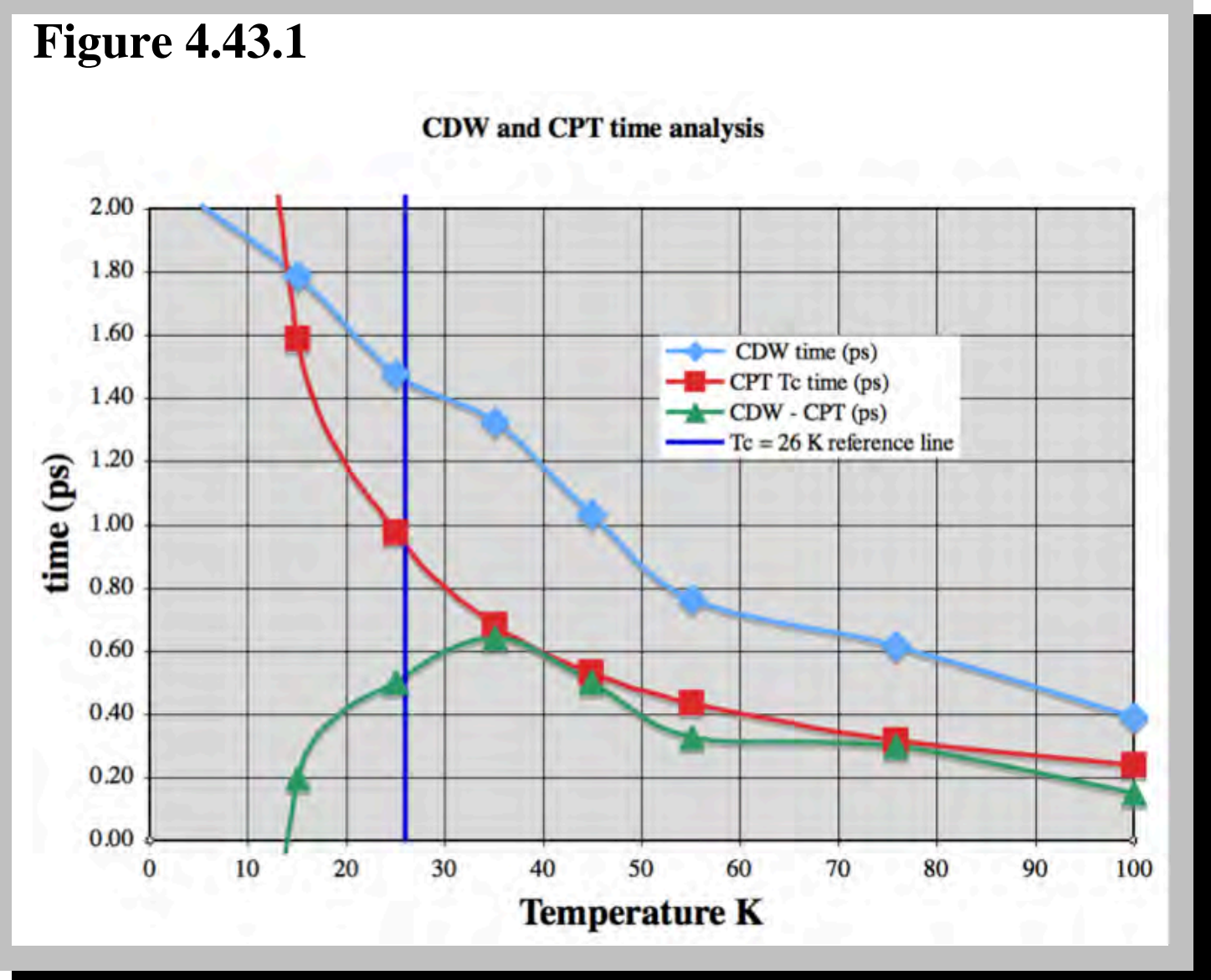

**Figure 4.43.1**

Space filling CPT theory, being a virtual particle concept may not always be expressed in CDW measurements, but in this case there is a peak in CDW-CPT in the upper 30K's.

(numerically similar to your intersection in Figure 4d) that may be a close indicator of your actual $T_c$ of 26 K.

The space filling CPT is the basis of the superconductor phenomenon that may or may not be expressed as CDW.

This CPT scaling is all consistent with Homes' Law.

This concept extends into cosmological dimensions.

In additional trisine model confirmation, it is noted that the spacing between $NbSe_2$ stripes ([133] Figure 2 A) is ~1 nm

This conforms to my stripe scaling Figure 2.3.1 with spacing

$$\frac{3B}{\sqrt{3}}$$

According to my scaling formula (similar to Homes law)

4 x 1 nm defined as $B$ as in equation 2.1.13a

$$E = k_b T_c = \frac{m_T v_{dx}^2}{2} = \hbar \frac{\pi}{time_{\pm}} = \frac{\hbar^2 K_B^2}{2 m_T}$$

would indicated a $T_c$ of 7.2 K for space filling CPT.

($m_T$ is a universal particle with mass = 110 x electron mass me)

Charge density waves CDW 4 x $NbSe_2$stripes may express the difference between CDW with underlying a more fundamental CPT.

In this case[133], charge density waves CDW are expressions of electron movement directly related to universal particle $m_t$ movement by the relationship $m_T = 110\ m_e$ with charge density wave CDW spacing/superconducting wave spacing = sqrt($m_T/m_e$), a relationship that is observed in this $NbSe_2$ superconductor but not all superconductors.

4.44. Robbert Dijkgraaf[134] provides a classical relationship based on string theory black hole dynamics.

$$F \Delta x = T_c \Delta S$$

This relationship is consistent with the following, noting that a de Broglie velocity $v_{dx}$ replaces the speed of light c.

$$F = \frac{m_T v_{dx}}{time_{\pm}}$$

The entropy relationship is the same as Dijkgraaf[134] with a removal of $\pi$ such that equation reads 2 versus $2\pi$

$$\Delta S = \frac{2 k_b v_{dx}}{\hbar} m_T \Delta x$$

The entropy has a Boltzmann constant value over all trisine dimensions recognizing the Heisenberg uncertainty relationship:

$$\frac{2 v_{dx}}{\hbar} m_T \Delta x = 1$$

This is all consistent with vacuum entropy equations 2.11.9a and 2.11.10 indicating singular entropy per cavity.

4.45. Newton's second law in conjunction with Hooke's law are the basis for presenting oscillatory motion[61]

$$Newtonian\ Force = ma = mv\frac{dv}{dx} = -kx$$

and let $$k = -m\frac{v}{x}\frac{dv}{dx} = -mH_U^2$$

From the third and fourth *Newtonian Force* terms we have:

$$\int mv\,dv + \int kx\,dx = 0$$

that integrates as follows with Total Energy($E$) as the integration constant:

$$\frac{1}{2}mv^2 + \frac{1}{2}kx^2 = E_{kinetic} + E_{potential} = \text{Total Energy}(E)$$

The maximum potential energy $E_{potential}$ at maximum $x$ and velocity $v$ at 0 represents the harmonic amplitude $A$ such that :

$$A^2 = \frac{2E}{k} = \frac{2mc^2}{k} = \frac{2mc^2}{-mH_U^2} = \frac{2c^2}{-H_U^2}$$

This equation compares nearly equivalent (2 vs 3 perhaps a *chain/cavity* relationship) eqn 2.11.14a.

$$R^2 = \frac{3c^2}{H_U^2}$$

This harmonic derivation may explain in part low angular anomalies[135] observed by the Planck spacecraft at first light as the universe bouncing against its own horizon with negative sign representing a 180 degree phase change.

4.46. A question logically develops from this line of thinking and that is "What is the source of the energy or mass density?". After the energy is imparted to the spacecraft (or other objects) passing through space, does the energy regenerate itself locally or essentially reduce the energy of universal field? And then there is the most vexing question based on the universal scaled superconductivity resonant hypothesis developed in this report and that is whether this phenomenon could be engineered at an appropriate scale for beneficial use.

One concept for realizing the capture of this essentially astrophysical concept, is to engineer an earth based congruent wave crystal reactor. Ideally, such a wave crystal reactor would be made in the technically manageable wave length (ultraviolet to microwave). Perhaps such coherent electromagnetic waves would interact at the $\hbar\omega/2$ energy level in the trisine CPT lattice pattern at the Schwinger limit with resultant virtual particle production. Conceptually a device as indicated in Figures 4.46.1, 4.46.2 and 4.46.3, may be appropriate where the green area represents reflecting, generating and polarizing surfaces of coherent electromagnetic radiation creating standing waves which interact in the intersecting zone defined by the trisine characteristic angle of (90 – 22.8) degrees and 120 degrees of each other. It is anticipated that Laguerre-Gaussian optical vortices and related superconducting modes will be spatially created in the vacuum in accordance with photon spatial positioning and total momentum (spin and angular) control, a possibility suggested by Hawton[85]. Schwinger pair production is anticipated at low electric fields because of geometrically defined permeability and permittivity associated with interfering laser beams (COVE project - Appendix N).

An alternative approach is made with the National Ignition Facility(NIF) at Livermore, CA. The question immediately arises whether the NIF could be run in a more continuous less powerful mode with selected laser standing wave beams of the trisine characteristic angle of (90 – 22.8) degrees and 120 degrees of each other with no material target and study the resultant interference pattern or lattice for virtual particles.

Above and beyond the purely scientific study of such a lattice, it would be of primary importance to investigate the critical properties of such a lattice as the basis of providing useful energy generating mechanisms for the good of society. The major conclusion of this report is that an interdisciplinary team effort is required to continue this effort involving physicists, chemists and engineers. Each has a major role in achieving the many goals that are offered if more complete understanding of the superconducting resonant phenomenon is attained in order to bring to practical reality and bring a more complete understanding of our universe.

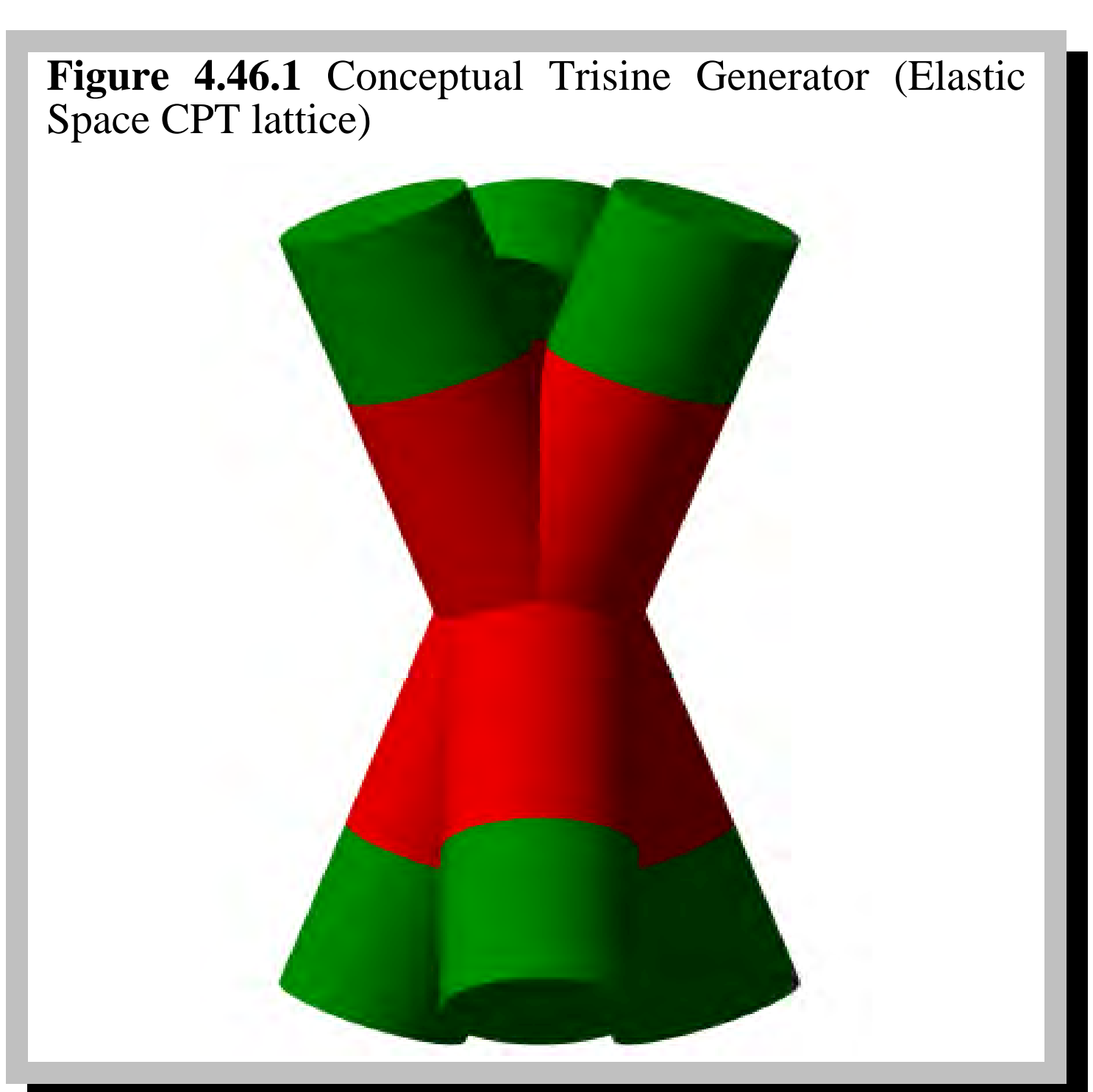

**Figure 4.46.1** Conceptual Trisine Generator (Elastic Space CPT lattice)

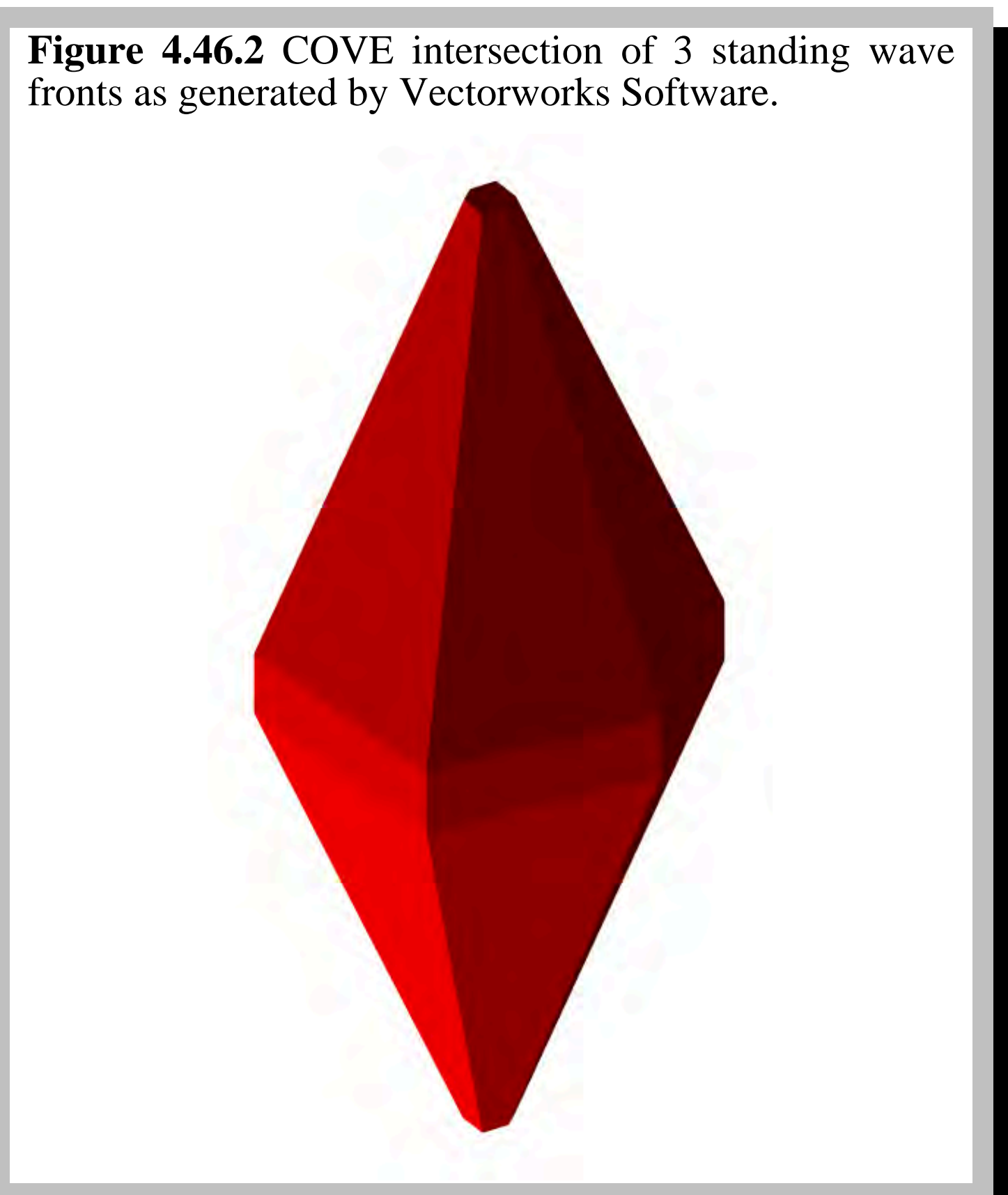

**Figure 4.46.2** COVE intersection of 3 standing wave fronts as generated by Vectorworks Software.

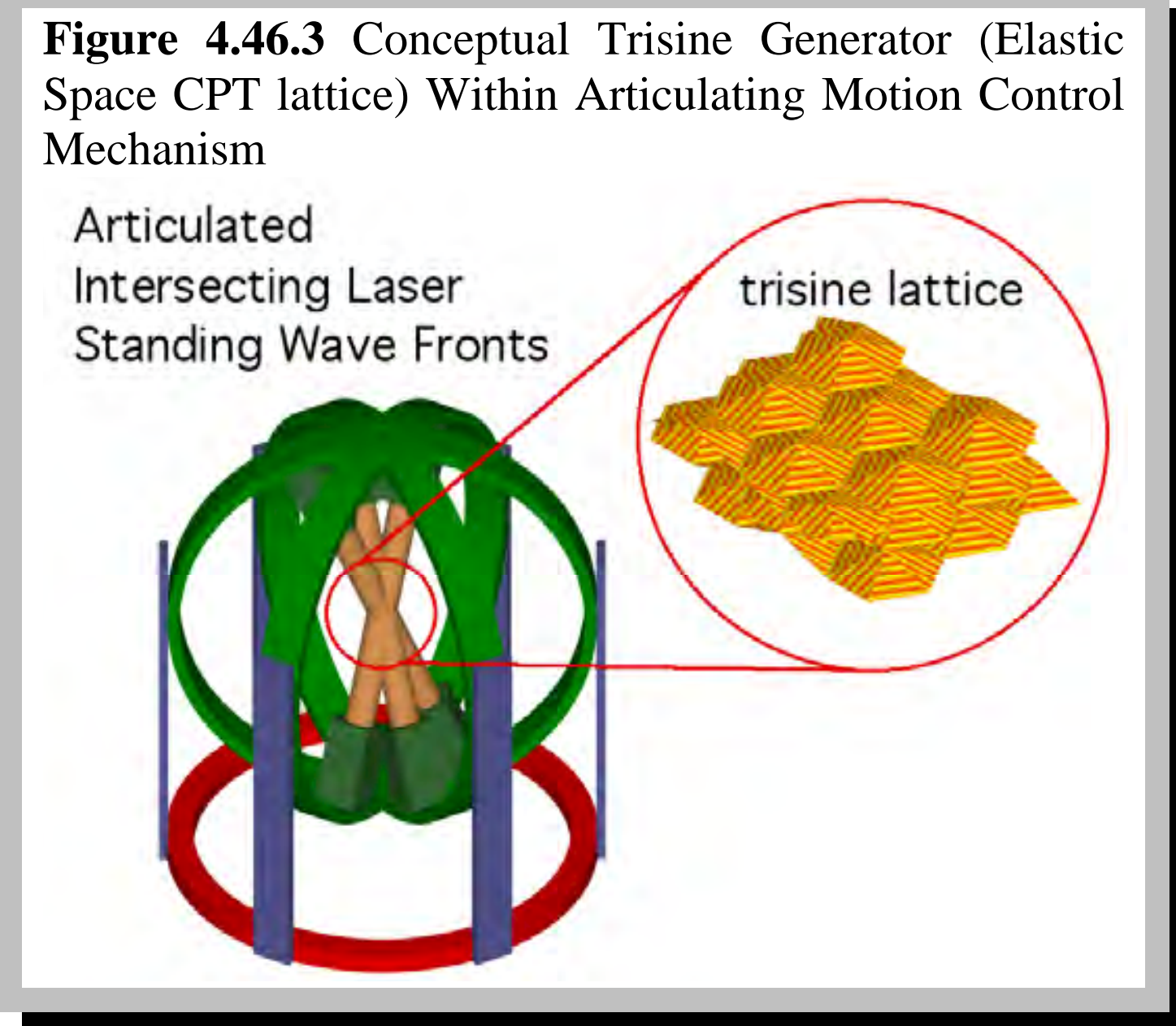

**Figure 4.46.3** Conceptual Trisine Generator (Elastic Space CPT lattice) Within Articulating Motion Control Mechanism

## 5. Variable And Constant Definitions

Fundamental Physical Constants are from National Institute of Standards (NIST).

'Present' values represent Universe at 13.7 billion years.

The $z$ factors $(1+z_R)^a$ or $(1+z_B)^b$ are presented for extrapolation of present values into the observational past consistent with tabular values in this paper. The two references $(1+z_R)^a$ or $(1+z_B)^b$ are presented representing both the universe Hubble radius of curvature $R_U$ and trisine cellular $B$ perspectives where $a/b = 1/3$.

| name | symbol | value | | units |
|---|---|---|---|---|
| Attractive Energy | $V\ or\ k_bT_c/\mathbb{C}$ | 5.58E-32 | (present) | $g\ cm^2 \sec^{-2} (erg)$ |
| Attractive Energy $z$ factors | $(1+z_R)^{2/3}$ or $(1+z_B)^2$ | | | *unitless* |
| Attractive Energy | $V_r\ or\ k_bT_{cr}/\mathbb{C}$ | 9.08E+21 | (present) | $g\ cm^2 \sec^{-2} (erg)$ |
| Attractive Energy $z$ factors | $(1+z_R)^{-2/3}$ or $(1+z_B)^{-2}$ | | | *unitless* |
| Black Body emission | $2E_v$ or $\sigma_S T_b^4$ | 3.14E-03 | (present) | $erg\ cm^{-2} \sec^{-1}$ |
| Black Body emission $z$ factors | $(1+z_R)^{4/3}$ or $(1+z_B)^4$ | | | *unitless* |
| Black Body emission reflective | $2E_{vr}$ or $\sigma_S T_{br}^4$ | 9.27E+84 | (present) | $erg\ cm^{-2} \sec^{-1}$ |
| Black Body emission reflective $z$ factors | $(1+z_R)^{-4/3}$ or $(1+z_B)^{-4}$ | | | *unitless* |
| Black Body energy/volume | $\rho_{BBenergy}$ | 4.20E-13 | (present) | $erg\ cm^{-3}$ |
| Black Body energy/volume | $\sigma_S T_b^4 (4/c)$ | 4.20E-13 | (present) | $erg\ cm^{-3}$ |
| Black Body energy/volume $z$ factors | $(1+z_R)^{4/3}$ or $(1+z_B)^4$ | | | *unitless* |
| Black Body energy/volume reflective | $\rho_{BBenergyr}$ | 1.24E+75 | (present) | $erg\ cm^{-3}$ |
| Black Body energy/volume reflective | $\sigma_S T_{br}^4 (4/c)$ | 1.24E+75 | (present) | $erg\ cm^{-3}$ |
| Black Body energy/volume reflective $z$ factors | $(1+z_R)^{-4/3}$ or $(1+z_B)^{-4}$ | | | *unitless* |
| Black Body mass/volume | $\rho_{BBmass}$ or $\sigma_S T_b^4 (4/c^3)$ | 4.67E-34 | (present) | $g\ cm^{-3}$ |
| Black Body mass/volume $z$ factors | $(1+z_R)^{4/3}$ or $(1+z_B)^4$ | | | *unitless* |
| Black Body mass/volume reflective | $\rho_{BBmassr}$ or $\sigma_S T_{br}^4 (4/c^3)$ | 1.38E54 | (present) | $g\ cm^{-3}$ |
| Black Body mass/volume reflective $z$ factors | $(1+z_R)^{-4/3}$ or $(1+z_B)^{-4}$ | | | *unitless* |
| Black Body momentum | $\rho_{BBmomentum}$ | 6.24E-26 | (present) | $g\ cm\ sec^{-1}$ |
| Black Body momentum | $4.96536456 k_b T_b / c$ | 6.24E-26 | (present) | $g\ cm\ sec^{-1}$ |
| Black Body momentum $z$ factors | $(1+z_R)^{-1/3}$ or $(1+z_B)^{-1}$ | | | *unitless* |
| Black Body momentum reflective | $\rho_{BBmomentumr}$ | 4.60E-04 | (present) | $g\ cm\ sec^{-1}$ |
| Black Body momentum reflective | $4.96536456 k_b T_{br} / c$ | 4.60E-04 | (present) | $g\ cm\ sec^{-1}$ |
| Black Body momentum reflective $z$ factors | $(1+z_R)^{1/3}$ or $(1+z_B)^1$ | | | *unitless* |
| Black Body photons/volume | $\rho_{BBphoton}$ | 4.12E+02 | (present) | $\#\ cm^{-3}$ |
| Black Body photons/volume | $8\pi\left(k_bT_b/(hc)\right)^3\cdot 2.404113806$ | 4.12E+02 | (present) | $\#\ cm^{-3}$ |
| Black Body photons/volume $z$ factors | $\left(1+z_R\right)^{3/3}$ or $\left(1+z_B\right)^3$ | | | *unitless* |
| Black Body photons/volume reflective | $\rho_{BBphotonr}$ | 1.65E+68 | (present) | $\#\ cm^{-3}$ |
| Black Body photons/volume reflective | $8\pi\left(k_bT_{br}/(hc)\right)^3\cdot 2.404113806$ | 1.65E+68 | (present) | $\#\ cm^{-3}$ |
| Black Body photons/volume reflective $z$ factors | $\left(1+z_R\right)^{-3/3}$ or $\left(1+z_B\right)^{-3}$ | | | *unitless* |
| Black Body photons/baryon | $\rho_{BBphoton}/\left(Avogadro\cdot\rho_U\right)$ | 1.07E+08 | (present) | $photon\ g^{-1}$ |
| Black Body photons/baryon $z$ factors | $\left(1+z_R\right)^0$ or $\left(1+z_B\right)^0$ | | | *unitless* |
| Black Body photons/baryon reflective | $\rho_{BBphotonr}/\left(Avogadro\cdot\rho_{Ur}\right)$ | 6.58E-07 | (present) | $photon\ baryon^{-1}$ |
| Black Body photons/baryon reflective $z$ factors | $\left(1+z_R\right)^0$ or $\left(1+z_B\right)^0$ | | | *unitless* |
| Black Body temperature $\left(T_b\right)$ | $\left(1/2\right)\left(1/\mathbb{C}\right)g_s^3 m_t v_\varepsilon^2\left(1/k_b\right)$ | 2.729 | (present) | kelvin |
| Black Body temperature $z$ factors | $\left(1+z_R\right)^{1/3}$ or $\left(1+z_B\right)^1$ | | | *unitless* |
| Black Body temperature $\left(T_{br}\right)$ reflective | $\left(1/2\right)\left(1/\mathbb{C}\right)g_s^3 m_t v_{\varepsilon r}^2\left(1/k_b\right)$ | 2.01E+22 | (present) | kelvin |
| Black Body temperature reflective $z$ factors | $\left(1+z_R\right)^{-1/3}$ or $\left(1+z_B\right)^{-1}$ | | | *unitless* |
| Black Body universe frequency | $\nu_{BBmax}$ | 1.60E+11 | (present) | $sec^{-1}$ |
| Black Body universe frequency | $2.821439372\,k_bT_b/h$ | 1.60E+11 | (present) | $sec^{-1}$ |
| Black Body universe frequency $z$ factors | $\left(1+z_R\right)^{1/3}$ or $\left(1+z_B\right)^1$ | | | *unitless* |
| Black Body universe frequency reflective | $\nu_{BB\max r}$ | 1.18E+33 | (present) | $sec^{-1}$ |
| Black Body universe frequency reflective | $2.821439372\,k_bT_{br}/h$ | 1.18E+33 | (present) | $sec^{-1}$ |
| Black Body universe frequency reflective $z$ factors | $\left(1+z_R\right)^{-1/3}$ or $\left(1+z_B\right)^{-1}$ | | | *unitless* |
| Black Body universe mass | $M_\nu$ or $V_U\sigma_sT_b^4\left(4/c^3\right)$ | 2.21E+52 | (present) | $g$ |
| Black Body universe mass $z$ factors | $\left(1+z_R\right)^{-5/3}$ or $\left(1+z_B\right)^{-5}$ | | | *unitless* |
| Black Body universe mass reflective | $M_{\nu r}$ or $V_{Ur}\sigma_sT_{br}^4\left(4/c^3\right)$ | 2.34E-100 | (present) | $g$ |
| Black Body universe mass reflective $z$ factors | $\left(1+z_R\right)^{5/3}$ or $\left(1+z_B\right)^5$ | | | *unitless* |
| Black Body universe viscosity | $\mu_{BB}$ | 1.49E-24 | (present) | $g\ cm^{-1}sec^{-1}$ |
| Black Body universe viscosity | $\sigma_sT_b^4\left(4/c^2\right)\lambda_{BB\max}$ | 1.49E-24 | (present) | $g\ cm^{-1}sec^{-1}$ |
| Black Body universe viscosity $z$ factors | $\left(1+z_R\right)^1$ or $\left(1+z_B\right)^3$ | | | *unitless* |
| Black Body universe viscosity reflective | $\mu_{BBr}$ | 5.95E+41 | (present) | $g\ cm^{-1}sec^{-1}$ |
| Black Body universe viscosity reflective | $\sigma_sT_{br}^4\left(4/c^2\right)\lambda_{BB\max r}$ | 5.95E+41 | (present) | $g\ cm^{-1}sec^{-1}$ |
| Black Body universe viscosity reflective $z$ factors | $\left(1+z_R\right)^{-1}$ or $\left(1+z_B\right)^{-3}$ | | | *unitless* |
| Black Body universe wave length max | $\lambda_{BBmax}$ | 1.06E-01 | (present) | $cm$ |
| Black Body universe wave length max | $hc/\left(4.96536456\cdot k_bT_b\right)$ | 1.06E-01 | (present) | $cm$ |
| Black Body universe wave length $z$ factors | $(1+z_R)^{-1/3}$ or $(1+z_B)^{-1}$ | | | *unitless* |
| Black Body universe wave length max reflective | $\lambda_{BB\max r}$ | 1.44E-23 | (present) | *cm* |
| Black Body universe wave length max reflective | $hc/(4.96536456 \cdot k_b T_{br})$ | 1.44E-23 | (present) | *cm* |
| Black Body universe wave length reflective $z$ factors | $(1+z_R)^{1/3}$ or $(1+z_B)^{1}$ | | | *unitless* |
| Boltzmann constant | $k_b$ | 1.380658120E-16 | | $g\ cm^2 \sec^{-2}\ Kelvin^{-1}$ |
| Bohr magneton | $\mu_B$ or $e\hbar/(2m_e c)$ | 9.27400949E-21 | | *erg/gauss (gauss* $cm^3$*)* |
| | | | | $g^{1/2}\ cm^{5/2} sec^{-1}$ |
| Bohr radius | $a_o$ or $\hbar^2/(m_e e^2)$ | 5.2917724924E-09 | | *cm* |
| | $\hbar c R_U cavity/(G m_t M_U)$ | | | |
| | $\hbar c R_{Ur} cavity_r/(G m_t M_{Ur})$ | | | |
| | $sherd^{1/2}\ \hbar c R_U cavity/(G_U m_t M_U)$ | | | |
| | $sherd_r^{1/2}\ \hbar c R_{Ur} cavity/(G_{Ur} m_t M_{Ur})$ | | | |
| Bose Einstein Condensate Survival Time | $BEC_{ST}$ *for 1 Mole* $(cm^3)$ *of 1* $g\ cm^{-3}$ *hydrogen BEC sec* | | | |
| Capacitance $(C_x)$ | *approach* $\varepsilon_x/(4\pi B)$ | 2.57E+13 | (present) | *cm* |
| Capacitance $(C_y)$ | $g_s \sqrt{3} approach\ \varepsilon_y/(8\pi B)$ | 2.57E+13 | (present) | *cm* |
| Capacitance $(C_z)$ | $\sqrt{3}\pi\ approach\ \varepsilon_z/(3\pi B)$ | 2.57E+13 | (present) | *cm* |
| Capacitance(C) $z$ factors | $(1+z_R)^{-2/3}$ or $(1+z_B)^{-2}$ | | | *unitless* |
| Capacitance $(C_{xr})$ | *approach* $\varepsilon_{xr}/(4\pi B_r)$ | 1.61E-40 | (present) | *cm* |
| Capacitance $(C_{yr})$ | $g_s \sqrt{3} approach_r\ \varepsilon_{yr}/(8\pi B_r)$ | 1.61E-40 | (present) | *cm* |
| Capacitance $(C_{zr})$ | $\sqrt{3}\pi\ approach_r\ \varepsilon_{zr}/(3\pi B_r)$ | 1.61E-40 | (present) | *cm* |
| Capacitance $(C_r)$ $z$ factors | $(1+z_R)^{2/3}$ or $(1+z_B)^{2}$ | | | *unitless* |
| Charge $(e)$ | $m_T^{1/2} l_T^{3/2} t_T^{-1} \alpha^{1/2}$ | 4.803206799E-10 | | $g^{1/2} cm^{3/2} sec^{-1}$ *(esu)* |
| | | 1.60218E-19 | | *coulomb* |
| Charge$^2$ $(e^2)$ | $m_T l_T^3 t_T^{-2} \alpha$ | 2.307077129E-19 | | $g\ cm^3 sec^{-2}$ |
| Cooper CPT conjugated pair | $\mathbb{C}$ | 2 | (constant) | *unitless* |
| Correlation length | $\ell$ | 60.3 | (present) | *cm* |
| Correlation length $z$ factors | $(1+z_R)^{1/3}$ or $(1+z_B)^{1}$ | | | *unitless* |
| COVE Area | $(c\ time_{\pm})^2$ | 7.87E+29 | (present) | $cm^2$ |
| COVE Area $z$ factors | $(1+z_R)^{-4/3}$ or $(1+z_B)^{-4}$ | | | *unitless* |
| COVE Length | $(c\ time_{\pm})$ | 8.87E+14 | (present) | *cm* |
| COVE Length $z$ factors | $(1+z_R)^{-2/3}$ or $(1+z_B)^{-2}$ | | | *unitless* |
| COVE Op frequency at in/out =1 | $1/(2\ time)$ | 1.69E-05 | (present) | $sec^{-1}$ |
| COVE Op frequency at in/out =1 $z$ factors | $(1+z_R)^{2/3}$ or $(1+z_B)^{2}$ | | | *unitless* |
| COVE Op wavelength at in/out =1 | $2c\ time$ | 1.77E+15 | (present) | $cm$ |
| COVE Op wavelength at in/out =1 $z$ factors | $(1+z_R)^{-2/3}$ or $(1+z_B)^{-2}$ | | | *unitless* |
| COVE Power at in/out =1 | $hc^3 time_{\pm}/(2cavity)$ | 1.68E+05 | (present) | $erg\ sec^{-1}$ |
| COVE Power at in/out =1 $z$ factors | $(1+z_R)^{1/3}$ or $(1+z_B)^{1}$ | | | *unitless* |
| COVE Power/area at in/out =1 | $power/area$ | 2.13E-25 | (present) | $erg\ sec^{-1}cm^{2}$ |
| COVE Power/area at in/out =1 $z$ factors | $(1+z_R)^{5/3}$ or $(1+z_B)^{5}$ | | | *unitless* |
| COVE Schwinger linear | $1/(B\ time_{\pm})$ | 1.53E-06 | (present) | $cm^{-1}\ sec^{-1}$ |
| COVE Schwinger linear $z$ factors | $(1+z_R)^{1}$ or $(1+z_B)^{3}$ | | | *unitless* |
| COVE Schwinger area | $1/(section\ time_{\pm})$ | 2.00E-08 | (present) | $cm^{-2}\ sec^{-1}$ |
| COVE Schwinger area $z$ factors | $(1+z_R)^{4/3}$ or $(1+z_B)^{4}$ | | | *unitless* |
| COVE Schwinger volume | $1/(cavity\ time_{\pm})$ | 2.15E-09 | (present) | $cm^{-3}\ sec^{-1}$ |
| COVE Schwinger volume $z$ factors | $(1+z_R)^{5/3}$ or $(1+z_B)^{5}$ | | | *unitless* |
| COVE Volume | $(c\ time_{\pm})^{3}$ | 6.99E+44 | (present) | $cm^{3}$ |
| COVE Volume $z$ factors | $(1+z_R)^{-6/3}$ or $(1+z_B)^{-6}$ | | | *unitless* |
| Critical magnetic field(internal) | $H_{c1}\ or\ \Phi_{\varepsilon}/(\pi\lambda^{2})$ | 2.17E-20 | (present) | $g^{1/2}\ cm^{-1/2}sec^{-1}$ |
| Critical magnetic field(internal) | $H_{c2}\ or\ 8\pi\Phi_{\varepsilon}/section$ | 1.24E-14 | (present) | $g^{1/2}\ cm^{-1/2}sec^{-1}$ |
| Critical magnetic field(internal) | $H_{c}\ or\ (H_{c1}H_{c2})^{1/2}$ | 1.64E-17 | (present) | $g^{1/2}\ cm^{-1/2}sec^{-1}$ |
| Critical magnetic fields(internal) $z$ factors | $(1+z_R)^{5/6}$ or $(1+z_B)^{15/6}$ | | | *unitless* |
| Critical magnetic field Reciprocity | $(H_{c1}H_{c1r})^{1/2}$ | 3.78E+13 | (constant) | $g^{1/2}\ cm^{-1/2}sec^{-1}$ |
| Critical magnetic field Reciprocity | $(H_{c2}H_{c2r})^{1/2}$ | 2.26E+19 | (constant) | $g^{1/2}\ cm^{-1/2}sec^{-1}$ |
| Critical magnetic field Reciprocity | $(H_{c}H_{cr})^{1/2}$ | 2.92E+16 | (constant) | $g^{1/2}\ cm^{-1/2}sec^{-1}$ |
| Critical magnetic field(internal) | $H_{cr1}\ or\ \Phi_{\varepsilon r}/(\pi\lambda_{r}^{2})$ | 2.30E+49 | (present) | $g^{1/2}\ cm^{-1/2}sec^{-1}$ |
| Critical magnetic field(internal) | $H_{cr2}\ or\ 8\pi\Phi_{\varepsilon r}/section_{r}$ | 1.24E+50 | (present) | $g^{1/2}\ cm^{-1/2}sec^{-1}$ |
| Critical magnetic field(internal) | $H_{cr}\ or\ (H_{cr1}H_{cr2})^{1/2}$ | 5.33E+49 | (present) | $g^{1/2}\ cm^{-1/2}sec^{-1}$ |
| Critical magnetic fields(internal) $z$ factors | $(1+z_R)^{-5/6}$ or $(1+z_B)^{-15/6}$ | | | *unitless* |
| Dirac's number - electron magnetic moment in Bohr magnetons | $g_d = g_e/2$ | 1.001160735884950 | | *unitless* |
| Displacement field (internal) $D_{fx}$ | $(\pi/g_s)4\pi\mathbb{C}e/approach$ | 1.06E-10 | (present) | $g^{1/2}\ cm^{-1/2}sec^{-1}$ |
| Displacement field (internal) $D_{fy}$ | $(4\pi/(3g_s))4\pi\mathbb{C}e/side$ | 1.22E-10 | (present) | $g^{1/2}\ cm^{-1/2}sec^{-1}$ |
| Displacement field (internal) $D_{fz}$ | $\left(\sqrt{3}/\left(2g_d^2\right)\right)4\pi\mathbb{C}e/section$ | 6.16E-12 | (present) | $g^{1/2}\ cm^{-1/2} sec^{-1}$ |
| Displacement field (internal) $D_f$ | $1/\left(1/D_{fx}+1/D_{fy}+1/D_{fz}\right)$ | 5.55E-12 | (present) | $g^{1/2}\ cm^{-1/2} sec^{-1}$ |
| Displacement field (internal) $z$ factors | $\left(1+z_R\right)^{2/3}$ or $\left(1+z_B\right)^2$ | | | *unitless* |
| Displacement field Reciprocity | $\left(D_{fx}D_{fxr}\right)^{1/2}$ | 4.25E+16 | (constant) | $g^{1/2}\ cm^{-1/2} sec^{-1}$ |
| Displacement field Reciprocity | $\left(D_{fy}D_{fyr}\right)^{1/2}$ | 4.91E+16 | (constant) | $g^{1/2}\ cm^{-1/2} sec^{-1}$ |
| Displacement field Reciprocity | $\left(D_{fz}D_{fzr}\right)^{1/2}$ | 2.48E+15 | (constant) | $g^{1/2}\ cm^{-1/2} sec^{-1}$ |
| Displacement field Reciprocity | $\left(D_{f}D_{fr}\right)^{1/2}$ | 2.24E+15 | (constant) | $g^{1/2}\ cm^{-1/2} sec^{-1}$ |
| Displacement field (internal) $D_{fxr}$ | $\left(\pi/g_s\right)4\pi\mathbb{C}e/approach_r$ | 1.71E+43 | (present) | $g^{1/2}\ cm^{-1/2} sec^{-1}$ |
| Displacement field (internal) $D_{fyr}$ | $\left(4\pi/\left(3g_s\right)\right)4\pi\mathbb{C}e/side_r$ | 1.98E+43 | (present) | $g^{1/2}\ cm^{-1/2} sec^{-1}$ |
| Displacement field (internal) $D_{fzr}$ | $\left(\sqrt{3}/\left(2g_s^2\right)\right)4\pi\mathbb{C}e/section_r$ | 9.99E+41 | (present) | $g^{1/2}\ cm^{-1/2} sec^{-1}$ |
| Displacement field (internal) $D_{fr}$ | $1/\left(1/D_{fx}+1/D_{fy}+1/D_{fz}\right)$ | 9.01E+41 | (present) | $g^{1/2}\ cm^{-1/2} sec^{-1}$ |
| Displacement field (internal) $z$ factors | $\left(1+z_R\right)^{-2/3}$ or $\left(1+z_B\right)^{-2}$ | | | *unitless* |
| Earth mass | $M_E$ | 5.979E27 | | $g$ |
| Earth radius | $R_E$ | 6.371315E08 | | $cm$ |
| Electric field (internal) $E_f$ | $1/\left(1/E_{fx}+1/E_{fy}+1/E_{fz}\right)$ | 2.54E-24 | (present) | $g^{1/2}\ cm^{-1/2} sec^{-1}$<br>$\left(volt/cm\right)/299.792458$ |
| Electric field (internal) $E_{fx}$ | $\mathbb{C}k_bT_c/\left(\mathbb{C}e2B\right)$ | 5.27E-24 | (present) | $g^{1/2}\ cm^{-1/2} sec^{-1}$ |
| Electric field (internal) $E_{fy}$ | $\mathbb{C}k_bT_c/\left(\mathbb{C}eP\right)$ | 6.08E-24 | (present) | $g^{1/2}\ cm^{-1/2} sec^{-1}$ |
| Electric field (internal) $E_{fz}$ | $\mathbb{C}k_bT_c/\left(\mathbb{C}eA\right)$ | 2.51E-23 | (present) | $g^{1/2}\ cm^{-1/2} sec^{-1}$ |
| Electric field (internal) $z$ factors | $\left(1+z_R\right)^{1}$ or $\left(1+z_B\right)^{3}$ | | | *unitless* |
| Electric field reciprocity | $\left(E_{fx}E_{fxr}\right)^{1/2}$ | 4.26E+16 | (constant) | $g^{1/2}\ cm^{-1/2} sec^{-1}$ |
| Electric field reciprocity | $\left(E_{fy}E_{fyr}\right)^{1/2}$ | 4.92E+16 | (constant) | $g^{1/2}\ cm^{-1/2} sec^{-1}$ |
| Electric field reciprocity | $\left(E_{fz}E_{fzr}\right)^{1/2}$ | 2.03E+17 | (constant) | $g^{1/2}\ cm^{-1/2} sec^{-1}$ |
| Electric field reciprocity | $\left(E_{f}E_{fr}\right)^{1/2}$ | 2.05E+16 | (constant) | $g^{1/2}\ cm^{-1/2} sec^{-1}$ |
| Electric field (internal) $E_{fr}$ | $1/\left(1/E_{fxr}+1/E_{fyr}+1/E_{fzr}\right)$ | 1.66E+56 | (present) | $g^{1/2}\ cm^{-1/2} sec^{-1}$<br>$\left(volt/cm\right)/299.792458$ |
| Electric field (internal) $E_{fxr}$ | $\mathbb{C}k_bT_{cr}/\left(\mathbb{C}e2B_r\right)$ | 3.44E+56 | (present) | $g^{1/2}\ cm^{-1/2} sec^{-1}$ |
| Electric field (internal) $E_{fyr}$ | $\mathbb{C}k_bT_{cr}/\left(\mathbb{C}eP_r\right)$ | 3.98E+56 | (present) | $g^{1/2}\ cm^{-1/2} sec^{-1}$ |
| Electric field (internal) $E_{fzr}$ | $\mathbb{C}k_bT_{cr}/\left(\mathbb{C}eA_r\right)$ | 1.64E+57 | (present) | $g^{1/2}\ cm^{-1/2} sec^{-1}$ |
| Electric field (internal) $z$ factors | $\left(1+z_R\right)^{-1}$ or $\left(1+z_B\right)^{-3}$ | | | *unitless* |
| Electron gyromagnetic-factor | $g_e$ | 2.0023193043617 | | *unitless* |

| | | | | | |
|---|---|---|---|---|---|
| Electron mass $(m_e)$ | $\alpha m_P \left(\Lambda_U^{1/6}/\cos(\theta)^{2/9}\right) l_P^{1/3}$ | | 9.10938975E-28 | | $g$ |
| Energy | $E_{cx}$ or $\left(m_T v_{cx}^2\right)/2$ or $k_B T_c$ | | 1.12E-31 | (present) | $g\ cm^2 sec^{-2}$ |
| Energy | $E_{cy}$ or $\left(m_T v_{cy}^2\right)/2$ | | 1.49E-31 | (present) | $g\ cm^2 sec^{-2}$ |
| Energy | $E_{cz}$ or $\left(m_T v_{cz}^2\right)/2$ | | 2.53E-30 | (present) | $g\ cm^2 sec^{-2}$ |
| Energy $z$ factors | $(1+z_R)^{2/3}$ or $(1+z_B)^2$ | | | | *unitless* |
| Energy reflective | $E_{cxr}$ or $\left(m_T v_{cxr}^2\right)/2$ or $k_B T_{cr}$ | | 1.82E+22 | (present) | $g\ cm^2 sec^{-2}$ |
| Energy reflective | $E_{cyr}$ or $\left(m_T v_{cyr}^2\right)/2$ | | 2.24E+22 | (present) | $g\ cm^2 sec^{-2}$ |
| Energy reflective | $E_{cyr}$ or $\left(m_T v_{cyr}^2\right)/2$ | | 4.11E+22 | (present) | $g\ cm^2 sec^{-2}$ |
| Energy reflective $z$ factors | $(1+z_R)^{-2/3}$ or $(1+z_B)^{-2}$ | | | | *unitless* |
| Euler number -natural log base | e | | 2.7182818284590 | | *unitless* |
| Euler-Mascheroni constant | *Euler* | | 0.5772156649015 | | *unitless* |
| Fine structure constant | $\hbar c/e^2$ | $1/\alpha$ | 137.03599936 | | *unitless* |
| | $(A/B)\sqrt{k_{mx}\varepsilon_x}/\varepsilon_{trisine}$ | $1/\alpha$ | 137.03599936 | | *unitless* |
| | $\sqrt{g_s}\,(\varepsilon/\varepsilon_{trisine})^2$ | $1/\alpha$ | 137.03599936 | | *unitless* |
| | $2\sqrt{k_{mz}\varepsilon_z}/\varepsilon_{trisine}$ | $1/\alpha$ | 137.03599936 | | *unitless* |
| Fine structure constant | $\hbar c/e^2$ | $1/\alpha$ | 137.03599936 | | *unitless* |
| | $(A_r/B_r)\sqrt{k_{mxr}\varepsilon_{xr}}/\varepsilon_{trisiner}$ | $1/\alpha$ | 137.03599936 | | *unitless* |
| | $\sqrt{g_s}\,(\varepsilon_r/\varepsilon_{trisiner})^2$ | $1/\alpha$ | 137.03599936 | | *unitless* |
| Fluxoid | $\Phi$ | | 2.0678539884E-7 | | $g^{1/2}cm^{3/2}sec^{-1}$ *(gauss* $cm^2$*)* |
| Fluxoid dielectric $(\varepsilon)$ mod | $\Phi_\varepsilon$ *or* $hv_\varepsilon/(\mathbb{C}e)$ | | 8.33E-13 | (present) | $g^{1/2}cm^{3/2}sec^{-1}$ *(gauss* $cm^2$*)* |
| Fluxoid dielectric $(\varepsilon)$ mod $z$ factors | $(1+z_R)^{1/6}$ or $(1+z_B)^{1/2}$ | | | | *unitless* |
| Fluxoid dielectric $(\varepsilon)$ mod reflected | $\Phi_{\varepsilon r}$ *or* $hv_{\varepsilon r}/(\mathbb{C}e)$ | | 5.13E-02 | (present) | $g^{1/2}cm^{3/2}sec^{-1}$ *(gauss* $cm^2$*)* |
| Fluxoid dielectric $(\varepsilon)$ mod reflected $z$ factors | $(1+z_R)^{-1/6}$ or $(1+z_B)^{-1/2}$ | | | | *unitless* |
| $\text{Force}_x$ | $m_T v_{dx}/time_\pm$ | | 5.06E-33 | (present) | $g\ cm/sec^2$ |
| $\text{Force}_y$ | $m_T v_{dy}/time_\pm$ | | 5.84E-33 | (present) | $g\ cm/sec^2$ |
| $\text{Force}_z$ | $m_T v_{dz}/time_\pm$ | | 2.41E-32 | (present) | $g\ cm/sec^2$ |
| Force | $m_t\sqrt{v_{dx}^2+v_{dy}^2+v_{dz}^2}/time_\pm$ | | 2.53E-32 | (present) | $g\ cm/sec^2$ |
| Force $z$ factors | $(1+z_R)^1$ or $(1+z_B)^3$ | | | | *unitless* |
| $\text{Force}_{xr}$ | $m_T v_{dxr}/time_{r\pm}$ | | 3.31E+47 | (present) | $g\ cm/sec^2$ |
| $\text{Force}_{yr}$ | $m_T v_{dyr}/time_{r\pm}$ | | 3.82E+47 | (present) | $g\ cm/sec^2$ |
| $\text{Force}_{zr}$ | $m_T v_{dzr}/time_{r\pm}$ | | 1.57E+48 | (present) | $g\ cm/sec^2$ |

| | | | | |
|---|---|---|---|---|
| Force$_r$ | $m_T\sqrt{v_{dxr}^2+v_{dyr}^2+v_{dzr}^2}\Big/time_{r\pm}$ | 1.65E+48 | (present) | $g\ cm/sec^2$ |
| Force$_r$ $z$ factors | $(1+z_R)^{-1}$ or $(1+z_B)^{-3}$ | | | *unitless* |
| Galactic diameter | $(24/\pi)(C\sigma/\mu H_U)$ or $(24/\pi)\left(Cc/\varepsilon^{1/2}H_U\right)$ | 1.10E+23 | (present) | cm |
| Galactic diameter $z$ factors | $(1+z_R)^{-5/6}$ or $(1+z_B)^{-15/6}$ | | | *unitless* |
| Hall quantum resistance | $h/e^2$ | 2.87206217786631E-08 | | $cm^{-1}sec$ |
| Hall quantum resistance (von Klitzing constant) | | 25812.80756 | (constant) | ohm |
| Homes' constant | ℧ | 191,537 | (constant) | $cm^5 sec^{-3}$ |
| Inductance$(L_x)$ | $\Phi_x\ time_\pm/(4\pi\mathbb{C}ev_{\varepsilon x})$ | 3.38E-05 | (present) | $cm^{-1}sec^2$ |
| Inductance$(L_y)$ | $\Phi_y\ time_\pm/(4\pi\mathbb{C}ev_{\varepsilon y})$ | 3.38E-05 | (present) | $cm^{-1}sec^2$ |
| Inductance$(L_z)$ | $\Phi_z\ time_\pm/(4\pi\mathbb{C}ev_{\varepsilon z})$ | 3.38E-05 | (present) | $cm^{-1}sec^2$ |
| Inductance$(L)$ $z$ factors | $(1+z_R)^{-2/3}$ or $(1+z_B)^{-2}$ | | | *unitless* |
| Inductance$(L_{xr})$ | $\Phi_{xr}\ time_{r\pm}/(4\pi\mathbb{C}ev_{\varepsilon xr})$ | 2.09E-58 | (present) | $cm^{-1}sec^2$ |
| Inductance$(L_{yr})$ | $\Phi_{yr}\ time_{r\pm}/(4\pi\mathbb{C}ev_{\varepsilon yr})$ | 2.09E-58 | (present) | $cm^{-1}sec^2$ |
| Inductance$(L_{zr})$ | $\Phi_{zr}\ time_{r\pm}/(4\pi\mathbb{C}ev_{\varepsilon zr})$ | 2.09E-58 | (present) | $cm^{-1}sec^2$ |
| Inductance$(L_r)$ $z$ factors | $(1+z_R)^{2/3}$ or $(1+z_B)^{2}$ | | | *unitless* |
| London penetration depth | $\lambda$ *or* $(1/g_s)\left((m_T v_\varepsilon^2/2)\,cavity\right)^{1/2}(\mathbb{C}e)$ | 3.50E+03 | (present) | *cm* |
| London penetration depth $z$ factors | $(1+z_R)^{-1/3}$ or $(1+z_B)^{-1}$ | | | *unitless* |
| London penetration depth | $\lambda_r$ *or* $(1/g_s)\left((m_T v_{\varepsilon r}^2/2)\,cavity_r\right)^{1/2}(\mathbb{C}e)$ | 2.67E-26 | (present) | *cm* |
| London penetration depth $z$ factors | $(1+z_R)^{1/3}$ or $(1+z_B)^{1}$ | | | *unitless* |
| Lamb shift | $\nu_{Lamb\ shift}$ | 1.0578330E09 | | $sec^{-1}$ |
| Mole of photons | *einstein* | | (constant) | *moles of photons* |
| Momentum vector | $p_x$ *or* $m_T v_{dx}$ *or* $\hbar K_B$ | 1.50E-28 | (present) | $g\ cm^{-1}sec^{-1}$ |
| Momentum vector | $p_y$ *or* $m_T v_{dy}$ *or* $\hbar K_B$ | 1.73E-28 | (present) | $g\ cm^{-1}sec^{-1}$ |
| Momentum vector | $p_z$ *or* $m_T v_{dz}$ *or* $\hbar K_A$ | 7.13E-28 | (present) | $g\ cm^{-1}sec^{-1}$ |
| Momentum $z$ factors | $(1+z_R)^{-1/3}$ or $(1+z_B)^{-1}$ | | | *unitless* |
| Momentum reflective vector | $p_{xr}$ *or* $m_T v_{dxr}$ *or* $\hbar K_{Br}$ | 6.04E-02 | (present) | $g\ cm^{-1}sec^{-1}$ |
| Momentum reflective vector | $p_{yr}$ *or* $m_T v_{dyr}$ *or* $\hbar K_{Br}$ | 6.97E-02 | (present) | $g\ cm^{-1}sec^{-1}$ |
| Momentum reflective vector | $p_{zr}$ *or* $m_T v_{dzr}$ *or* $\hbar K_{Ar}$ | 2.87E-01 | (present) | $g\ cm^{-1}sec^{-1}$ |
| Momentum $z$ factors | $(1+z_R)^{1/3}$ or $(1+z_B)^{1}$ | | | *unitless* |
| Neutral pi meson (pion) | $\pi^0$ | 2.406176E-25 | | *g* |
| Neutral pi meson (pion) in terms of trisine | $m_T\,(B/A)_t\,(g_s/g_d)$ | 2.407096E-25 | | *g* |
| Nuclear density | $m_T / cavity_{nucleus}$ | 2.34E14 | | $g\ cm^{-3}$ |
| Nuclear magneton | $e\hbar / \left(2m_p c\right)$ $\mu_p$ | 5.05078342E-24 | | *erg/gauss (gauss cm*$^3$*)* <br> $g^{1/2} cm^{5/2} sec^{-1}$ |
| Permeability | $k_m$ *or* $\left(1/\varepsilon\right) 1 / \left(1/\left(k_{mx}\varepsilon_x\right) + 1/\left(k_{my}\varepsilon_y\right) + 1/\left(k_{mz}\varepsilon_z\right)\right)$ | 6.73E+13 | (present) | *unitless* |
| Permeability | $k_{mx}$ | 2.00E+13 | (present) | *unitless* |
| Permeability | $k_{my}$ | 1.51E+13 | (present) | *unitless* |
| Permeability | $k_{mz}$ | 7.24E+13 | (present) | *unitless* |
| Permeability $z$ factors | $\left(1+z_R\right)^{-1/3}$ or $\left(1+z_B\right)^{-1}$ | | | *unitless* |
| Permeability | $k_{mr}$ *or* $\left(1/\varepsilon_r\right) 1 / \left(1/\left(k_{mxr}\varepsilon_{xr}\right) + 1/\left(k_{myr}\varepsilon_{yr}\right) + 1/\left(k_{mzr}\varepsilon_{zr}\right)\right)$ | 1.67E-13 | (present) | *unitless* |
| Permeability | $k_{mxr}$ | 4.97E-14 | (present) | *unitless* |
| Permeability | $k_{myr}$ | 3.73E-14 | (present) | *unitless* |
| Permeability | $k_{mzr}$ | 1.80E-11 | (present) | *unitless* |
| Permeability $z$ factors | $\left(1+z_R\right)^{1/3}$ or $\left(1+z_B\right)^{1}$ | | | *unitless* |
| Permittivity (dielectric) | $D_{fx} / E_{fx}$ or $\varepsilon_x$ | 2.00E+13 | (present) | *unitless* |
| Permittivity (dielectric) | $D_{fy} / E_{fy}$ or $\varepsilon_y$ | 2.00E+13 | (present) | *unitless* |
| Permittivity (dielectric) | $D_{fz} / E_{fz}$ or $\varepsilon_z$ | 2.46E+11 | (present) | *unitless* |
| Permittivity (dielectric) | $\left(1/\left(9g_s\right)\right) D_f / E_f$ or $\varepsilon$ | 2.40E+11 | (present) | *unitless* |
| | $1 / \left(1/\varepsilon_x + 1/\varepsilon_y + 1/\varepsilon_z\right)$ | 2.40E+11 | (present) | *unitless* |
| Permittivity (dielectric) $z$ factors | $\left(1+z_R\right)^{-1/3}$ or $\left(1+z_B\right)^{-1}$ | | | *unitless* |
| Permittivity (dielectric $\varepsilon_x$) | $D_{fxr} / E_{fxr}$ or $\varepsilon_{xr}$ | 4.97E-14 | (present) | *unitless* |
| Permittivity (dielectric $\varepsilon_y$) | $D_{fyr} / E_{fyr}$ or $\varepsilon_{yr}$ | 4.97E-14 | (present) | *unitless* |
| Permittivity (dielectric $\varepsilon_z$) | $D_{fzr} / E_{fzr}$ or $\varepsilon_{zr}$ | 6.10E-16 | (present) | *unitless* |
| Permittivity (dielectric $\varepsilon$) | $\left(1/\left(9g_s\right)\right) D_{fr} / E_{fr}$ or $\varepsilon_r$ | 5.95E-16 | (present) | *unitless* |
| Permittivity (dielectric $\varepsilon$) | $1 / \left(1/\varepsilon_{xr} + 1/\varepsilon_{yr} + 1/\varepsilon_{zr}\right)$ or $\varepsilon_r$ | 5.95E-16 | (present) | *unitless* |
| Permittivity (dielectric) $z$ factors | $\left(1+z_R\right)^{1/3}$ or $\left(1+z_B\right)^{1}$ | | | *unitless* |
| Planck reduced constant | $\hbar\left(h/2\pi\right)$ or $m_T (2B)^2 / \left(2\pi\ time_{\pm}\right)$ | 1.054572675E-27 | | $g\ cm^2 sec^{-1}$ |
| Planck reduced constant | $\hbar\left(h/2\pi\right)$ or $m_T (2B_r)^2 / \left(2\pi\ time_{r\pm}\right)$ | 1.054572675E-27 | | $g\ cm^2 sec^{-1}$ |
| Planck length $\left(l_P\right)$ | $\hbar^{1/2} G_U^{1/2} / c^{3/2}$ | 1.61624E-33 | (constant) | *cm* |
| Planck time $\left(t_P\right)$ | $\hbar^{1/2} G_U^{1/2} / c^{5/2}$ | 5.39121E-44 | (constant) | *sec* |
| Planck mass $\left(m_P\right)$ | $\hbar^{1/2} c^{1/2} / G_U^{1/2}$ | 2.17645E-05 | (constant) | *g* |
| Planck base speed $\left(s_P\right)$ | $m_P^0\ l_P^1 t_P^{-1} = \hbar^0 G^0 c^1$ | 2.99792E+10 | (constant) | $cm\ \sec^{-1}$ |
| Planck momentum $(p_P)$ | $m_P\ l_P t_P^{-1} = \left(\hbar c^3/G_U\right)^{1/2}$ | 6.52483E+05 | (constant) | $g\ cm\ \sec^{-1}$ |
| Planck energy $(E_P)$ | $m_P\ l_P^2 t_P^{-2} = \left(\hbar c^5/G_U\right)^{1/2}$ | 1.95610E+16 | (constant) | $g\ cm^2 \sec^{-2}\ (erg)$ |
| Planck force $(F_P)$ | $m_P l_P t_P^{-2} = c^4/G_U$ | 1.21027E+49 | (constant) | $g\ cm\ \sec^{-2}\ (dyne)$ |
| Planck density $(\rho_P)$ | $m_P l_P^{-3} = c^5/\left(\hbar G_U^2\right)$ | 5.15500E+93 | (constant) | $g\ cm^{-3}$ |
| Planck acceleration $(a_P)$ | $l_P t_P^{-2} = c^6/\left(\hbar G_U\right)$ | 1.03145E+97 | (constant) | $cm\ \sec^{-2}$ |
| Planck kinematic viscosity $(\nu_P)$ | $l_P^2 t_P^{-1} = \left(\hbar G_U/c\right)^{1/2}$ | 4.845244E-23 | (constant) | $cm^2\ \sec^{-1}$ |
| Planck absolute viscosity $(\mu_P)$ | $m_P\ l_P^{-1} t_P^{-1} = \left(c^9/\left(\hbar G_U^3\right)\right)^{1/2}$ | 2.49779E+71 | (constant) | $g\ cm^{-1}\ \sec^{-1}$ |
| Planck charge $(e)$ | $m_P^{1/2} l_P^{3/2} t_P^{-1} \alpha^{1/2} = \sqrt{\hbar c \alpha}$ | 4.80320E-10 | (constant) | $g^{1/2}\ cm^{3/2}\ \sec^{-1}$ |
| Proton mass $(m_p)$ | $(\hbar/c)(\pi/B_p)$ | 1.67262311E-24 | | $g$ |
| Proton radius | $B_p/\left(2\sin(\theta)\right)$ or $x_U/\left(8\sin(\theta)\right)$ | 8.59E-14 (8 E-14) | | $cm$ |
| Resonant CPT $time_{c\pm}$ | $time_{c\pm}$ | 2.96E+04 | (present) | $sec$ |
| Resonant CPT $time_{c\pm}$ $z$ factors | $(1+z_R)^{-2/3}$ or $(1+z_B)^{-2}$ | | | *unitless* |
| Resonant CPT $time_{cr\pm}$ | $time_{cr\pm}$ | 1.53E-49 | (present) | $sec$ |
| Resonant CPT $time_{cr\pm}$ $z$ factors | $(1+z_R)^{2/3}$ or $(1+z_B)^{2}$ | | | *unitless* |
| Saam's constant $(\Theta)$ | $3\pi^3 h^3/\left(6^{1/2} m_T^{3/2}(B/A)\right)$ | 5.89035133E-43 | | $erg^{3/2}\ cm^3$ |
| Stefan's constant $(\sigma_S)$ | $2\pi^5 k_b^4/\left(15c^2h^3\right)$ | 5.67040124E-05 | | $erg\ cm^{-2}\ sec^{-1}\ kelvin^{-4}$ |
| Superconductor gyro-factor | $g_s$ | 1.009804978188 | | *unitless* |
| Superconductor weak coupling constant | $2\mathrm{e}^{Euler}/\pi$ | 1.133865917 | | *unitless* |
| Superconductor conductance | $\sigma_{dc}\ \sqrt{3}\mathbb{C}^2 e^2/(hB)$ | 1.09E+07 | (present) | $\sec^{-1}$ |
| Superconductor conductance $z$ factors | $(1+z_R)^{1/3}$ or $(1+z_B)^{1}$ | | | *unitless* |
| Superconductor conductance | $\sigma_{dcr}\ \sqrt{3}\mathbb{C}^2 e^2/(hB_r)$ | 4.39E+33 | (present) | $\sec^{-1}$ |
| Superconductor conductance $z$ factors | $(1+z_R)^{-1/3}$ or $(1+z_B)^{-1}$ | | | *unitless* |
| Superconductor weak coupling constant | $2\mathrm{e}^{Euler}/\pi$ | 1.133865917 | | *unitless* |
| Temperature Fine Structure $(T_{å})$ | $g_d m_t \hbar c^3/\left(e^2 k_B\right)$ | 8.96E13 | | *kelvin* |
| | $g_d m_t c^2/(\alpha k_b)$ | | | |
| Temperature Universe $(T_a)$ | $m_t c^2/k_b$ | 6.53E+11 | | *kelvin* |
| | $G_U m_T M_U/(R_U k_b)$ | | | *kelvin* |
| | $G_{Ur} m_T M_{Ur}/(R_{Ur} k_b)$ | | | *kelvin* |
| Temperature Universe $(T_a)$ $z$ factors | $(1+z_R)^{0}$ or $(1+z_B)^{0}$ | | | *unitless* |
| Temperature Critical Black Body $(T_b)$ | $(1/2)(1/\mathbb{C})\, g_s^3 m_T v_\varepsilon^2 (1/k_b)$ | 2.729 | (present) | kelvin |
| Temperature Critical $(T_b)$ $z$ factors | $(1+z_R)^{1/3}$ or $(1+z_B)^{1}$ | | | *unitless* |
| Temperature Critical $\left(T_{br}\right)$ | $\left(1/2\right)m_T v_{\varepsilon r}^2/k_b$ | 2.01E+22 | (present) | *kelvin* |
| Temperature Critical $\left(T_{br}\right)$ $z$ factors | $\left(1+z_R\right)^{-1/3}$ or $\left(1+z_B\right)^{-1}$ | | | *unitless* |
| Temperature Critical $\left(T_c\right)$ | $\left(1/2\right)m_T v_{dx}^2/k_b$ | 8.11E-16 | (present) | *kelvin* |
| Temperature Critical $\left(T_c\right)$ $z$ factors | $\left(1+z_R\right)^{2/3}$ or $\left(1+z_B\right)^{2}$ | | | *unitless* |
| Temperature Critical Reciprocity | $\left(T_c T_{cr}\right)^{1/2}$ | 3.26E+11 | (constant) | *kelvin* |
| Temperature Phase $\left(T_{cr}\right)$ | $\left(1/2\right)m_T v_{dxr}^2/k_b$ | 1.29E38 | (present) | *kelvin* |
| | $m_T c^2 k_m^2/\left(2k_b\pi^2\right)$ | | | *kelvin* |
| Temperature Phase $\left(T_{cr}\right)$ $z$ factors | $\left(1+z_R\right)^{-2/3}$ or $\left(1+z_B\right)^{-2}$ | | | *unitless* |
| Temperature Dielectric $\left(T_d\right)$ | $m_T v_\varepsilon c/k_b$ | 1.33E+06 | (present) | *kelvin* |
| Temperature Phase $\left(T_d\right)$ $z$ factors | $\left(1+z_R\right)^{1/6}$ or $\left(1+z_B\right)^{3/6}$ | | | *unitless* |
| Temperature Dielectric $\left(T_{dr}\right)$ | $m_T v_{\varepsilon r} c/k_b$ | 1.62E16 | (present) | *kelvin* |
| Temperature Phase $\left(T_{dr}\right)$ $z$ factors | $\left(1+z_R\right)^{-1/6}$ or $\left(1+z_B\right)^{-3/6}$ | | | *unitless* |
| Temperature Momentum $(T_i)$ | $m_T v_{dx} c/k_b$ | 3.25E-02 | (present) | *kelvin* |
| | $\hbar K_B c/k_b$ | | | *kelvin* |
| | $m_T c^2/\left(\left(k_m\varepsilon_x\right)^{1/2}k_b\right)$ | | | *kelvin* |
| Temperature Momentum $(T_i)$ $z$ factors | $\left(1+z_R\right)^{1/3}$ or $\left(1+z_B\right)^{1}$ | | | *unitless* |
| Temperature Momentum $(T_{ir})$ | $m_T v_{dxr} c/k_b$ | 1.31E+25 | (present) | *kelvin* |
| Temperature Momentum $(T_{ir})$ $z$ factors | $\left(1+z_R\right)^{-1/3}$ or $\left(1+z_B\right)^{-1}$ | | | *unitless* |
| Temperature Resonant $\left(T_j\right)$ | $\left(m_T v_{dT}^4/4\right)\left(1/c^2\right)/k_b$ | 6.28E-40 | (present) | *kelvin* |
| | $m_T c^2\left(1/\left(k_m\varepsilon_x\right)+1/\left(k_m\varepsilon_y\right)+1/\left(k_m\varepsilon_z\right)\right)^2/\left(4k_b\right)$ | | | *kelvin* |
| Temperature Resonant $\left(T_j\right)$ $z$ factors | $\left(1+z_R\right)^{4/3}$ or $\left(1+z_B\right)^{4}$ | | | *unitless* |
| Temperature Resonant $\left(T_{jr}\right)$ | $\left(m_T v_{dTr}^4/4\right)\left(1/c^2\right)/k_b$ | 8.52E+64 | (present) | *kelvin* |
| | $m_T c^2\left(1/\left(k_{mr}\varepsilon_{xr}\right)+1/\left(k_{mr}\varepsilon_{yr}\right)+1/\left(k_{mr}\varepsilon_{zr}\right)\right)^2/\left(4k_b\right)$ | | | *kelvin* |
| Temperature Resonant $\left(T_{jr}\right)$ $z$ factors | $\left(1+z_R\right)^{-4/3}$ or $\left(1+z_B\right)^{-4}$ | | | *unitless* |
| Trisine base constant $\left(U\right)$ | $H_U/G_U$ | 3.46331E-11 | (constant) | $g\ cm^{-3}\sec$ |
| Trisine base length $\left(l_T\ or\ x_U\right)$ | $\hbar^1 m_T^{-1}c^{-1}=\hbar^{1/3}U^{-1/3}c^{-2/3}\cos\left(\theta\right)^{-1}$ | 3.51001E-13 | (constant) | *cm* |
| Trisine base time $\left(t_T\ or\ t_U\right)$ | $\hbar^{1/3}U^{-1/3}c^{-5/3}\cos\left(\theta\right)^{-1}$ | 1.17081E-23 | (constant) | *sec* |
| Trisine base mass $\left(m_T\ or\ m_U\right)$ | $\hbar^{2/3}U^{1/3}c^{-1/3}\cos\left(\theta\right)$ | 1.00218E-25 | (constant) | *g* |
| Trisine base speed $\left(s_T\right)$ | $m_T^0\ l_T^1 t_T^{-1}\ =\hbar^0 U^0 c^1$ | 2.99792E+10 | (constant) | $cm\ \sec^{-1}$ |
| Trisine base momentum $\left(p_T\right)$ | $m_T\ l_T t_T^{-1}\ =\hbar^{2/3}U^{1/3}c^{2/3}\cos\left(\theta\right)$ | 3.00447E-15 | (constant) | $g\ cm\ \sec^{-1}$ |
| Trisine base energy $\left(E_T\right)$ | $m_T l_T^2 t_T^{-2}\ =\hbar^{2/3}U^{1/3}c^{5/3}\cos\left(\theta\right)$ | 9.00718E-05 | (constant) | $g\ cm^2\sec^{-2}$ $\left(erg\right)$ |
| Trisine base force $(F_T)$ | $m_T l_T t_T^{-2} = \hbar^{1/3} U^{2/3} c^{7/3} \cos(\theta)^2$ | 2.56614E+08 | (constant) | $g\ cm\ \sec^{-2}$ $(dyne)$ |
| Trisine base density $(\rho_T)$ | $m_T l_T^{-3} = \hbar^{-1/3} U^{4/3} c^{5/3} \cos(\theta)^4$ | 2.31752E+12 | (constant) | $g\ cm^{-3}$ |
| Trisine base acceleration $(a_T)$ | $l_T t_T^{-2} = \hbar^{-1/3} U^{1/3} c^{8/3} \cos(\theta)$ | 2.56055E+33 | (constant) | $cm\ \sec^{-2}$ |
| Trisine base kinematic viscosity $(\nu_T)$ | $l_T^2 t_T^{-1} = \hbar^{1/3} U^{-1/3} c^{1/3} \cos(\theta)^{-1}$ | 1.05227E-02 | (constant) | $cm^2\ \sec$ |
| Trisine base absolute viscosity $(\mu_T)$ | $m_T\ l_T^{-1} t_T^{-1} = U c^2 \cos(\theta)^3$ | 2.43866E+10 | (constant) | $g\ cm^{-1}\ \sec^{-1}$ |
| Trisine base charge $(e)$ | $m_T^{1/2} l_T^{3/2} t_T^{-1} \alpha^{1/2} = \sqrt{\hbar c \alpha}$ | 4.80320E-10 | (constant) | $g^{1/2}\ cm^{3/2}\ \sec^{-1}$ |
| Trisine angle | $\theta$ or $\tan^{-1}(A/B)$ | 22.7979295 | | angular degrees |
| Trisine angle | $\theta$ or $\tan^{-1}(A_r/B_r)$ | 22.7979295 | | angular degrees |
| Trisine angle | or $\cos^{-1}\left(B\big/\left(A^2+B^2\right)^{1/2}\right)$ | 22.7979295 | | angular degrees |
| Trisine angle | or $\cos^{-1}\left(B_r\big/\left(A_r^2+B_r^2\right)^{1/2}\right)$ | 22.7979295 | | angular degrees |
| Trisine angle | or $\cos^{-1}\left(\left(1/(2K_B)\right)\Big/\left(\left(1/K_A\right)^2+\left(1/(2K_B)\right)^2\right)^{1/2}\right)$ | 22.7979295 | | angular degrees |
| Trisine angle | or $\cos^{-1}\left(\left(1/(2K_{Br})\right)\Big/\left(\left(1/K_{Ar}\right)^2+\left(1/(2K_{Br})\right)^2\right)^{1/2}\right)$ | 22.7979295 | | angular degrees |
| Trisine angular relationship | $\sin(\theta)\cos(\theta) = (B/A)\Big/\left((B/A)^2+1\right)$ | =0.357211055 | | *unitless* |
| Trisine area | *section* $2\sqrt{3}B^2$ | 1,693 | (present) | $cm^2$ |
| Trisine area | *approach* $\sqrt{3}AB$ | 356 | (present) | $cm^2$ |
| Trisine area | *side* $2AB$ | 411 | (present) | $cm^2$ |
| Trisine area | *trisine* $4\sqrt{3}B^2/\cos(\theta)$ | 3,670 | (present) | $cm^2$ |
| Trisine area $z$ factors | $(1+z_R)^{-2/3}$ or $(1+z_B)^{-2}$ | | | *unitless* |
| Trisine area reflected | $approach_r$ $\sqrt{3}A_r B_r$ | 2.19E-51 | (present) | $cm^2$ |
| Trisine area reflected | $side_r$ $2A_r B_r$ | 2.27E-50 | (present) | $cm^2$ |
| Trisine area reflected | $section_r$ $2\sqrt{3}B_r^2$ | 1.04E-50 | (present) | $cm^2$ |
| Trisine area reflected | *trisine* $4\sqrt{3}B_r^2/\cos(\theta)$ | 2.27E-50 | (present) | $cm^2$ |
| Trisine area reflected $z$ factors | $(1+z_R)^{2/3}$ or $(1+z_B)^2$ | | | *unitless* |
| Trisine unit | *cell* | | | *unitless* |
| Trisine constant | *trisine* $-\ln\left(2e^{Euler}/\pi - 1\right)$ | 2.010916597 | | *unitless* |
| Trisine density of states | $D(\xi_t)$ *or* $1/k_b T_c$ | 8.93E+30 | (present) | $sec^2 g^{-1} cm^{-2}$ |
| Trisine density of states $z$ factors | $(1+z_R)^{-2/3}$ or $(1+z_B)^{-2}$ | | | *unitless* |
| Trisine density of states(reflected) | $D(\xi_t)$ *or* $1/k_b T_{cr}$ | 5.51E-23 | (present) | $sec^2 g^{-1} cm^{-2}$ |
| Trisine density of states reflected $z$ factors | $(1+z_R)^{2/3}$ or $(1+z_B)^2$ | | | *unitless* |
| Trisine dimension | $A$ | 9.29 | (present) | $cm$ |
| Trisine dimension | $B$ | 22.11 | (present) | $cm$ |
| | $\pi l_T \left(c/v_{dx}\right)$ *or* $\pi l_T \sqrt{k_{mx}\varepsilon_x}$ | | | *cm* |
| Trisine dimension | $C$ | 23.98 | (present) | *cm* |
| Trisine dimension | $P$ | 38.29 | (present) | *cm* |
| Trisine dimensions *z* factors | $\left(1+z_R\right)^{-1/3}$ or $\left(1+z_B\right)^{-1}$ | | | *unitless* |
| Trisine length dimension reciprocity | $\left(AA_r\right)^{1/2}$ | 4.63E-13 | (constant) | *cm* |
| Trisine length dimension reciprocity | $\left(BB_r\right)^{1/2}$ | 1.12E-12 | (constant) | *cm* |
| Trisine dimension | $A_r$ | 2.31E-26 | (present) | *cm* |
| Trisine dimension | $B_r$ | 5.49E-26 | (present) | *cm* |
| | $\pi l_T \left(c/v_{dxr}\right)$ *or* $\pi l_T \sqrt{k_{mxr}\varepsilon_{xr}}$ | | | *cm* |
| Trisine dimension | $C_r$ | 5.96E-26 | (present) | *cm* |
| Trisine dimension | $P_r$ | 9.51E-26 | (present) | *cm* |
| Trisine dimensions *z* factors | $\left(1+z_R\right)^{1/3}$ or $\left(1+z_B\right)^{1}$ | | | *unitless* |
| Trisine frequency | $1/time$ *or* $v_{dx}/2B$ | 3.38E-05 | (present) | $sec^{-1}$ |
| Trisine frequency *z* factors | $\left(1+z_R\right)^{2/3}$ or $\left(1+z_B\right)^{2}$ | | | *unitless* |
| Trisine frequency reflective | $1/time_r$ *or* $v_{dxr}/2B_r$ | 5.48E+48 | (present) | $sec^{-1}$ |
| Trisine frequency reflective *z* factors | $\left(1+z_R\right)^{-2/3}$ or $\left(1+z_B\right)^{-2}$ | | | *unitless* |
| Trisine magneton | $e\hbar/\left(2m_T c\right)$ $\mu_t$ | 8.42151973E-23 | | *erg/gauss* |
| | | | | *gauss* $cm^3$ |
| | | | | $g^{1/2}cm^{5/2}sec^{-1}$ |
| Trisine size ratio | $\left(B/A\right)_t$ $2/sin(1\ radian)$ | 2.379146659 | | *unitless* |
| | $e^2/\left(\varepsilon v_{dx}\hbar\right)$ | | | |
| Trisine size ratio reflected | $\left(B_r/A_r\right)_t$ $2/sin(1\ radian)$ | 2.379146659 | | *unitless* |
| | $e^2/\left(\varepsilon_r v_{dxr}\hbar\right)$ | | | |
| Trisine spin moment | $\left(2/1\right)\left(\mathbb{C}e_{\pm}\hbar/2m_e v_{\varepsilon y}\right) = chain\ H_c$ | 1.72E-13 | (present) | *erg/gauss* |
| | | | | (*gauss* $cm^3$) |
| | | | | $g^{1/2}\ cm^{5/2}sec^{-1}$ |
| Trisine spin moment *z* factors | $\left(1+z_R\right)^{1/6}$ or $\left(1+z_B\right)^{3/6}$ | | | *unitless* |
| Trisine spin moment reflected | $\left(1/2\right)\left(\mathbb{C}e_{\pm}\hbar/2m_e v_{\varepsilon yr}\right) = chain_r\ H_{cr}$ | 8.56E-27 | (present) | *erg/gauss* |
| | | | | (*gauss* $cm^3$) |
| | | | | $g^{1/2}\ cm^{5/2}sec^{-1}$ |
| Trisine spin moment reflected $z$ factors | $(1+z_R)^{-1/6}$ or $(1+z_B)^{-3/6}$ | | | *unitless* |
| Trisine surface tension $(\sigma_T)$ | $k_b T_c / section$ | 6.61E-35 | (present) | $erg\ cm^{-2}$ |
| Trisine surface tension $(\sigma_T)$ $z$ factors | $(1+z_R)^{4/3}$ or $(1+z_B)^{4}$ | | | *unitless* |
| Trisine surface tension $(\sigma_{Tr})$ | $k_b T_{cr} / section_r$ | 1.74E+72 | (present) | $erg\ cm^{-2}$ |
| Trisine surface tension $(\sigma_{Tr})$ $z$ factors | $(1+z_R)^{-4/3}$ or $(1+z_B)^{-4}$ | | | *unitless* |
| Trisine surface tension seen $(\sigma_{cT})$ | $m_T c^2 / section$ | 5.32E-08 | (present) | $erg\ cm^{-2}$ |
| Trisine surface tension seen $(\sigma_{cT})$ $z$ factors | $(1+z_R)^{2/3}$ or $(1+z_B)^{2}$ | | | *unitless* |
| Trisine surface tension seen $(\sigma_{cTr})$ | $m_T c^2 / section_r$ | 8.64E+45 | (present) | $erg\ cm^{-2}$ |
| Trisine surface tension seen $(\sigma_{cTr})$ $z$ factors | $(1+z_R)^{-2/3}$ or $(1+z_B)^{-2}$ | | | *unitless* |
| Trisine relativistic mass | $m_r$ *or* $(1/2) m_T \left(v_{dx}^2 / c^2\right)$ | 1.24E-52 | (present) | $cm^3$ |
| Trisine relativistic mass $z$ factors | $(1+z_R)^{2/3}$ or $(1+z_B)^{2}$ | | | *unitless* |
| Trisine relativistic reflective mass | $m_{rr}$ *or* $(1/2) m_T \left(v_{dxr}^2 / c^2\right)$ | 20.2 | (present) | $cm^3$ |
| Trisine relativistic reflective mass $z$ factors | $(1+z_R)^{-2/3}$ or $(1+z_B)^{-2}$ | | | *unitless* |
| Trisine time | *time or* $2B / v_{dx}$ | 29,600 | (present) | *sec* |
| | or $t_T 2\pi k_{mx} \varepsilon_x$ | | | |
| Trisine time $z$ factors | $(1+z_R)^{-2/3}$ or $(1+z_B)^{-2}$ | | | *unitless* |
| Trisine time reflective | $time_r$ *or* $2B_r / v_{dxr}$ | 1.82E-49 | (present) | *sec* |
| | or $t_T 2\pi k_{mxr} \varepsilon_{xr}$ | | | |
| Trisine time reflective $z$ factors | $(1+z_R)^{2/3}$ or $(1+z_B)^{2}$ | | | *unitless* |
| Trisine time dimension reciprocity | $(time \cdot time_r)^{1/2}$ | 7.34E-23 | (constant) | *sec* |
| Trisine transformed mass | $m_T$ or $\hbar^{2/3} U^{1/3} c^{-1/3} \cos(\theta)$ | 1.00E-25 | (constant) | g |
| Majorana particle | or $\pi\hbar\ time_{\pm} / \left(2B^2\right)$ or $\pi\hbar / D_c$ | | (constant) | g |
| | or $\pi\hbar\ time_{r\pm} / \left(2B_r^2\right)$ or $\pi\hbar / D_{cr}$ | | (constant) | g |
| | or $(1/g_d)(2/3)(K_B K_P K_A)\ chain\ m_e$ | | (constant) | g |
| | or $(1/g_d)(2/3)(K_{Br} K_{Pr} K_{Ar})\ chain_r\ m_e$ | | (constant) | g |
| Trisine volume per cell | *cavity* $2\sqrt{3}AB^2$ | 15,733 | (present) | $cm^3$ |
| Trisine volume per cell $z$ factors | $(1+z_R)^{-1}$ or $(1+z_B)^{-3}$ | | | *unitless* |
| Trisine volume per chain | *chain or 2cavity/3* | 10,497 | | $cm^3$ |
| Trisine volume per chain $z$ factors | $(1+z_R)^{-1}$ or $(1+z_B)^{-3}$ | | | *unitless* |
| Trisine volume per cell reciprocity | $(cavity\ cavity_r)^{1/2}$ | *1.95E-36* | (constant) | $cm^3$ |
| Trisine volume per cell | $cavity_r$ $2\sqrt{3}A_r B_r^2$ | 2.41E-76 | (present) | $cm^3$ |
| Trisine volume per cell $z$ factors | $(1+z_R)^1$ or $(1+z_B)^3$ | | | *unitless* |
| Trisine volume per chain | *chain$_r$ or 2cavity$_r$/3* | 1.61E-76 | (present) | $cm^3$ |
| Trisine volume per chain $z$ factors | $(1+z_R)^1$ or $(1+z_B)^3$ | | | *unitless* |
| Trisine wave vector $(K_A)$ | $2\pi/A$ | 6.76E-01 | (present) | $cm^{-1}$ |
| Trisine wave vector $(K_B)$ | $2\pi/(2B)$ | 1.41E-01 | (present) | $cm^{-1}$ |
| Trisine wave vector $(K_C)$ | $4\pi/(3\sqrt{3}A)$ | 2.60E-01 | (present) | $cm^{-1}$ |
| Trisine wave vector $(K_{Dn})$ | $(3\pi^2\mathbb{C}/cavity)^{1/3}$ | 1.56E-01 | (present) | $cm^{-1}$ |
| Trisine wave vector $(K_{Ds})$ | $(4\pi^3\mathbb{C}/cavity)^{1/3}$ | 2.51E-01 | (present) | $cm^{-1}$ |
| Trisine wave vectors $z$ factors | $(1+z_R)^{1/3}$ or $(1+z_B)^1$ | | | *unitless* |
| Trisine wave vector $(K_{Ar})$ | $2\pi/A_r$ | 2.72E+26 | (present) | $cm^{-1}$ |
| Trisine wave vector $(K_{Br})$ | $2\pi/(2B_r)$ | 5.72E+25 | (present) | $cm^{-1}$ |
| Trisine wave vector $(K_{Cr})$ | $4\pi/(3\sqrt{3}A_r)$ | 1.05E+26 | (present) | $cm^{-1}$ |
| Trisine wave vector $(K_{Dnr})$ | $(3\pi^2\mathbb{C}/cavity_r)^{1/3}$ | 6.26E+25 | (present) | $cm^{-1}$ |
| Trisine wave vector $(K_{Ds})$ | $(4\pi^3\mathbb{C}/cavity_r)^{1/3}$ | 1.01E+26 | (present) | $cm^{-1}$ |
| Trisine wave vectors $z$ factors | $(1+z_R)^{-1/3}$ or $(1+z_B)^{-1}$ | | | *unitless* |
| Universe absolute viscosity | $\mu_U$ or $m_T v_{dx}/section$ | 8.85E-32 | (present) | $g\ cm^{-1}sec^{-1}$ |
| Universe absolute viscosity $z$ factors | $(1+z_R)^1$ or $(1+z_B)^3$ | | | *unitless* |
| Universe absolute viscosity reflected | $\mu_U$ or $m_T v_{cxr}/section_r$ | 5.78E48 | (present) | $g\ cm^{-1}sec^{-1}$ |
| Universe absolute viscosity reflected z factors | $(1+z_R)^{-1}$ or $(1+z_B)^{-3}$ | | | *unitless* |
| Universe absolute (seen) viscosity | $\mu_{Uc}$ or $m_T c/(2\Delta x^2)$ or $\mathbb{C}m_T c K_B^2$ | 1.21E-16 | (present) | $g\ cm^{-1}sec^{-1}$ |
| Universe absolute (seen) viscosity $z$ factors | $(1+z_R)^{2/3}$ or $(1+z_B)^{6/3}$ | | | *unitless* |
| Universe absolute (seen) viscosity | $\mu_{Ucr}$ or $m_T c/(2\Delta x_r^2)$ or $\mathbb{C}m_T c K_{Br}^2$ | 1.97E+37 | (present) | $g\ cm^{-1}sec^{-1}$ |
| Universe absolute (seen) viscosity $z$ factors | $(1+z_R)^{-2/3}$ or $(1+z_B)^{-6/3}$ | | | *unitless* |
| Universe area | $2\pi R_U^2$ | 3.17E+57 | (present) | $cm^2$ |
| Universe area $z$ factors | $(1+z_R)^{-2}$ or $(1+z_B)^{-6}$ | | | *unitless* |
| Universe area reflected | $2\pi R_{Ur}^2$ | 7.42E-103 | (present) | $cm^2$ |
| Universe area reflected $z$ factors | $(1+z_R)^2$ or $(1+z_B)^6$ | | | *unitless* |
| Universe coordinate age | $Age_U$ or $1/H_U$ | 4.32E17 | (present) | *sec* |
| Universe coordinate age $z$ factors | $(1+z_R)^{-1}$ or $(1+z_B)^{-3}$ | | | *unitless* |
| Universe coordinate reflected age | $Age_{Ur}$ or $1/H_{Ur}$ | 1.51E62 | (present) | *sec* |
| Universe coordinate reflected age *z* factors | $(1+z_R)^1$ or $(1+z_B)^3$ | | | *unitless* |
| Universe coordinate age at constant *G* | $Age_{UG}$ or $Age_U\sqrt{G_U/G}$ | 4.32E17 | (present) | *sec* |
| | $Age_U\sqrt{R_U c^2/(GM_U)}$ | | | |
| Universe coordinate age at constant *G* *z* factors | $(1+z_R)^{-1/2}$ or $(1+z_B)^{-3/2}$ | | | *unitless* |
| Universe coordinate age$_v$ at constant *G* | $Age_{UGv}$ | 4.32E17 | (present) | *sec* |
| | $Age_U\sqrt{R_U c^2/(G(M_U+M_v))}$ | | | |
| Universe coordinate age$_v$ at constant *G* *z* factors | | | | |
| | $(1+z_R)^{-1/2}$ or $(1+z_B)^{-3/2}$ $<(M_U = M_v = 4.44E31\ g)$ | | | *unitless* |
| | $(1+z_R)^{-2/3}$ or $(1+z_B)^{-2}$ $>(M_U = M_v = 4.44E31\ g)$ | | | *unitless* |
| Universe cosmological parameter | $\Lambda_U$ or $2H_U^2/c^2\cos(\theta)$ | 1.28958±.23E-56 | (present) | $cm^{-2}$ |
| Universe cosmological parameter *z* factors | $(1+z_R)^2$ or $(1+z_B)^6$ | | | *unitless* |
| Universe cosmological constant | $\Omega_{\Lambda_U}$ or $c^2\Lambda_U/(3H_U^2)$ | 0.72315 | (constant) | *unitless* |
| Universe cosmological constant | or $2/(3cos(\theta))$ | 0.72315 | (constant) | *unitless* |
| Universe diffusion coefficient | $D_c$ or $(2B)^2/time_{\pm} = 2\pi\hbar/m_T$ | 6.60526079E-2 | | $cm^2\ \sec^{-1}$ |
| | *or* $(\hbar^{1/3}U^{-1/3}c^{1/3})2\pi/\cos(\theta)$ | | (constant) | |
| Universe diffusion coefficient | $D_c$ or $(2B_r)^2/time_{r\pm} = 2\pi\hbar/m_T$ | 6.60526079E-2 | | $cm^2\ \sec^{-1}$ |
| | *or* $(\hbar^{1/3}U^{-1/3}c^{1/3})2\pi/\cos(\theta)$ | | (constant) | |
| Universe energy (seen) | $m_T c^2$ or $m_T l_T^2/t_T^2$ | 9.01E-05 | (constant) | $g\ cm^2\ sec^{-2}$ |
| Universe gravitational constancy | $G_U R_U$ or $3^{1/2}c/U$ | 1.4986653E+21 | (constant) | $cm^4 sec^{-2} g^{-1}$ |
| Universe gravitational constancy reflected | $G_{Ur}R_{Ur}$ or $3^{1/2}c/U$ | 1.4986653E+21 | (constant) | $cm^4 sec^{-2} g^{-1}$ |
| Universe gravitational energy density | $GM_U m_T/(R_U cavity)$ | 5.73E-09 | (constant) | $erg\ cm^{-3}$ |
| | or $sherd^{-1/2}G_U M_U m_T/(R_U cavity)$ | 5.73E-09 | (constant) | $erg\ cm^{-3}$ |
| | or $4\pi G\rho_U^2 c^2/H_U^2$ | 5.73E-09 | (constant) | $erg\ cm^{-3}$ |
| Universe gravitational energy density | $GM_{Ur}m_T/(R_{Ur}cavity_r)$ | 5.73E-09 | (constant) | $erg\ cm^{-3}$ |
| | or $4\pi G\rho_{Ur}^2 c^2/H_{Ur}^2$ | 5.73E-09 | (constant) | $erg\ cm^{-3}$ |
| Universe gravity scaling constant | $U$ or $H_U/G_U$ | 3.46E-11 | (constant) | $g\ sec\ cm^{-3}$ |
| Universe gravity scaling constant | or $H_{Ur}/G_{Ur}$ | 3.46E-11 | (constant) | $g\ sec\ cm^{-3}$ |
| Universe gravity scaling constant | or $4m_t^3 c/(\pi\hbar^2)$ | 3.46E-11 | (constant) | $g\ sec\ cm^{-3}$ |
| Universe gravity scaling constant | or $4\pi M_U/(V_U H_U)$ | 3.46E-11 | (constant) | $g\ sec\ cm^{-3}$ |
| Universe gravity scaling constant | or $4\pi M_{Ur}/(V_{Ur}H_{Ur})$ | 3.46E-11 | (constant) | $g\ sec\ cm^{-3}$ |
| Universe gravity scaling constant | or $4\pi m_t/(cavity\ H_U)$ | 3.46E-11 | (constant) | $g\ sec\ cm^{-3}$ |
| Universe gravity scaling constant | or $4\pi m_t/(cavity_r\ H_{Ur})$ | 3.46E-11 | (constant) | $g\ sec\ cm^{-3}$ |
| Universe gravitational parameter | $G_U$ or $R_U c^2/M_U$ | 6.67384E-8 | (present) | $cm^3 sec^{-2} g^{-1}$ |
| | or $G\left(M_{Upresent}/M_U\right)^{1/2}$ | 6.67384E-8 | (present) | $cm^3 sec^{-2} g^{-1}$ |
| | or $Euler\sqrt{3}\pi^3\hbar^4/\left(4m_T^5\ cavity\ c^2\right)$ | 6.67384E-8 | (present) | $cm^3 sec^{-2} g^{-1}$ |
| Universe gravitational parameter $z$ factors | $(1+z_R)^1$ or $(1+z_B)^3$ | | | *unitless* |
| Universe gravitational parameter | $G_{Ur}$ or $R_{Ur} c^2/M_{Ur}$ | 4.36E+72 | (present) | $cm^3 sec^{-2} g^{-1}$ |
| | or $G\left(M_{Upresent}/M_{Ur}\right)^{1/2}$ | 4.36E+72 | (present) | $cm^3 sec^{-2} g^{-1}$ |
| | or $Euler\sqrt{3}\pi^3\hbar^4/\left(4m_T^5\ cavity_r\ c^2\right)$ | 4.36E+72 | (present) | $cm^3 sec^{-2} g^{-1}$ |
| Universe gravitational parameter $z$ factors | $(1+z_R)^{-1}$ or $(1+z_B)^{-3}$ | | | *unitless* |
| Universe Newtonian gravitation | $G$ or $\left(sherd/sherd_{present}\right)^{1/2} G_U$ | 6.67384E-8 | (constant) | $cm^3 sec^{-2} g^{-1}$ |
| Universe Newtonian gravitation | $G$ or $\left(sherd_r/sherd_{present}\right)^{1/2} G_{Ur}$ | 6.67384E-8 | (constant) | $cm^3 sec^{-2} g^{-1}$ |
| Universe Hubble parameter | $H_U$ or $2^{-1/2}\cos(\theta)^{1/2}\Lambda_U^{1/2}c$ | 2.31E-18 | (present) | $sec^{-1}$ |
| Universe Hubble parameter | or $(B/A)\left(1/\sqrt{3}\right)(v_{dx}/v_c)(1/time)$ | 2.31E-18 | (present) | $sec^{-1}$ |
| | | 71.23 | (present) | *km/(sec Mpc)* |
| Universe Hubble parameter $z$ factors | $(1+z_R)^1$ or $(1+z_B)^3$ | | | *unitless* |
| Universe Hubble parameter | $H_{Ur}$ or $2^{-1/2}\cos(\theta)^{1/2}\Lambda_{Ur}^{1/2}c$ | 1.51E+62 | (present) | $sec^{-1}$ |
| Universe Hubble parameter | or $(B_r/A_r)\left(1/\sqrt{3}\right)(v_{dxr}/v_c)(1/time_{\pm r})$ | 1.51E+62 | (present) | $sec^{-1}$ |
| Universe Hubble parameter $z$ factors | $(1+z_R)^{-1}$ or $(1+z_B)^{-3}$ | | | *unitless* |
| Universe Hubble parameter | $H_{UG}$ or $H_U\left(G/G_U\right)^{1/2}$ | 2.31E-18 | (present) | $sec^{-1}$ |
| Universe Hubble parameter $z$ factors | $(1+z_R)^{1/2}$ or $(1+z_B)^{3/2}$ | | | *unitless* |
| Universe Hubble parameter | $H_{UGr}$ or $H_{Ur}\left(G/G_{Ur}\right)^{1/2}$ | 1.87E+22 | (present) | $sec^{-1}$ |
| Universe Hubble parameter $z$ factors | $(1+z_R)^{-1/2}$ or $(1+z_B)^{-3/2}$ | | | *unitless* |
| Universe inflation number | $\#_U$ or $V_U m_t/(m_e cavity)$ or $M_U/m_e$ | 3.32E+83 | (present) | *unitless* |
| Universe inflation number $z$ factors | $(1+z_R)^{-2}$ or $(1+z_B)^{-6}$ | | | *unitless* |
| Universe inflation number | $\#_{Ur}$ or $V_{Ur} m_t/(m_e cavity_r)$ or $M_{Ur}/m_e$ | 7.77E-77 | (present) | *unitless* |
| Universe inflation number $z$ factors | $(1+z_R)^2$ or $(1+z_B)^6$ | | | *unitless* |
| Universe kinematic viscosity (seen) | $\nu_{Uc}$ or $\mu_{Uc}/\rho_U$ | 1.39E13 | (present) | $cm^2 sec^{-1}$ |
| Universe kinematic viscosity (seen) $z$ factors | $(1+z_R)^{-1/3}$ or $(1+z_B)^{-1}$ | | | *unitless* |
| Universe kinematic viscosity (seen) | $\nu_{Ucr}$ or $\mu_{Ucr}/\rho_{Ur}$ | 4.73E-14 | (present) | $cm^2 sec^{-1}$ |
| Universe kinematic viscosity (seen) $z$ factors | $(1+z_R)^{1/3}$ or $(1+z_B)^1$ | | | *unitless* |
| Universe kinematic viscosity | $\nu_U$ or $\left(\pi A/B\right) x_U^2/t_U$ | 1.39-02 | (constant) | $cm^2 sec^{-1}$ |
| Universe kinematic viscosity | *or* $\mu_U/\rho_U$ | 1.39-02 | (constant) | $cm^2 sec^{-1}$ |
| Universe kinematic viscosity | $\mu_{Ur}/\rho_{Ur}$ | 1.39-02 | (constant) | $cm^2 sec^{-1}$ |
| Universe mass density $\rho_U$ | $m_T/cavity$ or $2H_U^2/\left(8\pi G_U\right)$ | 6.38E-30 | (present) | $g$ $\text{cm}^{-3}$ |
| Universe number density $n$ | $\mathbb{C}/cavity$ | 1.27E-04 | (present) | # $\text{cm}^{-3}$ |
| Universe mass and number density $z$ factors | $\left(1+z_R\right)^1$ or $\left(1+z_B\right)^3$ | | | *unitless* |
| Universe mass density $\rho_{Ur}$ | $m_T/cavity_r$ or $2H_{Ur}^2/\left(8\pi G_{Ur}\right)$ | 4.16E50 | (present) | $g$ $\text{cm}^{-3}$ |
| Universe number density $n_r$ | $\mathbb{C}/cavity_r$ | 8.30E+75 | (present) | # $\text{cm}^{-3}$ |
| Universe mass and number density $z$ factors | $\left(1+z_R\right)^{-1}$ or $\left(1+z_B\right)^{-3}$ | | | *unitless* |
| Universe mass (no radiation) | $M_U$ or $\rho_U V_U$ | 3.02E56 | (present) | $g$ |
| Universe (no radiation) mass $z$ factors | $\left(1+z_R\right)^{-2}$ or $\left(1+z_B\right)^{-6}$ | | | *unitless* |
| Universe (no radiation) mass reflective | $M_{Ur}$ or $\rho_{Ur} V_{Ur}$ | 2.34E-100 | (present) | $g$ |
| Universe (no radiation) mass reflective $z$ factors | $\left(1+z_R\right)^2$ or $\left(1+z_B\right)^6$ | | | *unitless* |
| Universe pressure | $p_U$ or $k_b T_c/cavity$ | 7.11E-36 | (present) | $g\ cm^{-1} sec^{-2}$ |
| Universe pressure $z$ factors | $\left(1+z_R\right)^{5/3}$ or $\left(1+z_B\right)^5$ | | | *unitless* |
| Universe pressure reflective | $p_{Ur}$ or $k_b T_{cr}/cavity_r$ | 7.53E+97 | (present) | $g\ cm^{-1} sec^{-2}$ |
| Universe pressure reflective $z$ factors | $\left(1+z_R\right)^{-5/3}$ or $\left(1+z_B\right)^{-5}$ | | | *unitless* |
| Universe Hubble radius of curvature | $R_U$ or $\sqrt{3}c/H_U$ | 2.25E28 | (present) | $cm$ |
| Universe Hubble radius $z$ factors | $\left(1+z_R\right)^{-1}$ or $\left(1+z_B\right)^{-3}$ | | | *unitless* |
| Universe Hubble reflected radius | $R_{Ur}$ or $\sqrt{3}c/H_{Ur}$ | 3.44E-52 | (present) | $cm$ |
| Universe Hubble reflected radius $z$ factors | $\left(1+z_R\right)^1$ or $\left(1+z_B\right)^3$ | | | *unitless* |
| Universe Hubble surface constant | $M_U/R_U^2$ | 0.599704 | | $g\ cm^{-2}$ |
| Universe Hubble surface constant reflective | $M_{Ur}/R_{Ur}^2$ | 0.599704 | | $g\ cm^{-2}$ |
| Universe Hubble volume | $V_U$ or $\left(4\pi/3\right) R_U^3$ | 4.74E85 | (present) | $cm^3$ |
| Universe Hubble volume $z$ factors | $\left(1+z_R\right)^{-3}$ or $\left(1+z_B\right)^{-9}$ | | | *unitless* |
| Universe Hubble volume reflective | $V_{Ur}$ or $\left(4\pi/3\right) R_{Ur}^3$ | 1.70E-154 | (present) | $cm^3$ |
| Universe Hubble volume reflective $z$ factors | $\left(1+z_R\right)^3$ or $\left(1+z_B\right)^9$ | | | *unitless* |
| Universe Reynolds number | $R_{eU}$ or $\rho_U v_{dx} cavity/\left(\mu_U section\right)$ | 1.00 | (constant) | *unitless* |
| Universe Reynolds number | $R_{eUr}$ or $\rho_{Ur} v_{dxr} cavity_r/\left(\mu_{Ur} section_r\right)$ | 1.00 | (constant) | *unitless* |
| Universe sherd | $\left(M_{Upresent}/m_t\right)/\left(M_U/m_t\right)$ | 1 | (present) | *unitless* |
| Universe sherd | $\left(V_{Upresent}/cavity_{present}\right)/\left(V_U/cavity\right)$ | 1 | (present) | *unitless* |
| Universe sherd $z$ factors | $(1+z_R)^{-2}$ or $(1+z_B)^{-6}$ | | | *unitless* |
| Universe *sherd*$_r$ | $(M_{Upresent}/m_T)/(M_{Ur}/m_T)$ | 4.27E+159 | (present) | *unitless* |
| Universe *sherd*$_r$ | $(V_{Upresent}/cavity_{present})/(V_{Ur}/cavity_r)$ | 4.27E+159 | (present) | *unitless* |
| Universe *sherd*$_r$ $z$ factors | $(1+z_R)^{2}$ or $(1+z_B)^{6}$ | | | *unitless* |
| Universe $z_R$ factor | $R_{Upresent}/R_U = 2.25E28\ cm/R_U$ | | | *unitless* |
| Universe $z_B$ factor | $B_{present}/B = 22.1\ cm/B$ | | | *unitless* |
| Universe Total volume | $V_U\ sherd$ | 4.74E85 | (present) | $cm^3$ |
| Universe Total volume $z$ factors | $(1+z_R)^{-1}$ or $(1+z_B)^{-3}$ | | | *unitless* |
| Universe Total volume | $V_{Ur}\ sherd_r$ | 7.27E+05 | (present) | $cm^3$ |
| Universe Total volume $z$ factors | $(1+z_R)^{1}$ or $(1+z_B)^{3}$ | | | *unitless* |
| Velocity of light | $c$ or $x_U/t_U$ | 2.997924580E10 | | $cm\ sec^{-1}$ |
| Velocity of light | $H_U R_U/\sqrt{3}$ | 2.997924580E10 | | $cm\ sec^{-1}$ |
| Velocity of light | $H_{Ur} R_{Ur}/\sqrt{3}$ | 2.997924580E10 | | $cm\ sec^{-1}$ |
| Velocity of light dielectric $\varepsilon$ modified | $v_\varepsilon\ or\ c/\sqrt{\varepsilon}\ or\ 2v_{dT}\sqrt{k_m}$ | 6.12E+04 | (present) | $cm\ sec^{-1}$ |
| Velocity of light dielectric $\varepsilon$ modified | $v_{\varepsilon x}$ or $c/\sqrt{\varepsilon_x}$ | 6.70E+03 | (present) | $cm\ sec^{-1}$ |
| Velocity of light dielectric $\varepsilon$ modified | $v_{\varepsilon y}$ or $c/\sqrt{\varepsilon_y}$ | 6.70E+03 | (present) | $cm\ sec^{-1}$ |
| Velocity of light dielectric $\varepsilon$ modified | $v_{\varepsilon z}$ or $c/\sqrt{\varepsilon_z}$ | 6.05E+04 | (present) | $cm\ sec^{-1}$ |
| Velocity of light dielectric $\varepsilon$ modified $z$ factors | $(1+z_R)^{1/6}$ or $(1+z_B)^{1/2}$ | | | *unitless* |
| Velocity dielectric $\varepsilon$ Reciprocity | $(v_{\varepsilon x} v_{\varepsilon xr})^{1/2}$ | 3.00E+10 | (constant) | $cm\ sec^{-1}$ |
| Velocity dielectric $\varepsilon$ Reciprocity | $(v_{\varepsilon x} v_{\varepsilon xr})^{1/2}$ | 3.00E+10 | (constant) | $cm\ sec^{-1}$ |
| Velocity dielectric $\varepsilon$ Reciprocity | $(v_{\varepsilon x} v_{\varepsilon xr})^{1/2}$ | 2.71E+11 | (constant) | $cm\ sec^{-1}$ |
| Velocity dielectric $\varepsilon$ Reciprocity | $(v_{\varepsilon x} v_{\varepsilon xr})^{1/2}$ | 2.74E+11 | (constant) | $cm\ sec^{-1}$ |
| Velocity (reflective $v_{\varepsilon x}$) | $v_{\varepsilon xr}$ or $(c/v_{\varepsilon x})c$ or $c/\sqrt{\varepsilon_r}$ | 7.61E+16 | (present) | $cm\ sec^{-1}$ |
| Velocity (reflective $v_{\varepsilon y}$) | $v_{\varepsilon yr}$ or $(c/v_{\varepsilon y})c$ or $c/\sqrt{\varepsilon_{xr}}$ | 6.59E+16 | (present) | $cm\ sec^{-1}$ |
| Velocity (reflective $v_{\varepsilon z}$) | $v_{\varepsilon zr}$ or $(c/v_{\varepsilon z})c$ or $c/\sqrt{\varepsilon_{yr}}$ | 1.60E+16 | (present) | $cm\ sec^{-1}$ |
| Velocity (reflective $v_{\varepsilon}$) | $v_{\varepsilon r}$ or $(c/v_{\varepsilon})c$ or $c/\sqrt{\varepsilon_{zr}}$ | 7.44E+15 | (present) | $cm\ sec^{-1}$ |
| Velocity (reflective $v_{\varepsilon}$) $z$ factors | $(1+z_R)^{-1/6}$ or $(1+z_B)^{-1/2}$ | | | *unitless* |
| Velocity (de Broglie) | $v_{dx}$ or $\hbar K_B/m_T$ or $c/\sqrt{\varepsilon_x k_{mx}}$ | 1.49E-03 | (present) | $cm\ sec^{-1}$ |
| Velocity (de Broglie) | $v_{dy}$ or $\hbar K_P/m_T$ or $c/\sqrt{\varepsilon_y k_{my}}$ | 1.73E-03 | (present) | $cm\ sec^{-1}$ |
| Velocity (de Broglie) | $v_{dz}$ or $\hbar K_A/m_T$ or $c/\sqrt{\varepsilon_z k_{mz}}$ | 7.11E-03 | (present) | $cm\ sec^{-1}$ |
| Velocity (de Broglie) | $v_{dT}$ or $\sqrt{v_{dx}^2+v_{dy}^2+v_{dz}^2}$ or $c/\sqrt{4\varepsilon k_m}$ | 7.47E-03 | (present) | $cm\ sec^{-1}$ |
| Velocity (de Broglie) $z$ factors | $(1+z_R)^{1/3}$ or $(1+z_B)^1$ | | | *unitless* |
| Velocity (de Broglie ) Reciprocity | $(v_x v_{xr})^{1/2}$ | 3.00E+10 | (constant) | $cm\ sec^{-1}$ |
| Velocity (de Broglie ) Reciprocity | $(v_y v_{yr})^{1/2}$ | 3.46E+10 | (constant) | $cm\ sec^{-1}$ |
| Velocity (de Broglie ) Reciprocity | $(v_z v_{zr})^{1/2}$ | 1.43E+11 | (constant) | $cm\ sec^{-1}$ |
| Velocity (de Broglie ) Reciprocity | $(v_{dT} v_{dTr})^{1/2}$ | 1.50E+11 | (constant) | $cm\ sec^{-1}$ |
| Velocity (reflective $v_{dx}$ ) | $v_{dxr}$ or $(c/v_{dx})c$ or $c/\sqrt{\varepsilon_{xr} k_{mxr}}$ | 6.02E+23 | (present) | $cm\ sec^{-1}$ |
| Velocity (reflective $v_{dy}$ ) | $v_{dyr}$ or $(c/v_{dy})c$ or $c/\sqrt{\varepsilon_{yr} k_{myr}}$ | 5.21E+23 | (present) | $cm\ sec^{-1}$ |
| Velocity (reflective $v_{dz}$ ) | $v_{dzr}$ or $(c/v_{dz})c$ or $c/\sqrt{\varepsilon_{zr} k_{mzr}}$ | 1.26E+23 | (present) | $cm\ sec^{-1}$ |
| Velocity (reflective $v_{dT}$ ) | $v_{dTr}$ or $\sqrt{v_{dxr}^2+v_{dyr}^2+v_{dzr}^2}$ or $c/\sqrt{4\varepsilon_r k_{mr}}$ | 8.06E+23 | (present) | $cm\ sec^{-1}$ |
| Velocity (reflective) $z$ factors | $(1+z_R)^{-1/3}$ or $(1+z_B)^{-1}$ | | | *unitless* |
| Zeta function $\zeta(n)$ | $\zeta(n)=\sum_n^\infty 1/n^2$ | | | *unitless* |
| Zeta function $\zeta(3/2)$ | | 2.61237530936309 | | *unitless* |
| Zeta function $\zeta(2)$ | | 1.64493406684823 | $\pi^2/6$ | *unitless* |
| Zeta function $\zeta(5/2)$ | | 1.34148725725091 | | *unitless* |
| Zeta function $\zeta(3)$ | | 1.20205690117833 | | *unitless* |
| Zeta function $\zeta(7/2)$ | | 1.12673386618586 | | *unitless* |
| Zeta function $\zeta(4)$ | | 1.08232323371114 | $\pi^4/90$ | *unitless* |
| Zeta function $\zeta(9/2)$ | | 1.05470751076145 | | *unitless* |
| Zeta function $\zeta(5)$ | | 1.03692775485763 | | *unitless* |

## Appendix A. Trisine Universal Number $m_T$ And *B/A* Ratio Derivation

BCS approach [2]

$$\Delta_o^{BCS} = \frac{\hbar\omega}{\sinh\left(\frac{1}{D(\in_T)V}\right)} \qquad \text{(A.1)}$$

Kittel approach [4]

$$\Delta_o = \frac{2\hbar\omega}{e^{\frac{1}{D(\in_T)V}} - 1} \qquad \text{(A.2)}$$

Let the following relationship where BCS (equation A.1) and Kittel (equation A.2) approach equal each other define a particular superconducting condition.

$$k_b T_c = \begin{cases} \dfrac{e^{Euler}}{\pi} \dfrac{\hbar\omega}{\sinh\left(\frac{1}{D(\in_T)V}\right)} \\ \dfrac{\hbar\omega}{e^{\frac{1}{D(\in_T)V}} - 1} \end{cases} \qquad \text{(A.3)}$$

$$where \quad \frac{e^{Euler}}{\pi} = 1.133865917$$

Define:

$$\frac{1}{D(\in_T)V} = trisine \qquad \text{(A.4)}$$

Then:

$$trisine = -\ln\left(\frac{2e^{Euler}}{\pi} - 1\right) = 2.01091660 \qquad \text{(A.5)}$$

Given:

$$\frac{2m_T k_b T_c}{\hbar^2} = \begin{Bmatrix} \dfrac{(KK)_{\hbar\omega} e^{Euler}}{\pi \sinh(trisine)} \\ \dfrac{(KK)_{\hbar\omega}}{e^{trisine} - 1} \end{Bmatrix} = K_B^2 \qquad \text{(A.6)}$$

Ref: [2,4]

And within Conservation of Energy Constraint ($|n| < \infty$):

$$\left(\frac{1}{g_s}\right)\left(K_B^{2+n} + K_C^{2+n}\right) = \left(K_{Ds}^{2+n} + K_{Dn}^{2+n}\right) \qquad \text{(A.7)}$$

And within Conservation of Momentum Constraint ($|n| < \infty$):

$$(g_s)\left(K_B^{1+n} + K_C^{1+n}\right) = \left(K_{Ds}^{1+n} + K_{Dn}^{1+n}\right) \qquad \text{(A.8)}$$

And $(KK)$ from table A.1:

$$(KK) = K_C^2 + K_{Ds}^2 \qquad \text{(A.9)}$$

Also from table A.1:

$$\frac{2\Delta_o^{BCS}}{k_b T_c} = \frac{2\pi}{e^{Euler}} = \frac{2K_{Ds}}{K_B} \overset{.065\%}{\approx} 3.527754 \qquad \text{(A.10)}$$

Also:

$$\frac{\Delta_o}{k_b T_c} = 2 \qquad \text{(A.11)}$$

Also from table A.1:

$$trisine = \frac{K_B^2 + K_P^2}{K_P K_B} \qquad \text{(A.12)}$$

**Table A.1** Trisine Wave Vector Multiples $(KK)$ vs. $K_B K_B$

| | | $\dfrac{(KK)e^{Euler}}{\pi \sinh(trisine)}$ | |
|---|---|---|---|
| $K$ | $\dfrac{(KK)}{K_B K_B}$ | $\dfrac{(KK)}{K_B K_B}\dfrac{1}{e^{trisine}-1}$ | $\dfrac{K}{K_B}$ |
| $K_{Dn}$ | 1.1981 | 0.1852 | 1.0947 |
| $K_P$ | 1.3333 | 0.2061 | 1.1547 |
| $K_{Ds}$ | 3.1132 | 0.4813 | 1.7646 |
| $K_C$ | 3.3526 | 0.5185 | 1.8316 |
| $K_A$ | 22.6300 | 3.4999 | 4.7587 |
| $(K_C + K_{Ds})/2^{1/2}$ | 6.4687 | 1.0000 | 3.5962 |

But also using the following resonant condition in equation A.13 and by iterating equations A.7, A.8 and A.13.

$$k_b T_c = \begin{Bmatrix} \dfrac{\hbar^2 K_B^2}{2m_T} \\ \dfrac{\hbar^2}{\dfrac{m_e m_T}{m_e + m_T}(2B)^2 + \dfrac{m_T m_p}{m_T + m_p}(A)^2} \end{Bmatrix} \qquad \text{(A.13)}$$

This trisine model converges at:

$$m_T / m_e = 110.122753426 \text{ where } \alpha m_t / m_e \sim \frac{4}{5}$$

$$B/A = 2.379760996 \sim 2\csc\left(\frac{180}{\pi}\right) \sim \frac{\text{neutral pion}\left(\pi^0\right)}{m_t} \qquad \text{(A.14)}$$

## Appendix B. Debye Model Normal And Trisine Reciprocal Lattice Wave Vectors

This appendix parallels the presentation made in ref [4 pages 121 - 122].

$$N = \begin{Bmatrix} \left(\dfrac{L}{2\pi}\right)^3 \dfrac{4\pi}{3} K_{Dn}^3 \\ \left(\dfrac{L}{2\pi}\right)^3 K_{Ds}^3 \end{Bmatrix} = \begin{Bmatrix} \text{Debye Spherical Condition} \\ \text{Debye Trisine Condition} \end{Bmatrix} \qquad \text{(B.1)}$$

where:

$$K_{Ds} = \left(K_A K_B K_P\right)^{\frac{1}{3}} \tag{B.2}$$

$$\frac{dN}{d\omega} = \begin{bmatrix} \dfrac{cavity}{2\pi^2} K_{Dn}^2 \dfrac{dK_{Dn}}{d\omega} \\ \dfrac{cavity}{8\pi^3} 3K_{Ds}^2 \dfrac{dK_{Ds}}{d\omega} \end{bmatrix} = \left\{ \begin{matrix} \text{Debye Spherical Condition} \\ \text{Debye Trisine Condition} \end{matrix} \right\} \tag{B.3}$$

$$K = \frac{\omega}{v} \tag{B.4}$$

$$\frac{dK}{d\omega} = \frac{1}{v} \tag{B.5}$$

$$\frac{dN}{d\omega} = \left\{ \begin{matrix} \dfrac{cavity}{2\pi^2} \dfrac{\omega^2}{v^3} \\ \dfrac{3\cdot cavity}{8\pi^3} \dfrac{\omega^2}{v^3} \end{matrix} \right\} = \left\{ \begin{matrix} \text{Debye spherical} \\ \text{Debye trisine} \end{matrix} \right\} \tag{B.6}$$

$$N = \left\{ \begin{matrix} \dfrac{cavity}{6\pi^2} \dfrac{\omega^3}{v^3} \\ \dfrac{cavity}{8\pi^3} \dfrac{\omega^3}{v^3} \end{matrix} \right\} = \left\{ \begin{matrix} \dfrac{cavity}{6\pi^2} K_{Dn}^3 \\ \dfrac{cavity}{8\pi^3} K_{Ds}^3 \end{matrix} \right\} = \left\{ \begin{matrix} \text{Debye Spherical Condition} \\ \text{Debye Trisine Condition} \end{matrix} \right\} \tag{B.7}$$

Given that $N = 1$ Cell

$$\left\{ \begin{matrix} K_{Dn} \\ K_{Ds} \end{matrix} \right\} = \left\{ \begin{matrix} \left(\dfrac{6\pi^2}{cavity}\right)^{\frac{1}{3}} \\ \left(\dfrac{8\pi^3}{cavity}\right)^{\frac{1}{3}} \end{matrix} \right\} = \left\{ \begin{matrix} \text{Debye Spherical Condition} \\ \text{Debye Trisine Condition} \end{matrix} \right\} \tag{B.8}$$

## Appendix C. One Dimension And Trisine Density Of States

This appendix parallels the presentation made in ref [4 pages 144-155].

Trisine

$$\frac{2K_C^3}{\left(\dfrac{2\pi}{L}\right)^3} = N \tag{C.1}$$

One Dimension

$$\xi = \frac{\hbar^2}{2m_T} N^2 \frac{\pi^2}{\left(2B\right)^2} \tag{C.2}$$

Now do a parallel development of trisine and one dimension density of states.

$$N = \begin{bmatrix} \dfrac{2K_C^3 \cdot cavity}{8\pi^3} \\ \left(\dfrac{2m_T}{\hbar^2}\right)^{\frac{1}{2}} \dfrac{2B}{\pi} \xi^{\frac{1}{2}} \end{bmatrix} = \begin{bmatrix} \text{Trisine Dimensional Condition} \\ \text{One Dimensional Condition} \end{bmatrix} \tag{C.3}$$

$$\frac{dN}{d\xi} = \left\{ \begin{matrix} \dfrac{2\cdot cavity}{8\pi^3} \left(\dfrac{2m_T}{\hbar^2}\right)^{\frac{3}{2}} \dfrac{3}{2} \xi^{\frac{1}{2}} \\ \left(\dfrac{2m_T}{\hbar^2}\right)^{\frac{1}{2}} \dfrac{2B}{\pi} \dfrac{1}{2} \xi^{-\frac{1}{2}} \end{matrix} \right\} = \left\{ \begin{matrix} \text{Trisine Dimensional Condition} \\ \text{One Dimensional Condition} \end{matrix} \right\} \tag{C.4}$$

$$\frac{dN}{d\xi} = \left\{ \begin{matrix} \dfrac{2\cdot cavity}{8\pi^3} \left(\dfrac{2m_T}{\hbar^2}\right)^{\frac{3}{2}} \dfrac{3}{2} \xi^{\frac{1}{2}} \\ \left(\dfrac{2m_T}{\hbar^2}\right)^{\frac{1}{2}} \dfrac{2B}{\pi} \dfrac{1}{2} \xi^{-\frac{1}{2}} \end{matrix} \right\} = \left\{ \begin{matrix} \text{Trisine Dimensional Condition} \\ \text{One Dimensional Condition} \end{matrix} \right\} \tag{C.5}$$

$$\frac{dN}{d\xi} = \left\{ \begin{matrix} 2\dfrac{3}{2} \dfrac{cavity}{8\pi^3} \left(\dfrac{2m_T}{\hbar^2}\right)^{\frac{3}{2}} \left(\dfrac{2m_T}{\hbar^2}\right)^{-\frac{1}{2}} K_C \\ \left(\dfrac{2m_T}{\hbar^2}\right)^{\frac{1}{2}} \dfrac{2B}{\pi} \dfrac{1}{2} \left(\dfrac{2m_T}{\hbar^2}\right)^{\frac{1}{2}} \dfrac{1}{K_B} \end{matrix} \right\} = \left\{ \begin{matrix} \text{Trisine Dimensional Condition} \\ \text{One Dimensional Condition} \end{matrix} \right\} \tag{C.6}$$

$$\frac{dN}{d\xi} = D\left(\xi\right) = \left\{ \begin{matrix} \dfrac{3}{8} \dfrac{cavity}{\pi^3} \left(\dfrac{2m_T}{\hbar^2}\right) K_C \\ \dfrac{2m_T}{\hbar^2 K_B^2} \end{matrix} \right\} = \left\{ \begin{matrix} \text{Trisine Dimensional Condition} \\ \text{One Dimensional Condition} \end{matrix} \right\} \tag{C.7}$$

Now equating trisine and one dimensional density of states results in equation C.8.

$$\begin{aligned} K_C &= \frac{8\pi^3}{3\cdot cavity \cdot K_B^2} \\ &= \frac{4\pi}{3\sqrt{3}A} \end{aligned} \tag{C.8}$$

$\xi$ = energy

$D\left(\xi\right)$ = density of states

## Appendix D. Ginzburg-Landau Equation And Trisine Structure Relationship

This appendix parallels the presentation made in ref [4 Appendix I].

$$\left|\psi\right|^2 = \frac{\alpha}{\beta} = \frac{\mathbb{C}}{cavity} \tag{D.1}$$

$$\frac{\alpha^2}{2\beta} = \frac{k_b T_c}{chain} = \frac{\hbar^2 K_B^2}{2m_T \cdot chain} = \frac{H_c^2}{8\pi} \tag{D.2}$$

$$\lambda = \frac{m_T v_\varepsilon^2}{4\pi\left(\mathbb{C}\cdot e\right)^2 \left|\psi\right|^2} = \frac{m_t v_\varepsilon^2 \beta}{4\pi q^2 \alpha} = \frac{m_T v_\varepsilon^2 cavity}{4\pi\left(\mathbb{C}\cdot e\right)^2 \mathbb{C}} \tag{D.3}$$

$$\beta = \frac{\alpha^2 2m_T chain}{2\hbar^2 K_B^2} = \frac{\alpha^2 chain}{2k_b T_c} \tag{D.4}$$

$$\alpha = \beta \frac{\mathbb{C}}{cavity} = \frac{\alpha^2 2m_T chain}{2\hbar^2 K_B^2} \frac{\mathbb{C}}{cavity} = \frac{\alpha^2 chain}{2k_b T_c} \frac{\mathbb{C}}{cavity} \quad \text{(D.5)}$$

$$\alpha = k_b T_c \frac{cavity}{chain} = \frac{\hbar^2 K_B^2}{2m_T} \frac{cavity}{chain} \quad \text{(D.6)}$$

$$\ell^2 \equiv \frac{\hbar^2}{2m_e \alpha} = \frac{\hbar^2}{2m_e} \frac{2m_T \cdot chain}{\hbar^2 K_B^2 \cdot cavity} = \frac{m_T}{m_e} \frac{1}{K_B^2} \frac{chain}{cavity} \quad \text{(D.7)}$$

$$H_{c2} = \frac{\phi_\varepsilon \cdot 2 \cdot 4\pi}{section} = \frac{\phi_\varepsilon}{\pi \ell^2} \frac{m_T}{m_e} \frac{cavity}{chain} \quad \text{(D.8)}$$

## Appendix E. Heisenberg Uncertainty within the Context of De Broglie Condition

Given Heisenberg uncertainty:

$$\Delta p_x \Delta x \geq \frac{h}{4\pi} \quad \text{(E.1)}$$

And the assumption that mass 'm' is constant then:

$$m \Delta v \Delta x \geq \frac{h}{4\pi} \quad \text{(E.2)}$$

Given de Broglie Condition:

$$p_x \; x = h \quad \text{(E.3)}$$

And again assuming constant mass 'm':

$$mvx = h \quad \left(m_t v_{dx} 2B = h\right) \quad \text{(E.4)}$$

Rearranging Heisenberg Uncertainty and de Broglie Condition:

$$m \geq \frac{h}{4\pi} \frac{1}{\Delta v \Delta x} \quad \text{and} \quad m = h \frac{1}{vx} \quad \text{(E.5)}$$

This implies that:

$$m \geq m \quad \text{(E.6)}$$

but a constant mass cannot be greater than itself, therefore:

$$m = m \quad \text{(E.7)}$$

therefore:

$$h \frac{1}{vx} = \frac{h}{4\pi} \frac{1}{\Delta v \Delta x} \quad \text{(E.8)}$$

And:

$$vx = 4\pi \Delta v \Delta x \quad \text{(E.9)}$$

Both the Heisenberg Uncertainty and the de Broglie condition are satisfied. Uncertainty is dimensionally within the de Broglie condition. Schrödinger in the field of quantum mechanics further quantified this idea.

## Appendix F. Equivalence Principle In The Context Of Work Energy Theorem

Start with Newton's second law

$$F = \frac{dp}{dt} \quad \text{(F.1)}$$

and for constant mass

$$F = \frac{dp}{dt} = \frac{d(mv)}{dt} = m \frac{dv}{dt} \quad \text{(F.2)}$$

then given the 'work energy theorem'

$$F = \frac{dp}{dt} = \frac{d(mv)}{dt} = m \frac{dv}{dt} \frac{ds}{ds} = mv \frac{dv}{ds}$$

$$\text{where } \frac{ds}{dt} = v \quad \text{and therefore: } Fds = mvdv \quad \text{(F.3)}$$

where $ds$ is test particles incremental distance along path '$s$'

Now given Newton's second law in relativistic terms:

$$F = \frac{dp}{dt} = \frac{d}{dt} \frac{mv}{\sqrt{1 - \frac{v^2}{c^2}}} \quad \text{(F.4)}$$

let Lorentz transform notation as '$b$' :

$$b = \frac{1}{\sqrt{1 - \frac{v^2}{c^2}}} \quad \text{(F.5)}$$

therefore:

$$F = \frac{dp}{dt} = \frac{d\left(mvb\right)}{dt} \quad \text{(F.6)}$$

Now transform into a work energy relationship with a $K$ factor of dimensional units 'length/mass'

$$a\left(\frac{1}{K}\right) ds = mbvdv \quad \text{(F.7)}$$

acceleration '$a$' can vary and the path '$s$' is arbitrary within the dimensional constraints of the equation.
Define acceleration '$a$' in terms of change in Volume '$V$' as:

$$\frac{d^2V}{dt^2} \frac{1}{V} = \text{constant} \quad \text{(F.8)}$$

and incremental path 'ds' as:

$$ds = 2(area)dr = 2(4\pi r^2)dr \quad \text{(F.9)}$$

'*area*' is the surface of the spherical volume '*V*'
Now work-energy relationship becomes:

$$\frac{d^2V}{dt^2} \frac{1}{V} \frac{1}{K} 2(4\pi r^2)dr = mbvdv \quad \text{(F.10)}$$

Now integrate both sides:

$$\frac{d^2V}{dt^2} \frac{1}{V} \frac{1}{K} 2(\frac{4}{3}\pi r^3) = -\left(mc^2 - mv^2\right)b \quad \text{(F.11)}$$

for $v << c$ then $b \sim 1$

and $V = \frac{4}{3}\pi r^3$

then:

$$-\frac{d^2V}{dt^2}\frac{1}{V}\frac{2}{K} = \left(\frac{mc^2}{V} - \frac{mv^2}{V}\right) \quad \text{(F.12)}$$

and:

$$-\frac{d^2V}{dt^2}\frac{1}{V} = \frac{K}{2}\left(\frac{mc^2}{V} - \frac{mv^2}{V}\right) \quad \text{(F.13)}$$

and

$$-\frac{d^2V}{dt^2}\frac{1}{V} = \frac{K}{2}\left(\frac{mc^2}{V} - \frac{mv}{time}\left(\frac{1}{area_x} + \frac{1}{area_y} + \frac{1}{area_z}\right)\right) \quad \text{(F.14)}$$

This is the same as the Baez-Bunn narrative [144] interpretation except for the negative signs on momentum. Since momentum *'mv'* is a vector and energy 'mv$^2$ ' is a scalar perhaps the relationship should be expressed as:

$$-\frac{d^2V}{dt^2}\frac{1}{V} = \frac{K}{2}\begin{pmatrix} \frac{mc^2}{V} \pm \frac{mv}{area_x time} \\ \pm \frac{mv}{area_y time} \\ \pm \frac{mv}{area_z time} \end{pmatrix} \quad \text{(F.15)}$$

or perhaps in terms of spherical surface area '$area_r$'

$$-\frac{d^2V}{dt^2}\frac{1}{V} = \frac{K}{2}\left(\frac{mc^2}{V} \pm \frac{mv}{area_r time}\right) \quad \text{(F.16)}$$

Lets look again at the equation F.7 and define another path *'s'*:

$$ds = 16\pi r dr \qquad a = \frac{d^2r}{dt^2} = \text{constant} \quad \text{(F.17)}$$

then

$$a\left(\frac{2}{K}\right)8\pi r dr = mbvdv \quad \text{(F.18)}$$

Then integrate:

$$a\left(\frac{2}{K}\right)4\pi r^2 = -\left(mc^2 - mv^2\right)b \quad \text{(F.19)}$$

Given that:

$$K = \frac{8\pi G}{c^2} \quad \text{(F.20)}$$

and solve equation F.19 for *'a'*

$$a = -\frac{Gm}{r^2}\ \frac{1}{\mathrm{b}} \quad \text{(F.21)}$$

or

$$a = -\frac{Gm}{r^2}\sqrt{1-\frac{\mathrm{v}^2}{\mathrm{c}^2}} \quad \text{(F.22)}$$

Does this equation correctly reflect the resonant gravitational acceleration such as a satellite orbiting the earth??

Of course this equation reduces to the standard Newtonian gravitational acceleration at 'v << c'

$$a = -\frac{Gm}{r^2} \quad \text{(F.23)}$$

The equation F.7 would appear to be general in nature and be an embodiment of the equivalence principle. Various candidate accelerations *'a'* and geometric paths *'ds'* could be analyzed within dimensional constraints of this equation.

A simple case defined by a path *'ds/$C_t$'* may be appropriate for the modeling the Pioneer deceleration anomaly.

$$Ma\frac{1}{C_t}ds = F\frac{1}{C_t}ds = mbvdv \quad \text{(F.24)}$$

Where a 'thrust' constant ($C_t$) is associated with object of mass (M) moving on curvature *'ds'*.

### Appendix G. Koide Constant for Leptons - tau, muon and electron

Koide Constant for Lepton masses follows naturally from nuclear condition as defined in paragraph 1.2.

$$\begin{aligned} \mathrm{m}_\tau &= \frac{2}{3}\hbar c\left(K_{Ds} + K_{Dn}\right)\frac{1}{c^2} \\ &= \frac{2}{3}\hbar c\left(K_B + K_C\right)\frac{g_s}{c^2} \\ &= 1.777\ GeV/c^2 \quad (1.77684(17)) \\ \mathrm{m}_\mu &= \frac{1}{3}\frac{m_e}{m_T}\frac{\hbar^2}{2m_T}\left(K_{Ds}^2 + K_{Dn}^2\right)\frac{g_s}{c^2} \\ &= \frac{1}{3}\frac{m_e}{m_T}\frac{\hbar^2}{2m_T}\left(K_B^2 + K_C^2\right)\frac{1}{c^2} \\ &= .1018\ GeV/c^2 \quad (.1056583668(38)) \end{aligned} \quad \text{(G1)}$$

$$\mathrm{m}_e = \frac{B}{A} cos\left(\theta\right) \frac{m_e}{m_T}\left(g_s - 1\right)\hbar c\left(K_{Ds} + K_{Dn}\right)\frac{1}{c^2}$$

$$= \frac{B}{A} cos\left(\theta\right) \frac{m_e}{m_T}\left(g_s - 1\right)\hbar c\left(K_B + K_C\right)\frac{g_s}{c^2}$$

$$= .0005129\ GeV / c^2 \quad \left(.000510998910(13)\right)$$

$$\frac{\mathrm{m}_\tau + \mathrm{m}_\mu + \mathrm{m}_e}{\left(\sqrt{\mathrm{m}_\tau} + \sqrt{\mathrm{m}_\mu} + \sqrt{\mathrm{m}_e}\right)^2} = \frac{chain}{cavity} = \frac{2}{3}$$

## Appendix H. Active Galactic Nuclei (AGN) and Variable Star RRc Tidal Dimension (T) in Oscillating Dark Energy Matter Medium (U)

$$\frac{GM_{AGN}M_{test}}{R_{AGN}} - \frac{GM_{AGN}M_{test}}{R_{AGN} + T} = \frac{GM_U M_{test}}{R_U} - \frac{GM_U M_{test}}{R_U - B} \quad \text{(H.1)}$$

$$\frac{M_{AGN}}{R_{AGN}} - \frac{M_{AGN}}{R_{AGN} + T} = \frac{M_U}{R_U} - \frac{M_U}{R_U - B} \quad \text{(H.2)}$$

$$M_{AGN}T \frac{1}{R_{AGN}\left(R_{AGN} + T\right)} = M_U B \frac{1}{R_U\left(R_U + B\right)} \quad \text{(H.3)}$$

$$\frac{M_{AGN}}{R_{AGN}^2} T = \frac{M_U}{R_U^2} B \approx 0.6\,B \qquad \text{eqn } 2.11.34 \quad \text{(H.4)}$$

$$T \approx 0.6\ \frac{R_{AGN}^2}{M_{AGN}} B \approx 0.6\ \frac{R_{AGN}^3}{M_{AGN}} \frac{B}{R_{AGN}} \approx 0.6\ \frac{1}{\rho_{AGN}} \frac{B}{R_{AGN}} \quad \text{(H.5)}$$

Active Galactic Nuclei (AGN) or any other interstellar gaseous medium (including planetary atmosphere) Oscillates with Dark Energy Medium (U).

## Appendix I. Dimensional Analysis of $\hbar$, $G$, $c$, $H$

These calculations are presented in context of historical appreciation of the Planck Units. Mead[141] deals with the mass(m), length(l), time(t) dimensionality associated with universal constants (*h,G,c*) known to Max Planck in 1899. The Hubble parameter(*H*) was conceptually unknown at that time. These relationships are independent of the particular (*m, l, t*) units used (in this case(cgs) gram, centimeter, second)

As a later extension of this concept, George Lemaitre conceptually interpreted Albert Einstein's field equations in terms of universe expansion (1927)[142] which was verified by Edwin Hubble (1929).

The observed Hubble expansion value *H* has changed over the years and has narrowed over the last decade with WMAP and Planck spacecrafts and galactic formation changes with redshift (*z*) particularly with the newly commissioned dark matter survey in Chile.

In this context, there is justification to add the Hubble constant '*H*' to the physics dimensional analysis conversation.

$$\hbar,\ G,\ c,\ H$$

The dimensional analysis concept provides a mechanism for arrangement of these constants in order to proved investigative leads for continued experimental work.

Let *H* be a fundamental constant representative of an ubiquitous expanding universe (including a cosmological parameter) on equal par with *G*, *c* and $\hbar$ In terms mass(*m*), length(*l*), and time(*t*), these constants are represented by their dimensional and observed numerical values:

| | | | |
|---|---|---|---|
| $\hbar\ m^1 l^2 t^{-1}$ | = | $1.05E-27$ | $g^1 cm^2 sec^{-1}$ |
| $G\ m^{-1} l^3 t^{-2}$ | = | $6.67E-8$ | $g^{-1} cm^3 sec^{-2}$ |
| $c\ m^0 l^1 t^{-1}$ | = | $3.00E10$ | $g^0 cm^1 sec^{-1}$ |
| $H\ m^0 l^0 t^{-1}$ | = | $2.31E-18$ | $g^0 cm^0 sec^{-1}$ |

Below, a (*m, l, t*) dimensional analysis of these relationships of these constants is executed, one of these relationships being the Planck units.

Within this context, Planck units do not warrant special consideration above others.

Planck units are an expression of the limited information of Max Planck's era (circa 1900). Based on this (*m, l, t*) dimensional analysis (below dotted line), the following $H/G$ (*m, l, t*) dimensional relationships appear to be best, because they conform to physical nuclear measurements and could be considered as a base to investigate thermodynamics and dissipative processes.

| | | | | |
|---|---|---|---|---|
| $H/G$ | = | $m^1 l^{-3} t^1$ | 3.46E-11 | $g\ cm^{-3} sec$ |
| $H/G\ m$ | = | $\hbar^{2/3}\left(\frac{H}{G}\right)^{1/3} c^{-1/3}$ | 1.09E-25 | $g$ |
| $H/G\ l$ | = | $\hbar^{1/3}\left(\frac{H}{G}\right)^{-1/3} c^{-2/3}$ | 3.23E-13 | $cm$ |
| $H/G\ t$ | = | $\hbar^{1/3}\left(\frac{H}{G}\right)^{-1/3} c^{-5/3}$ | 1.08E-23 | $sec$ |

Further, there is a subsequent relationship to charge $\left(e\right)$ and the fine structure constant $\left(\alpha\right)$

$$\left(H/G\ m\right)^{1/2}\left(H/G\ l\right)^{3/2}\left(H/G\ t\right)^{-1} e^{-1} = \alpha^{-1/2}$$

and

$$\left(H/G\ l\right)\left(H/G\ t\right)^{-1} = speed\ of\ light(c)$$

This same condition exits for Planck units.

$$\left(Planck\ m\right)^{1/2}\left(Planck\ l\right)^{3/2}\left(Planck\ t\right)^{-1} e^{-1} = \alpha^{-1/2}$$

and

$$\left(Planck\ l\right)\left(Planck\ t\right)^{-1} = speed\ of\ light(c)$$

The nuclear $\left(H/G\ m,\ H/G\ l,\ H/G\ t\right)$ dimensional result (1.09E-25 g, 3.23E-13 cm, 1.08E-23 sec) (generally consistent with trisine condition 1.2) based in part on a cosmological $H$ appears to be incongruous, but the $\left(H/G\ m,\ H/G\ l,\ H/G\ t\right)$ dimensional analysis assumes a *H/G* constant ratio which may reflect much higher respective values of $H$ and $G$ at the Big Bang when nuclear particles were synthesized. $\dot{G}/G$ studies are projected to resolve this question.

The Planck unit $\left(Planck\ m,\ Planck\ l,\ Planck\ t\right)$ dimensional result (2.18E-5 g, 1.61E-33 cm, 5.39E-44 sec) does not share a similar physical dimensional perspective except perhaps at the very beginning of the universe (trisine condition 1.1).

Linear extrapolation of currently available $\dot{G}/G$ values indicate that $G$ has changed less that 1 percent in universe 13.7 billion year age. But the $G$ relationship with Universe Age may not be linear but *~1/Uage*. Gravity($G$) studies of early universe are required.

These universal values with appropriate $2\pi$ phase factor allow basis for trisine lattice scaling as follows:

$$2\pi\ m_T c\ H/G\ l = m_T v_{dx} 2B = m_T \left(2B\right)^2 / time = h$$

Dimensional analysis indicates the total set combinations worthy of investigation:

$$\begin{array}{lcll}
\hbar^{a_{11}} G^{a_{21}} c^{a_{31}} H^{a_{41}} & = & m^1 l^0 t^0 & m \\
\left(m^1 l^2 t^{-1}\right)^{a_{11}} \left(m^{-1} l^3 t^{-2}\right)^{a_{21}} \left(m^0 l^1 t^{-1}\right)^{a_{31}} \left(m^0 l^0 t^{-1}\right)^{a_{41}} & = & m^1 l^0 t^0 & m \\
\hbar^{a_{12}} G^{a_{22}} c^{a_{32}} H^{a_{42}} & = & m^0 l^1 t^0 & l \\
\left(m^1 l^2 t^{-1}\right)^{a_{12}} \left(m^{-1} l^3 t^{-2}\right)^{a_{22}} \left(m^0 l^1 t^{-1}\right)^{a_{32}} \left(m^0 l^0 t^{-1}\right)^{a_{42}} & = & m^0 l^1 t^0 & l \\
\hbar^{a_{13}} G^{a_{23}} c^{a_{33}} H^{a_{43}} & = & m^0 l^0 t^1 & t \\
\left(m^1 l^2 t^{-1}\right)^{a_{13}} \left(m^{-1} l^3 t^{-2}\right)^{a_{23}} \left(m^0 l^1 t^{-1}\right)^{a_{33}} \left(m^0 l^0 t^{-1}\right)^{a_{43}} & = & m^0 l^0 t^1 & t
\end{array}$$

with dimensionless constants

$$\begin{array}{cccc}
a_{11} & a_{21} & a_{31} & a_{41} \\
a_{12} & a_{22} & a_{32} & a_{42} \\
a_{13} & a_{23} & a_{33} & a_{43}
\end{array}$$

required to maintain expression mass($m$), length($l$), and time($t$) dimensionality.

Solve these three equation's dimensionless constants for mass($m$) necessarily eliminating one column parameter.

$$\begin{array}{rrlrrlrrlrrlcr}
+ & 1 & a_{11} & - & 1 & a_{21} & + & 0 & a_{31} & + & 0 & a_{41} & = & 1 \\
+ & 2 & a_{11} & + & 3 & a_{21} & + & 1 & a_{31} & + & 0 & a_{41} & = & 0 \\
- & 1 & a_{11} & - & 2 & a_{21} & - & 1 & a_{31} & - & 1 & a_{41} & = & 0
\end{array}$$

Solve these three equation's dimensionless constants for length($l$) necessarily eliminating the same one column parameter from above.

$$\begin{array}{rrlrrlrrlrrlcr}
+ & 1 & a_{12} & - & 1 & a_{22} & + & 0 & a_{32} & + & 0 & a_{42} & = & 0 \\
+ & 2 & a_{12} & + & 3 & a_{22} & + & 1 & a_{32} & + & 0 & a_{42} & = & 1 \\
- & 1 & a_{12} & - & 2 & a_{22} & - & 1 & a_{32} & - & 1 & a_{42} & = & 0
\end{array}$$

Solve these three equation's dimensionless constants for time($t$) necessarily eliminating the same one column parameter from above.

$$\begin{array}{rrlrrlrrlrrlcr}
+ & 1 & a_{13} & - & 1 & a_{23} & + & 0 & a_{33} & + & 0 & a_{43} & = & 0 \\
+ & 2 & a_{13} & + & 3 & a_{23} & + & 1 & a_{33} & + & 0 & a_{43} & = & 0 \\
- & 1 & a_{13} & - & 2 & a_{23} & - & 1 & a_{33} & - & 1 & a_{43} & = & 1
\end{array}$$

Selected four(4) groups three(3) need be selected for an exact solution in accordance with combination theory ${}_4C_3 = 4$.

$$\begin{array}{cccc}
0 & a_{21} & a_{31} & a_{41} \\
0 & a_{22} & a_{32} & a_{42} \\
0 & a_{23} & a_{33} & a_{43}
\end{array} \quad \text{Group One(1)}$$

$$\begin{array}{cccc}
a_{11} & 0 & a_{31} & a_{41} \\
a_{12} & 0 & a_{32} & a_{42} \\
a_{13} & 0 & a_{33} & a_{43}
\end{array} \quad \text{Group Two(2)}$$

$$\begin{array}{cccc}
a_{11} & a_{21} & 0 & a_{41} \\
a_{12} & a_{22} & 0 & a_{42} \\
a_{13} & a_{23} & 0 & a_{43}
\end{array} \quad \text{Group Three(3)}$$

$$\begin{array}{cccc}
a_{11} & a_{21} & a_{31} & 0 \\
a_{12} & a_{22} & a_{32} & 0
\end{array} \quad \text{Group Four(4)}$$

$a_{13}$ $a_{23}$ $a_{33}$ 0

group one(1) corresponds to

$$\hbar^0 G^{a_{21}} c^{a_{31}} H^{a_{41}} = m^1 l^0 t^0 = m$$
$$\hbar^0 G^{a_{22}} c^{a_{32}} H^{a_{42}} = m^0 l^1 t^0 = l$$
$$\hbar^0 G^{a_{23}} c^{a_{33}} H^{a_{43}} = m^0 l^0 t^1 = t$$

where:

$$m = \hbar^0 G^{-1} c^3 H^{-1} = 1.75\text{E+}56 \quad g$$
$$l = \hbar^0 G^0 c^1 H^{-1} = 1.30\text{E+}28 \quad cm$$
$$t = \hbar^0 G^0 c^0 H^{-1} = 4.32\text{E+}17 \quad sec$$

trial $l/t = c$ 3.00E+10 $cm\ sec^{-1}$

trial $e = m^{1/2} l^{3/2} t^{-1}$ = 4.51E+52 $g^{1/2} cm^{3/2} sec^{-1}$

group two(2) corresponds to

$$\hbar^{a_{11}} G^0 c^{a_{31}} H^{a_{41}} = m^1 l^0 t^0 = m$$
$$\hbar^{a_{12}} G^0 c^{a_{32}} H^{a_{42}} = m^0 l^1 t^0 = l$$
$$\hbar^{a_{13}} G^0 c^{a_{33}} H^{a_{43}} = m^0 l^0 t^1 = t$$

where:

$$m = \hbar^1 G^0 c^{-2} H^1 = 2.71\text{E-}66 \quad g$$
$$l = \hbar^{-5/2} G^0 c^{1/2} H^{1/2} = 7.29\text{E+}63 \quad cm$$
$$t = \hbar^0 G^0 c^0 H^{-1} = 4.32\text{E+}17 \quad sec$$

trial $l/t$ = 1.69E+46 $cm\ sec^{-1}$

trial $e = m^{1/2} l^{3/2} t^{-1}$ = 2.37E+45 $g^{1/2} cm^{3/2} sec^{-1}$

group three(3) corresponds to

$$\hbar^{a_{11}} G^{a_{21}} c^0 H^{a_{41}} = m^1 l^0 t^0 = m$$
$$\hbar^{a_{12}} G^{a_{22}} c^0 H^{a_{42}} = m^0 l^1 t^0 = l$$
$$\hbar^{a_{13}} G^{a_{23}} c^0 H^{a_{43}} = m^0 l^0 t^1 = t$$

where:

$$m = \hbar^{3/5} G^{-2/5} c^0 H^{1/5} = 1.44\text{E-}17 \quad g$$
$$l = \hbar^{1/5} G^{1/5} c^0 H^{-3/5} = 5.64\text{E+}03 \quad cm$$
$$t = \hbar^0 G^0 c^0 H^{-1} = 4.32\text{E+}17 \quad sec$$

trial $l/t$ = 1.30E-14 $cm\ sec^{-1}$

trial $e = m^{1/2} l^{3/2} t^{-1}$ = 3.71E-21 $g^{1/2} cm^{3/2} sec^{-1}$

group four(4) corresponds to Planck units

$$\hbar^{a_{11}} G^{a_{21}} c^{a_{31}} H^0 = m^1 l^0 t^0 = m$$
$$\hbar^{a_{12}} G^{a_{22}} c^{a_{32}} H^0 = m^0 l^1 t^0 = l$$
$$\hbar^{a_{13}} G^{a_{23}} c^{a_{33}} H^0 = m^0 l^0 t^1 = t$$

where:

$$m = \hbar^{1/2} G^{-1/2} c^{1/2} H^0 = 2.18\text{E-}5 \quad g$$
$$l = \hbar^{1/2} G^{1/2} c^{-3/2} H^0 = 1.61\text{E-}33 \quad cm$$
$$t = \hbar^{1/2} G^{1/2} c^{-5/2} H^0 = 5.39\text{E-}44 \quad sec$$

trial $l/t$ = 3.00E+10 $cm\ sec^{-1}$

trial $e = m^{1/2} l^{3/2} t^{-1}$ = 5.62E-09 $g^{1/2} cm^{3/2} sec^{-1}$

Alpha($\alpha$) modification to achieve charge(e)

$$e = m^{1/2} l^{3/2} t^{-1} \alpha^{1/2} = 4.80\text{E-}10 \quad g^{1/2} cm^{3/2} sec^{-1}$$

The other exact solution alternative is to combine two constants so that all four $\hbar$, *G*, *c*, *H* can be arranged in groups of three including $\hbar/H$, $c/H$, $\hbar/G$, $c/G$, $c/\hbar$ and $H/G$ and procedurally solved as the other groups above.

group $\hbar/H$

$$\hbar/H = m^1 l^2 t^0 = 4.56\text{E-}10 \quad g\ cm^2$$
$$\hbar/H\ m = G^{-2/3} (\hbar/H)^{1/3} c^{4/3} = 4.36\text{E+}15 \quad g$$
$$\hbar/H\ l = G^{1/3} (\hbar/H)^{1/3} c^{-2/3} = 3.24\text{E-}13 \quad cm$$
$$\hbar/H\ t = G^{1/3} (\hbar/H)^{1/3} c^{-5/3} = 1.08\text{E-}23 \quad sec$$
$$\hbar/H\ l/t = c = 3.00\text{E+}10 \quad cm\ sec^{-1}$$
$$\hbar/H\ e = m^{1/2} l^{3/2} t^{-1} = 1.13\text{E+}12 \quad g^{1/2} cm^{3/2} sec^{-1}$$

group $c/H$

$$c/H = m^0 l^1 t^0 = 1.30\text{E+}28 \quad cm$$
$$c/H\ m = G^{-1/3} \hbar^{2/3} (c/H)^{-1/3} = 1.09\text{E-}25 \quad g$$
$$c/H\ l = G^0 \hbar^0 (c/H)^1 = 1.30\text{E+}28 \quad cm$$
$$c/H\ t = G^{-1/3} \hbar^{-1/3} (c/H)^{5/3} = 1.73\text{E+}58 \quad sec$$
$$c/H\ l/t = \ = 7.48\text{E-}31 \quad cm\ sec^{-1}$$
$$c/H\ e = m^{1/2} l^{3/2} t^{-1} = 2.81\text{E-}29 \quad g^{1/2} cm^{3/2} sec^{-1}$$

group $\hbar/G$

$$\hbar/G = m^2 l^1 t^1 = 1.58\text{E-}20 \quad g\ cm\ sec$$
$$\hbar/G\ m = H^0 (\hbar/G)^{1/2} c^{1/2} = 2.18\text{E-}05 \quad g$$
$$\hbar/G\ l = H^{-1} (\hbar/G)^0 c^1 = 1.30\text{E+}28 \quad cm$$
$$\hbar/G\ t = H^{-1} (\hbar/G)^0 c^0 = 4.32\text{E+}17 \quad sec$$
$$\hbar/G\ l/t = c \quad 3.00\text{E+}10 \quad cm\ sec^{-1}$$
$$\hbar/G\ e = m^{1/2} l^{3/2} t^{-1} = 1.59\text{E+}22 \quad g^{1/2} cm^{3/2} sec^{-1}$$

group $c/G$

$$c/G = m^1 l^{-2} t^1 = 4.49\text{E+}17 \quad g\ cm^{-2} sec$$
$$c/G\ m = H^0 \hbar^{1/2} (c/G)^{1/2} = 2.18\text{E-}05 \quad g$$
$$c/G\ l = H^{-1/2} \hbar^{1/4} (c/G)^{-1/4} = 4.58\text{E-}03 \quad cm$$
$$c/G\ t = H^{-1} \hbar^0 (c/G)^0 = 4.32\text{E+}17 \quad sec$$

$$c/G\ l/t = \quad 1.06\text{E-}20 \quad cm\ sec^{-1}$$
$$c/G\ e = m^{1/2}l^{3/2}t^{-1} = 3.34\text{E-}24 \quad g^{1/2}cm^{3/2}sec^{-1}$$

group $c/\hbar$

$$c/\hbar = m^{-1}l^{-1}t^{0} = 2.84\text{E+}37 \quad g^{-1}\ cm^{-1}$$
$$c/\hbar\ m = H^{1/2}G^{-1/4}\left(c/\hbar\right)^{-3/4} = 7.68\text{E-}36 \quad g$$
$$c/\hbar\ l = H^{-1/2}G^{1/4}\left(c/\hbar\right)^{-1/4} = 4.58\text{E-}03 \quad cm$$
$$c/\hbar\ t = H^{-1}G^{0}\left(c/\hbar\right)^{0} = 4.32\text{E+}17 \quad sec$$
$$c/\hbar\ l/t = \quad 1.06\text{E-}20 \quad cm\ sec^{-1}$$
$$c/\hbar\ e = m^{1/2}l^{3/2}t^{-1} = 1.99\text{E-}39 \quad g^{1/2}cm^{3/2}sec^{-1}$$

group $H/G$

$$H/G = m^{1}l^{-3}t^{1} = 3.46\text{E-}11 \quad g\ cm^{-3}sec$$
$$H/G\ m = \hbar^{2/3}\left(H/G\right)^{1/3}c^{-1/3} = 1.09\text{E-}25 \quad g$$
$$H/G\ l = \hbar^{1/3}\left(H/G\right)^{-1/3}c^{-2/3} = 3.23\text{E-}13 \quad cm$$
$$H/G\ t = \hbar^{1/3}\left(H/G\right)^{-1/3}c^{-5/3} = 1.08\text{E-}23 \quad sec$$
$$H/G\ l/t = c \quad 3.00\text{E+}10 \quad cm\ sec^{-1}$$
$$H/G\ e = m^{1/2}l^{3/2}t^{-1} = 5.62\text{E-}09 \quad g^{1/2}cm^{3/2}sec^{-1}$$

Alpha($\alpha$) modification to achieve charge(e)

$$H/G\ e \quad m^{1/2}l^{3/2}t^{-1}\alpha^{1/2} = 4.80\text{E-}10 \quad g^{1/2}cm^{3/2}sec^{-1}$$

Mead's[141] Heisenberg uncertainty logic can be applied to (group *H/G*) mass(m), length(l) and time(t) which are more intuitively physical than the Planck units (group 4) and both conform to charge definition(e) and the speed of light(c) while all of the other groups do not.

(what Planck length is physically significant at 1.61E-33 cm?)

(what Planck mass is physically significant at 2.18E-5 g?)

(what Planck time is physically significant at 5.39-44 sec?)

It is necessary to include the $cos(\theta)$ *or* $(\pi/4)^{1/3}$ factor in group $H/G$ to conform with the rest of this report while maintaining speed of light charge relationships and also conformance with universe density $\rho_U$ and expansion $H_U$.

$$H/G = \frac{8\pi}{2}\frac{\rho_U}{H_U} = U$$

group $H/G$ mod

$$H/G = m^{1}l^{-3}t^{1}$$
$$= 3.46\text{E-}11 \quad g\ cm^{-3}sec$$
$$H/G \text{ mod } m = \hbar^{2/3}\left(H/G\right)^{1/3}c^{-1/3}cos\left(\theta\right)$$
$$= 1.00\text{E-}25 \quad g$$
$$H/G \text{ mod } l = \hbar^{1/3}\left(H/G\right)^{-1/3}c^{-2/3}/cos\left(\theta\right)$$
$$= 3.51E-13 \quad cm$$
$$H/G \text{ mod } t = \hbar^{1/3}\left(H/G\right)^{-1/3}c^{-5/3}cos\left(\theta\right)$$
$$= 1.17E-23 \quad sec$$
$$H/G \text{ mod } l/t = c$$
$$= 3.00\text{E+}10 \quad cm\ sec^{-1}$$
$$H/G \text{ mod } e = m^{1/2}l^{3/2}t^{-1}$$
$$= 5.62\text{E-}09 \quad g^{1/2}cm^{3/2}sec^{-1}$$

Alpha($\alpha$) modification to achieve charge(e)

$$H/G \text{ mod } e \quad m^{1/2}l^{3/2}t^{-1}\alpha^{1/2} = 4.80\text{E-}10 \quad g^{1/2}cm^{3/2}sec^{-1}$$

The $H/G$ mod mass(m) x $cos(\theta)$ = 1.00E-25 g conforms to a particle size between the electron and proton that has an alternate solution for $H/G$ mod mass(m) x $cos(\theta)$ and identity with $m_t$ in the following equation

$$energy = k_b T_c = \frac{\hbar^2 K_B^2}{2m_T} = \frac{\hbar^2}{\frac{m_e m_T}{m_e + m_T}\left(2B\right)^2 + \frac{m_T m_p}{m_T + m_p}\left(A\right)^2}$$

where

$$\theta = \tan^{-1}\left(A/B\right)_T$$

The $H/G$ mod mass(m) x $cos(\theta)$

= trisine mass $\left(m_T\right)$ 1.00E-25 g

as per equations 2.1.53 and 2.1.53r:

$$m_T = \frac{3}{4}\frac{m_e}{g_s}\left(\frac{cavity}{\left(\Delta x\ \Delta y\ \Delta z\right)} - \frac{m_p}{m_e}\right) \qquad (2.1.53)$$

$$m_T = \frac{3}{4}\frac{m_e}{g_s}\left(\frac{cavity_r}{\left(\Delta x_r\ \Delta y_r\ \Delta z_r\right)} - \frac{m_p}{m_e}\right) \qquad (2.1.53r)$$

The $H/G$ mod length(l) $/cos(\theta)$

= trisine length $\left(l_T\right)$ 3.51E-13 cm

conforms to nuclear size and Compton wavelength $\hbar/\left(mc\right)$ = 3.51E-13 cm.

The $H/G$ mod time(t) $/cos(\theta)$

= trisine time $\left(t_T\right)$ = 1.17E-23 sec

conforms to typical nuclear particle decay time. It is noted the characteristic $cos(\theta)$ has a value very nearly equal to $(\pi/4)^{1/3}$.

(The Standard Model predicts a top quark mean lifetime to be roughly 5E–25 sec about 20 times shorter than the timescale for strong interactions -WIKI)

Max Planck's 1899 statement is retained with the *H/G* units

"These necessarily retain their meaning for all times and for all civilizations, even extraterrestrial and non-human ones, and can therefore be designated as 'natural units.'"

These $H/G$ mod mass(m), length(l) and time(t) may be an alternative way based on standard model physics to quantify the Hubble parameter ($H$) and support Mead's premise of the connection between gravity and fundamental length.

## Appendix J. CMBR and Universe Luminosity Equivalence

CMBR Black Body radiation equals Milky Way luminous Black Body radiation (with reasonable dimensional assumptions)

FOR PRESENT UNIVERSE CONDITIONS

CMBR Black Body radiation equals universe luminous Black Body radiation AT ANY UNIVERSE AGE as solar type luminescent objects being 3.7 Percent of universe density $\rho_U$ (equation 2.11.16) as presently nearly indicated (4.6 percent) by Wilkinson Microwave Anisotropy Probe (WMAP)

IF UNIVERSE DENSITY IS ADIABATICALLY EXPANDED AT THE 4/3 POWER.

(This is consistent with CMBR temperature $T_b$ adiabatic expansion as indicated in equations 2.7.3a-c)

With the universe dimensionally described over 100's of orders of magnitude, this one percent dimensional relationship (F = 3.7 percent vs 4.6 percent WMAP)

OVER THE UNIVERSE AGE $1/H_U$ should be placed in the appendix of astrophysical discussion.

Calculations are as follows:

(integrated Black Body relationships are used-no /Hz)

(full spherical relationships are used-no steradians)

0.1367 — *watt/(cm² AU² solar mass)*

1,367,000 — *erg/(cm² sec AU² solar mass unit)*
sunBlackBody

1.50E+13 — *cm/AU*

3.06E+32 — energy total from the sun *erg*
*SunE*

4.00E+11 — stars in Milky Way
*MWstars*
visually observed and estimated

1.99E+33 — solar mass *(g)*
*Ms*

1.22E+44 — energy total from the Milky Way *(erg)*
*MWstars x SunE*

3.67E+22 — average distance*(MWR)* of Milky Way stars*(cm)* 38,600 *light years*
(a reasonable parameter which makes the model work)

3.85E-24 — Milky Way density *(g/cm³)* assumed sphere

2.04E+18 — average Milky Way luminosity over spherical area defined by earth distance from sun
*MWstars SunE (cm/AU)²/MWR²*

9.10E-02 — Milky Way luminosity at earth *(erg/sec/cm²)*

6.67E-08 — Newton *(G cm³ sec⁻² g⁻¹)*

2.31E-18 — Hubble parameter *(sec⁻¹)*
71.2 *km/(sec million parsec)*

2.25E+28 — universe visible radius cm
$R_U = sqrt(3)c/H_U$ equation 2.11.6

6.38E-30 — university mass density *g/cm³*
$\rho_U = \left(2/(8\pi)\right)\left(H_U^2/G_U\right)$ equation 2.11.16

3.02E+56 — university mass *(g)*
$M_U = \rho_U \left(4\pi/3\right) R_U^3$ equation 2.11.15

1.52E+23 — number of universe solar mass units
$M_U$ */solar mass* $= M_n$

0.037 — luminous fraction of the universe
*F*

3.40E-03 — universe luminosity at earth erg/cm2/sec
$F\, M_n\, sunBlackBody\, (cm/AU)^2/R_U^2 = UL$

4.54E-13 — universe luminosity density at earth *erg/cm³*
*ULc/4*

3.41E-03 — CMBR at $T_b = 2.7\ K$ Black Body universe emission *erg/(cm² sec)*
$\left(\pi^2/\left(15\hbar^3c^3\right)\right)\left(k_bT_b\right)^4$

4.55E-13 — CMBR at $T_b = 2.7\ K$ energy density at earth *erg/cm³*
$\left(1/4\right)\left(\pi^2/\left(15\hbar^3c^2\right)\right)\left(k_bT_b\right)^4$

1.00 — ratio CMBR/universe luminescent BlackBody Radiation

and also:

1.00 — ratio CMBR to Milky Way luminescent BlackBody Radiation

## Appendix K. Hydrogen Bose Einstein Condensate as Dark Matter

A hypothesis is that the spectroscopically undetectable dark matter is actually Big Bang nucleosynthetic adiabatically produced (extremely cold nanokelvin to femtokelvin) colligative Bose Einstein Condensate(BEC) hydrogen helium at density on the order of 1 g/cm$^3$ or 1 mole/cm$^3$ or Avogadro/cm$^3$. This is substantiated by the following established Bose Einstein Condensate relationship

$$k_b T_c = \left( \frac{\rho_{dark\ matter\ paticle\ density}}{\xi\left(\frac{3}{2}\right)^{\frac{2}{3}}} \right) \frac{h^2}{2\pi m_p \left(a m_p\right)^{\frac{2}{3}}}$$

that is derived from published form

$$k_b T_c = \left( \frac{n_{dark\ matter\ paticle\ density}}{\xi\left(\frac{3}{2}\right)^{\frac{2}{3}}} \right) \frac{h^2}{2\pi m_p}$$

by a multiplier factor

$$\frac{\left(a m_p\right)^{\frac{2}{3}}}{\left(a m_p\right)^{\frac{2}{3}}}$$

For the universe present condition

$T_c$ = 8.11E-16 and $\rho_{dark\ matter\ particle\ density}$ = 1 g/cm$^3$

then $a$ = 5.22E25 superimposed states with origins in Big Bang nucleosynthesis (not presently observed under laboratory conditions)

and the hydrogen BEC volume is:

$$\frac{a m_p}{\rho_{dark\ matter\ particle\ density}} = 87.3 \text{ cm}^3$$

or about 4.4 cm on a side.

This relationship reverts to equation 2.10.6 when $a$ = 1 and mass = $m_T$.

The exact nature of hydrogen at these cold temperatures may also have metallic hydrogen, frozen hydrogen and/or superconducting properties.

Fact:

The observed amount of hydrogen in the universe is observed spectroscopically as neutral hydrogen through the 21 cm line assuming a Beers law.

Beers law assumption:

"Assume that particles may be described as having an area, alpha, perpendicular to the path of light through a solution (or space media), such that a photon of light is absorbed if it strikes the particle, and is transmitted if it does not."

"Expressing the number of photons absorbed by the (concentration c in) slab (in direction z) as dIz, and the total number of photons incident on the slab as Iz, the fraction of photons absorbed by the slab is given by:"

dIz/Iz = -alpha c dz

In other words, the fraction of photons absorbed when passing through an observed slab is equal to the total opaque area of the particles in the slab.

In other words, larger particles with the same aggregate volume concentration in a supporting medium will not attenuate as many incident photons.

Issue one:

If the Big Bang nucleosynthetic hydrogen were in the form of adiabatically produced very cold chunks, their aggregate total opaque area would be much smaller than a diffuse hydrogen gas resulting in many orders of magnitude lower Beers law photon attenuation rates. Mean free paths are of multi galactic length dimensions. (These BEC's would not be spectroscopically detected with present tools)

Issue two:

Given these primeval (through adiabatic Big Bang expansion) hydrogen dense (1 g/cc) BEC's large chunks, say 10's of kilometers in diameter, The parallel to antiparallel hydrogen (proton electron spin) configuration resulting in 21 cm line would occur randomly in the BEC chunk hydrogens at the known 2.9E-15 /sec rate. Transmittance accordance with Beer's law would indicate that the 22 cm radiation from the BEC center would be attenuated more than that from the BEC near surface resulting in a lower than reality hydrogen spatial density reading obtained by a distant observer.

Issue three:

The hydrogen dense (1 g/cc) BEC chunks should be stable relative to higher temperature of a intergalactic media

(ionized gas). There is the question whether Intergalactic Medium (IGM) composed of 'hot' atoms would 'sublimate' any BECs that were adiabatically produced at the initial Big Bang. In order to address this question, assume the IGM is totally ionized at the given universe critical density of ~1 proton mass per meter$^3$. (It is very clear that this ionized IGM was not there in the beginning but ejected into space from stars etc.) Now take a 1 meter$^3$ hydrogen BEC. It would have $100^3$ x (Avogadro number) hydrogen atoms in it. If there were no relative motion between the BEC and the hot atoms, the BEC would not degrade. If there was relative motion between the BEC and the hot atoms, the BEC could travel $100^3$ x (Avogadro number) meters or 6.023E29 meters through space before 'melting'. The length 6.023E29 meters is larger than the size of the visual universe (2.25E26 meter). There could have been multi kilometer sized hydrogen BEC's at the adiabatic expansion still remaining at the present time and at nearly that size. Also, as a BEC, it would not have a vapor pressure. It would not sublime. The hydrogen BECs perhaps are still with us.

Issue four:

Over time, the BEC would absorb the CMBR microwave background and heat up, but how much?

Black Body Background radiation at 2.732 K emits at power density of 0.00316 erg/cm$^2$/sec (stefan's constant x 2.7324) with wavelength at black body max at 0.106 cm (1.60531E+11 Hz)

The 1S-2S characteristic BEC hydrogen transition is at 243 nm (1.23E+15 Hz) This 1S-2S transition would have to be initiated before subsequent Rydberg transitions.

This 243 nm (1.23E+15 Hz) absorption is way out on the CMBR tale and absorbing in a band width of ~1E6 Hz with a natural line width of 1.3 Hz. [90]

Calculations indicate the einsteins absorbed at this characteristic hydrogen BEC frequency at 243 nm (band width of ~1E6 Hz with a natural line width of 1.3 Hz) from CMBR would not substantially affect hydrogen BEC chunks over universe life 13.7 billion years.

Issue five:

Even though, spectroscopically undetectable these BECs would be at the same abundance (relative concentration) as predicted by nucleosynthetic models and would be considered baryonic matter. How flexible are nucleosynthetic models as to baryonic mass generation (keeping the abundance (relative concentration to say observed helium) the same? The presently spectroscopically undetectable BECs would in principle not be spectroscopically detectable all the way back to 13.7 billion year beginning and not observable in the WMAP data.

Issue six:

These hydrogen dense (1 g/cc) BECs could be the source media for star formation (~5 percent of universe density) by gravitational collapse and as such, their residual (~25 percent of universe density) presence would be gravitationally detected by observed galactic rotation features and gravitational lensing characteristics

Issue seven:

A reference to Saha equation indicates an equilibrium among reactants in a single phase prior to recombination

hydrogen + photon <> proton + electron

There may have been a non equilibrium condition between this phase and with another nucleosynthetic adiabatically produced solid cold BEC hydrogen phase prior to and continuing on through the recombination event and not spectroscopically observed by WMAP. (such multiphase non equilibrium reactions are important, particularly in $CO_2$ distribution in air phase and associated $H_2CO_3$, $HCO_3^-$ and $CO_3^=$ in water phase)

| | | |
|---|---|---|
| CMBR Temperature: | 3,000 | K |
| Radiated power area density: | 4.593E+09 | erg/cm$^2$/sec |
| Radiated power: | 4.593E+09 | erg/sec |
| radiation mass density | 6.819E-22 | g/cm$^3$ |
| pressure=1/3 energy/volume | 2.043E-01 | dyne/cm$^2$ |
| photon number density | 5.477E+11 | photon/cm$^3$ |
| photon energy density | 6.401E-01 | erg/cm$^3$ |
| Most probable energy | 1.764E+14 | /sec |
| hydrogen BEC | | |
| absorbed energy | 1.301E-12 | erg/cm$^3$ |
| absorbed freq | 1.234E+15 | /sec |
| absorbed wavelength | 2.430E-05 | cm |

Ref: [90]

Calculations indicate the einsteins absorbed by BEC at the characteristic hydrogen BEC frequency 1.234E+15 /sec (way out on the CMBR 3,000 K tail) (extremely narrow absorption band width of ~1E6 Hz with a natural line width of 1.3 Hz) would not substantially degrade extremely cold hydrogen dense (1 g/cc) BEC chunks at universe Big Bang age of 376,000 years. Indeed the $BEC_{ST}$ at 376,000 years is 1.05E8 years. This is an estimate realizing that CMBR temperature changes with time. $BEC_{ST}$ keeps ahead of the universe age $Age_{UG}$ back to introductory condition 1.9 and again from introductory conditions 1.4 to 1.1.

If it assumed the BEC absorption band with approaches the nature line width of 1.3 Hz from 1E6 Hz, it is evident that the $BEC_{ST}$ would survive through all of the universe nucleosynthetic events. It is noteworthy that under the present CMBR temperature $T_b$, the hydrogen BECs would survive essentially forever (>1E300 years – the upper limit of the current computational ability).

| *Conditions Defined In The Introduction* | $T_b$ *kelvin* | *BEC* $T_c$ *kelvin* | $BEC_{ST}$ *(seconds) at 1E6 Hz bandwidth* | $BEC_{ST}$ *(seconds) at 1.3 Hz bandwidth* | $Age_{UG}$ *(seconds)* | $Age_{UGv}$ *(seconds)* |
|---|---|---|---|---|---|---|
| 1.11 | 2.73E+00 | 8.11E-16 | >1.00E+300 | >1.00E+300 | 4.32E+17 | 4.32E+17 |
| | 1.00E+01 | 1.09E-14 | >1.00E+300 | >1.00E+300 | 6.16E+16 | 6.16E+16 |
| | 2.50E+01 | 6.81E-14 | >1.00E+300 | >1.00E+300 | 1.56E+16 | 1.56E+16 |
| | 5.00E+01 | 2.72E-13 | >1.00E+300 | >1.00E+300 | 5.52E+15 | 5.51E+15 |
| | 1.00E+02 | 1.09E-12 | 6.12E+263 | 4.70E+269 | 1.95E+15 | 1.95E+15 |
| | 2.50E+02 | 6.81E-12 | 3.47E+109 | 3.34E+115 | 4.93E+14 | 4.91E+14 |
| | 5.00E+02 | 2.72E-11 | 1.50E+58 | 1.66E+64 | 1.74E+14 | 1.73E+14 |
| | 1.00E+03 | 1.09E-10 | 3.69E+32 | 3.44E+38 | 6.16E+13 | 6.08E+13 |
| | 2.00E+03 | 4.36E-10 | 5.64E+19 | 4.83E+25 | 2.20E+13 | 2.12E+13 |
| | 2.50E+03 | 6.81E-10 | 1.63E+17 | 1.29E+23 | 1.56E+13 | 1.51E+13 |
| 1.10 | 3.00E+03 | 9.80E-10 | 3.31E+15 | 2.55E+21 | 1.19E+13 | 1.14E+13 |
| | 1.00E+04 | 1.09E-08 | 3.28E+09 | 2.54E+15 | 1.94E+12 | 1.73E+12 |
| | 1.50E+04 | 2.45E-08 | 4.54E+08 | 3.50E+14 | 1.06E+12 | 8.96E+11 |
| | 1.00E+05 | 1.09E-06 | 7.28E+06 | 5.61E+12 | 6.16E+10 | 3.21E+10 |
| | 1.50E+05 | 2.45E-06 | 4.38E+06 | 3.34E+12 | 3.37E+10 | 1.50E+10 |
| | 1.00E+06 | 1.09E-04 | 5.52E+05 | 4.23E+11 | 1.95E+09 | 3.69E+08 |
| | 1.50E+06 | 2.45E-04 | 3.63E+05 | 2.79E+11 | 1.06E+09 | 1.65E+08 |
| 1.9 | 1.18E+07 | 1.52E-02 | 4.54E+04 | 3.50E+10 | 4.80E+07 | 2.70E+06 |
| 1.8 | 2.94E+08 | 9.41E+00 | 1.82E+03 | 1.40E+09 | 3.87E+05 | 4.35E+03 |
| 1.7 | 5.48E+08 | 3.27E+01 | 9.76E+02 | 7.51E+08 | 1.52E+05 | 1.25E+03 |
| 1.6 | 9.24E+08 | 9.30E+01 | 5.77E+02 | 4.45E+08 | 6.94E+04 | 4.41E+02 |
| 1.5 | 1.45E+09 | 2.29E+02 | 3.68E+02 | 2.84E+08 | 3.53E+04 | 1.79E+02 |
| 1.4 | 1.69E+11 | 3.11E+06 | 3.16E+00 | 2.44E+06 | 2.81E+01 | 1.32E-02 |
| 1.3 | 5.49E+13 | 3.28E+11 | 9.72E-03 | 7.51E+03 | 4.79E-03 | 1.25E-07 |
| 1.2 | 9.07E+14 | 8.96E+13 | 5.92E-04 | 4.54E+02 | 7.14E-05 | 4.57E-10 |
| 1.1 | 9.19E+15 | 9.20E+15 | 5.81E-05 | 4.48E+01 | 2.21E-06 | 4.45E-12 |

For general consistency with the rest of the report, the data is also presented in seconds.

| *Conditions Defined In The Introduction* | $T_b$ *kelvin* | *BEC* $T_c$ *kelvin* | $BEC_{ST}$ *(seconds) at 1E6 Hz bandwidth* | $BEC_{ST}$ *(seconds) at 1.3 Hz bandwidth* | $Age_{UG}$ *(years)* | $Age_{UGv}$ *(years)* |
|---|---|---|---|---|---|---|
| 1.11 | 2.73E+00 | 8.11E-16 | >1.00E+300 | >1.00E+300 | 1.37E+10 | 1.37E+10 |
| | 1.00E+01 | 1.09E-14 | >1.00E+300 | >1.00E+300 | 1.95E+09 | 1.95E+09 |
| | 2.50E+01 | 6.81E-14 | >1.00E+300 | >1.00E+300 | 4.95E+08 | 4.94E+08 |
| | 5.00E+01 | 2.72E-13 | >1.00E+300 | >1.00E+300 | 1.75E+08 | 1.75E+08 |
| | 1.00E+02 | 1.09E-12 | 6.12E+263 | 4.70E+269 | 6.18E+07 | 6.17E+07 |
| | 2.50E+02 | 6.81E-12 | 3.47E+109 | 3.34E+115 | 1.56E+07 | 1.56E+07 |
| | 5.00E+02 | 2.72E-11 | 1.50E+58 | 1.66E+64 | 5.52E+06 | 5.49E+06 |
| | 1.00E+03 | 1.09E-10 | 3.69E+32 | 3.44E+38 | 1.95E+06 | 1.93E+06 |
| | 2.00E+03 | 4.36E-10 | 5.64E+19 | 4.83E+25 | 6.98E+05 | 6.73E+05 |
| | 2.50E+03 | 6.81E-10 | 1.63E+17 | 1.29E+23 | 4.95E+05 | 4.79E+05 |
| 1.10 | 3.00E+03 | 9.80E-10 | 3.31E+15 | 2.55E+21 | 3.77E+05 | 3.62E+05 |
| | 1.00E+04 | 1.09E-08 | 3.28E+09 | 2.54E+15 | 6.15E+04 | 5.49E+04 |
| | 1.50E+04 | 2.45E-08 | 4.54E+08 | 3.50E+14 | 3.36E+04 | 2.84E+04 |
| | 1.00E+05 | 1.09E-06 | 7.28E+06 | 5.61E+12 | 1.95E+03 | 1.02E+03 |
| | 1.50E+05 | 2.45E-06 | 4.38E+06 | 3.34E+12 | 1.07E+03 | 4.75E+02 |
| | 1.00E+06 | 1.09E-04 | 5.52E+05 | 4.23E+11 | 6.18E+01 | 1.17E+01 |
| | 1.50E+06 | 2.45E-04 | 3.63E+05 | 2.79E+11 | 3.36E+01 | 5.24E+00 |
| 1.9 | 1.18E+07 | 1.52E-02 | 4.54E+04 | 3.50E+10 | 1.52E+00 | 8.55E-02 |
| 1.8 | 2.94E+08 | 9.41E+00 | 1.82E+03 | 1.40E+09 | 1.23E-02 | 1.38E-04 |
| 1.7 | 5.48E+08 | 3.27E+01 | 9.76E+02 | 7.51E+08 | 4.82E-03 | 3.97E-05 |
| 1.6 | 9.24E+08 | 9.30E+01 | 5.77E+02 | 4.45E+08 | 2.20E-03 | 1.40E-05 |
| 1.5 | 1.45E+09 | 2.29E+02 | 3.68E+02 | 2.84E+08 | 1.12E-03 | 5.67E-06 |
| 1.4 | 1.69E+11 | 3.11E+06 | 3.16E+00 | 2.44E+06 | 8.91E-07 | 4.18E-10 |
| 1.3 | 5.49E+13 | 3.28E+11 | 9.72E-03 | 7.51E+03 | 1.52E-10 | 3.96E-15 |
| 1.2 | 9.07E+14 | 8.96E+13 | 5.92E-04 | 4.54E+02 | 2.26E-12 | 1.45E-17 |
| 1.1 | 9.19E+15 | 9.20E+15 | 5.81E-05 | 4.48E+01 | 7.01E-14 | 1.41E-19 |

These numbers are then expressed graphically following Figure K1.

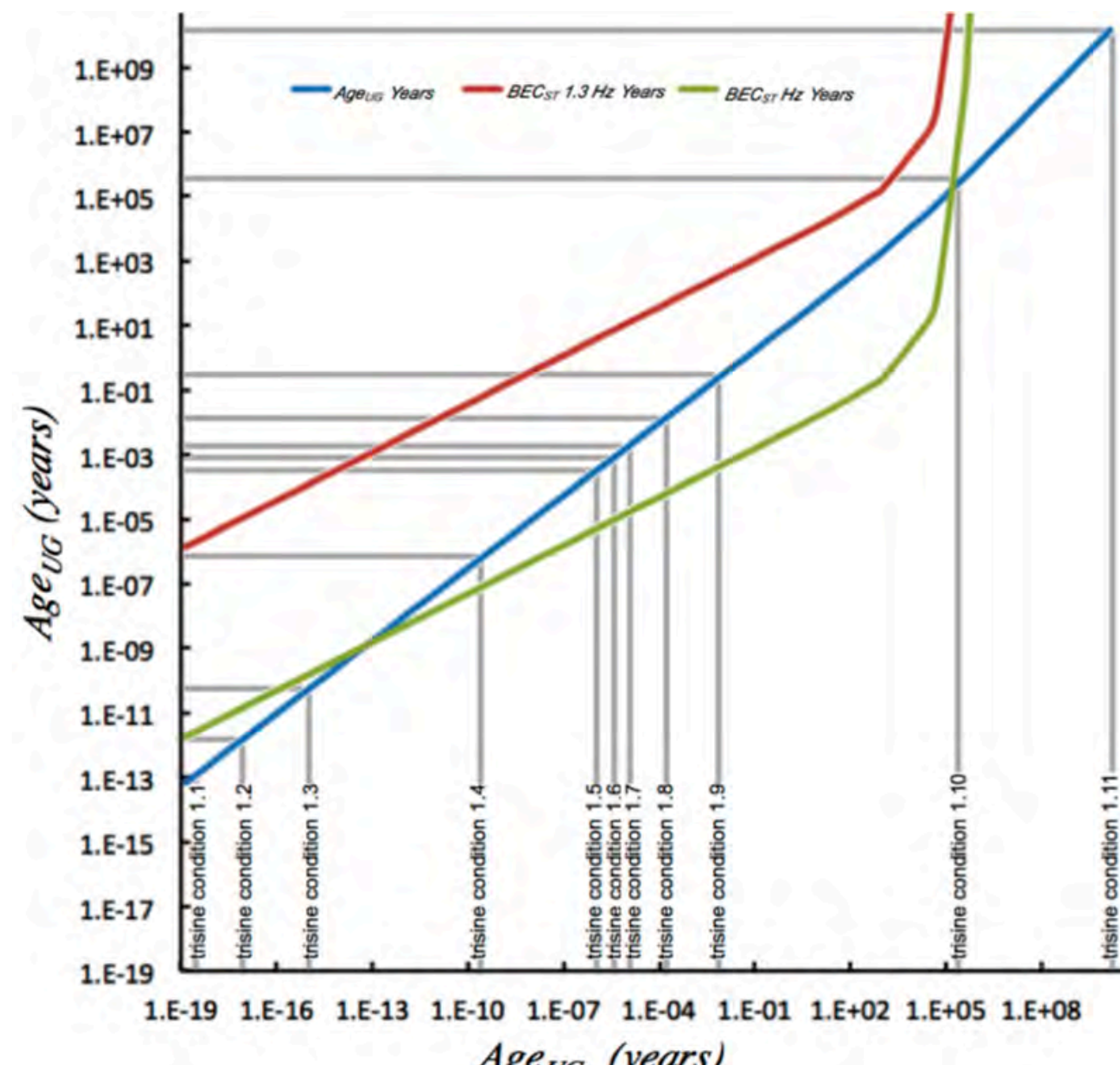

The 3,000 K CMBR at 376,000 years is hot but has a very low heat capacity. It would be like touching a red to white hot space shuttle tile with your finger and not getting burned.

How would these cold hydrogen BECs coming through the recombination event be observed in Baryonic Acoustic Oscillations (BAO) (which does have a measurable BEC? dark matter component) in the context that BECs would have different masses ($m$) than currently modeled gaseous hydrogen?

$$BAO\ frequency_{BEC} \sim \sqrt{1/m_{BEC}}$$

and

$$BAO\ frequency_{gaseous\ hydrogen} \sim \sqrt{1/m_{gaseous\ hydrogen}}$$

but

$$\sqrt{1/m_{gaseous\ hydrogen}} \neq \sqrt{1/m_{BEC}}$$

Can 'first three minute' nucleosynthesis dynamics 'precipitate out' these cold hydrogen BECs?

The permittivity permeability model as presented in this paper allows a mechanism to do such.

Issue eight

A gravitationally unstable system loses energy (for star formation)

Assuming that the universe is a closed system that conserves energy, then where does this gravitational energy go?

Issue nine

Superconducting particles in the universe vacuum should not clump due to magnetic field expulsion characteristic. This nonclumping is observed in dark matter observations.

Hypothesis:

The energy goes to an endothermic phase change (like solid to gas) and more specifically the change from Big Bang nucleosynthetic hydrogen Bose Einstein Condensate(H-BEC) to hydrogen gas that with this energy transfer gaseous hydrogen is thermodynamically able to gravitationally collapse to stars.

This assumes ~30 percent of the universe is H-BEC 'dark matter' that provides a large heat sink for a small portion that endothermically absorbs energy through 'sublimation' made available for gravitationally collapsing star forming H-gas.

Hubble and Spitzer have observed early (z = 5.7) galaxies with a distinct red color into the infrared [109]. Could this galaxy red color be due to red shifted characteristic 1S-2S Hydrogen Bose Einstein Condensate(HBEC) at 243 nanometer red shifted

(1+5.7) x 243 nanometer = 1.63 micro meter?

(in the graphically expressed beginning red detection 'within error' of [109] Figure 3 ERS-1, 2, 3, 4). Galactic formation may have this z adjusted red detection at all epochs.

Big Bang nucleosynthetic HBEC (presently expressed as dark matter) was a robust active galactic formation source and may have contributed to the observed 13 billion year old galactic red as a portion(~4%) of HBEC was destroyed and then proceeded up the Rydberg scale before gravitational collapse in the galactic formation process all observed as a distant reddish glow.

This reddish glow should be characteristic of all early galaxies in the process formation from Big Bang nucleosynthetic HBEC.

Is this the missing link to understanding dark matter as Hydrogen Bose Einstein Condensate(HBEC)?

Figure K2 Dark Matter as Baryonic Hydrogen

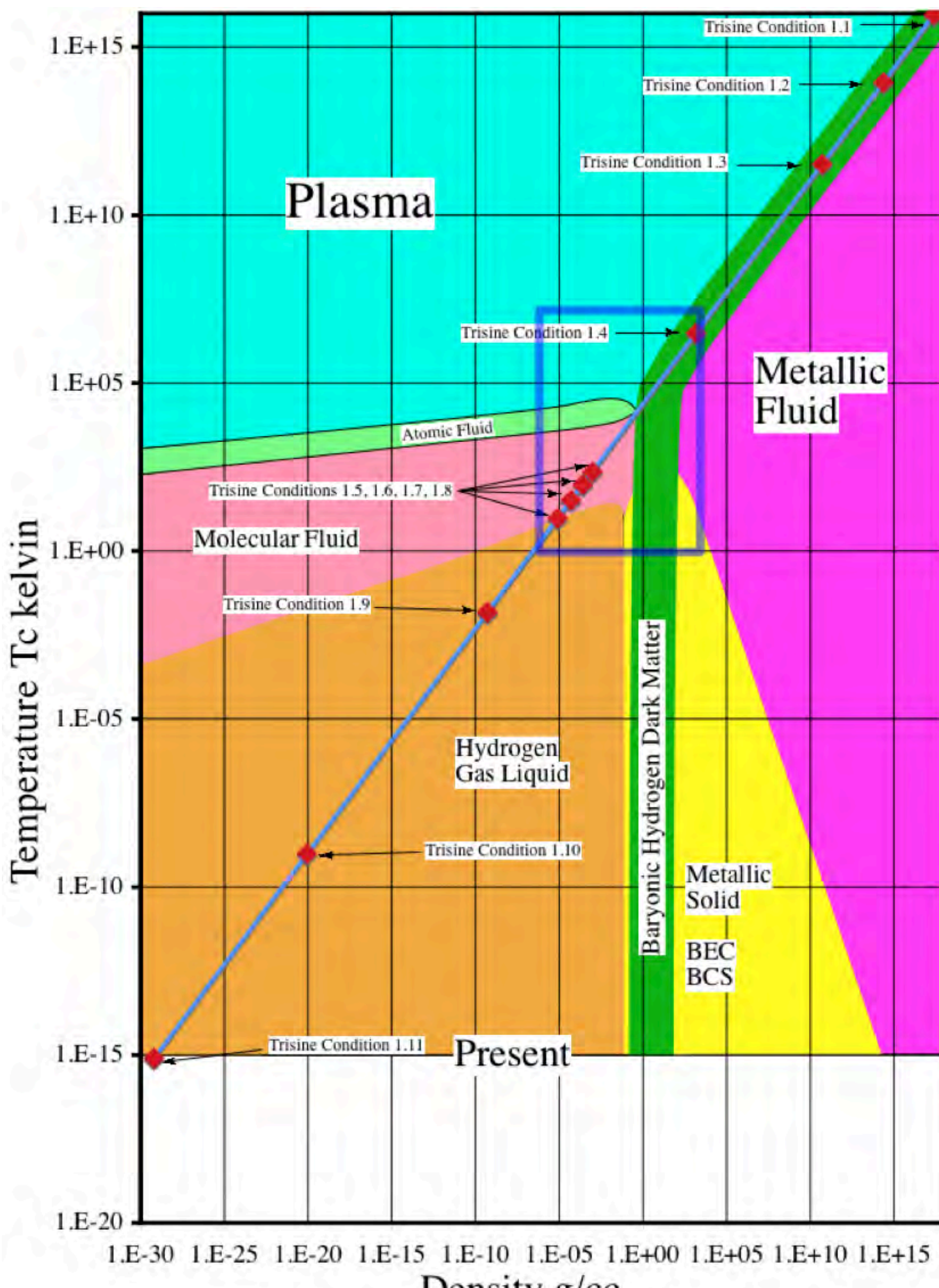

Figure K2 graphically presents these thoughts in terms of a expanded hydrogen phase diagram. The dark blue insert box represents available laboratory hydrogen phase data is available. Extended phase relationships are extrapolated from this available experimental data.

A separate colder $(T_c)$ phase than CMBR $(T_b)$ is present from the universe beginning (initially as subatomic particles) and separates after Condition 3 as baryonic metallic (BEC and or BCS) hydrogen and retains this form (dark green) until the present time. The separate colder $(T_c)$ phase provides the trapping $(kT_c)$ mechanism for this baryonic metallic (BEC and or BCS) hydrogen. This baryonic metallic (BEC and or BCS) hydrogen is considered to be dark matter. It is not presently observed spectroscopically but only gravitationally.

The phase density Figures 4.42.3 (Blue line) has a slope correlation to Figure K2. The magnitude difference can be explained in terms of particle coalesence dynamics as presented in Appendix L and in particular equation L2.14.

Question?

Does transition-state theory as written in physical chemistry books and detailed in wiki Ref 1:

https://en.wikipedia.org/wiki/Transition_state_theory

enter into Big Bang Nucleosynthesis BBN

particularly with the introduction of redshift z and the colligative properties of BBN plasma as suggested by

Ref 2:

Closing in on the Border Between Primordial Plasma and Ordinary Matter

http://www.bnl.gov/bnlweb/pubaf/pr/PR_display.asp?prID=1446

"the different phases exist under different conditions of temperature and density, which can be mapped out on a “phase diagram,” where the regions are separated by a phase boundary akin to those that separate liquid water from ice and from steam. But in the case of nuclear matter, scientists still are not sure where to draw those boundary lines.

RHIC is providing the first clues." ?

The general reaction expression given by the transition-state theory

$$-\frac{d\left[A^{'}\right]}{dt}=\frac{k_bT}{h}\left[A^{'}\right]\left[B^{'}\right]$$

$$\left[A^{'}\right]=\left[A_0^{'}\right]\left[B_0^{'}\right]e^{\frac{-k_bTt}{h}}$$

Now assume T is reaction temperature that scales as

$CMBR\ T_o(z+1)^1$

and

t is universe time 1/H that scales as

$t_o(z+1)^{-3}$

then generally redshifted z transition-state theory is:

$$\left[A^{'}\right]=\left[A_0^{'}\right]\left[B_0^{'}\right]e^{\frac{-k_bT_o(z+1)^1t_o(z+1)^{-3}}{h}}$$

and by reduction

$$\left[A^{'}\right]=\left[A_0^{'}\right]\left[B_0^{'}\right]e^{\frac{-k_bT_ot_o(z+1)^{-2}}{h}}$$

and specifically at two redshifts $z_1$ and $z_2$

$$\left[A^{'}\right]=\left[A_0^{'}\right]\left[B_0^{'}\right]e^{\frac{-k_bT_ot_o(z_1+1)^{-2}}{h}}$$

$$\left[A^{'}\right]=\left[A_0^{'}\right]\left[B_0^{'}\right]e^{\frac{-k_bT_ot_o(z_2+1)^{-2}}{h}}$$

and the ratio for two reaction temperatures $T_1$ and $T_2$ replacing temperature $T_o$ (universe time to scaling is the same and nuclide reactant standard free energies are the same at $T_1$ and $T_2$ redshift $z$ scaling)

$$\frac{\left[A_1^{'}\right]}{\left[A_2^{'}\right]}=\frac{\left[A_0^{'}\right]\left[B_0^{'}\right]}{\left[A_0^{'}\right]\left[B_0^{'}\right]}e^{-k\frac{T_ot_o(z_1+1)^{-2}}{T_ot_o(z_2+1)^{-2}}}$$

Taking the BBN z at 1 minute (192,000) and 3 minute (134,000)

(This is somewhat arbitrary but is in the time frame of Weinberg's book The First Three Minutes [100])

and normalizing into k = 1

$$\frac{\left[A_1^{'}\right]}{\left[A_2^{'}\right]}=e^{-k\frac{T_1 1.92^{-2}}{T_2 1.34^{-2}}}$$

$$\frac{\left[A_1^{'}\right]}{\left[A_2^{'}\right]}=e^{-k\frac{T_1 1.92^{-2}}{T_2 1.34^{-2}}}$$

for $\frac{T_1}{T_2}=1$ then $\frac{\left[A_1^{'}\right]}{\left[A_2^{'}\right]}=.61$

for $\frac{T_1}{T_2}=.8$ then $\frac{\left[A_1^{'}\right]}{\left[A_2^{'}\right]}=.68$

and .68/.61 = 1.11.

(millions of other redshifted $z$ time $t$ and temperature $T$ conditions could be made to make the same point)

The different reaction temperatures $T_1/T_2=1$ are achieved by colligative properties as suggested by Ref 2 analogous to boiling point raising without changing the nuclide reaction chemistry.

In BBN, the reaction products would be the same (standard model quarks, gluons etc) but their mass amounts would be altered.

The Ref 2 BNL RHIC experiments hint at this colligative mechanism, something that may be missed in conventional BBN calculations.

Observed H, He and Li abundances (ratios) would be the same. The universe H He masses would be different.

This is all posed as a question as to the possibility as to universe baryonic mass originating in BBN as greater than the conventionally held value at ~4%.

## Appendix L. Model Development For Classical Fluid Shear and Fluid Drop Modification

The Starry Night a painting by Vincent Van Gogh
A fluid vision of the Universe

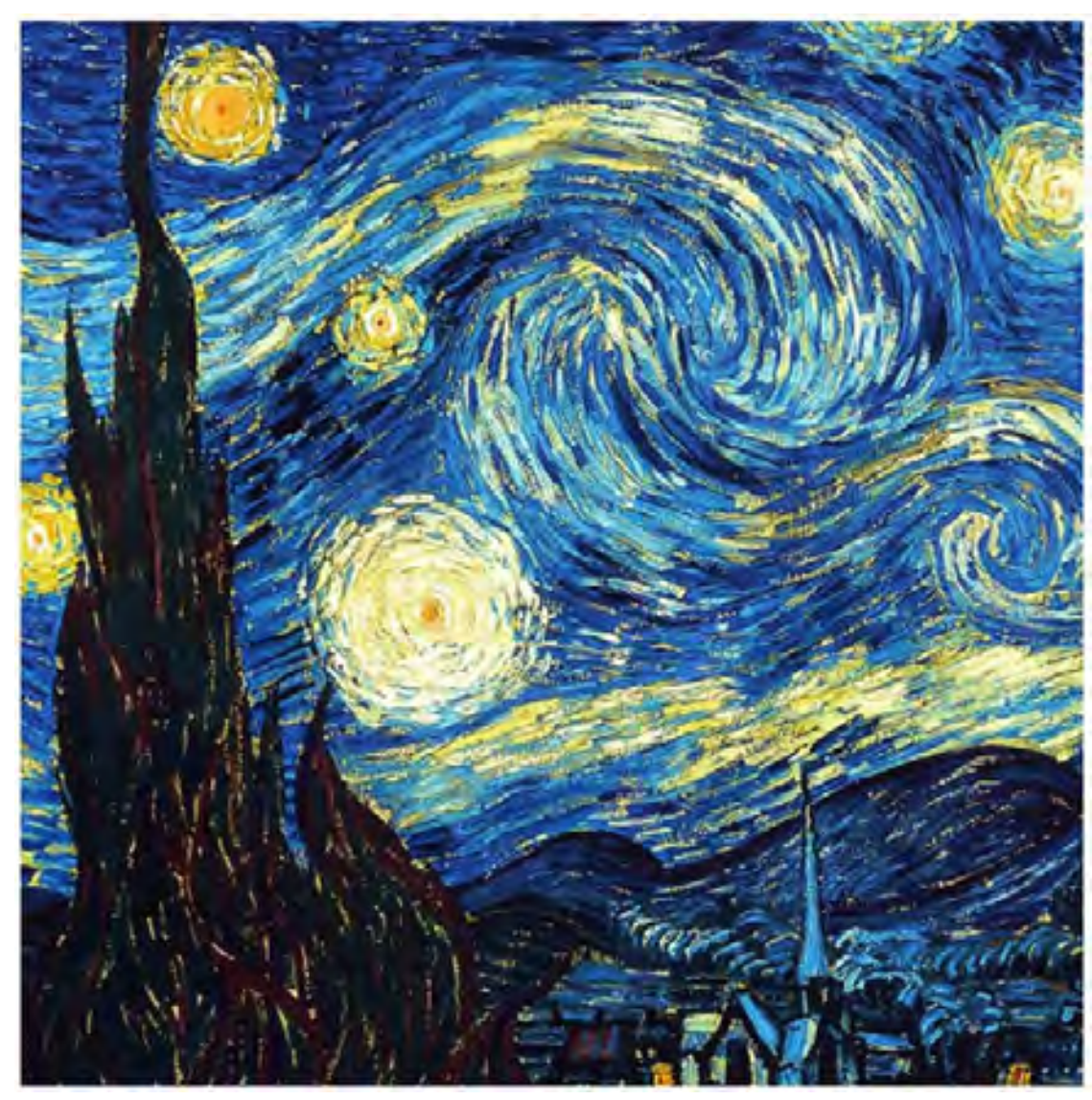

The following fluid work was primarily done in the 1980's as a theoretical framework for industrial separator design. As such it was done in the English (ft lb sec) system that was prevalent at that time although dimensionality dictates any other system such as cgs or SI may be used with equal applicability with the understanding that:

$$F = M\,a$$

$$lb_{force} = \frac{lb_{force}}{gravity_{acceleration}} gravity_{acceleration}$$

$$\partial = \rho g$$

$$\frac{lb_{force}}{ft^3} = \frac{lb_{force}}{ft^3 gravity_{acceleration}} gravity_{acceleration}$$

Variable Definition

Fluid Viscosity $\left(\mu_1 \;\; lb\ sec/ft^2\right)$

Interfacial Tension $\left(\sigma \;\; ft\ lb/ft^2\right)$

Fluid Specific Weight $\left(\partial_1 \;\; lb/ft^3\right)$

Second Phase Specific Weight $\left(\partial_2 \;\; lb/ft^3\right)$

Gravity Acceleration $\left(g \;\; ft/sec^2\right)$

### L1. Introduction

The figure one display of the shear problem and following equations L1.1 and L1.2 are paraphrased from reference 1 and used as an initial starting point for this derivation or model development for the quantitative influence of shear ($G$) on droplet size ($D$) in various hydraulic functional units (HFUs) such as pipes, valves, pumps, nozzles etc.

**Figure L1.** Shear forces along parallel planes of an elemental volume of fluid

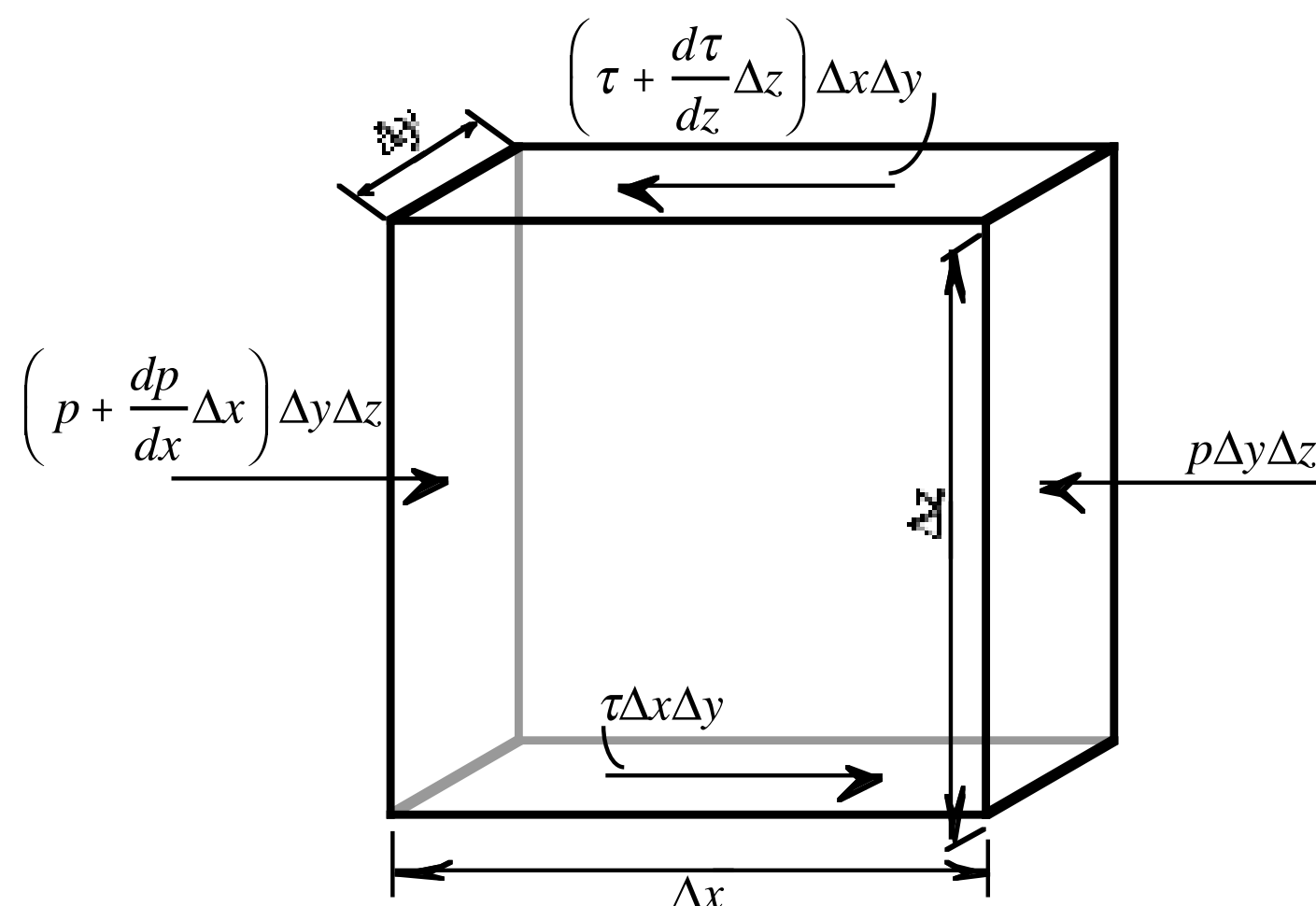

As illustrated in Figure L1 the fluid shear $(G)$ is derived below by equating the forces acting on a cube of fluid in shear in one direction to those in the opposite direction, or

$$p\ \Delta y\ \Delta z + \left[\tau + \frac{d\tau}{dz}\Delta z\right]\Delta x\ \Delta y = \tau\ \Delta x\ \Delta y + \left[p + \frac{dp}{dx}\Delta x\right]\Delta y\ \Delta z \qquad \text{L1.1}$$

Here $p$ is the pressure, $\tau$ the shear intensity, and $\Delta x$, $\Delta y$, and $\Delta z$ are the dimensions of the cube. It follows that:

$$\frac{d\tau}{dz} = \frac{dp}{dx} \qquad \text{L1.2}$$

The power expended, or the rate at which work is done by the couple $(\tau\ \Delta x\ \Delta y)$, equals the torque $(\tau\ \Delta x\ \Delta y)\Delta z$ times the angular velocity $dv/dz$. Therefore the power $(P)$ consumption per volume $(V)$ of fluid is as follows:

$$\frac{P}{V}=\frac{(\tau\ \Delta x\ \Delta y)\Delta z}{\Delta x\ \Delta y\ \Delta z}\frac{dv}{dz}=\tau\frac{dv}{dz} \qquad \text{L1.3}$$

Defining $\tau=\mu\ dv/dz$ and $G=dv/dz$

then

$P/V=\ \mu(dv/dz)^2=\mu G^2$ or as follows:

$$G^2=\frac{P}{\mu V} \qquad \text{L1.4}$$

## L2. Basic Model Development

Define the list of variables.

*n* = number of second phase fluid droplets per volume

*V* = volume of n second phase fluid droplets

*Vol* = volume of first phase supporting fluid

*V/Vol* = concentration*(C)* of second phase fluid droplets in first phase supporting fluid

*D* = diameter of individual second phase fluid droplets

*S* = surface area of n second phase fluid droplets

*t* = reactor residence time

From these variables the following surface, volume and droplet relationships logically follow.

$$n=\frac{6V}{\pi D^3}=\frac{S}{\pi D^2} \qquad \text{L2.1}$$

$$D=\frac{6V}{S} \quad \text{and} \quad S=\frac{6V}{D} \qquad \text{L2.2}$$

$$V=\frac{SD}{6} \qquad \text{L2.3}$$

As per the Smoluchowski droplet collision relationship as derived circa 1917, verified through the years and typically presented in reference 1. Droplet collision rate is directly proportional to fluid shear (*G*) in units of /sec.

$$\frac{1}{Vol}\frac{dn}{dt}=\frac{4}{3}\frac{n^2d^3G_1}{Vol^2}\frac{1}{2} \qquad \text{L2.4}$$

The 1/2 factor is introduced so that particle or drop collisions are not counted twice.

Now using the defined relationships between volume, surface area and droplet size, derive and expression for change of droplet surface area as a function of second phase fluid concentration (*C*), fluid shear (*G*) and droplet diameter (*D*).

$$\begin{aligned}\frac{1}{Vol}\frac{dS}{dt}&=\frac{2}{3\pi}\frac{S^2dG_1}{Vol^2} \\ &=\frac{2}{3\pi}\frac{S^2 6VG_1}{SVol^2}=\frac{4}{\pi}\frac{SVG_1}{Vol^2} \\ &=\frac{4}{\pi}\frac{6V^2G_1}{Vol^2D}=\frac{24}{\pi}\frac{C^2G_1}{D}\end{aligned} \qquad \text{L2.5}$$

In terms of change in surface area per second phase fluid volume, equation L2.5 is changed by factor *C* = *V/Vol*.
The Smoluchowski droplet collision relationship is now expressed as:

$$\frac{1}{V}\frac{dS}{dt}=\frac{24}{\pi}\frac{CG_1}{D} \qquad \text{L2.6}$$

Equation L2.6 represents the increased second phase fluid surface area due to second phase fluid droplets colliding with each other, but it is well known that under conditions of shear there is a tendency for second phase fluid droplets to break up. An extreme case would be addition of second phase fluid to first phase supporting fluid in a blender and the emulsification of the second phase fluid therein.

The power (*P*) to create this second phase fluid surface may be expressed in terms of the Gibb's interfacial tension (energy/area) $\sigma$ as in equation L2.7

$$P=\sigma\frac{dS}{dt} \qquad \text{L2.7}$$

The well known (reference 1) dissipative power ($P$) relationship is expressed in equation L2.8 in terms of first phase supporting fluid absolute viscosity $(\mu)$ and first phase supporting fluid shear $(G_1)$.

$$P = \mu_1 V G_1^2 \qquad \text{L2.8}$$

Now equating 2.7 and 2.8

$$\mu_1 V G_1^2 = \sigma \frac{dS}{dt} \qquad \text{L2.9}$$

Now again a relationship is established in equation L2.10 for change in surface area (S) per second phase fluid volume (V) per time (t) from equation L2.9.

$$\frac{1}{V}\frac{dS}{dt} = \frac{\mu_1 G_1^2}{\sigma} \qquad \text{L2.10}$$

Now assume an equilibrium condition under which the coalescing of droplets equals the breakup of droplets and this is expressed by subtracting equation L2.10 from equation L2.6 and set equal to zero.

$$\frac{1}{V}\frac{dS}{dt} = \frac{24}{\pi}\frac{CG_1}{D} - \frac{\mu_1 G_1^2}{\sigma} = 0 \qquad \text{L2.11}$$

Now solving equation 2.11 for $D$. This represents the steady state droplet diameter under a particular shear condition.

$$D = \frac{24}{\pi}\frac{C\sigma}{\mu_1 G_1} \qquad \text{L2.12}$$

Equation 2.12 is the steady state condition, but it may be useful to establish the time dependent change in droplet diameter by solving the differential equation L2.11.

$$S = 6\frac{V}{D} \qquad dS = -6\frac{V}{D^2}dD \qquad \text{L2.13}$$

Therefore:

$$\frac{dD}{dt} = \frac{4CG_1 D}{\pi} - \frac{\mu G_1^2 D^2}{6\sigma} \qquad \text{L2.13}$$

A heat and mass transfer function can be added to L2.13 for evaporation and condesation affects.

Solution of equation L2.13 is presented in equation L2.14.

$$D_a = \frac{\beta D_i \exp(\beta\ t)}{\alpha D_i \left(\exp(\beta\ t) - 1\right) + \beta} \qquad \text{L2.14}$$

Where $D_i$ is the initial droplet size and $D_a$ is final droplet size,

And where:

$$\text{Constant}(\alpha) = \frac{\mu_1 G_1^2}{6\sigma} \qquad \text{L2.15}$$

$$\text{and} \quad \text{Constant}(\beta) = \frac{4C_i G_1}{\pi}$$

And where equations L2.12 and L2.13 are equivalent to:

$$D_a = \frac{\beta}{\alpha} \quad \text{and} \quad \frac{dD}{dt} = \beta D - \alpha D^2 \qquad \text{L2.16}$$

With equation L2.12 an equilibrium droplet diameter can be found and/or with equation L2.14 a time dependent droplet diameter can be found, if second phase fluid concentration (C), second phase fluid/first phase supporting fluid interfacial tension, first phase supporting fluid absolute viscosity $(\mu_1)$ and first phase supporting fluid shear $(G_1)$ are known. From this relationship it can be seen that second phase fluid droplet size ($D$) is directly proportional to second phase fluid concentration ($C$) (the more second phase fluid drops to collide or coalesce to form bigger drops), directly proportional to interfacial tension $(\sigma)$ (the more interfacial tension $(\sigma)$ the more difficult to break up drops or decrease interfacial tension $(\sigma)$ by adding surfactant results in ease in drop break up), inversely proportional to first phase supporting fluid viscosity $(\mu_1)$ (the more viscous the supporting fluid, the more power is transmitted to the second phase fluid drop), and inversely proportional to first phase supporting fluid shear ($G$) (the more shear the smaller drop produced as in the blender analogy).

All these parameters are measurable except fluid shear ($G$)

which can be solved by equation L2.8 by introducing conventional head loss ($H$), flow rate ($Q$) and specific gravity $(\delta)$.

$$G_1^2 = \frac{P}{\mu_1 V} = \frac{Q\delta_1 H}{\mu_1 V} = \frac{V\!/\!t\ \delta H}{\mu_1 V} = \frac{Vol\!/\!t\ \delta H}{\mu_1 Vol} = \frac{\delta_1}{\mu_1}\frac{H}{t} \qquad \text{L2.17}$$

Now presenting the standard Darcy Weisbach head loss (for pipe) and minor loss (for valves, elbows etc.) formulas (reference 2) and where f is the Moody friction factor(which can be explicitly calculated in turbulent and laminar regions via reference 4), g is gravity acceleration, $L$ is pipe length, $K$ is unitless factor for particular valve, elbow etc., $R_h$ is hydraulic radius ($R_h$ = d/4 for pipe with diameter ($d$)) and $v$ is fluid velocity.

$$H = f\frac{L}{d}\frac{v^2}{2g} \quad \text{and} \quad H = K\frac{v^2}{2g} \qquad \text{L2.18}$$

Now letting v = L/t then

$$\frac{H}{t} = f\frac{1}{d}\frac{v^3}{2g} \quad \text{and} \quad \frac{H}{t} = \frac{K}{L}\frac{v^3}{2g} \qquad \text{L2.19}$$

Now combining equations L2.17 and L2.19 then fluid shear in the pipe or fitting can be calculated as in equation L2.20.

$$G_1^2 = \frac{\delta_1}{\mu_1}\frac{H}{t} = \frac{\delta_1}{\mu_1} f\frac{1}{d}\frac{v^3}{2g} \qquad \text{L2.20}$$

and

$$G_1^2 = \frac{\delta_1}{\mu_1}\frac{H}{t} = \frac{\delta_1}{\mu_1}\frac{K}{L}\frac{v^3}{2g}$$

Now all of the relationships are presented in order to calculate a droplet size as a function of head loss. The Reynolds number $(R_e)$ reference is presented in equation L2.21.

$$R_e = \frac{\delta_1 v 4 R_h}{g\mu_1} \text{ where } R_h \text{ is the hyraulic radius} \qquad \text{L2.21}$$

The following equation L2.22 represents the classical Stokes law relationship (with two phase specific weight difference $(v_s)$) to overflow rate with the principle of overflow rate (flow rate $Q$/(effective area ($A$))) as generally used to size separation devices by equating droplet settling/rising velocity $(v_s)$ to overflow rate.

$$v_s = \frac{Q}{A} = \frac{\Delta\delta D^2}{18\mu_1} \qquad \text{L2.22}$$

## L3. Model Development Including Second Phase Viscosity Contribution to Interfacial Tension

$$H_D = \frac{32\mu_2 D v_a}{\gamma_2 D^2} \qquad v_c = 2v_a \qquad \text{L3.1}$$

$$H_D = \frac{v_c - v_a}{D} = G_2 \qquad \text{L3.2}$$

And therefore:

$$\frac{2v_a - v_a}{D} = \frac{v_a}{D} = G_2 \qquad \text{L3.3}$$

And:

$$H_D = \frac{32\mu_2 G_o}{\gamma_2} \qquad \text{L3.4}$$

For a sphere with diameter(D), the interfacial tension $(\sigma_{\mu_2})$ is:

$$H_D = \frac{6}{\pi D^3}\frac{Energy}{\gamma_2} = \frac{6}{\gamma_2}\frac{Energy}{\pi D^2}\frac{1}{D} = \frac{6}{\gamma_2}\frac{\sigma_{\mu_2}}{D} \qquad \text{L.3.5}$$

For a six sided cube with each side (D), the interfacial tension $(\sigma_{\mu_2})$ is:

$$H_D = \frac{1}{D^3}\frac{Energy}{\gamma_2} = \frac{6}{\gamma_2}\frac{Energy}{6D^2}\frac{1}{D} = \frac{6}{\gamma_2}\frac{\sigma_{\mu_2}}{D} \qquad \text{L3.6}$$

And then combine (L3.5 or L 3.6) and L 3.4 and therefore:

$$H_D = \frac{32\mu_2 G_2}{\gamma_2} = \frac{6}{\gamma_2}\frac{\sigma_{\mu_2}}{D} \qquad \text{L3.7}$$

And now solve equation L3.7 for the interfacial tension $\left(\sigma_{\mu_2}\right)$:

$$\sigma_{\mu_2} = \frac{16}{3}\mu_2 G_2 D \qquad \text{L3.8}$$

And now from equation L1.4:

$$G_1^2 = \frac{P}{\mu_1 V} \text{ and } G_2^2 = \frac{P}{\mu_2 V} \qquad \text{L3.9}$$

Or by rearranging equation 2.9:

$$G_2 = \sqrt{\frac{\mu_1}{\mu_2}} G_1 \qquad \text{L3.10}$$

Now combine equations 3.8 and 3.10:

$$\sigma_{\mu_2} = \frac{16}{3}\mu_2 \sqrt{\frac{\mu_1}{\mu_2}} G_1 D \qquad \text{L3.11}$$

Now extending the basic model developed in equation 2.13 section 2 for the addition of the equivalent second phase viscosity:

$$\frac{dD}{dt} = \frac{4CG_1D}{\pi} - \frac{\mu_1 G_1^2 D^2}{6\left(\sigma + \sigma_{\mu_2}\right)} \qquad \text{L3.12}$$

Now substituting equation 3.11 into equation 3.12:

$$\frac{dD}{dt} = \frac{4CG_1D}{\pi} - \frac{\mu_1 G_1^2 D^2}{6\left(\sigma + \frac{16}{3}\mu_2 \sqrt{\frac{\mu_1}{\mu_2}} G_1 D\right)} \qquad \text{L3.13}$$

## L4. Charge Related Surface Tension

$$\sigma_{\pm} = \Psi_o (8RT\varepsilon\varepsilon_o(1000c))^{\frac{1}{2}} \sinh\left(\frac{Z\Psi_o F}{2RT}\right) \qquad \text{L3.14}$$

= charge related surface tension

Where:

$\Psi_o$ = potential (volt) (joule/coulumb) L3.15

Where:

$R$ = Gas Constant $\left(8.314 \text{ joule mol}^{-1} \text{ K}^{-1}\right)$ L3.16

Where:

$T$ = Temperature (degree Kelvin) L3.17

Where:

$\varepsilon$ = dielectric constant of first phase supporting fluid L3.18

Where:

$\varepsilon_o$ = permittivity of free space $\left(8.854 \text{ x } 10^{-12} \text{ coulumb volt}^{-1} \text{ meter}^{-1}\right)$ L3.13

Where:

$c$ = molar electrolyte concentration L3.19

Where:

$Z$ = ionic charge L3.20

Where:

$F$ = Faraday constant $\left(96485 \text{ coulumb mole}^{-1}\right)$ L3.21

## Appendix M. Classical Phase Separation

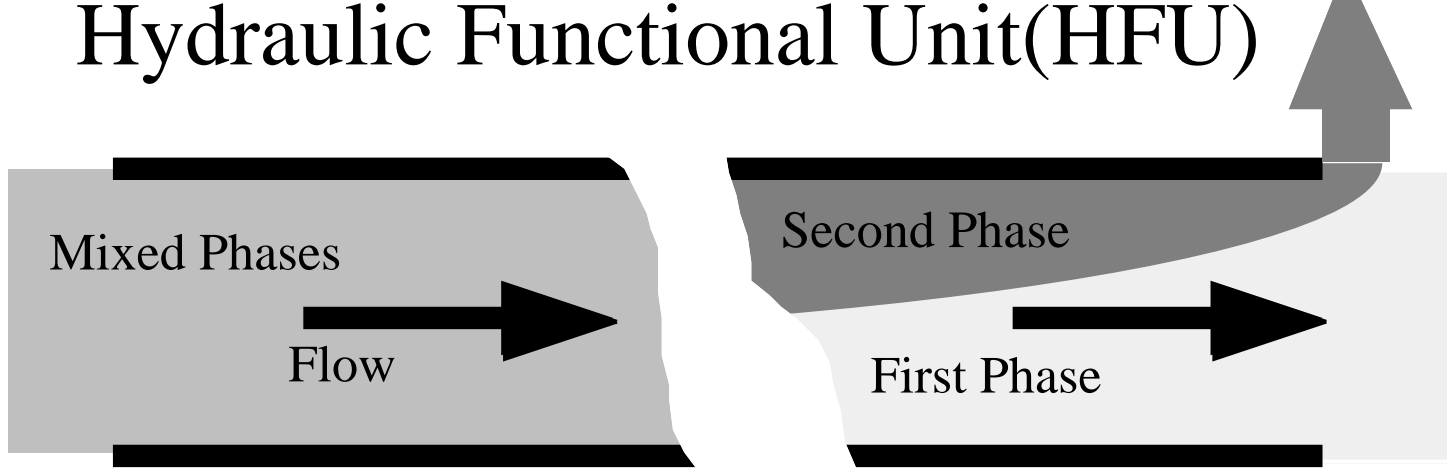

This work was primarily done in the 1980's as a theoretical framework for industrial separator design. As such it was done in the English (ft lb sec) system that was prevalent at that time although dimensionality dictates any other system such as cgs or SI may be used with equal applicability with the understanding that:

$$F = M\,a$$

$$lb_{force} = \frac{lb_{force}}{gravity_{acceleration}}\, gravity_{acceleration}$$

$$\partial = \rho g$$

$$\frac{lb_{force}}{ft^3} = \frac{lb_{force}}{ft^3\, gravity_{acceleration}}\, gravity_{acceleration}$$

Variable Definition

Fluid Viscosity $(\mu_1)$
Fluid Specific Weight $(\partial_1)$
Second Phase Specific Weight $(\partial_2)$
Gravity Acceleration $(g)$

## M1. PIPE HFU

**Figure** M1.1

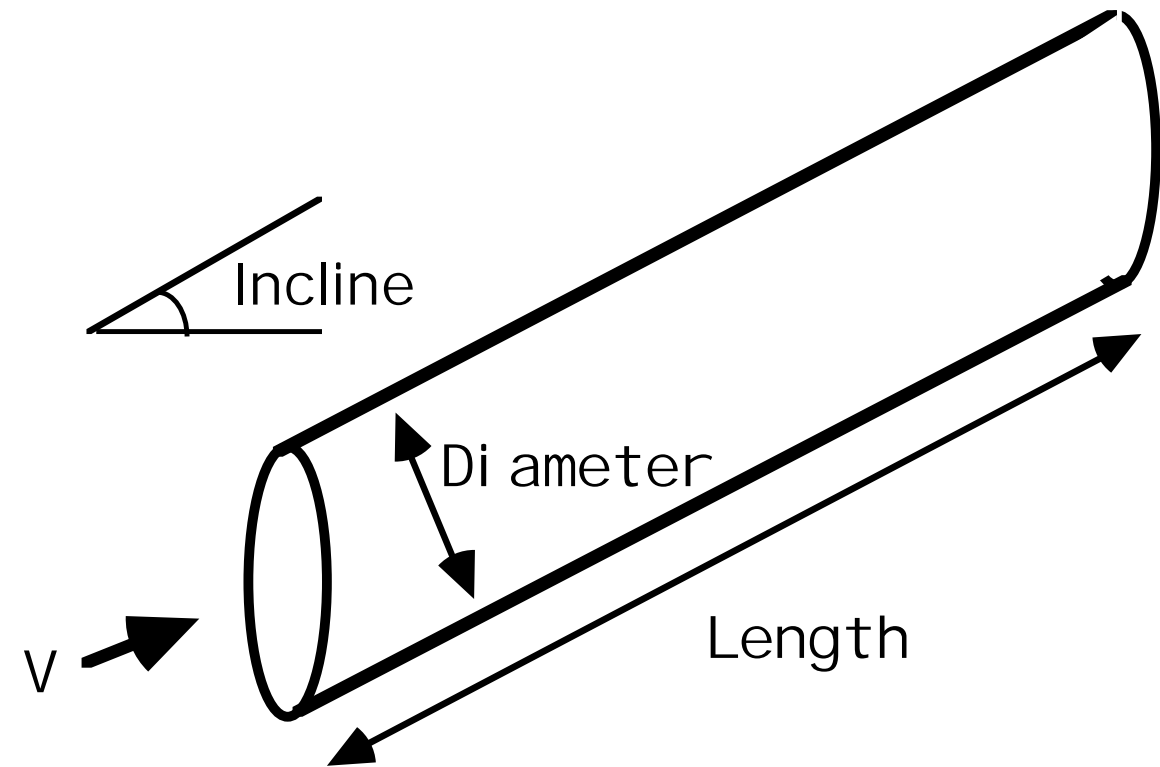

$$Approach\ Velocity(V) = \frac{4\ Flowrate}{\pi\ Diameter^2} \qquad \text{(M1.1)}$$

$$Internal\ Velocity(V_i) = V \qquad \text{(M1.2)}$$

$$AreaFactor(A_F) = \frac{Diameter}{Length\ \cos(Incline)} \qquad \text{(M1.3)}$$

$$Time(t) = \frac{Length}{V} \qquad \text{(M1.4)}$$

$$Hydraulic\ Radius(R_h) = \frac{Diameter}{4} \qquad \text{(M1.5)}$$

$$Head\ Loss\ Per\ Time\left(\frac{H}{t}\right) = \frac{fV^3}{8gR_h} \qquad \text{(M1.6)}$$

$$Roughness\ to\ Diameter\ Ratio(r) = \frac{Roughness}{Diameter} \qquad \text{(M1.7)}$$

## M2. OTHER HFU

**Figure** M2.1

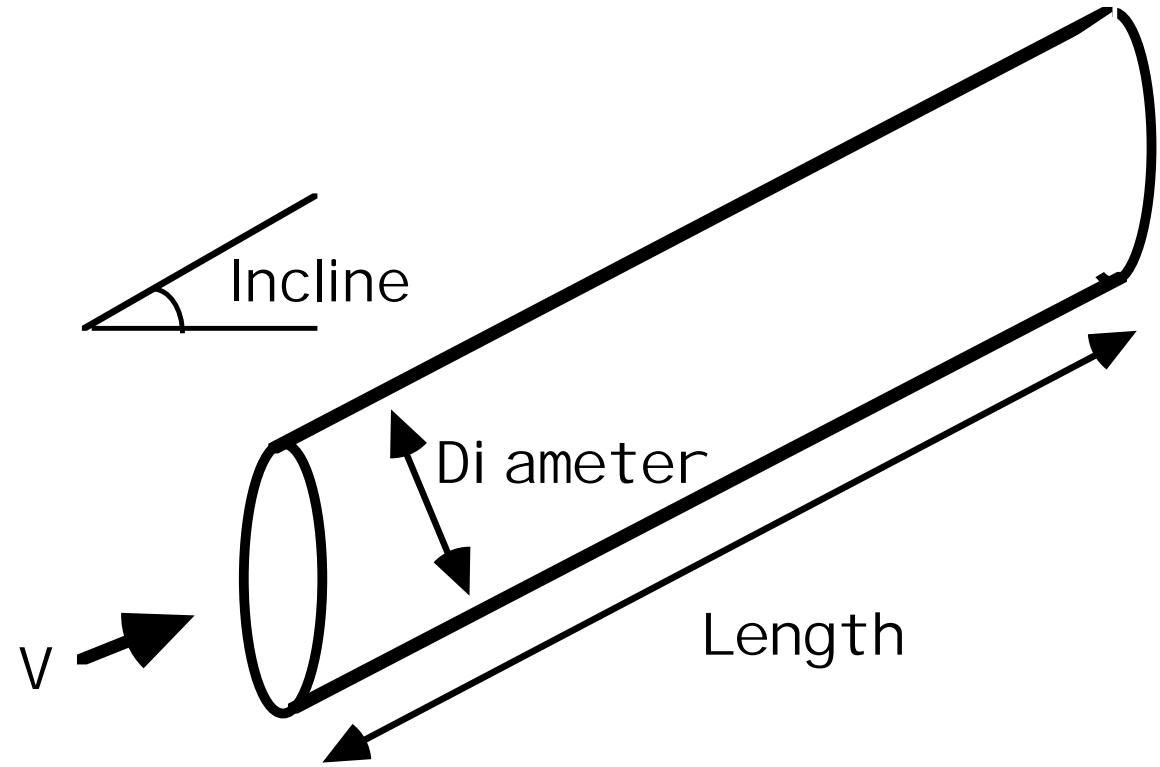

$$Approach\ Velocity(V) = \frac{4\ Flowrate}{\pi\ Diameter^2} \quad \text{(M2.1)}$$

$$Internal\ Velocity(V_i) = V \quad \text{(M2.2)}$$

$$AreaFactor(A_F) = \frac{Diameter}{Length\ \cos(Incline)} \quad \text{(M2.3)}$$

$$Time(t) = \frac{Length}{V} \quad \text{(M2.4)}$$

$$Hydraulic\ Radius(R_h) = \frac{Diameter}{4} \quad \text{(M2.5)}$$

$$Head\ Loss\ Per\ Time\left(\frac{H}{t}\right) = \frac{KV^3}{2\ g\ Length} \quad \text{(M2.6)}$$

## M3. PUMP HFU

**Figure** M3.1

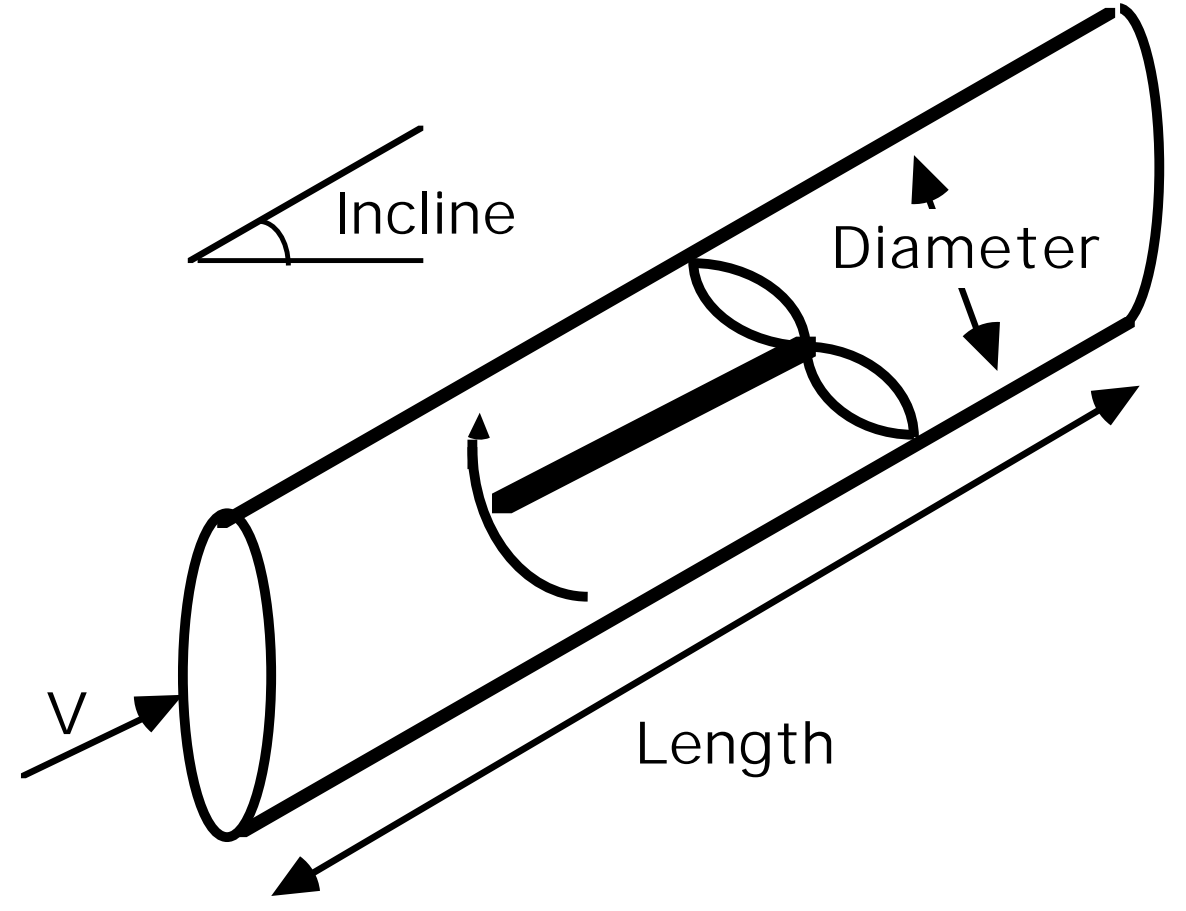

$$SpaceVelocity(V) = \left(\frac{8gR_h}{f + C_F}\frac{H}{t}\right)^{1/3} \quad \text{(M3.1)}$$

$$Internal\ Velocity(V_i) = V \quad \text{(M3.2)}$$

$$AreaFactor(A_F) = \frac{Diameter}{Length\ \cos(Incline)} \quad \text{(M3.3)}$$

$$Time(t) = \frac{Length}{V} \quad \text{(M3.4)}$$

$$Hydraulic\ Radius(R_h) = \frac{Diameter}{4} \quad \text{(M3.5)}$$

$$Head\ Loss\ Per\ Time\left(\frac{H}{t}\right) = \frac{(1 - E_F)(Pumphead + Head)}{t} \quad \text{(M3.6)}$$

$$Roughness\ to\ Diameter\ Ratio(r) = \frac{Roughness}{Diameter} \quad \text{(M3.7)}$$

## M4. VORTEX HFU

**Figure** M4.1

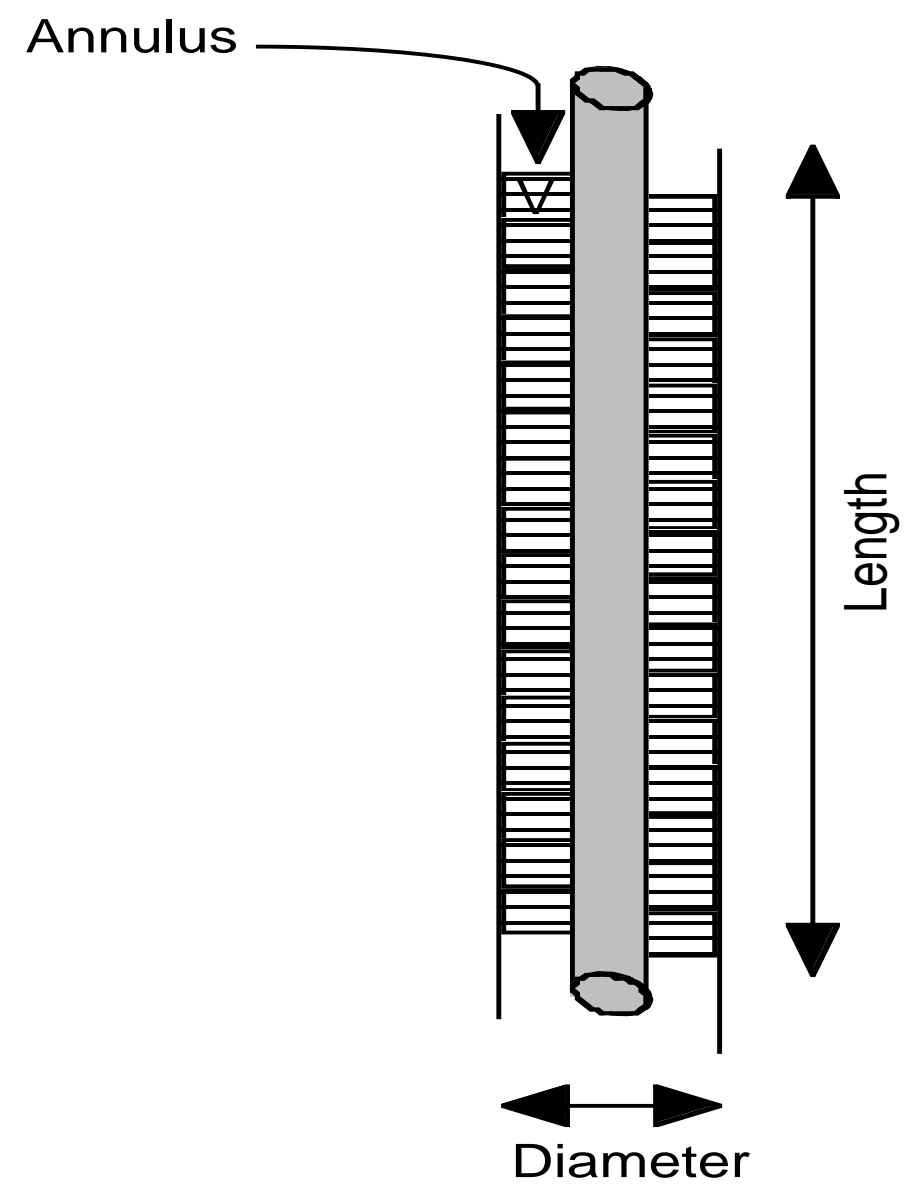

*Approach Area* = (M4.1)

$$\frac{\pi}{4}\left(Diameter^2 - \left(Diameter - 2Annulus\right)^2\right)$$

$$Approach\ Velocity(V_A) = \frac{Flowrate}{Approach\ Area} \quad (M4.2)$$

*Tangential Velocity* $(V_T)$ (M4.3)

$$= \frac{Flowrate}{K\ \pi(Diameter)(Annulus)}$$

$$Internal\ Velocity(V_i) = V_A + V_T \quad (M4.4)$$

$$Area\ Factor(A_F) = \frac{Annulus}{V_i t} \quad (M4.5)$$

$$Time(t) = \frac{\left(Approach\ Area\right)\ Length}{Flowrate} \quad (M4.6)$$

$$Hydraulic\ Radius(R_h) = \frac{Diameter}{4} \quad (M4.7)$$

*Tangential gravity* $\left(g_T\right)$ = (M4.8)

$$g + \frac{V_T^2}{\dfrac{Diameter}{2} - \dfrac{Annulus}{3}}$$

$$Head\ Loss\ Per\ Time\left(\frac{H}{t}\right) = \frac{fV_A^3}{8g_T R_h} \quad (M4.9)$$

*Roughness to Diameter Ratio* $(r)$ = (M4.10)

$$\frac{Roughness}{Diameter}$$

$$Vortex\ Reynolds\ Number = \frac{\partial_1 V_A 4R_h}{g_T \mu_1} \quad (M4.11)$$

## M5. MIXER HFU

**Figure** M5.1

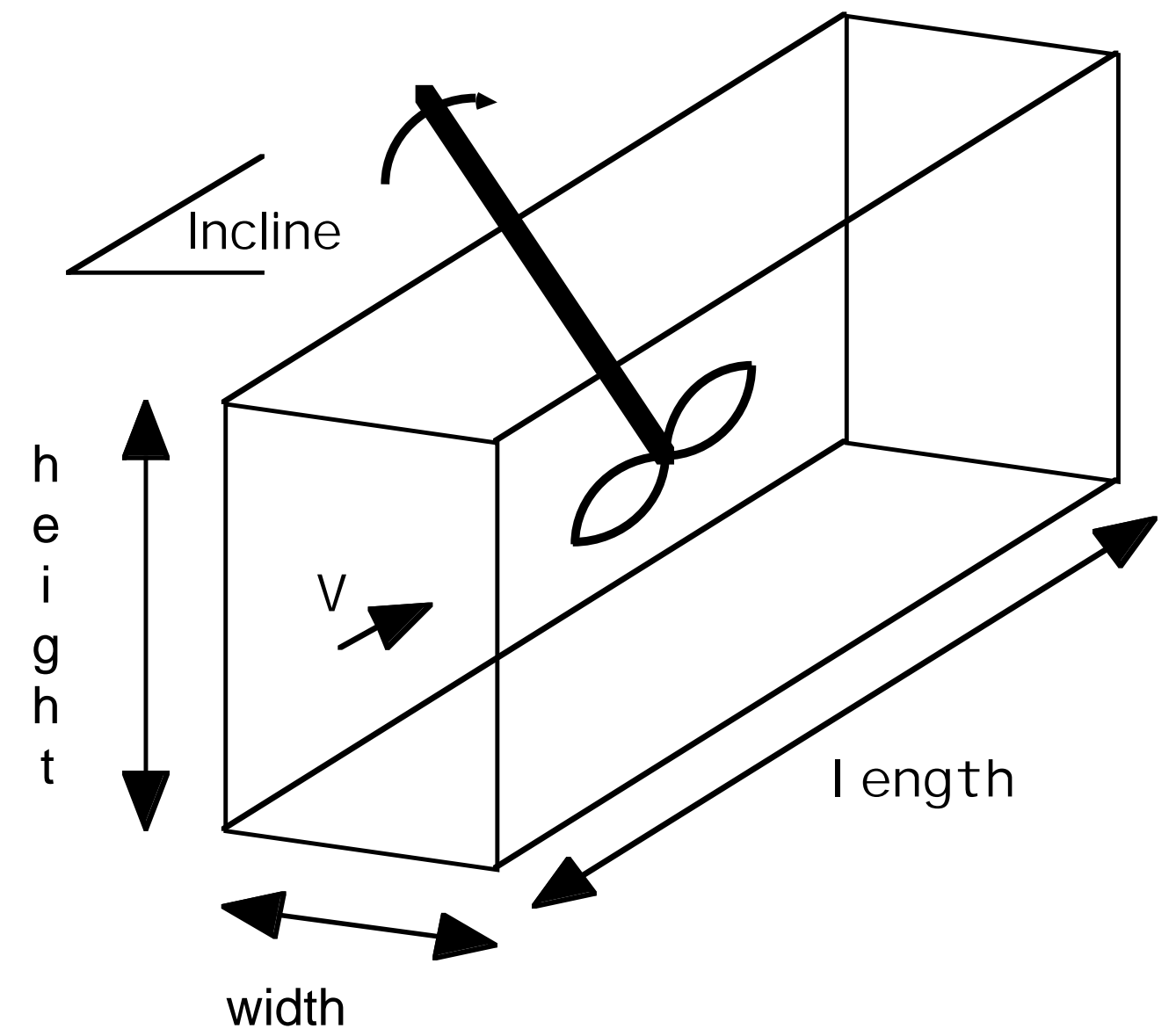

$$Internal\ Velocity(V_i) = V \quad (M5.1)$$

$$AreaFactor(A_F) = \frac{Height}{LengthCOS(Incline)} \quad (M5.2)$$

$$Time(t) = \frac{Length}{V} \quad (M5.3)$$

$$Hydraulic\ Radius(R_h) = \frac{Width\ Height}{2(Width + Height)} \quad (M5.4)$$

$$Head\ Loss\ Per\ Time\left(\frac{H}{t}\right) = \frac{SHead}{\partial_1 Height\ Width\ Length} \quad \text{(M5.5)}$$

$$SpaceVelocity(V) = \left(\frac{8gR_h}{f + C_F}\frac{H}{t}\right)^{1/3} \quad \text{(M5.6)}$$

$$Roughness\ to\ Diameter\ Ratio(r) = \frac{Roughness}{Diameter} \quad \text{(M5.7)}$$

## M6. COALESCER HFU

**Figure** M6.1

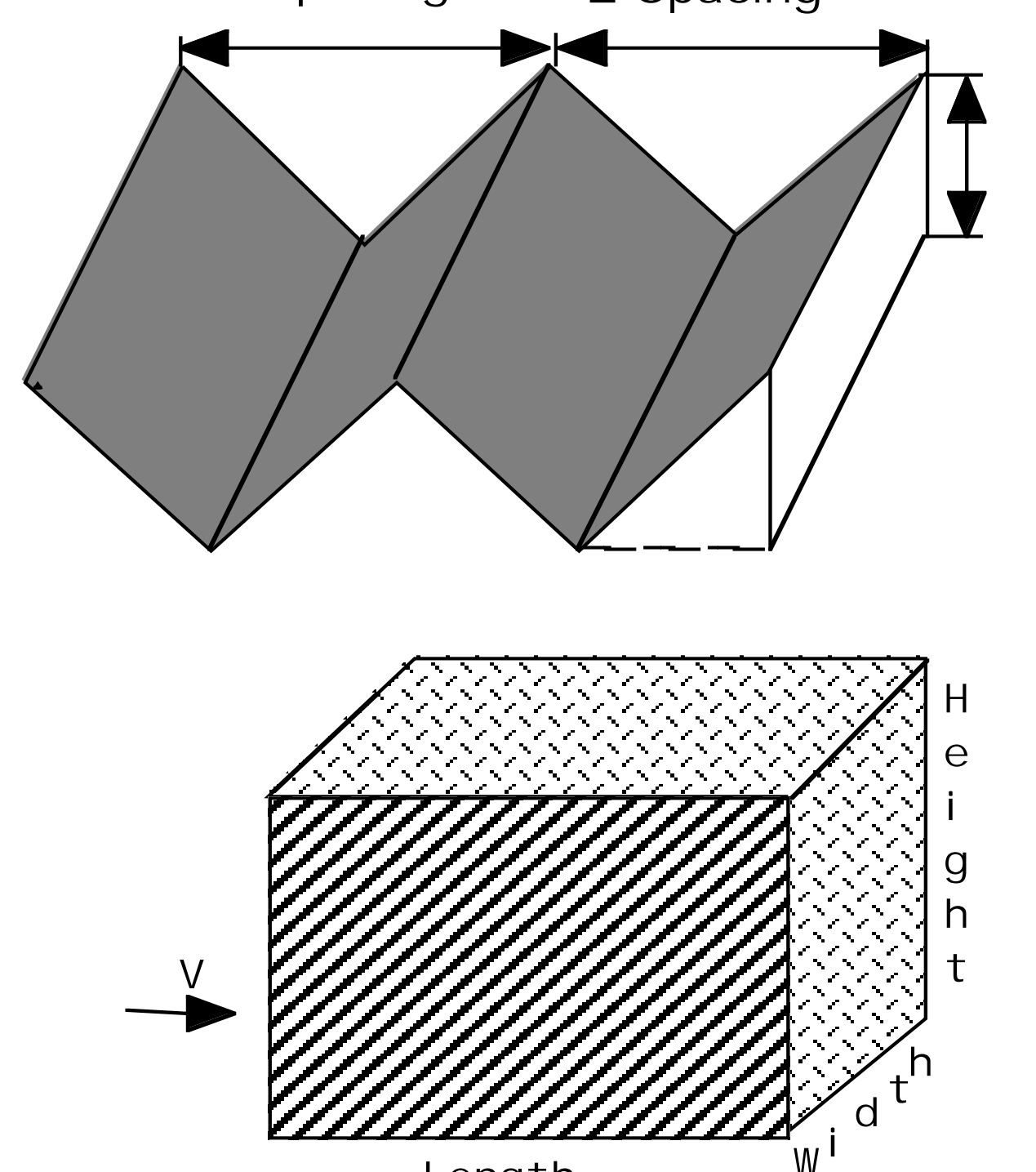

$$Approach\ Velocity(V) = \frac{Flowrate}{Height\ Width} \quad \text{(M6.1)}$$

$$Internal\ Velocity(V_i) = \frac{V}{Porosity\ \cos(Angle)} \quad \text{(M6.2)}$$

$$Area\ Factor(A_F) = \frac{Height}{Length\ \cos(Incline)} \quad \text{(M6.3)}$$

$$Time(t) = \frac{Length}{V} \quad \text{(M6.4)}$$

$$B_D = \frac{2\ Spacing\ \cos(Angle)}{\sin(2Angle)} \quad \text{(M6.5)}$$

$$A_D = \frac{2\ Spacing}{\sin(2\ Angle)} \quad \text{(M6.6)}$$

$$Head\ Loss\ Per\ Time\left(\frac{H}{t}\right) = \frac{fVV_i^2}{8gR_h} \quad \text{(M6.7)}$$

$$Hydraulic\ Radius(R_h) = \frac{1}{\frac{2}{A_D} + \frac{2}{B_D}} \quad \text{(M6.8)}$$

## M7. PARALLEL PLATE SEPARATOR HFU

**Figure** M7.1

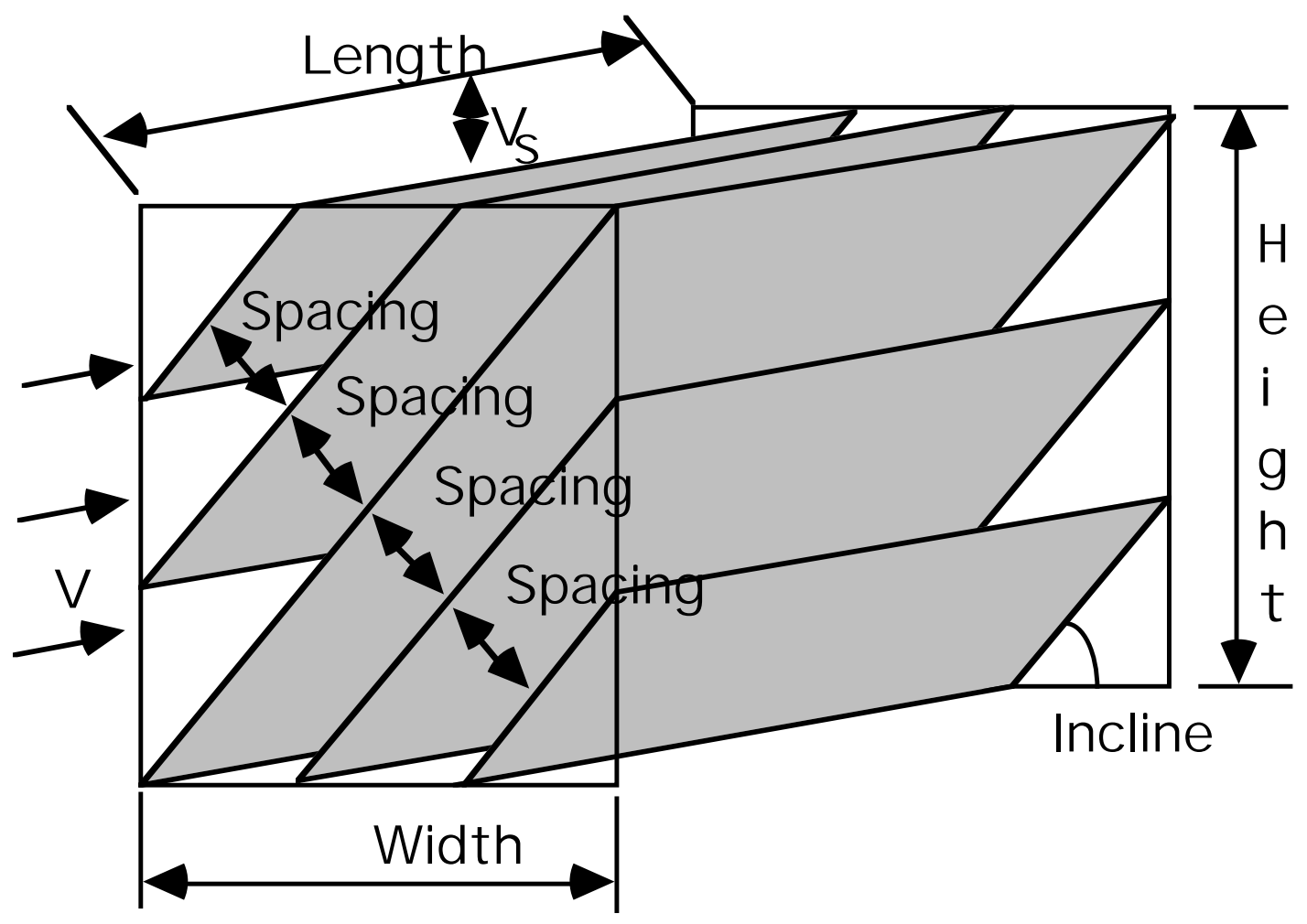

$$Approach\ Velocity(V) = \frac{Flowrate}{Height\ Width} \quad \text{(M7.1)}$$

$$Internal\ Velocity(V_i) = \frac{V}{Porosity} \quad \text{(M7.2)}$$

$$Area\ Factor(A_F) = \frac{Spacing}{Length \cos(Incline)} \quad \text{(M7.3)}$$

$$Time(t) = \frac{Length}{V} \quad \text{(M7.4)}$$

$$Head\ Loss\ Per\ Time\left(\frac{H}{t}\right) = \frac{fV^3}{8gR_h} \quad \text{(M7.5)}$$

$$Hydraulic\ Radius(R_h) = \frac{Spacing}{2} \quad \text{(M7.6)}$$

## M8. CURVED ELEMENT SEPARATOR

**Figure** M8.1

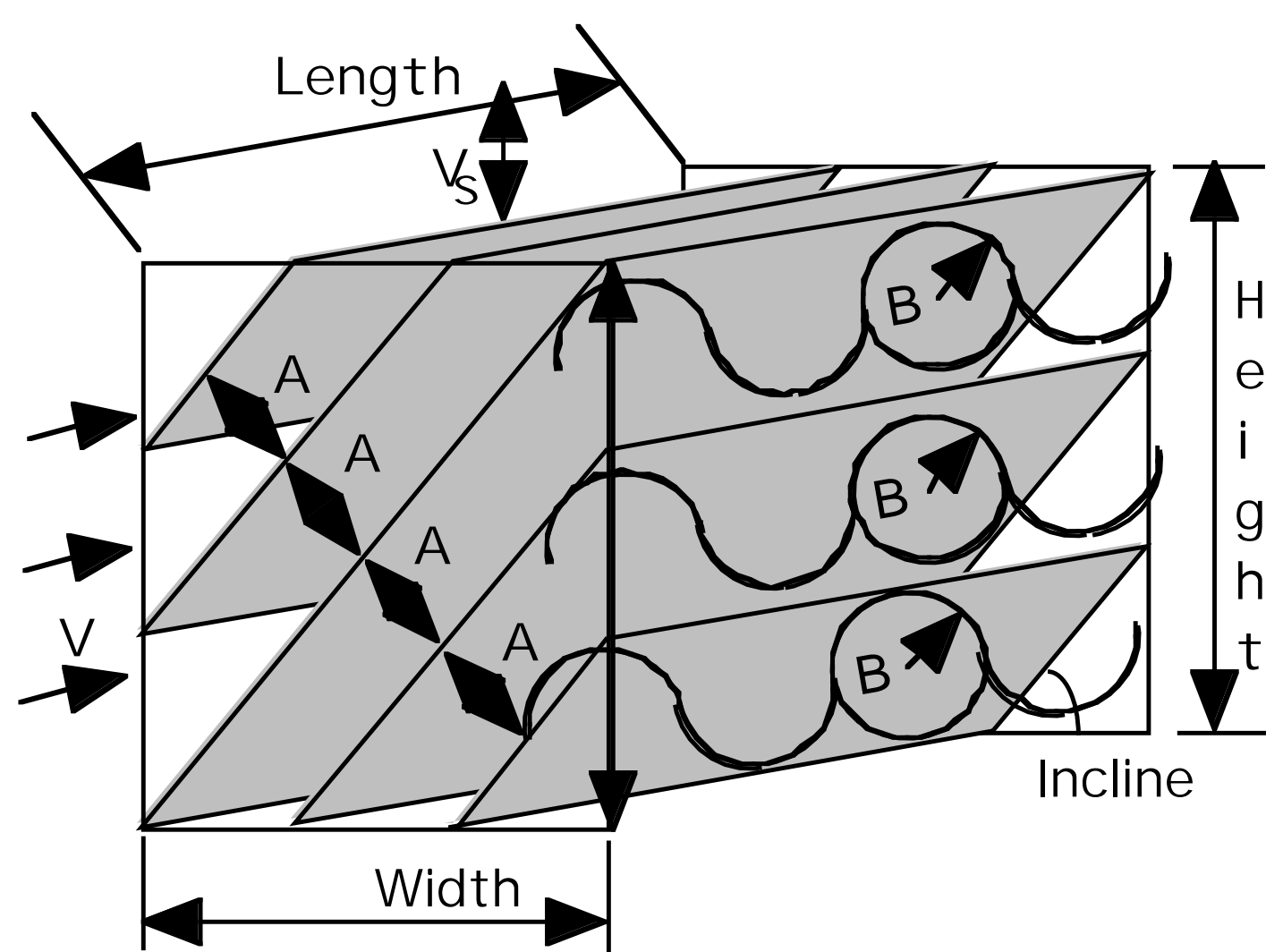

$$Approach\ Velocity(V) = \frac{Flowrate}{Height\ Width} \quad \text{(M8.1)}$$

$$Internal\ Velocity(V_i) = \frac{V}{Porosity\ \cos(Angle)} \quad \text{(M8.2)}$$

$$Area\ Factor(A_F) = \frac{A}{Length \cos(Incline)} \quad \text{(M8.3)}$$

$$Time(t) = \frac{Length}{V} \quad \text{(M8.4)}$$

$$Head\ Loss\ Per\ Time\left(\frac{H}{t}\right) = \frac{fVV_i^2}{8gR_h} \quad \text{(M8.5)}$$

$$Hydraulic\ Radius(R_h) = \frac{1}{\frac{2}{A} + \frac{2}{B}} \quad \text{(M8.6)}$$

## M9. OPEN CHANNEL HFU

**Figure** M9.1

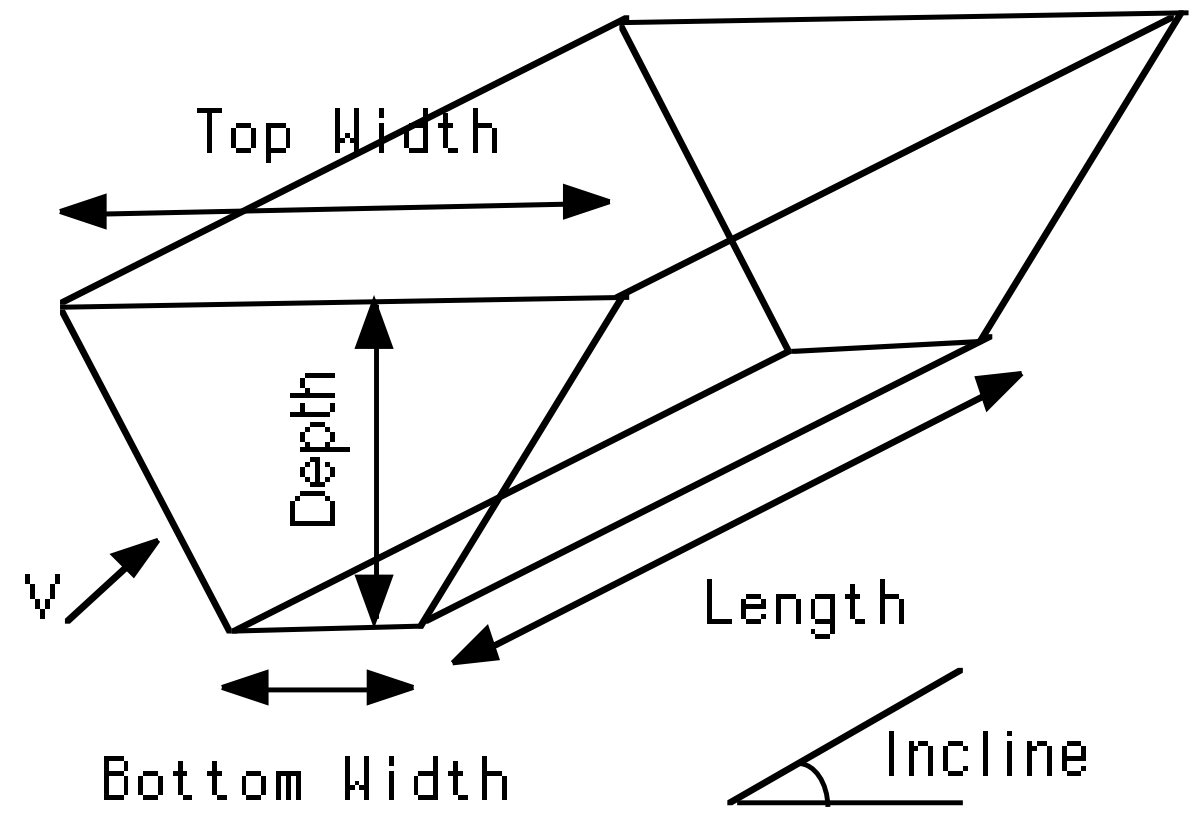

$$X_F = \frac{Depth}{Bottom\ Width} \quad \text{(M9.1)}$$

$$e_F = \frac{Top\ Width - Bottom\ Width}{2} \quad \text{(M9.2)}$$

$$Z_F = \frac{e_F}{Depth} \quad \text{(M9.3)}$$

$$Wetted\ Perimeter(P) = \quad \text{(M9.4)}$$

$$Bottom\ Width + 2(e_F^2 + Depth_F^2)^{\frac{1}{2}}$$

$$Approach\ Area = \left(\frac{1}{X_F} + Z_F\right) Depth^2 \quad \text{(M9.5)}$$

$$Approach\ Velocity(V) = \frac{Flowrate}{Approach\ Area} \quad \text{(M9.6)}$$

$$Internal\ Velocity(V_i) = V \quad \text{(M9.7)}$$

$$Area\ Factor(A_F) = \frac{Depth}{Length \cos(Incline)} \quad \text{(M9.8)}$$

$$Time(t) = \frac{Length}{V} \quad \text{(M9.9)}$$

$$Head\ Loss\ Per\ Time\left(\frac{H}{t}\right) = \frac{fV^3}{8gR_h} \quad \text{(M9.10)}$$

$$Hydraulic\ Radius(R_h) = \frac{A}{P} \quad \text{(M9.11)}$$

$$Roughness\ to\ Diameter\ Ratio(r) = \frac{Roughness}{Height} \quad \text{(M9.12)}$$

## M10. CLOSED CHANNEL HFU

**Figure** M10.1

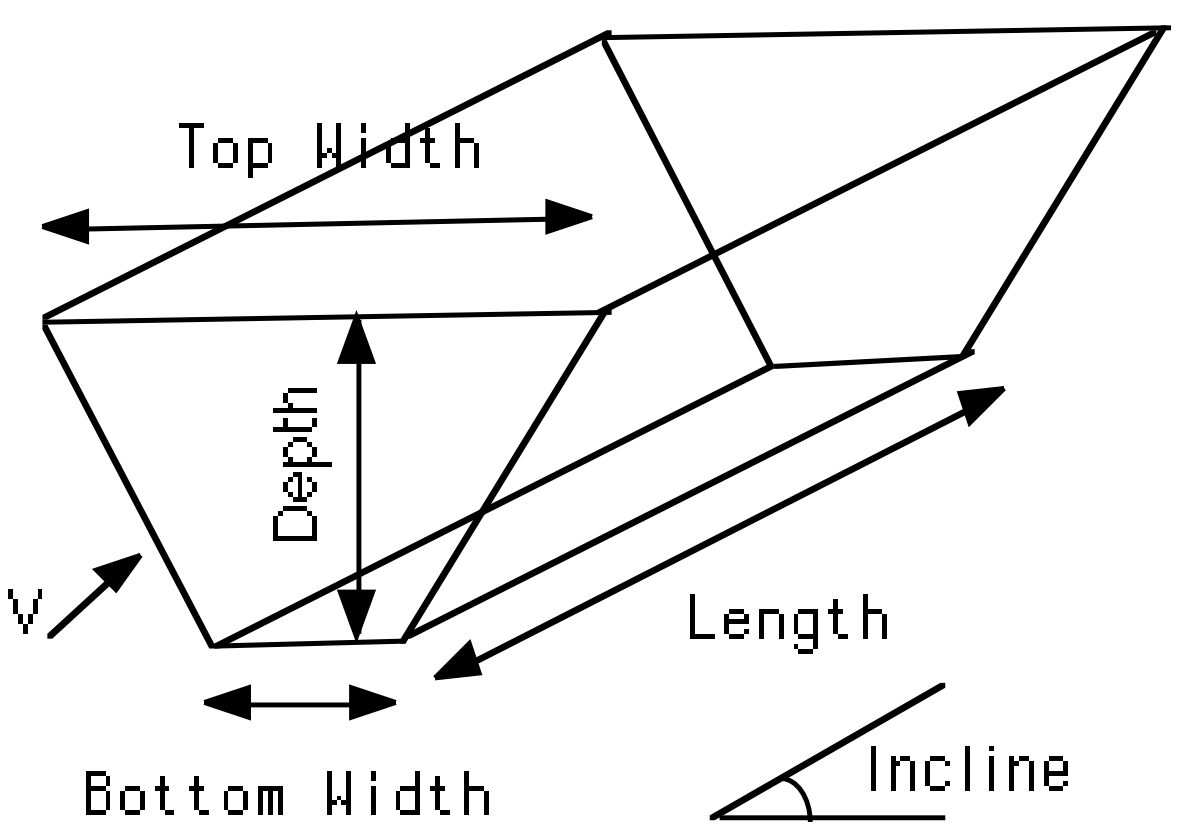

$$X_F = \frac{Depth}{Bottom\ Width} \quad \text{(M10.1)}$$

$$e_F = \frac{Top\ Width - Bottom\ Width}{2} \quad \text{(M10.2)}$$

$$Z_F = \frac{e_F}{Depth} \quad \text{(M10.3)}$$

$$Wetted\ Perimeter(P) = \quad \text{(M10.4)}$$

$$2e_F + 2Bottom\ Width + 2(e_F^2 + Depth_F^2)^{\frac{1}{2}}$$

$$Approach\ \text{Area} = \left(\frac{1}{X_F} + Z_F\right)Depth^2 \quad \text{(M10.5)}$$

$$Approach\ Velocity(V) = \frac{Flowrate}{Approach\ \text{Area}} \quad \text{(M10.6)}$$

$$Internal\ Velocity(V_i) = V \quad \text{(M10.7)}$$

$$Area\ Factor(A_F) = \frac{Depth}{Length\cos(Incline)} \quad \text{(M10.8)}$$

$$Time(t) = \frac{Length}{V} \quad \text{(M10.9)}$$

$$Head\ Loss\ Per\ Time\left(\frac{H}{t}\right) = \frac{fV^3}{8gR_h} \quad \text{(M10.10)}$$

$$Hydraulic\ Radius(R_h) = \frac{A}{P} \quad \text{(M10.11)}$$

$$Roughness\ to\ Diameter\ Ratio(r) = \quad \text{(M10.12)}$$

$$\frac{Roughness}{Height}$$

## M11. PARTICLE DENSITY

**Figure** M11.1 Third Phase Drop Coated with Second Phase

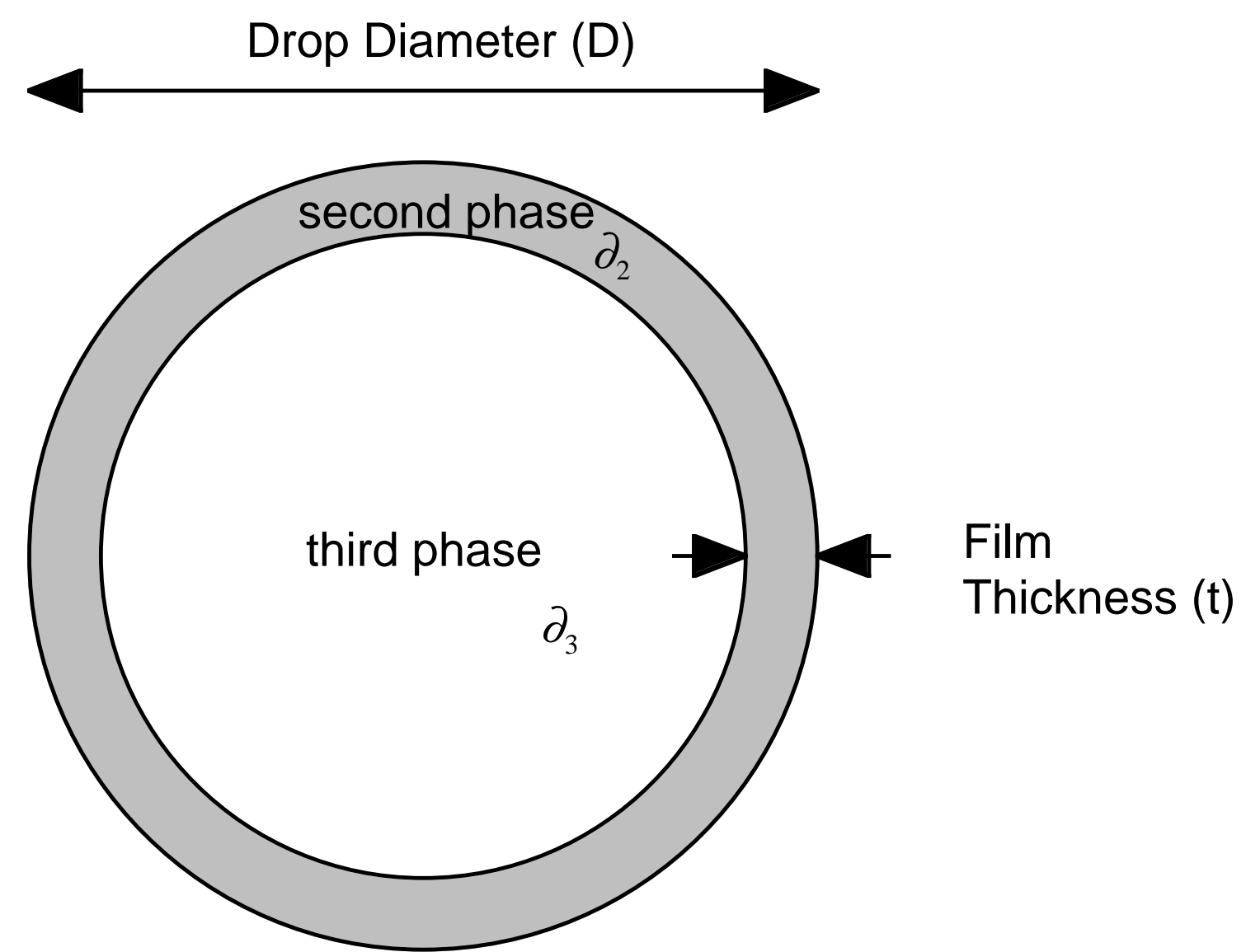

HFU Inlet Concentration $(C_i)$

$$C_i = C_2 + C_3 \quad \text{(M11.1)}$$

Particle Film Thickness (t)

$$t = \left(1 - \frac{C_3}{C_2 + C_3}\right)\frac{D}{2} \quad \text{(M11.2)}$$

Particle Density $(\partial_D)$ on addition of a third gaseus phase with density $(\partial_3)$

$$\partial_D = \frac{\partial_2 \left(D-2t\right)^3 + \partial_3 D^3}{D^3} \qquad \text{(M11.3)}$$

## M12. PARTICLE COALESCENCE

Figure M12.1

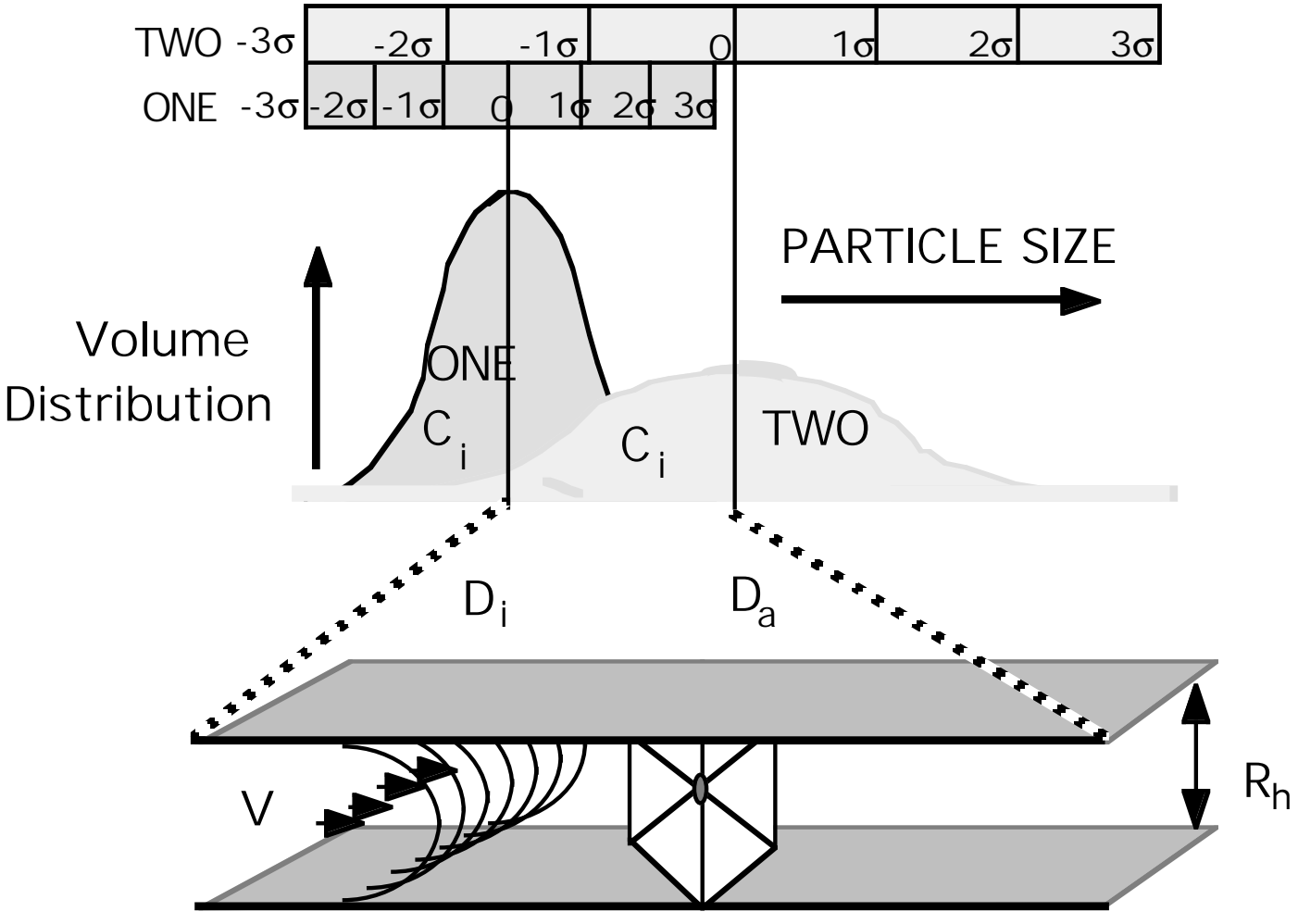

HFU Reynolds Number $\left(R_e\right)$

$$R_e = \frac{\partial_1 V_i 4 R_h}{g \mu_1} \qquad \text{(M12.1)}$$

Churchill interim factor $\left(E\right)$ in terms of $R_e$ and roughness factor $\left(r\right)$

$$\mathrm{E} = \left(-2.457 \ln\left(\left(\frac{7}{R_e}\right)^{.9} + .27r\right)\right)^{16} \qquad \text{(M12.2)}$$

Churchill Flow Friction Factor $\left(f\right)$

$$f = 8\left(\left(\frac{8}{R_e}\right)^{12} + \left(\left(\frac{37530}{R_e}\right)^{16} + E\right)^{-1.5}\right)^{\frac{1}{12}} \qquad \text{(M12.3)}$$

Fluctuation Thermal Kinetic Shear $\left(G_k\right)$

$$G_k = \frac{8 k_b T}{\mu_1 D^3} \qquad \text{(M12.4)}$$

Friction Factor Shear $\left(G_f\right)$

$$G_f = \sqrt{\frac{\partial_1 H}{\mu_1 t}} \qquad \text{(M12.5)}$$

Distance $\left(R\right)$ between Particles

$$R = \left(\frac{3}{2C_i}\right)^{1/3} D \qquad \text{(M12.6)}$$

Coalesced Particle Diameter $\left(D_a\right)$

$$D_a = \frac{\delta D_i \exp\left(\delta\, t\right)}{\alpha D_i \left(\exp\left(\delta\, t\right) - 1\right) + \delta} \qquad \text{(M12.7)}$$

$$\text{Constant}(\alpha) = \frac{\mu_1 G^2}{6 \beta_e} - \frac{2 F_s G}{3 \pi R_h} \qquad \text{(M12.8)}$$

$$\text{Constant}(\delta) = \frac{4 F_p C_i G}{\pi} \qquad \text{(M12.9)}$$

Surface Factor $\left(F_s\right)$ and Sphere Factor $\left(F_p\right)$

$$F_s = \exp\left(\frac{-12 Zp}{\pi \partial_D G}\right) \qquad \text{(M12.10)}$$

$$F_p = \exp\left(\frac{-6 Zp}{\pi \partial_D G}\right) \qquad \text{(M12.11)}$$

Gibbs Surface Tension $\left(\beta\right)$.

$$\beta = \frac{\mathit{Energy}}{\mathit{Interfacial\ Area\ between\ Phases}\ 1 \,\&\, 2} \qquad \text{(M12.12)}$$

Effective Surface Tension $\left(\beta_e\right)$

$$\beta_e = \frac{16}{3}\mu_2\sqrt{\frac{\mu_1}{\mu_2}}GD + \beta \qquad \text{(M12.13)}$$

## M13. SIMILITUDE PARTICLE

Generalized HFU/Particle Similitude Assumption

$$HFU\ Shear(G) = Particle\ Shear\left(\frac{v_{sr}}{D_r}\right) \qquad \text{(M13.1)}$$

Similitude Particle Settling/Rising Velocity $\left(v_{sr}\right)$

$$v_{sr} = \frac{\left|\partial_1 - \partial_D\right| D_r^2}{18\mu_1} \qquad \text{(M13.2)}$$

Similitude Particle $\left(D_r\right)$

$$D_r = \frac{2}{3}\frac{a\mu_1 G}{\left|\partial_1 - \partial_D\right|} \qquad \text{(M13.3)}$$

where $a$ = Toppler Factor
with typical value of 5

## M14. SETTLING/RISING PARTICLE

Particle Settling Velocity $\left(v_s\right)$

$$v_s = A_F v_i \qquad \text{(M14.1)}$$

Particle Drag Coefficient $\left(C_d\right)$

$$C_d = \frac{24}{R_{eP}} + \frac{3}{R_{eP}^{1/2}} + .34 \qquad \text{(M14.2)}$$

Particle Reynolds Number $\left(R_{ep}\right)$

$$R_{eP} = \frac{\partial_1 v_s D_c}{g\mu_1} \qquad \text{(M14.3)}$$

Settling/Rising Particle $\left(D_c\right)$

$$D_c = \frac{3C_d\partial_1 v_s^2}{4\left|\partial_1 - \partial_D\right| g} \qquad \text{(M14.4)}$$

## M15. PARTICLE REMOVAL

Figure M15.1

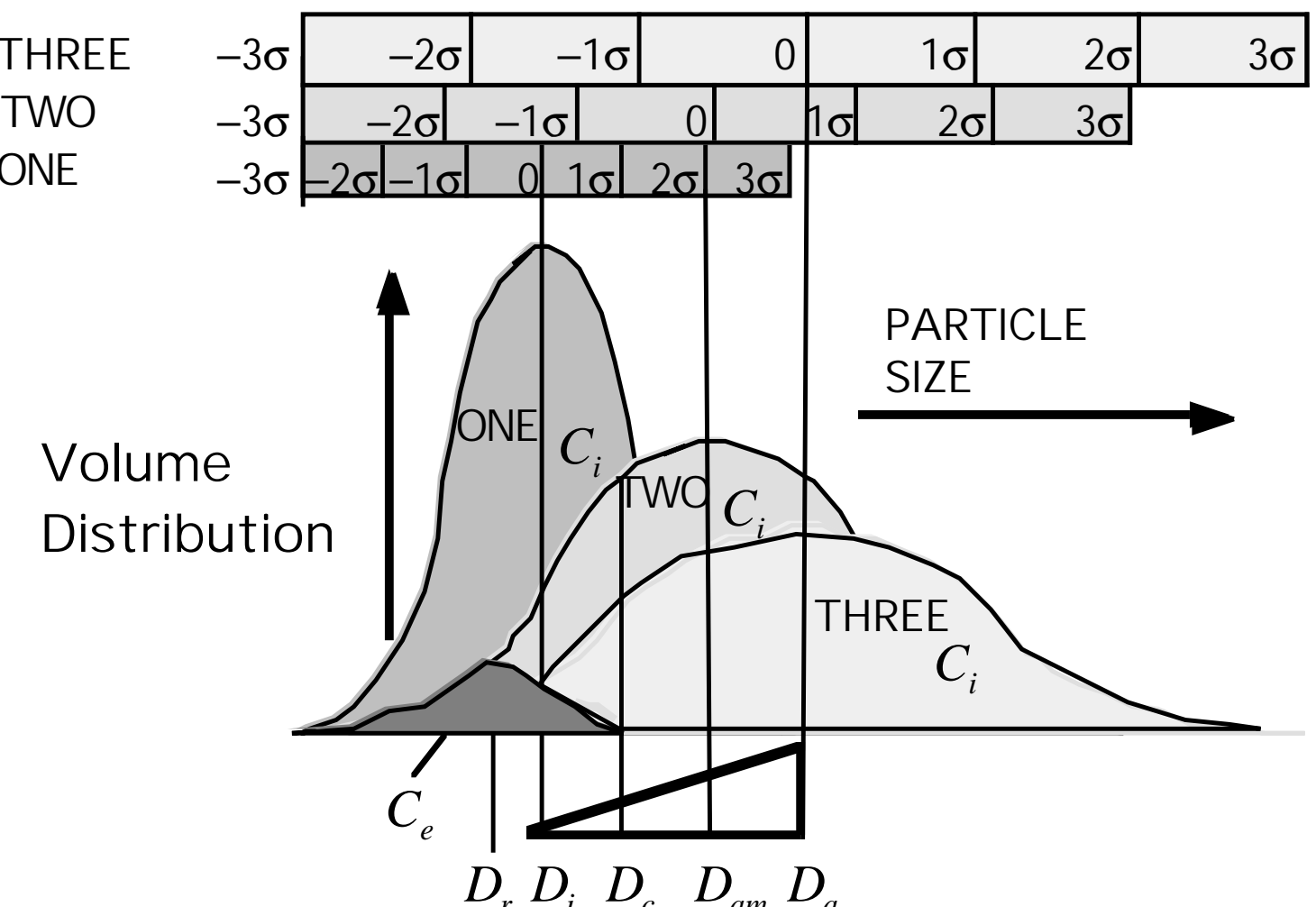

Mean Coalesced Particle Diameter $\left(D_{am}\right)$

$$D_{am} = \frac{1}{3}D_i + \frac{2}{3}D_a \qquad \text{(M15.1)}$$

Gaussian Influent Particle Distribution $\left(f\left(D\right)\right)$

$$f(D) = \frac{1}{\sqrt{2\pi}}\exp\left(\frac{-1}{2}\left(\frac{D_{am} - D}{D_{am}}\right)^2\right) = \frac{1}{\sqrt{2\pi}}\exp\left(\frac{-\sigma^2}{2}\right) \qquad \text{(M15.2)}$$

Influent concentration $\left(C_i\right)$

$$C_i = \int_0^\infty D^n f(D)\,dD \qquad \text{(M15.3)}$$

Camp's Particle Effluent Stokes Trial 1 $\left(C_{e1}\right)$ and Similitude Trial 2 $\left(C_{e2}\right)$ Effluents.

$$C_{e1} = \int_0^{D_C}\left(1 - \frac{D^2}{D_c^2}\right)D^n f(D)\,dD \qquad \text{(M15.4)}$$

$$C_{e2} = \int_0^{aD_s} \left(1 - \frac{D^2}{(aD_s)^2}\right) D^n f(D)\, dD \tag{M15.5}$$

$1.2 < a < 2.8$

| n | Distribution | n | Distribution |
|---|---|---|---|
| 0 | Volume | 3 | Number |
| 1 | Area | 4 | Reciprocal Diameter |
| 2 | Diameter | 6 | Reciprocal Volume |

**Figure** M15.2

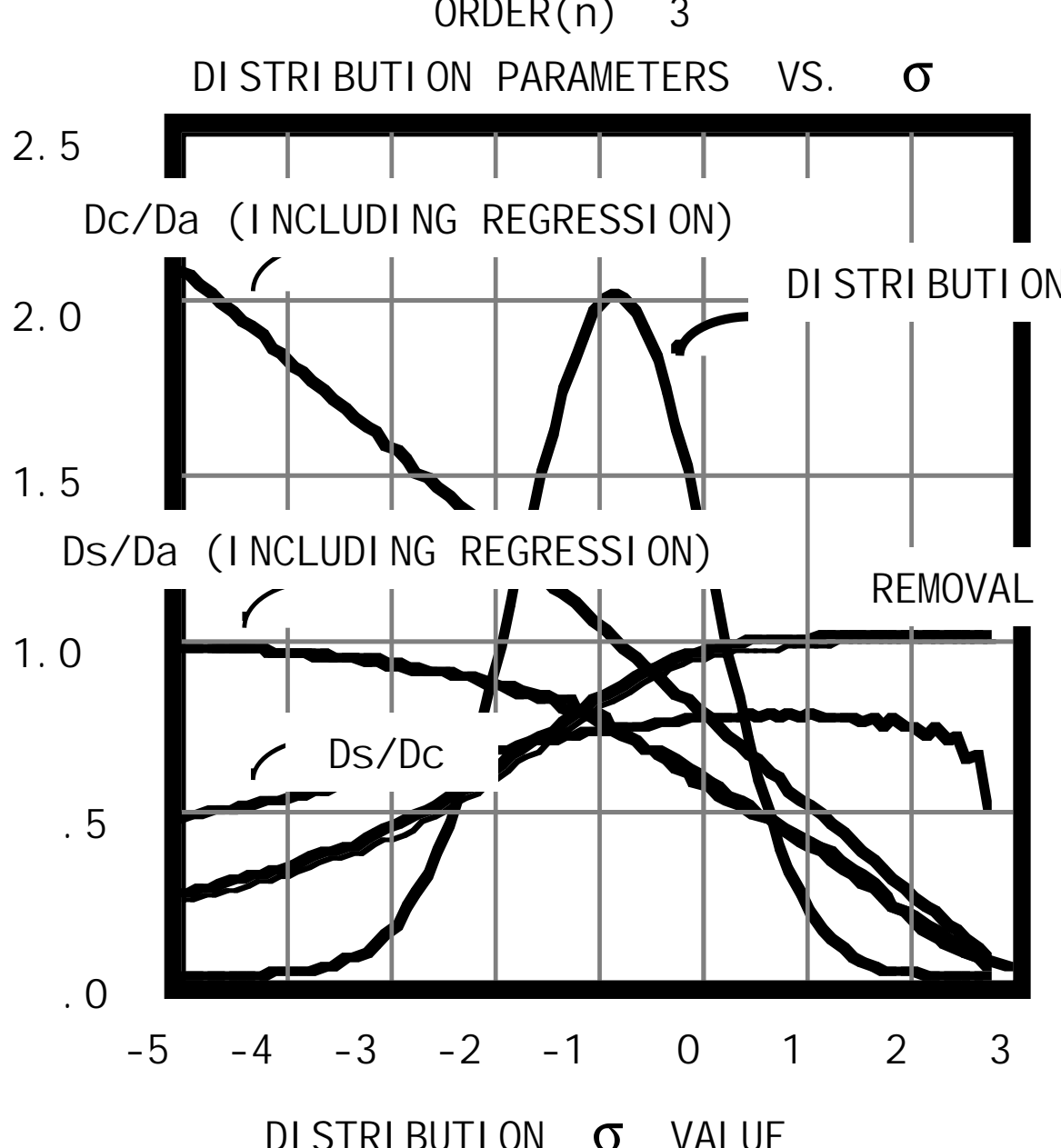

HFU Effluent $(C_e)$

$$C_e = \text{Maximum}(C_{e1}, C_{e2}) \tag{M15.6}$$

## Appendix N. Authority for Expenditure (Project Cost Estimate)

| | | | |
|---|---|---|---|
| AFE number | 51507 | | |
| Project Name | Three Dimensional Laser Energetics<br>Relating Trisine Geometry to the CPT theorem | Date | May 15, 2007 |
| Location | Corpus Christi, Texas | | |
| Description | Apparatus and Support Facilities for 10 years | | |

|  |  |  |  |  | Completed Cost |
|---|---|---|---|---|---|
| PROJECT - INTANGIBLE COSTS | | | | | |
| Project Physicists | | | | | $60,000,000 |
| Project Chemists | | | | | $60,000,000 |
| Project Engineers | | | | | $60,000,000 |
| Technicians Draftsmen Electronic Electrical Specifications | | | | | $60,000,000 |
| Salaries management and office staff | | | | | $20,000,000 |
| Architectural Services | | | | | $1,500,000 |
| Landscaping | | | | | $500,000 |
| Maintenance | | | | | $1,000,000 |
| Office Library Supplies | | | | | $2,000,000 |
| Library Reference Searches | | | | | $3,000,000 |
| Patent Trademark Attorney Services | | | | | $1,500,000 |
| Legal Services | | | | | $5,000,000 |
| Financial Services | | | | | $500,000 |
| Travel | | | | | $5,000,000 |
| Publication Services | | | | | $1,000,000 |
| Marketing | | | | | $500,000 |
| Infrastructure maintainance | | | | | $10,000,000 |
| Utilities | | | | | $1,000,000 |
| Security | | | | | $1,000,000 |
| | | | | | $0 |
| SubTotal | | | | | $293,500,000 |
| Contingencies | 50.00% | | | | $146,750,000 |
| PROJECT - INTANGIBLE COSTS | | | | | $440,250,000 |
| | | | | | |
| PROJECT - TANGIBLE COSTS | | | | | |
| Vibration Attenuation foundation | | | | | $5,000,000 |
| Laser Reactor Area | 5000 | sq ft | $5,000 | per sq ft | $25,000,000 |
| Assembly Area | 10000 | sq ft | $500 | per sq ft | $5,000,000 |
| Office Laboratory Library Visualization Space | 20000 | sq ft | $500 | per sq ft | $10,000,000 |
| Office Laboratory Library Visualization Equipment Furniture | | | | | $25,000,000 |
| Office Laboratory Computers CAD Software Plotters Displays Printers LAN | | | | | $10,000,000 |
| Heating Air Conditioning | | | | | $5,000,000 |
| Heavy Equipment movement and placement trollleys and lifts | | | | | $10,000,000 |
| Test Benches and Equipment Storage | | | | | $15,000,000 |
| Test and Measurement Equipment | | | | | $25,000,000 |
| Faraday and Optical Cage | | | | | $25,000,000 |
| Three Dimensional Positioning System | | | | | $5,000,000 |
| Laser Generators | | | | | $100,000,000 |
| Laser Optics, Polarizers and Rectifiers | | | | | $20,000,000 |
| Vacuum Enclosure for Laser and Optics | | | | | $50,000,000 |
| Vacuum Pumping Equipment | | | | | $5,000,000 |
| Cryogenic Equipment | | | | | $50,000,000 |
| Data Sensors and Acquisition Equipment | | | | | $50,000,000 |
| Computers and Servers for Analysing and Storing Data | | | | | $100,000,000 |
| Laser Interference Reactor | | | | | $100,000,000 |
| PROJECT - TANGIBLE COSTS | | | | | $640,000,000 |
| TOTAL PROJECT COSTS | | | | | $1,080,250,000 |
| | | | | | |
| LEASE EQUIPMENT | | | | | |
| truck and sedan Vehicle | | | | | $200,000 |
| Out Sourcing Services | | | | | $500,000,000 |
| | | | | | $0 |
| | | | | | $0 |
| TOTAL LEASE EQUIPMENT | | | | | $500,200,000 |
| | | | | | |
| TOTAL COST | | | | | $1,580,450,000 |